\documentclass[a4paper,12pt]{these}
\usepackage[greek,english]{babel}
\usepackage{isolatin1}
\usepackage{graphics}
\usepackage{graphicx}
\usepackage{fancyheadings}
\usepackage{amsmath}
\usepackage{amsbsy}
\usepackage{pifont}
\usepackage{float}
\usepackage{subfigure}
\usepackage{amssymb}
\usepackage{delarray}
\usepackage{rotating}
\usepackage{isolatin1}
\usepackage{makeidx}
\usepackage{color}
\usepackage[sf,hang,sf]{caption}
\usepackage[multiple]{footmisc}
\makeindex
\input psfig.sty
\input axodraw.sty
\DeclareGraphicsExtensions{.png,.pdf,.eps}

\DeclareMathAlphabet{\mathbfit}{T1}{cmr}{bx}{it}
\DeclareMathAlphabet{\bbox}{T1}{cmr}{bx}{it}
\DeclareMathAlphabet{\bbbox}{T1}{cmr}{bx}{n}

  {\begin{displaymath}%
      \everymath={\displaystyle}%
      \begin{array}{l}}%
  {\end{array}\end{displaymath}}

\def\addcontentsline#1#2#3{%
  \addtocontents{#1}{\protect\contentsline{#2}{#3}{\sffamily\thepage}}}

\renewcommand{\chaptermark}[1]% 
{\markboth{#1}{}}
\renewcommand{\sectionmark}[1]% 
{\markright{  \thesection\  #1}}

\newenvironment{myitem}{%
\begin{itemize}}{\end{itemize}}

\newcommand{\ce}[1]{Eq.~(\ref{#1})}

\newcommand{\cf}[1]{{\sffamily Fig.~\ref{#1}}}
\newcommand{\ct}[1]{{\sffamily Tab.~(\ref{#1})}}
\newcommand{\x}{\bar{x}}
\newcommand{\y}{\bar{y}}
\newcommand{\z}{\bar{z}}

\renewcommand{\footnote}[1]{\footnotemark\footnotetext{\ #1}}

\newcommand{\bfsf}[1]{{\bfseries \sffamily #1}}

\newfont{\police}{cmr7}
\newfont{\ssf}{cmss10}
\newfont{\ssfi}{cmssi10}

\newcommand{\ep}{\varepsilon}

\newcommand{\vect}[1]{\vec{#1}}

\newcommand{\eqs}[1]{\begin{equation} \begin{split} #1\end{split} \end{equation} }

\newcommand{\ks}[1]{#1 \!\!\! \slash } 

\newcommand{\ga}{\gamma^5}
\newcommand{\gmu}{\gamma^\mu}
\newcommand{\gnu}{\gamma^\nu}
\newcommand{\gmup}{\gamma^{\mu'}}

\newcommand{\grho}{\gamma^\rho}

\newcommand{\grhop}{\gamma^{\rho'}}

\newcommand{\gtaup}{\gamma^{\tau'}}
\newcommand{\gz}{\gamma^0}

\newcommand{\ie}{{\it i.e.}}
\newcommand{\eg}{{\it e.g.}}
\newcommand{\etal}{{\it et al.}}

\newcommand{\nn}{\nonumber}

\relax
\def\frac#1#2{{#1\over#2}}
\def\<{\langle}\def\>{\rangle}
\newcount\parenthesis \parenthesis=0 \newcount\n
\def\({\global\advance\parenthesis by1\left(}
\def\){\global\advance\parenthesis by-1\right)}
\def\{{\global\advance\parenthesis by1\left\lbrace}
\def\}{\global\advance\parenthesis by-1\right\rbrace}
\def\[{\relax} % dummy parenthesis
\def\]{\relax} % dummy parenthesis
\def\Loop#1\Repeat{\global\n=0\global\let\body=#1\iterate}
\def\iterate{\body\let\next=\iterate\else\let\next=\relax\fi\next}
\def\ldd{\ifnum\n<\parenthesis\global\advance\n by1
\left.\nulldelimiterspace=0pt\mathsurround=0pt}
\def\rdd{\ifnum\n<\parenthesis\global\advance\n by1
\right.\nulldelimiterspace=0pt\mathsurround=0pt}

\newcount\exacount\exacount=0\font\caps=cmcsc10
\def\Istrut{\vrule height11pt depth4pt width0pt}
\def\TRIexa#1#2#3#4{\global\advance\exacount by1\par\filbreak
{\offinterlineskip
  \vbox{\hrule\hbox to\hsize{\Istrut\vrule
      \hbox to 8mm{\hfil\caps\the\exacount\hfil}\vrule
      \quad\rm#1\hfill\vrule
      \hbox to 32mm{\hfill{\caps Mode: }{\tt #2}\hfill}\vrule
      \hbox to 32mm{\hfill{\caps Tolerance: }{\tt #3}\hfill}\vrule}
    \hrule\hbox to\hsize{\Istrut\vrule\hfill#4\hfill\vrule}\hrule}
}\nobreak}
\usepackage{array}
\usepackage{longtable}
\usepackage{calc}
\usepackage{multirow}
\usepackage{hhline}
\usepackage{ifthen}

\newlength{\gnumericTableWidth}
\newlength{\gnumericTableWidthComplete}

\title {}

\newcommand{\clearemptydoublepage}{\newpage{\pagestyle{empty}\cleardoublepage}}

\begin{document}

\pagestyle{empty}

\thispagestyle{empty}
%\vspace*{-3.5cm}
%\begin{center}
%\centerline{\includegraphics[height=2.75cm]{blason_coulb.png}}
%\vspace*{-.2cm}
%\sc \large University of Liege\\
%
%Faculty of Sciences -- Fundamental Theoretical Physics 

%{\large F}ACULTY OF {\large S}CIENCES \\
%\vspace*{.5cm}

%\large  Group of Fundamental Theoretical Physics 
%\end{center}
%\vspace*{-1.5cm}

\begin{center}
\sffamily

\bf\sffamily
{\Huge Q}\LARGE{UARKONIUM} \Huge{P}\LARGE{RODUCTION}  \LARGE{AT} 

\vspace*{.2cm}
\Huge{H}\LARGE{IGH}-\Huge{E}\LARGE{NERGY}
{\Huge H}\LARGE{ADRON} {\Huge C}\LARGE{OLLIDERS} 
\vspace*{0.5cm}
 \Large

A Systematic Gauge-Invariant Approach to Relativistic 
Effects of $J/\psi$, $ \psi'$ and $ \Upsilon$ Production \\

\vspace*{2.5cm}
%\large 
\sc \normalsize Thesis by \\

\Large  \sc Jean-Philippe Lansberg \\

\vspace*{.5cm}
\normalsize University of Li\`ege -- Fundamental Theoretical Physics
\end{center}

\vspace*{.5cm}

\begin{center}
%\large  
\sc

\vspace*{1.5cm}
\centerline{\includegraphics[height=3.5cm]{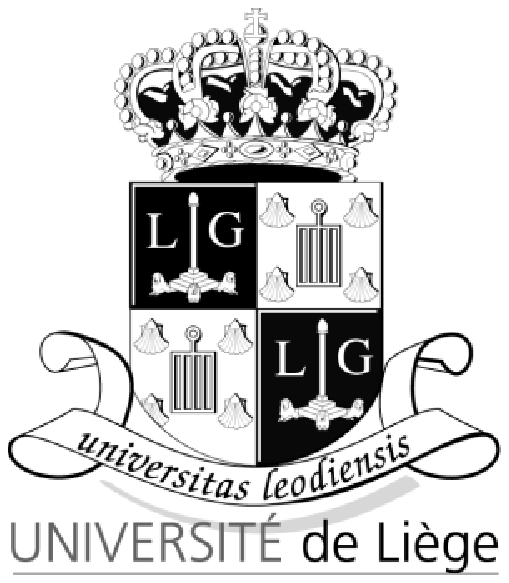}}
\sc \large University of Li\`ege\\
Li\`ege, Belgium
\vspace*{1.5cm}
%Faculty of Sciences \\
%Fundamental Theoretical Physics 

Thesis presented in fulfilment of the requirements for \\
 the Degree of  Doctor of Science

%\vfill
\vspace*{2.5cm}
\normalsize
Academic year 2004-2005\\
--Defended on April 22, 2005--
%{ \large University of Liege}
\end{center}

\rmfamily
\newpage
%\thispagestyle{empty}
%\

\vspace*{2cm}\sffamily

{\bfseries Thesis Jury:}\\

\begin{flushleft}
\begin{tabular}{ll}
\ {\bfseries President:} &\\
Yves Lion                & (Li\`ege)  \\
\\
\ {\bfseries External Members:}&\\
Fran\c cois Le Diberder  & (IN2P3 \&  LAL, Orsay) \\
James W. Stirling        & (CPT \& IPPP, Durham) \\
\\
\ {\bfseries Internal Members:} &\\
Maxim V. Polyakov           & (PTF, Li\`ege)\\
Lech Szymanowski         & (PTF, Li\`ege  \& SINS, Warsaw) \\

\\
\ {\bfseries Advisor:} &\\
Jean-Ren\'e Cudell       & (PTF, Li\`ege) \\
\end{tabular}

\vfill

\underline{ \ \hspace{8cm} \ } \\\vspace*{.5cm}

ISBN: 2-87456-004-9   \hspace*{.5cm}D/2005/8886/11\\
Printed by ``Les \'editions de l'Universit\'e de Li\`ege'', Li\`ege, Belgique, 2005\\\vspace*{.2cm}

\underline{ \ \hspace{8cm} \ } \\

\vspace*{3cm}

\end{flushleft}

\rmfamily
\newpage

\chapter*{}
\ \

\vspace{10cm}
\begin{flushright}
{ \large \sffamily \itshape 
A tout ces gens,

qui ne me doivent rien,

\vspace*{1mm} et \`a qui je dois \'enorm\'ement.}
\end{flushright}
\vspace*{1cm}
\begin{flushleft}
\textgreek{Afier'wnw ep'ishs aut'hn thn ergas'ia \\
stis Mo'uses mou, \\ pareljontik'es, 
\\paro'uses \\
kai mellontik'es.}
\end{flushleft}

\newpage

\chapter*{}
\ \

\vspace{13cm}
\begin{flushright}
{ \sffamily \itshape You see, in this world,\\
there's two kinds of people\\
... my friend:\\ 
Those with loaded guns, \\
and those who dig...you dig!\\

\vspace{0.5cm}
The man with no name (Clint Eastwood),\\
{\em The Good, the Bad and the Ugly}
}
\end{flushright}
\newpage

%--------------------------------------------------------------------
\clearemptydoublepage

\sffamily
\pagestyle{empty}
\chapter*{\sffamily Remerciements -- Acknowledgments}
\rmfamily
\thispagestyle{empty}
Je voudrais tout d'abord remercier ma m\`ere, ma marraine et mon parrain pour leur
soutien constant, tant mat\'eriel que moral, depuis ces longues ann\'ees. 
Inutile de pr\'eciser que cela aurait \'et\'e singuli\`erement plus compliqu\'e sans leur pr\'esence.
Je tiens aussi \`a remercier les autres membres de ma famille qui m'ont tous, \`a un moment ou 
\`a un autre, donn\'e le petit coup de pouce n\'ecessaire quand il le fallait.

Ensuite, bien \'evidemment, je tiens \`a remercier Jean-Ren\'e Cudell tant pour l'encadrement 
scientifique qui m'a permis de mener \`a bien ce travail de longue haleine, que pour l'aide,
un peu moins scientifique, mais tout aussi pr\'ecieuse, que furent ses innombrables relectures 
de mes \'ecrits (lettres, e-mails, m\'emoires et cette th\`ese), ses conseils administratifs, ses 
suggestions et soutiens pour mes demandes de financements, etc. Cela fait maintenant plus de six ans 
que je suis rentr\'e pour la premi\`ere fois dans son bureau, o\`u j'appris que ``la Chromodynamique
Quantique n'\'etait qu'une matrice \`a la place d'un nombre dans le facteur de phase de la fonction 
d'onde'' (sic); si ma motivation pour la recherche est intacte, c'est en 
grande partie gr\^ace \`a lui.

Mes plus sinc\`eres remerciements vont \'egalement au chef de notre groupe, Joseph Cugnon, pour m'avoir
accord\'e le poste de chercheur IISN qui m'a permis de mener \`a bien cette th\`ese, pour m'avoir
autoris\'e \`a prendre part \`a de nombreux s\'ejours scientifiques qui furent pour moi une source
fondamentale de motivations et qui, maintenant, me permettent de jouir d'une certaine 
perspective sur mes recherches futures gr\^ace aux contacts que j'ai pu nouer avec d'autres 
physiciens. Je le remercie aussi de m'avoir confi\'e
une partie de l'organisation de la conf\'erence en D\'ecembre dernier, 
l'\'edition des proceedings, ainsi que 
d'autres t\^aches dans le cadre de notre groupe. Cela m'a permis de d\'ecouvrir pleinement des 
aspects de la vie scientifique auxquels normalement nous n'avons pas l'occasion de go\^uter
en tant que doctorants.

J'aimerais en outre remercier Laurent et Voica -- {\it la mia sorella} -- pour la relecture 
attentive de ma th\`ese, mais bien au-del\`a de cette relecture, pour nos innombrables 
discussions \'electroniques, tout au long de ma th\`ese. J'esp\`ere les avoir aid\'es
et divertis lors de moments moins agr\'eables autant qu'eux ont pu le faire pour moi.
Pour leur amiti\'e, je voudrais aussi remercier Muriel, Fr\'ed\'eric, Philippe et Nicolas. 

Next, I woud like to thank Yura Kalinovsky for his everlasting good mood but also for his
scientific help concerning this work. Without his method to tackle this ``bloody'' (sic) integral
appearing in the chapter 6, my task would have been certainly more awkward. I also thank him and Pedro
Costa to have shared their results with me.

I would like also to express my sincere acknowledgments to the members of the jury for having accepted 
to join it, in spite of their overbooked agenda. I am deeply grateful to them for that. I would like
also to underline the constant kindness and attention of Lech Szymanowski.

I would like finally to thank all the physicists whom I met and who did their best to answer 
my questions.

Enfin, je r\'eserve quelques mots  de remerciements pour mes coll\`egues de bureau, Michel et Aurore; 
pour Alice et C\'edric qui 
remplacent avantageusement les ordinateurs du 4/03, les \'etudiants en physique que j'ai eu le plaisir
de c\^otoyer, dont Christophe et Gr\'egory, le groupe des ``ing\'enieurs'', St\'ephane et Fran\c cois 
(``ce sont quand m\^eme de bons acteurs !''), Marie-Paule Bi\'emont pour sa pr\'ecieuse aide 
logistique, Michel Bawin pour nos discussions Linuxiennes ainsi 
que toutes les personnes qui se sont int\'eress\'ees \`a un moment ou \`a un autre \`a mon travail. 

Merci \`a tous !\thispagestyle{empty}

\pagestyle{fancy}
\sffamily
\clearemptydoublepage
\tableofcontents 
\rmfamily
\clearemptydoublepage
\chapter{Introduction}

\section{History}

\subsection{$J/\psi$ and the November revolution}\index{jpsi@$J/\psi$}

\index{SLAC}
Simultaneously discovered in November 1974 by Ting {\it et al.}~\cite{Aubert:1974js} 
at the Brookhaven National Laboratory\footnote{BNL} and by Richter 
{\it et al.}~\cite{Augustin:1974xw} at the Stanford Linear Accelerator\footnote{SLAC}, 
the $J/\psi$ directly stands out from the crowd of other particles 
by its peculiar name composed of two symbols: $J$ by Ting in comparison\footnote{See 
however the left part of~\cf{fig:spear_psi_decay}~\dots} to the electric 
current $j_\mu(x)$, and $\psi$ because of the typical pattern of its decay in $\pi^+ \pi^-$ at SPEAR 
(see~\cf{fig:spear_psi_decay}).\index{BNL}

\begin{figure}[H]
\centerline{\includegraphics[width=3.0cm]{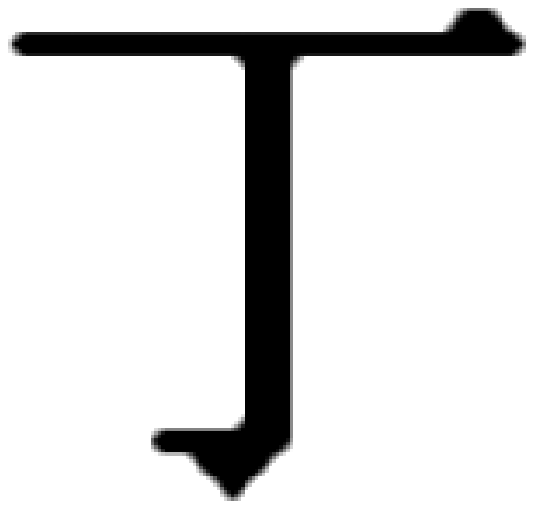} \ \includegraphics[width=4.5cm]
{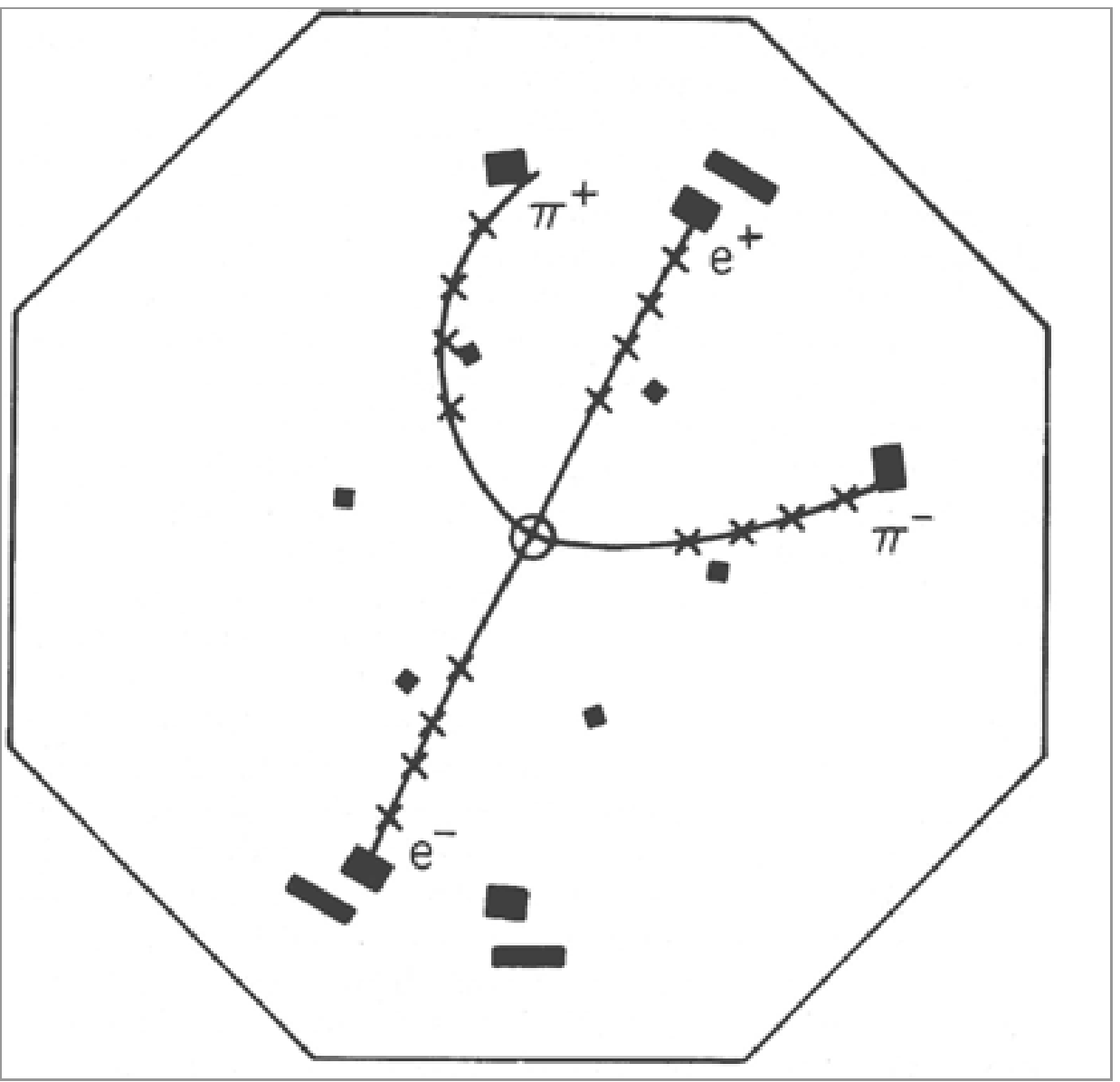}}
\caption{Left: the sound [d\={\i}ng] in the Chinese alphabet; Right: typical Mark I (SPEAR) 
event display for $J/\psi$ decay (from~\cite{Lynch:1975ze}).}
\label{fig:spear_psi_decay}
\end{figure}

Ting's experiment was based on the high-intensity proton beams of 
the Alternating Gradient Synchrotron (AGS) working at the energy of 30 GeV, 
which bombarded a fixed target with the consequence of producing showers of 
particles detectable by the appropriate apparatus. More precisely, they used 
a beryllium target and a beam with a momentum in the laboratory frame of 19-20 GeV. 
Taking advantage of a double-arm spectrometer, they detected a strong peak 
in the energy distribution of the produced electron-positron pairs at a value of 3.1 GeV 
as can be seen on~\cf{fig:psi_discovery_ting}. 
On the other hand, Richter's experiment used the electron-positron 
storage ring SPEAR, whose center-of-momentum energy could be tuned at the desired value. With 
the Mark I detector, they discovered a sharp enhancement of the production
cross section in different channels:  $e^+ e^-$, $\mu^+ \mu^-$,
$\pi^+ \pi^-$,\dots The resulting cross sections are shown in~\cf{fig:psi_discovery_richter}.

\begin{figure}[H]
\centerline{\includegraphics[width=8cm]{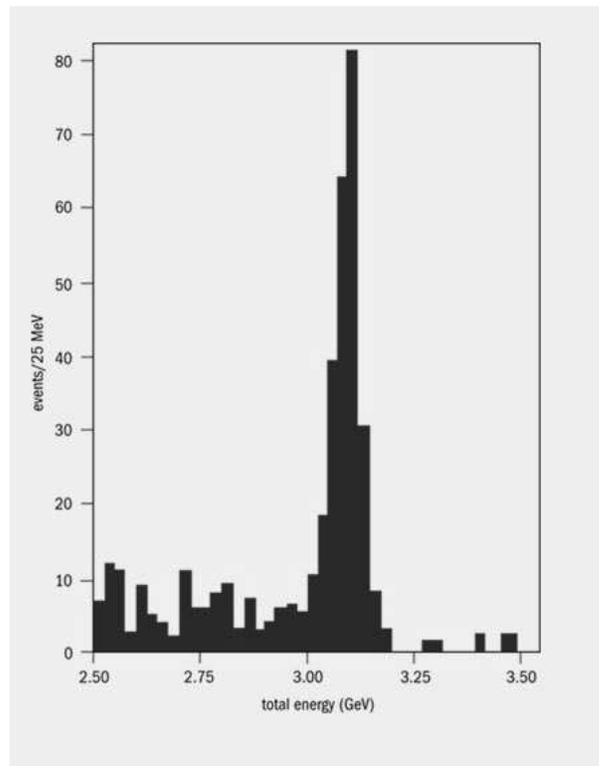}}
\caption{Energy distribution of the produced electron-positron pairs for Ting's experiment
(from~\cite{cern_courier}).}
\label{fig:psi_discovery_ting}
\end{figure}

In the following weeks, the Frascati group (Bacci~\etal~\cite{Bacci:1974za}) 
confirmed the presence of this new particle 
whose mass was approximately 3.1 GeV. The confirmation was so fast that it was actually 
published in the same issue of Physical Review Letters, {\it ie.} vol. 33, no. 23, 
issued the second of December 1974. In the meantime --ten days later according Richter's 
Nobel Lecture--, his group discovered another resonant state with a slightly higher mass 
which was called\footnote{We shall also make us of the name $\psi(2S)$.} $\psi'$.

In 1976, Ting and Richter were awarded the Nobel prize for this simultaneous discovery. 
Indeed, considering the state of comprehension of particle physics at that time, this discovery was a
 shock for many. During the two years between the discovery and the award of the Nobel prize, an 
intense research activity in this domain took place. Were these new states the sign
of a new generation of quarks ? Was the quark model to be finally trusted? 

An important fact to bear in mind  here is that further studies 
based on the on- and off-peak production of muons and 
on forward/backward asymmetry in the observed angular distribution promptly established 
that the quantum numbers of the $J/\psi$ were the same as those of the photon , 
\ie~$1^{--}$. In a sense, 
this is not surprising at all since it is much more likely to produce a resonance with the same quantum
number as the photon in $e^+ e^-$ colliding experiments than other resonances. 

\begin{figure}[H]
\centerline{\includegraphics[height=8cm,angle=90]{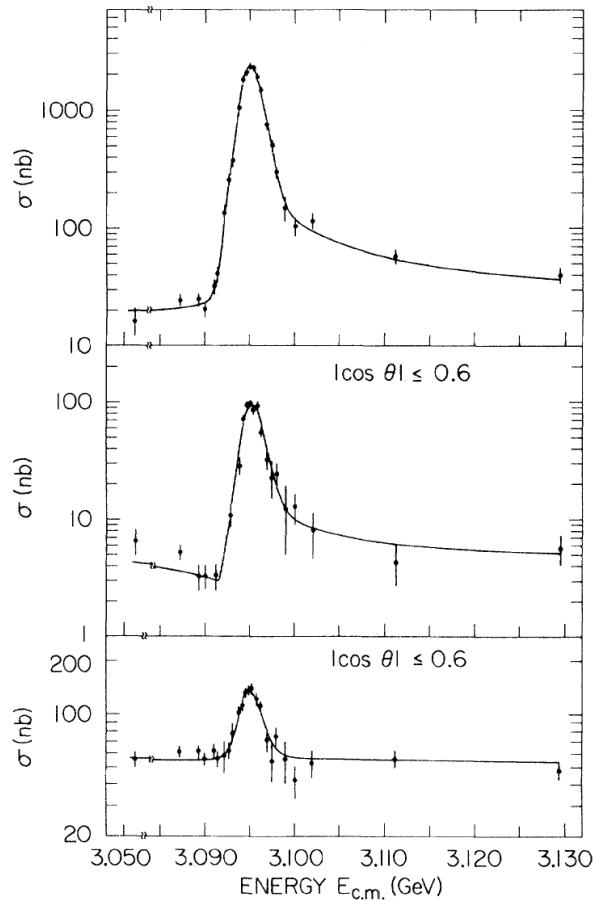}}
\caption{Various pair production cross sections as functions of the center-of-momentum energy 
(top: hadrons; middle: $\pi^+\pi^-$, $\mu^+\mu^-$ and $K^+K^-$; bottom: $e^+e^-$)~\cite{Augustin:1974xw}.}
\label{fig:psi_discovery_richter}
\end{figure}

Moreover, since the following ratio
\eqs{R= \frac{\hbox{cross section for $e^+ e^- \to$ hadrons}}{\hbox{cross section for 
$e^+ e^- \to \mu^+\mu^-$}}} was much larger on-resonance than off, it was then clear that the
$J/\psi$ did have direct hadronic decays. The same conclusion held for the $\psi'$ as well. 
The study of multiplicity in pion decays
indicated that $\psi$ decays were restricted by a specific selection rule, called $G$-parity 
conservation, known to hold only for hadrons. Consequently, $J/\psi$ and $\psi'$ entered 
the family of hadrons with isospin 0 and $G$-parity -1.\index{G@$G$-parity}

Particles with charge conjugation $C$ different from -1, \ie~with quantum 
numbers  different from those of the photon, were found later. Indeed, they were only produced by decay of 
$\psi''$ and the detection of the radiated photon during the (electromagnetic)
decay was then required. The first to achieve this task and discover a new state 
was the DASP collaboration~\cite{Braunschweig:1975ac} based at DESY 
(Deutsches Elektronen-Synchrotron), Hamburg, 
working at an $e^+e^-$ storage ring called DORIS. This new particle, named\footnote{The name chosen 
here reveals that physicists already had at this time an interpretation of this state as bound-state
of quarks $c$. Indeed, the symbol $P$ follows from the classical spectroscopic notation for
a $\ell=1$ state, where $\ell$ stands for the angular momentum quantum number.} 
$P_c$, had a mass of approximately 3.500 GeV. At SPEAR other resonances at 
3.415, 3.450 and 3.550 GeV were discovered, the 3.500 GeV state was confirmed. 
Later, these states were shown to be $C=+1$.

Coming back to the ratio $R$, let us now analyse the implication in the framework of the 
quark model postulated by Gell-Mann\footnote{The name ``quark'' was chosen by Gell-Mann. 
He found this word in the James Joyce's novel {\it Finnegans Wake} where the following
enigmatic sentence appears: 'Three quarks for Muster Mark'. This name comes from 
the standard English verb {\it quark}, meaning ``to caw, to croak'', and also from the dialectal
verb {\it quawk}, meaning ``to caw, to screech like a bird''.} and Zweig in 1963. 
In 1974 at the London conference, 
Richter presented~\cite{Richter:1974vz} the experimental situation as in~\cf{fig:R-richter}.

\begin{figure}[H]
\centerline{\includegraphics[width=7cm]{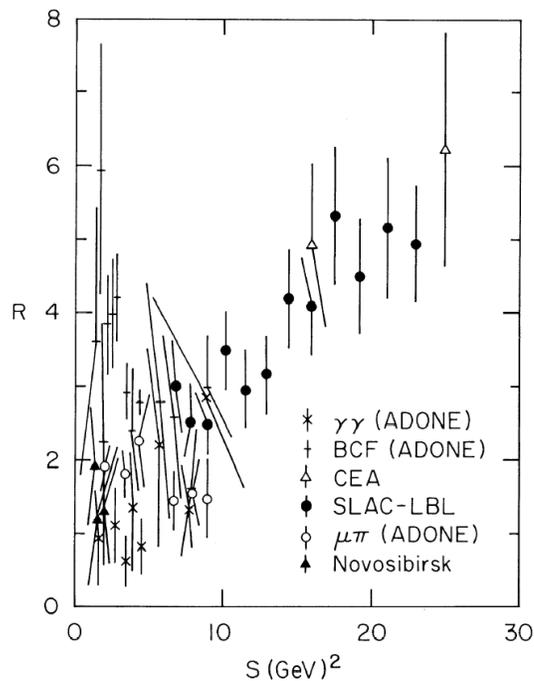}}
\caption{Experimental status of $R$ as of July 1974 (from~\cite{Richter:1974vz}).}
\label{fig:R-richter}
\end{figure} 

In the framework of the Gell-Mann-Zweig quark model with 
three quarks,\index{quark! Gell-Mann-Zweig model}
a plateau was expected with a value of $2/3$ or $2$ if the quark were considered as 
``coloured'' -- a new concept introduced recently then --. The sign of a new quark -- as the 
charm quark first proposed by Bjorken and Glashow in 1964~\cite{Bjorken:1964gz}-- would 
have been another plateau from a certain energy (roughly its mass) with a 
height depending on its charge. Retrospectively, one cannot blame anyone for not having seen
this ``plateau'' on this kind of plots. Quoting Richter, ``the situation that prevailed in 
the Summer of 1974'' was ``vast confusion''.

This charm quark  was also
theoretically needed following the work of Glashow, Iliopoulos and Maiani~\cite{Glashow:1970gm},
which suggested that a symmetry between leptons and quarks should exist in order to cancel the anomaly in
weak decays. And following this symmetry proposal, the charm quark 
was expected to exist and to have an electric charge $2/3$. $R$ was therefore 
to be $10/3$ in the coloured quark model, still not obvious in~\cf{fig:R-richter}. This explains 
why the discovery of such sharp and easily interpreted resonances in November 1974 
was a revolution in particle physics.\index{November revolution}

It became quite rapidly obvious that the $J/\psi$ was the lowest-mass $c\bar c$ system 
with the same quantum numbers as photons -- explaining why  it was so much produced 
compared to some other members of its system. These  $c\bar c$  bound states were named ``charmonium'', 
firstly by Appelquist, De R\'ujula,
Politzer and Glashow~\cite{Appelquist:1974yr}\footnote{To be complete, the initial title of the preprint
``Charmonium spectroscopy'' was immediately refused by the senior editor Pasternak 
of Physical Review Letters
because of ``frivolous new terminology''. A consensus eventually emerged thanks to Glashow: the term
was accepted in the text but not in the title!}, in analogy with positronium, 
which has a similar bound-state level structure.

At that time, the charm had nevertheless always been found hidden, that is in charm-anti-charm 
bound states. On the other hand, it became quickly clear that the mass of the lightest possible
state constituted of a charm quark and a light quark, thus an explicit charmed meson, named $D$, 
was to lie\index{D@$D$} around 1900 MeV. A more precise lower bound, 1843~MeV, was inferred 
from the fact that the $\psi'$ ($m_{\psi'}=3684$ MeV) was too narrow or, in other words, 
had too long a lifetime to be above \index{psip@$\psi'$} the threshold to produce a pair 
of $D$'s, that is $2 m_D$. The upper bound was, on the other hand, deduced from the point
where $R$ started to rise before reaching the plateau with 4 active quark flavours.

In order to study these $D$ mesons, the investigations were based on the assumption that 
the $D$ was to decay weakly and preferentially into strange quarks. The weak character of 
the decay motivated physicists to search for parity-violation signals. The 
first resonance attributed to $D^0$ meson was found in $K^+\pi^+$ decay by Goldhaber 
{\it et al.}~\cite{Goldhaber:1976xn} in 1976. A little later, $D^+$ and $D^-$ were also 
discovered as well as an excited state, $D^*$, with a mass compatible with the 
decay $\psi'''\to D^0 D^*$. And, finally, the most conclusive evidence 
for the discovery of charmed meson was the observation of parity violation in $D$ 
decays~\cite{Wiss:1976gd}. To complete the picture of the charm family, the first charmed baryon
was discovered during the same year~\cite{Knapp:1976qw}. The quarks were not anymore just a way to 
interpret symmetry in masses and spins of particles, they had acquired an existence. 

\subsection{The truth: a third generation}

\index{SLAC}
In 1975, another brand new particle was unmasked at SLAC by Perl \etal~\cite{Perl:1975bf},
for which he was awarded the Nobel prize twenty years later. This was the first particle of a third 
generation of quarks and leptons, the $\tau$, a {\it lepton}\footnote{in 
contradiction \index{tau@$\tau$} with the etymology because it was {\it heavier} 
than most of known {\it baryons} at that time.}. Following the 
standard model, which was more and more a reference trusted by 
most, two other quarks were expected to exist. Their discovery 
would then be the very proof of the theory that was developed since the sixties. 
Two names for the fifth quark were already chosen: ``beauty'' 
and ``bottom'', and in both cases represented by the letter $b$; the sixth quark was as well 
already christened with the letter $t$ for the suggested names\footnote{The 
name ``truth'' was given as a reference to the fact that its discovery should be the proof
of the truth of the standard model.}
 ``true'' or ``top''.\index{quark!$t$} \index{quark!$b$}

The wait was not long: two years. After a false discovery\footnote{Nonetheless,  
this paper~\cite{Hom:1976cv} first suggested the notation $\Upsilon$ for any ``onset of high-mass 
dilepton physics''.} of a resonance at 6.0 GeV, a new dimuon resonance similar \index{upsi@$\Upsilon$}
to $J/\psi$ and called $\Upsilon$ was brought to light at Fermilab, thanks to the FNAL 
proton synchrotron accelerator, by 
Herb \etal~\cite{Herb:1977ek}, with a mass of 9.0 GeV; as for charmonia, the first radial excited 
state ($\Upsilon(2S)$) was directly found~\cite{Innes:1977ae} thereafter. Various confirmations of these 
discoveries were not long to come. The $3S$ state was then found~\cite{Ueno:1978vr} at Fermilab 
as well as evidence that the $4S$ state was lying above the threshold for the production of $B$ mesons. 
The latter was confirmed at the Cornell $e^+e^-$ storage ring  with the CLEO detector. The 
first evidence for $B$ meson and, thus, for unhidden $b$ quark, was also brought
 by the CLEO collaboration~\cite{Bebek:1980zd} 
in 1980. One year later, ``the first evidence for baryons with naked beauty''({\it sic}) was reported by 
CERN physicists~\cite{Basile:1981wr}\footnote{Here is an unambiguous sign that European editors
had a much softer editorial approach concerning titles of articles\dots}.\index{CERN}
\index{CLEO}

Another decade was needed for the discovery of the sixth quark which was definitely christened
 ``top''. Indeed, in 1994, the CDF Collaboration observed it at the Tevatron collider in \index{Tevatron}
Fermilab~\cite{Abe:1994st}. This accelerator -- which reaches now a c.m. energy of 1960 GeV -- 
and this collaboration will accompany us throughout this thesis, but for a different signal.
\index{CDF} Unfortunately, due to its very short lifetime, this quark cannot bind with 
its antiquark to form the toponium.\index{toponium} To conclude this historical prelude, we give 
the spectra of the $c\bar c$ and $b\bar b$ systems as well as two tables summing up the
characteristics of the observed states as of today.
\begin{table}[H]
\centering
\begin{tabular}{|c||c|c|c|c|}
\hline Meson & $n ^{2S+1}L_J$ & $J^{PC}$& Mass (GeV)& $\Gamma_{\mu\mu}$ (keV)\\
\hline \hline 
$\eta_c$  &  $1\ ^1S_0$  &  $0^{-+}$ & 2.980 & N/A\\
$J/\psi$  &  $1\ ^3S_1$  &  $1^{--}$ & 3.097 & 5.40\\
$\chi_{c0}$, $\chi_{c1}$, $\chi_{c2}$  &  $1\ ^3P_{0,1,2}$  &  $0^{++}$,$1^{++}$,$2^{++}$ & 
3.415,3.511,3.556 & N/A\\
$h_c$  &  $1\ ^1P_0$  &  $1^{+-}$ & 3.523 & N/A\\
$\eta_c(2S)$  &  $2\ ^1S_0$  &  $0^{-+}$ & 3.594 & N/A\\
$\psi'$          & $2\ ^3S_1$  & $1^{--}$ & 3.686 & 2.12\\
\hline
\end{tabular}
\caption{Properties of charmonia (cf.~\cite{pdg}).}\label{table:sum_charmonium}
\end{table}

\begin{figure}[H]
\centering\includegraphics[width=15cm]{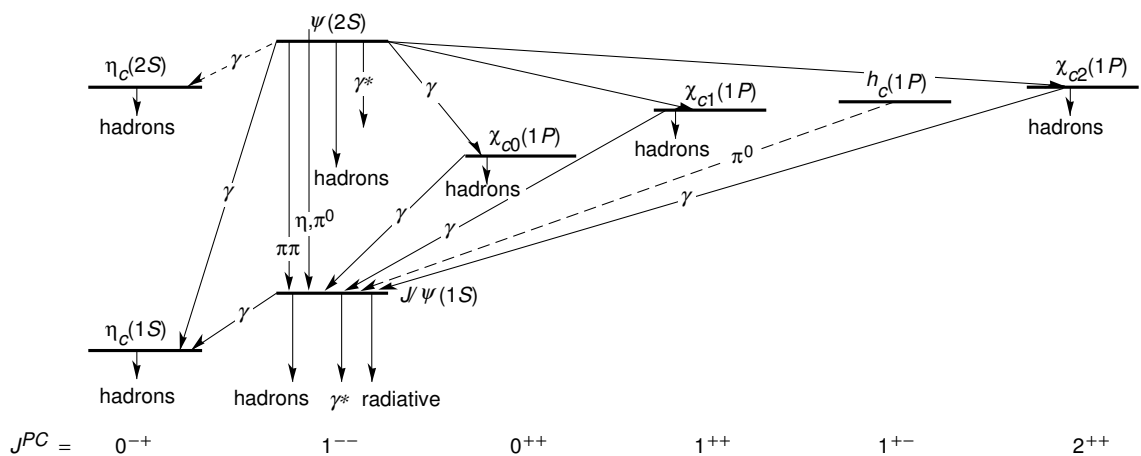}
\caption{Spectrum and transitions of the charmonium family (from~\cite{pdg}).}
\label{fig:charm_spectrum}
\end{figure}
\begin{figure}[H]
\centering\includegraphics[width=15cm]{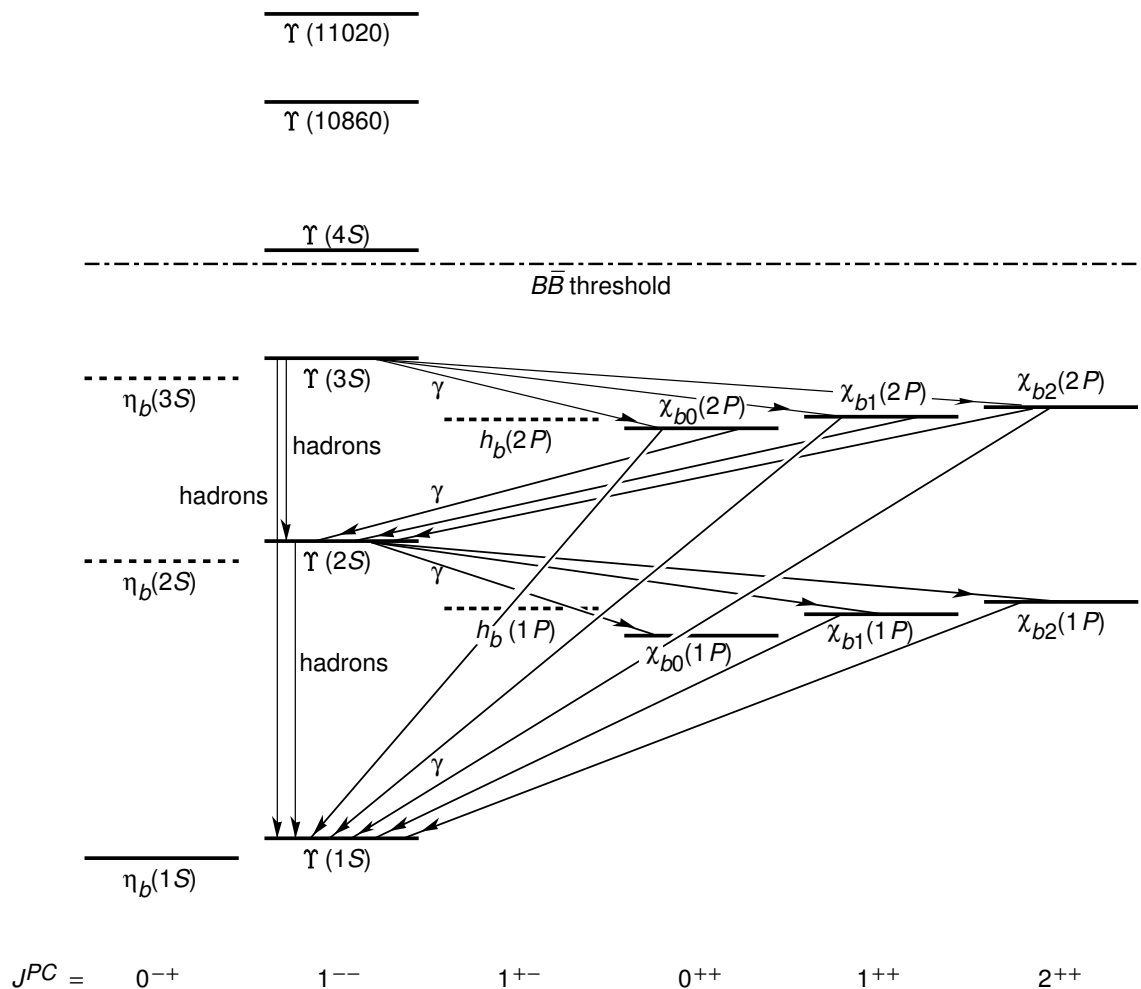}
\caption{Spectrum and transitions of the bottomonium family (from~\cite{pdg}).}
\label{fig:bottom_spectrum}
\end{figure}

\begin{table}[h]
\centering
\begin{tabular}{|c||c|c|c|c|}
\hline Meson & $n ^{2S+1}L_J$ & $J^{PC}$& Mass (GeV)& $\Gamma_{\mu\mu}$ (keV)\\
\hline \hline 
$\Upsilon(1S)$   & $1 ^3S_1$  & $1^{--}$ & 9.460 & 1.26\\
$\chi_{b0}$, $\chi_{b1}$ , $\chi_{b2} (1P)$  &  $1\ ^3P_{0,1,2}$  &  $0^{++}$,$1^{++}$,$2^{++}$ & 
9.860,9.893,9.913 & N/A\\
$\Upsilon(2S)$   &  $2\ ^3S_1$  & $1^{--}$ & 10.023 & 0.32\\
$\chi_{b0}$, $\chi_{b1}$, $\chi_{b2}$ (2P) &  $1\ ^3P_{0,1,2}$  &  $0^{++}$,$1^{++}$,$2^{++}$ & 
10.232,10.255,10.269 & N/A\\
$\Upsilon(3S)$   &  $3\ ^3S_1$ & $1^{--}$ & 10.355 & 0.48\\
\hline
\end{tabular}
\caption{Properties of bottomonia (cf.~\cite{pdg}) .}\label{table:sum_bottomonium}
\end{table}

\section{Motivations}

\subsection{Why study quarkonium physics in the high-energy regime ? }

The fact that $J/\psi$ was the first particle with a $c$ quark to be discovered is not
an accident: it is in fact of fundamental importance in particle physics.
Let us go back to its discovery: as we have said, two totally different experiments
were involved in it. The first was a fixed-target experiment, whose main advantage was to scan all
the energy range below the c.m. energy of the collision. Its main disadvantage
was to produce many other particles than the resonance which it looked for. The second experiment was 
taking place at an $e^+e^-$ collider with totally opposite characteristics, one given energy and a 
clean environment surrounding the expected resonance. In a sense, the $J/\psi$ discovery is
as if the Higgs boson had been discovered simultaneously at LEP and at the Tevatron\dots
\index{Tevatron} \index{LEP}

The discovery in a fixed-target experiment shows that the 
 $J/\psi$ leptonic decay is significant, making it highly detectable compared to particles
only detected through their hadronic decays. $J/\psi$ is in fact 
so highly detectable that it was not a major problem to have bunches of other hadrons 
produced during the same collision. 

Let us explain why this leptonic decay is significant.
Firstly, the open charm channel (decay to $D$-mesons) is kinematically blocked 
since $m_{J/\psi} \leq 2 m_D$; secondly, colour prevents decays in a single gluon; thirdly, 
the Landau-Yang theorem~\cite{LandauYang} or the $G$-parity conservation forbids its\index{G@$G$-parity}
decay into two gluons. Therefore, three gluon emissions are required and, as the gluon emission rate from
heavy quarks is reduced\footnote{This comes from the fact that the scale of the process is of the order of
$m_Q$ and makes $\alpha_s$ relatively low. In fact, this results 
in the Zweig rules.\index{Zweig rules}}, 
one is not too surprised to get a total width of $91 \pm 3$ keV,
much smaller than that of usual hadronic decays. Moreover, this value also includes leptonic 
channels. On the other hand, leptonic decays are not subject to all these restrictions: a single photon
decay is allowed, making them competitive with hadronic decays. More precisely, they account for
one third of the total width.\index{Landau-Yang theorem}

This significant branching ratio into leptons, especially into muons, is an incredible chance, like 
a metal detector to find a needle in a haystack. In high-energy hadron colliders, the number
of background hadrons becomes annoyingly large and complicates the detection of particles with
constrained kinematical characteristics. However, for muons, this is a 
completely different story, as they can go through the calorimeters without interaction
and be seen  in dedicated detectors. These can measure precisely their
charge, momentum and direction, determining therefore their track. It is in turn 
possible to discriminate precisely which dimuon, out of the complete set
of the detected muons and anti-muons, results from the decay
of a $J/\psi$.  The latter being a significant source of dimuons, it is
therefore an ideal probe to understand high-energy hadron collisions where many other studies
are  prevented due to a high background.

Concerning its discovery with an $e^+e^-$ collider, it results partly from the fact that
the rather low $J/\psi$ mass was accessible 
to the available experiments whereas had it been lower, say 1.5-2 GeV, $J/\psi$ would perhaps have been 
lost among excited states of strange baryons.

The reasonably low mass of the $J/\psi$ makes it highly producible, 
with an easily reachable threshold in energy. This explains why it is
abundantly produced at the Tevatron, even at relatively large \index{Tevatron}
transverse momentum. This qualifies it, once more, as a good probe to study mechanisms associated 
with its production. Furthermore, its  production at large transverse momentum,
\ie~in a range where perturbative Quantum Chromodynamics (pQCD) is likely to be applicable, offers
us a way to test the application of QCD. This statement that we have a sufficiently 
high scale to use pQCD will follow us all along this thesis.

To complete the picture, the $\Upsilon$'s, which are heavier, 
will be a useful benchmark; even though they are less produced due to their higher mass, 
they set a higher scale, for which pQCD should apply better, even under unfavourable 
kinematical conditions. Their quarks being heavier, 
their velocity inside the meson is expected to be much smaller and, consequently, relativistic effects 
to be less important. This lowers theoretical uncertainties and constrains 
more the quantities to be tested.

To be more definite, quarkonium production was thought, till the middle of the nineties, 
to be a good way to extract the gluon distribution function, both in 
electron-proton collisions where the modelisations are
more reliable, and in proton-proton collisions where two gluon pdf's enter the calculation.
As we shall see, the mechanisms of production of these quarkonia are not yet understood and
do not provide us with a powerful tool to achieve the extraction of these pdf's.

\index{parton distribution function}

\subsection{The eighties: the time for predictions}

\subsubsection{ The Colour-Singlet Model}\index{colour-singlet!model|(}
This model is\footnote{before the inclusion of fragmentation contributions.} 
the most natural application of QCD to heavy quarkonium production in the high-energy regime. 
It takes its inspiration in the factorisation theorem of QCD~\cite{facto}\footnote{proven for 
some definite cases, {\it e.g.} Drell-Yan process.}
where the hard part is calculated by the strict application pQCD and the soft part
is factorised in a universal wave function. This model is meant to describe the
production not only of $J/\psi$, $\psi(2S)$, $\Upsilon(1S)$, $\Upsilon(2S)$ and $\Upsilon(3S)$, \ie~the 
$^3S_1$ states, but also the singlet $S$ states $\eta_c$ and $\eta_b$ as well as the 
$P$ ($\chi$) and $D$ states.\index{factorisation}\index{wave function}

Its greatest quality resides in its predictive power as the only input, apart form the pdf, 
 namely
the wave function, can be determined from data on decay processes or by application
of potential models. Nothing more is required. It was first applied to 
hadron colliders~\cite{CSM_hadron}, then to electron-proton colliders~\cite{Berger:1980ni}. 
The cross sections for $^3S_1$ states were then calculated, as well as also for $\eta$ and $\chi$, 
for charmonium and for bottomonium. These calculations were compared to ISR and FNAL \index{ISR} 
data from $\sqrt{s}=27$ GeV to $\sqrt{s}=63$ GeV for which the data extended to 6 GeV 
for the transverse momentum. Updates~\cite{Halzen:1984rq,Glover:1987az} of the model
to describe collisions at the CERN $p\bar p$ collider ($\sqrt{s}=630$ GeV) were then presented. 
At that energy, the possibility that the charmonium be produced from the decay of a beauty 
hadron was becoming competitive. Predictions for Tevatron energies were also made~\cite{Glover:1987az}.
\index{Tevatron}

In order to introduce the reader to several concepts and quantities that will be useful
throughout this work, let us proceed with a detailed description of this model.

\subsubsection{The Model as of early 90's}

It was then  based on several approximations or postulates:
\begin{itemize}
\item If we decompose the quarkonium production in two steps, first the creation of 
two {\it on-shell} heavy quarks ($Q$ and $\bar Q$) and then their binding to make 
the meson, one {\it postulates the factorisation} of these two processes.
\index{factorisation}
\item As the scale of the first process is approximately $M^2+p_T^2$, one  
considers it as a {\it   perturbative} one. One supposes that its 
cross section be computable with Feynman-diagram methods.
\item As we consider only bound states of heavy quarks (charm and bottom quarks), 
their velocity in the meson must be small. One therefore supposes that the meson be created 
with its 2 constituent quarks {\it   at rest} in the meson frame. This 
is {\it the static approximation.} \index{static!approximation}
\item One finally assumes that the colour and the spin of the $Q\bar Q$ pair do not 
change during the binding. Besides, as physical states are colourless, one requires the pair 
 be produced in a {\it   colour-singlet state}. This explains the name Colour-Singlet Model (CSM).
\index{colour-singlet!state}
\end{itemize}

In high-energy hadronic collisions, the leading contribution comes from a gluon 
fusion process; as the energy of the collider increases, the initial parton momentum fraction $x_i$ needed 
to produce the quarkonium decreases to reach the region in $x$ where the number of gluons 
becomes much larger than the number of quarks. One has then only six Feynman diagrams  
for the $^3S_1$ states production associated with a gluon\footnote{This is 
the dominant process when the transverse momentum of the meson
is non-vanishing.} (see \cf{fig:CSM_box}).

\begin{figure}[h]
\centerline{\includegraphics[width=12cm]{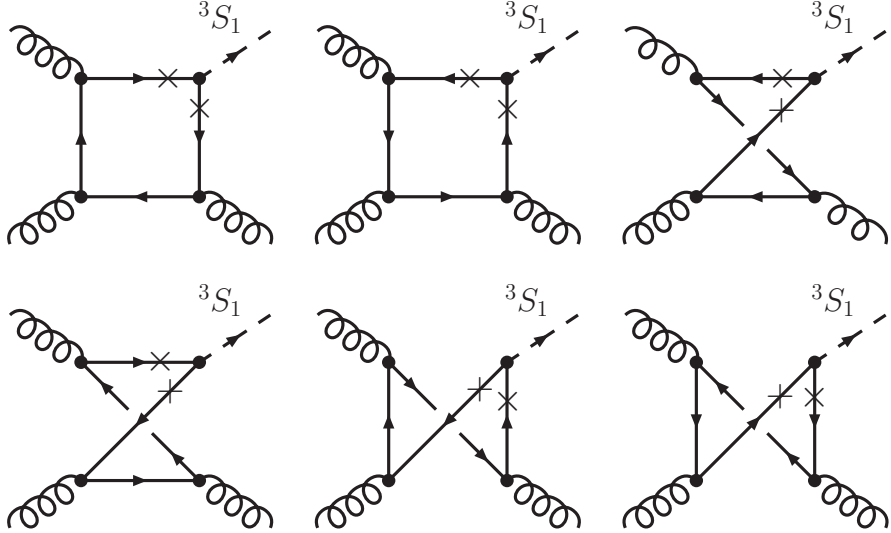}}
\caption{The 6 diagrams for $gg \to \!\!\ ^3S_1 g$ at LO within the CSM.}
\label{fig:CSM_box}
\end{figure} 

One usually starts with  ${\cal M}(p)$, the perturbative amplitude to produce 
the heavy quark pair on-shell with relative momentum $p$ and in a configuration similar 
to the one of the meson. To realize the latter constraint, 
one introduces a projection operator\footnote{In fact, this amounts to associate a $\gamma^5$ 
matrix to pseudoscalars, $\gamma^\mu$ to vectors, etc.}; the amplitude ${\cal M}(p)$ is then 
simply calculated with the usual Feynman rules.\index{projection operator}

The amplitude to produce the meson is thence given by 
\begin{equation}\label{eq:amp_CSM_1}
{\cal A}=\int \Phi(\vect{p}) {\cal M}(p)  \delta(2p^0) dp, 
\end{equation}
where $\Phi(\vect{p})$ is the usual Schr\"odinger wave-function.\index{wave function}

Fortunately, one does not have to carry out the integration thanks to the {\it the static approximation}
which amounts to considering the first non-vanishing term of $\cal A$ when the perturbative
part $\cal M$ is expanded in $p$. For $S$-wave, this gives

\begin{equation}\label{eq:psim2}
\int \Phi(\vect{p}) {\cal M}(p) \delta(2p^0)dp \simeq \left.{\cal M}\right|_{p=0} 
\left.\Psi\right|_{\vect{x}=0},
\end{equation}
where $\Psi$ is the wave-function in coordinate space,  and $\left.\Psi\right|_{\vect{x}=0}$ 
or $\Psi(0)$ is its value at the origin. For $P$-waves, $\Psi(0)$ is zero, and the second term
in the Taylor serie must be considered; this makes appear $\Psi'(0)$.  $\Psi(0)$ (or  $\Psi'(0)$)  is the 
non-perturbative input, which is also present in the leptonic decay width from 
which it can be extracted.

If the perturbative part ${\cal M}(p)$ is calculated at the leading order in $\alpha_s$, 
this will be referred to as the Leading Order CSM (LO CSM)~\cite{CSM_hadron,Berger:1980ni}. 
\index{colour-singlet!model|)}

\subsubsection{The Colour Evaporation Model}\index{colour evaporation model|(}

This model was initially introduced in 1977~\cite{Fritzsch:1977ay,Halzen:1977rs} and was revived 
in 1996 by Halzen~\etal~\cite{CEM}, following a prescription by Buchm\"uller 
and Hebecker~\cite{Buchmuller:1995qa} to explain rapidity gap production in deep-inelastic scattering.
\index{rapidity gap}

The main characteristic of this model is the total rejection of the role of colour. Contrarily 
to the CSM, the heavy quark pair produced by the perturbative interaction is not assumed to 
be in a colour-singlet state. One simply considers that the colour and the spin of the 
asymptotic $Q\bar Q$ state is randomised by numerous soft interactions occurring after its production,
and that, as a consequence, it is not correlated with the quantum numbers of the pair right after 
its production.

A first consequence of this statement is that the production of a $^3S_1$ state by one gluon
is possible, whereas in the CSM it was  forbidden solely by colour conservation. In addition,
the probability that the $Q\bar Q$ pair eventually be in a colour-singlet state is therefore 
$\frac{1}{9}$, which gives the total cross section to produce a quarkonium:
\begin{equation}\label{eq:CEM1}
\sigma_{onium}=\frac{1}{9}\int_{2m_Q}^{2m_{\bar qQ}} dm \frac{d\sigma_{Q\bar Q}}{dm}.
\end{equation}
This amounts to integrating the cross sections of production of $Q\bar Q$ from the threshold $2m_Q$ up
to the threshold to produce two charm or beauty mesons (designated here by $\bar qQ$) 
and one divides by 9 to get the probability of having a colour-singlet state.

The procedure to get the cross section for a definite state, for instance a $J/\psi$, is tantamount
to ``distributing'' the cross sections among all states:
\begin{equation}\label{eq:CEM2}
\sigma_{J/\psi}=\rho_{J/\psi}\sigma_{onium}.
\end{equation}
The natural value for $\rho_{J/\psi}$, as well as for the other states in that approximation, is the 
inverse of the number of quarkonium lying between $2m_c$ and $2m_D$. This can be refined by 
factors arising from spin multiplicity arguments
or by the consideration of the mass difference between the produced  and the final
states. These are included in the Soft-Colour-Interactions approach (SCI)~\cite{Edin:1997zb}.
\index{soft-colour interactions}

By construction, this model is unable to give information about the polarisation of \index{polarisation}
the quarkonium produced, which is a key test for the other models~\cite{yr}. Furthermore,
nothing but fits can determine the values to input for $\rho$, except perhaps the SCI approach
which relies on Monte-Carlo simulation, but this does not help to understand the underlying
mechanisms. Considering production ratios for charmonium states within the simplest
approach for spin, we should have for instance $\sigma[\eta_c]:\sigma[J/\psi] = 
1:3$ and $\sigma[\chi_{c0}]:\sigma[\chi_{c1}]:\sigma[\chi_{c2}] = 
1:3:5$, whereas deviations from the predicted ratio for $\chi_{c1}$ 
and $\chi_{c2}$ have been observed. Moreover, it is unable to describe the observed variation 
-- from one process to another -- of the production ratios for charmonium states. For example, 
the ratio of the cross sections for $\chi_c$ and $J/\psi$ differs significantly in 
photoproduction and hadroproduction, whereas for the CEM these number are strictly constant. 

All these arguments make us think that despite its simplicity and its 
phenomenological reasonable grounds, this model is less reliable than the CSM. It is also
instructive to point out that the invocation of reinteractions after the $Q\bar Q$ pair 
production contradicts factorisation, which is albeit required when
the pdf are used. However, as we shall see now, the CSM has undergone in the nineties 
a lashing denial from the data.\index{colour evaporation model|)}

\subsection{The nineties: the time for disillusions}

\subsubsection{$\psi'$ anomaly}\index{psip@$\psi'$!anomaly}

In 1984, Halzen \etal~\cite{Halzen:1984rq} noticed that charmonium production from 
a weak $B$ decay  could
be significant at high energy -- their prediction were made for $\sqrt{s}=540$ GeV -- 
and could even dominate 
at high enough $p_T$. This can be easily understood having in mind that the $B$ meson is produced
by fragmentation\index{fragmentation} of a $b$ quark -- the latter dresses with a 
light quark --. And to produce only one $b$ quark at high  $p_T$ is not so burdensome; 
the price to pay is only to put one quark propagator off-shell, instead of two for $gg \to \psi g$. 

This idea was confirmed by the calculations of Glover \etal~\cite{Glover:1987az}, 
which were used as a benchmark by the UA1 Collaboration~\cite{Albajar:1990hf}. After the introduction
of a $K$ factor of 2, the measurements agreed with the predictions; however the $p_T$ slope 
 was not compatible with $b$-quark fragmentation simulations.

\index{production!prompt|(}
From 1993, the CDF collaboration undertook an analysis 
of $\psi$\footnote{In the following, $\psi$ stands for both $J/\psi$ and $\psi'$.} production. 
They managed to extract unambiguously the {\it prompt} component of the signal (not from $B$ 
decay) using a Silicon Vertex Detector (SVX)~\cite{CDF7997a}\footnote{The details of the 
analysis are given in the following chapter.}.\index{silicon vertex detector}

The preliminary results showed an unexpectedly large {\it prompt} component. 
For the $\psi'$, the {\it prompt} cross section was orders of magnitude above the predictions of
the LO CSM (compare the data to the dashed curve on \cf{fig:dsdpt_braaten-94} (b)) . 
This problem was then referred to as the $\psi'$ anomaly.  For  
$J/\psi$, the discrepancy was smaller (\cf{fig:dsdpt_braaten-94} (a)), but it was conceivably blurred 
by the fact that a significant part of the production was believed to come from $\chi_c$ 
radiative feed-down, but no experimental results were there to confirm this hypothesis.

\begin{figure}[h]
\centering
\mbox{\subfigure[$J/\psi$]{\includegraphics[width=8cm]{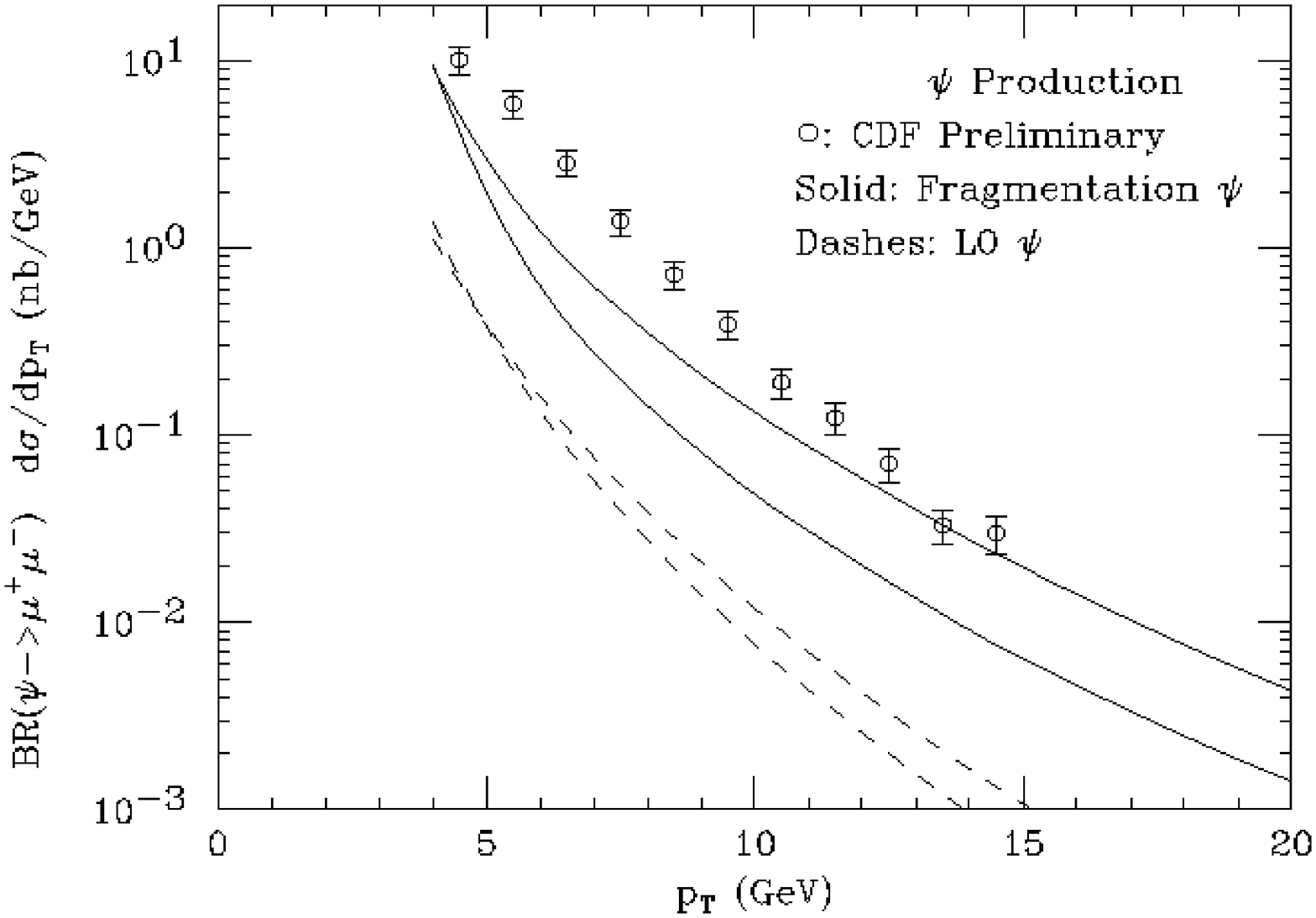}}\quad\quad
      \subfigure[$\psi'$]{\includegraphics[width=8cm]{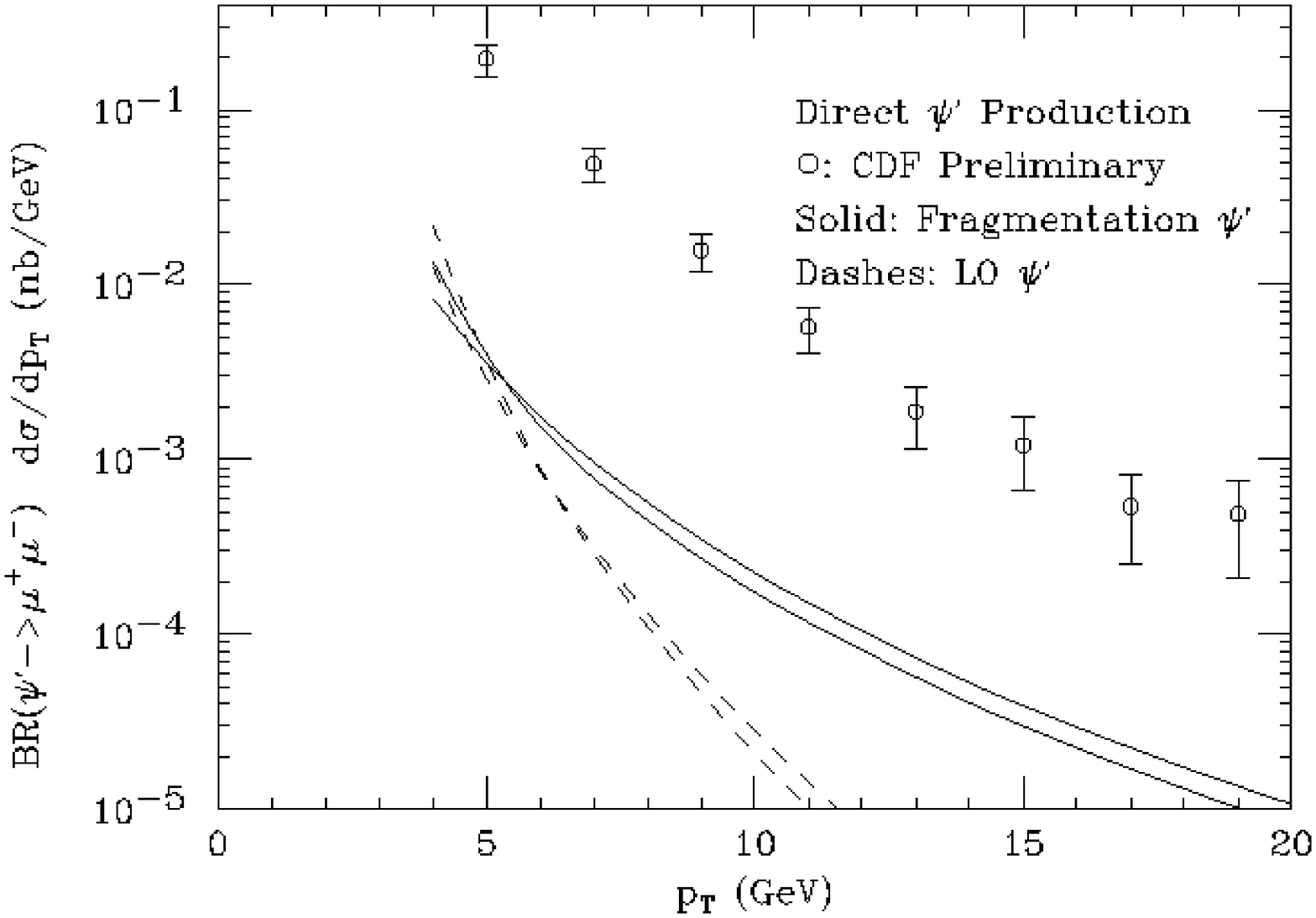}}}
\caption{ Cross sections obtained by Braaten \etal~for LO CSM (dashed curves) 
of  $J/\psi$ (a) and $\psi'$ (b), as well as for CSM fragmentation contribution (solid curves). In 
each case, the two curves depict the extremum values obtained by varying parameters such as $m_c$ and 
the different scales: $\mu_R$, $\mu_F$, $\mu_{frag}$ (from~\cite{Braaten:1994xb}).}
\label{fig:dsdpt_braaten-94}
\end{figure}

\index{production!prompt|)}

\subsubsection{Fragmentation within the CSM}\index{fragmentation|(}\index{Tevatron|(}

The prediction from the LO CSM for {\it prompt} $\psi$ being significantly below 
these preliminary measurements
by CDF, Braaten and Yuan~\cite{Braaten:1993rw} pointed out in 1993 that gluon fragmentation 
processes, even though of higher order in $\alpha_s$, were to prevail over the LO CSM for $S$-wave 
mesons at large $p_T$, in same spirit as the production from $B$ was dominant as it came from 
a fragmentation process. Following this paper, with Cheung and Fleming~\cite{Braaten:1993mp}, 
they considered the fragmentation of $c$ quark into a $\psi$ in $Z_0$ decay. From this
calculation, they extracted the corresponding fragmentation function. In another 
paper~\cite{Braaten:1994kd}, 
they considered gluon fragmentation into $P$-wave mesons. All the tools were then at hand 
for a full prediction of the prompt component of the $J/\psi$ and $\psi'$ at the Tevatron.
This was realised simultaneously by Cacciari and Greco \cite{Cacciari:1994dr} and by
Braaten \etal~\cite{Braaten:1994xb}. We shall discuss their results after a digression 
about the formalism.\index{Tevatron}

To understand how the fragmentation process is factorised from the hard interaction and
how it can be so enhanced that it becomes larger than the LO terms, let us consider the
$\eta_c$ production as was done in~\cite{Braaten:1993rw}. Since its charge conjugation is +1, 
it can be produced in association with two gluons. The LO CSM diagram is depicted 
in~\cf{fig:etac_fragm_LO_CSM} (a), whereas one of
the NLO CSM diagram is on depicted in~\cf{fig:etac_fragm_LO_CSM} (b). 

\begin{figure}[h]
\centering
\mbox{\subfigure[$J/\psi$]{\includegraphics[width=5cm]{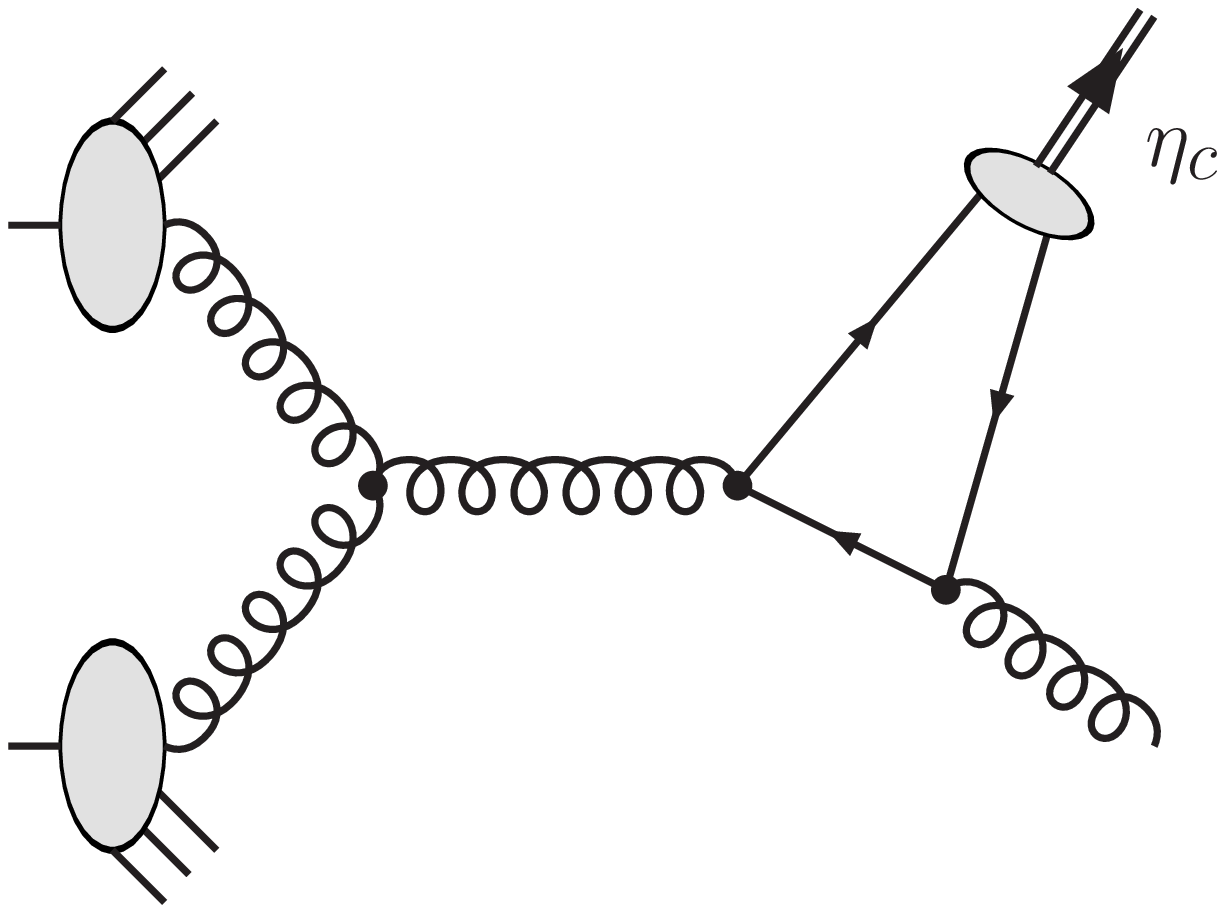}}\quad\quad
      \subfigure[$\psi(2S)$]{\includegraphics[width=6cm]{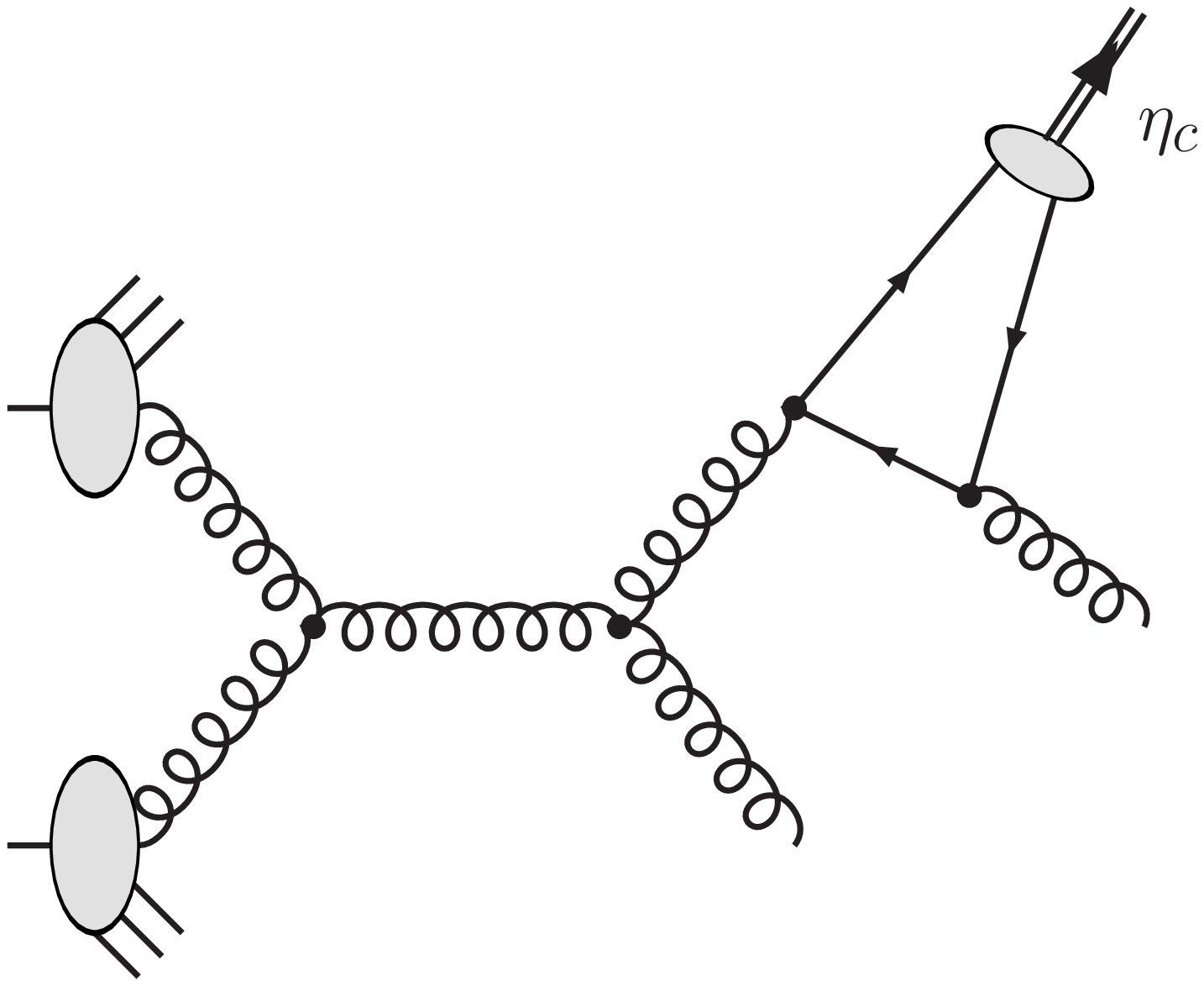}}}
\caption{The LO CSM diagram for $\eta_c$ production in gluon fusion process (a), one of
the NLO CSM diagram for the same process (b) .}
\label{fig:etac_fragm_LO_CSM}
\end{figure}

Normally, the two virtual gluons of the NLO process (\cf{fig:etac_fragm_LO_CSM} (b)) are 
off their mass shell by  $\sim p_T$, as well as the virtual gluon and the virtual quark
in the LO process (\cf{fig:etac_fragm_LO_CSM} (a)). The NLO diagram is then solely 
 suppressed by one power of 
$\alpha_s$ at the chosen scale, which can be $p_T$. However, in some regions of  phase
space, namely when the off-shell gluon attached to the $c\bar c$ is soft 
-- with an off-shellness of the order of $m_c$ --, this diagram is enhanced by a factor
of $\frac{p^2_T}{m^2_c}$ compared to the LO one. At sufficiently large values of $p_T$, 
this enhancement can overcome the $\alpha_s$ suppression. This behaviour can in fact be
traced back to the impossibility of large relative momentum configurations for the constituents of the
bound state.

In this region, this NLO contribution can factorised: the first factor is the 
amplitude $gg\to gg^*$ where the off-shell  and the on-shell gluons in the final state have 
a high $p_T$ and where the same off-shell gluon has a small off-shellness $q^2 \ll p^2_T$;
the second factor is the propagator of the gluon, $1/q^2$, and the third corresponds to
the fragmentation function $g^* \to \eta_c g$. Taking into consideration 
the restriction on the relative momentum
within the bound state, we can therefore express the differential cross section to produce the
$\eta_c$ with a momentum $P$ as:
\eqs{
d \sigma_{\eta_c}(P) \simeq \int^1_0 dz d \sigma_g\left(\frac{P}{z}\right) D_{g_\to\eta_c}(z),
}
$d \sigma_g\left(\frac{P}{z}\right)$ being the differential cross section for $gg\to gg^*$ where 
the momentum of $g^*$ is $\frac{P}{z}$ and $D_{g_\to\eta_c}(z)$ is the 
gluon fragmentation function.\index{fragmentation!function}

\index{fragmentation!scale}
It is now possible to generalise this to all quarkonia. To take into account collinear production 
of fragmenting gluons from a higher energy parton, a fragmentation scale $\mu_{frag}$ is 
introduced which is meant to include logarithms of $p_T/m_Q$ 
appearing in these collinear emissions ($\ln (p_T/m_Q)=\ln (p_T/\mu_{frag}) +\ln (\mu_{frag}/m_Q)$). 
Hence, to all orders in $\alpha_s$, we have the following fragmentation cross section for a quarkonium
$\cal Q$:
\eqs{
\sigma_{{\cal Q}}(P) \simeq \sum_i\int^1_0 dz d \sigma_i(\frac{P}{z},\mu_{frag}) D_{i_\to{\cal Q}}(z,\mu_{frag}).
}
The scale is as usual chosen to avoid large logarithms of $p_T/\mu_{frag}$ in 
$\sigma_i(\frac{P}{z},\mu_{frag})$, that is $\mu_{frag}\simeq p_T$. The summation of the corresponding
large logarithms of $\mu_{frag}/m_Q$ appearing in the fragmentation function is realised via
 an evolution equation~\cite{Curci:1980uw,Collins:1981uw}.\index{evolution equation}

\begin{figure}[H]
\centering
\includegraphics[width=8.5cm,clip=true]{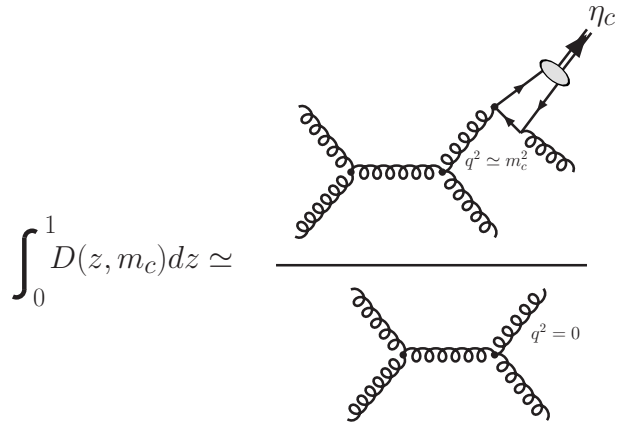}
\caption{Illustration of how to obtain $\int^1_0 D(z,m_c)$.}
\label{fig:etac_fragm_ratio_CSM}
\end{figure}

The interesting point raised  by Braaten and Yuan is that the fragmentation functions can be 
calculated perturbatively in $\alpha_s$ at the scale $\mu_{frag}=2m_Q$. For instance, in the case of gluon
fragmentation into an  $\eta_c$, the trick is to note that $\int^1_0 dz D_z(z,m_c)$ is the 
ratio to the rates for the well-known $gg\to gg$ process and
$gg\to \eta_c gg$ (\cf{fig:etac_fragm_ratio_CSM}). After some manipulations, 
the fragmentation function can be obtained from this ratio
by identifying the integrand in the $z$ integral. This gives :
\eqs{
D_{g_\to\eta_c}(z,2m_c)= \frac{1}{6} \alpha^2_s(2m_c) \frac{|\psi(0)|^2}{m_c^3} [3z-2z^2+2(1-z)\ln(1-z)].
}

The other fragmentation functions of a given parton $i$ into a given quarkonium $\cal Q$, 
$D_{i_\to{\cal Q}}(z,\mu_{frag})$, were obtained in the same spirit. For 
the Tevatron, the differential cross section versus $p_T$ of 
various CSM fragmentation processes are plotted 
in \cf{fig:dsdpt_contrib_braaten-94}.

\begin{figure}[H]
\centering
\includegraphics[width=10cm]{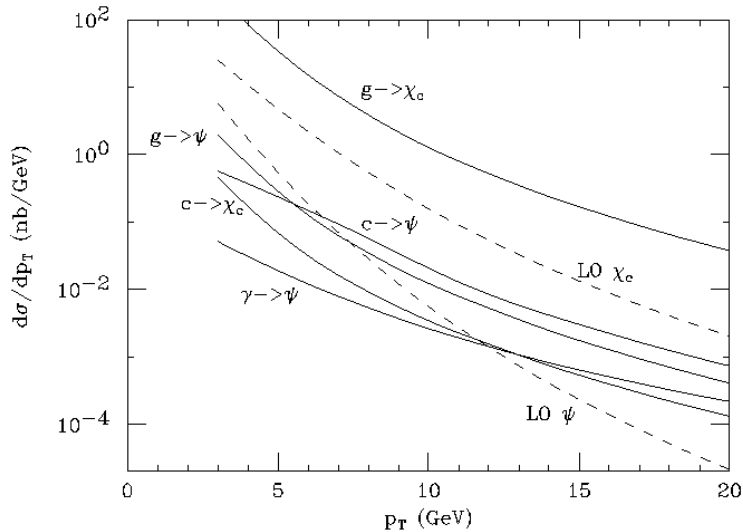}
\caption{Differential cross section versus $p_T$ of various CSM fragmentation processes for $J/\psi$ 
to be compared with the LO contributions (from~\cite{Braaten:1994xb}).}
\label{fig:dsdpt_contrib_braaten-94}
\end{figure}

 The {\it prompt} component of the $J/\psi$ and the direct component of the $\psi'$ could 
in turn be obtained and compared with the preliminary data of CDF (see the solid curves
in plots in \cf{fig:dsdpt_braaten-94} above). For the $J/\psi$, the previous disagreement 
was reduced and could be accounted for by the theoretical and experimental 
uncertainties; on the other hand, for the $\psi'$, the disagreement
continued to be dramatic. The situation would be clarified by the extraction
of the direct component for $J/\psi$, for which theoretical uncertainties are reduced 
and are similar to those for the $\psi'$.

The CDF collaboration undertook the disentanglement of the direct $J/\psi$ signal~\cite{CDF7997b}. 
They searched 
for $J/\psi$ associated with the photon emitted during this radiative decay: 
the result was a direct cross section 30 times above expectations from LO CSM plus fragmentation. 
This was the confirmation that the CSM was not the suitable model for heavy quarkonium production
in high-energy hadron collisions.

\begin{figure}[H]
\centering
\includegraphics[width=10cm]{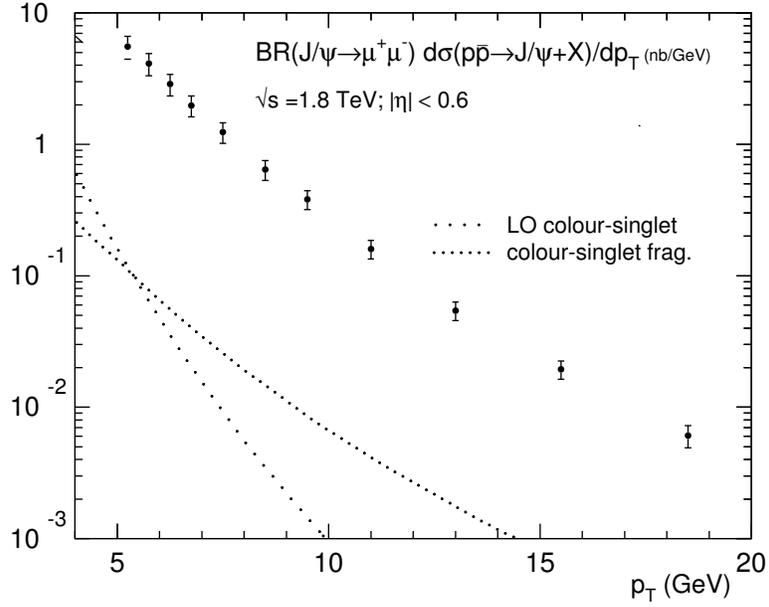}
\caption{Differential cross section versus $p_T$ of the CSM (fragmentation and LO) production 
to be compared with the direct production of $J/\psi$ from CDF (Adapted from~\cite{Kramer:2001hh}).}
\label{fig:dsdpt_CSM_vs_CDF}
\end{figure}

\index{colour-singlet!model|)}
\index{fragmentation|)}
\subsubsection{NRQCD: successes and doubts\dots}\index{NRQCD|(}\index{colour-octet mechanism|(}
\index{Fock state|(}

In 1992, Bodwin~\etal~considered new Fock state contributions in order to
cancel exactly the IR divergence in the light hadron decays of $\chi_{c1}$ at LO. This decay
proceeds via two gluons, one real and one off-shell, which gives the 
light-quark pair forming the hadron. When the first gluon becomes soft, the decay width diverges.
The conventional treatment~\cite{barbieri:div_p_wave}, which amounts to regulating the divergence 
by an infrared cut-off
identified with the binding energy of the bound state, was not satisfactory: it supposed
a logarithmic dependence of $\Psi'(0)$ upon the binding energy. They looked at
this divergence as being a clear sign that the factorisation was violated in the CSM.

Their new Fock states for $\chi_c$ were \eg~a gluon plus a $c\bar c$ pair, 
in a $^3 S_1$ configuration and in 
a {\it colour-octet} state. The decay of this Fock state occurred through
the transition of the coloured pair into a gluon (plus the other
gluon already present in the Fock state as a spectator). 
This involves a new phenomenological parameter, $H_8$, which is related to 
the probability that the $c\bar c$ pair of the $\chi_c$ be in 
a colour-octet $S$-wave state. The key point of this procedure is that a 
logarithmic dependence on a new scale $\Lambda$ -- a typical momentum scale
for the light quark -- appears naturally within the effective field theory Non-Relativistic
Quantum Chromodynamic (NRQCD)~\cite{Caswell:1985ui,Bodwin:1994jh}.

This effective theory is based on a systematic expansion in both
$\alpha_s$ and $v$, which is the quark velocity within the bound state.  For charmonium, 
$v_c^2\simeq 0.23$ and for bottomonium $v_b^2\simeq 0.08$. One of the main novel features
of this theory is the introduction of dynamical gluons in the Fock state decomposition of the physical
quarkonium states. In the case of $S$-wave orthoquarkonia ($^3S_1$), we schematically have, in the Coulomb
gauge:
\eqs{\label{eq:NRQCD_decomp1}
|{\cal Q}_Q\rangle=&{\cal O}(1) | Q\bar Q [\ ^3S_1^{(1)}]\rangle
+{\cal O}(v) | Q\bar Q [\ ^3P_J^{(8)}g]\rangle
+{\cal O}(v^2) | Q\bar Q [\ ^1S_0^{(8)}g]\rangle\\
&+{\cal O}(v^2) | Q\bar Q [\ ^3S_1^{(1,8)}gg]\rangle
+{\cal O}(v^2) | Q\bar Q [\ ^3D_J^{(1,8)}gg]\rangle
+\dots
}
whereas for  $P$-wave orthoquarkonia ($^3P_J$), the decomposition is as follows
\eqs{\label{eq:NRQCD_decomp2}
|{\cal Q}_{Q_J}\rangle=&{\cal O}(1) | Q\bar Q [\ ^3P_J^{(1)}]\rangle
+{\cal O}(v) | Q\bar Q [\ ^3S_1^{(8)}g]\rangle
+\dots
}
In these two formulae, the superscripts (1) and (8) indicate the colour state of the $Q \bar Q$ pair.
The ${\cal O}(v^n)$ factors indicate the order in the velocity expansion at which the corresponding
Fock state participates to the creation or the annihilation of quarkonia. This follows from the 
{\it velocity scaling rules} of NRQCD (see \eg~\cite{Bodwin:1994jh}).\index{velocity scaling rules}

In this formalism, it is thus possible to demonstrate, in the limit of large quark mass, the 
factorisation between the short-distance -- and perturbative -- contributions and 
the hadronisation of the $Q \bar Q$, described by non-perturbative
matrix elements defined within NRQCD. For instance, the differential cross section for the production of
a quarkonium $\cal Q$ associated with some other hadron $X$ reads
\eqs{\label{eq:facto_nrqcd}
d \sigma ({\cal Q}+X)=\sum d \hat\sigma (Q\bar Q[^{2S+1}L_J^{(1,8)}]+X)
\langle{\cal O}^{\cal Q}[^{2S+1}L_J^{(1,8)}] \rangle,
}
where the sum stands for $S$, $L$, $J$ and the colour.

The long-distance matrix element (LDME) $\langle{\cal O}^{\cal Q}[^{2S+1}L_J^{(1,8)}] \rangle$ 
takes account of the transition between the $Q \bar Q$ pair and the final physical state ${\cal Q}$.
Its power scaling rule comes both from the suppression in $v$ of the Fock state component
$[^{2S+1}L_J^{(1,8)}]$ in the wave function of $\cal Q$ and
from the scaling of the NRQCD interaction  responsible for the transition. \index{wave function}

Usually, one defines ${\cal O}^{\cal Q}[^{2S+1}L_J^{(1,8)}]$ as 
the production operator that creates and annihilates a point-like $Q \bar Q$ pair in the specified state. 
This has the following general expression 
\eqs{\label{eq:def_op_prod}
{\cal O}^{\cal Q}[^{2S+1}L_J^{(1,8)}]&=\chi^\dagger K \psi\left(\sum_X\sum_{J_z}|{\cal Q}+X\rangle
\langle{\cal Q}+X|\right)  \psi^\dagger K \chi\\
&=\chi^\dagger K \psi (a^\dagger_{\cal Q} a_{\cal Q}) \psi^\dagger K \chi,}
where\footnote{The second line of~\ce{eq:def_op_prod} is nothing but a short way of 
expressing this operator.} $\psi$ and $\chi$ are Pauli spinors and the matrix $K$ 
is  a product of colour, spin and covariant derivative factors.
These factors can be obtained from the NRQCD lagrangian~\cite{Bodwin:1994jh}. For instance,
${\cal O}^{\cal Q} [\ ^3S_1^{(1)}]=\chi^\dagger \vect \sigma\psi 
  (a^\dagger_{\cal Q} a_{\cal Q})\psi^\dagger \vect \sigma\chi$. However some transitions
require the presence of chromomagnetic ($\Delta L=0$ and $\Delta S=\pm1$) and chromoelectric
($\Delta L=\pm1$ and $\Delta S=0$) terms for which the expressions of $K$ are more complicated.

On the other hand, the general scaling rules relative to the LDME's~\cite{Kramer:2001hh} give:
\eqs{
\langle{\cal O}^{\cal Q}[^{2S+1}L_J^{(1,8)}] \rangle=v^{3+2L+2E+4M},
}
where $E$ and $M$ are the minimum number of chromoelectric and chromomagnetic transitions for the 
$Q\bar Q$ to go from the state $[^{2S+1}L_J^{(1,8)}]$ to the dominant quarkonium ${\cal Q}$ Fock
state.

The idea of combining fragmentation as the main source of production with allowed
transitions between a $\chi_c$ to a $^3S_1$ in a colour-octet state, had already been applied 
by Braaten and Yuan in the above mentioned paper~\cite{Braaten:1994kd}.
Indeed, similar formulae can be written for fragmentation functions~\cite{Braaten:1994vv}:
\eqs{
D_{g\to {\cal Q}}(z,\mu)=\sum d_{[^{2S+1}L_J^{(1,8)}]}(z,\mu) 
\langle{\cal O}^{\cal Q}[^{2S+1}L_J^{(1,8)}] \rangle, 
}
where $d_{[\cdot]}(z,\mu)$ accounts for short-distance contributions 
and does not depend on which  $\cal Q$ is involved.
But since the theoretical predictions for prompt $J/\psi$ production did not disagree dramatically 
with data and since there was no possible $\chi_c$ decay to $\psi'$, a possible 
enhancement of the $\chi_c$ cross section by colour-octet mechanism (COM) was not seen as a key-point
both for $J/\psi$ and $\psi'$ production.

However, in the case of $J/\psi$ and $\psi'$ production, COM could still matter in a different 
manner: fragmentation of a gluon
into a $^3P_J^{(8)}$ is possible with solely one gluon emission and fragmentation into a $^3S_1^{(8)}$
requires no further gluon emission 
(at least in the hard process described by $d_{[\cdot]}(z,\mu)$). Concerning
the latter process, as two chromoelectric transitions are required 
for the transition $|c\bar cgg\rangle$ to
$|c\bar c\rangle$, the associated LDME $\langle{\cal O}^{\psi}[^{3}S_1^{(8)}] \rangle$ was 
expected to scale as $m_c^3 v^7$. In fact, $d_{[^{3}S_1^{(8)}]}$, the contribution 
to the fragmentation function of the short-distance process $g\to\!\!\ ^{3}S_1^{(8)}$ was already known
since the paper of Braaten and Yuan~\cite{Braaten:1994kd} and could be used here, 
as the hard part $d_{[\cdot]}(z,\mu)$ 
of the fragmentation process is independent of the quarkonium. 

In a key paper, Braaten and Fleming~\cite{Braaten:1994kd} combined everything together 
to calculate, for the Tevatron, the fragmentation rate of a 
gluon into an octet $^3S_1$ that subsequently evolves into a $\psi'$. They obtained,
with\footnote{This corresponds to a suppression of 25 compared to the ``colour-singlet matrix
element'' $\langle{\cal O}^{\psi(2S)}[^{3}S_1^{(1)}]\rangle$, which scales as $m_c^3 v^3$. This is 
thus in reasonnable agreement with a $v^4$ suppression.}  
$\langle{\cal O}^{\psi(2S)}[^{3}S_1^{(8)}] \rangle=4.2 \times 10^{-3}$ GeV$^3$, 
a perfect agreement with the CDF preliminary data: compare
these predictions in~\cf{fig:dsdpt_COM_vs_CDF} with the previous ones in~\cf{fig:dsdpt_braaten-94} (b).

\begin{figure}[h]
\centering
\includegraphics[width=10cm]{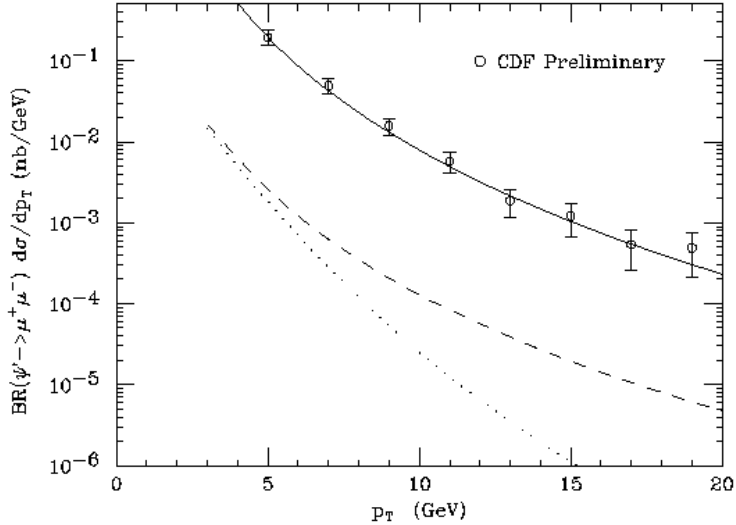}
\caption{Differential cross section versus $p_T$ of the CSM (dotted: LO; dashed: fragmentation and LO) 
production and of the COM fragmentation (solid curve) to be compared with CDF 
preliminary direct production of $\psi'$ (from~\cite{Braaten:1994kd}).}
\label{fig:dsdpt_COM_vs_CDF}
\end{figure}

A straightforward and unavoidable consequence of this solution to the $\psi'$ anomaly was
raised by Cho and Wise~\cite{Cho:1994ih}: the $\psi'$, produced in the latter way,
was 100\% transversally polarised. A suggested test of this prediction was also given, \ie~the 
measurement of
the lepton angular distribution in $\psi'\to \ell^+ \ell^-$, which should behave as
\eqs{
\frac{d\Gamma}{d \cos \theta} (\psi'\to \ell^+ \ell^-) \propto (1+\alpha \cos^2 \theta),
}
with $\alpha$=1 for 100\% transversally polarised particles. Further details are given in 
section~\ref{sec:polarisation}.\index{polarisation}

Following these studies, a complete survey on the colour-octet mechanism was made in two papers
by Cho and Leibovich~\cite{Cho:1995vh,Cho:1995ce}. The main achievements of these papers
were the calculation of colour-octet $P$-state contributions to $\psi$, the first prediction for direct 
colour-octet $J/\psi$ production through \eg~the $\chi_c$ feed-down calculation, the first
predictions for $\Upsilon$ and the complete set of $2\to 3$ parton processes like $ij\to c \bar c +k$,
all this in agreement with the data.

\index{polarisation}
A few year later, however, the CDF collaboration carried out the study of polarisation of 
prompt $\psi$~\cite{Affolder:2000nn}. The outcome was $\alpha=0$ for $\psi'$, or even $\alpha<0$ when
$p_T$ becomes large. For $J/\psi$, the same behaviour showed up, whereas comparison with theory is still
difficult because of the $\chi_c$ feed-down. In any case, no sign of strong transverse polarisation
was seen, in total contradiction with all predictions of the COM.

\index{NRQCD|)}\index{colour-octet mechanism|)}

\subsubsection{Summary}

To sum up, we have seen that the LO CSM  gives  smaller contributions than the 
fragmentation processes for the CSM, 
where the heavy quark pair produced in the hard part of the process has strictly 
the same quantum numbers as the final physical state. On the other hand, in order 
to solve problems connected with the factorisation 
for $\chi_c$ decay, one is driven to introduce new contributions from higher Fock states.

In the context of production, this allows a given physical state to come from a heavy quark
pair with different quantum numbers, \eg~in a colour-octet state. For the case of fragmentation,
even though these higher Fock-state contributions are suppressed, they involve lower-order 
contributions in $\alpha_s$ and are shown to dominate for intermediate and large momentum transfer.
\index{Fock state|)}

All these statements can be easily understood by analysing the large $P_T$ behaviour of the 
differential cross section, $\frac{d\sigma }{dP^2_T}$, from typical diagrams relevant for 
the considered processes. For the LO CSM, at large $P_T$, the two internal lines not linked 
with the bound state are off-shell by~$\sim P^2_T$ and make the partonic differential 
cross section behave like $\frac{d\sigma}{dP^2_T}\sim \frac{1}{P^8_T}$,
as depicted in~\cf{fig:summary_CSM_COM_pt_depend} (a). 

To what concerns fragmentation, it behaves like
a single particle production process, thus like $\frac{d\sigma}{dP^2_T}\sim \frac{1}{P^4_T}$, 
if we consider that, at large $P_T$, the effects of the quarkonium mass become negligible.
Even though fragmentation processes (\cf{fig:summary_CSM_COM_pt_depend} (b)) are of higher order 
in  $\alpha_s$ ($\alpha_s^5$ instead of  $\alpha_s^3$), 
they scale like $\frac{P^4_T}{(2m_c)^4}$ compared to LO CSM and dominate over the latter 
for $P_T\gg 2m_c$.

The inclusion of colour-octet channels can be analysed with similar arguments.  For the 
$^3S_1^{(8)}$ channel, the only difference comes from the powers of $\alpha_s$ and $v$. 
No similar estimation for power counting of the $P_T$ dependence can be done for the 
other ${\cal O}(v^4)$ channels~\cite{Cho:1995ce}, however the gluon $t$-channel dominance
argument~\cite{Kramer:2001hh} provides us with the scaling $\frac{1}{P^6_T}$. Indeed these 
channels are significant at moderate 
$P_T$ of the order of $2m_c$, as one can conclude from the complete analysis made in~\cite{Cho:1995ce}.

On the other hand, the statement that the polarisation of $\psi$ produced at large $P_T$ be 
entirely transverse -- at variance with experimental data -- follows simply and only 
from the on-shellness of the fragmenting gluon and from the approximate
spin symmetry of NRQCD.\index{polarisation}

Considering that no solution of the  polarisation issue has been  proposed so far, we 
have decided to undertake an analysis of the non-static effect of the LO CSM, 
with the first hope to state that the static approximation\index{static!non-static}
was not viable for previously unknown reasons, but also with the goal to attain new insights 
about some other 
unknown effects linked with quarkonium production.
As we shall see, it seems that the static approximation is not problematic. 
However, we have realised than unconstrained contributions 
could be introduced in the description of the production of the quarkonium in 
association with a gluon -- as it should be at non vanishing transverse momentum. 
This introduction can bring the theoretical predictions
in agreement with experimental measurements, for the cross section and for the polarisation.
\index{Tevatron|)}

\begin{figure}[h]
\centering
\includegraphics[height=20cm]{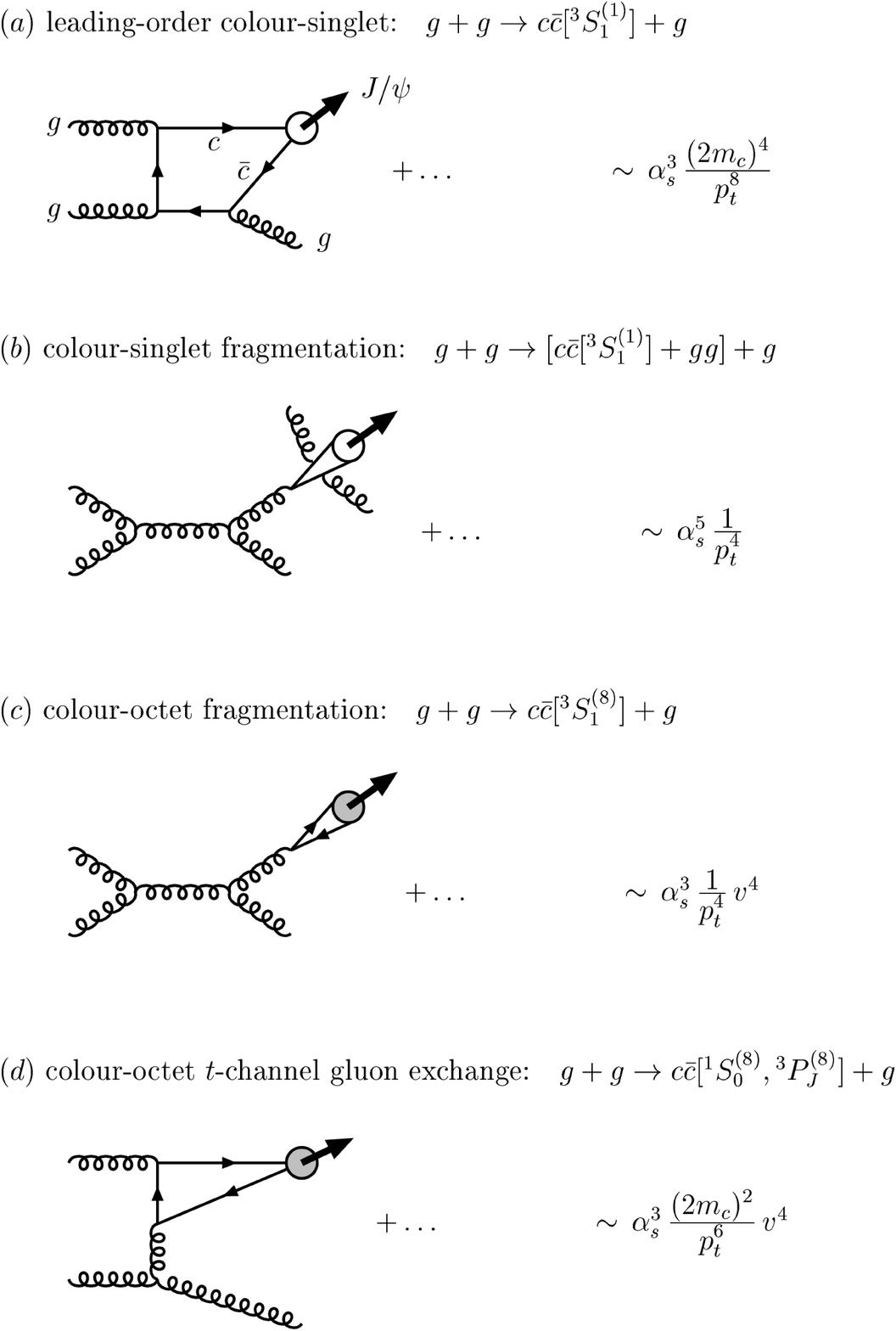}
\caption{Summary of the different processes arising in the direct production of $\psi$ within the 
CSM and the COM and their $p_T$ dependence (from Kr\"amer~\cite{Kramer:2001hh}).}
\label{fig:summary_CSM_COM_pt_depend}
\end{figure}

\chapter{Experimental results}\label{ch:experim}

\section{Foreword}

Before broaching a review of the most recent results on high-energy hadroproduction
of $J/\psi$, $\psi'$ and $\Upsilon$'s, let us remind the reader
of the various kinds of collisions and colliders in which these particles can be created. 

For most of the following discussion, we shall limit ourselves to  high-energy collider experiments, 
because most of the models that we shall consider in this thesis 
are  valid in the perturbative regime of QCD, \ie~when the coupling constant $\alpha_s$ is
clearly less than 1; this is likely to be the case for reactions with c.m. energy larger than,
say,  5 GeV. 
Furthermore, the factorisation hypothesis relies on the assumption of having a sufficiently
inclusive reaction, although the quarkonium is actually detected. This is another reason to
impose a large energy. Concerning fixed target experiments, for which factorisation
is uncertain, we shall not include them in our survey, even though heavy quarkonium 
production in that regime is certainly of great interest.

Limiting ourselves to high-energy collisions, the most recent published results for quarkonium
hadroproduction come from two accelerators:
\begin{enumerate}
\item \bfsf{The Tevatron} at Fermilab which -- as stated by its name -- runs at TeV energy with 
proton-antiproton collisions. For Run I (``Run IA'' 1993-94 and ``Run IB'' 1995-96), the 
energy in c.m. was 1.8 TeV. For Run II, it has been increased to 1.96 TeV. The experimental 
analyses for this Run are still being carried out.
\item \bfsf{RHIC} at BNL running at 200 GeV for the $J/\psi$ study with proton-proton collisions.
\end{enumerate}\index{Tevatron} \index{RHIC}\index{BNL}

There exist also data from electron-proton collision from \bfsf{HERA} at DESY running 
with electron-proton beams with proton energy of 820 GeV and electrons energy of 27.6 GeV.

Results for $e$-$p$ collisions come for the most part from two collaborations, namely 
ZEUS and H1. Two classes of processes can be distinguished depending \index{HERA}
on the off-shellness of the photon emitted by the scattering electron. If the photon is on-shell, 
we are dealing with photoproduction, otherwise, we deal with electroproduction. 
Results on photoproduction are compatible with NLO calculations within
CSM~\cite{Kramer:1995nb}. Occasionally, we shall mention these in the following discussion.

Finally, three $e^+e^-$ collider are yielding data on quarkonium at present time:
\begin{enumerate}
\item \bfsf{LEP} at CERN running from around $Z_0$ mass up to around 210 GeV with 
electron-positron collisions.\index{CERN} \index{LEP}
\item $B$ factories, \bfsf{PEP II} at SLAC and \bfsf{KEK B} at KEK running \index{KEK} \index{SLAC}
mainly at the $\Upsilon (4S)$ (10.58 GeV) 
resonance with electron-positron collisions.
\end{enumerate}
However, quarkonium physics at LEP is very different from that at $B$ factories. 
Due to its large c.m. energy, LEP can significantly produce $\psi$ through
$\gamma \gamma$ collisions which can be selected via kinematical constraints. From
a theoretical point of view, this is rather interesting since this way of producing
$\psi$ is similar to those at work in hadroproduction, although potentially better
understood. On the other hand, to what concerns $B$ factories results, we think that the energy 
is such that the reactions are not sufficiently inclusive to be straightforwardly 
compared to reactions that take place at HERA, RHIC or the Tevatron. However, it is very 
important to keep in mind that there exist serious disagreements between measurements 
and theoretical predictions
of both CSM and COM. Sometimes they are thought to come from missing contributions 
at end-point which arise when re-summation techniques are applied; this is indeed the 
sign that some kinds of ``exclusive'' effects become important at these low energies. For further
information, the interested reader may have a look at~\cite{yr}.

Concerning hadroproduction, the above-mentioned deviations between theoretical predictions 
 and measurements of the CDF collaboration could be actually considered 
as a confirmation of the prior UA1 analysis~\cite{Albajar:1990hf}. 
The deviations were confirmed by D$\emptyset$ later on~\cite{Abachi:1996jq}. For the Tevatron, 
only CDF analysis will be presented in the following. Concerning RHIC, we shall expose the
analysis for $J/\psi$ in 200 GeV $pp$ collisions from the PHENIX collaboration~\cite{Adler:2003qs}.
\index{PHENIX}

It is worth pointing out here that high-energy  $p$-$\bar p$ and $p$-$p$ collisions 
give similar results for the same kinematics, due to the small contribution of 
valence quarks.\index{quark!valence}

\section{Different type of heavy quarkonium production}\label{sec:genmeson}

As we have already explained, the detection of quarkonia proceeds via the identification
of their leptonic decay products. We give in~\ct{table:dimuon} the relative decay widths into
muons.

\begin{table}[H]
\centering
\begin{tabular}{|l|l|}
\hline Meson & $\Gamma(\mu^+\mu^-)/\Gamma(total)$ \\
\hline \hline 
$J/\psi $        &  $0.0588 \pm 0.0010$  \\
$\psi(2S) $      &  $0.0073 \pm 0.0008$  \\
$\Upsilon(1S)$   &  $0.0248 \pm 0.0006$  \\
$\Upsilon(2S)$   &  $0.0131 \pm 0.0021$  \\
$\Upsilon(3S)$   &  $0.0181 \pm 0.0017$  \\
\hline
\end{tabular}
\caption{Table of branching ratios in dimuons~\cite{pdg}.}\label{table:dimuon}
\index{branching ratio} 
\end{table}

It is interesting to note that the CDF analysis that lead to the problem of quarkonium
hadroproduction 
was in fact not devoted to $\psi$ production but to $b$-quark production through
its (weak) decay into  $\psi$~\cite{CDF7997a}. The value of this decay 
was already quite well known~\cite{CLEO} at that time and a measurement of the 
proportion of $\psi$'s produced through this decay was expected to give 
the total number of $b$ quarks produced.

Their study succeeded and they extracted the percentage of $\psi$'s produced through 
$b$ decays and of $\psi$'s {\it promptly} produced. The latter was unexpectedly
large. Indeed, the production through $b$ decays was expected to behave like
a fragmentation process and to dominate at moderate and large $p_T$. This was not the case
and the slight disagreement between total production predictions~\cite{Glover:1987az}
and UA1 measurements~\cite{Albajar:1990hf}
was blown away by a striking incompatibility between prompt $\psi$ production predictions
and CDF data. A factor 60 of discrepancy
was for instance found for $\psi'$ (see \cf {fig:dsdpt_braaten-94}~(b)).\index{production!prompt} 

In this case, no higher excited states  was expected to contribute to its production
(see \cf{fig:charm_spectrum}), 
contrarily to the case of the $J/\psi $ which can be the product of a $\chi_c$ ($P$-state)
electromagnetic decay. In order
to get a complete picture of the problem, the CDF collaboration decided to proceed to a study of the direct
production of this meson~\cite{CDF7997b}. This involved the detection of the photon emitted during the
radiative decay,\index{chi@$\chi_c$} 
\begin{equation}
\chi_c \to J/\psi+ \gamma .
\end{equation}
They then found  a direct production 30 times above theoretical predictions.
\index{production!direct}

Briefly, the problem of direct $J/\psi$ production separation comprises three steps:
\begin{myitem}
\item muon detection;
\item elimination of $J/\psi$ produced by hadrons containing $b$ quarks ;
\item elimination of $\chi_c$ radiative-decay production;
\end{myitem}

The following table summarises the different processes to be discussed and the quantities linked.
\begin{table}[H]
\begin{center}
{\footnotesize
\begin{tabular}{|c|l|c|c|c|}
\hline $1^{st}$ step & $2^{nd}$ step &$3^{rd}$ step & Type & \begin{tabular}{c}Associated
\\ quantity \end{tabular}\\
\hline \hline 
& $c\bar c \to J/\psi$ &$ -$          &  Direct prod.
& $ F(b \!\!\!/, \chi\!\!\!/ ,\psi\!\!\!/ (2S))^{J/\psi}$\\$p\bar p \to c\bar c + X$ & $c\bar c \to \chi_c$ &  $\chi_c \to J/\psi+ \gamma$ & 
\begin{tabular}{c} \hbox{Prompt prod. by }  \\ \hbox{decay of } $\chi_c$ \end{tabular}
 & $ F(b \!\!\!/)_\chi^{J/\psi}$ \\ & $c\bar c \to \psi(2S)$ &  $\psi(2S) \to J/\psi+ X$ & 
\begin{tabular}{c} 
\hbox{Prompt prod. by}  \\\hbox{decay of } $\psi(2S)$ \end{tabular}
 & $ F(b \!\!\!/)_{\psi(2S)}^{J/\psi}$ \\
\hline 
$\begin{array}{c}
p\bar p \to \bar bc +X\\  
\bar bc\to\bar c  c + \ell^-+\bar\nu_\ell
\end{array}$  
&$\begin{array}{ll}
\\ c\bar c \to J/\psi \\
c\bar c \to \chi_c
\end{array}$
&  $\begin{array}{cl} \\ - \\
\chi_c \to J/\psi+ \gamma
\end{array}$
& $\begin{array}{ll}
\\-\\ -
\end{array}$
& $\begin{array}{c}\\ f_b (or F_b) \\
\end{array}$
\\
etc \dots& & & &\\
\hline
\end{tabular}
}
\caption{Different processes involved in $J/\psi$ production accompanied by quantities
used in the following discussion.}\label{table:jpsi00}
\end{center}
\end{table}
Let us explain the different fractions that appear in the latter table: 
\begin{itemize}
\item $ F(b \!\!\!/, \chi\!\!\!/,\psi\!\!\!/ (2S) )^{J/\psi}$ is the {\it prompt}  fraction of $J/\psi$ 
that do not come $\chi$, neither from $\psi(2S)$, \ie~the direct fraction.
\item $F(b \!\!\!/)_{\psi(2S)}^{J/\psi}$ is the {\it prompt} fraction of $J/\psi$ that
come from $\psi(2S)$. 
\item $F(b \!\!\!/)_\chi^{J/\psi}$ is the {\it prompt} fraction of $J/\psi$ that
come from $\chi$.
\item $F_b$ (or $f_b$) is the {\it non-prompt} fraction or equally the fraction that come for $b$ quarks.
\end{itemize}

Concerning $\psi(2S)$,  due to the absence of stable higher excited states likely to decay into it, we have
the following summary for the different processes to be discussed and the quantities linked:
\begin{table}[H]
\begin{center}
{\footnotesize
\begin{tabular}{|c|l|c|c|c|}
\hline $1^{st}$ step & $2^{nd}$ step &$3^{rd}$  step & Type & \begin{tabular}{c}Associated
\\ quantity \end{tabular}\\
\hline \hline 
 &  & &  
& \\
 $p\bar p \to c\bar c + X$  & $c\bar c \to \psi(2S)$ &  $-$ & 
Direct/Prompt prod.
 & $ F(b \!\!\!/)^{\psi(2S)}$\\    &  &   & 
 &  \\

\hline 
$\begin{array}{c}
p\bar p \to \bar bc +X\\  
\bar bc\to\bar c  c + \ell^-+\bar\nu_\ell
\end{array}$  
&$ c\bar c \to \psi(2S)$
&   $-$ 
& $-$
& $f_b (or F_b)$
\\
etc \dots& & & &\\
\hline
\end{tabular}
}
\caption{Different processes involved in $\psi'$ production accompanied by quantities
used in the following discussion.}\label{table:psi200}
\end{center}
\end{table}
Let us explain the different fractions that appear in the latter table: 
\begin{itemize}
\item $F(b \!\!\!/)^{\psi(2S)}$ is the {\it prompt}  fraction of $\psi(2S)$, \ie~the direct production.
\item $F_b$ (or $f_b$) is the {\it non-prompt} fraction or equally the fraction that come for $b$ quarks.
\end{itemize}

Considering the similarities between the charmonium and bottomonium systems 
(see \cf{fig:bottom_spectrum} for the spectrum), a substantive part of the 
discussion will be also devoted to it. As can be seen in \ct{table:dimuon},
the leptonic branching ratio are also relatively high. The detection and 
the analysis of the bottomonia are therefore carried out in the same fashion. 
For the extraction of the direct production, the $b$-quark feed-down is 
obviously not relevant, only the decays from stable higher resonances
of the family are to be considered.\index{upsi@$\Upsilon$}

All the quantities useful for the bottomonium discussion are summarised in the following table: 
\begin{table}[H]
\begin{center}
{\footnotesize
\begin{tabular}{|c|l|c|c|c|}
\hline $1^{st}$ step & $2^{nd}$ step &$3^{rd}$ step & Type & \begin{tabular}{c}Associated
\\ quantity \end{tabular}\\
\hline \hline 
$p\bar p \to b\bar b + X$ & $b\bar b \to \Upsilon(nS)$ &$-$          &  
Direct prod.& $F_{direct}^{\Upsilon(nS)}$\\
&  $b\bar b \to \chi_b$  & $\chi_b \to  \Upsilon(nS)+ \gamma$  & 
\begin{tabular}{c} \hbox{Prod. by }  \\ \hbox{decay of } $\chi_b$ \end{tabular}
 & $F_{\chi_b}^{\Upsilon(nS)}$ \\    & $b\bar b \to \Upsilon(n'S)$ &  $\Upsilon(nS) \to \Upsilon(n'S)+ X$  & 
\begin{tabular}{c} 
\hbox{Prod. by}  \\\hbox{decay of } $\Upsilon(n'S)$ \end{tabular}
 &  $F_{\Upsilon(n'S)}^{\Upsilon(nS)}$ \\
\hline
\end{tabular}
}
\caption{Different processes involved in $\Upsilon(nS)$ ($n=1,2,3$) production accompanied by quantities
used in the following discussion.}\label{table:upsi00}
\end{center}
\end{table}
Let us explain the different fractions that appear in the latter table: 
\begin{itemize}
\item  $F_{direct}^{\Upsilon(nS)}$ is the direct  fraction of $\Upsilon(nS)$.
\item  $F_{\chi_b}^{\Upsilon(nS)}$ is the fraction of $\Upsilon(nS)$ that come from $\chi_b$. 
\item $F_{\Upsilon(n'S)}^{\Upsilon(nS)}$ is the  fraction of $\Upsilon(nS)$ that come from a higher $\Upsilon(n'S)$.
\end{itemize}

\section[CDF analysis for $\psi$ production cross sections]{CDF analysis for $\psi$ production cross sections\protect\footnote{Some plots shown in the following
are preliminary. These are displayed only for an illustrative purpose. 
The discussion would have been the same if the corresponding final plots had been
available. The plots relative to final observables are however all published ones.}\protect\footnote{In the following, $\psi$ stands for both $J/\psi$ and $\psi'$.}}

\label{sec:CDF_psi_prod}\index{CDF|(}

\subsection{Succinct description of the parts of the detector used in this analysis}

Without entering  too much into the details, let us recall the main elements of
the detector used for this analysis.

\begin{itemize}
\item A superconducting solenoid;
\item A central muon detector (CMU: central muon system (or central muon chambers)). It 
covers the region\footnote{$\eta \equiv  - \ln(\tan(\theta/2))$ where $\theta$ is the polar angle 
of the particle in the c.m. frame.} $|\eta| < 0.6$;\index{pseudorapidity}
\item The silicon vertex detector (SVX) which among other purposes enables to detect a second vertex 
resulting from the decay of a beauty hadron in a $\psi$. This detector directly surrounds the beam;
\index{silicon vertex detector}
\item Two calorimeters: an electromagnetic one  and a hadronic 
one. They cover the whole azimuthal angle $\phi$ (cf. \cf{fig:schem1})
and spread till pseudorapidity $\eta$ of 4.2 in absolute value;
\item A central tracking chamber (CTC) lying in the induction of the magnet. It enables the 
determination of the transverse momenta of the particles. 
It covers a pseudorapidity range of $|\eta| < 1.1$.
\end{itemize}

\subsection{$\psi$ total production} \label{subsec:psi_detect}

Muons are selected by a three-level trigger:
\index{trigger}
\begin{itemize}
\item The \bfsf{level 1} requires two track segments in the CMU, separated by at least 
$5^\circ$ in azimuth (cf. \cf{fig:schem1});
\item The \bfsf{level 2} requires that at least one of the muon track segments
is matched in $\phi$ to a track found in the CTC by the central fast tracker,
a hardware track-finding processor. These two tracks are then considered 
to come from a single particle;
\item The \bfsf{level 3} requires a pair of oppositely charged muons after 
full track reconstruction.
\end{itemize}

Besides this selection, muons are to satisfy other criteria:

\begin{itemize}
\item The CTC track is extrapolated to the CMU chambers and is required to lie within
3 $\sigma$ of the CMU hits, where $\sigma$ is the uncertainty in the extrapolated 
position due to multiple scatterings;
\item A nonzero energy deposit  is required in the calorimeter tower situated
in front of the muon-chamber segment;
\item To remain in the region of good trigger efficiency, both muons are required
to have $p_T>2.0$ GeV and one muon $p_T>2.8$ GeV. The two muon tracks are fitted with the 
requirement that they originate from a common point;
\item To remain in a region of good acceptance to $\psi$ decays, the $\psi$ candidate
is required to satisfy $|\eta|<0.6$ and $p_T> 5$ GeV.
\item The invariant mass of the pair must lie within the region $ 3057 \hbox { MeV} < 
M(pair) < 3137 \hbox { MeV}$ 
in order to be considered as coming from a $J/\psi$ and $ 3646 \hbox { MeV} < 
M(pair) < 3726 \hbox { MeV}$ to be considered as coming from a $\psi'$. 
\end{itemize}

\begin{figure}[H]
\centering\includegraphics[width=7.5cm]{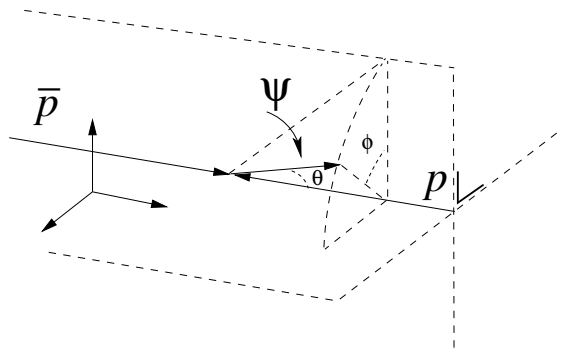}
\caption{General drawing defining the angles.}
\label{fig:schem1}
\end{figure}

\begin{figure}[H]
\centering\includegraphics[width=13cm]{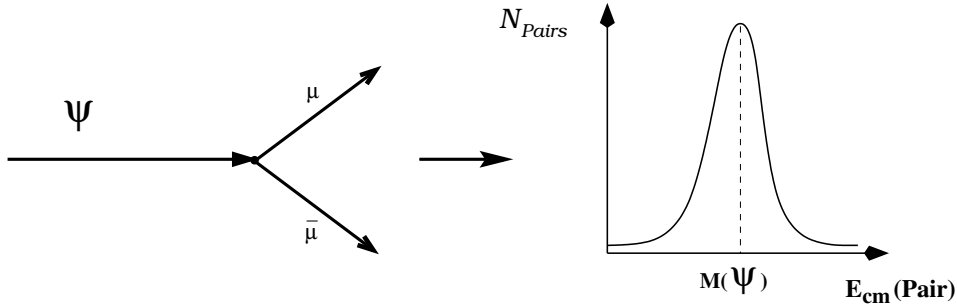}
\caption{Invariant mass distribution of the muon pair in which the $\psi$ decays.  
The c.m. energy amounts to $\psi$ mass.}
\label{fig:schem2}
\end{figure}

On the other hand, 60\% of the muon tracks also have hits in the SVX which measures
only $r$ and $\phi$ and the resolution of their track impact parameter is
$(13 + 40/p_T) \mu$m, where $p_T$ (expressed in GeV) is the track momentum 
transverse to the beam line.

The sample of $p\bar p$ collisions amounts to $17.8 \pm 0.6$ pb$^{-1}$ 
at $\sqrt{s}=1.8$ TeV. For $J/\psi$, the considered sample consists however of
 $15.4 \pm 0.6$ pb$^{-1}$  of integrated luminosity\footnote{This difference
is due to the subtraction of data taken during a period of reduced
level 3 tracking efficiency. These data have however been taken into account
for $\psi(2S)$ after a correction derived from the $J/\psi$ sample.}. \index{luminosity}

The number of $\psi$ candidate is therefore determined by fitting
the mass distribution of the muons in the c.m. frame
(cf. \cf{fig:schem2} and \cf{fig:reson} (a) and (b)) after subtraction of the  
noise, which is evaluated as follows.

Various corrections are evaluated by Monte-Carlo simulation. They incorporate deviations
due to processes as $\psi \to \mu^+\mu^-\gamma$. Muon momenta are smeared in order
to simulate the finite detector resolution. As the measurement of the muon energy 
is a function of $p_T$, data are fitted in separate $p_T$ bins.

Concerning the trigger efficiency $\epsilon_{\rm trigger}$, it is accounted for by
weighting the event with $1/\epsilon_{\rm trigger}$, which depends on $p_T$.

\begin{figure}[h]
\centering
\mbox{\subfigure[$J/\psi$]{\includegraphics[height=7cm]{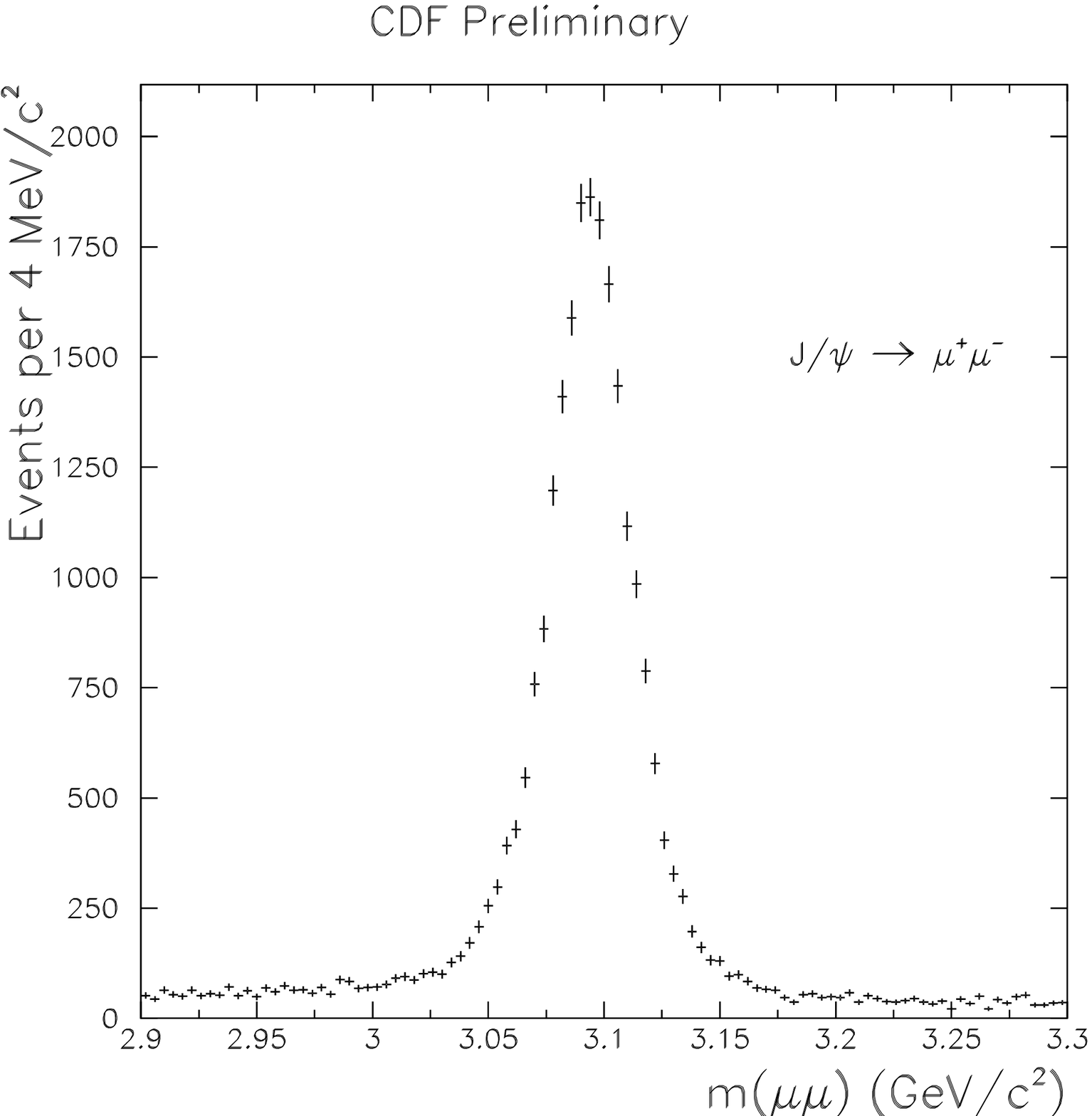}}\quad\quad
      \subfigure[$\psi'$]{\includegraphics[height=7cm]{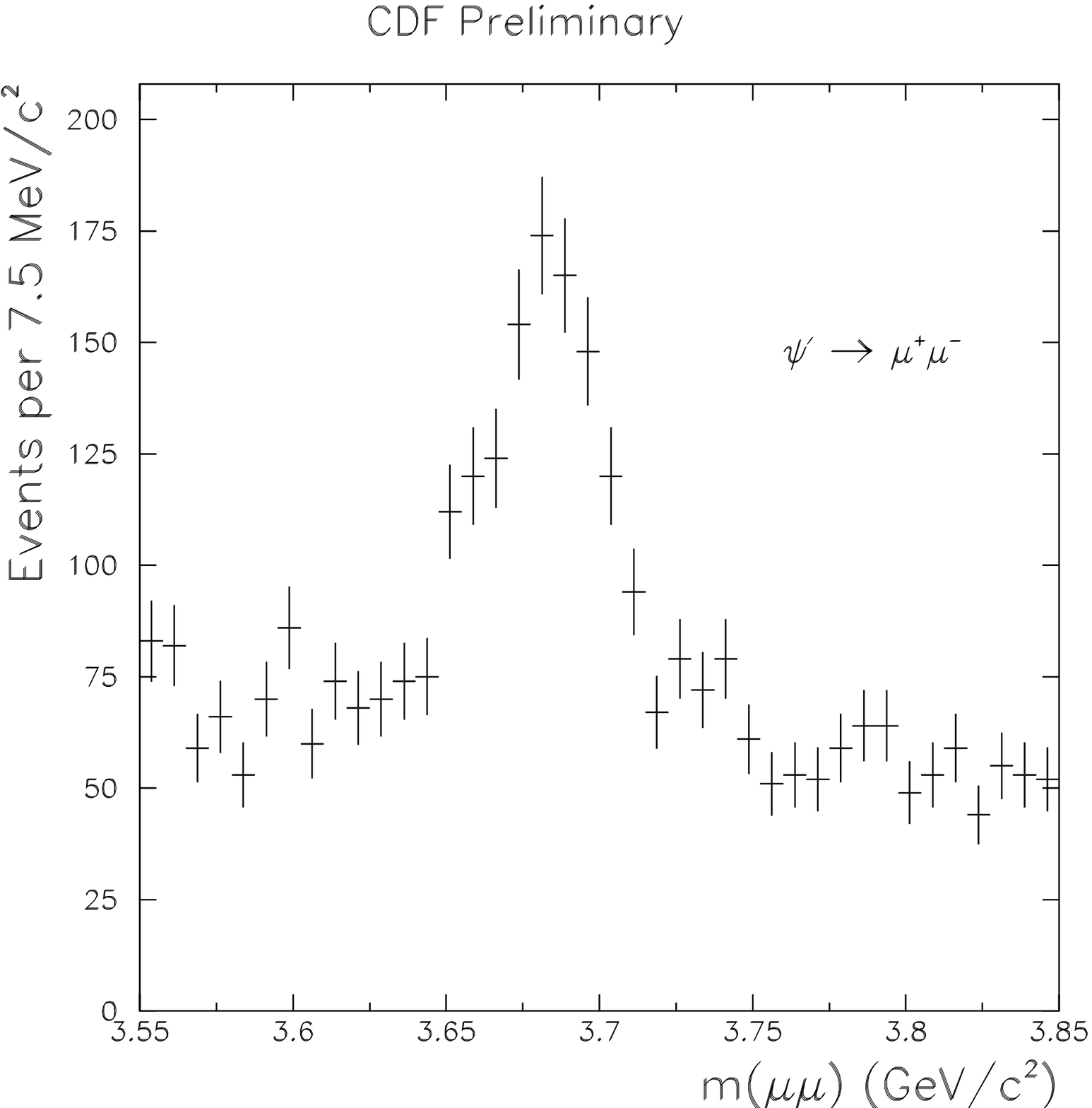}}}
\caption{Resonance peak of $J/\psi$ (a) and $\psi'$ (b).}
\label{fig:reson}
\end{figure}

The mass distribution is then fitted to the signal shape fixed
by simulation and to a linear background. The fit yields the mass of
the particle and a background estimate; from this, it is
easy to find back the number of particles which has decayed
and from the branching ratio, the number of produced particles.

For the present study, the fits were reasonably good for each $p_T$-bin,
the $\chi^2$ per degree of freedom ranging from 0.5 to 1.5. The measured width
of the mass peak
was from 17 MeV to 35 MeV for $p_T$ from 5 to 20 GeV.

Approximatively, 22100 $J/\psi$ candidates and 800 $\psi'$ candidates above
a background of 1000 events were observed.

The acceptance was also determined  by Monte-Carlo simulation. $\psi$'s
were generated with uniform distribution in $p_T$, $\eta$ and $\phi$. A simulated
detector was used and the kinematic cuts already discussed were applied to the 
generated events. This leads to an acceptance of 9\% at $p_T=5$ GeV which rises
to a plateau of 28 \% for $p_T>14$ GeV. It is important to note that the acceptance
depends on the $\psi$ polarisation. By simulation, the uncertainty on this polarisation
is shown to produce a  15\% variation of the acceptance at $p_T=$ 5 GeV and 5\% at 20 
GeV for prompt production, whereas the non-prompt production variation is from 6\% to 3\% 
for the same transverse momenta.

The efficiency of the CMU segment reconstruction is measured to be $97.2 \pm 1.2 \%$ 
using dimuon events recorded with a single muon trigger. For CTC efficiency reconstruction,
it is determined by a Monte-Carlo technique and is $96.4 \pm 2.8\%$ for both CTC tracks.
The efficiency of the CMU-CTC matching requirements is estimated from the number of 
$J/\psi$ before and after the matching requirements, to be $90.5 \pm 1.0 \%$.

Systematics on the efficiency of the three triggers are evaluated by varying the 
functional form of the trigger efficiency. The $J/\psi$ efficiency is then $97.0 \pm 0.2 \%$
and the  $\psi'$ one is $92.3 \pm 0.2 \%$.

The $\psi$ differential cross section is experimentally defined as 
\begin{equation} \frac{d\sigma(\psi)}{dp_T}\cdot{\cal B}(\psi \to \mu^+\mu^-)=
\frac{N(\psi)}{\epsilon \cdot \int{\cal L}dt\cdot \Delta p _T},
\end{equation}
where $N(\psi)$ is the number of $\psi$ candidates in the $\Delta p_T$ bin 
with efficiency corrections applied, and  $\epsilon$ ($\neq \epsilon_{trigger}$) 
is the product of the acceptances (kinematic and detector-related), of the
reconstruction efficiencies and of the corrections due to selection cuts.

The integrated cross sections were then measured to be 
\begin{equation}
\sigma(J/\psi)\cdot{\cal B}(J/\psi \to \mu^+\mu^-)= 17.4 \pm 0.1 (stat.)^{+2.6}
_{-2.8}(syst.) \ { \rm nb}
\end{equation}

\begin{equation}
\sigma(\psi(2S))\cdot{\cal B}(\psi(2S) \to \mu^+\mu^-)= 0.57 \pm 0.04 (stat.)^
{+0.08}_{-0.09}(syst.) \ { \rm nb}.
\end{equation}

\ 

\noindent where $\sigma(\psi)\equiv \sigma(p\bar p\to \psi X, p_T(\psi)>5 \hbox{ GeV},
|\eta(\psi)|<0.6)$. The evolution of $d\sigma \times {\cal B}$ as a function of $p_T$ is
shown in \cf{fig:sectioneff1}.

\begin{figure}[H]
\centering
\mbox{\subfigure[$J/\psi$]{\includegraphics[height=7cm]{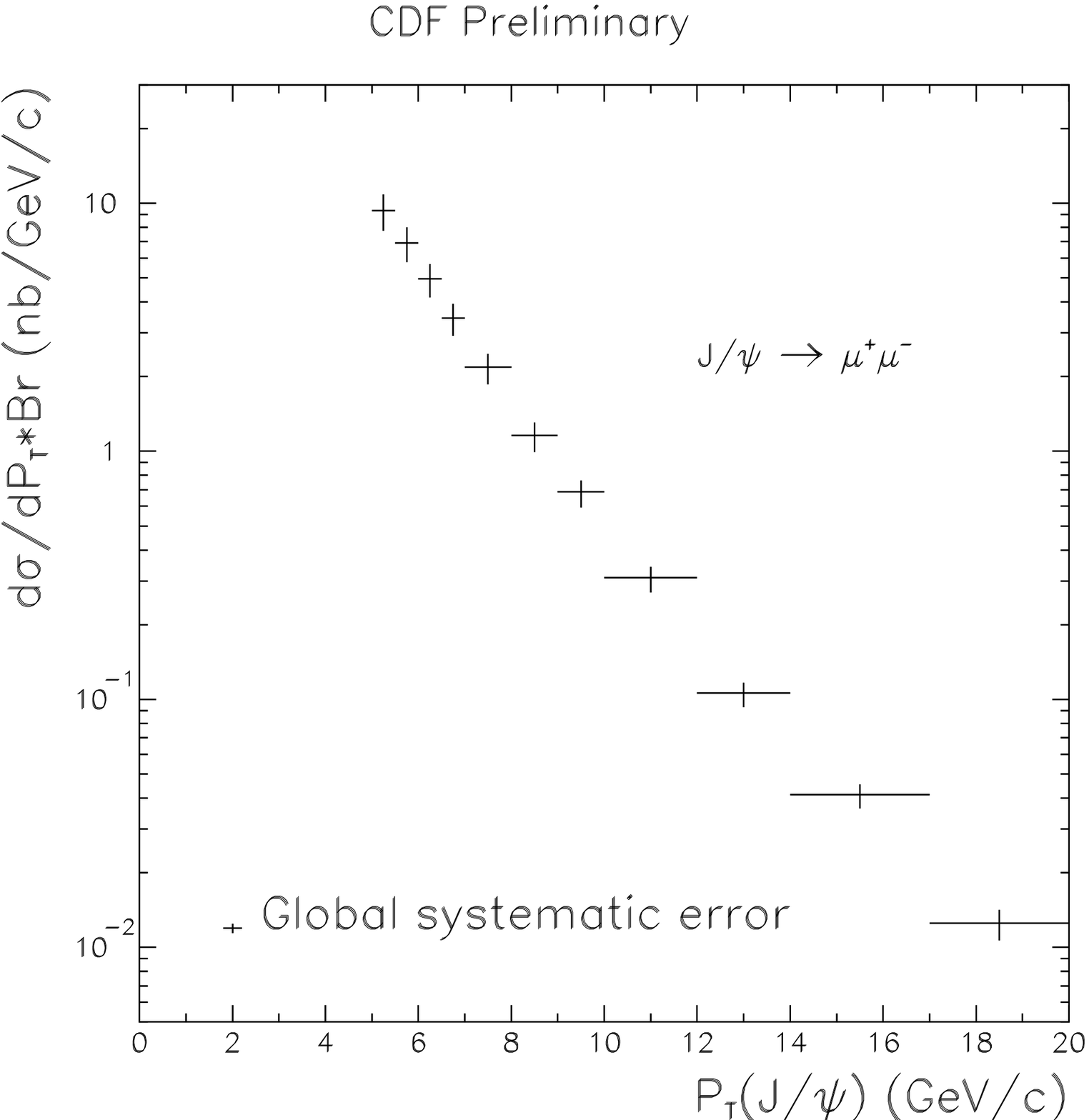}}\quad\quad
      \subfigure[$\psi'$]{\includegraphics[height=7cm]{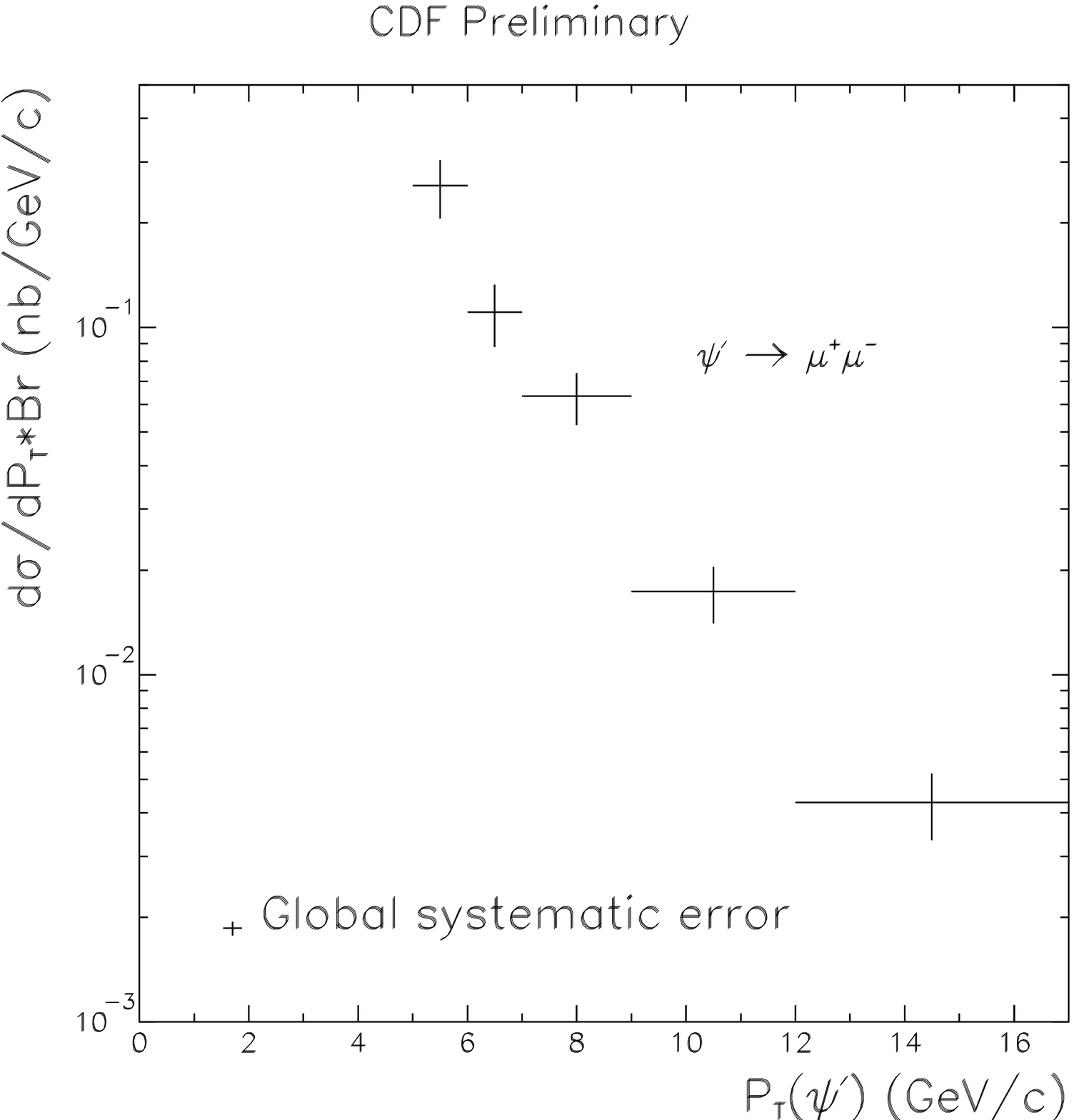}}}
\caption{$\frac{d\sigma}{dp_T}{\cal B}$ (nb/GeV) as a function of $p_T $ for $J/\psi$ (a) 
and for $\psi'$ (b).}
\label{fig:sectioneff1}
\end{figure}

\subsection{Disentangling prompt charmonia}\label{subsec:psipr}

As already seen, prompt $\psi$'s are the one that do not come from the decay of $B$ mesons.
Their production pattern has the distinctive feature compared to non-prompt one that
there exists a measurable distance between the $B$ production vertex and its decay into
charmonium.

To proceed, it is necessary to use the SVX, whose resolution is $40 \mu$m, whereas 
$B$ lifetime is $c\tau_B\approx 450\mu$m. Muons are constrained to come from the 
same point which is called the secondary vertex, as opposed to the 
primary vertex , which is the collision point
of protons. Then the projection of the distance between these two vertices on the
$\psi$ momentum, $L_{xy}$, can be evaluated. It is converted into a proper time equivalent quantity
using the formula, $ c\tau = \frac{L_{xy}}{{p_T(\psi)\over m(\psi)}\cdot F_{corr}}$, where
$F_{corr}$ is a correction factor, estimated by Monte-Carlo simulations, which links 
the $\psi$ boost factor  $\beta_T\gamma$ to this of the $B$~\cite{CDF7193}.

The prompt component of the signal is parametrised by  $c\tau=0$ (a single vertex), 
the component coming from $B$ decay is represented by an exponential whose lifetime is  
$\tau_b$ and is convoluted
with the resolution function.

The $c\tau$ distribution (cf. \cf{fig:ctau1}) is fitted in each $p_T$-bin with an unbinned 
log-likelihood function. The noise is allowed to vary within the normalisation uncertainty extracted
from the sidebands. The fraction of $\psi$ coming from $b$, $f_b(p_T)$, is displayed as a function
of $p_T$ in \cf{fig:prompt_result} (a).

\begin{figure}[h]
\centering
 \mbox{\subfigure[$J/\psi$]{\includegraphics[height=7cm]{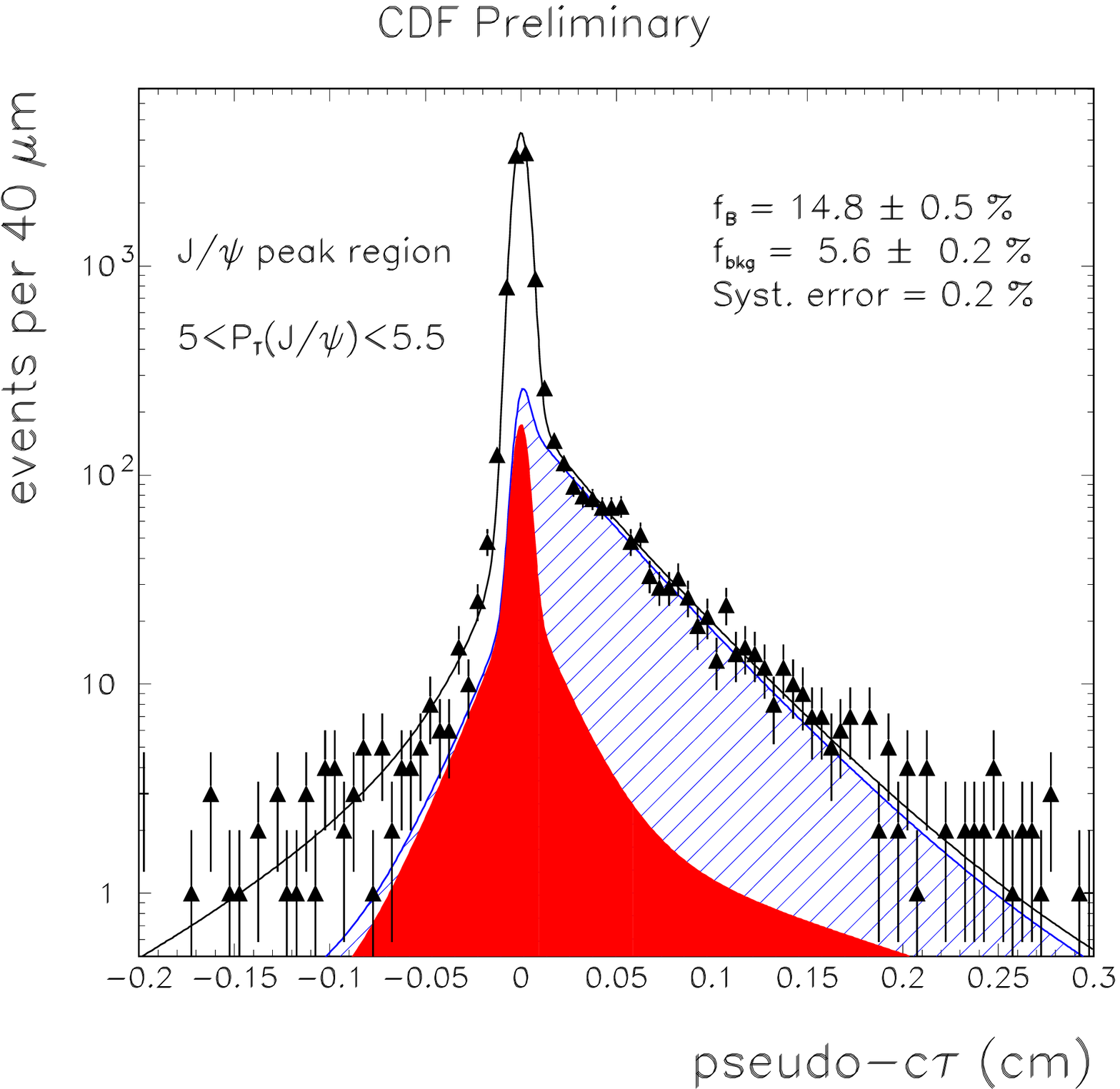}}\quad
      \subfigure[$\psi'$]{\includegraphics[height=7cm]{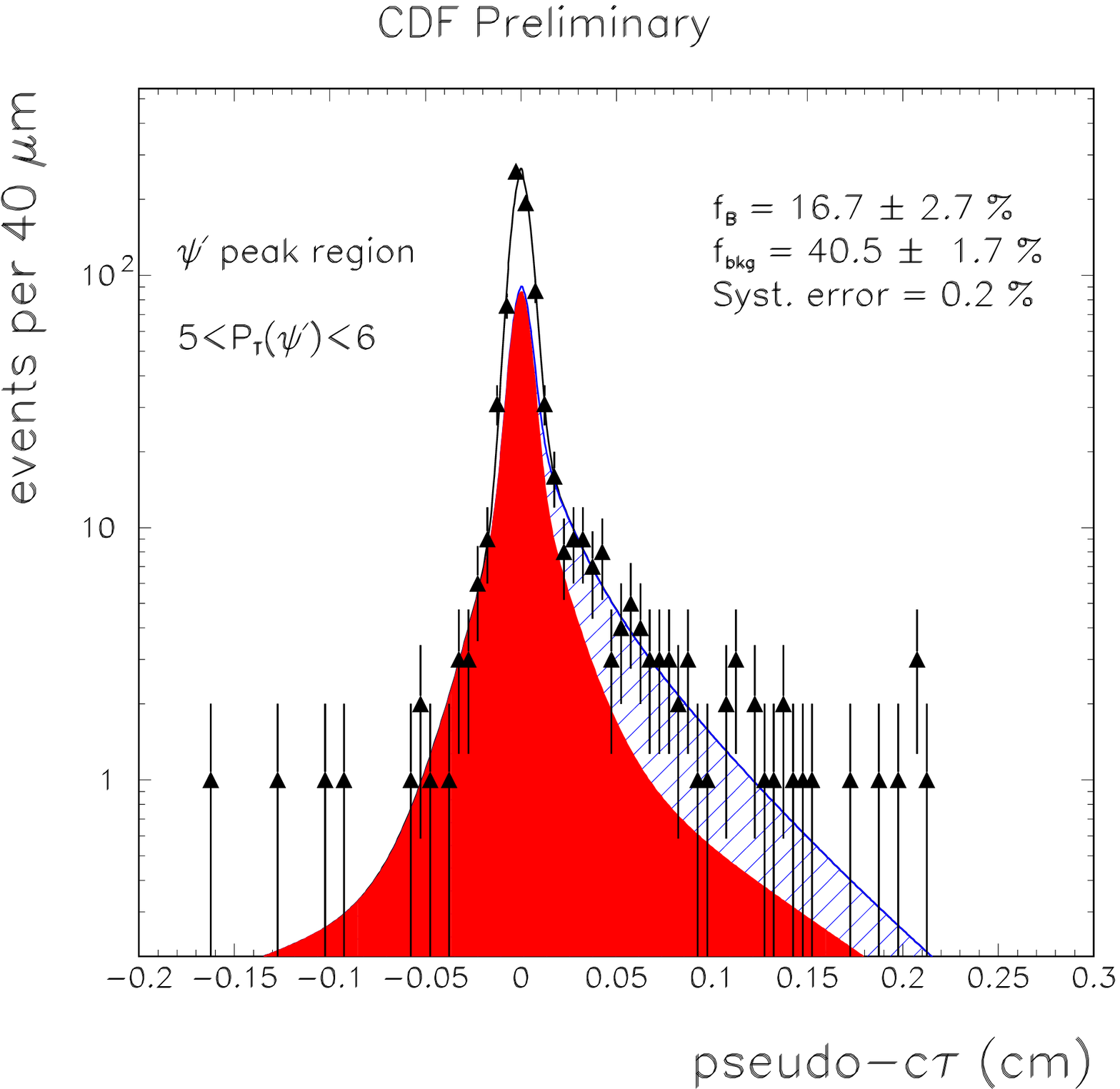}}}
\caption{$c\tau$ distribution for $J/\psi$ (a) and for $\psi'$ (b). 
The solid region is the noise contribution (of null lifetime), the white region (just below the peak)
give the null-lifetime signal: the prompt production, and finally the hashed region (mainly on 
the right side) is due to $B$ component over background.}
\label{fig:ctau1}
\end{figure}

\begin{figure}[h]
\centering
 \mbox{\subfigure[$f_b(p_T)$]{\includegraphics[height=7cm]{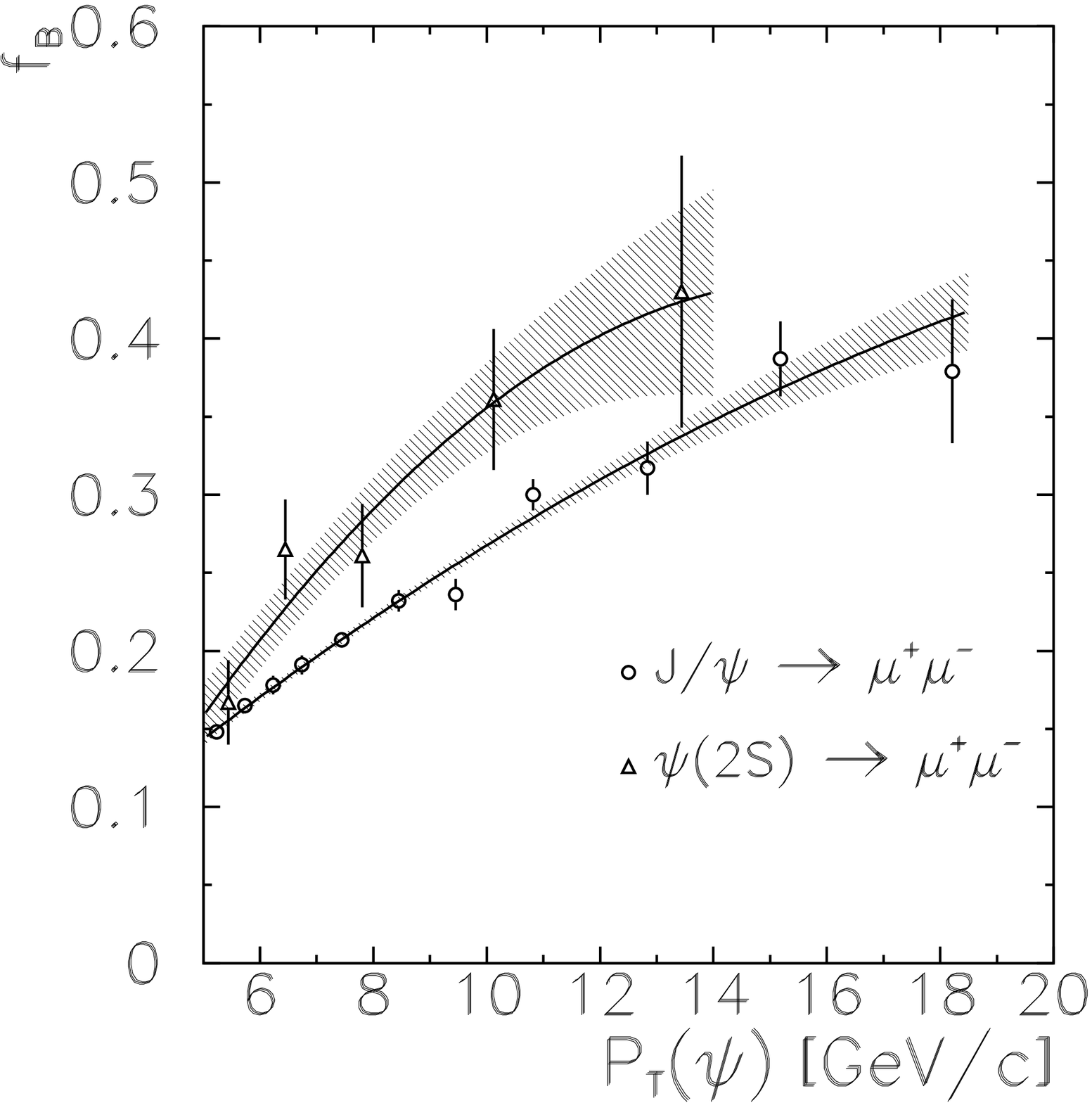}}\quad
      \subfigure[$\frac{d\sigma}{dp_T}{\cal B}$]{\includegraphics[height=7cm]
{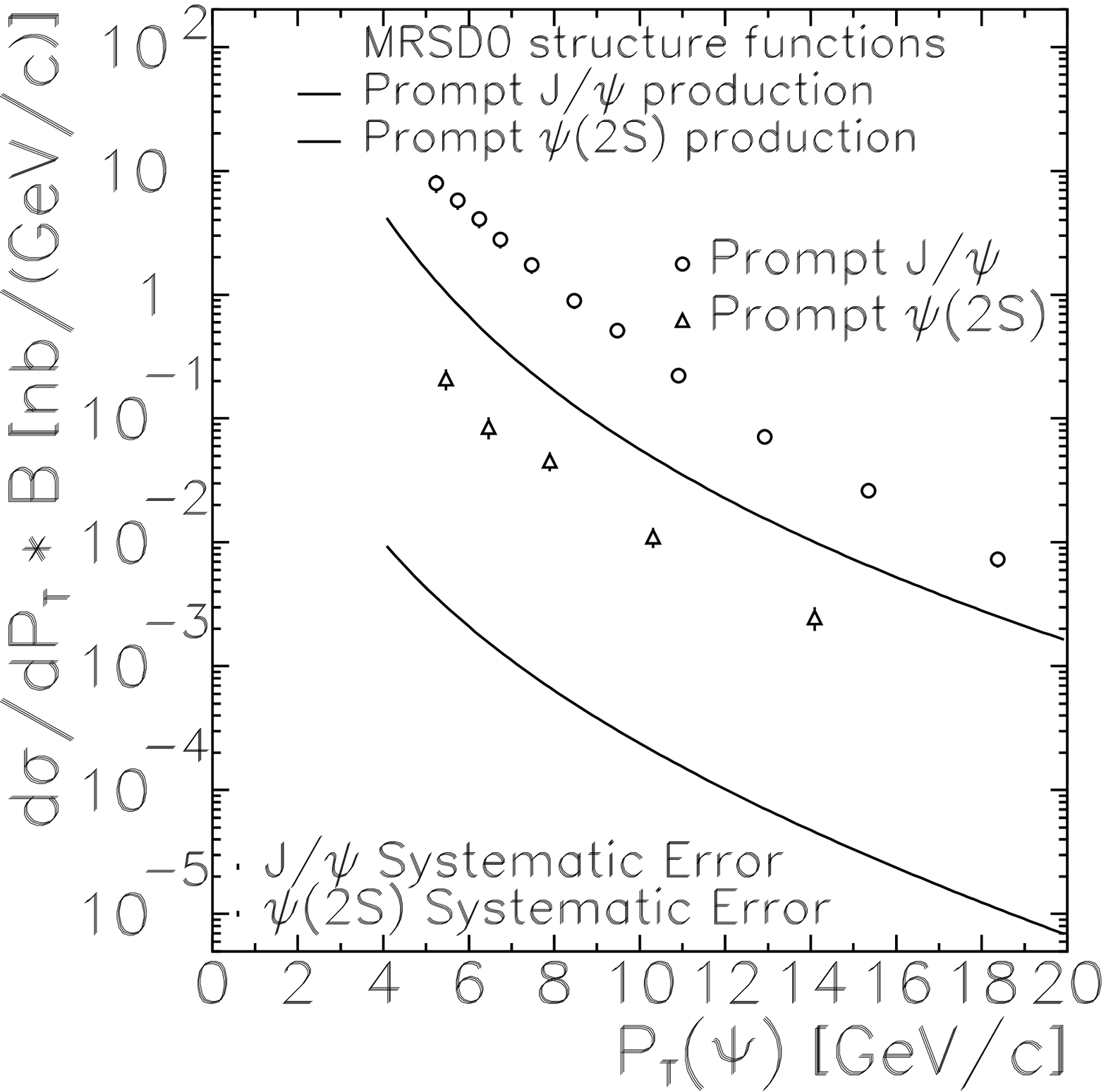}}}
 \caption{(a) Fraction of $\psi$ from $B$ decay as a function of $p_T$;
(b) $\frac{d\sigma}{dp_T}{\cal B}$ from the prompt component of $\psi$~\cite{CDF7997a}. 
The lines are the theoretical expectations based on the CSM (LO + 
fragment.)\cite{Braaten:1994xb,Cacciari:1994dr}. }
\label{fig:prompt_result}
\end{figure}

Uncertainties are evaluated as follows: a different choice of the fitting procedure implies
a mean relative variation of $\pm 0.9 \%$, this is included in the systematics on  $f_b$.
In the same fashion, a variation of the lifetime of $B$'s of 1 $\sigma$ modifies $f_b$ by $0.7 \%$.

The $\psi$ production cross section from $B$ decays is thus extracted by 
multiplying\footnote{In order to reduce statistical fluctuations, $f_b$ is fitted by a parabola
weighted by the observed shape of the cross section~\cite{CDF7193}.} $f^{fit}_b(p_T)$
by the inclusive $\psi$ cross section. Multiplying the latter by $(1-f^{fit}_b(p_T))$, 
we obviously get the prompt-production cross section (cf. \cf{fig:prompt_result} (b)).

\index{production!direct|(}

\subsection{Disentangling the direct production of $J/\psi$}
\label{subsec:psidir}

The problem here is to subtract the $J/\psi$ coming from $\chi_c$ decay, assuming 
that this is the only source of prompt $J/\psi$ besides the direct production after
subtraction of $ F(b \!\!\!/)_{\psi(2S)}^{J/\psi}$, the {\it prompt} fraction of $J/\psi$ that
come from $\psi(2S)$. The latter is obtained from the $\psi(2S)$ cross section from the previous
section and from Monte-Carlo simulation of the decays $\psi(2S)\to~J/\psi X$ where $X=\pi\pi$, $\eta$
and $\pi^0$. The delicate point here is the detection of the photon emitted during the 
radiative decay of the $\chi_c$. 

The events used in the analysis are collected by a dimuon trigger with same requirements as prior
except than the lower bound on $p_T$ is 4 GeV instead of 5 GeV.

With this selection, a sample of 34367 $J/\psi$ is then obtained from which $32642\pm 185$ 
is the number of real $J/\psi$ when the estimated background is removed. 

The requirements are as follow:

\begin{itemize}
\item an energy deposition of at least 1 GeV in a cell of the central electromagnetic calorimeter;
\item a signal in the fiducial volume of the strip chambers (CES);
\item the absence of charged particles pointing to the photon candidate cell (the no-track cut).
\end{itemize}

The direction of the photon is determined from the location of the signal in the strip chambers and from 
the event interaction point. All combinations of the $J/\psi$ with all photon candidates 
that have passed these tests are made and the invariant-mass difference defined as  
$\Delta M = M(\mu^+\mu^-\gamma)-M(\mu^+\mu^-)$
can then be evaluated and plotted (see~\cf{fig:deltaM}).

\begin{figure}[h]
\centering\includegraphics[height=12cm]{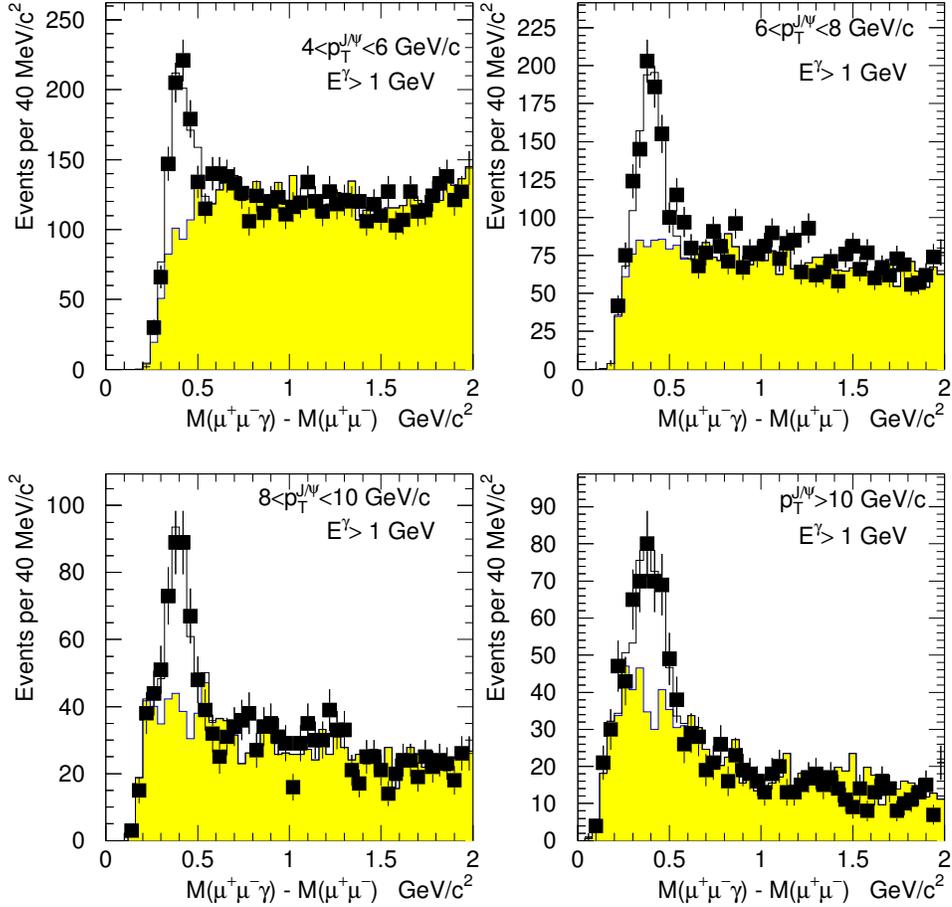}
\caption{Number of events as a function of the mass difference after the selection procedure 
for the different $p_T$-bins. The points are the data; the solid region is Monte-Carlo 
simulation of the background. The solid line is a fit resulting from the background added 
to the contribution expected from a gaussian distribution~\cite{CDF7997b}.}
\label{fig:deltaM}
\end{figure}

As expected, a clear peak from $\chi_c$ decays is visible near $\Delta M= 400$ MeV. Yet, distinct 
signals for $\chi_{c,1}$ and $\chi_{c,2}$ are not resolved as the two states are separated by 45.6~MeV 
and as the mass resolution of the detector is predicted to be respectively 50 and 55~MeV.

With regard to the background, its shape is simulated by replacing all 
charged particles (different from muons) by $\pi^0$, $\eta$ and $K_S^0$ with probabilities 
proportional to $4:2:1$. The detector response which results from the decay of these 
$\pi^0$, $\eta$ and $K_S^0$ is obtained using a Monte-Carlo simulation. The
$\Delta M$ distribution for the background is consequently obtained by applying on these simulated events 
the above-mentioned procedure as for real $\chi_c$ reconstruction. A simple test, 
which consists in trying to reproduce with this Monte-Carlo simulation the sideband signal, where 
there should not be any $\chi_c$ signal, was carried out successfully. Indeed imposing
an equal number of events, the shape of the two distribution agree quite well 
(see \cf{fig:test-bckgd-simul}).

\begin{figure}[h]
\centering\includegraphics[height=7cm]{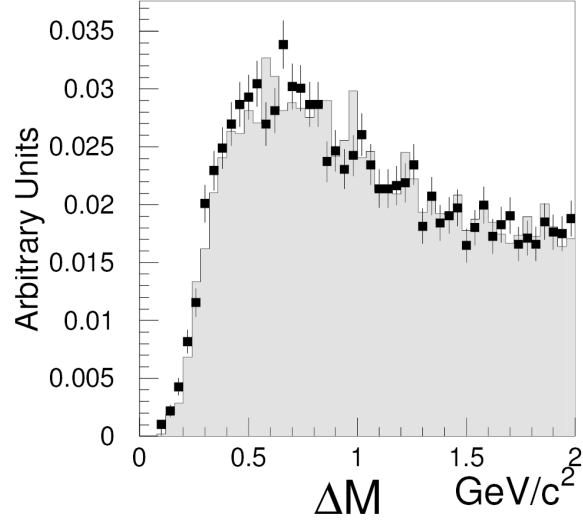}
\caption{Comparison between the $\Delta M$ distribution for dimuons in the $J/\psi$ sidebands, 
and the corresponding one predicted by the Monte-Carlo calculation; the two distributions are 
normalised to equal area and the vertical scale is arbitrary~\cite{CDF7997b}.}
\label{fig:test-bckgd-simul}
\end{figure}

Eventually, the $\Delta M$ distribution obtained from the data is fitted with a 
gaussian and with the background fixed by the procedure explained above but with a free normalisation. 
The parameter of the gaussian then lead to the number of signal events:  $1230 \pm 72$ $\chi_c$.

The analysis of the direct $J/\psi$ signal is done within four $p_T$-bins:  
$ 4 < p_T^{J/\psi}<6$ ,  $ 6 < p_T^{J/\psi} < 8$ ,  $8 < p_T^{J/\psi}< 10$ and  $  p_T^{J/\psi}> 10$ GeV. 
The proportion of  $J/\psi$ produced by $\chi_c$ decay is given by:
\begin{equation}
F_\chi^{J/\psi} = \frac{N^\chi}{N^{J/\psi}\cdot A^\gamma_{J/\psi} \cdot \epsilon^\gamma_{trk}
\cdot \epsilon^\gamma_{env}}
\end{equation}
where $ A^\gamma_{J/\psi}$ is the probability to reconstruct the photon once the $J/\psi$ is found
(or the photon acceptance),
$\epsilon^\gamma_{trk}$ is the efficiency of the no-track-cut and $\epsilon^\gamma_{env}$ the efficiency
to reconstruct the photon in the presence of additional energy deposited in the calorimeter. 
The latter quantities as well as the acceptance of the photon are evaluated with the help 
of a Monte-Carlo simulation in the spirit of
the previous section. For instance, this acceptance is $0.146 \pm 0.002(stat.)$ for
$p_T^{J/\psi} > 4.0$ GeV, whereas  $\epsilon^\gamma_{trk}$ amounts to $ 97.9^{+2}_{-5}\%$  
and $\epsilon^\gamma_{env}$ to $ 96.5^{+3.5}_{-4.5}\%$.

The total systematic uncertainty is estimated to be  $\pm 18.9\%$ and is correlated in 
the four bins~\cite{CDF7997b}.

For $p_T^{J/\psi} > 4.0$ GeV and $|\eta^{J/\psi}|<0.6$,  the fraction of $J/\psi$ from $\chi_c$ is then
\begin{equation}
F_{\chi}^{J/\psi} =27.4 \% \pm 1.6\%(stat.) \pm 5.2\% (syst.)
\end{equation}

The last step now is the disentanglement of the prompt $\chi_c$ production, that is the determination of 
$F(b \!\!\!/)_\chi^{J/\psi} $. Let us here draw the reader's attention to the fact that by
selecting prompt $J/\psi$ (cf. section~\ref{subsec:psipr}), $J/\psi$ produced by {\it non-prompt} $\chi_c$
have also been eliminated; it is thus necessary to remove only the prompt production by $\chi_c$ decay, and
nothing else. This necessitates the knowledge of $F(b \!\!\!/)_\chi^{J/\psi}$, the {\it prompt} fraction of $J/\psi$ that come from $\chi$.

The latter is calculated as follows:
\eqs{
F(b \!\!\!/)_\chi^{J/\psi}=\frac{N^\chi-N_b^\chi}{(N^{J/\psi}-N_b^{J/\psi})\cdot A^\gamma_{J/\psi} 
\cdot \epsilon^\gamma_{trk}\cdot \epsilon^\gamma_{env} }=F_\chi^{J/\psi}\frac{1-F^\chi_b}{1-F^{J/\psi}_b},
}
where $N_b^\chi$, $N_b^{J/\psi}$ are the number of reconstructed $\chi_c$ and $J/\psi$ from $b$'s, 
 $F_b^\chi$, $F_b^{J/\psi}$ are the corresponding fractions.

$F_b^{J/\psi}$ (or $f_b$) has been  extracted in the section~\ref{subsec:psipr}; $F_b^\chi$
 is obtained in same way and is $17.8 \% \pm 0.45\%$ for $p_T>4.0$ GeV. Consequently,
\begin{equation}
F(b \!\!\!/)_\chi^{J/\psi} =29.7 \% \pm 1.7\%(stat.) \pm 5.7\% (syst.)
\end{equation}
and its evolution as a function of $p_T$ is shown in \cf{fig:fraction_direct}. It is also 
found that the fraction of directly produced  $J/\psi$ is
\begin{equation}\label{eq:jpsi_dir_prod_rate}
 F_{direct}^{J/\psi}=64 \% \pm 6 \%, 
\end{equation} 
and is almost constant from 5 to 18 GeV in $p_T$. We therefore conclude that direct production is the 
principal contribution to $J/\psi$ (see \cf{fig:fraction_all} ).

\begin{figure}[h]
\centerline{\includegraphics[width=14cm]{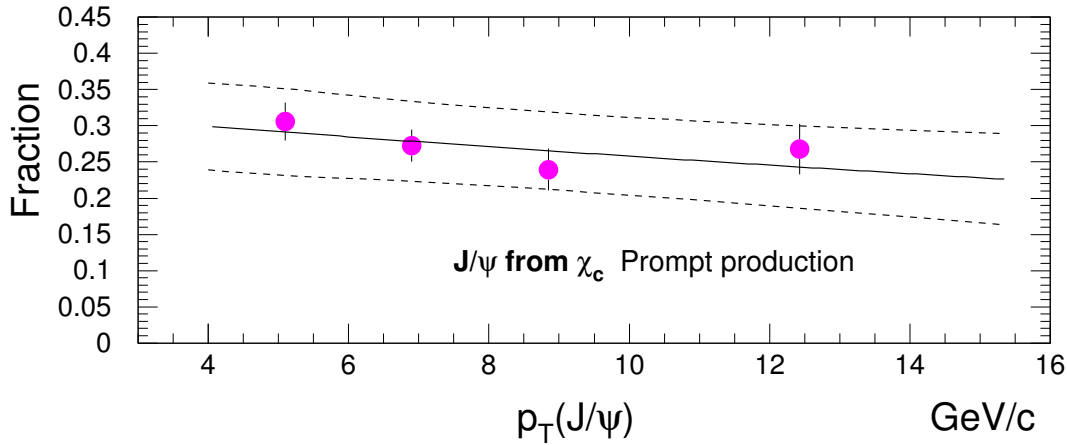}}
\caption{ The fraction of $J/\psi$ from $\chi_c$ as a function of $p_T$ with 
the contribution of $b$'s removed. The error bars correspond to statistical 
uncertainty.The solid line is the parametrisation of the fraction. The dashed 
lines show the upper and lower bounds corresponding to the statistical and systematic
uncertainties combined~\protect\cite{CDF7997b}.}
\label{fig:fraction_direct}
\end{figure}

\begin{figure}[h]
\centerline{\includegraphics[width=9.5cm]{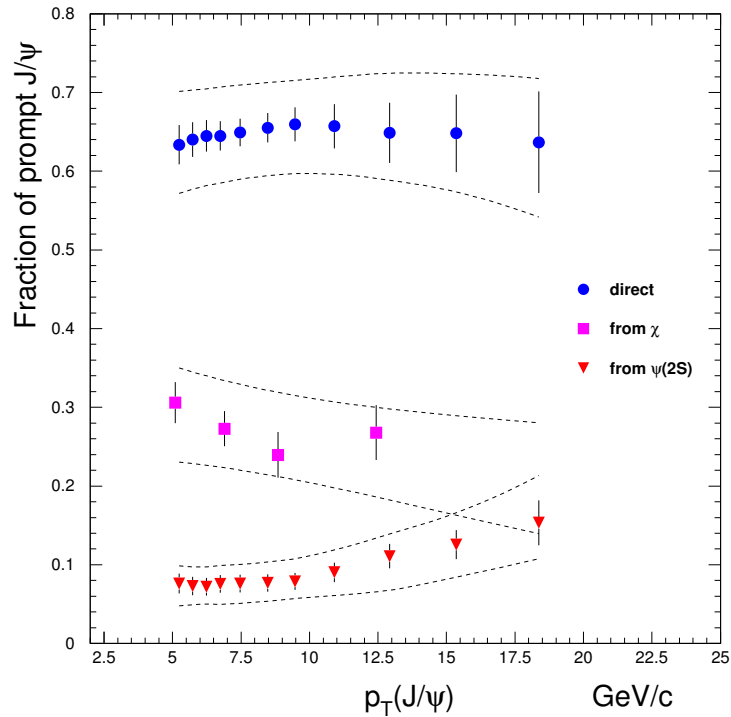}}
\caption{Fractions of $J/\psi$ with the contribution of
$b$'s removed~\protect\cite{CDF7997b}.}
\label{fig:fraction_all}
\end{figure}

In order to get the cross section of direct $J/\psi$ production, it sufficient now
to extract the contribution of $\psi'$ obtained by Monte-Carlo simulation and of $\chi_c$ obtained
by multiplying the cross section of prompt production by the factor $F(b \!\!\!/)_\chi^{J/\psi}$, 
which is a function of $p_T^{J/\psi}$. The different cross sections are displayed in~\cf{fig:direct}.

\begin{figure}[H]
\centerline{\includegraphics[width=9.5cm]{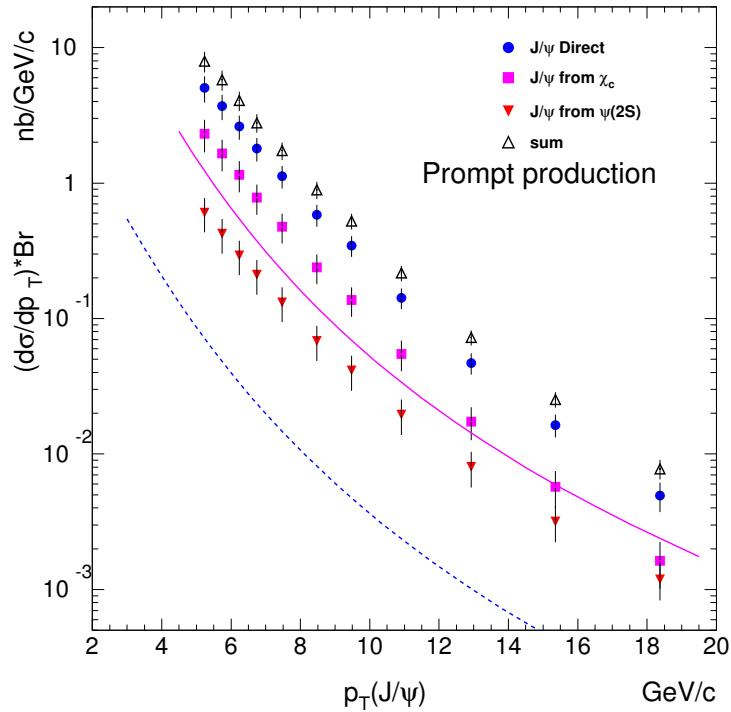}}
\caption{ Differential cross section for prompt  production of $J/\psi\to \mu^- \mu^+$ as 
a function of $p_T$. The dashed line gives the prediction of the CSM 
(LO+Fragmentation)~\cite{Braaten:1994xb,Cacciari:1994dr}  
and the  solid lines that of  the CSM + preliminary COM calculations for $J/\psi$ from 
$\chi_c$ (see~\cite{CDF7997b}), 
to be compared respectively with $\bullet$ and $\blacksquare$.}
\label{fig:direct}
\end{figure}

\index{production!direct|)}\index{CDF|)}

\section{CDF measurement of the $\Upsilon$ production cross sections}\label{sec:CDF_upsi_prod}
\index{upsi@$\Upsilon$|(}\index{CDF|(}

As  $\Upsilon$ states are similar
to $\psi$ states for which data are numerous and in contradiction with predictions, their study
is just as good a point of comparison. Furthermore, many approximations performed in the models
 are better for $b \bar b$ states. Firstly, there is no expected decay
from $t$ quarks; secondly, the scale set by the quark mass is higher
and therefore perturbation methods should be more suitable; thirdly, fragmentation would be
significant only for large $p_T$. Classical production mechanisms based on LO pQCD 
can thus be considered as the only source of production; finally, the velocity of the heavy quarks 
within the meson being smaller, relativistic effects should be smaller, rendering the predictions
of non-relativistic models more reliable.

In  this section, we shall sum up the procedure and the results on
 $\Upsilon$ production in $p\bar p$ at $\sqrt{s}=1.8$ TeV. These results were exposed 
in two letters~\cite{CDF7595,Acosta:2001gv}, and we shall mainly focus on the second one, 
which considered data collected in 1993-95 and 
corresponding to an integrated luminosity of $77 \pm 3$ pb$^{-1}$.

The parts of the detector used are the same as for the $\psi$'s, with an additional muon 
detector, called central muon upgrade (CMP) outside the CMU. It covers the same rapidity
range, \ie~$|\eta| < 0.6$, but it is located behind an additional steel shielding of
$\sim 2.4$ interaction lengths. This increases its signal purity, although some muons from the CMU
could have too few energy to reach it.

\subsection{$\Upsilon$ total production} \label{subsec:upsi_detect}

To what concerns muons, they are selected by a three level trigger:
\index{trigger}
\begin{itemize}
\item The \bfsf{level 1} requires two track segments in the CMU ( with efficiencies 
$\epsilon_1(p_T=1.5 \hbox{ GeV})=40 \%$ -- $\epsilon_1(p_T>3.0 \hbox{ GeV})=93 \%$);
\item The \bfsf{level 2} requires that at least one of the muon-track segments
be matched in $\phi$ (within 5$^\circ$) to a track found in the CTC by the central fast tracker. 
These two tracks are then considered 
to come from a single particle (the efficiency is here 
$\left<\epsilon_1(p_T>2.5 \hbox{ GeV})\right>=95 \%$);
\item The \bfsf{level 3} performs a 3D reconstruction and requires that muons pairs have
an invariant mass in the region from 8.5 to 11.5 GeV.
\end{itemize}

Besides the selection, muons are to satisfy other criteria to isolate $\Upsilon$ resonances:

\begin{itemize}
\item Identification by CMU of both muons is required as well as at least one identification
by the CMP;
\item The momentum of each muons is determined using the CTC information combined with 
the constraint that they come from the beam line (no secondary vertex possible here);
\item Only muons with $p_T>2.2$ GeV were selected;
\item Each muon-chamber track is required to match  the extrapolation of a CTC to within
3 $\sigma$ in $r-\phi$ and 3.5 $\sigma$ in $z$ ($\sigma$ is the calculated uncertainty due to
multiple scatterings and detection uncertainties, exactly as for the $\psi$'s.);
\item Muons are required to have opposite charge and the rapidity \index{rapidity} of
the reconstructed pair is to be in the CMU fiducial rapidity region $|y|<0.4$.
\item The transverse momentum of the pair is finally restricted to be between 0 and 20 GeV.
\end{itemize}

The mass distribution obtained by CDF is shown in \cf{fig:upsi-reson}.
\begin{figure}[H]
\centerline{\includegraphics[width=8cm]{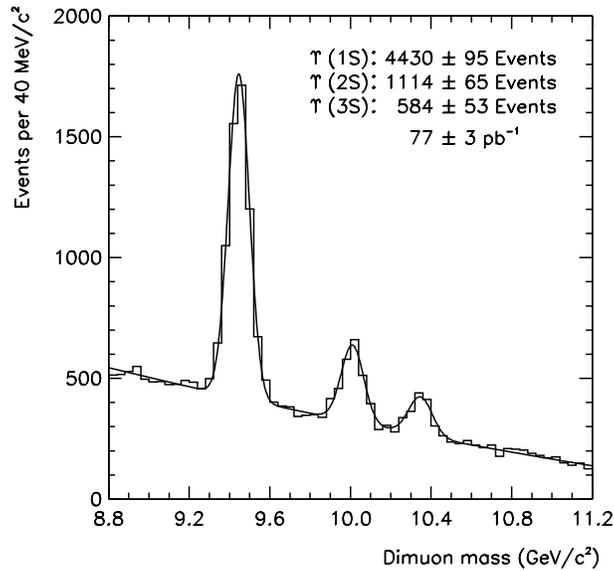}}
\caption{ Resonance peaks of  $\Upsilon(1S)$, $\Upsilon(2S)$, $\Upsilon(3S)$. The histogram represents
the data and the solid curve a gaussian fit to each resonance plus a quadratic background~\cite{Acosta:2001gv}.}
\label{fig:upsi-reson}
\end{figure}

The differential cross section times the branching ratio in dimuon is experimentally obtained as follow:
\eqs{
\left<\frac{d^2\sigma(\Upsilon)}{dp_T dy}\right>_{|y|<0.4} {\cal B}(\Upsilon\to \mu^+ \mu^-)=
\frac{N_{fit}}{A (\int {\cal L}dt)\Delta p_T \Delta y \epsilon}
}
where $N_{fit}$ is the number of $\Upsilon$ extracted from the gaussian fit 
on the signal for each $p_T$-bins similar to \cf{fig:upsi-reson}, $A$ is the 
geometric and kinematic acceptance determined by Monte-Carlo simulation, 
$\epsilon$ represents the various efficiency corrections detailed 
in~\cite{Acosta:2001gv}. $A$ is comparable 
for the three states and varies from $17\%$ to $24\%$ as a function of $p_T$.

Systematic uncertainties on $\frac{d^2\sigma}{dp_T dy}$ arise from ${\cal L}$ ($4.1\%$), 
the level 1 and 2 trigger-efficiency correction $\epsilon_{trig}$ ($4\%$) and from 
the remaining efficiency corrections ($6\%$). Uncertainties are also due to the fact
that  the $\Upsilon(2S)$ and $\Upsilon(3S)$ polarisations are unknown. They were evaluated by 
running the simulation with extreme value of polarisation for these states (longitudinal 
or transverse). For the $\Upsilon(1S)$, as the polarisation is measured (cf.
section \ref{sec:polarisation}), the limits obtained are used as extrema for this variation 
and it is seen that the uncertainties are then negligible and no systematic error from 
this source is then to be added.

\begin{figure}[H]
\centering\includegraphics[height=8.5cm]{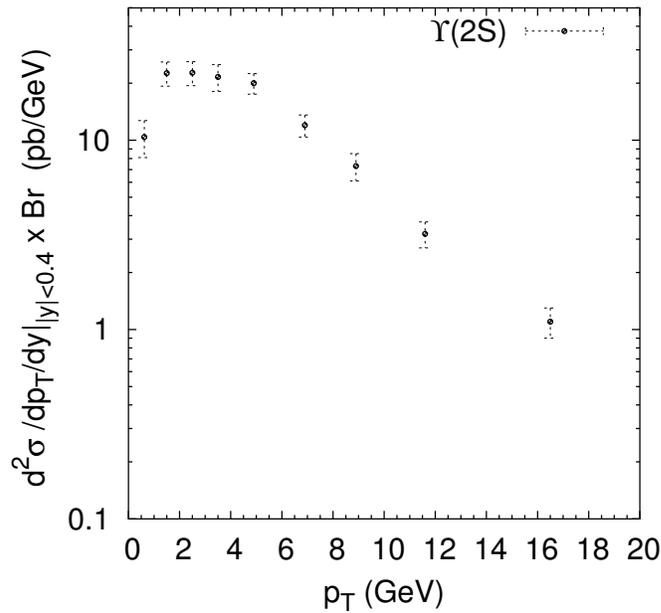}
 \caption{Differential cross section of  $\Upsilon(1S)\to \mu^- \mu^+$ as a function of $p_T$ for 
$|y|<0.4$.}
\label{fig:dsdpt_upsi1s}
\end{figure}

\begin{figure}[H]
\centering\includegraphics[height=8.5cm]{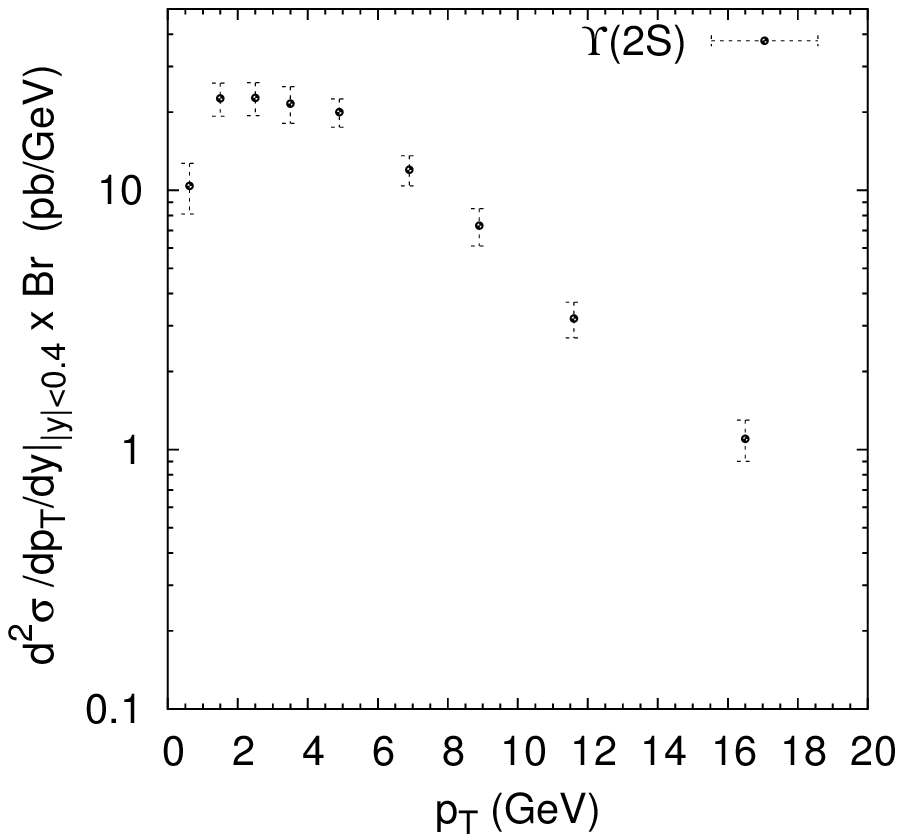}
 \caption{Differential cross section of  $\Upsilon(1S)\to \mu^- \mu^+$ as a function of $p_T$ for 
$|y|<0.4$.}
\label{fig:dsdpt_upsi2s}
\end{figure}

\begin{figure}[H]
\centering\includegraphics[height=8.5cm]{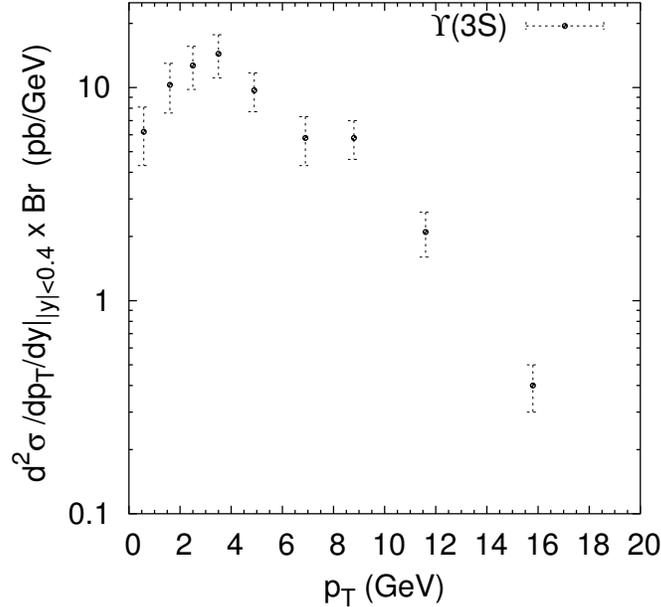}
 \caption{Differential cross section of  $\Upsilon(1S)\to \mu^- \mu^+$ as a function of $p_T$ for 
$|y|<0.4$.}
\label{fig:dsdpt_upsi3s}
\end{figure}

\subsection{Disentangling direct production of $\Upsilon(1S)$}\label{subsec:upsidir}

The analysis~\cite{Affolder:1999wm} presented here is for the most part the same as 
described in section~\ref{subsec:psidir}.
It is based on 90 pb$^{-1}$ of data collected during the 1994-1995 run. The measurement 
has been constrained to the range $p_T>8.0$ GeV because the energy of the photon emitted during the decay 
of $\chi_b$ decreases with $p_T$ and ends up to be too small for the photon to be detected
properly. In the same spirit, analysis relative to $\Upsilon(2S)$ has not been carried out, once
again because of the lower energy of the radiative decay. Concerning $\Upsilon(3S)$, except for the
unobserved $\chi_b(3P)$ (which is nevertheless supposed to be below the $B\bar B$ threshold),
no $P$-states are supposed to be parents. 

The event used  are collected by a dimuon trigger (details are given in~\cite{Affolder:1999wm}) 
with the following requirements, which slightly differ from section~\ref{subsec:upsi_detect}:

\begin{itemize}

\item Two oppositely charged muons candidates observed in the muon chambers,
\item Both with $p_T > 2.0$ GeV,
\item The pair must have $p_T > 8.0$ GeV and $|\eta|<0.7$,
\item In order to be considered as coming from a $\Upsilon(1S)$, the invariant mass of the pair must lie within the region: 
\eqs{9300 \hbox { MeV} < M(pair) < 9600 \hbox { MeV}.}  
\end{itemize}

With this selection, a sample of 2186 $\Upsilon(1S)$ candidates is then obtained from which $1462\pm 55$ 
is the estimated number of $\Upsilon(1S)$ after subtraction of the background. 
The resulting mass distribution is plotted 
in~\cf{fig:upsi_from_chi_mass_inv}. 

In this sample, photons likely to come from $\chi_b$ decay are selected as for the $J/\psi$ case, except
for the energy deposition in the central electromagnetic calorimeter, which is lowered to 0.7 GeV.

\begin{figure}[h]
\centering\includegraphics[height=9cm]{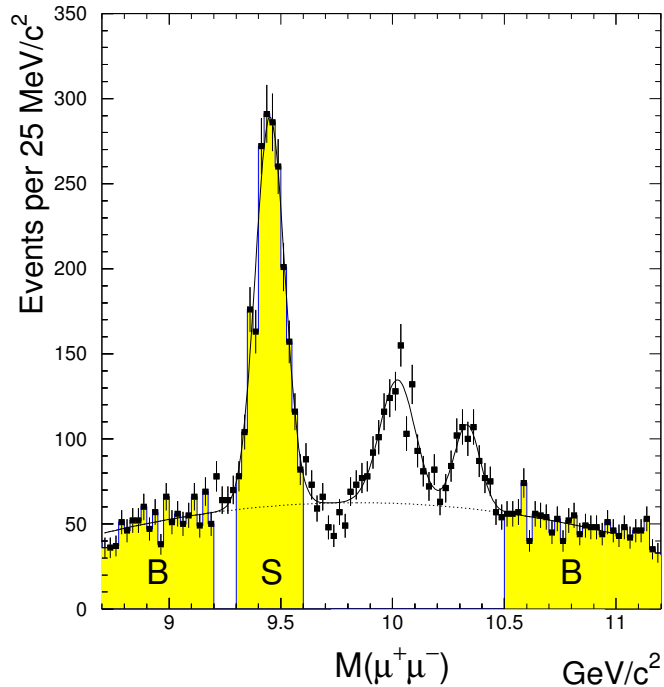}
\caption{Muon pairs invariant mass distribution after the selection for $p_T > 8.0$~GeV. The dotted
line is the fit to the background, the solid line is the fit to the data; the region S is the signal
region and the region B is the sideband region.}
\label{fig:upsi_from_chi_mass_inv}
\end{figure}

The direction of the photon is determined from the location of the signal in the strip chambers and from 
the event-interaction point. All combinations of the $\Upsilon(1S)$ with all photon candidates 
that have passed these tests are made and the invariant-mass difference still defined 
by $\Delta M = M(\mu^+\mu^-\gamma)-M(\mu^+\mu^-)$
can then be evaluated. It is plotted in~\cf{fig:upsi_from_chi_deltaM}.

\begin{figure}[h]
\centering\includegraphics[height=8cm]{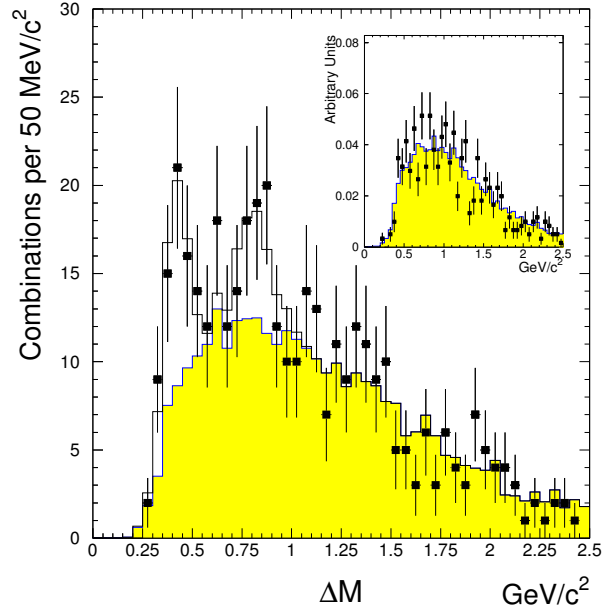}
\caption{Number of events as a function of the mass difference after the selection procedure. 
The points are the data; the solid region is a Monte-Carlo 
simulation of the background. The solid line is a fit resulting from the background added 
to the contribution expected from two gaussian distributions~\cite{Affolder:1999wm}. The inset
shows the comparison between the background of $\Delta M$ extracted from the sidebands and with
the Monte-Carlo simulation.}
\label{fig:upsi_from_chi_deltaM}
\end{figure}

The background is evaluated as in the $J/\psi$ analysis. A fit to the data then yields the number
of signal events: $35.3 \pm 9.0$ for $\chi_b(1P)$ and $28.5 \pm 12.0$ for $\chi_b(2P)$. 
The analysis of the direct $\Upsilon(1S)$ 
signal is done within one single $p_T$-bin, as opposed to 
$J/\psi$ case. The proportion of  $\Upsilon(1S)$ produced by $\chi_b$ decay is given by:
\begin{equation}
F_\chi^{J/\psi} = \frac{N^{\chi_b}}{N^{\Upsilon}\cdot A^\gamma_{\Upsilon} \cdot \epsilon^\gamma},
\end{equation}
where $N^{\chi_b}$ and $N^{\Upsilon}$ are the numbers of reconstructed 
$\chi_b$ and $\Upsilon(1S)$ respectively, $ A^\gamma_{\Upsilon}$ is the probability to 
reconstruct the photon once the $\Upsilon$ is found,
$\epsilon^\gamma$ is the efficiency of the isolation cuts. The latter quantities, as well as 
the acceptance of the photon, are evaluated with the help of simulations in the same spirit as that 
described in the previous section. For instance, this acceptance is $0.142 \pm 0.004 (stat.)$ for
$\chi_b(1P)$ and  $0.284 \pm 0.006 (stat.)$ for $\chi_b(2P)$, the difference between the two
coming mainly from the lower energy of the photon for $\chi_b(2P)$ decays. The efficiencies 
$\epsilon^\gamma$ obtained are $0.627 \pm 0.013(stat.)$ for $\chi_b(1P)$ and  
$0.651 \pm 0.0013(stat.)$ for $\chi_b(2P)$. The efficiency for inclusive $\Upsilon(1S)$ is
supposed to be the same as for the sub-sample considered.

The systematic uncertainties on $ F_{\chi_b}^{\Upsilon(1S)} $ associated with the 
variation of the shape of the $p_T$ spectrum and of the decay angular distribution 
is $\pm 13 \%$ for $\chi_b(1P)$ and  
$\pm 9 \%$ for $\chi_b(2P)$. The uncertainty from $N^{\chi_b}$ is $\pm 7 \%$ for $\chi_b(1P)$ and  
$\pm 9 \%$ for $\chi_b(2P)$, this comes from the background 
of the $\Delta M$ distribution and, from the uncertainty on the resolution 
of the gaussian signals in this 
same distribution. Other uncertainties are also considered, see \cite{Affolder:1999wm} for more
details. Assuming the independence of all these uncertainties, they are combined 
in a total systematic uncertainty:~$\pm 16.4 \%$ for $\chi_b(1P)$ and  
$\pm 13.7 \%$ for $\chi_b(2P)$.

For $p_T^{\Upsilon(1S)} > 8.0$ GeV and $|\eta^{\Upsilon(1S)}|<0.7$,  
the fractions of $\Upsilon(1S)$ from $\chi_b(1P)$ and  from  $\chi_b(2P)$ are 
\eqs{
F_{\chi_b(1P)}^{\Upsilon(1S)} &=27.1 \% \pm 6.9\%(stat.) \pm 4.4\% (syst.),\\
F_{\chi_b(2P)}^{\Upsilon(1S)} &=10.5 \% \pm 4.4\%(stat.) \pm 1.4\% (syst.).
}

In order to get the direct production of $\Upsilon(1S)$, one must still evaluate
the feed-down from the $S$-waves $\Upsilon(2S)$ and $\Upsilon(3S)$. This is done using
Monte-Carlo simulations of these decays normalised to the production cross sections
discussed in section~\ref{sec:CDF_upsi_prod}. It is found that for $p_T^{\Upsilon(1S)} > 8.0$ GeV
the fraction of $\Upsilon(1S)$ from $\Upsilon(2S)$ and $\Upsilon(3S)$ are
respectively
\eqs{
F_{\Upsilon(2S)}^{\Upsilon(1S)} &=10.1 \%^{+7.7}_{-4.8}\%,\\
F_{\Upsilon(3S)}^{\Upsilon(1S)} &=0.8 \%^{+0.6}_{-0.4}\%.
}

Concerning the unobserved $\chi_b(3P)$, a maximal additional contribution 
is taken into account by supposing than all the $\Upsilon(3S)$ are due to 
$\chi_b(3P)$ and from theoretical expectation for the  decay of this state, 
a relative rate of $\Upsilon(1S)$ from $\chi_b(3P)$ can obtained. This rate
is less than $6\%$.

In conclusion, the fraction of directly produced $\Upsilon(1S)$ is obtained from
\eqs{
F_{direct}^{\Upsilon(1S)} =1- F_{\chi_b(1P)}^{\Upsilon(1S)}-F_{\chi_b(2P)}^{\Upsilon(1S)}-
F_{\Upsilon(2S)}^{\Upsilon(1S)}-F_{\Upsilon(3S)}^{\Upsilon(1S)},
}
for which the uncertainties come from that of cross sections and of 
branching ratio. It is eventually found that 
\eqs{
F_{direct}^{\Upsilon(1S)} &=50.9 \% \pm 8.2\%(stat.) \pm 9.0\% (syst.).
}
\index{upsi@$\Upsilon$|)}

%***************************************************************************
%***************************************************************************
%***************************************************************************
%***************************************************************************
%                            POLARISATION
%***************************************************************************
%***************************************************************************
%***************************************************************************
%***************************************************************************

\section{Polarisation at the Tevatron}
\label{sec:polarisation} \index{polarisation|(}\index{spin-alignment|see{polarisation}}

As the considered bound states, $\psi$ and $\Upsilon$ are massive spin-1 particles, 
they have three polarisations. In addition to measurements of their 
momentum after their production, the experimental set-up is sufficiently refined
to provide us with a measurement of their spin alignment through an analysis of 
the angular distribution of the dimuon pairs from the decay.

The CDF collaboration has carried out two analyses, one for the $\psi$ 
states~\cite{Affolder:2000nn} -- for which the feed-down
from $b$ decay has been subtracted, but not the indirect component in the case of the 
$J/\psi$ -- and another for $\Upsilon(nS)$~\cite{Acosta:2001gv}.

In the following, we shall proceed to a brief outline of the analysis 
and give its main results.

\subsection{Polarisation parameters}

The polarisation state of the quarkonium can be deduced from the angular dependence of
its decay into $\mu^{+}\mu^{-}$. Taking the spin quantisation axis along  the quarkonium 
momentum direction in the $p\bar{p}$ c.m. frame, we define $\theta$ as the angle  
between the $\mu^{+}$  direction in the quarkonium rest frame  
and the quarkonium  direction in the lab frame (see \cf{fig:pol_def_theta}). 
Then the normalised angular distribution $I(\cos \theta)$ is 
given\footnote{For a derivation, see the Appendix A of~\cite{Cropp:2000zk}.} by
\eqs{
  I(\cos \theta) = \frac{3}{2(\alpha+3)}(1+\alpha \ \cos^2\theta)
 \label{def_alpha}
}
where the interesting quantity is 
\eqs{
\alpha=\frac{\frac{1}{2}\sigma_T-\sigma_L}{\frac{1}{2}\sigma_T+\sigma_L}.
}
$\alpha=0$ means that the mesons are unpolarised, $\alpha=+1$ corresponds to
a full transverse polarisation and $\alpha=-1$ to a longitudinal one.

As the expected behaviour is biased by muons cuts -- for instance there exists 
a severe reduction of the acceptance  as $\theta$ approaches 0 and 180 degrees, due 
to the $p_T$ cuts on the muons --, the method followed by CDF was to
compare measurements, not with a possible $(1+\alpha \ \cos^2\theta)$ distribution, 
but with distributions obtained after simulations of quarkonium decays taking account  
the geometric and kinematic acceptance of the detector as well as the reconstruction efficiency.

\begin{figure}[H]
\centerline{\includegraphics[width=5cm]{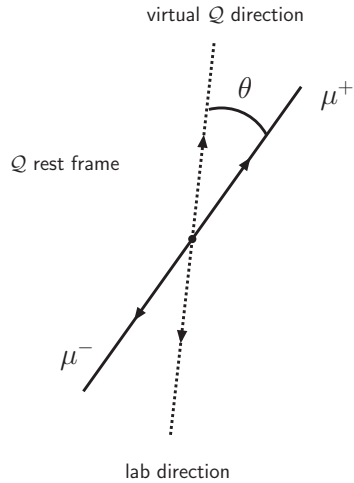}}
\caption{ Definition of the angle $\theta$ used in the polarisation analysis of a quarkonium
${\cal Q}$.}
\label{fig:pol_def_theta}
\end{figure}

The parts of the detector used are the same as before, with the additional central muon upgrade (CMP) 
outside the CMU. 

\subsection{Study of the $\psi$'s} \label{subsec:psi_detect_pol}

We give here the results relative to the CDF analysis published in 2000~\cite{Affolder:2000nn}.
The data used correspond to an integrated luminosity of 110 pb$^{-1}$ collected
between 1992 and 1995. The three-level trigger is similar as before. Beside this on-line selection, 
muons are to satisfy other criteria to isolate $\psi$ resonances:

\begin{itemize}
\item Identification by CMU of both muons is required as well as a matching of the
corresponding tracks within the CMU with one of the CTC tracks. This reduces the rate from 
sources such as $\pi/K$ meson decay-in-flight;
\item These matches are required to pass a maximum $\chi^2$ cut of 9 and
 12 (for 1 degree of freedom) in the $\phi$ and $z$ views respectively~\cite{Affolder:2000nn};
\item Only muons with $p_T>2$ GeV were selected;
\end{itemize}

\subsubsection{Disentangling prompt production}

The method followed here is on many aspects similar to the one exposed in section~\ref{subsec:psipr}.
As before, prompt events have $c\tau$ consistent with zero whereas $b$-quark decays have
 an exponential $c\tau$ distribution, all this with a smearing due to  the detector resolution.
The $c\tau$  distribution is fitted to obtain the relative
fractions of prompt and $b$-decay production.
 Results are that the measured fraction of $J/\psi$ mesons which come from $b$-hadron decay,
$F^{J/\psi}_b$,  increases from $(13.0\pm 0.3)\%$ at $p_T = 4 $ GeV to 
$(40\pm 2)\%$ at 20 GeV and for $\psi'$ mesons, 
$F^{\psi'}_b$ is  $(21\pm2)\%$ at 5.5 GeV and $(35\pm4)\%$  at  20 GeV.

The following study is then based on two samples
a short-lived one dominated by prompt production, and a long-lived one dominated by $b$-decays. 
The definitions used to separate the samples are the following  $-0.1\leq c\tau \leq 0.013[0.01]$ cm
for the  short-lived sample, and for the long-lived sample $0.013[0.01]\leq c\tau \leq 0.3$ cm, 
for the $J/\psi$ $[\psi']$ analyses respectively\footnote{The boundary between the two $c\tau$ regions
has been determined separately for the $J/\psi$ and for the $\psi'$.
Depending on $p_T^{\psi}$, the prompt fraction in the short-lived sample ranges from 85\%[86\%]
 to 96\%[95\%],and the $B$-decay fraction in the long-lived sample ranges from 83\%[86\%] to
98\%[91\%], for $J/\psi$ $[\psi']$ respectively.}.

Within a 3-standard-deviation mass window around the $J/\psi$ peak, 
the data sample is of 180 000 $J/\psi$ events. In order to study the effect of $p_T$, 
the data are divided into seven $p_T$-bins from 4 to 20 GeV.
Because the number of $\psi'$  events is lower, data for $\psi'$  
are divided into three  $p_T$-bins from 5.5 to 20 GeV.

The $J/\psi$ sample 
is divided into three sub-samples where the analysis is executed independently: 
the short-lived and long-lived SVX samples described above, and a third sample 
(the CTC sample) in which neither muon has SVX information and no 
$c\tau$ measurement is made. 

For $\psi'$, both muons are required to be reconstructed in the SVX and the number
of signal events is found to $1855\pm65$ in a 3-standard-deviation mass window
around the $\psi'$ mass. The sample is further divided
into two sub-samples based on the $c\tau$ distribution instead of 3 for $J/\psi$. 
Once again, because the number of events is lower than in the $J/\psi$ case,
10 bins are used in $|\cos \theta |$ instead of 20 in $\cos \theta $ for $J/\psi$.
Similarly to the latter, the number of events in each bin is obtained by fitting
the mass distribution each time.

\subsubsection{$J/\psi$ polarisation measurement}

The $J/\psi$ polarisation is determined through a fit of  the measured muonic
$\cos \theta$ distributions to a set of Monte-Carlo templates. 
%\cite{rjcthesis}
Implementing the experimental constraints with  detector and trigger simulations, 
these templates are generated by processing simulated samples of 
$J/\psi \rightarrow \mu^+ \mu^-$ decays. 

\begin{figure}[H]
\centerline{\includegraphics[width=14cm]{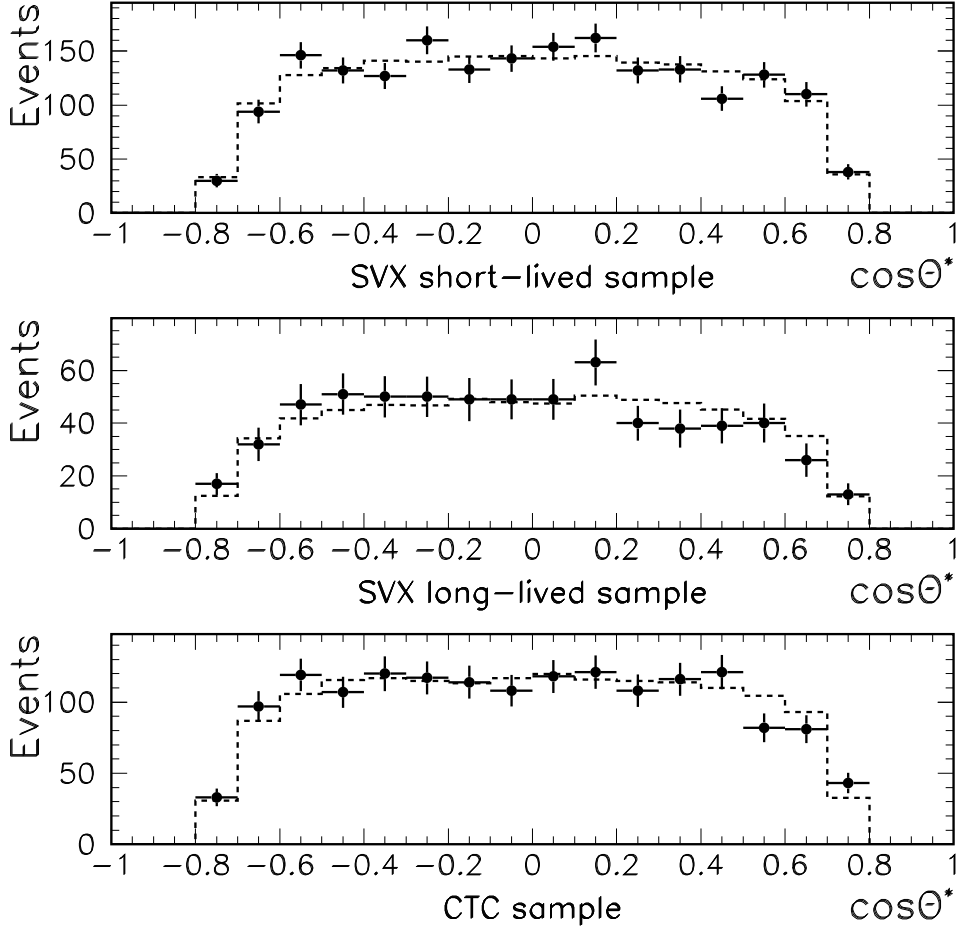}}
\caption{The points represent the data (with sideband signal subtracted); the dashed
lines is the fit of the $\cos \theta$  distributions in
 the 12-15 GeV bin. The three plots correspond to the three samples defined in the text.}
\label{fig:det_costheta_jpsi}
\end{figure}

The polarisation is obtained using
a $\chi^2$ fit of the data to a weighted sum of transversely 
polarised and longitudinally polarised templates. The weight obtained with the fit
provides us with the polarisation. Explanations relative to
 procedure used can be found in \cite{Affolder:2000nn} and \cite{Cropp:2000zk}. 

Note however that the polarisation is measured in each $p_T$ bin and that separate polarisation 
measurements for direct $J/\psi$ production and for production via $\chi_c$ and 
$\psi'$ decays was found to be unfeasible. Let us recall here that $\chi_c$ and 
$\psi'$ were shown to account for 36$\pm$6\% of the prompt production 
(cf.~\ce{eq:jpsi_dir_prod_rate}) and to be mostly constant in the considered $p_T$ range.

For  $J/\psi$, the fit is performed simultaneously on
 the SVX short-lived, SVX long-lived and CTC samples, with two
fit parameters: $\alpha_{Prompt}$ and $\alpha_{from B}$. As observed above,
the long-lived and short-lived samples are not pure, this is of course taken into
account.  The $B$-decay fraction in the CTC sample is assumed to be the same as in
 the SVX sample. This seems reasonable as the main difference
between the two samples resides primarily in the $z$ position of the primary vertex. 
The fit in the $p_T$ range 
12-15 GeV  is shown in \cf{fig:det_costheta_jpsi}.

Except in the lowest $p_T$ bins, the systematic uncertainties are much smaller than 
the statistical one. The values obtained for $\alpha_{Prompt}$ and $\alpha_{from B}$ 
are given in \ct{tab:res_pol_psi1s} and plotted with a comparison to NRQCD \index{NRQCD}
predictions in \cf{fig:evol_alpha_jpsi}.

\begin{table}[H]
\begin{center}
\begin{tabular}{|c|c|c|c|} \hline
$p_T$ bin (GeV) & Mean $P_T$  (GeV) & $\alpha_{Prompt}$ & $\alpha_{from B}$ \\ \hline \hline
$4-5$   & 4.5  & $0.30\pm0.12\pm0.12$    & $-0.49\pm0.41\pm0.13$   \\ 
$5-6$   & 5.5  & $0.01\pm0.10\pm0.07$    & $-0.18\pm0.33\pm0.07$  \\ 
$6-8$   & 6.9  & $0.178\pm0.072\pm0.036$ & $0.10\pm0.20\pm0.04$  \\
$8-10$  & 8.8  & $0.323\pm0.094\pm0.019$ & $-0.06\pm0.20\pm0.02$  \\
$10-12$ & 10.8 & $0.26\pm0.14\pm0.02$    & $-0.19\pm0.23\pm0.02$  \\
$12-15$ & 13.2 & $0.11\pm0.17\pm0.01$    & $0.11\pm^{0.31}_{0.28}\pm0.02$  \\
$15-20$ & 16.7 & $-0.29\pm0.23\pm0.03$   & $-0.16\pm^{0.38}_{0.33}\pm0.05$  \\ \hline
\end{tabular}
\caption{Fit results for $J/\psi$ polarisation, with
statistical and systematic uncertainties.}
\label{tab:res_pol_psi1s}
\end{center}
\end{table}

\begin{figure}[H]
  \centerline{\includegraphics[width=16cm]{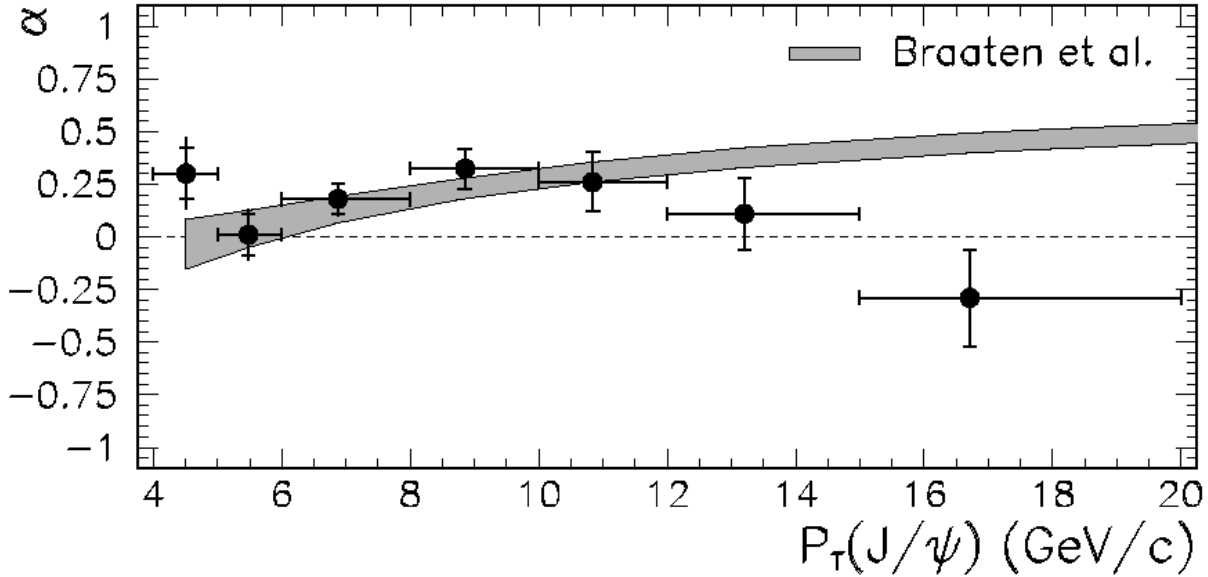}}
\caption{$\alpha_{Prompt}$ for  $J/\psi$ fitted  for $|y^{J/\psi}| < 0.6$.
 Full error bars denote statistical and systematic uncertainties added in quadrature; ticks
denote statistical errors alone.
 The shaded band shows a  NRQCD-factorisation prediction \protect\cite{Braaten:1999qk} which 
includes the contribution from $\chi_c$ and $\psi'$ decays.  }
\label{fig:evol_alpha_jpsi}
\end{figure}

\subsubsection{$\psi'$ polarisation measurement}

The procedure to obtain $\alpha_{Prompt}$ and $\alpha_{from B}$ is similar. Weighted
simulations of the angular distribution are fitted to the data and the weight obtained gives the 
polarisation. The $|\cos \theta |$ distributions in the two $c\tau$ sub-samples 
are fitted simultaneously. As  there is no expected radiative decay from 
higher excited charmonia, we are in fact dealing with direct production.  
The fit of the angular distribution in the three $p_T$-bins is shown on \cf{fig:det_costheta_psi2s}.

\begin{figure}[H]
\centerline{\includegraphics[width=14cm]{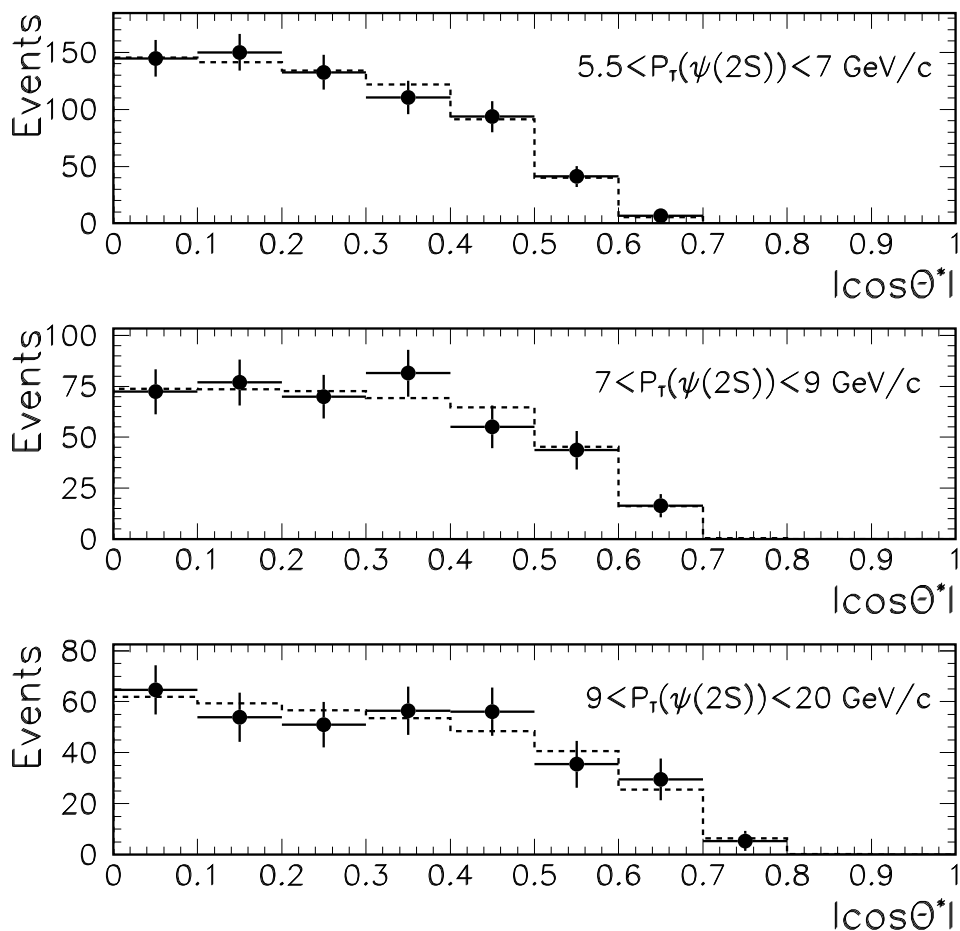}}
\caption{The points represent the $\psi'$ data for the short-lived sample only; the dashed
lines is the fit to the $|\cos \theta|$  distributions. The three plots correspond to the 
three samples in $p_T$ defined in the text. Note the extension of the acceptance 
range in $|\cos \theta|$ as $p_T$ increases.}
\label{fig:det_costheta_psi2s}
\end{figure}

Anew, the systematic uncertainties~\cite{Affolder:2000nn} are much smaller than 
the statistical uncertainties.  The values obtained for $\alpha_{Prompt}$ and $\alpha_{from B}$ 
are given in \ct{tab:res_pol_psi2s} and plotted with a comparison to NRQCD \index{NRQCD}
predictions in \cf{fig:evol_alpha_psi2s}.

\begin{figure}[H]
  \centerline{\includegraphics[width=16cm]{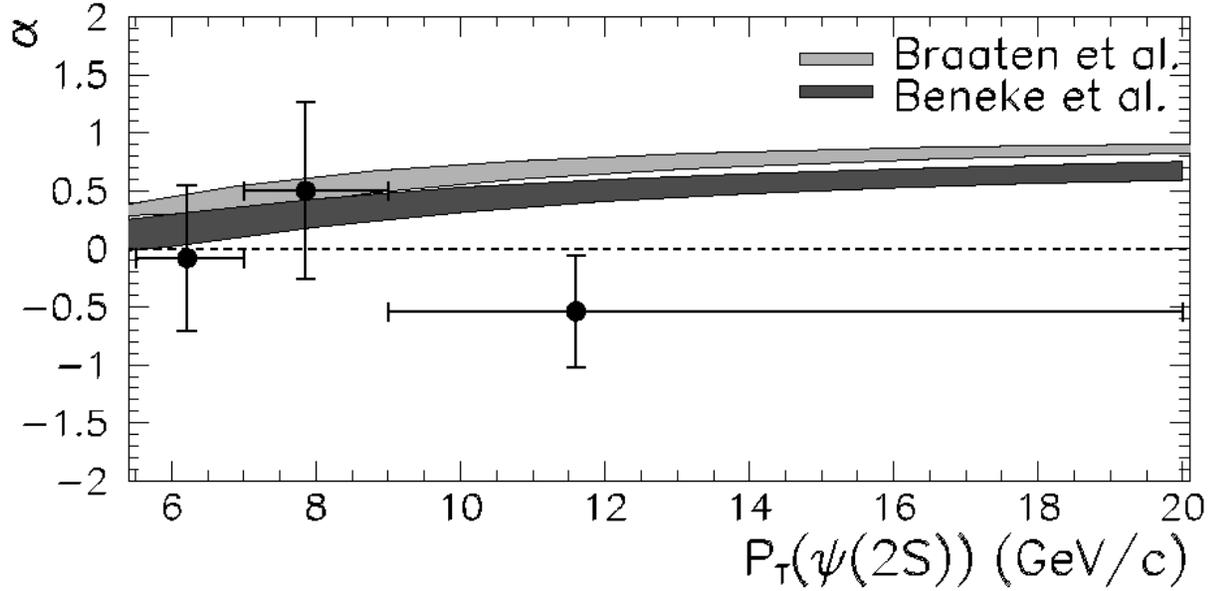}}
\caption{$\alpha_{Prompt}$ for  $\psi'$ fitted  for $|y^{\psi'}| < 0.6$.
 Error bars denote statistical and systematic uncertainties added in quadrature.
Shaded bands show two NRQCD-factorisation predictions \protect\cite{Beneke:1996yw,Braaten:1999qk}. }
\label{fig:evol_alpha_psi2s}
\end{figure}

\begin{table}[H]
\begin{center}
\begin{tabular}{|c|c|c|c|} 
\hline
$p_T$ bin (GeV) & Mean $p_T$  (GeV) & $\alpha_{Prompt}$ & $\alpha_{from B}$ \\ \hline\hline
$5.5-7.0$  & 6.2  & $-0.08\pm0.63\pm0.02$ & $-0.26\pm1.26\pm0.04$ \\
$7.0-9.0$  & 7.9  & $ 0.50\pm0.76\pm0.04$ & $-1.68\pm0.55\pm0.12$ \\
$9.0-20.0$ & 11.6 & $-0.54\pm0.48\pm0.04$ & $ 0.27\pm0.81\pm0.06$ \\ \hline
\end{tabular}
\caption{Fit results for $\psi(2S)$ polarisation, with
statistical and systematic uncertainties.}
\label{tab:res_pol_psi2s}
\end{center}
\end{table}

\subsection{Study of the $\Upsilon(1S)$} \label{subsec:upsi_detect_pol}

As the procedure is quite similar as of the $\psi$ study, we just present here the results.
The measurements are made in the region of $p_T$ from 0 to 20 GeV and with $|y|<0.4$ and the 
data are separated in four $p_T$-bins~\cite{Acosta:2001gv}.

Using the same technique as before, the longitudinal polarised fraction is 
obtained. \cf{fig:costheta_upsi1s} shows the $\cos \theta $ 
distribution with the two templates used,  and the resulting fit. 

\begin{figure}[H]
\centerline{\includegraphics[width=8cm]{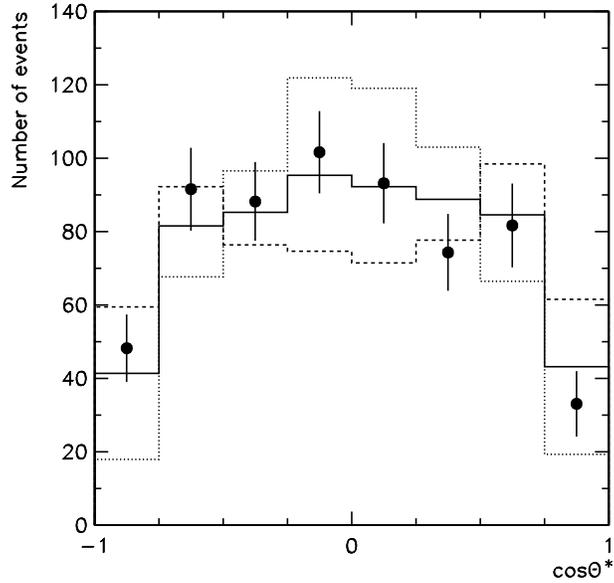}}
\caption{ Uncorrected $\cos \theta$ distribution for $|y^{\Upsilon(1S)}| < 0.4$ and for 
the highest $p_T$-bin. The points are the data; the solid line, the result giving the weight
of the transversal template (dashed line) and the longitudinal template 
(dotted line)~\protect\cite{Acosta:2001gv}.}
\label{fig:costheta_upsi1s}
\end{figure}

In \ct{tab:res_pol_upsi1s} are given 
the results for $\alpha$ and the same values are plotted in \cf{fig:cdf_ups1s_pol_nrqcd}
in comparison with a theoretical prediction from NRQCD~\cite{Braaten:2000gw}. In conclusion,
$\Upsilon(1S)$ are shown to be mostly unpolarised.

\begin{figure}[H]
\centerline{\includegraphics[width=8cm]{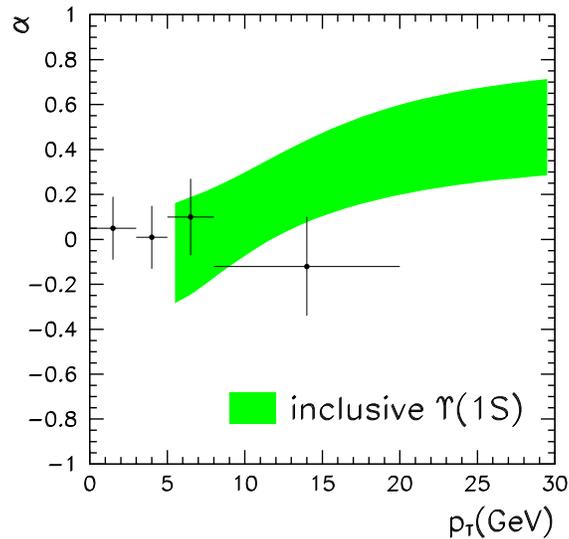}}
\caption{ $\alpha$ for  $\Upsilon(1S)$ fitted  for $|y^{\Upsilon(1S)}| < 0.4$.
The theoretical band represents the NRQCD prediction~\protect\cite{Braaten:2000gw}.}
\label{fig:cdf_ups1s_pol_nrqcd}
\end{figure}

\begin{table}[H]
\center \begin{tabular}{|c|c|} 
\hline
$p_T$ bin (GeV) & $\alpha$ \\ \hline\hline
$0.0-3.0$  & $+0.05\pm0.14$  \\
$3.0-5.0$  &  $+ 0.01\pm0.14$ \\
$5.0-8.0$ & $+0.10\pm0.17$ \\ 
$8.0-20.0$ & $-0.12\pm0.22$ \\ 
\hline
\end{tabular}
\caption{Fit results for $\Upsilon(1S)$ polarisation.}
\label{tab:res_pol_upsi1s}
\end{table}
\index{CDF|)}
\index{polarisation|)}

\section{PHENIX analysis for $J/\psi$ production cross sections}\index{PHENIX|(}

In this section, we present the first measurements of $pp\to J/\psi+ X$ at RHIC obtained
by the PHENIX experiment~\cite{Adler:2003qs} at a c.m. energy of $\sqrt{s}=200$ GeV. The analysis
was carried out by detecting either dielectron or dimuon pairs. The data were taken  during
the run at the end of 2001 and at the beginning of 2002. The data 
amounted to 67 nb$^{-1}$ for $\mu^+\mu^-$  and\footnote{The difference results from different
cuts on the extrapolated vertex position.} 82 nb$^{-1}$ for $e^+e^-$.
The $B$-decay feed-down is estimated to contribute less than 4\% at $\sqrt{s}= 200$ GeV
and is not studied separately.

Concerning the dielectron study,  it relied on:
\begin{itemize}
\item Two central spectrometer arms covering $|\eta| \leq 0.35$ and $\Delta\phi =90^\circ$;
\item An electromagnetic calorimeter (EMC);
\item A ring imaging \v Cerenkov detector (RICH) with a rejection threshold for pions of
4.7~GeV.\index{Cerenkov@\v Cerenkov}
\end{itemize}
whereas for muons, were used:
\begin{itemize}
\item Two muon-spectrometer arms covering $\Delta\phi=360^\circ$, of which
only one was used ($1.2 \leq \eta \leq 2.2$);
\item A muon-identifier system (MuID);
\item A muon-tracker system (MuTr).
\end{itemize}

Concerning dielectrons, the trigger requirement was either  a deposition of 0.75 GeV in a
$2\times2$ pile of EMC towers or 2.1 GeV in a $4\times4$ pile. Electron candidates had
to be associated with a RICH ring (with at least 2 photo-tubes hit) and with an EMC
hit (within $\pm 4 \sigma$ for the position). Furthermore, a correlation between
the energy deposition in the EMC, $E$, and the momentum of the reconstructed track, $p$, was
required through the condition $0.5\leq E/p \leq 1.5$. Finally, to avoid pion signals from the RICH,
a 5 GeV cut-off for the electrons was imposed.

On the other hand,  the muon trigger required two deeply penetrating track seeds (``roads'') in two
distinct azimuthal quadrants of the MuID~\cite{Sato:2002dv}. Then these seeds were matched to
the cluster hits in the MuTr stations. A fit with energy-loss correction of the MuID and MuTr 
hit positions and of the collision vertex finally gave the momentum.

The net yield of $J/\psi$ within the region $ 2.80 \hbox { GeV} < 
M(pair) < 3.40 \hbox { GeV}$ was found to be $46.0 \pm 7.4$ for electrons, whereas
for muons, it was $65.0 \pm 9.5$ $J/\psi$  within the region 
$ 2.71 \hbox { GeV} < M(pair) < 3.67 \hbox { GeV}$. The mass distributions 
obtained are shown in~\cf{fig:phenix_mass_dist} for both
dimuon and dielectron selections. As an indication of the background, a like-sign pair 
distribution is shown by dashed lines.

\begin{figure}[H]
\centerline{\includegraphics[width=12cm]{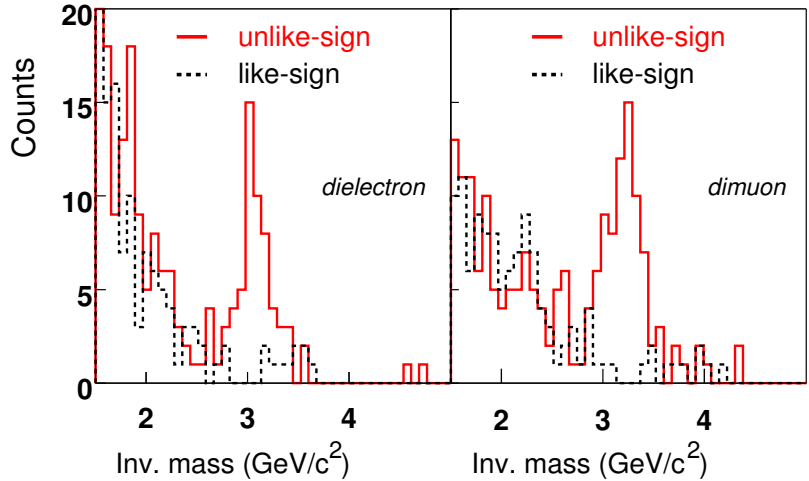}}
\caption{The invariant mass distribution for dielectron (left) and dimuon (right) pairs.
The solid lines are for unlike-sign pairs and the dashed lines for like-sign pairs 
(from~\cite{Adler:2003qs}).}
\label{fig:phenix_mass_dist}
\end{figure}

The cross sections were in turn extracted using 
\eqs{\label{eq:cross_section_PHENIX}
{\cal B} \frac{d^2\sigma}{dy dp_T}=\frac{N_{J/\psi}}{(\int {\cal L} dt) \Delta y \Delta p_T}
\frac{1}{\epsilon_{bias}\epsilon_{trig}} \frac{1}{A\epsilon_{rec}},
}
where $N_{J/\psi}$ is the measured $J/\psi$ yield, $\int {\cal L} dt$ the integrated luminosity,
$\epsilon_{bias}$ is the minimum bias trigger which is used to measure the luminosity, 
$\epsilon_{trig}$ is the trigger efficiency, $A$ the acceptance for $J/\psi$ reconstruction and
finally $\epsilon_{rec}$ its efficiency.

We shall not describe here the method to evaluate the systematics, we simply reproduce {\it verbatim} 
the table given in~\cite{Adler:2003qs}. Note however that the analysis of errors did not 
take into account possible polarisation effects.

\begin{table}[H]
\begin{center}
  \begin{tabular}{|c||c|c|} 
\hline
   Quantity  & $e^+e^-$ &  $\mu^+\mu^-$ \\ \hline\hline
   Yield & $\pm$ 5\% & $\pm$ 5\% \\
   $A \epsilon_{rec}$ & 0.026-0.010 $\pm$ 13\% & 0.038 - 0.017 $\pm$ 13\% \\
   $\epsilon_{trig}$ & (2$\times$2) 0.87-0.90 $\pm$ 5\% &  \\
                     & (4$\times$4) 0.30-0.74 $\pm$ 36\% &   \\
   $\epsilon_{bias}$ & 0.75 $\pm$ 3\% & 0.75 $\pm$ 3\% \\
   $\int{{\cal L}dt}$ & 82 nb$^{-1}$ $\pm$ 9.6\% & 67 nb$^{-1}$ $\pm$ 9.6\% \\
\hline
   Total & $\pm$ 15\%(sys) $\pm$ 10\%(abs) & $\pm$ 14\%(sys) $\pm$10\%(abs) \\
\hline
  \end{tabular}
\caption{Table of quantities and their systematic-error estimates. Ranges are given
for $p_T$ dependent quantities.  For the $\mu^+\mu^-$ case, the values of
$A \epsilon_{rec}$ and $\epsilon_{trig}$ are combined. The absolute 
cross-section normalisation uncertainty from $\epsilon_{bias}$ and 
$\int{{\cal L}dt}$ is kept separate and is labeled ``(abs)'' (taken from~\cite{Adler:2003qs}).
\label{tab:systematic_error} }
\end{center}
\end{table}

From~\ce{eq:cross_section_PHENIX}, the cross section as a function
of $p_T$ is obtained for the two analyses. For $p_T$ larger than two 2 GeV, it is compared 
in~\cf{fig:dsdpt_PHENIX} with LO CSM and COM without fragmentation but with $\chi_c$ feed-down. These
theoretical predictions are from Nayak~\etal~\cite{Nayak:2003jp}.

\begin{figure}[H]
\centerline{\includegraphics[width=16cm]{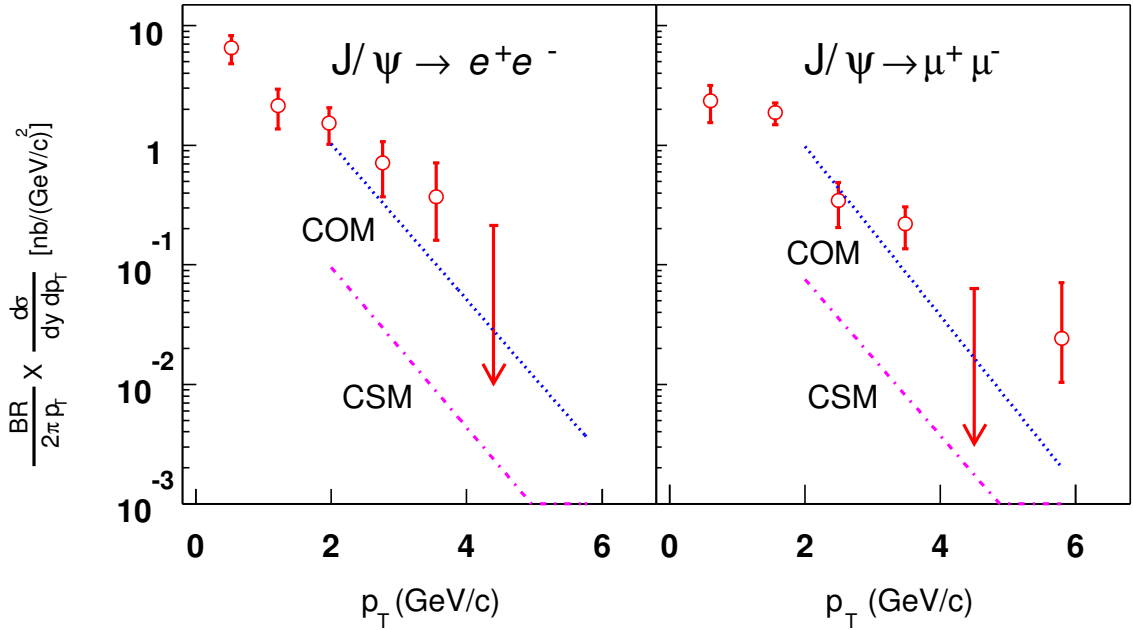}}
\caption{The differential cross section measured to be compared with theoretical predictions
from CSM and COM (without fragmentation but with $\chi_c$ feed-down) of~\cite{Nayak:2003jp}
(adapted from~\cite{Adler:2003qs}).}
\label{fig:dsdpt_PHENIX}
\end{figure}
\index{PHENIX|)}

\clearemptydoublepage
\chapter{Building a new model}\label{ch:build_model}

\section{Foreword}

We first present a study~\cite{Lansberg:2002cz} of the theoretical 
uncertainties of the LO CSM as an indication of the uncertainties
of all theoretical model. As we shall see, these are far from being 
sufficient to explain the discrepancies with experimental measurements. 

These theoretical uncertainties can arise from errors on the leptonic
decay width determination. The latter is used to determine the wave function at the origin,
which is the non-perturbative input of the model. Similar inputs
exist for the COM and they are associated with the probability for the $Q\bar Q$ pair to eventually
evolve into a given physical state. These probabilities are obtained from fits and are  
subject to similar uncertainties. 

Another main source of uncertainty is the parton distribution function. In the case at hand, only
gluons are actually involved, as the energy of the colliding hadrons is very large, and the
required momentum fraction of partons consequently low.

After the presentation of this study, we shall expose the guidelines that we shall follow in order to 
build a new model; in a sense, the model serves as a {\it non-static extension} of the CSM, but it could
also be seen as a generalisation of the COM for $Q\bar Q g$ states with the gluon having a large
momentum, whereas, in NRQCD, the momentum of the gluon associated with the heavy quark pair 
in the considered Fock state is necessarily lower than $Mv$ where $v$ is the typical velocity
of the quarks within the bound state of mass $M$.\index{Fock state} \index{static!non-static}
\index{colour-octet mechanism}

\section{Theoretical uncertainties of the LO CSM}
%%%%%%%%%%%%%%%%%%%%%%%%%%%%%%%%%%%%
\subsection{Wave function}\index{LO CSM|(}\index{wave function|(}

At LO in $\alpha_s$, the leptonic decay width reads:
\begin{equation}\label{eq:CSM_Gamma}
\Gamma(^3S_1 \to \ell \bar\ell)= \frac{64 \pi}{9} 
\frac{\alpha^2 e^2_q|\Psi(0)|^2}{M^2},
\end{equation}
where $\alpha$ is the fine structure constant and $e_q$ the quark charge in units of $e$.
This is one of the consequences of the factorisation postulate.\index{factorisation}

We thus find that the error on  $\Gamma_{^3S_1\to\mu\mu}$ introduces an error of at least 
$10 \%$  on the cross section.

\begin{center}
\begin{tabular}{|c|c|c|}
\hline Meson & $|\Psi(0)|^2 \pm \sigma_{|\Psi(0)|^2}$ & Relative error  \\
\hline \hline 
$J/\psi $        &  $0.041 \pm 0.0042$ GeV$^3$ & $10 \%$	\\
$\psi(2S) $      &  $0.024 \pm 0.0025$ GeV$^3$ & $10 \%$	\\
$\Upsilon(1S)$   &  $0.397 \pm 0.015  $  GeV$^3$ & $4 \%$ 	\\
$\Upsilon(2S)$   &  $0.192 \pm 0.030  $  GeV$^3$ & $16 \%$	\\
$\Upsilon(3S)$   &  $0.173 \pm 0.032  $  GeV$^3$ & $18 \%$ \\
\hline
\end{tabular}
\end{center}

Besides, one may consider the Colour Singlet Model as a level of approximation
of the Non Relativistic Quantum Chromodynamics (NRQCD) (already exposed in the first chapter). Indeed
this formalism describes the production of quarkonia by two main classes of processes, the first
when the quark pair is perturbatively produced within a colour-singlet configuration -- it can then
have exactly the same quantum number as the meson, as in the CSM, or be in a different spin
configuration--, the second class is when the quark pair is perturbatively produced within a colour-octet 
configuration.\index{NRQCD}

In fact, as already mentioned, NRQCD predicts that there is an infinite 
number of Fock-state contributions to the 
 production of a quarkonium ${\cal Q}$. Practically, one is driven to truncate the series; this
is quite natural in fact since most of the contributions are known to be suppressed by factor
of ${\cal O}(\frac{v^2}{c^2})$, where $v$ is the quark velocity in the bound state.

For definiteness, the latest studies accommodating the production rate of $^3S_1$ states retain
only the colour singlet state with the same quantum numbers as the bound state and colour 
octet $P$-wave states and singlet $S$-wave. In this context, the CSM 
can be thought as a further approximation to the NRQCD formalism.

However, in this formalism, factorisation tells us that each contribution is the product of 
a perturbative part and a non-perturbative matrix element, giving, roughly speaking, the probability
that the quark pair perturbatively produced will evolve into the considered physical bound state. If one 
transposes this to the CSM, this means the wave function at the origin corresponds to
this non-perturbative element. This seems reasonable since the wave function squared is also
a probability.

Yet, one has to be cautious because one actually links production processes with decay processes. In 
NRQCD, two different matrix elements are defined for the ``colour singlet'' production and decay, and
they are likely to be different and independent.

The only path left to recover the CSM is the use of a further approximation, the 
vacuum saturation approximation. The latter tells us that the matrix element for the 
decay is linked to the one for the production. This enables us to relate the wave function
at the origin appearing in \ce{eq:CSM_Gamma} to the colour-singlet NRQCD matrix element
for production. This gives:\index{vacuum saturation approximation}
\eqs{\left< {\cal O}_1^\psi(^3S_1)\right>=18 \left|\psi(0) \right|^2 +{\cal O}(v^4).
}

The conclusion that could be drawn within NRQCD is that the extraction of non-perturbative
input for production from the one for decay is polluted by factors of ${\cal O}(v^4)$.
\index{wave function|)}

\subsection{Parton distribution functions}\index{parton distribution function|(}

Let us now analyse the uncertainties due to the parton distribution 
functions (pdf). As there are two gluons coming from the initial 
hadronic state, these can induce non-negligible uncertainties on 
the cross section, either directly or via the associated $\alpha_s(Q^2)$.

For coherence, we have limited ourselves 
to LO pdf's. As a consequence, we have not felt the necessity to refine
the study to the point of using the very last pdf on the market. This choice is further
justified because we want to have an estimate of the effects due to extreme-case choices.
Here are different plots obtained for a set of 12 pdf's (see the legend of
Fig.~\ref{fig:CSM_variation_pdf}) coded in PDFLIB~\cite{pdflib} which illustrate these effects:
\begin{figure}[H]
\centerline{\includegraphics[width=10cm]{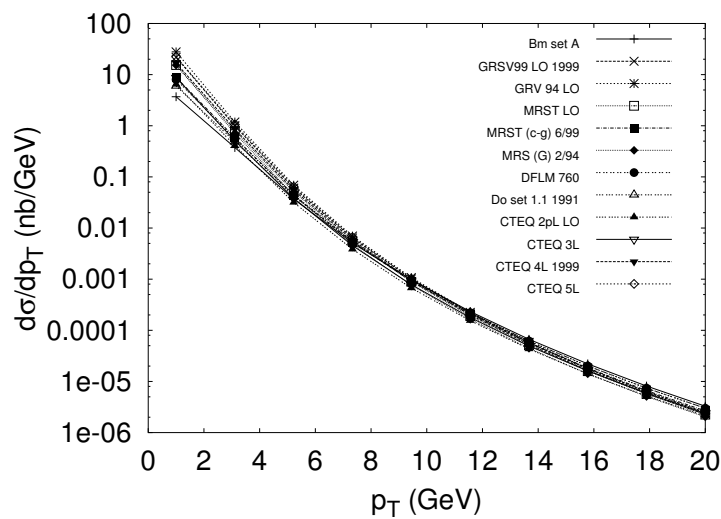}}
\caption{Variation of the cross sections obtained with various 
pdf's~\cite{pdflib} due to the pdf's only.}
\label{fig:CSM_variation_pdf}
\end{figure} 
%%%%%%%%%%%%%%%%%%%%%%%%
\begin{figure}[H]
\centerline{\includegraphics[width=10cm]{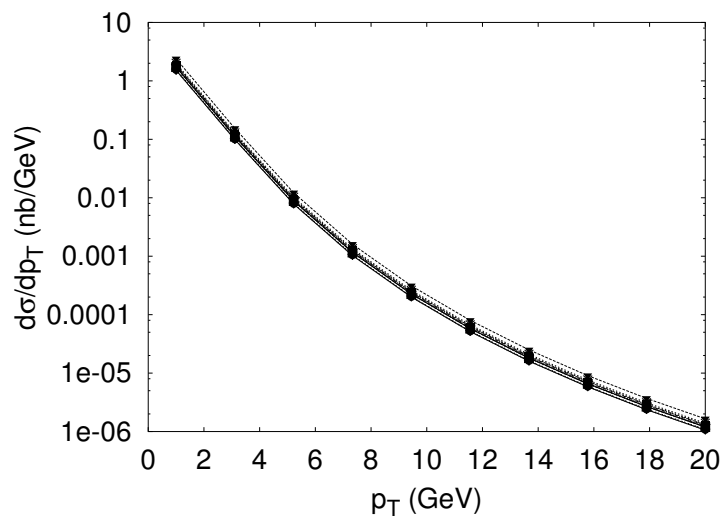}}
\caption{Variation of the cross sections due to $\alpha_s$ only.}
\label{fig:CSM_variation_alpha}
\end{figure} 
%%%%%%%%%%%%%%%%%%%%%%%%
\begin{figure}[H]
\centerline{\includegraphics[width=10cm]{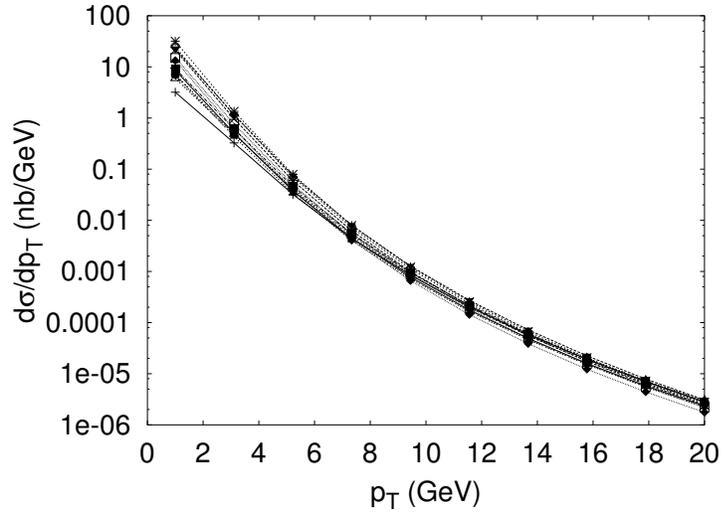}}
\caption{Variation of the cross sections obtained with various 
pdf's~\cite{pdflib} due to $\alpha_s$ and the pdf's.}
\label{fig:CSM_variation_all}
\end{figure} 
%%%%%%%%%%%%%%%%%%%%%%%%
\begin{figure}[H]
\centerline{\includegraphics[width=10cm]{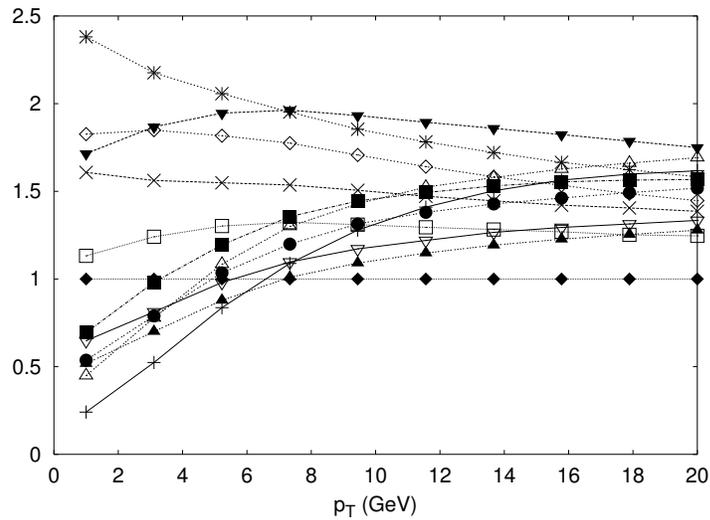}}
\caption{Ratio of the cross sections obtained with various 
pdf's~\cite{pdflib} to the 
cross section obtained with MRS(G) 2-94.}
\label{fig:CSM_variation_alphas}
\end{figure} 
%%%%%%%%%%%%%%%%%%%%%%%%
\begin{figure}[H]
\centerline{\includegraphics[width=10cm]{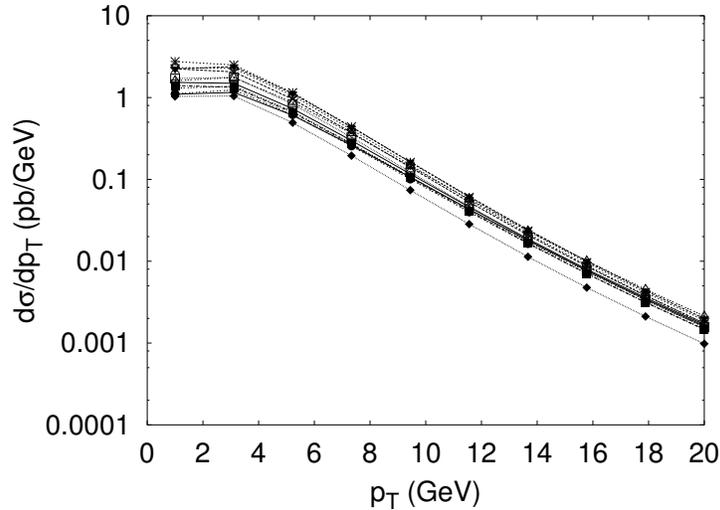}}
\caption{Variation of the cross sections obtained with various 
pdf's~\cite{pdflib} for $\Upsilon(3S)$.}
\label{fig:CSM_variation_ups3S}
\end{figure} 

We thus found that the overall factor resulting from these 3 sources 
of uncertainties is at most 2-3 and roughly constant for different values of $p_T$.
Hence, the discrepancy of the CSM with data is not due to conventional theoretical errors.
Furthermore, these could not change polarisation results either, for which data seem 
to contradict NRQCD.

\index{parton distribution function|)}
\index{LO CSM|)}

\section{Assumptions for a new model}

As we have already stated, no solution of the polarisation issue has been proposed so far, while we 
just have seen that conventional theoretical uncertainties are under control. We therefore propose
to go beyond the traditional treatment which is to consider the heavy quarkonia as purely 
non-relativistic states with the consequence of neglecting {\it a priori} any effects that may come from 
the internal dynamics of quarks in the bound state. This approximation is sensible and supported
by arguments such as the success of potential models, with instantaneous interactions. Nevertheless, 
it has 
never been demonstrated that non-static effects are systematically negligible and that 
the internal dynamics as well as the quark off-shellness in the meson do not provide any 
sizable modification of their description.

In order to tackle internal dynamics, we shall describe the heavy quarkonia with a phenomenological
3-particle vertex, made of a Dirac structure inferred from the quantum numbers of 
the considered quarkonium and
a vertex function inspired by the Bethe-Salpeter formalism. 
The description and the justification of the different steps leading to 
this 3-particle vertex is made in the next chapter. 

Nevertheless, as we shall see in chapter~\ref{ch:vertex_derivation}, the introduction
of vertex functions breaks gauge invariance in production processes associated with a photon
or a gluon. Our solution to this problem is the introduction of new 4-particles vertices
restoring the invariance. A complete derivation of their characteristics will be undertaken.
Gauge-invariance restoration leaves some freedom to add autonomous 4-particle vertices, which are gauge invariant alone. 

In chapter~\ref{ch:application}, we shall then apply our approach to the determination 
of the imaginary part of the amplitude for $gg\to\!\ ^3S_1g$. This amplitude is the expected
main contribution for $\psi$ and $\Upsilon$ production at the Tevatron and RHIC at low $P_T$. This
will enable us to calculate the various cross sections for these. In turn, we shall introduce
the above-mentioned autonomous contributions that will allow us to fit the data rather well.
A further and complete agreement with the data, namely with polarisation, will then be 
accomplished by a combination of our prediction with the colour octet $^3S_1$ fragmentation. Indeed,
we eventually obtain a polarisation parameter roughly constant and compatible with an unpolarised
production.

\clearemptydoublepage
\chapter{Two-particle description of heavy quarkonia}\label{ch:normalisation}
In this chapter, we shall introduce
the  quantities required to study processes involving heavy quarkonia, 
such as decay and production mechanisms. 
In field theory, all the information
needed can be parameterised by a vertex function, which describes
the coupling of the bound state to the quarks and contains the information about the size of 
the bound state, the amplitude of probability for given quark configurations and
the normalisation of the bound-state wave functions.\index{vertex function}\index{wave function}

The vertex function has close links with the classical non-perturbative wave function
introduced in quantum mechanics. We shall first consider those, and give their characteristics
if one thinks that the 
quarkonia are purely non-relativistic bound-states. This will show us 
also that the mass of the quark is actually a poorly  constrained parameter, 
which can be chosen with some freedom.\index{quark!mass}

Then, we shall start the description of our approach to build the 
quarkonia vertex functions. Along the way, we shall explain why 
we do not rely on the Bethe-Salpeter Equation~\cite{Salpeter:1951sz} (BSE), 
usually used to constrain 
the properties of vertex functions, for example, to relate the mass of the bound 
state  to the mass of its constituent quarks. Beside problems with gauge invariance,
all predictions coming from BSE
are realised within Euclidean space (see \eg~\cite{Maris:2003vk,Burden:1996nh,Ivanov:1998ms})
and we shall show that the continuation to Minkowsky space is in general problematic.
We shall make the choice to remain in Minkowsky space, at variance
with other phenomenological models (see \eg~\cite{Ivanov:2000aj,Ivanov:2003ge}).
The price to pay for not using BSE will be additional uncertainties due to the functional
dependence of the vertex. Fortunately, most of these can be cancelled if one fixes
the normalisation from the study of the leptonic decay width. 

We shall thus explain only in few words how in principle from a given choice for the vertex functions, we
may be able to fix some of the parameters by the use of the BSE.

\section{The charmonium model}

As soon as the $J/\psi$ and the $\psi'$ were discovered and the community of particle 
physicists started to convince itself that they were composed of a $c$ and a $\bar c$ quark,
Applequist and Politzer~\cite{Appelquist:1975ya} proposed that these states 
should share some properties with positronium states made of $e^+$ and $e^-$. 
Their statement was mainly motivated by the discovery of asymptotic
freedom in QCD~\cite{Gross:1973id,Politzer:1973fx,Gross:1973ju,Politzer:1974fr}.

The identification of the $J/\psi$ and $\psi'$ as ``ortho-charmonium''~\footnote{$S$-wave ($n=0,\ell=0$),
$J$=1.} was then straightforward. Concerning the decay, as the $c$-quark mass sets a sufficiently high
scale compared to $\Lambda_{QCD}$, a formula similar to that for positronium 
-- derived from the QED perturbative expansion -- was therefore believed to be 
sufficient, with the natural replacements of $\alpha$ by $\alpha_{s}$ and $m_e$ by $m_c$. In the 
latter formula, the use of a coulombic wave function gave the right order of magnitude for the
leptonic decay width $\Gamma_{\ell\ell}$~\cite{Buchmuller:1992zf}.\index{wave function}

As the Bohr radius ($(\frac{2}{3} \alpha_{s} m_c)^{-1}$) of these states was of 
the order of 1 fm (a typical hadron size), they could not be purely coulombic.
A strong binding force operating at large distance was then added. It resulted
in a coulombic potential with a linear rising term preventing the freedom of quarks.

This gave birth to what is referred to as the charmonium model~\cite{Eichten:1974af} whose potential
was initially of the form:\index{charmonium model}\index{potential}
\eqs{\label{eq:pot_charm_model}
V(r)=-\frac{4}{3} \frac{\alpha_{s}}{r} + k r,
}
where $k$ is the string tension.

From the Schr\"odinger equation \index{Schr\"odinger equation}
\eqs{\label{eq:schro_eq}
\left[ -\frac{1}{m_c}\Delta+V(r) \right] \Psi_{n\ell m}(r)= E_{n\ell} \Psi_{n\ell m}(r),
}
the wave functions $\Psi_{n\ell m}(r)$ and the binding energies $E_{n\ell}$ were then  calculated, 
the corresponding mass spectrum  being then built using $M_{n\ell}(c\bar c)=2 m_c+E_{n\ell}$.
\index{wave function}

\subsection{Refinements}

Later on, in order to solve the entire spectroscopy of the charmonium and bottomonium 
family, other forms for the potential were proposed: 
\begin{itemize}
\item The ``Cornell'' (COR) potential~\cite{cornell_potential} 
\eqs{\label{eq:pot_COR}
V(r)=-\frac{k}{r} + \frac{r}{a^2},
}
with\footnote{Comparing this form to the one of \ce{eq:pot_charm_model}, we see that
$\alpha_s=0.38$  makes $k=\frac{4\alpha_s}{3}$ be $0.52$.} $k=0.52$, 
$a=2.34$ GeV$^{-1}$~\footnote{$\frac{1}{a^2}$ here plays the role of the string tension
$k$ of \ce{eq:pot_charm_model}.} and $m_c=1.84$ GeV or  $m_b=5.18$ GeV.
\item The ``Buchm\"uller \& Tye'' (BT) potential~\cite{Buchmuller:1980su} such that
\eqs{\label{eq:pot_BT}
V(r)&\sim k r \ \ (r_{crit}\leq r),\\
V(r)&\sim \frac{1}{r\ln(1/\Lambda_{QCD}^2r^2)}\left[1+{\cal O}\left(\frac{1}{\ln(1/\Lambda_{QCD}^2r^2)}\right)\right],
}
where $k$ is again the string tension, 
$\Lambda_{QCD}$ is the QCD running scale,
$r_{crit}$ is fixed by $\frac{1}{r_{crit}^2\Lambda^2_{QCD,\overline{MS}}}=100$ 
(see~\cite{Buchmuller:1980su})
 and $m_c=1.48$ GeV or $m_b=4.88$ GeV.
\item The ``Logarithmic'' (LOG) potential~\cite{Quigg:1977dd} such that
\eqs{\label{eq:pot_LOG}
V(r)=-0.6635 \hbox{ GeV } + (0.733 \hbox{ GeV}) \log(r \cdot 1 \hbox{ GeV }),
}
with $m_c=1.5$ GeV or $m_b=4.906$ GeV.

\item the ``Power-law'' (POW) potential~\cite{Martin:1980jx}
\eqs{\label{eq:pot_POW}
V(r)=-8.064 \hbox{ GeV } + (6.898 \hbox{ GeV}) (r \cdot 1 \hbox{ GeV })^{0.1},
}
with $m_c=1.8$ GeV or $m_b=5.174$ GeV.

\end{itemize}

For illustrative purpose, we  show in~\cf{fig:wave_funct_from_pot} some wave 
functions obtained by Kopeliovich~\etal~\cite{Kopeliovich:2003cn} 
from~Eqs.(\ref{eq:schro_eq}--\ref{eq:pot_POW}). 
As one may see, \index{quark!mass}
even though the values of the charm quark mass, $m_c$, differ substantially from
one potential to another, the wave functions do not differ much for large $r$. On the 
other hand, they differ by 30 \% at the origin, value which is a key input 
in the static approximation. Usually, it is extracted from the leptonic decay width but one is free to rely
on these determinations from the resolution of the Schr\"odinger equation.

Please note also the presence of a node in the wave function for the first radial excitation ($\psi'$)
as should be from general theorems of quantum mechanics.  
This node at around 0.4 fm produces the suppression of photoproduction of $\psi'$ 
compared to $J/\psi$~\cite{Kopeliovich:1991pu}.\index{wave function}

\begin{figure}[H]
\centering
\mbox{\subfigure[$J/\psi$]{\includegraphics[width=6cm]{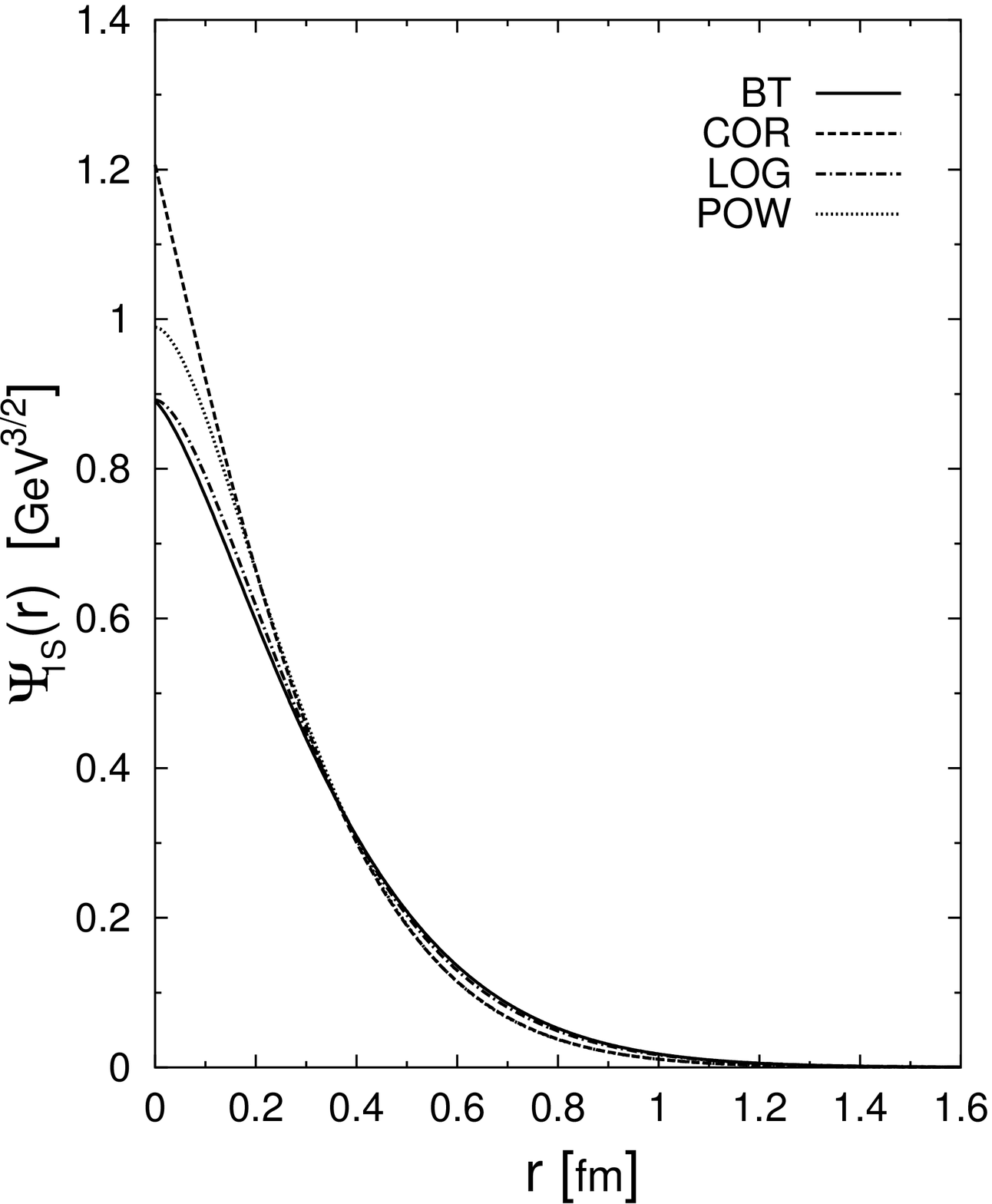}}\quad\quad
      \subfigure[$\psi(2S)$]{\includegraphics[width=6cm]{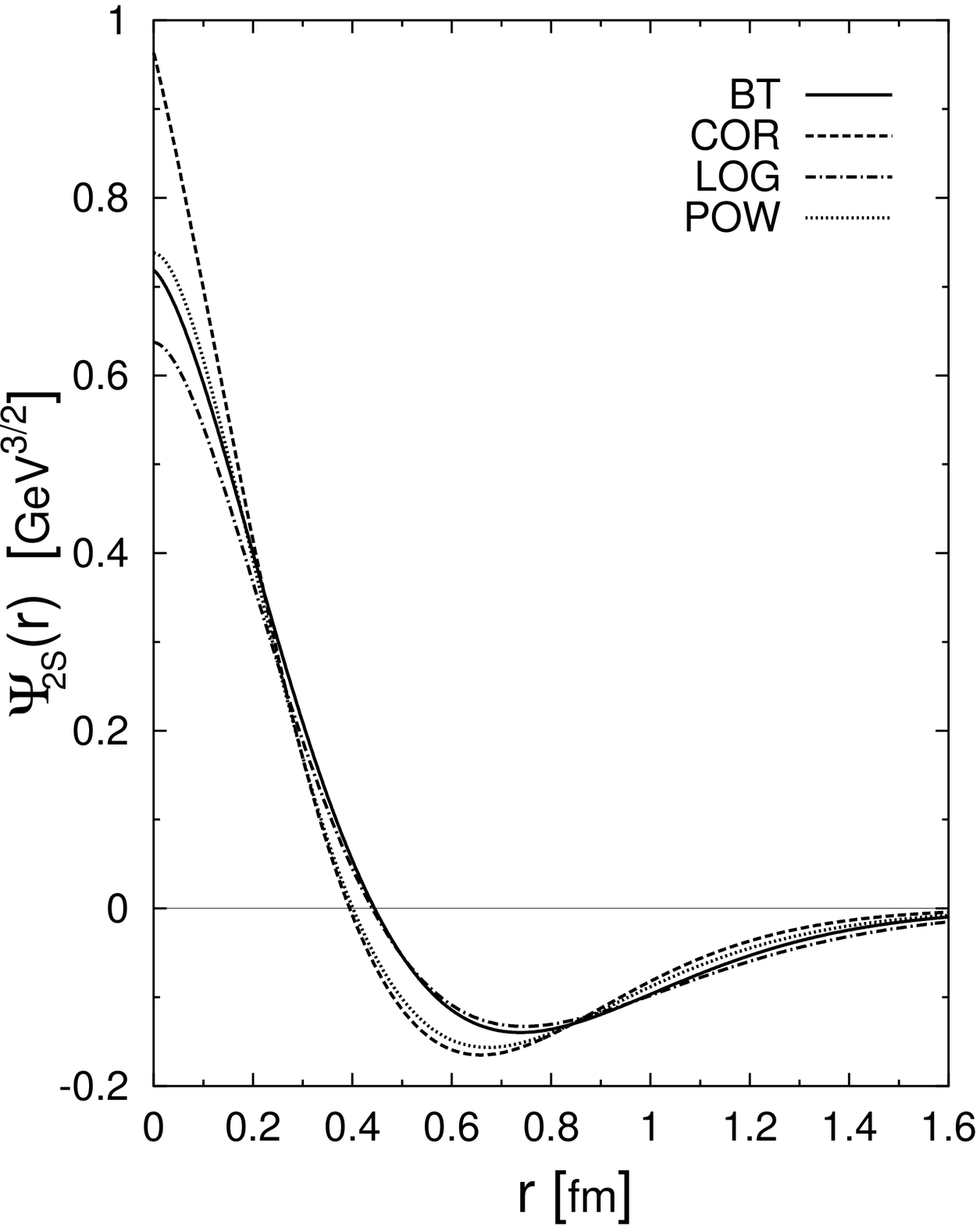}}}
\caption{The radial wave function $\Psi_{n\ell}(r)$ for the $J/\psi$ ($1S$) and
$\psi'$  ($2S$) states obtained by Kopeliovich \etal~\cite{Kopeliovich:2003cn} 
for four potential: BT, COR, LOG and POW.}
\label{fig:wave_funct_from_pot}
\end{figure}

\section{Phenomenological approach to the vertex function}

\subsection{Wave function vs. vertex function}\index{vertex function|(}\index{wave function|(}

We have seen in the previous section how one can calculate the wave function for the quarkonia. 
One thing is however lacking, that is how to include this information in a calculation involving Feynman
diagrams. The usual procedure can be summarised as follows:

\begin{itemize}
\item the quarks are to be on-shell within the quarkonia and therefore have to be produced on-shell;
\item the Schr\"odinger wave function provides with the binding-probability amplitude ; 

\item the quantum numbers of the bound state come uniquely from the quarks configuration. The parity $P$ 
and charge conjugation $C$ are therefore given by $P=(-)^{L+1}$ and $C=(-)^{L+S}$ where $L$ and $S$
are given by the corresponding eigenvalues of the Schr\"odinger equation.
\end{itemize}\index{charge conjugation}\index{parity conjugation}

From this, it follows that the amplitude for producing a bound state of quantum numbers $J,L,S$ is
\begin{equation}\label{eq:opproject1}
{\cal M}_{tot}= \sum_{L_z,S_z} \langle LL_z; SS_z |JJ_z \rangle
\sum_{\lambda, \bar\lambda} \int \frac{d^3\vec p}{(2\pi)^3} {\cal M}_{\lambda, \bar\lambda}(\vect p)
{\Psi_{\lambda, \bar\lambda}^{L_zS_z}(\vect p )}^*,
\end{equation}
with $\vect p$, the relative momentum between the two quarks, $\lambda, \bar\lambda$, 
the spin projections of the quarks along the $z$-axis (quarkonium direction), 
${\cal M}_{\lambda, \bar\lambda}(\vect p)$, 
the amplitude for producing the heavy-quark pair
and ${\Psi_{\lambda, \bar\lambda}^{L_zS_z}(\vect p )}$, the wave function of the bound state.

The important point here is that the latter reads: 
\begin{equation}\label{eq:opproject2}
{\Psi_{\sigma \bar\sigma}^{L_zS_z}(\vect p)}^* = \Psi_{LL_z}^* (\vect p) \frac{1}{2m} \bar
v(-\vect p,\bar\sigma) \bar{\cal P}_{SS_z}(\vect p,-\vect p) u(\vect p,\sigma),
\end{equation}
where $\Psi_{LL_z} (\vect p)$ is the non-relativistic wave function, $\bar v$ and $u$ are the spinors
for the two on-shell  heavy quarks -- as stated in the assumptions -- and
$\bar{\cal P}_{SS_z}$ is {\it a spin-projection operator}\index{spin-projection operator}
 whose  explicit expression is~\cite{Kuhn:1979bb} 
\begin{eqnarray} 
\bar{\cal P}_{SS_z}(\vect p_1,\vect p_2) &=& \frac{1}{2m}
\sum_{\lambda_1,\lambda_2}
\langle \frac{1}{2}\lambda_1,\frac{1}{2}\lambda_2 | SS_z \rangle
v(\vect p_2,\lambda_2)
\bar u(\vect p_1,\lambda_1) \nonumber \\
&=& \frac{1}{2m} \frac{\ks p_2 - m}{\sqrt{p_{2,0}+m}}\bar{\Pi}_{SS_z}
\frac{\gamma^0+1}{2\sqrt{2}}  \frac{\ks p_1 + m}{\sqrt{p_{1,0}+m}}
 \\ \label{eq:opproject3} 
\hbox{ where } \ \ \bar\Pi_{SS_z} &=& 
\left[
\begin{array}{cc}
-\gamma_5 & (S=0) \\
\gamma^\mu \ep_\mu^*(S_z) & (S=1) \\
\end{array} 
\right. \label{eq:opproject4}
\end{eqnarray}

The expressions in \ce{eq:opproject4} are what we shall use for our vertex, for respectively
a pseudoscalar particle and a vector particle. But we are going further by lifting  
the constraint that the quark be on their mass shell. {\it This generalises structureless vertices
of the usual pQCD to bound states.}\index{wave function|)}

\subsection{Our choice for the vertex function}

It has been shown elsewhere (see \eg~\cite{Burden:1996nh}) that other structures than $\gamma^\mu$ Dirac
matrix for vector bound state are suppressed by one order of magnitude 
in the case of light mesons, and that this suppression is increasing for the meson $\phi$ ($s \bar s$). 
We have therefore reasons to believe that their effect are even more negligible in heavy 
quarkonia.\index{phi@$\phi$}

As a consequence, the following phenomenological Ansatz for the vector-meson vertex, 
inspired by~\ce{eq:opproject4}, is likely to be sufficient for our purposes. It
reads:
\begin{eqnarray}\label{vf}
  V_{\mu}(p,P) &=& \Gamma(p,P) \gamma_\mu,
\end{eqnarray}
with $P$ the total momentum of the bound state, $P=p_{1}-p_{2}$, and $p$ the
relative one, $p=(p_{1}+p_{2})/2$ as drawn in \cf{fig:BS_vertex_phenoa}. This Ansatz
amounts to multiplying the {\it point} vertex (corresponding to a structureless particle) 
by a function, $\Gamma(p,P)$.

\begin{figure}[H]
\centering
\includegraphics[width=12.0cm]{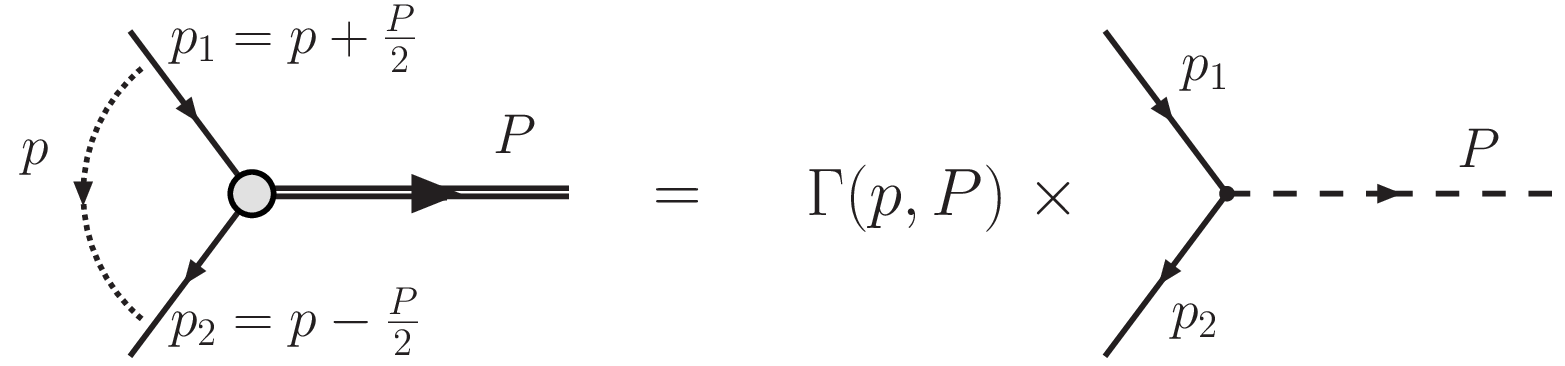}
\caption{Phenomenological vertex obtained by multiplying a {\it point} vertex, representing a structureless
particle, by a vertex function (or form factor).}
\label{fig:BS_vertex_phenoa}
\end{figure}

The function $\Gamma$, which we shall call vertex function, 
can be chosen in different ways. Two simple Ans\"atze are commonly used 
in the context of Bethe-Salpeter Equation. We shall consider both. 
They correspond to two extreme choices at large distances: a dipolar 
form which decreases gently with its argument, and a gaussian form: \index{vertex function!dipole}
\index{vertex function!gaussian}
\eqs{
\Gamma_0(p,P)=\frac{N}{(1-\frac{p^2}{\Lambda^2})^2} 
,}
and
\eqs{
\Gamma_0(p,P)=N e^{\frac{p^2}{\Lambda^2}} 
,}
both with a free size parameter $\Lambda$. As $\Gamma(p,P)$ in principle
depends on $p$ and $P$, we may shift the variable and use $p^2-\frac{(p.P)^2}{M^2}$ instead
of $p^2$ which has the advantage of reducing to $-|\vect p|^2$ in the rest frame of the bound state.  
We shall refer to it as the {\it vertex function with shifted argument}. In the latter
cases, we have -- in the rest frame --, 

\eqs{\Gamma(p,P)=\frac{N}{(1+\frac{|\vect p|^2}{\Lambda^2})^2}} 
and 
\eqs{\Gamma(p,P)=N e^{\frac{-|\vect p|^2}{\Lambda^2}}.}

\subsubsection{Excited states}\index{excited states}

As one knows, 
the number of nodes in the wave function, in whatever space, increases 
with the principal quantum number $n$. This simple feature can be used
to differentiate between $1S$ and $2S$ states.

We thus simply  have to determine the position of the node of the wave function in momentum space.
To what concerns the vertex function, working in the meson rest frame, \index{node}
the node comes through a
prefactor, $1-\frac{|\vect p|}{a_{node}}$, which multiplies
the vertex function for the $1S$ state. Explicitly, this gives, 
\eqs{\Gamma_{2S}(p,P)=N'\left(1-\frac{|\vect p|}{a_{node}}\right)
\frac{1}{(1+\frac{|\vect p|^2}{\Lambda^2})^2}} 
and 
\eqs{\label{eq:gam2s_exp}
\Gamma_{2S}(p,P)=N'\left(1-\frac{|\vect p|}{a_{node}}\right) e^{\frac{-|\vect p|^2}{\Lambda^2}}.}

In order to determine the node position in momentum space, we can use two methods. The first is
to fix $a_{node}$  from its known value in position space, \eg~from potential studies, and to Fourier-transform the vertex function. In the case of a 
gaussian form, this can be carried out analytically. 

The second method is to impose the following relation\footnote{inspired from the orthogonality
between the $1S$ and $2S$ wave functions.} between the $1S$ and $2S$ vertex functions:
\eqs{\label{eq:ortho_1S_2S}
\int|\vect p|^2 d|\vect p| e^{\frac{-|\vect p|^2}{\Lambda^2}} \left(1-\frac{|\vect p|}{a_{node}}\right) e^{\frac{-|\vect p|^2}{\Lambda^2}}
=0.
}
The two methods give compatible results. The 
introduction of a second node to describe the $3S$ states
is similar.\index{vertex function|)}

\section{Normalising: the leptonic decay width}\index{leptonic decay width|(}\index{normalisation|(}

To fix the normalisation parameter $N$, we shall use the leptonic 
decay width. Besides, as the calculation is similar to the one for the production 
processes, we shall give a detailed explanation of this calculation as a pedagogical 
introduction to the production calculations which are the core of this work (see 
chapter~\ref{ch:application}).

The width in term of the decay amplitude $\cal M$, is given by
\eqs{\label{eq:lept_width}
\Gamma_{\ell\ell}=\frac{1}{2 M}\frac{1}{(4\pi^2)}\int \left|\bar{\cal M}\right|^2 d_2(PS),
}
where $d_2(PS)$ is the two-particle phase space~\cite{barger}.
\index{two-particle phase space}

The amplitude is obtained as usual via Feynman rules, for which we shall
use our vertex function at the meson-quark-antiquark vertex. At leading order,
the square of the amplitude is obtained from the cut-diagram drawn in \cf{fig:decay_diag_1}.

\begin{figure}[h]
\centering{\mbox{\includegraphics[width=14cm]{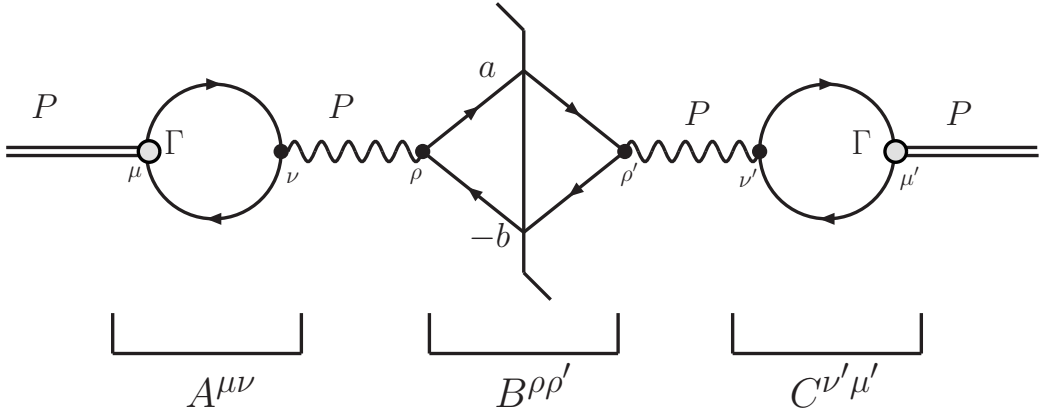}}}
\caption{Feynman diagram for $^3S_1\to \ell \bar \ell$.}\label{fig:decay_diag_1}.
\end{figure}

In terms of the sub-amplitudes $A^{\mu\nu}$, $B^{\mu\nu}$ and $C^{\mu\nu}$ defined in
\cf{fig:decay_diag_1}, we have\footnote{In the Feynman gauge. The calculation can be shown
to be gauge-invariant, though.} :\index{gauge!Feynman}
\eqs{
\label{eq:decomp_ampl_inv_decay}
\int \left|\bar{\cal M}\right|^2 d_2(PS)=\frac{1}{3} 
\Delta_{\mu\mu'}
A^{\mu\nu}\left(\frac{-ig_{\nu\rho}}{M^2}\right)B^{\rho\rho'}(\frac{-ig_{\rho'\nu'}}{M^2})
C^{\nu'\mu'},
}
where the factor $\Delta_{\mu\nu}=(g_{\mu\nu}-\frac{P_\mu P_{\nu'}}{M^2})=
\sum_i \ep_{i,\mu} \ep^\star_{i,\mu'}$ results from the sum over polarisations 
of the meson and the factor $\frac{1}{3}$ accounts for the averaging on these initial polarisations.

%**************************
\subsection{First sub-amplitude}

From the Feynman rules and using the vertex functions discussed above, we have (see \cf{fig:decay_diag_1a})
\eqs{\label{eq:Amunu_start}
iA^{\mu\nu}&=-3e_Q \int\! \frac{d^4k}{(2 \pi)^4}
{\rm Tr} \left((i \Gamma(p,P)\gmu)  \frac{i(\ks k -\frac{1}{2}\ks P +m)}{(k-\frac{P}{2})^2-m^2+i\ep}
(i e \gnu)  \frac{i(\ks k +\frac{1}{2}\ks P +m)}{(k+\frac{P}{2})^2-m^2+i\ep}
\right)\\
 &=-3e_Q\int\! \frac{d^4k}{(2 \pi)^4}
 \Gamma(p,P)
\frac{g^{\mu\nu} (M^2+4m^2-4k^2)+8 k^\mu k^\nu-2P^\mu P^\nu}
{((k-\frac{P}{2})^2-m^2+i\ep))((k+\frac{P}{2})^2-m^2+i\ep))},
}
$e_Q$ is the heavy-quark charge, $-1$ comes for the fermionic loop, $3$ is the colour factor.

\begin{figure}[h]
\centering{\mbox{\includegraphics[width=6cm]{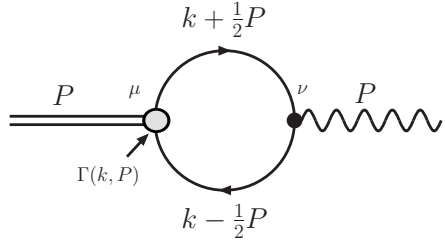}}}
\caption{Feynman diagram for $^3S_1\to \gamma$.}\label{fig:decay_diag_1a}
\end{figure}

Since the relative momentum is $k$, we shall have in the bound-state rest frame,
for the dipolar form of the vertex function as a function of $k$ ,
\eqs{
\Gamma(p,P)=\frac{N}{(1+\frac{|\vect k|^2}{\Lambda^2})^2}
,}
and for the gaussian form, 
\eqs{
\Gamma(p,P)=N e^{\frac{-|\vect k|^2}{\Lambda^2}}
.}
The expressions for the vertex functions of excited states immediately follow from the above discussion 
and are not written down.

We can easily motivate our choice of the argument of the vertex function, 
beside its simple relation with that of wave functions. Its main virtue is to regularise
the integration in all spatial directions. In the meson-rest frame, 
the $k^0$ integral can be done by standard residue techniques, and the remaining $\vect k$
integral is guaranteed to converge. If one uses $k^2$ instead of $|\vect k|^2$  as an argument, 
it is possible to show that along the light cone, one obtains logarithmic divergences,
which would presumably need to be renormalised. Besides, the $k^0$ integral then also 
becomes dependent on the singularity structure of the vertex function. These two reasons
make us prefer the shifted vertex in our calculation.

A notable simplification 
can be obtained by guessing the tensorial form of $A^{\mu\nu}$. Indeed, current
conservation (gauge invariance) for the photon can be expressed as:
\index{integration by residues}
\index{current conservation}\index{gauge!invariance}

\eqs{\label{eq:GI_Amunu}
A^{\mu\nu}P_\nu=0.
}
It is equivalent to 
\eqs{
iA^{\mu\nu}=iF (g^{\mu\nu}-\frac{P^\mu P^\nu}{M^2})=iF\Delta^{\mu\nu},
}
the coefficient $F$ being $\frac{A^\mu_{\ \mu}}{3}$ since $A^\mu_{\ \mu}= A^{\mu\nu}g_{\nu\mu}=3F$.
This assumption can be easily verified in the bound-state rest frame $P=(M,0,0,0)$ 
for which \ce{eq:GI_Amunu} reduces
to $A^{\mu0}M=0\Rightarrow A^{\mu0}=0$. We shall check this at the end of this calculation.

It is therefore sufficient to compute the following quantity, where we set $|\vect k|^2\equiv K^2$:
\eqs{
iA^\mu_{\ \mu}=-3e_Q&\int_0^\infty 4 \pi K^2 dK \Gamma(-K^2)\times \\
&\int_{-\infty}^\infty \frac{dk_0}{(2 \pi)^4}
\frac{4 (-2(k^2-\frac{M^2}{4})+4m^2)}
{((k-\frac{P}{2})^2-m^2+i\ep)((k+\frac{P}{2})^2-m^2+i\ep))}.
}

\subsubsection{Integration on $k_0$}

Let us first integrate on $k_0$ by residues. Defining $E=\sqrt{K^2+m^2}$, we determine
easily the position of the pole in $k_0$ (still in the bound-state rest frame)
\eqs{
&(k+\frac{P}{2})^2-m^2+i\ep=k^2_0+k_0 M+\frac{M^2}{4}-K^2-m^2+i\ep=\\&
(k_0+\frac{M}{2})^2-E^2+i\ep=((k_0+\frac{M}{2})+E-i\ep')((k_0+\frac{M}{2})-E+i\ep')=\\
&(k_0-(\frac{-M}{2}+E+i\ep'))(k_0-(\frac{-M}{2}+E-i\ep')),
}
and 
\eqs{
&(k-\frac{P}{2})^2-m^2+i\ep=k^2_0-k_0 M+\frac{M^2}{4}-K^2-m^2+i\ep=\\&
(k_0-\frac{M}{2})^2-E^2+i\ep=((k_0-\frac{M}{2})+E-i\ep')((k_0-\frac{M}{2})-E+i\ep')=\\
&(k_0-(\frac{M}{2}-E+i\ep'))(k_0-(\frac{M}{2}+E-i\ep')).
}

To calculate the integral on $k_0$, we choose the contour as drawn in \cf{fig:k0_plane}.
Two poles $-E-\frac{M}{2}$ and $-E+\frac{M}{2}$ are located in the upper half-plane and 
the two others in the lower
half-plane. We shall  therefore have two residues to consider. The contribution of the 
contour ${\cal C}_R$ vanishes as $R$ tends to $\infty$. 

\eqs{
&iA^\mu_{\ \mu}=\frac{-3e_Q}{\pi^3} \int_0^\infty K^2 dK \Gamma(-K^2)\times \\
&\int_{-\infty}^\infty 
\frac{dk_0(-2k_0^2+2K^2+\frac{M^2}{4}+4m^2)}
{(k_0-(\frac{-M}{2}-E+i\ep))(k_0-(\frac{-M}{2}-E-i\ep))
(k_0-(\frac{M}{2}-E+i\ep))(k_0-(\frac{M}{2}+E-i\ep))}\\
&=\frac{-3e_Q}{\pi^3} \int_0^\infty K^2 dK \Gamma(-K^2)\times\\
&2i\pi\left[
\frac{-2(-E-\frac{M}{2})^2+\frac{M^2}{2}+2E^2+2m^2}{(-2E)(-M)(-2(E+\frac{M}{2}))}\right.
+\left.
\frac{-2(-E+\frac{M}{2})^2+\frac{M^2}{2}+2E^2+2m^2}{M(-2(E-\frac{M}{2}))(-2E)}
\right]\\
&=\frac{-3e_Q}{\pi^3} \int_0^\infty K^2 dK \Gamma(-K^2)\frac{2i\pi}{4ME}
\underbrace{\left[-\frac{2EM+2m^2}{E+\frac{M}{2}}+\frac{2EM+2m^2}{E-\frac{M}{2}}\right]}
_{\frac{M(4E^2+2m^2)}{E^2-\frac{M^2}{4}}}.
}

\begin{figure}[ht]
\centering{\mbox{\includegraphics[width=10 cm]{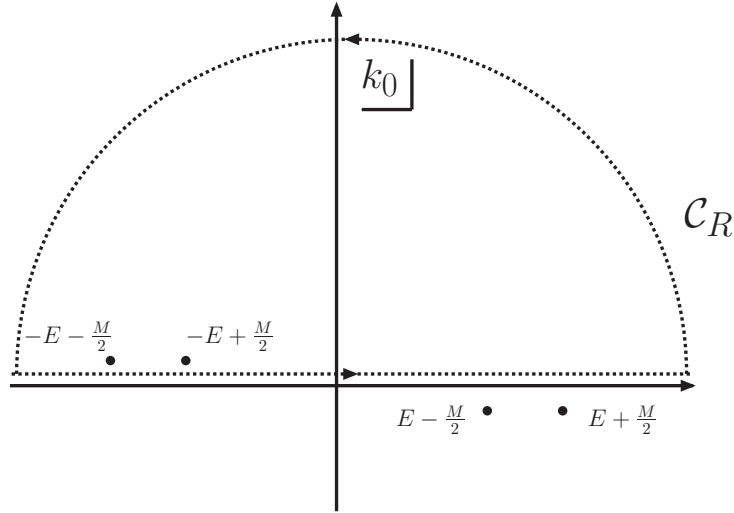}}}
\caption{Illustration of the contour chosen to integrate on $k_0$.}\label{fig:k0_plane}.
\end{figure}

\subsubsection{Setting the constituent quark mass}\index{quark!mass}

This way of proceeding is correct only if the poles of the upper-plane are on the left
 and those of the lower-plane are on the right\footnote{If they cross, the integral
acquires a discontinuity from the pinch and becomes complex.}.

This crossing would first occur between the points $-E+\frac{M}{2}$ and $E-\frac{M}{2}$. 
Solving for $E$, we have:
\eqs{
-E+\frac{M}{2}=E-\frac{M}{2} &\Rightarrow E=\frac{M}{2}=\sqrt{K^2+m^2} \\
                             &\Rightarrow  \frac{M^2}{4}=K^2+m^2 \\
 &\Rightarrow    \frac{M^2}{4}-m^2=K^2>0.
}
The crossing is therefore impossible when $M<2m$. Therefore, to get
a coherent description of all charmonia (resp. bottomonia) below the open charm (resp. beauty) 
threshold, we shall set the quark mass high enough to avoid crossing for all of these.  This sets
$m_c$ to 1.87 GeV ($m_D$) and $m_b$ to 5.28 GeV  ($m_B$). Considering the variety of results
obtained from potential model, this seems to be a sensible choice.\index{B@$B$}\index{D@$D$}

\subsubsection{Integration on $K$}

Putting	it all together, we get 
\eqs{
A^\mu_{\ \mu}=\frac{-3 e_Q}{\pi^2} \int_0^\infty dK \frac{K^2\Gamma(-K^2)}{\sqrt{K^2+m^2}}
\frac{(2K^2+3m^2)}{(K^2+m^2-\frac{M^2}{4})}.
}
One is left with the integration on $K$ for which we define the integral $I$ depending 
on the vertex function whose normalisation is pulled off,
\eqs{
I(\Lambda,M,m)\equiv\int_0^\infty  \frac{dK K^2}{\sqrt{K^2+m^2}}
\frac{\Gamma(-K^2)}{N}
\frac{(2K^2+3m^2)}{(K^2+m^2-\frac{M^2}{4})},
}
$I$ is a function of $\Lambda$ through the vertex function $\Gamma(-K^2)$ 
and is not in general computable analytically.
In the following we shall leave it as is and express 
$A^{\mu\nu}$ as:
\eqs{
A^\mu_{\ \mu}=\frac{-3 e_Q}{\pi^2} N I(\Lambda,M,m) \Rightarrow A^{\mu\nu}=\frac{- e_Q}{\pi^2} N I(\Lambda,M,m) \Delta^{\mu\nu}.
}

\subsubsection{Check of the assumption on $A^{\mu\nu}$}

As said before, we can test our assumption of\index{gauge!invariance}
gauge invariance by checking that $A^{\mu0}=0$. For $A^{i0}$, this is 
straightforward as the integrand is odd. This is a bit less trivial 
for $A^{00}$. Let us go back to \ce{eq:Amunu_start} and write down
$A^{00}(K)$ omitting the integration on $K$ and neglecting the constant factors:

\eqs{\label{eq:A00}
&A^{00}(K)\propto\int_{-\infty}^\infty dk_0
\frac{ (M^2+4m^2-4k^2)+8 k_0 k_0-2 M^2}
{((k-\frac{P}{2})^2-m^2+i\ep)((k+\frac{P}{2})^2-m^2+i\ep)}\\
&\! \propto \!\!
\int_{-\infty}^\infty 
\frac{dk_0(k_0^2+E^2-\frac{M^2}{4})}
{(k_0-(\frac{-M}{2}-E+i\ep))(k_0-(\frac{-M}{2}-E-i\ep))
(k_0-(\frac{M}{2}-E+i\ep))(k_0-(\frac{M}{2}+E-i\ep))}\\
& \propto
\left[
\frac{E^2+\frac{M^2}{4}+EM-\frac{M^2}{4}+E^2}{(-2E)(-M)(-2(E+\frac{M}{2}))}\right.
+\left.
\frac{E^2+\frac{M^2}{4}-EM-\frac{M^2}{4}+E^2}{M(-2(E-\frac{M}{2}))(-2E)}
\right]\propto
\left[\frac{-1}{M}+\frac{1}{M}\right]=0.
}

%**************************
\subsection{Second sub-amplitude}
From the Feynman rules, we have
\eqs{
B^{\rho\rho'}&= (ie)^2 \int d_2(PS) {\rm Tr} (\grho(-\ks a+m_\ell)\grhop (\ks b+m_\ell))\\
&=(ie)^2 \int d_2(PS) 4(a.b g^{\rho\rho'}-a^\rho b^{\rho'}-a^{\rho'} b^{\rho}
+g^{\rho\rho'} m_\ell^2).
}
where $a$ and $b$ are the momenta of the lepton and antilepton ($a+b=P$), and $m_\ell$ their mass. 
As $m_\ell$ is much smaller than $M$ and $m$, we shall neglect it.

To perform the integration on the two-particle phase space, we shall use the following 
well-known relations~\cite{barger}
\eqs{\int d_2(PS) a^\alpha b^\beta=&\frac{\pi M^2}{24}\sqrt{\lambda}
\left(g^{\alpha \beta}\lambda +2 \frac{P^\alpha P^\beta}{M^2}\left[1+\frac{a^2}{M^2}+
\frac{b^2}{M^2}-2 \frac{(a^2-b^2)^2}{M^4}\right]\right)\\
\simeq&\frac{\pi M^2}{24}\left(g^{\alpha\beta}+2 \frac{P^\alpha P^\beta}{M^2}\right),\\
\int d_2(PS) a.b=&\int d_2(PS)a^\alpha b^\beta g_{\alpha\beta}  =\frac{\pi M^2}{24}\left(4+2\right)=\frac{\pi M^2}{4}.}
where $\lambda=
\lambda(1,a^2/M^2,b^2/M^2)=\lambda(1,m_\ell^2/M^2,m_\ell^2/M^2)=1-2 m_\ell^2/M^2\simeq 1$ if we
neglect the lepton mass.

Therefore, we have
\eqs{
B^{\rho\rho'}=& (ie)^2 \left[\pi M^2 g^{\rho \rho'}-8 \frac{\pi M^2}{24}\left(g^{\rho\rho'}+2 
\frac{P^\rho P^{\rho'}}{M^2}\right) \right]=(ie)^2 \frac{2\pi}{3} M^2\underbrace{\left[g^{\rho\rho'}- 
\frac{P^\rho P^{\rho'}}{M^2}\right]}_{\Delta^{\rho\rho'}}.
}
%**************************                                                                                                  
\subsection{Third sub-amplitude}

We simply have 
\eqs{
C^{\nu'\mu'}=(A^{\nu'\mu'})^\dagger
=A^{\nu'\mu'}.
}

%**************************
\subsection{Results}

Now that all quantities in \ce{eq:decomp_ampl_inv_decay} are determined, we can combine them. 
Thence, we obtain\footnote{Recall that the projector $\Delta_{\mu\nu}$ satisfies 
$\Delta_{\mu\nu}\Delta^{\mu\nu}=3$ and $\Delta_{\mu\nu}\Delta^{\mu\nu'}=\Delta_{\nu}^{\ \nu'}$.}:
\eqs{&\int \left|\bar{\cal M}\right|^2 d_2(PS)=\frac{1}{3}
\Delta_{\mu\mu'}\frac{-1}{M^4}\left(\frac{ e_Q}{\pi^2} N I(\Lambda,M,m)\right)^2\left((ie)^2 \frac{2\pi}{3} M^2\right)
  \Delta^{\mu\nu}g_{\nu\rho}
\Delta^{\rho\rho'}g_{\rho'\nu'}
\Delta^{\nu'\mu'}.
}

The leptonic decay width eventually reads from~\ce{eq:lept_width}
%($\Gamma_{\ell\ell}=\frac{1}{2 M}\frac{1}{(4\pi^2)}\int \left|\bar{\cal M}\right|^2 d_2(PS)$)
:
\eqs{\Gamma_{\ell\ell}=N^2 \frac{e^2}{12 \pi M^3}  \left(\frac{e_Q}{\pi^2} I(\Lambda,M,m)\right)^2.}

\subsubsection{Numerical results}

The value of $N$ is in practice obtained by replacing $\Gamma_{\ell\ell}$ by its measured value,
 $e_Q$ by $\frac{2e}{3}$ for $c$ quarks and $\frac{-e}{3}$ for $b$ quarks
and, finally, by introducing the value obtained for $I$ for the 
chosen value of $\Lambda$. The formulae to be used
for the different quarkonia are shown in \ct{tab:det_N}.

\begin{table}[h]
\begin{center}
\begin{tabular}{|c|c|c|}\hline
Meson &$\Gamma_{\ell\ell}$(keV)& $N$\\
\hline\hline
$J/\psi$      & $5.40$ &$\sqrt{\frac{(5.4 \times10^{-6}) 27 \pi^5 M_{J/\psi}^3}{e^4 
I(\Lambda,M_{J/\psi},m_c)^2}}$ \\
$\psi'$       & $2.12$ &$\sqrt{\frac{(2.12 \times10^{-6}) 27 \pi^5 M_{\psi'}^3}{e^4 
I(\Lambda,M_{\psi'},m_c)^2}}$\\
$\Upsilon(1S)$& $1.31$ &$\sqrt{\frac{(1.31 \times10^{-6}) 108 \pi^5 M_{\Upsilon(1S)}^3}{e^4 
I(\Lambda,M_{\Upsilon(1S)},m_b)^2}}$ \\
$\Upsilon(2S)$& $0.57$ &$\sqrt{\frac{(0.57 \times10^{-6}) 108 \pi^5 M_{\Upsilon(2S)}^3}{e^4 
I(\Lambda,M_{\Upsilon(2S)},m_b)^2}}$ \\
$\Upsilon(3S)$& $0.48$ &$\sqrt{\frac{(0.48 \times10^{-6}) 108 \pi^5 M_{\Upsilon(3S)}^3}{e^4 
I(\Lambda,M_{\Upsilon(3S)},m_b)^2}}$ \\
\hline
\end{tabular}
\caption{Formulas to be used to determine $N$.}\label{tab:det_N}
\end{center}
\end{table}

We sketch some plots of $N$ for the $J/\psi$ and for the $\Upsilon(1S)$ to show its dependence 
on $\Lambda$ and $m_Q$ and the value it actually takes. The values obtained are also given in tables 
in Appendix~\ref{sec:res_norm_tab} and will be used in the production calculations 
presented in the following.

We have chosen the range of size of the mesons, \ie~$\Lambda$, following BSE 
studies~\cite{costa} and other 
phenomenological models~\cite{Ivanov:2000aj,Ivanov:2003ge}, where
 the commonly accepted values for the $J/\psi$ are from $1$ to $2.4$ 
GeV, and for the $\Upsilon(1S)$ from $3.0$ to $6.5$ GeV.

For illustration, we show in~\ct{tab:res_norm_meth2} several values of $\Lambda$ 
obtained with BSE~\cite{costa}, accompanied with the vertex-function normalisation.

\begin{table}[h]
\begin{center}
\begin{tabular}{|c||c|}
\hline $m_c$ (GeV) & Vertex functions \\ 
\hline 
\begin{tabular}{c}
     \\
     \\
1.6  \\
1.7  \\
1.87 \\
\end{tabular}
&
\begin{tabular}{c|c}
Gaussian& Dipole \\
\hline
\begin{tabular}{c|c} 
$N$     & $\Lambda$ (GeV)   \\
 1.59   &  2.34       \\
  2.28   &  2.08    \\
 3.31   &  1.97       \\
\end{tabular}
&
\begin{tabular}{c|c} 
$N$     & $\Lambda$ (GeV)  \\
 1.33   &   2.87      \\
 1.87   &   2.54     \\
 2.76   &   2.31       \\
\end{tabular}         \\
\end{tabular} \\
\hline
\end{tabular}
\end{center}
\caption{Set of values for $m_c$, $\Lambda$ and $N$ obtained for the $J/\psi$ within BSE approach~\cite{costa}.}
\label{tab:res_norm_meth2}
\end{table}

To what concerns the $\psi'$, we give the results for $\Lambda$=1.8 GeV, $m_c=1.87$ GeV 
 and a gaussian vertex function (\ct{tab:N_psip}). 
The constraint for the node position (\ce{eq:ortho_1S_2S}) gives $a_{node}=1.46 $ GeV for
these values of $\Lambda$ and $m_c$.
The normalisation $N'=16.36$  will then be the value that we shall 
retain for the numerical applications. Note that the normalisation is strongly
dependent on the node position; it is even possible to make the 
decay width vanish, with the consequence of obtaining an infinite value for $N$. 

\begin{table}[h]
\begin{center}
\begin{tabular}{|c|c|}\hline
$a_{node}$ (GeV) &$N'$\\
\hline\hline
0.8     & 2.02   \\
1.46    & 16.36  \\
2.0     & 4.12   \\
\hline
\end{tabular}
\caption{Normalisation for the $\psi'$ as a function of $a_{node}$.}\label{tab:N_psip}
\end{center}
\end{table}

\begin{figure}[H]
\centering{\mbox{\includegraphics[width=12 cm]{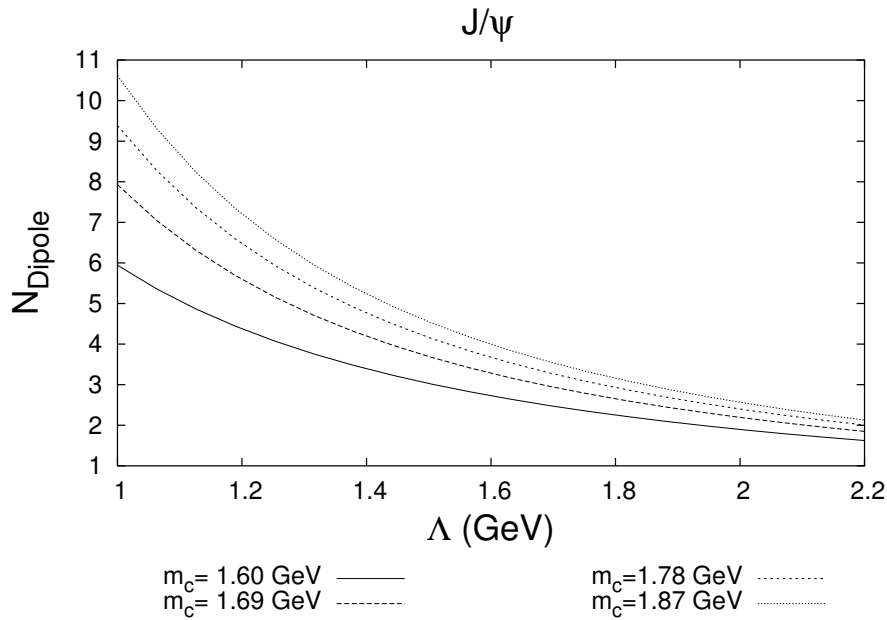}}}
\caption{Normalisation for a dipolar form for $J/\psi$ as a function of $\Lambda$.}
\label{fig:N_dip_jpsi_lam}
\end{figure}

\begin{figure}[H]
\centering{\mbox{\includegraphics[width=12 cm]{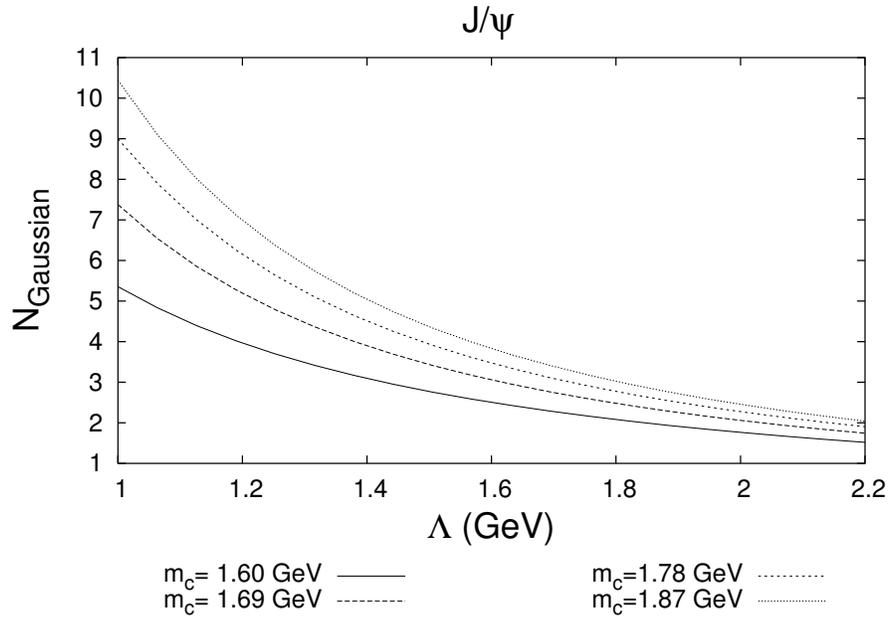}}}
\caption{Normalisation for a gaussian form for $J/\psi$ as a function of $\Lambda$.}
\label{fig:N_exp_jpsi_lam}
\end{figure}

\begin{figure}[H]
\centering{\mbox{\includegraphics[width=12 cm]{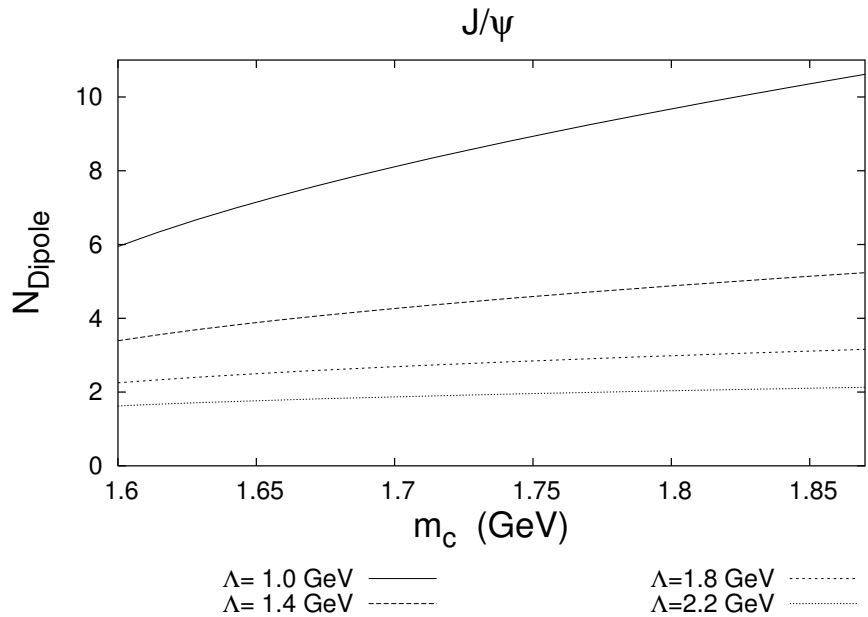}}}
\caption{Normalisation for a dipolar form for $J/\psi$ as a function of the 
$c$-quark mass.}\label{fig:N_dip_jpsi_mq}
\end{figure}

\begin{figure}[H]
\centering{\mbox{\includegraphics[width=12 cm]{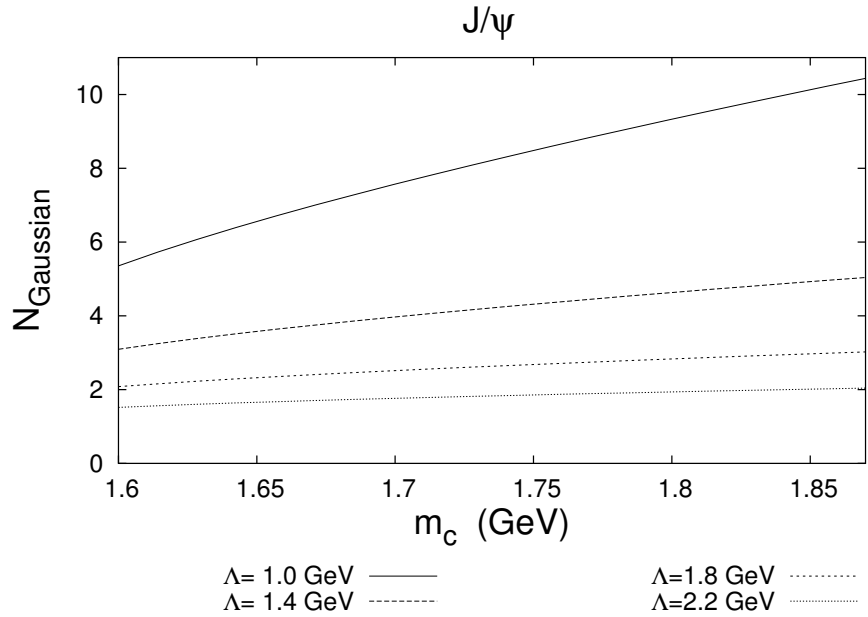}}}
\caption{Normalisation for a gaussian form for $J/\psi$ as a function of the 
$c$-quark mass.}\label{fig:N_exp_jpsi_mq}
\end{figure}

\begin{figure}[H]
\centering{\mbox{\includegraphics[width=12 cm]{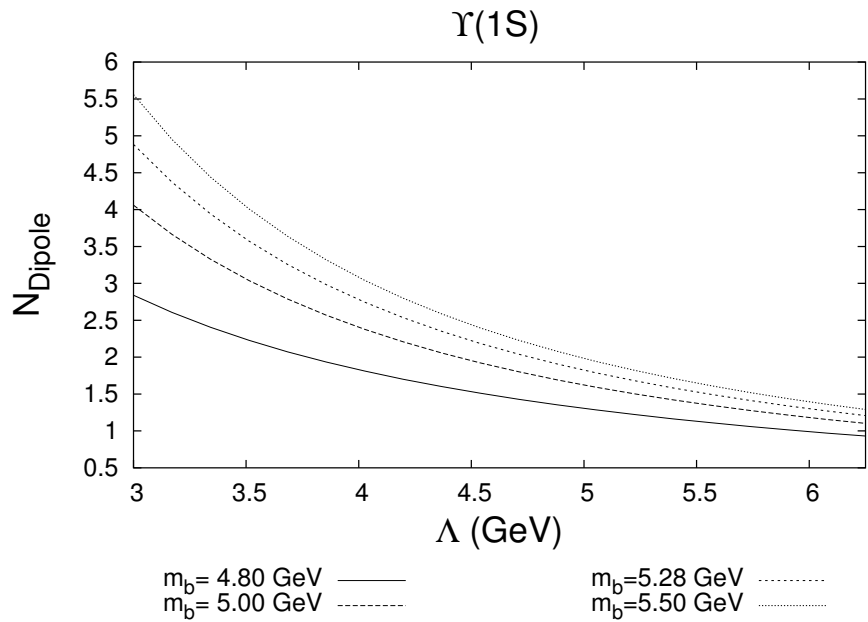}}}
\caption{Normalisation for a dipolar form for $\Upsilon(1S)$  as a function of $\Lambda$.}
\label{fig:N_dip_upsi_lam}
\end{figure}

\begin{figure}[H]
\centering{\mbox{\includegraphics[width=12 cm]{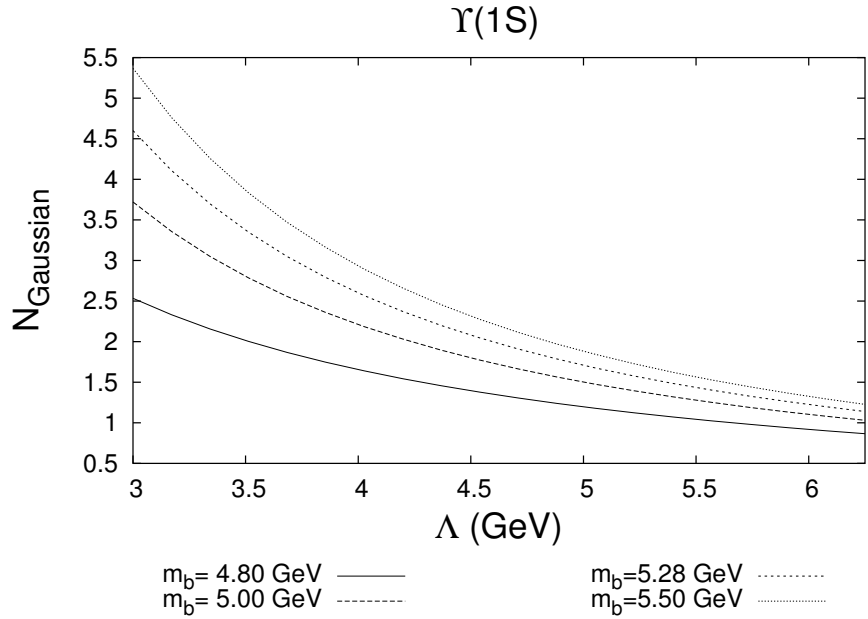}}}
\caption{Normalisation for a gaussian form for $\Upsilon(1S)$ as a function of $\Lambda$.}
\label{fig:N_exp_upsi_lam}
\end{figure}

\begin{figure}[H]
\centering{\mbox{\includegraphics[width=12 cm]{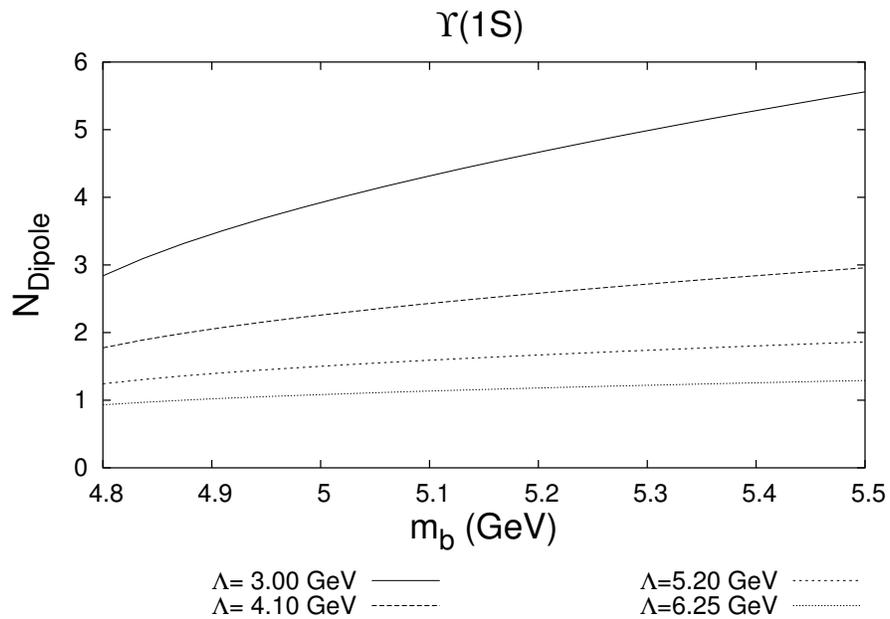}}}
\caption{Normalisation for a dipolar form for $\Upsilon(1S)$ as a function of the 
$b$-quark mass.}\label{fig:N_dip_upsi_mq}
\end{figure}

\begin{figure}[H]
\centering{\mbox{\includegraphics[width=12 cm]{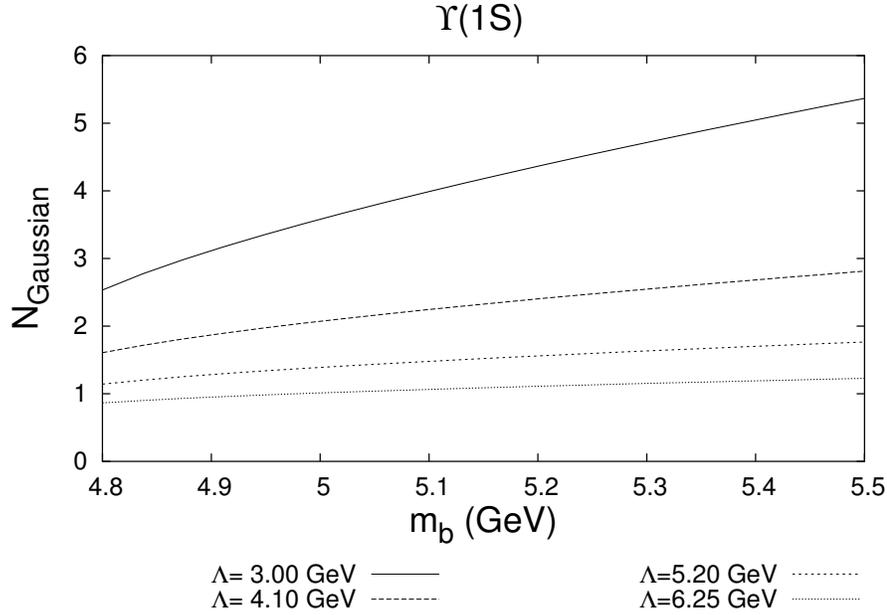}}}
\caption{Normalisation for a gaussian form for $\Upsilon(1S)$ as a function of the 
$b$-quark mass.}\label{fig:N_exp_upsi_mq}
\end{figure}

\index{leptonic decay width|)}

\section{Another point of view}

In appendix~\ref{sec:norm_alt_method}, we give some details about
 another method for determining the parameters of the vertex function. The latter relies
on a Wick rotation of the relative momentum in order to work in 
Euclidean space. This procedure is absolutely common among the community of 
physicists which employ the Bethe-Salpeter equation and the Dyson-Schwinger equation 
to obtain the mass, the normalisation and 
the decay width of mesons. 

We have been unable to justify this Wick rotation, especially in the case of the 
gaussian vertex function which reaches infinite values when $|k_0|^2\to \infty$ and 
$|{\cal I}m \ k_0| > |{\cal R}e \ k_0|$. However, as one can see in~\ct{tab:res_norm_meth2}, 
the results are  similar to the ones obtained by our method and considering that 
this more classical method also gives $\Lambda$, $m$ and $N$, 
we have thought useful for the reader to sketch some of
its results. This can be seen as an {\it a posteriori} 
justification for the values of the parameters that we used, such as 
$\Lambda=1.8$~GeV for $m_c$=1.87~GeV.
\index{Wick rotation} \index{Euclidean space}

\index{normalisation|)}

\clearemptydoublepage
\chapter{Bound states and gauge invariance}\label{ch:vertex_derivation}
\index{gauge!invariance|(}

As already mentioned in the previous chapter,
in order to tackle processes involving bound-states, 
we shall phenomenologically generalise the Feynman rules for point-like particles 
to bound states by introducing vertex functions or form factors, keeping by hypothesis
the same Dirac structure as for the corresponding point-like particle.

However, before using this formalism, one needs to deal with gauge invariance.
Indeed, if we consider the production of a bound state associated with
a photon or a gluon\footnote{This depends of course on the bound-state content: quark, 
lepton,\dots},
this may couple with the constituent before the binding but also after, even though the
bound state is neutral for the considered interaction. In the framework of Bethe-Salpeter 
equations (BSE)~\cite{Salpeter:1951sz}, one has to go back
to the details of the interactions occurring in the bound state. We have to take into consideration
which interactions occur and which type of possible diagram we shall have. 
This renders this study cumbersome~\cite{Gross:1987bu,Drell:1971vx}.

In the approach we follow, which we  refer to as ``phenomenological'',
gauge invariance is of course also broken. The vertex functions get different 
values for different kinematical
configurations and the different diagrams, contributing at a given order in the expansion
parameter, have different weights. As one knows, only the sum of the whole set of diagrams at a given 
order provides gauge-invariant results. Therefore gauge invariance is 
broken\footnote{To the exception of the decay process of the previous chapter.}.

The aim of this chapter is to introduce the reader to our  method to restore gauge invariance. 
We shall also choose a phenomenological point of view. Indeed our method 
is based on the introduction of
effective new 4-particle vertices as a supplement to the 3-particle ones.
We shall also show what constraints these effective new vertices have to satisfy.

First we shall present our derivation for a pseudoscalar particle, produced in association with a photon.
Compared to our final application to heavy quarkonium production (\eg~$J/\psi$ and $\Upsilon$), 
this brings two main simplifications to the discussion: the absence of colour factors and a 
single vector particle, namely the photon. 
Thence we shall give a similar derivation for vector-meson production with a gluon. 
Finally, we shall state that, as there is an unconstrained degree of freedom in this restoration, 
we are free to add any suitable contribution to the amplitude. This will be accompanied 
with the derivation of the  general form for these. 

As said above, we consider the production of a bound state ${\cal Q}$ associated with
a photon or a gluon, the simplest processes are therefore $q\bar q \to{\cal Q} \gamma$
or $q\bar q \to{\cal Q}  g$. We shall consider these here and this will be used in
the calculation of the imaginary part of the amplitude for $gg \to Vg$ as done in the next chapter.

\section{The simpler case of  $q\bar q \to X_0 \gamma$}

\begin{figure}[h]
\centering
\mbox{\subfigure[Direct]{\includegraphics[height=5.0cm]{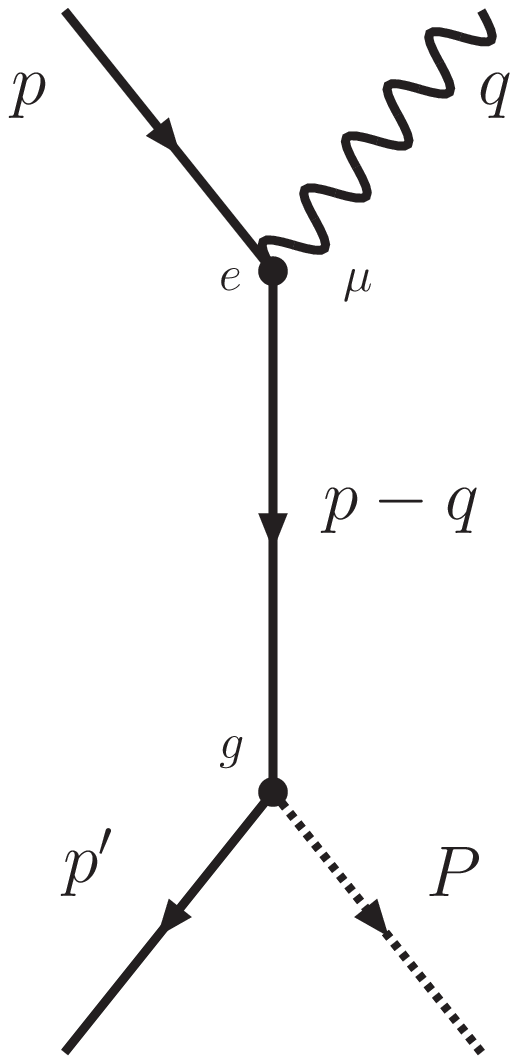}}\quad
      \subfigure[Crossed]{\includegraphics[height=5.0cm]{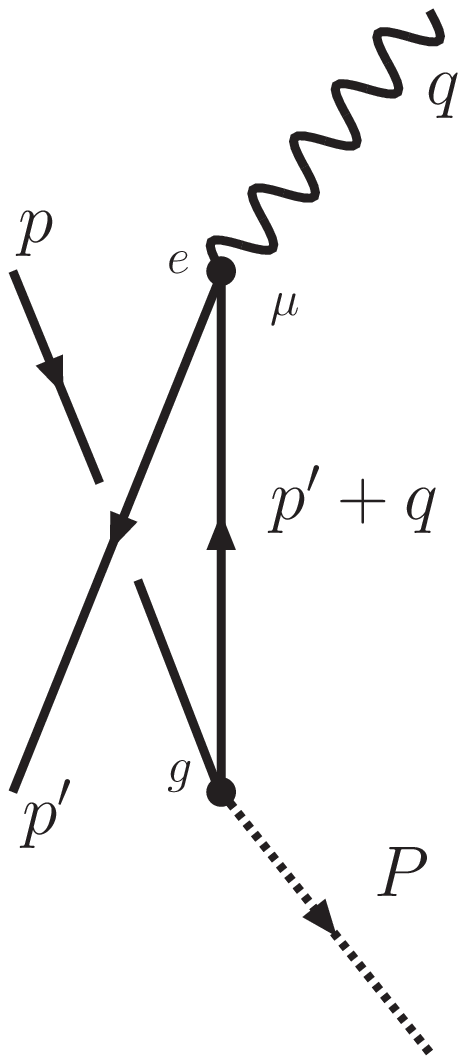}}}
\caption{Feynman diagrams for the leading order contributions for $q\bar q \to X_0 \gamma$.}
\label{fig:qqX0gam}
\end{figure}

We start from the usual amplitude at leading order (LO) for $q\bar q \to X_0 \gamma$, 
where $X_0$ stands for a pseudoscalar structureless (point) particle and $\gamma$ 
for a photon: 
\eqs{\label{eq:ampl_5_gen}
{\cal M} &=-i\,g\,e\,\underbrace{\bar u(p')}_{\bar v(-p')}V_{point}^\mu \ep^\star_\mu(q) u(p)\\
&= -i\,g\,e\,\bar u(p')\left(\frac{\ga 
((\ks p -\ks q)+m) \gmu}{(p-q)^2-m^2}+ \frac{\gmu((\ks p' +\ks q)+m)\ga}
{(p'+q)^2-m^2}\right) \ep^\star_\mu(q) u(p),}
where the momenta are defined as in Fig.~\ref{fig:qqX0gam}, $e$ is the QED charge and 
$g$ is the $q\bar q X_0$ coupling. To emphasise the different Dirac structures appearing 
in it, we can rewrite it as:
\eqs{\label{eq:vmu_5_gen}
\bar u(p')V_{point}^\mu \ep^\star_\mu(q) u(p)=\bar u(p')\left( \alpha \ga (\ks p -\ks q) \gmu 
+\beta \ga\gmu +\delta \gmu (\ks p' +\ks q)\ga
+\lambda \gmu \ga\right) \ep^\star_\mu(q) u(p).}
The factors $\alpha,\dots,\lambda$ are functions  of scalar products of momenta of the problem
and take the following values in this case:
\begin{center}
\boxed{\parbox{\linewidth}{
\eqs{\alpha_{point}\equiv\frac{1}{(p-q)^2-m^2}=P_1^{-1},\  \beta_{point}\equiv\frac{m}{(p-q)^2-m^2}=mP_1^{-1},\\
\delta_{point}\equiv\frac{1}{(p'+q)^2-m^2}=P_2^{-1},  \ \lambda_{point}\equiv\frac{m}{(p'+q)^2-m^2}=mP_2^{-1}.}
}}
\end{center}

As announced, should we generalise this amplitude in order to describe the production of
a bound state, the introduction of vertex functions would break gauge invariance. To restore the
latter, we may add to these two diagrams  a new contribution, constructed from 
a new vertex. This new vertex would account then for missing contributions, akin to direct interactions 
between the photon and the bound state, $X_0$. We shall call it $V_4^\mu$.

In turn, this novel vertex has to satisfy constraints, \eg~to have
the same symmetries as the sum of the 
previous contributions (which we shall refer to as the {\it point} ones). 
To enforce this symmetry constraint, we shall 
merely impose it on the partial amplitude built only from this 
effective vertex.

A first constraint comes from charge conjugation which can be assimilated to the crossing of 
the quark and the antiquark and reads (denoting the transformation $\cal T$): 
\eqs{\label{eq:sym_constr_5}
V^\mu (p,p',q,P,m)={\cal T}(V^\mu (p,p',q,P,m))=\gz V^{\mu\dagger}(-p',-p,q,P,-m)\gz.} 
Indeed, returning to the usual form for the amplitude (where we omitted the spinors $u$,
the polarisation vector $\ep$ and the other factors),
\eqs{V_{point}^\mu= \frac{\ga ((\ks p -\ks q)+m) \gmu}{(p-q)^2-m^2}+ 
\frac{\gmu((\ks p' +\ks q)+m)\ga}{(p'+q)^2-m^2},}
and using the following well-known relations, 
$(\ga)^\dagger=\ga$, $\gmu\ga=-\ga\gmu$, $(\gmu)^\dagger=\gz \gmu\gz$, we get:
\eqs{{\cal T} (V^\mu)=\gz V^{\mu\dagger}(-p',-p,q,P,-m)\gz&= \frac{\gmu ((-\ks p' -\ks q)-m) (-\ga)}
{(-p'-q)^2-m^2}+ 
\frac{(-\ga)((-\ks p +\ks q)-m)\gmu}{(-p+q)^2-m^2}\\
&=V^\mu (p,p',q,P,m).}

To sum up, the operation ${\cal T}(\cdot)= \gz (\cdot)^\dagger(-p',-p,q,P,-m)\gz$ 
should have no effect on the sought-after gauge-invariance restoring sub-amplitude:
\eqs{V_4^\mu(p,p',q,P,m)= \gz V_4^{\mu\dagger}(-p',-p,q,P,-m)\gz.}

In the following, instead of deriving directly the effective vertex restoring gauge invariance in cases
where the $q\bar q X_0$ vertex factors are simply the {\it point} ones multiplied by a 
function (this procedure
is illustrated in \cf{fig:BS_vertex_phenob}), we prefer to derive explicit and general constraints 
expressing how gauge invariance would be broken and consequently how to restore it taking into 
account the above symmetry.

\begin{figure}[H]
\centering
\includegraphics[width=12.0cm]{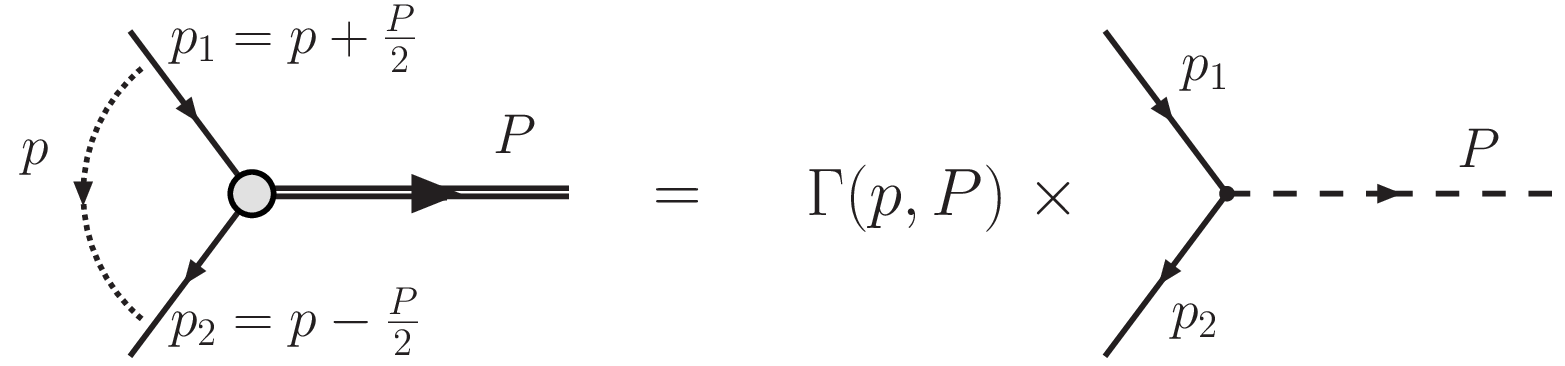}
\caption{Phenomenological vertex obtained by multiplying a {\it point} vertex, representing a structureless
particle, by a vertex function (or form factor).}
\label{fig:BS_vertex_phenob}
\end{figure} 

These explicit and general constraints will actually relate the parameters 
$\alpha,\dots,\lambda$. Having these constraints at hand, 
it will then be easier to construct our effective contribution from the general expression of 
Eq.~(\ref{eq:vmu_5_gen}). We shall simply input in \ce{eq:vmu_5_gen} $\alpha,\dots,\lambda$ 
fulfilling the obtained relations between them.

\subsection{Symmetry: general constraints}\label{subsec:sym_gen_case_pion}

In order to extract such relations for the symmetry of~\ce{eq:sym_constr_5}, we shall first 
rewrite the amplitude~\ce{eq:vmu_5_gen} by ordering the Dirac structures to insure
immediate identifications of the factors of the latter structures. 

\eqs{\label{eq:vmu_5_ordering}
V^\mu=& \alpha \ga (\ks p -\ks q) \gmu + \beta \ga\gmu +\delta \gmu (\ks p' +\ks q)\ga
+\lambda \gmu \ga\\=&
-\alpha \ga \gmu\ks p +2 \alpha \ga p^\mu-\alpha \ga \gmu \ks q-2\alpha \ga q^\mu
\beta \ga\gmu -\delta \ks p'\gmu\ga\\&+2\delta p'^\mu \ga+\delta\gmu\ks q\ga
+\lambda \gmu \ga\\ \overset{(*)}{=}&m \alpha\gmu\ga + 2\alpha \ga p^\mu
-\alpha \gmu \ga \ks q - 2 \alpha \ga q^\mu - \beta \gmu \ga -m \delta \gmu \ga 
+2\delta \ga p'^\mu \\&- \delta \gmu \ga \ks q + \lambda \gmu \ga.
}
The equality $(*)$ is obtained by means of Dirac equations for the on-shell quark
and antiquark: $\ks p  u(p)=m  u(p)$ and $\bar u(p') \ks p'= m \bar u(p')$ (or
$\bar v(-p') (-\ks p')= -m \bar v(-p')$).

By inspection of Eq.~(\ref{eq:vmu_5_ordering}), we see that we only have three independent 
Dirac structures: $\ga$, $\gmu\ga$ and $ \gmu \ga \ks q$. We have however to 
keep in mind that $p$, $p'$ and $q$ are independent, and that the identification
must be true for all their values. 

Schematically, we shall write:

\begin{table}[H]
\begin{center}
{\small\begin{tabular}{|c||c|}
\hline Structures & Factors  \\
\hline \hline 
$\ga$             &  $2\alpha p^\mu -2\alpha q^\mu + 2 \delta p'^\mu$	\\
$\gmu \ga$        &  $m \alpha -\beta -m\delta +\lambda$	\\
$\gmu \ga \ks q$  &  $-\alpha-\delta$ 	\\
\hline
\end{tabular}}
\end{center}
\end{table}

Now, we shall apply the transformation ${\cal T}(\cdot)=\gz (\cdot)^\dagger(-p',-p,q,V,-m)\gz
=\gz (\cdot)_{1\to2}^\dagger \gz$ to
the different structures and their coefficients. For the structures, we have

\eqs{ 
\ga &\to \gz (\ga)^\dagger \gz= -\ga \\
\gmu\ga &\to \gz (\gmu\ga)^\dagger \gz=\gz (\ga)^\dagger \gz \gmu \gz \gz= -\ga \gmu = \gmu \ga\\
\gmu\ga \ks q &\to \gz (\gmu\ga\ks q)^\dagger \gz=
\gz \gz \ks q\gz (\ga)^\dagger\gz \gmu\gz\gz
=\ks q \gmu \ga = \gmu \ga \ks q+2q^\mu \ga,}
and for the coefficients\footnote{The indices 1 and 2 stand for the two sets of variables
$(p,p',q,V,m)$ and $(-p',-p,q,V,-m)$ for which these factors are to be evaluated, {\it e.g.} 
$\alpha_1=\alpha(p,p',q,V,m)$ and $\alpha_2=\alpha(-p',-p,q,V,-m)$.}:
\eqs{
2\alpha_1 p^\mu -2\alpha_1 q^\mu + 2 \delta_1 p'^\mu &\to 
-2\alpha_2p'^\mu-2\alpha_2q^\mu -2 \delta_2 p^\mu\\
m \alpha_1 -\beta_1 -m\delta_1 +\lambda_1 &\to 
-m \alpha_2 -\beta_2 +m\delta_2 +\lambda_2\\
-\alpha_1-\delta_1 &\to -\alpha_2-\delta_2} 

Finally, we get,
\begin{table}[H]
\begin{center}
{\small\begin{tabular}{|c||c|c|}
\hline Structures & Factors 1 & Factors 2\\
\hline \hline 
$\ga$             &  $2\alpha_1 p^\mu -2\alpha_1 q^\mu + 2 \delta_1 p'^\mu$&
$2\alpha_2p'^\mu+2\alpha_2q^\mu +2 \delta_2 p^\mu-(\alpha_2+\delta_2)2q^\mu$\\
$\gmu \ga$        &  $m \alpha_1 -\beta_1 -m\delta_1 +\lambda_1$&
$-m \alpha_2 -\beta_2 +m\delta_2 +\lambda_2$	\\
$\gmu \ga \ks q$  &  $-\alpha_1-\delta_1$&$ -\alpha_2-\delta_2$	\\
\hline
\end{tabular}}
\end{center}
which gives 4 different identities:
\eqs{
\alpha_1 &=\delta_2 \\
\delta_1&=\alpha_2 \\
m \alpha_1 -\beta_1 -m\delta_1 +\lambda_1&=
-m \alpha_2 -\beta_2 +m\delta_2 +\lambda_2 \\
-\alpha_1-\delta_1&=-\alpha_2-\delta_2}
\end{table}
In short, we have
\begin{center}
\boxed{\parbox{\linewidth}{
\eqs{
\alpha_1 &=\delta_2 \\
\delta_1&=\alpha_2  \\
-\beta_1+\lambda_1&=-\beta_2 +\lambda_2}
}}
\end{center}

These three last equations contain all the constraints in terms of $\alpha,\dots,\lambda$,
coming from \ce{eq:sym_constr_5}. The first thing to notice is that the {\it point} case 
obviously fulfils these conditions. Indeed, noticing that 
$P_1\overset{1\to2}{\longrightarrow}P_2$,
\eqs{
1)\ &\delta_{point,1}=P_2^{-1}\Rightarrow \delta_{point,2}=P_1^{-1}=\alpha_{point,1},\\
2)\ &\alpha_{point,1}=P_1^{-1} \Rightarrow \alpha_{point,2 }=P_2^{-1}=\delta_{point,1},\\
3)\ &-\beta_{point,1}+\lambda_{point,1}=-m P_1^{-1} +m P_2^{-1}  \\&\Rightarrow 
-\beta_{point,2}+\lambda_{point,2}=-(-m)P_2^{-1}+(-m)P_1^{-1}=\lambda_{point,1}-\beta_{point,1}.
}
The most important result is that we now have a means to deduce in a  straightforward 
way whether or not given contributions, for instance the generalisation of these {\it point} 
contributions to 
bound state ones, have the symmetry required.

\subsection{Gauge invariance: general constraints}\label{subsec:gau_inv_gen_case_pion}

$\cal M$, the amplitude under consideration in~\ce{eq:ampl_5_gen}
involves $\ep^\star_\mu(q)$, the polarisation vector of the photon.
Being a four-vector, the latter possesses four degrees of freedom, one is removed by the
orthogonality condition:
\eqs{\ep^\star_\mu(q) q^\mu=0,}
three remain, whereas we know that photons have only two. Supposing the photon
travels in the $z$ direction, one possible choice for these two polarisations could be
\eqs{ 
\ep_x^\mu=(0,1,0,0) \hbox{ and }  \ep_y^\mu=(0,0,1,0)
}

The third degree of freedom can be expressed by noticing that any new vector obtained
by a shift in the photon momentum direction $q$ is also valid:
\eqs{
\ep'^{\mu}=\ep^\mu_{x,y}+\chi q^\mu,
}
A unique definition of the polarisation
vectors is then only achieved by imposing a further condition upon $\chi$. This condition
is equivalent to fixing the gauge (see \eg~\cite{gross}), or equivalently,
two different values of $\chi$ correspond to two different gauges
\footnote{Let $\Lambda$ be the gauge-transformation parameter such that 
$A'^\mu =A^\mu+\partial^\mu \Lambda(x)$. The wave function transforms like $\psi'=e^{i \Lambda(x)}\psi$ 
and
the polarisation vector like $\ep^\mu e^{ikx}=\ep^\mu e^{ikx}+q^\mu \int \Lambda(x) e^{ikx} dx$. We
may then identify $\chi$ with $\int \Lambda(x) e^{ikx} dx$.}
.

Gauge invariance of an amplitude $\cal M $ involving $n$ external photons is therefore 
realised by
the absence of $\chi$ in its expression. Hence, as $\chi$ appears solely in association
with\footnote{$\mu$ is the index of the photon for which we look at gauge-invariance 
or current conservation.}
$q^\mu$, {\it a  necessary condition for a gauge-invariant result} resides
in the current-conservation condition: \index{current conservation|see{gauge invariance}}

\eqs{
V_{\mu_1\dots\mu_n}q_i^{\mu_i}=0, \, \forall i,
}
defining $V_{\mu_1\dots\mu_n}$  by  ${\cal M}= V_{\mu_1\dots\mu_n}
\underset{i}{\Pi}\ep^{\mu_i}(q_i)$.

To what concerns $\cal M$ of ~\ce{eq:ampl_5_gen}, the requested condition reads
\eqs{
V^\mu q_\mu=0.
}

We shall perform the same ordering as before but for
the amplitude $V^\mu$ contracted with $q_\mu$, the momentum of the photon.  
As the contraction  vanishes, it will be in turn expressed in terms of 
identities for the coefficients $\alpha,\beta,\delta,\lambda$.

\eqs{0=V^\mu q_\mu=& \alpha \ga (\ks p -\ks q) \ks q + \beta \ga\ks q +\delta \ks q 
(\ks p' +\ks q)\ga+\lambda \ks q \ga \\
=& -\alpha \ga \ks q\ks p +2 \alpha \ga p.q-\alpha \ga q^2+
\beta \ga\ks q -\delta \ks p'\ks q\ga\\&+2\delta p'.q \ga+\delta q^2\ga
-\lambda \ga\ks q
\\=& -m \alpha\ga \ks q + 2\alpha  p.q \ga
-\alpha q^2 \ga + \beta  \ga \ks q+m \delta \ga \ks q 
+2\delta p'.q \ga + \delta q^2\ga -\lambda  \ga \ks q
}

Finally we get,
\begin{table}[H]
\begin{center}
{\small\begin{tabular}{|c||c|c|}
\hline Structures & Factors \\
\hline \hline 
$\ga$        &$2\alpha p.q -\alpha q^2 +\delta q^2+2\delta p'.q  $\\
$\ga \ks q$  &$-m \alpha +\beta +m \delta -\lambda$\\
\hline
\end{tabular}}
\end{center}
\end{table}

And we can write --the two factors (right column of the last table)  being identically zero--

\begin{center}
\framebox{\parbox{\linewidth}{
\eqs{\label{eq:gi_pion_1}
-\alpha P_1 +\delta P_2=0}
\eqs{\label{eq:gi_pion_2}
\beta-\lambda= m(\alpha-\delta)}
}}
\end{center}

We may check that the {\it point} case 
follows these relations. This is straightforward since

\eqs{
1)\ &-\alpha_{point} P_1 +\delta_{point} P_2=-P_1^{-1} P_1 +P_2^{-1} P_2=0,\\
2)\ &\beta_{point}-\lambda_{point}=m P_1^{-1}-m P_2^{-1}= m(\alpha_{point}-\delta_{point}).
}

\subsection{Restoring gauge invariance for the ``phenomenological approach''}
\label{subsec:gau_inv_classical_case_pion}

Having at our disposal the latter constraints, we are now ready to construct an effective
4-particle vertex that will restore gauge invariance for the cases where a bound-state vertex function 
enters the calculation.

The values of $\alpha,\beta,\delta,\lambda$ for our ``phenomenological 
approach'', where we introduce a 3-particle vertex function $\Gamma$ ($V_3= \Gamma \ga$), 
are easily shown to be:
\eqs{ \alpha=\frac{\Gamma_1}{P_1},\ \beta=\frac{m\Gamma_1}{P_1},
\delta=\frac{\Gamma_2}{P_2},\lambda=\frac{m\Gamma_2}{P_2},}
where $\Gamma_1=\Gamma(2p-q)$ and $\Gamma_2=\Gamma(2p'+q)$. This amounts to
the following modifications compared to the point case:
\eqs{\alpha=\alpha_{point}\times \Gamma_1,\ \beta=\beta_{point}\times \Gamma_1,\
\delta=\delta_{point}\times\Gamma_2,\ \lambda=\lambda_{point}\times\Gamma_2.}

We directly see that \ce{eq:gi_pion_2} is automatically satisfied 
by this choice contrarily to \ce{eq:gi_pion_1} for $\Gamma_1\neq \Gamma_2$. 
This is actually true as long as $\beta=m\alpha$ and $\lambda=m\delta$.
{\it We have thus expressed through one single relation {\rm(\ce{eq:gi_pion_1})} involving 
one single Dirac structure {\rm($\gamma^5$)} the breaking of gauge invariance.} 

The fact that \ce{eq:gi_pion_2} is satisfied shows us that either the effective 4-particle 
vertex should not contribute to this equation -- no 
term in $\ga \gmu $ at all (before contraction with $q^\mu$)~--, or that its 
contribution should be transverse, \ie~should vanish when contracted with $q^\mu$.

Concerning the crossing symmetry, let us first analyse the constraint $\alpha_1=\delta_2$:
the symmetry of the vertex function, $\Gamma_1 \overset{1\to2}{\longrightarrow}\Gamma_2$, 
and $\alpha_{point,1}=\delta_{point,2}$ imply that the latter also holds 
in the  ``phenomenological approach''. The same is true for the other symmetry relations 
as can easily be seen.

Let us turn back to the derivation of the new vertex.
For simplicity, we shall first introduce terms that explicitly do not 
contribute to~\ce{eq:gi_pion_2}. 
On the other hand, Eq.~(\ref{eq:gi_pion_1}) is not satisfied by our approach and 
corrections from the 4-particle vertex should then appear in this equation, 
\ie~with a $\ga$ Dirac matrix for sole Dirac structure. This implies that it should be of the general
form $k^\mu\ga $. The vertex should as well satisfy the 
symmetry relation $V_4^\mu (p,p',q,V,m)={\cal T}(V_4^\mu)=\gz V_4^\dagger(-p',-p,q,V,-m)\gz$.

\subsubsection{First solutions}

Setting $k^\mu$ to $(p+p')^\mu$ in $k^\mu \ga $ , defining  $\eta$ as a multiplicative factor, we can take
$2\eta(p+p')^\mu \ga$ which can be rewritten as:
\eqs{2\eta(p+p')^\mu \ga=\eta(2p'^\mu\ga-m\gmu\ga
+2p^\mu\ga-\ga \gmu m)=\eta(\gmu \ks p' \ga +\ga \ks p \gmu).}

As we have seen, under the transformation $\cal T$, $\ga$ changes to 
$-\ga$ and $\eta_1(p+p')$ will give $-\eta_2(p'+p)$, therefore, we have to 
impose that 
\eqs{\label{eq:eta_sym_pion_1}\eta_1=\eta_2.}

Concerning gauge invariance, a similar equation to~Eq.~(\ref{eq:gi_pion_1}), with this new term
added and with the values of $\alpha$ and $\delta$ for the ``phenomenological approach'', reads
\eqs{\label{eq:gi_pion_1b}
0=-\alpha P_1 +\delta P_2+\eta(p+p').q \ga
=-\frac{\Gamma_1}{P_1}P_1+\frac{\Gamma_2}{P_2}P_2+2\eta(p+p').q \ga,}
Solving for $\eta$, this determines the effective vertex:
\eqs{\label{eq:v_eff_pion_1}
V_{4}^\mu=2\eta(p+p')^\mu \ga =\frac{\Gamma_1-\Gamma_2}{(p+p').q}\ga(p+p')^\mu
}

We also see that $\eta$ satisfies the symmetry relation Eq.~(\ref{eq:eta_sym_pion_1})
as $\Gamma_1$ transforms into $\Gamma_2$.

The similar vertex 
\eqs{\label{eq:v_eff_pion_2}
V_{4}^\mu= \frac{\Gamma_1-\Gamma_2}{(p-p').q}\ga(p-p')^\mu
}
will suit also. Contracted with $q^\mu$, it indeed gives 
$(\Gamma_1-\Gamma_2)\ga$ as required to guarantee
gauge invariance. The symmetry constraint requires $\eta$ to
be odd in $\cal T$. This is also verified as
$\eta=\frac{\Gamma_1-\Gamma_2}{2(p-p').q}$.

If we consider a virtual photon, \ie with $q^2\neq 0$, we can also choose
\eqs{\label{eq:v_eff_pion_3}
V_{4}^\mu= \frac{\Gamma_1-\Gamma_2}{q^2}\ga q^\mu,
}
nevertheless $V_{4}^\mu \ep(q)_\mu=0$ due to $\ep_\mu(q) q^\mu=0$.

\subsubsection{Singularity structure}

Now that we have seen that different choices are possible, let us turn to
another issue, \ie~the singularity structure of the 4-particle vertex. Indeed
the above mentioned vertices have all different denominators, with a different
singularity structure, whereas we shall demand that the gauge-invariance restoring contribution does not 
introduce new singularities in the problem. The easiest way to insure this
is to limit our choice to vertices
with the same denominators as the ones already in the problem.

Of course any combination of the previous solutions is also correct. Thus any
structure that would  not contribute to~\ce{eq:gi_pion_1} and \ce{eq:gi_pion_2} and with
the correct symmetry can be added to them. This freedom will be discussed 
thoroughly for the  $q\bar q \to V_0 g$ case
(see section~\ref{sec:auton_contrib}).

Since all structures similar to $\frac{\Gamma_1-\Gamma_2}{k.q}\ga k^\mu$ will satisfy
the gauge-invariance constraint, it is sufficient to concentrate on its 
symmetry. Since the odd character of $(\Gamma_1-\Gamma_2)$ is compensated by
$\ga$, it is obvious that any combination of the vectors of the problem
will yield a correct $k$ (since $k$ is both in the numerator and in the
denominator). 

If we choose $k^\mu=p^\mu (p'.q)+p'^\mu (p.q)$, we get
\eqs{\label{eq:v_eff_pion_4}
V_{4}^\mu=(\Gamma_1-\Gamma_2) \frac{p^\mu (p'.q)+p'^\mu (p.q)}{ 2(p'.q)(p.q)}\ga= 
(\Gamma_1-\Gamma_2)\left(\frac{p^\mu }{2(p.q)}+\frac{p'^\mu }{2(p'.q)}\right)
}

Except for the term $q^2$, the denominators of this solution are equivalent to the
propagators of the problem. For $q^2=0$, they do not bring any new singularities in the 
problem. But we look for a more general case\dots

A slight different choice for $k^\mu$ gives:
\eqs{\label{eq:v_eff_pion_4b}
V_{4}^\mu=&(\Gamma_1-\Gamma_2) \frac{(2p-q)^\mu (2p'.q+q^2)+(2p'+q)^\mu (2 p.q-q^2)}
{ 2(2p'.q+q^2)(2 p.q-q^2)}\ga\\=& 
(\Gamma_1-\Gamma_2)\left(\frac{(-2p+q)^\mu }{2P_1}+\frac{(2p'+q)^\mu}{2P_2}\right)\ga
}

The great advantage of this vertex, contrarily to Eq.~(\ref{eq:v_eff_pion_4}), is that 
it does not introduce new singularities in the problem for any value of $q^2$. It can therefore 
be used in model calculation for pion structure functions where the probing
photon is off-shell~\cite{pdf_pion,gpd_pion}.

%*************************************************************************
%*************************************************************************
%*************************************************************************
%*************************************************************************

\section{Derivation of an effective vertex for $q\bar q \to V_0 g$}

\begin{figure}[H]
\centering
\mbox{\subfigure[Direct]{\includegraphics[height=5.0cm]{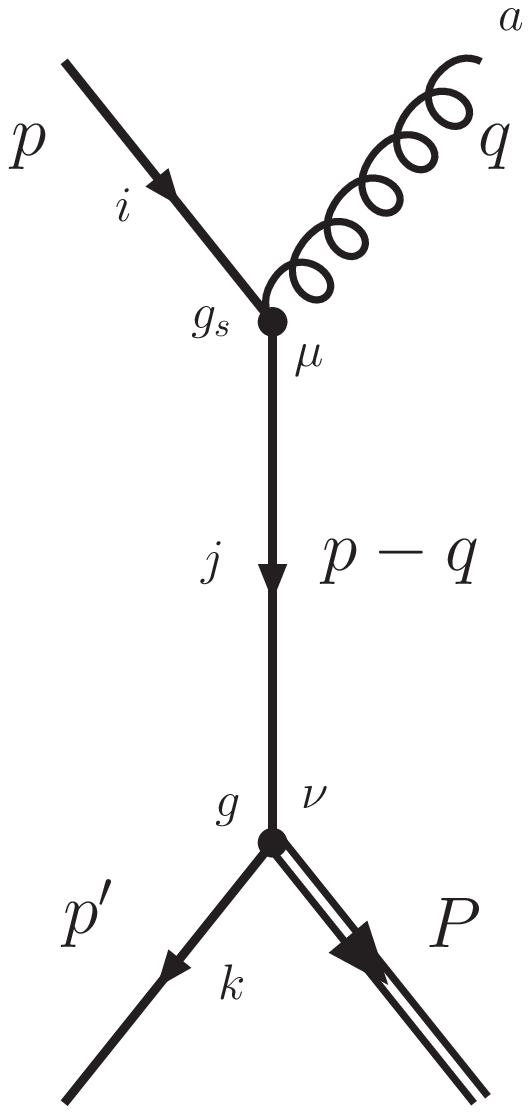}}\quad
      \subfigure[Crossed]{\includegraphics[height=5.0cm]{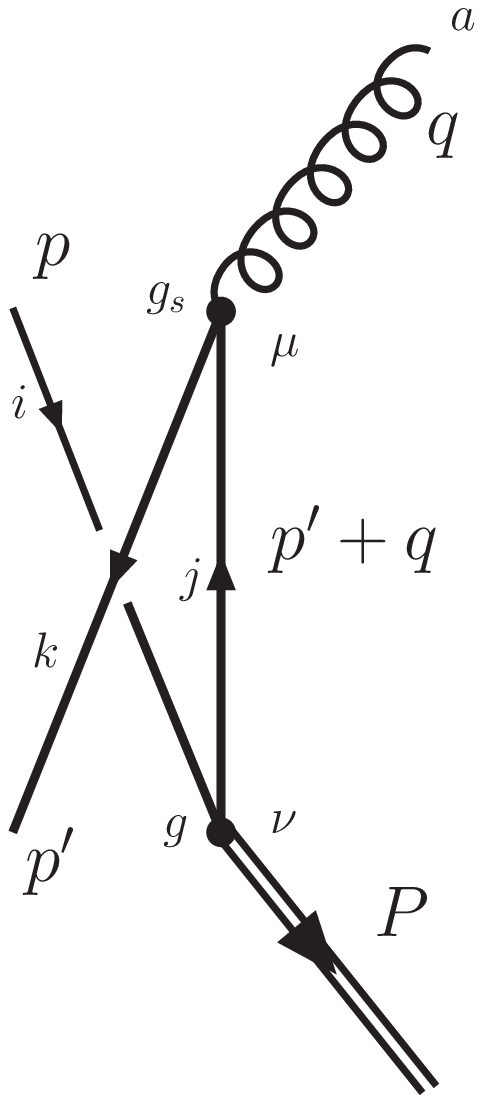}}}
\caption{Feynman diagrams for the leading order contribution of $q\bar q \to V_0 g$.}
\label{fig:qqV0g}
\end{figure}

We shall here deal with a vector particle and with a gluon
field. We start from the amplitude at LO which we rewrite in a simplified form comparable to
\ce{eq:ampl_5_gen}: 
\eqs{\label{eq:ampl_v_gen}
{\cal M}&=-i\,g\,g_s\,T^a_{ji} \delta_{kj}
\underbrace{\bar u(p')}_{\bar v(-p')}V^{\mu\nu} \ep^\star_\mu(q) \ep^\star_\nu(P) u(p) \\
&=-i\,g\,g_s\,T^a_{ki} 
\bar u(p')\left(\frac{\gnu ((\ks p -\ks q)+m) \gmu}{(p-q)^2-m^2}+ 
\frac{\gmu((\ks p' +\ks q)+m)\gnu}{(p'+q)^2-m^2}\right) \ep^\star_\mu(q) \ep^\star_\nu(P)u(p),}
where the momenta are defined as in Fig.~\ref{fig:qqV0g}, $g_s$ is the QCD coupling,
$g$ the $q\bar q V^0$ coupling, $T^a$ the colour matrices 
%(see appendix~\ref{ch:conventions})
 and $\delta_{kj}$ the colour factor for a singlet state. 
To emphasise the different structures\footnote{Note that these are not necessarily independent.} 
appearing in it, we can rewrite it as  :
\eqs{\label{eq:vmu_v_gen}
\bar u(p')V^{\mu\nu} \ep^\star_\mu(q) \ep^\star_\nu(P)u(p)=
\bar u(p')
&\left( \alpha \gnu (\ks p -\ks q) \gmu + \beta \gnu\gmu +\right.\\
& \ \left.\delta \gmu (\ks p' +\ks q)\gnu 
+\lambda \gmu \gnu+\kappa g^{\mu\nu} \right) \ep^\star_\mu(q) \ep^\star_\nu(P) u(p).}
Note that we have added a new structure $g^{\mu\nu}$ to get a more general discussion in the following.

The factors $\alpha,\dots,\kappa$, then take the following values in the QCD perturbative 
case for point-like particles:
\begin{center}
\boxed{\parbox{\linewidth}{
\eqs{\label{eq:val_point_vect}
\alpha_{point}=\frac{1}{(p-q)^2-m^2}\equiv P_1^{-1},&\  \beta_{point}=\frac{1}{(p-q)^2-m^2}\equiv m P_1^{-1},\\
\delta_{point}=\frac{1}{(p'+q)^2-m^2}\equiv P_2^{-1},&  \ \lambda_{point}=\frac{m}{(p'+q)^2-m^2}\equiv m P_2^{-1}, \\
\kappa_{point}&=0 ,}
}}
\end{center}
with the short notations for the propagator denominators ($p^2=m^2,p'^2=m^2,q^2=0$),
\eqs{P_1^{-1}=\frac{1}{(p-q)^2-m^2}=\frac{-1}{2p.q} \hbox{ and} \ P_2^{-1}=\frac{1}{(p'+q)^2-m^2}=\frac{1}{2p'.q}.}

Similarly to the pseudoscalar case, we work out a first constraint from the quark
crossing symmetry, which reads: 
\eqs{\label{eq:sym_const_vect}V^{\mu\nu} (p,p',q,P,m)=-\gz V^{\mu\nu\dagger}(-p',-p,q,P,-m)\gz.} 
Indeed, returning to the usual form for the amplitude (where we omitted the spinors $u$,
the polarisation vectors $\ep$ and the other factors),
\eqs{V^{\mu\nu}= \frac{\gnu ((\ks p -\ks q)+m) \gmu}{(p-q)^2-m^2}+ 
\frac{\gmu((\ks p' +\ks q)+m)\gnu }{(p'+q)^2-m^2},}
and using the following well-known relations
, $\gmu\gnu =-\gnu \gmu+2g^{\mu\nu}$, $(\gmu)^\dagger=\gz \gmu\gz$, we get:
\eqs{{\cal T} (V^{\mu\nu})=
\gz V^{\mu\nu \dagger}(-p',-p,q,P,-m)\gz&= \frac{\gmu ((-\ks p' -\ks q)-m) \gnu}
{(-p'-q)^2-m^2}+ 
\frac{\gnu ((-\ks p +\ks q)-m)\gmu}{(-p+q)^2-m^2}\\
&=-V^{\mu\nu} (p,p',q,P,m).}

Note the change of sign compared to the pseudoscalar case (\ce{eq:sym_constr_5}).

\subsection{Symmetry: general constraints}\label{subsec:sym_gen_case_vect}

As for the pseudoscalar case, we shall now extract from \ce{eq:sym_const_vect} relations between
the parameters $\alpha,\dots,\kappa$. Using the relations 
$\gmu \ks q \gnu=2 \gmu q^\nu - \gmu \gnu \ks q=2 \gmu q^\nu - 2 g^{\mu \nu} \ks q+\gnu \gmu \ks q$,
\eqs{\label{eq:vmunu_ordering}
V^{\mu\nu}&= \alpha \gnu  (\ks p -\ks q) \gmu + \beta \gnu \gmu +\delta \gmu (\ks p' +\ks q)\gnu 
+\lambda \gmu \gnu +\kappa g^{\mu\nu}
\\&=2 \alpha \gnu  (p-q)^\mu +\alpha \gmu \gnu (\ks p -\ks q) 
-2 \alpha g^{\mu\nu} (\ks p -\ks q)- \beta \gmu \gnu +2 \beta g^{\mu\nu} +\lambda \gmu \gnu 
\\ &+2 \delta p'^\mu \gnu -\delta \ks p' \gmu \gnu +2\delta  \gmu q^{\nu} - \delta \gmu \gnu \ks q
+\kappa g^{\mu\nu}
\\&\overset{(*)}{=} 2\alpha \gnu (p-q)^\mu + m \alpha \gmu\gnu-\alpha \gmu \gnu \ks q 
-2 m \alpha g^{\mu\nu}+2\alpha g^{\mu\nu} \ks q-\beta \gmu \gnu +2  \beta g^{\mu\nu}
\\ &+\lambda\gmu\gnu +2 \delta p'^\mu \gnu -m \delta \gmu\gnu
+2\delta \gmu q^{\nu}-\delta \gmu\gnu \ks q+\kappa g^{\mu\nu}.
}
The equality $(*)$ is obtained by means of Dirac equations for the on-shell quark
and antiquark: $\ks p  u(p)=m u(p)$ and $\bar u(p') \ks p'= m u(p')$.

By inspection of Eq.~(\ref{eq:vmunu_ordering}), we see that we have to deal with six  
Dirac structures: $\gmu $, $\gnu $, $\gmu\gnu $, $ \gmu \gnu  \ks q$, $g^{\mu\nu}$ and 
$g^{\mu\nu}\ks q$;  we also have to keep in mind that $p$, $p'$ and $q$ are 
independent, and that the identification is expected to be true for all their values. 

Schematically, to help the identification, we shall write:

\begin{center}
{\small\begin{tabular}{|c||c|}
\hline Structures & Factors  \\
\hline \hline 
$\gnu $             &  $2\alpha p^\mu -2\alpha q^\mu + 2 \delta p'^\mu$	\\
$\gmu $             &  $2 \delta q^\nu$                                               \\
$\gmu \gnu $        &  $m \alpha -\beta -m\delta +\lambda$	\\
$\gmu \gnu  \ks q$  &  $-\alpha-\delta$ 	\\
$g^{\mu\nu}$        &  $-2 m \alpha +2 \beta +\kappa$ \\
$g^{\mu\nu}\ks q$   &  $2\alpha$ \\
\hline
\end{tabular}}
\end{center}

Now, we shall apply the transformation $-{\cal T}(\cdot)=-\gz (\cdot)^\dagger(-p',-p,q,P,-m)\gz$ to
the different structures and their coefficients. We obtain
\eqs{ \gnu  &\to -\gz (\gnu )^\dagger \gz= -\gnu,\\
\gmu  &\to -\gz (\gmu )^\dagger \gz= -\gmu,  \\
\gmu\gnu  &\to 
-\gnu  \gmu = \gmu \gnu -2 g^{\mu\nu},\\
\gmu\gnu  \ks q &\to 
-\ks q \gnu \gmu =\gnu \ks q\gmu  - 2q^\nu \gmu  = \gmu \gnu  \ks q-2 g^{\mu\nu} \ks q , \\
g^{\mu\nu}&\to -g^{\mu\nu},
\\  g^{\mu\nu}\ks q&\to -g^{\mu\nu}\ks q,}
and for the coefficients\footnote{The minus sign of the
transformation $-\gz (\cdot)_{1\to 2}^{\dagger}\gz$
is already taken into account in the transformation laws for the structures.}:
\eqs{
2\alpha_1 p^\mu -2\alpha_1 q^\mu + 2 \delta_1 p'^\mu &\to 
-2\alpha_2 p'^\mu -2\alpha_2 q^\mu - 2 \delta_2 p^\mu,   \\ 
2 \delta_1 q^\nu &\to 2 \delta_2 q^\nu,                 
\\
m \alpha_1 -\beta_1 -m\delta_1 +\lambda_1 &\to   
-m \alpha_2 -\beta_2 +m\delta_2 +\lambda_2,          \\
-\alpha_1-\delta_1 &\to  -\alpha_2-\delta_2,             \\
-2 m \alpha_1 +2 \beta_1 +\kappa_1 &\to  
2 m \alpha_2 +2 \beta_2 +\kappa_2,                     \\
2\alpha_1 &\to 2 \alpha_2.
}

Finally, we get,
\begin{center}
{\small\begin{tabular}{|c||c|c|}
\hline Structures & Factors 1 & Factors 2\\
\hline \hline 
$\gnu $             &  $2\alpha_1 p^\mu -2\alpha_1 q^\mu + 2 \delta_1 p'^\mu$	
&$2\alpha_2 p'^\mu -2\delta_2 q^\mu + 2 \delta_2 p^\mu$\\
$\gnu $             &  $2 \delta_1 q^\nu$                                               
&$2 \alpha_2 q^\nu$\\
$\gmu \gnu $        &  $m \alpha_1 -\beta_1 -m\delta_1 +\lambda_1$	
&$-m \alpha_2 -\beta_2 +m\delta_2 +\lambda_2$\\
$\gmu \gnu  \ks q$  &  $-\alpha_1-\delta_1$ 	
&$-\alpha_2-\delta_2$\\
$g^{\mu\nu}$        &  $-2 m \alpha_1 +2 \beta_1 +\kappa_1$ 
&$-2 m \delta_2 -2 \lambda_2 -\kappa_2$\\
$g^{\mu\nu}\ks q$   &  $2\alpha_1$ 
&$2\delta_2$\\
\hline
\end{tabular}}
\end{center}
which gives the following 8 identities relating $\alpha,\dots,\kappa$ out of which
5 are independent: 
\eqs{
2\alpha_1 &=2\delta_2, \\
-2\alpha_1&=-2\delta_2, \\
2\delta_1&=2\alpha_2, \\
2\delta_1&=2\alpha_2,  \\
m \alpha_1 -\beta_1 -m\delta_1 +\lambda_1&=
-m \alpha_2 -\beta_2 +m\delta_2 +\lambda_2, \\
-\alpha_1-\delta_1&=-\alpha_2-\delta_2,\\
-2m\alpha_1+2\beta_1+\kappa_1&=-2m\delta_2-2\lambda_2-\kappa_2\\
2\alpha_1&=2\delta_2.
}

These relations reduce to 4 conditions:

\begin{center}
\boxed{\parbox{\linewidth}{\eqs{
\alpha_1 &=\delta_2, \\
\delta_1&=\alpha_2, \\
-\beta_1+\lambda_1&=-\beta_2 +\lambda_2,\\
2(\beta_1+\lambda_2)&=-(\kappa_1+\kappa_2)
}}}
\end{center}

Let us check that the {\it point} case 
fulfils these conditions. 
\eqs{
1)\ &\delta_{point,1}=P_2^{-1}\Rightarrow \delta_{point,2}=P_1^{-1}=\alpha_{point,1},\\
2)\ &\alpha_{point,1}=P_1^{-1} \Rightarrow \alpha_{point,2 }=P_2^{-1}=\delta_{point,1},\\
3)\ &-\beta_{point,1}+\lambda_{point,1}=-m P_1^{-1} +m P_2^{-1}  \\&\Rightarrow 
-\beta_{point,2}+\lambda_{point,2}=-(-m)P_2^{-1}+(-m)P_1^{-1}=\lambda_{point,1}-\beta_{point,1},\\
4)\ &-(\kappa_{point,1}+\kappa_{point,2})=0=2(\beta_{point,1}+\lambda_{point,2})=m P_1^{-1}+(-m)P_1^{-1}.
}
We have as in the pseudoscalar case a means to deduce in a straightforward 
way whether or not given contributions, for instance the generalisation of these {\it point} 
contributions to bound state ones, have the symmetry required.

\subsection{Gauge invariance: general constraints}\label{subsec:gau_inv_gen_case_vect}

If we now turn to gauge invariance, we shall perform the same ordering but for
the amplitude $V^{\mu\nu}$ contracted with $q_\mu$, the momentum of the gluon. 
Current conservation implies that the contraction should then vanish. This annulment
is in turn expressed in terms of identities for the coefficients 
$\alpha,\beta,\delta,\lambda,\kappa$.

\eqs{\label{eq:GI_vect_1}
0=V^{\mu\nu} q_\mu=& \alpha \gnu  (\ks p -\ks q) \ks q + \beta \gnu \ks q +\delta \ks q 
(\ks p' +\ks q)\gnu +\lambda \ks q \gnu +\kappa q^\nu \\
=& -\alpha \gnu  \ks q\ks p +2 \alpha \gnu  p.q-\alpha \gnu  q^2+
\beta \gnu \ks q -\delta \ks p'\ks q\gnu \\&+2\delta p'.q \gnu +\delta q^2\gnu 
+2\lambda q^\nu-\lambda \gnu \ks q+\kappa q^\nu
\\=& -m \alpha\gnu\ks q  + 2\alpha  p.q \gnu
-\alpha q^2 \gnu  + \beta  \gnu  \ks q-2 m \delta q^\nu +m \delta \gnu \ks q\\ 
&+2\delta p'.q \gnu  + \delta q^2\gnu 
+2\lambda q^\nu-\lambda  \gnu  \ks q+\kappa q^\nu
}

We are left with 3 independent  structures, $\gnu $, $\gnu  \ks q$ and $q^\nu$ with 
the following coefficients:

\begin{center}
{\small\begin{tabular}{|c||c|c|}
\hline Structures & Factors \\
\hline \hline 
$\gnu $        &$2\alpha p.q -\alpha q^2 +2\delta p'.q +\delta q^2$ \\
$\gnu  \ks q$  &$-m \alpha +\beta +m \delta -\lambda$\\
$q^\nu$        &$-2m\delta+2 \lambda +\kappa$ \\ 
\hline
\end{tabular}}
\end{center}

This gives
\begin{center}
\boxed{\parbox{\linewidth}{
\eqs{\label{eq:gi_vect_1}
-\alpha P_1 +\delta P_2=0}
\eqs{\label{eq:gi_vect_2}
\beta-\lambda= m(\alpha-\delta)}
\eqs{\label{eq:gi_vect_3}
\kappa= 2(m\delta-\lambda)}
}}
\end{center}

The {\it point} case follows these relations. Indeed,

\eqs{
1)\ &-\alpha_{point} P_1 +\delta_{point} P_2=-P_1^{-1} P_1 +P_2^{-1} P_2=0\\
2)\ &\beta_{point}-\lambda_{point}=m P_1^{-1}-m P_2^{-1}= m(\alpha_{point}-\delta_{point})\\
3)\ &2(m\delta_{point}-\lambda_{point})=2(m P_2^{-1}-m P_2^{-1})=0=\kappa_{point}
}

\subsection{Restoring gauge invariance for the ``phenomenological approach''}
\label{subsec:gau_inv_classical_case_vect}

Let us construct an effective 4-particle vertex restoring gauge invariance, broken by
the ``phenomenological approach''. If we introduce a 3-particle vertex function 
$\Gamma$ ($V_3^\nu =\Gamma \gnu$), the values of $\alpha,\beta,\delta,\lambda$  are :
\eqs{ \alpha=\frac{\Gamma_1}{P_1},\ \beta=\frac{m\Gamma_1}{P_1},
\delta=\frac{\Gamma_2}{P_2},\lambda=\frac{m\Gamma_2}{P_2},\kappa=0.}
where $\Gamma_1=\Gamma((2p-q)^2)$ and $\Gamma_2=\Gamma((2p'+q)^2)$.
\eqs{\alpha=\alpha_{point}\times \Gamma_1,\ \beta=\beta_{point}\times \Gamma_1,\
\delta=\delta_{point}\times\Gamma_2,\ \lambda=\lambda_{point}\times\Gamma_2.}

From the point of view of gauge invariance, we directly see that 
\ce{eq:gi_vect_2} and \ce{eq:gi_vect_3} are automatically satisfied 
by our  phenomenological 3-point vertex contrarily to \ce{eq:gi_vect_1} at 
least as long as $\Gamma_1\neq \Gamma_2$. Concerning the symmetry under $-\cal T$, 
it can easily be seen that it is also satisfied.

The single relation \ce{eq:gi_vect_1} derived from coefficient identification of one single Dirac 
structure ($\gnu$) will be also our starting point to restore gauge invariance 
in this case. However we shall proceed bearing in mind not to add any contribution to
\ce{eq:gi_vect_2} and \ce{eq:gi_vect_3}, {\it i.e.} no term in $\gnu \gmu $ at all (before contraction with
$q^\mu$), or only a transverse one (vanishing due to the contraction).

\subsubsection{First solution}

Starting from contribution whose Dirac structure is $\gnu$, we can choose
the following one (with the multiplicative factor $\eta$):
\eqs{2\eta(p+p')^\mu \gnu=\eta(2p'^\mu\gnu -m\gmu\gnu 
+2p^\mu\gnu -\gnu  \gmu m)=\eta(\gmu \ks p' \gnu  +\gnu  \ks p \gmu) .}

As we have seen, under the transformation $-{\cal T}(\cdot)$, $\gnu $ changes to 
$-\gnu $ and $\eta_1(p+p')$ changes\footnote{The minus sign of $-{\cal T}(\cdot)$ is considered only the
transformation of the structure.} to $-\eta_2(p'+p)$, therefore, we have to 
impose that 
\eqs{\label{eq:eta_sym_vect_1}\eta_1=\eta_2.}

Concerning gauge invariance, the introduction of this new contribution adds a new term 
to~\ce{eq:gi_vect_1}:
\eqs{\label{eq:gi_vect_1b}
0=-\alpha P_1 +\delta P_2+2\eta(p+p').q 
=-\frac{\Gamma_1}{P_1}P_1+\frac{\Gamma_2}{P_2}P_2+2\eta(p+p').q.}

Solving for $\eta$, we have:
\eqs{\label{eq:v_eff_vect_1}
V_{4}^{\mu\nu}=2\eta(p+p')^\mu \gnu  =\frac{\Gamma_1-\Gamma_2}{(p+p').q}\gnu (p+p')^\mu,
}
$\eta$ being $\frac{\Gamma_1-\Gamma_2}{2(p+p').q}$. We in turn see that $\eta$ satisfies 
the symmetry relation Eq.~(\ref{eq:eta_sym_vect_1}) as $\Gamma_1$ transforms into $\Gamma_2$.

The similar vertex 
\eqs{\label{eq:v_eff_vect_2}
V_{4}^{\mu\nu}= \frac{\Gamma_1-\Gamma_2}{(p-p').q}\gnu (p-p')^\mu
}
also satisfies all the criteria. Contracted with $q^\mu$, it indeed gives 
$(\Gamma_1-\Gamma_2)\gnu $ as required to obtain
gauge invariance. The symmetry constraint would then require $\eta$ to
be odd. This is also verified as
$\eta=\frac{\Gamma_1-\Gamma_2}{2(p-p').q}$.

As in the pseudoscalar case, if we consider a virtual gluon, 
\ie~with $q^2\neq 0$, we might have also chosen
\eqs{\label{eq:v_eff_vect_3}
V_{4}^{\mu\nu}= \frac{\Gamma_1-\Gamma_2}{q^2}\gnu  q^\mu,
}
which does not contribute due to  $\ep_\mu(q) q^\mu=0$.

\subsubsection{Generalising\dots }

Similarly to the pseudoscalar case, where $\frac{\Gamma_1-\Gamma_2}{k.q}\ga  k^\mu$
was a possible gauge-invariance restoring vertex for any $k$, we have here 
$\frac{\Gamma_1-\Gamma_2}{k.q}\gnu  k^\mu$ .

If we choose $k^\mu=p^\mu (p'.q)+p'^\mu (p.q)$, we get
\eqs{\label{eq:v_eff_vect_4}
V_{4}^{\mu\nu}&=(\Gamma_1-\Gamma_2) \frac{p^\mu (p'.q)+p'^\mu (p.q)}{ 2(p'.q)(p.q)}\gnu = 
(\Gamma_1-\Gamma_2)\left(\frac{p^\mu }{2(p.q)}+\frac{p'^\mu }{2(p'.q)}\right)\gnu
\\&=(\Gamma_1-\Gamma_2)\left(\frac{p^\mu }{P_1}-\frac{p'^\mu }{P_2}\right)\gnu
}

In this case, this choice contains the same propagators as the {\it point} case:
it cannot introduce new singularities in the problem. As before, there exist also 
choices of $k$ to make appear the denominator of the problem when $q^2\neq 0$,
\ie~for an off-shell gluon. If we choose $k^\mu=(2p-q)^\mu (2p'.q+q^2)+(2p'+q)^\mu (2p.q-q^2)$, we get
\eqs{\label{eq:v_eff_vect_4b}
V_{4}^{\mu\nu}&=(\Gamma_1-\Gamma_2) \frac{(2p-q)^\mu (2p'.q+q^2)+(2p'+q)^\mu (2p.q-q^2)}
{ 2(2p'.q+q^2)(2p.q-q^2)}\gnu\\&=
\frac{(\Gamma_1-\Gamma_2)}{2}\left(\frac{(2p-q)^\mu }{2p.q-q^2}+\frac{(2p'+q)^\mu }{2p'.q+q^2}\right)\gnu
=(\Gamma_1-\Gamma_2)\left(\frac{(2p-q)^\mu }{2P_1}-\frac{(2p'+q)^\mu }{2P_2}\right)\gnu,
}
suitable for an off-shell gluon.

The choice of~\ce{eq:v_eff_vect_4} associated
with the contributions derived from our phenomenological 3-point vertex will be what we
shall refer to as the Gauge-Invariant Phenomenological Approach (GIPA). The latter will be
considered as our benchmark for the study of non-static effects and will
be studied in details when applied to the calculation of the quarkonium cross section
at the Tevatron and at RHIC in chapter~\ref{ch:application}.\index{GIPA}

\section{Autonomous contributions for $q\bar q \to V_0 g$}\label{sec:auton_contrib}

Following the discussion for $q\bar q \to V_0 g$, we may generalise the vertex by adding 
new contributions that are  gauge-invariant. This is in fact equivalent to what
is implicitly done when passing from one gauge-invariance restoring contribution to another .

For this reason we shall derive here a generic form for a gauge-invariant contribution that
can be added to the amplitude for the bound state. We shall nevertheless restrain 
ourselves to certain class of contributions that satisfy 
reasonable constraints, namely

\begin{itemize}
\item They should vanish for a structureless bound state, that is
 when $\Gamma_1 =\Gamma_2$.
\item The colour factors, as well as the coupling constant from QCD, are to be the same 
as in the corresponding perturbative contribution with the same external particles.
\item Once again, no new singularities can be brought into the problem.
\item The new contributions should have the same symmetries under quark crossing as the 
structureless ones.
\end{itemize}

As we are dealing with two vectors particles and two of spin 1/2, we 
have have to have the following generic form for the amplitude 
\eqs{\bar u(p')V^{\mu\nu} \ep^\star_\mu(q)  \ep^\star_\nu(P) u(p),}
$\ep^\star_\mu(q)$ is the gluon helicity vector and $\ep^\star_\nu(P)$ that of the vector meson.

The only independent tensorial structures that we can introduce are  made of Dirac matrices 
and of the momenta of the problem. From momentum conservation, we might choose: $P$, $q$ 
and $p+p'$ (recall that $p-p'=q+P$). Therefore, $V^{\mu\nu}$ could be constructed from
a product like the following
\eqs{\label{eq:generic_form_vmunu_nonGI_qqVg}
V^{\mu\nu}= (\Gamma_1-\Gamma_2)\left(  \begin{matrix} \gmu \\ P^\mu \\ q^\mu \\ 
(p+p')^\mu \end{matrix}\right) \otimes 
\left(  \begin{matrix} \gnu \\ P^\nu \\ q^\nu \\ (p+p')^\nu \end{matrix}\right),
}
and with $g^{\mu\nu}$, but this simple combination does not insure gauge-invariance.

The latter can be insured by use of the projector 
${\cal P}_{\ \mu'}^\mu=g_{\ \mu'}^\mu-\frac{P^\mu q_{\mu'}}{q.P}$. 
For $g^{\mu\nu}$, this gives $V^{\mu\nu}= (\Gamma_1-\Gamma_2)(g^{\mu\nu}-\frac{P^\mu q^\nu}{q.P})$.  
As the index of the bound-state $\nu$ is not affected by the gauge constraints, let us consider 
the factors in \ce{eq:generic_form_vmunu_nonGI_qqVg}  with index $\mu$ first:

\eqs{
{\cal P}_{\ \mu'}^{\mu}\left(  \begin{matrix} \gmup \\ P^{\mu'} \\ q^{\mu'} \\ (p+p')^{\mu'} 
\end{matrix}\right) =
\left(  \begin{matrix} \gmu-\frac{P^\mu \ks q}{q.P} \\ 0 \\ q^\mu \\ (p+p')^\mu -\frac{P^\mu  
(p+p').q}{q.P}\end{matrix}\right)
}

We can drop $q^\mu$ as the $\mu$ index is the one of the gluon and $\ep(q).q=0$. We do the same 
for $P^\nu$  ($\ep(P).P=0$) and finally for $P^\mu$. Indeed because the gluon is real 
and thus transverse, we have  $\ep(q).k=0$ for all time-like vectors, $k$, like $q+P$ 
--which reduce to $(\sqrt{s},0,0,0)$ in the CM frame--. $\ep(q).(q+P)=0$ 
and $\ep(q).q=0$ obviously give $\ep(q).P=0$. We therefore have 

\eqs{
V^{\mu\nu}= (\Gamma_1-\Gamma_2)\left(  \begin{matrix} \gmu \\ 0 \\ 0\\ (p+p')^\mu \end{matrix}\right) 
\otimes \left(  \begin{matrix} \gnu \\ 0 \\ q^\nu \\ (p+p')^\nu \end{matrix}\right).
}

This drives us to the generic contribution :
\eqs{\label{eq:general_form_vmunu_contrib_qqVg}
  V^{\mu\nu}=(\Gamma_1-\Gamma_2)(\alpha_1 \gnu+\alpha_2 q^\nu+\alpha_3 (p+p')^\nu ) (\beta_1 \gmu+\beta_3 
(p+p')^\mu)+\xi g^{\mu\nu}
}

Please remember that the latter expression is without terms that 
explicitly do not contribute due to contraction with helicity vectors of the bound-state and 
of the gluon. $V^{\mu\nu}$ in \ce{eq:general_form_vmunu_contrib_qqVg} is not meant to verify 
current conservation as is. We would rather write to be complete:

\eqs{\label{eq:general_form_vmunu_full_qqVg}
V^{\mu\nu}=&(\Gamma_1-\Gamma_2)
((\alpha_1\gnu + \alpha_2P^\nu + \alpha_3q^\nu + \alpha_4(p+p')^\nu) \\
&(\beta_1(\gmu-\frac{P^\mu \ks q}{q.P}) +\beta_2 q^\mu +\beta_3 ((p+p')^\mu -\frac{P^\mu  (p+p').q}{q.P})
+\xi (g^{\mu\nu}-\frac{P^\mu q^\nu}{q.P}))}

As long as the coefficients $\alpha,\beta,\gamma,\kappa,\lambda,\xi$ are not singular anywhere, 
the second condition that we required is satisfied. To what concerns the crossing symmetry, first
let us recall that $\Gamma_1-\Gamma_2$ changes signs 
since $\Gamma_1 \overset{1\to2}{\longrightarrow}\Gamma_2$, the rest of the expression
must then also change sign. Under $-\cal{T}(\cdot)$, 
we have:
\begin{eqnarray}
\gmu P^\nu            \to& - \gmu P^\nu         \hspace*{3.75cm}      
\gmu q^\nu            &\to - \gmu q^\nu               \nn\\
\gmu (p+p')^\nu       \to& + \gmu (p+p')^\nu       \hspace*{0.5cm}     
(p+p')^\mu (p+p')^\nu &\to - (p+p')^\mu (p+p')^\nu    \nn\\
P^\mu q^\mu           \to& - P^\mu q^\mu            \hspace*{2.5cm}    
P^\mu (p+p')^\nu      &\to + P^\mu (p+p')^\nu         \nn\\
\gmu \gnu             \to& + \gmu \gnu -2 g^{\mu\nu}  \hspace*{2.75cm}  
g^{\mu\nu}            &\to - g^{\mu\nu}               
%}
\end{eqnarray}
the other combinations are trivially obtained from these. As we required them to change sign, to cancel
that of $\Gamma_1-\Gamma_2$, we have to impose that $\gmu (p+p')^\nu$  as well as
$q^\mu (p+p')^\nu$ have coefficients changing sign contrary to  $(p+p')^\mu (p+p')^\nu$,
whose coefficient can be a constant.

On the other hand, we have to rule out $\gmu \gnu$, the term $-2 g^{\mu\nu}$
cannot be cancelled by the terms proportional to $g^{\mu\nu}$. Only more complicated combination
starting from a more general product:
\eqs{
V^{\mu\nu}= (\Gamma_1-\Gamma_2)\left(  \begin{matrix} \gmu \\ P^\mu \\ q^\mu \\ 
(p+p')^\mu \end{matrix}\right) \otimes 
\left(\begin{matrix}  I \\ \ks q \\ \ks P\\ 
(\ks p+\ks p')\end{matrix}\right) \otimes
\left(  \begin{matrix} \gnu \\ P^\nu \\ q^\nu \\ (p+p')^\nu \end{matrix}\right),
}
would allow to have terms including $\gmu \gnu$.

For a first approach, we shall retain here only vertices with a constant multiplicative factor. 
This enables us to drop some terms in \ce{eq:general_form_vmunu_contrib_qqVg} which eventually becomes:

\eqs{\label{eq:general_form_vmunu_contrib_qqVgb}
V^{\mu\nu}=(\Gamma_1-\Gamma_2)(&\alpha \, \gmu q^\nu + \beta \, (p+p')^\mu (p+p')^\nu +
 \xi \, g^{\mu\nu}).
}

\subsection{Our Feynman rules}

Considering the process $q \bar q \to V_0 g$ in the GIPA,
where the $q \bar q V_0$ vertex is $\Gamma(p_{rel}^2,P)\gmu$ ($p_{rel}$ is the relative
momentum of the quark and the antiquark and $P$ the momentum of $V_0$), the choice given
by~\ce{eq:v_eff_vect_4} is written explicitly in the first line of~\ct{tab:feyn_rules}.

To what concerns autonomous contributions, we limit ourselves to the ones where the free multiplicative
factor is a constant. The Feynman rules for these three possible choices 
derived in~\ce{eq:general_form_vmunu_contrib_qqVgb} are also given in~\ct{tab:feyn_rules}.

\begin{table}[H]
\begin{tabular}{|l|c|}
\hline &\\
\includegraphics[height=3.0cm,clip=true]{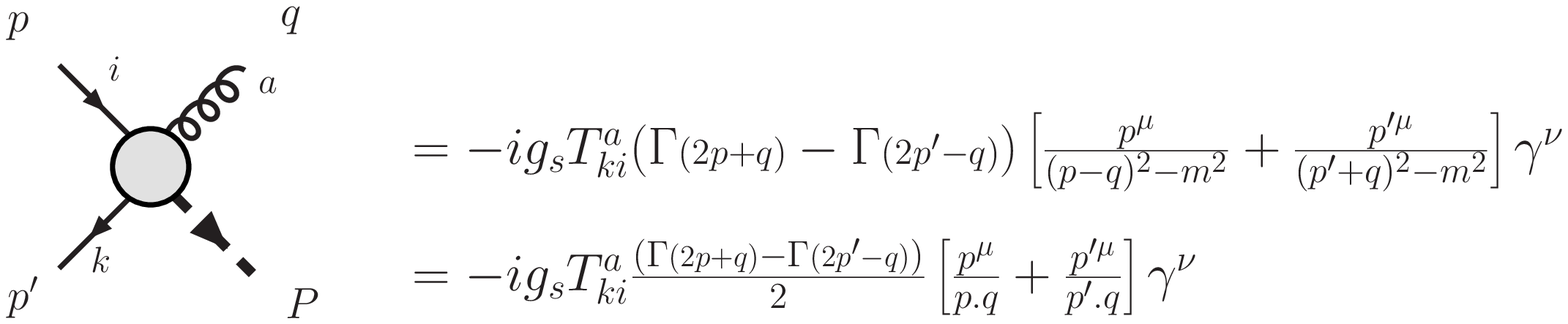}& 
\parbox{2.5cm}{ cf.\ce{eq:v_eff_vect_4}\\\\\\\\} \\
\includegraphics[height=3.0cm]{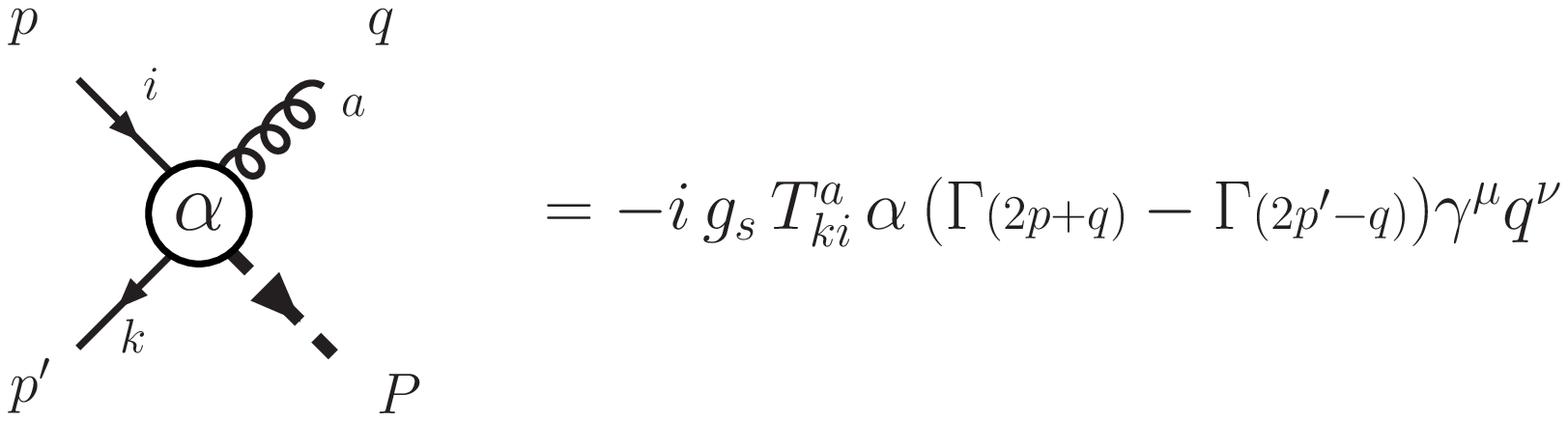}& \parbox{2.5cm}{ First term \\in \ce{eq:general_form_vmunu_contrib_qqVgb} \\\\\\\\\\ }\\
\includegraphics[height=3.0cm]{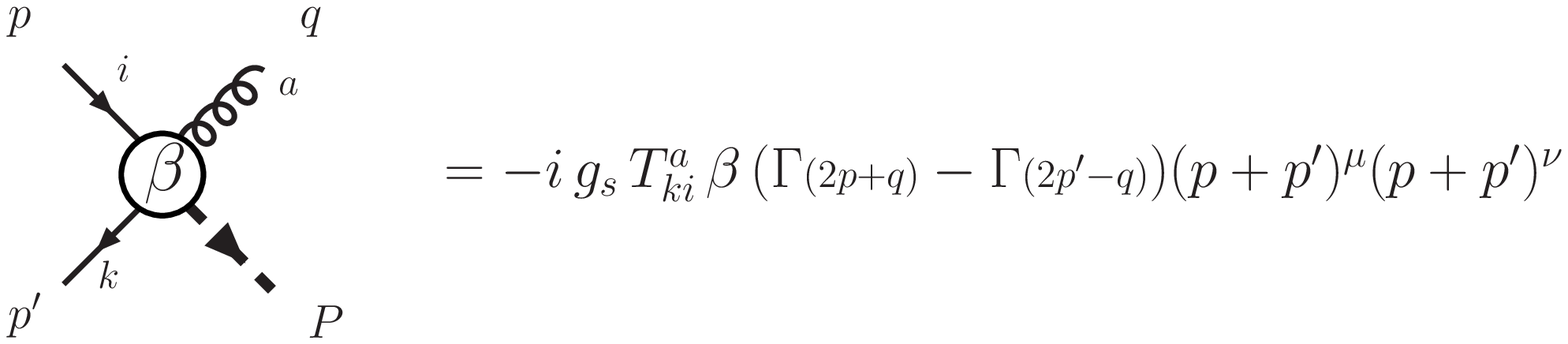}&  \parbox{2.5cm}{ Second term \\in \ce{eq:general_form_vmunu_contrib_qqVgb}\\\\\\\\\\ }\\
\includegraphics[height=3.0cm]{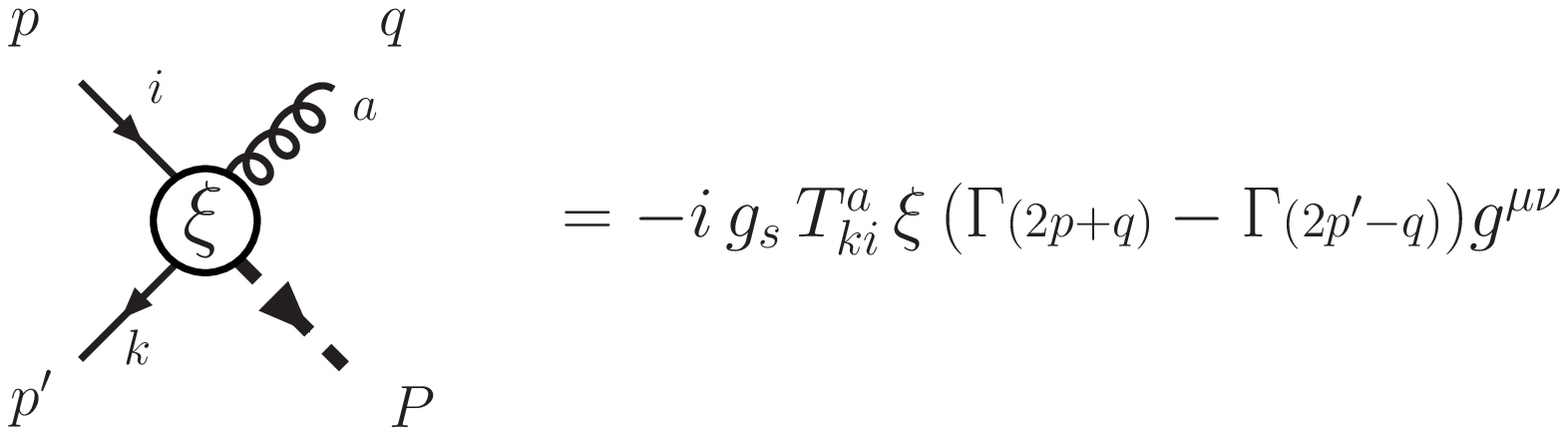}& \parbox{2.5cm}{Third term \\in \ce{eq:general_form_vmunu_contrib_qqVgb}     \\\\\\\\\\ }\\
\hline
\end{tabular}
\caption{Feynman rules for one choice of gauge-invariance restoring vertex (first line) 
and for the three possible choices of autonomous contributions with a constant multiplicative factor.}
\label{tab:feyn_rules}
\end{table}

\index{gauge!invariance|)}

\clearemptydoublepage

\chapter{$J/\psi$ and $\Upsilon$ production  at RHIC and at the Tevatron}\label{ch:application}

We shall now calculate the inclusive polarised cross sections for $pp\to \psi X$ (or
$p\bar p\to \psi X$) and $p\bar p\to \Upsilon X$ (or $pp\to \Upsilon X$)  
in the experimental conditions of the Tevatron collider based at Fermilab and RHIC at BNL. 
As already mentioned, the Tevatron is a collider with two beams of equal energy, 
one of protons and one of antiprotons. The energy in center of momentum frame 
amounts\footnote{In run I where the data were taken.} to 1.8 TeV. There exist two main detectors operating
on this collider,  CDF and D$\emptyset$. We shall here solely consider 
the results coming from the CDF collaboration as their analysis
reckon, oppositely to that of D$\emptyset$, on a central magnet which enables to identify the charges of
the detected particles. 
As a consequence, their results are slightly more reliable. Concerning
RHIC, we shall focus on the PHENIX analysis for $pp$ collisions at $\sqrt{s}=200$~GeV. 
\index{Tevatron} \index{RHIC}

The main study of the CDF analysis, as seen in chapter~\ref{ch:experim}, has been centered
on $J/\psi$, $\psi'$ and the $\Upsilon(nS)$. For the latter, the direct production 
has been extracted only for  $\Upsilon(1S)$. The $P_T$ range was from 5 to 20 GeV and the pseudorapidity
range from -0.6 to 0.6 for $\psi$ and from -0.4 to 0.4 $\Upsilon$. PHENIX studied only 
total $J/\psi$ production in the central region $|y| \leq  0.35$ and in the forward one 
$-2.2\leq y\leq-1.2$. The $P_T$ range was much smaller, from 0.4 GeV to 5.5~GeV.

In those conditions (very high energy and hadron beams), we have already explained that that the main 
partonic contribution will be gluon fusion. 
Therefore, one can easily convince oneself that {\it a priori} the LO sub-process to produce 
$J/\psi$ and $\Upsilon$ (generically denoted ${\cal Q}$) at medium transverse momentum 
should be $gg\to {\cal Q} g$, where 
the final state gluon is required to provide with the desired transverse momentum.

We first explain the kinematics, underlining the link between the hadronic cross section
and the partonic one. The latter is then expressed in terms of helicity amplitudes, which will enable us to
calculate the polarisation parameter $\alpha_{pol}$ measured by CDF (see section~\ref{sec:polarisation}).

\section{Generalities on the kinematics}

The link between the partonic cross sections and the hadronic ones relies on the following general formula:
\begin{equation}
\label{eq:Edsdp3}
E\frac{d^3\sigma}{dP^3}=\int_0^1 dx_1 dx_2 \ G_1(x_1) \ G_2(x_2)\ \frac{\hat s}{\pi}\frac{d\sigma}{d\hat t}
\delta(\hat s +\hat t +\hat u -M^2),
\end{equation}
where $P=(E,\vect P)$ is the momentum of the meson in the c.m. frame 
of the colliding hadrons. Note on the way that these variables are 
experimentally accessible. The functions $G_i$ are the parton distribution functions, which give
the probability that a parton $i$ take a momentum fraction $x_i$ off the parent hadron.

In the c.m. frame of the colliding hadrons, introducing the rapidity $y$, the transverse momentum 
$\vect P_T=(P_x,P_y)$ and the transverse energy\footnote{$P_T\equiv|\vect P_T|$} $E_T=\sqrt{P_T^2+M^2}$, we
may decompose $P$ as follows (satisfying the on-shell constraint $P^2=M^2$):
\begin{equation}
\label{eq:def_p}
P=(\sqrt{M^2+P_x^2+P_y^2}\cosh y,P_x,P_y, \sqrt{M^2+P_x^2+P_y^2}\sinh y)=
(E_T\cosh y,\vect P_T, E_T\sinh y),
\end{equation}
with
\eqs{y=\frac{1}{2}\ln\left(\frac{E+P_z}{E-P_z}\right)=\tanh^{-1}\left
(\frac{P_z}{E}\right).} 

Now, let us express the measure $d^3P$ in terms of the 
variables $P_T$ and $y$, {\it i.e.} $d^3P=dP_z \, d\vect P_T=dP_z \, d\phi \, P_T\, dP_T$.
As we want to keep $P_T$, we shall only express the $z$-component in terms of $y$, in order 
to get $\frac{dP_z}{dy}$:
\eqs{P_z=E_T\sinh y \Rightarrow \left|\frac{dP_z}{dy}\right|=E_T\cosh y\equiv E,}
which enables us to rewrite the integration element as:
\begin{equation}\label{eq:dsigmadpt_a}
E\frac{d^3\sigma}{dP^3}=\frac{d^3\sigma}{dy P_T d P_T d\phi},
\end{equation}

As the integrand is symmetric in $\phi$, we obtain the double-differential 
cross section upon  $P_T$ and $y$
from \ce{eq:Edsdp3}:
\eqs{
\frac{d\sigma}{dydP_T}=\int_0^1 dx_1 dx_2 G_1(x_1) G_2(x_2) 2 \hat s  P_T\frac{d\sigma}{d\hat t}
\delta(\hat s+\hat t+\hat u-M^2).
}
When there might be an ambiguity, the Mandelstam variables relative 
to the partonic process are discriminated from the hadronic ones by circumflexes. To simplify the
notations, especially when dealing uniquely with partonic processes, we shall sometimes 
drop the circumflexes, as in the next section.

At this stage, we can perform the integration upon $x_2$ (or $x_1$) using the delta function,
which gives us the jacobian:
\eqs{\frac{1}{\left|\frac{d(\hat s +\hat t +\hat u -M^2)}{dx_2}\right|}.
}

To evaluate the latter, we have to express $\hat s$, $\hat t$ and $\hat u$ in terms of $x_1$, 
$x_2$, $P_T$ and $y$.  If the momenta of the
initial gluons are $k_1$ and $k_2$ and that of the final ones is $k_4$ (see~\cf{fig:kin_k1k2k3k4}),
we have:
\eqs{
\hat s= (k_1+k_2)^2=(q+P)^2,\\
\hat t=(k_1-P)^2=(k_2-q)^2,\\
\hat u=(k_1-q)^2=(k_2-P)^2.
}

\begin{figure}[h]
\centering{\mbox{\includegraphics[width=5cm]{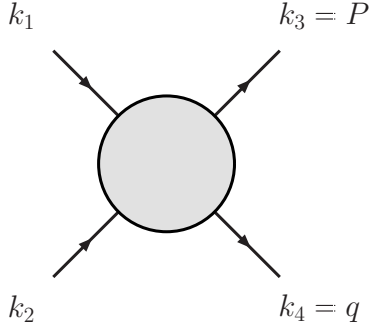}}}
\caption{Definition of $k_1$, $k_2$, $k_3$ ($P$) and $k_4$ ($q$).}
\label{fig:kin_k1k2k3k4}
\end{figure}

Defining the momenta of the two hadrons as $p_1$ and $p_2$, we may always choose
the axis such as to suppress their transverse components (along the $x$ and $y$ axes for
definiteness). For massless initial hadrons, we then have in their c.m. frame
\eqs{
p_1&=(\frac{\sqrt{s}}{2},0,0,\frac{\sqrt{s}}{2}),\\
p_2&=(\frac{\sqrt{s}}{2},0,0,-\frac{\sqrt{s}}{2}).
}

Now, by definition of the momentum fractions 
$x_1$ and $x_2$:
\eqs{
\hat s=(k_1+k_2)^2=\frac{s}{4}(x_1+x_2,0,0,x_1-x_2)^2=s x_1x_2,
}
and 
\begin{equation}
\begin{split}
\hat t = \, & (k_1-P)^2=\underbrace{k_1^2}_{0}+\underbrace{P^2}_{M^2}-2\underbrace{k_1}_{x_1p_1}\cdot P \\
= & M^2-2x_1\frac{\sqrt{s}}{2}(E-P_z).
\end{split}
\end{equation}

From \ce{eq:def_p}, we have $(E-P_z)=E_T (\cosh y-\sinh y)=E_T e^{-y}$ and
\begin{equation}
\hat t = M^2-x_1e^{-y}\sqrt{s}E_T.
\end{equation}
Similarly, we get
\begin{equation}
\hat u =M^2-x_2e^{y}\sqrt{s}E_T.
\end{equation}
Writing down $\hat s +\hat t +\hat u -M^2=s x_1 x_2 +M^2-x_1e^{-y}\sqrt{s}E_T-x_2e^{y}\sqrt{s}E_T$ and 
solving for $x_2$, we obtain:
\begin{equation}\label{eq:x1x2}
x_2=\frac{x_1E_T\sqrt{s}e^{-y}-M^2}{\sqrt{s}(\sqrt{s} x_1-E_Te^{y})}.
\end{equation}
For completeness, solving for $x_1$ we have
\begin{equation}\label{eq:x2x1}
x_1=\frac{x_2E_T\sqrt{s}e^{y}-M^2}{\sqrt{s}(\sqrt{s} x_2-E_Te^{-y})}.
\end{equation}
We also easily compute the jacobian of the $\delta$ integration:
\eqs{
\frac{1}{\left|\frac{d(\hat s +\hat t +\hat u -M^2)}{dx_2}\right|}
=\frac{1}{\sqrt{s}(\sqrt{s}x_1-E_T e^{y})}.
}
As $x_2$ is now a function of $x_1$, we are left with a one-dimensional integration which 
is restricted by
\eqs{
0\le x_2\le 1.
}
Supposing that\footnote{The conditions $\hat t\leq0$ and $\hat u \leq 0$ give 
$x_1\geq \frac{M^2e^y}{\sqrt{s}E_T}$ and $x_2\geq \frac{M^2e^{-y}}{\sqrt{s}E_T}$ which
are always less restrictive than $x_1\geq \frac{E_Te^y}{\sqrt{s}}$ and $x_2\geq \frac{E_Te^{-y}}{\sqrt{s}}$
for $P_T\neq 0$. Furthermore, as long as $|y|$ is not of the order of $\ln(\frac{\sqrt{s}}{P_T})$ 
($|y|=3.8$ at $P_T=40$ GeV), $x_1\geq \frac{E_Te^y}{\sqrt{s}}$ is always less restrictive 
than~\ce{eq:low_bound_x_1}.} $x_1\ge\frac{E_T e^{y}}{\sqrt{s}}$, 
we get the following condition:
\eqs{\label{eq:low_bound_x_1}
x_1\ge \frac{E_T\sqrt{s}e^{y}-M^2}{\sqrt{s}(E_T e^{-y}-\sqrt{s})}\equiv x_1^{min}.
}
Therefore, the double differential cross section upon $P_T$ and $y$ takes the following form:
\eqs{
\frac{d\sigma}{dydP_T}=\int_{x_1^{min}}^1 dx_1 \frac{2 \hat s P_T G_1(x_1) G_2(x_2(x_1))}
{\sqrt{s}(\sqrt{s}x_1-E_T e^{y})}\frac{d\sigma}{d\hat t}.
}

The last step is now to relate the partonic differential cross section $\frac{d\sigma}{d\hat t}$ to
the amplitude calculated from our model. To this end, 
we use the well-known formula:
\begin{equation} \label{eq:sigmadt}
\frac{d\sigma}{d\hat t}=\frac{1}{16\pi \hat s^2}|\overline{\cal M}|^2,
\end{equation}
where $|\overline{\cal M}|^2$ is the squared amplitude averaged on the initial spin and colour degrees
of freedom and summed over the final ones for an unpolarised cross section, or 
averaged only for the colour for polarised cross sections.

The calculation of the latter amplitude within the framework of our model is the subject the following section.

\section{Amplitude for $gg\to \!\!\ ^3S_1  g$}

In this section, we shall give our procedure to calculate
the imaginary of the amplitude for $gg\to \!\!\ ^3S_1  g$ in 
 our formalism through Cutkovsky rules. We shall consider this amplitude
 at leading order in $\alpha_s$.

First, we further develop the kinematics to avoid any ambiguity and we
derive expressions for the polarisation vectors of all particles. Then
we write down the expressions for the helicity amplitudes. 
The cutting rules impose further constraints on the internal kinematics, which we give. We then
organise the calculation by classifying the integrals that enter the amplitude.

\begin{figure}[h]
\centering{\mbox{\includegraphics[width=5.0cm]{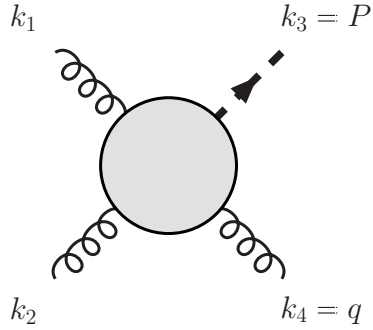}}}
\caption{Kinematics of the process $g g \rightarrow \!\!\ ^3S_1 g$.}
\label{fig:kin_ggVg}
\end{figure}

\subsection{Further kinematics and polarisation vectors}\label{subsec:polvec}

The momenta of the particles are as shown in \cf{fig:kin_ggVg}. 
The on-shellness conditions are then:
$k_{1}^{2}=k_{2}^{2}=k_{4}^{2}=0$,
$k_{3}^{2}=M^{2}$. The Mandelstam  variables are defined as usual (dropping the circumflexes)
\begin{eqnarray}\label{invars}
 && s = (k_{1}+k_{2})^{2} = 2 k_{1}.k_{2} = (k_{3}+k_{4})^{2}= M^{2} + 2 k_{3}.k_{4}\,, \\
\label{invart}
 && t = (k_{1}-k_{3})^{2} = -2 k_{1}.k_{3} + M^{2} = (k_{2}-k_{4})^{2} = - 2 k_{2}.k_{4}\,, \\
\label{invaru}
 && u = (k_{1}-k_{4})^{2} = - 2 k_{1}.k_{4}= (k_{2}-k_{3})^{2} = -2 k_{2}.k_{3} + M^{2}\, , \\
 && s+t+u = M^{2} \, .
\end{eqnarray}
and all scalar products, $k_{i}.k_{j}$, are functions of the latter:
\begin{eqnarray}
&& k_{1}.k_{2} = \frac{s}{2}\,,  k_{1}.k_{3} = \frac{M^{2}-t}{2}  \,,
k_{1}.k_{4} = -\frac{u}{2},\\
&& k_{2}.k_{3} = \frac{M^{2}-u}{2}  \,, k_{2}.k_{4} = -\frac{t}{2} \, ,
k_{3}.k_{4} = \frac{s-M^{2}}{2}  \, .
\end{eqnarray}

In order to compute polarised cross sections, we have to introduce explicitly 
the polarisation vectors for the vector meson and for the gluons. It is however 
sufficient to determine for all polarisations the scalar product between these 
vectors and the other vectors of the problem, {\it i.e.} $k_1,\dots,k_4$. 

Let $ \ep_i\equiv\ep(k_i)$ be the polarisation vector of the different particles
considered in the present case, we shall perform a Sudakov decomposition
of these along the momenta of the problem and along 
a vector $\epsilon_T$ ($\epsilon_T.\epsilon_T =-1$) orthogonal 
to the plane of the scattering.

A generic helicity vector $\ep^h_{i}$, where $i$ is the particle label 
($i=1,..4$), $h$ the helicity state ($T_1$ or $T_2$ for
the gluons and $T_1$, $T_2$ or $L$ for the vector meson), then becomes:

\eqs{\label{eq:pol-decomp}
\ep^{h}_{i}=a^{h}_{i} k_1+b^{h}_{i} k_2+c^{h}_{i} k_4+d^{h}_{i} \epsilon_T.
}

We first impose that each vector be orthogonal to the 
momentum of the particle it describes: 
\eqs{\label{eq:pol-orthk}k_i.\ep^h_{i}=0,}
for all polarisation $h$.

Then we require that vectors of different helicities ($h\neq h'$) and of one given particle  
be orthogonal to each other: 
\eqs{\label{eq:pol-orthe}\ep^h_{i}.\ep^{h'}_{i}=0.}
We in turn impose the correct normalisations, 
\eqs{(\ep_{i}^{h})^2=-1 .}

The previous conditions are valid for all polarisations. In the case of 
the transverse ones, we shall introduce this transverse property by requiring 
that only spatial components remain. In other words, the scalar product between
transverse polarisations and a vector of the generic form $v=(v_0,0,0,0)$ should vanish. 

In the c.m.~frame  (of the gluons), $k_1+k_2$ or
$k_3+k_4$ have the latter property since $k_1+k_2=k_3+k_4=(\sqrt{s},0,0,0)$, and thus the 
condition for transverse polarisations becomes:
\eqs{\label{eq:pol-orthtr}
(k_1+k_2).\ep_{i}^{T}=(k_3+k_4).\ep_{i}^{T}=0,\\
}
Taking into account the conditions of \ce{eq:pol-orthk}, we then derive
\eqs{ \label{eq:pol-orthtr2}
\ep_{1}^T.k_2=0,\\
\ep_{2}^T.k_1=0,\\
\ep_{3}^T.k_4=0,\\
\ep_{4}^T.k_3=0.
}

Here, we are free to choose $\ep_{i}^{T_1}$ to be a vector orthogonal to all the others, 
namely
\eqs{
 \ep_{1}^{T_1}=\ep_{2}^{T_1}=\ep_{3}^{T_1}=\ep_{4}^{T_1}=\epsilon_T,
}
since this satisfies all conditions cited above.

Having fixed $\ep_{i}^{T_1}$, we shall determine $\ep_{i}^{T_2}$ imposing $\ep_{i}^{T_1}.\ep_{i}^{T_2}=0$. 
We have:

\eqs{
a_{1}^{T_2}=a_{2}^{T_2}=a_{3}^{T_2}=a_{4}^{T_2}=\sqrt{\frac{s}{u t}}\frac{t}{s}, \\
b_{1}^{T_2}=b_{2}^{T_2}=-b_{3}^{T_2}=-b_{4}^{T_2}=\sqrt{\frac{s}{u t}}\frac{u}{s}, \\
c_{1}^{T_2}=c_{2}^{T_2}=\sqrt{\frac{s}{u t}},\\
c_{3}^{T_2}=c_{4}^{T_2}=\sqrt{\frac{s}{u t}}\frac{(t-u)}{t+u}, \\
d_{1}^{T_2}=d_{2}^{T_2}=d_{3}^{T_2}=d_{4}^{T_2}=0. \\
}

Finally, requiring $\ep_{3}^{T_1}.\ep_{3}^L=0=\ep_{3}^{T_2}.\ep_{3}^L$, we fully determine $\ep_{3}^L$:
\eqs{
a_{3}^L=\frac{1}{M}, b_{3}^L=\frac{1}{M},c_{3}^L=-\frac{1}{M}\frac{s+M^2}{s-M^2},d_{3}^L=0.
}

Thus, the complete expressions of the polarisation vectors are the following:
\eqs{\label{eq:pol-final}
\ep_{1}^{T_1}&=\ep_{2}^{T_1}=\ep_{3}^{T_1}=\ep_{4}^{T_1}=\epsilon_T,\\
\ep_{1}^{T_2}&=\sqrt{\frac{1}{s t u}}\left(t k_1+ u k_2+ s k_4\right)=\ep_{2}^{T_2},\\
\ep_{3}^{T_2}&=\sqrt{\frac{1}{s t u}}\left(t k_1-u k_2+\left(\frac{s(t-u)}{t+u}\right) k_4\right)
=\ep_{4}^{T_2},\\
\ep_{3}^{L}  &= \frac{1}{M}\left(k_1+k_2-\left(\frac{s+M^2}{s-M^2}\right) k_4\right),\\
}

\subsubsection{Compatibility with the standard choice}

We shall show here that our result can give in a non-covariant form the standard choice for
instance taken in~\cite{Cho:1995vh,Cho:1995ce}. Let us do that for the longitudinal 
meson polarisation vector.
We have with the notations explained above:
\eqs{
\ep_{3}^{L}  &= \frac{1}{M}\left(k_1+k_2-\left(\frac{s+M^2}{s-M^2}\right) k_4\right)
}
whereas the standard  choice used in~\cite{Cho:1995vh,Cho:1995ce} is in the meson rest frame
($k_3=(M,0,0,0,)$):
\eqs{
\ep_{3}^{L}  &= (0,0,0,1),\\
}

First we rewrite $\ep_{3}^{L}$ in terms of $k_3$, this gives 
\eqs{
\ep_{3}^{L}  &= \frac{1}{M}\left(k_3-\frac{2M^2}{s-M^2} k_4\right).
}
Choosing the $z$-axis direction to be the same as $-\vect k_4$, we have 
$k_4=(k_{4,0},0,0,-|k_{4,z}|)$, which, for an on-shell gluon, gives $k_4=(k_{4,0},0,0,-k_{4,0})$.
We are thus left to determine $k_{4,0}$, which is given by
\eqs{
s=(k_3+k_4)^2=M^2+0+2 k_3.k_4= M^2+2 M k_{4,0}\ \Rightarrow 
k_{4,0}=\frac{s-M^2}{2 M}
}

Finally, we have 
\eqs{
\ep_{3}^{L} = \frac{1}{M}((M,0,0,0,)-(M,0,0,-M))= (0,0,0,1).
}

\subsection{Expressions for the amplitude}

As we shall consider only its imaginary part, the amplitude of the process
can be diagrammatically presented by a product of two scattering amplitudes
$g g \rightarrow q \bar{q}$ and $q \bar{q} \rightarrow {\cal Q} g$ where we  put 
the quark lines with momenta $\ell+k_{1}$ and $\ell-k_{2}$ (\cf{fig:ampl_ggqq_qqVg})
on their mass shell, so that they satisfy the relations: $(\ell+k_{1})^2=m^{2}$, $(\ell-k_{2})^2=m^{2}$,
or equivalently we replace their propagators by $\delta$ functions:
\eqs{
\frac{1}{(\ell+k_{1})^{2}-m^{2}} 
\frac{1}{(\ell-k_{2})^{2}-m^{2}} \to&\, \frac{1}{2} (2\pi)^{2} \delta^{+}((\ell+k_{1})^{2}-m^{2} ) 
\delta^{-}((\ell-k_{2})^{2}-m^{2})  \\
=&\, \frac{1}{2} (2\pi)^{2} \delta( 2 \ell.(k_{1}+k_{2})) \delta(\ell^{2}+2 \ell.k_{1}-m^{2})
\times\\\times&\,
\theta(\ell^0+k_1^0)\theta(-\ell^0+k_2^0),
\label{cutt}
}
where the factor $\frac{1}{2}$ comes from $\Im m{\cal A}= \frac{1}{2} {\rm Disc} ({\cal A})$, 
and the discontinuity is obtained with the introduction of the above-mentioned $\delta$ functions.

At leading order in $\alpha_s$ ($g_s$), we have three diagrams for $g g \rightarrow q \bar{q}$ and
also three for $q \bar{q} \rightarrow {\cal Q} g$, in  our framework. The six 
diagrams\footnote{Compared to notations used for the derivation of the gauge-invariance restoring
vertices (chapter~\ref{ch:vertex_derivation}), we have 
$p\to \ell+k_1$, $p'\to \ell-k_2$, $q\to k_4$ and $P\to k_3$.} are drawn in~\cf{fig:ampl_ggqq_qqVg}.

\begin{figure}[h]
\centering{\mbox{\includegraphics[width=15cm]{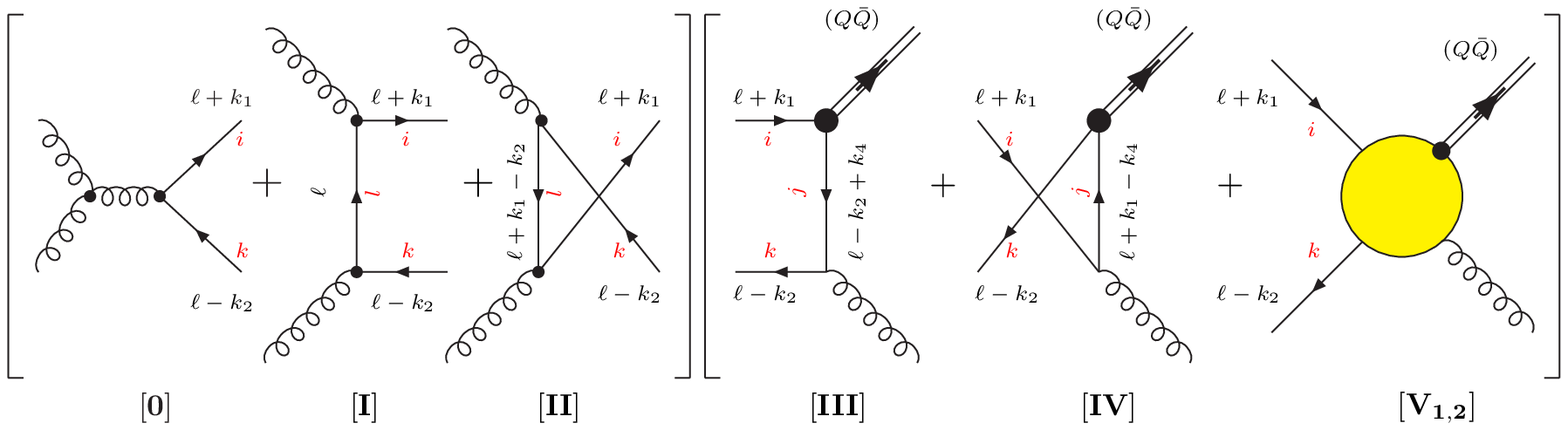}}}
\caption{Different contributions to the imaginary part of the amplitude
for $q \bar{q} \rightarrow {\cal Q} g$.}
\label{fig:ampl_ggqq_qqVg}
\end{figure}

\newpage

These scattering amplitudes have the following explicit forms, where $p$, $q$, $r$ and $s$ 
are the helicity states of the particles,
\begin{eqnarray}
{ \bf [O]}\!\!\!&=&\!\!\!  
(-g_s) f^{abd}\bar u(\ell+k_1)
\left[(k_1-k_2)^\tau g^{\mu\nu}+(k_1+2k_2)^\mu g^{\nu\tau}-(2k_1+k_2)^\nu g^{\tau\mu}\right]\times
\nonumber\\ &\times&
\frac{-i g_{\tau \tau'} \delta_{d d'}}{(k_1+k_2)^2} (-i g_s T_{ik}^{d'})\gtaup
u(\ell+k_1)\ep_{\mu}^p(k_1)\ep_{\nu}^q(k_2)\\
\!\!\!&=&\!\!\!  
g_s^2 \frac{f^{abd} T_d}{s} 
\bar u(\ell-k_2)
\left[(\ks k_1-\ks k_2) g^{\mu\nu}+(k_1+2k_2)^\mu\gnu-(2k_1+k_2)^\nu \gmu\right]
u(\ell+k_1)\times\nonumber\\
&\times&\ep_{\mu}^p(k_1)\ep_{\nu}^q(k_2), \\
{ \bf [I]}\!\!\!&=&\!\!\! 
(-i g_s)^2 T_{il}^aT_{lk}^b  \bar u(\ell-k_2)
\frac{i\left[ \gamma^{\mu} \left(
    \ks{\ell} + m \right) \gamma^{\nu} \right]}{\ell^{2}-m^{2}}  u(\ell+k_1)
\ep_{\mu}^p(k_1)\ep_{\nu}^q(k_2), \\
{ \bf [II]}\!\!\!&=&\!\!\! 
(-i g_s)^2 T_{il}^bT_{lk}^a  \bar u(\ell-k_2)
\frac{i\left[ \gamma^{\nu} \left(
    ( \ks{\ell}+\ks{k}_{1}-\ks{k}_{2}) + m \right) \gamma^{\mu} \right]}
{ ( \ell+k_{1}-k_{2} )^{2}-m^{2}} 
u(\ell+k_1)\ep_{\mu}^p(k_1)\ep_{\nu}^q(k_2), \\
{ \bf [III]}\!\!\!&=&\!\!\! 
(-i g_s T_{kj}^c)  (-i \Gamma_{1}\delta^{ji}) \bar u(\ell-k_2)
\frac{i\left[ \gamma^{\sigma} \left(
    ( \ks{\ell}-\ks{k}_{2}+\ks{k}_{4}) + m \right) \gamma^{\rho} \right]}
{ ( \ell-k_{2}+k_{4} )^{2}-m^{2}}
 u(\ell+k_1)\times\nonumber\\
&\times&\ep^{r\star}_{\rho}(k_3)\ep^{s\star}_{\sigma}(k_4) , \\
{ \bf [IV]}\!\!\!&=&\!\!\! 
(-i \Gamma_{2}\delta^{kj})(-i g_s T_{ji}^c)  \bar u(\ell-k_2)
\frac{i\left[ \gamma^{\rho} \left(
    ( \ks{\ell}+\ks{k}_{1}-\ks{k}_{4}) + m \right) \gamma^{\sigma} \right]}
{ ( \ell+k_{1}-k_{4} )^{2}-m^{2}}
u(\ell+k_1)\times\nonumber\\
&\times&\ep^{r\star}_{\rho}(k_3)\ep^{s\star}_{\sigma}(k_4) , \\
{ \bf [V]}\!\!\!&=&\!\!\!{ \bf [V_{1}]}  + { \bf [V_{2}]} \, \\
{ \bf [V_{1}]}\!\!\!&=&\!\!\!
-i \left( \Gamma_{1}-\Gamma_{2} \right)(-i g_s)T_{ki}^c \bar u(\ell-k_2)
\gamma^{\rho}
 \frac{i(\ell^{\sigma}+k_{1}^{\sigma}) }{( \ell+k_{1}-k_{4} )^{2}-m^{2}}
u(\ell+k_1)\times\nonumber\\
&\times&\ep^{r\star}_\rho(k_3)\ep^{s\star}_{\sigma}(k_4) , \\
{ \bf [V_{2}]}\!\!&=&\!\!
-i \left( \Gamma_{1}-\Gamma_{2} \right) (-i g_s)T_{ki}^c \bar u(\ell-k_2)
\gamma^{\rho}
\frac{-i(\ell^{\sigma}-k_{2}^{\sigma}) }{( \ell-k_{2}+k_{4} )^{2}-m^{2}}
 u(\ell+k_1)\times\nonumber\\
&\times&\ep^{r\star}_{\rho}(k_3)\ep^{s\star}_{\sigma}(k_4) .
\end{eqnarray}

Before going further, let us draw the reader's attention to some points concerning these amplitudes:
\begin{itemize}
\item Gauge invariance imposes that\index{gauge!invariance}
$({\bf [0]}+{\bf [I]}+{\bf [II]})\ep^\nu(k_2)k_1^\mu=0$ and not simply
$({\bf [0]}+{\bf [I]}+{\bf [II]})k_1^\mu=0$. The same remark holds also for $k_2$. Indeed, using the
Dirac equation for the on-shell quarks, we have
$({\bf [0]}+{\bf [I]}+{\bf [II]})k_1^\mu=i g_s^2 [T^a,T^b] \bar u(\ell+k_1) \frac{k_2^\nu}{s} \ks k_1
u(\ell-k_2)$ which is orthogonal to $\ep_\nu(k_2)$. Their product therefore vanishes.
\item Gauge invariance for the right part is insured by the introduction of 
the diagrams\index{gauge!invariance}
 ${ \bf [V_{1}]}$ and ${ \bf [V_{2}]}$ that come from the choice of the following 
gauge-invariance restoring vertex derived in section~\ref{subsec:gau_inv_classical_case_vect}, 
\ce{eq:v_eff_vect_4} (see also \ct{tab:feyn_rules})
\end{itemize}
\eqs{
(\Gamma_1-\Gamma_2)\left(\frac{p^\mu }{P_1}-\frac{p'^\mu }{P_2}\right)\gnu
\rightarrow (\Gamma_1-\Gamma_2)\left(\frac{\ell^\sigma+k_1^\sigma }
{( \ell+k_{1}-k_{4} )^{2}-m^{2}}+\frac{-(\ell^\sigma 
-k_2^\sigma)}{( \ell-k_{2}+k_{4} )^{2}-m^{2}}\right)\grho
}
\begin{itemize}
\item We have split the contribution ${ \bf [V]}$ in two parts as they contain two different
propagators.
\item We have chosen to omit the $\frac{1}{\sqrt{3}}$ factor in the colour factor 
for the quark-antiquark-meson vertex. Of course, the same choice has been made during the
normalisation procedure (cf.~\ce{eq:Amunu_start}).
\item The contributions of the triple-gluon vertex ${\bf [0]}$ will vanish when contracted
with the right part of the diagram due to Furry's theorem. Indeed, a loop with an odd number of
vertices of vectors particles vanishes. We have nevertheless checked numerically
that it was the case, also for the gauge-invariance restoring vertex ${ \bf [V]}$.\index{Furry's theorem}
\item This set of contribution is what we shall refer to as the Gauge-Invariant 
Phenomenological Approach (GIPA).\index{GIPA}

\end{itemize}

The vertex functions depend on the relative momentum of the quarks at the vertices as follows
\begin{eqnarray}
    \Gamma_{1} = \Gamma\left( \ell+\frac{1}{2}\left(k_{1}-k_{2}+k_{4} \right)\right) \, , \,\,
    \Gamma_{2} = \Gamma\left( \ell+k_{1}-\frac{1}{2}k_{4}\right) \, .
\end{eqnarray}

Multiplying the amplitudes from the right with the ones from the left 
and taking the trace in Dirac matrix space and in colour space, we define the following 
quantities\footnote{Note that the following quantities are defined 
so that they contain neither the denominator of
the quarks propagators nor the meson vertex function.}:
\eqs{
\mbox{tr}^{pqrs}_{I-III}=&-g_s^3\mbox{Tr}\left[T^aT^bT^c\right]
\mbox{Tr} \left[
 \gamma^{\mu} \left( \ks{\ell} + m \right) \gamma^{\nu} 
(   ( \ks{\ell}-\ks{k}_{2} ) + m  )
\right.\hspace{3.5cm}\ \\&\left.\hspace*{-0cm}\times
 \gamma^{\sigma} \left( ( \ks{\ell}-\ks{k}_{2}+\ks{k}_{4}) + m \right) \gamma^{\rho} 
(  ( \ks{\ell}+\ks{k}_{1}) + m  )
\right]\ep_{\mu}^p(k_1)\ep_{\nu}^q(k_2)\ep^{r\star}_{\rho}(k_3)\ep^{s\star}_{\sigma}(k_4)
}
\eqs{
\mbox{tr}^{pqrs}_{I-IV}=&
-g_s^3\mbox{Tr}\left[T^aT^bT^c\right]
\mbox{Tr} \left[
 \gamma^{\mu} \left( \ks{\ell} + m \right) \gamma^{\nu} 
(   ( \ks{\ell}-\ks{k}_{2} ) + m  )
\right.\hspace{2.2cm} \ \\&\left.\times
 \gamma^{\rho} \left( ( \ks{\ell}+\ks{k}_{1}-\ks{k}_{4}) + m \right) \gamma^{\sigma} 
(  ( \ks{\ell}+\ks{k}_{1}) + m  )
\right]  \ep_{\mu}^p(k_1)\ep_{\nu}^q(k_2)\ep^{r\star}_{\rho}(k_3)\ep^{s\star}_{\sigma}(k_4)
} 
\eqs{
\mbox{tr}^{pqrs}_{II-III}=&
-g_s^3\mbox{Tr}\left[T^bT^aT^c\right]
\mbox{Tr} \left[
 \gamma^{\nu} \left(( \ks{\ell}+\ks{k}_{1}-\ks{k}_{2}) + m \right) \gamma^{\mu} 
(   ( \ks{\ell}-\ks{k}_{2} ) + m  )
\right.\hspace{0cm}\ \\&\left.\times
 \gamma^{\sigma} \left( ( \ks{\ell}-\ks{k}_{2}+\ks{k}_{4}) + m \right) \gamma^{\rho} 
(  ( \ks{\ell}+\ks{k}_{1}) + m  )
\right]\ep_{\mu}^p(k_1)\ep_{\nu}^q(k_2)\ep^{r\star}_{\rho}(k_3)\ep^{s\star}_{\sigma}(k_4)
} 
\eqs{
\mbox{tr}^{pqrs}_{II-IV}=&
-g_s^3\mbox{Tr}\left[T^bT^aT^c\right]
\mbox{Tr} \left[
 \gamma^{\nu} \left(( \ks{\ell}+\ks{k}_{1}-\ks{k}_{2}) + m \right) \gamma^{\mu} 
(   ( \ks{\ell}-\ks{k}_{2} ) + m  )
\right.\hspace{0cm}\ \\&\left.\times
 \gamma^{\rho} \left( ( \ks{\ell}+\ks{k}_{1}-\ks{k}_{4}) + m \right) \gamma^{\sigma} 
(  ( \ks{\ell}+\ks{k}_{1}) + m  )
\right]\ep_{\mu}^p(k_1)\ep_{\nu}^q(k_2)\ep^{r\star}_{\rho}(k_3)\ep^{s\star}_{\sigma}(k_4)
}
\eqs{
\mbox{tr}^{pqrs}_{I-V_1}=&
-g_s^3\mbox{Tr}\left[T^aT^bT^c\right]
\mbox{Tr} \left[
 \gamma^{\mu} \left(\ks{\ell} + m \right) \gamma^{\nu} 
(   ( \ks{\ell}-\ks{k}_{2} ) + m  )
\right.\hspace{3.5cm}\ \\&\left.\times
\gamma^{\rho}
(  ( \ks{\ell}+\ks{k}_{1}) + m  )
\right]
(\ell^{\sigma}+k_{1}^{\sigma})\ep_{\mu}^p(k_1)\ep_{\nu}^q(k_2)\ep^{r\star}_{\rho}(k_3)\ep^{s\star}_{\sigma}(k_4)
} 
\eqs{
\mbox{tr}^{pqrs}_{I-V_2}=&
-g_s^3\mbox{Tr}\left[T^aT^bT^c\right]
\mbox{Tr} \left[
 \gamma^{\mu} \left(\ks{\ell} + m \right) \gamma^{\nu} 
(   ( \ks{\ell}-\ks{k}_{2} ) + m  )
\right.\hspace{3.5cm}\ \\&\left.\times
 \gamma^{\rho}
(  ( \ks{\ell}+\ks{k}_{1}) + m  )
\right]
(-)(\ell^{\sigma}-k_{2}^{\sigma})\ep_{\mu}^p(k_1)\ep_{\nu}^q(k_2)\ep^{r\star}_{\rho}(k_3)\ep^{s\star}_{\sigma}(k_4)
} 
\eqs{
\mbox{tr}^{pqrs}_{II-V_1}=&
-g_s^3\mbox{Tr}\left[T^bT^aT^c\right]
\mbox{Tr} \left[
 \gamma^{\nu} \left( ( \ks{\ell}+\ks{k}_{1}-\ks{k}_{2}) + m \right) \gamma^{\mu} 
(   ( \ks{\ell}-\ks{k}_{2} ) + m  )
\right.\hspace{1cm}\ \\&\left.\times
\gamma^{\rho}
(  ( \ks{\ell}+\ks{k}_{1}) + m  )
\right]
(\ell^{\sigma}+k_{1}^{\sigma})\ep_{\mu}^p(k_1)\ep_{\nu}^q(k_2)\ep^{r\star}_{\rho}(k_3)\ep^{s\star}_{\sigma}(k_4)
}
\eqs{
\mbox{tr}^{pqrs}_{II-V_2}=&
-g_s^3\mbox{Tr}\left[T^bT^aT^c\right]
\mbox{Tr} \left[
 \gamma^{\nu} \left( ( \ks{\ell}+\ks{k}_{1}-\ks{k}_{2}) + m \right) \gamma^{\mu} 
(   ( \ks{\ell}-\ks{k}_{2} ) + m  )\right.\hspace{1cm}\ \\&\left.\times
 \gamma^{\rho}
(  ( \ks{\ell}+\ks{k}_{1}) + m  )
\right]
(-)(\ell^{\sigma}-k_{2}^{\sigma})\ep_{\mu}^p(k_1)\ep_{\nu}^q(k_2)\ep^{r\star}_{\rho}(k_3)\ep^{s\star}_{\sigma}(k_4)
}

Using the latter quantities, it is now straightforward to rewrite the different contributions
to the imaginary of $gg\to \!\!\ ^3S_1 g$ by including the integration over the internal quarks momenta
constrained by the two $\delta$ functions coming from the cutting rules which is
designated by 
$d_{PS}= \frac{d^{4} \ell}{8\pi^2} \delta( 2 \ell.(k_{1}+k_{2})) 
\delta(\ell^{2}+2 \ell.k_{1}-m^{2}) \theta(\ell^0+k_1^0)\theta(-\ell^0+k_2^0)$ in the following,

\eqs{
{ \bf [I]}  { \bf [III]} (p,q,r,s)
 &= \int d_{PS}\Gamma_{1} \mbox{tr}^{pqrs}_{I-III}
\frac{1}{\ell^{2}-m^{2}} 
\frac{1}{ ( \ell-k_{2}+k_{4} )^{2}-m^{2}} 
\\ % }
{ \bf [I]}  { \bf [IV]}(p,q,r,s)
 & = \int d_{PS}\Gamma_{2} \mbox{tr}^{pqrs}_{I-IV}
\frac{1}{\ell^{2}-m^{2}} 
\frac{1}{ ( \ell+k_{1}-k_{4} )^{2}-m^{2}} 
\\ 
{ \bf [II]}  { \bf [III]}(p,q,r,s)
&= \int d_{PS}\Gamma_{1} \mbox{tr}^{pqrs}_{II-III}
\frac{1}{ ( \ell+k_{1}-k_{2} )^{2}-m^{2}} 
\frac{1}{ ( \ell-k_{2}+k_{4} )^{2}-m^{2}} 
\\ % }
{ \bf [II]}  { \bf [IV]}(p,q,r,s)
&= \int d_{PS}\Gamma_{2} \mbox{tr}^{pqrs}_{II-IV}
\frac{1}{ ( \ell+k_{1}-k_{2} )^{2}-m^{2}} 
\frac{1}{ ( \ell+k_{1}-k_{4} )^{2}-m^{2}} 
\\ % }
{ \bf [I]}  { \bf [V_{1}]}(p,q,r,s)
& = \int d_{PS}\left( \Gamma_{1}-\Gamma_{2}\right) \mbox{tr}^{pqrs}_{I-V_1}
\frac{1}{\ell^{2}-m^{2}} \frac{1}{( \ell+k_{1}-k_{4} )^{2}-m^{2}}
\\ %}
{ \bf [I]}  { \bf [V_{2}]} (p,q,r,s)
&= \int d_{PS}
\left( \Gamma_{1}-\Gamma_{2}\right) \mbox{tr}^{pqrs}_{I-V_2}
\frac{1}{\ell^{2}-m^{2}} 
\frac{1}{( \ell-k_{2}+k_{4} )^{2}-m^{2}} 
\\ %}
{ \bf [II]} { \bf [V_{1}]}(p,q,r,s)
&= \int d_{PS}\left( \Gamma_{1}-\Gamma_{2}\right) \mbox{tr}^{pqrs}_{II-V_1}
\frac{1}{ ( \ell+k_{1}-k_{2} )^{2}-m^{2}}
\frac{1}{( \ell+k_{1}-k_{4} )^{2}-m^{2}}
\\ %}
{ \bf [II]} { \bf [V_{2}]} (p,q,r,s)
&=\int d_{PS}
\left( \Gamma_{1}-\Gamma_{2}\right) \mbox{tr}^{pqrs}_{II-V_2}
\frac{1}{ ( \ell+k_{1}-k_{2} )^{2}-m^{2}}
\frac{1}{( \ell-k_{2}+k_{4} )^{2}-m^{2}}
}

The full polarised amplitude of $g g \to\!\!\ ^3S_1 g$ is given by summing all these expressions.

%%%%%%%%%%%%%%%%%%%%%%%%%%%%%%%%%%%%%%%%%%%%%%%%%%%%%%
\subsection{Integration over internal kinematics}

To extract  explicit forms for the conditions constraining the integration phase space,
we shall decompose the vector $\ell$ along three independent light-cone vectors $k_1$,
$k_2$, $k_4$ and $\epsilon_{T}$
\begin{eqnarray} \label{l}
    \ell = \x k_{1} + \y k_{2} + \z k_{4} + \lambda \epsilon_{T}.
\end{eqnarray}
In terms of these new variables $(\x,\y,\z,\lambda)$, we can rewrite the element of integration
$d^{4}\ell$ as
\begin{eqnarray}
    d^{4}\ell = \frac{1}{2} \sqrt{stu} \,\,  d\x \, d\y \, d\z\, d\lambda
=\frac{1}{4} \sqrt{stu} \, d\x \, d\y \, d\z\,\frac{1}{\sqrt{\lambda^{2}}}
\left[d\lambda^{2}\left.(\cdot)\right|_{\sqrt{\lambda}=\lambda^{2}}
+d\lambda^{2}\left.(\cdot)\right|_{\sqrt{\lambda}=-\lambda^{2}} \right]
 \, .
\end{eqnarray}

From \ce{l} and  \ce{invars}-\ce{invaru},  we  find
\begin{eqnarray}\label{eq:scal_prod_ell}
   \ell^{2}   &=& (\x\y)s - (\x\z)u - (\y\z)t +\lambda^{2}\epsilon_{T}^{2}, \\
  \ell.k_{1} &=& \frac{1}{2} (\y s -\z u), \\
  \ell.k_{2} &=& \frac{1}{2} (\x s -\z t), \\
  \ell.k_{3} &=& \frac{1}{2} (\x (s+u)+\y (s+t) -\z (u+t)), \\
  \ell.k_{4} &=& - \frac{1}{2} (\x u + \y t ),
  \end{eqnarray}
and
\begin{eqnarray}
\delta( 2 \ell.(k_{1}+k_{2})) = \frac{1}{(u+t)} \, \delta \left(
\z - \frac{s}{u+t} (\x+\y)\right).
\end{eqnarray}
The integration over $\z$ can thus be done.
The second $\delta$-function in \ce{cutt} allows to integrate over $\lambda^2$:
\begin{eqnarray}
  \delta ( (\ell+k_{1})^{2}-m^{2}) =
\delta \left(
\lambda^{2}  +
\left[
\frac{s u}{u+t} (\x^{2}+\x)
+\frac{s t}{u+t} (\y^{2}-\y) +m^{2}
  \right]
\right)
\end{eqnarray}
We have for $d^{4} \ell$:
\begin{eqnarray}
\int  d^{4}  \ell (\cdot)
&=& \frac{1}{4} \frac{\sqrt{s t u}}{u+t} \int \,\, (d\x d\y)
d \z\delta \left(
\z - \frac{s}{u+t} (\x+\y)\right)
\frac{d(\lambda^{2})}{\sqrt{\lambda^{2}}}
\delta \left(
\lambda^{2} +
\left[
..  \right]
\right) (\cdot) \nonumber \\
&=&
\frac{1}{4} \frac{\sqrt{s t u}}{u+t} \,\,
\int (d\x d\y) \, \, \frac{1}{\sqrt{-\frac{su}{u+t}\left[\x^{2}+\x+
\frac{t}{u}(\y^{2}-\y) + \frac{u+t}{s u} m^{2}\right]}}(\cdot)|_{\dots}\nonumber
\\&=&\frac{1}{4} \frac{\sqrt{s t u}}{u+t} \,\,
\int (d\x d\y) \, \, \frac{1}{\sqrt{\lambda^2}}(\cdot)|_{\dots}
\,  \, .
\label{mese}
\end{eqnarray}

We can rewrite the latter square root ({\it i.e. }$\lambda$),
\begin{eqnarray}
    \sqrt{\frac{-su}{u+t}\left[ \x^{2}+\x+
\frac{t}{u}(\y^{2}-\y) + \frac{u+t}{s u} m^{2}\right]} =
\sqrt{\frac{-su}{u+t}(\x-\x_{1}(\y))
(\x-\x_{2}(\y))}
\end{eqnarray}
so that the condition $\x_{2}(\y) \leq \x \leq \x_{1}(\y)$
giving the limits of integration over $\x$ appears clearly:
\begin{eqnarray}%
   \x_{1,2}(\y) = - \frac{1}{2} \pm \frac{1}{2}
\sqrt{
1-4 \frac{t}{u} \left( \y^{2}-\y + \frac{u+t}{s t} m^{2} \right)
 } \, .
\end{eqnarray}
From this expression we also have a constraint upon the $\y$-integration. 
Similarly, the latter square root can be rewritten: 
\begin{eqnarray}%
    \sqrt{1-4 \frac{t}{u} \left( \y^{2}-\y + \frac{u+t}{s t} m^{2} \right)} =
\sqrt{-4 \frac{t}{u} (\y-\y_{1})(\y-\y_{2})} .
\end{eqnarray}
We see that if $t/u \geq 0$, then the expression is real for $\y_{2} \leq \y \leq \y_{1}$, where
\begin{eqnarray}%
    \y_{1,2}  = \frac{1}{2} \pm \frac{1}{2}\sqrt{\frac{u+t}{t} \, \frac{s-4m^{2}}{s}}
= \frac{1}{2} \pm \frac{1}{2} \frac{\sqrt{(s-M^{2})(s-4m^{2}) }}{\sqrt{-st} }
\end{eqnarray}

From the boundary values $\y_1$ and $\y_2$, we also infer that $s\geq M^{2}$ and $s\geq4m^{2}$.
The integral from \ce{mese} then becomes:
\begin{eqnarray}%
\int d^{4} \ell \, \, (\cdot) &=&
\frac{1}{4} \frac{\sqrt{s t u}}{u+t} \,\,
\int_{\y_{2}}^{\y_{1}} d \y
\int_{\x_{2}(\y) }^{\x_{1}(\y)} d \x \,
\frac{1}{\sqrt{-\frac{su}{u+t}(\x-\x_{1}(\y))(\x-\x_{2}(\y)) }}
 (\cdot)|_{\dots} \nonumber\\
&=&
\frac{1}{4} \frac{\sqrt{s t u}}{u+t} \,\,
\int [ d\x d  \y  ] \,
\frac{1}{\sqrt{-\frac{su}{u+t}(\x-\x_{1}(\y))(\x-\x_{2}(\y)) }}
 (\cdot)|_{\dots}
\end{eqnarray}

%%%%%%%%%%%%%%%%%%%%%%%%%%%%%%%%%%%%%%%%%%%%%%%%%%%%%%
\subsection{Organising the integration}
In the same fashion, we rewrite the propagators in terms of the new variables of integration
$(\x,\y,\z,\lambda)$:
\begin{eqnarray}%
 \frac{1}{(\ell^{2}-m^{2})} &=& \frac{u+t}{s u}\, \, \frac{1}{\x - \frac{t}{u} \y}
= -\frac{u+t}{s t}\, \, \frac{1}{\y - \frac{u}{t} \x}  \, , \\
 \frac{1}{((\ell+k_{1}-k_{2})^{2}-m^{2})} &=&
- \frac{u+t}{s u}\, \, \frac{1}{\x - \frac{t}{u}\y + \frac{u+t}{u}}
= \frac{u+t}{s t}\, \, \frac{1}{\y - \frac{u}{t}\x - \frac{u+t}{t}} \, ,\\
  \frac{1}{((\ell-k_{2}+k_{4})^{2}-m^{2})} &=&
- \frac{1}{u}\, \, \frac{1}{\x + \frac{t}{u} (\y-1)}
= \frac{1}{t}\, \, \frac{1}{\y -1  + \frac{u}{t} \x}
\, ,  \\
 \frac{1}{((\ell+k_{1}-k_{4})^{2}-m^{2})} &=&
 \frac{1}{u}\, \, \frac{1}{\x + \frac{t}{u}\y+1}
=  \frac{1}{t}\, \, \frac{1}{\y + \frac{u}{t}(\x+1)}
\, , \\
\end{eqnarray}

Concerning the vertex factors $\Gamma_{i}$, as discussed previously, 
we shall employ either the dipolar form or the gaussian form both with the shifted
argument $p^2_{i,sh}=p_i^2-\frac{(p_i.k_3)^2}{M^2}$ instead of $p_i^2$:
The relative
momentum squared $p_{i}^2$ is calculated from the momenta at the meson vertex
\eqs{
p^2_1&= \frac{-u \x-t (\y-1)}{2}+m^2-\frac{M^2}{4}=-\frac{s-4 m^2+(2\x+1)u+(2\y-1)t}{4}\\
p^2_2&= \frac{u (\x+1)+t \y}{2}+m^2-\frac{M^2}{4} =-\frac{s-4 m^2-(2\x+1)u-(2\y-1)t}{4}
}
and we therefore obtain the following expressions for the  argument of the vertex functions 
at the considered relative momenta:

\begin{eqnarray}%
  p^2_{1,sh} &=& \left(\frac{s-4m^2+(2\x+1)u+(2\y-1)t}{4}+\frac{(u\x+t(\y-1))^2}{4M^2}\right)\\
  p^2_{2,sh} &=& \left(\frac{s-4m^2-(2\x+1)u-(2\y-1)t}{4}+\frac{(u(\x+1)+t\y-M^2)^2}{4M^2}\right)
\nn\end{eqnarray}
and 
\eqs{
\Gamma_{i} = \frac{N}{(1+\frac{p^2_{i,sh}}{\Lambda^2})^2} \text{ or }
\Gamma_{i} = N \exp(-\frac{p^2_{i,sh}}{\Lambda^2}).
}

All integrals that we have to calculate can be recast in the following forms:
\begin{eqnarray}
 \int d_{PS} \!&&\!\!\!\!\frac{1}{(\ell^{2}-m^{2})}  \frac{1}{((\ell-k_{2}+k_{4})^{2}-m^{2})}
\Gamma_{i}\times [\ldots]
 = - \frac{1}{8 \pi^{2}} \frac{\sqrt{s t u}}{4 (u+t)}  \nonumber \hspace{2cm}\\&&
\frac{u+t}{su^2}
\int [d\x d\y] \frac{1}{\sqrt{\lambda^2}}
\frac{1}{\x - \frac{t}{u} \y} \, \frac{1}{\x + \frac{t}{u} (\y-1)}
\, \Gamma_{i}\times[\ldots]
\end{eqnarray}
\begin{eqnarray}
\int d_{PS} \!&&\!\!\!\!\frac{}{(\ell^{2}-m^{2})} \frac{1}{((\ell+k_{1}-k_{4})^{2}-m^{2})}\Gamma_{i}\times[\ldots] =
\frac{1}{8 \pi^{2}} \frac{\sqrt{s t u}}{4 (u+t)} \nonumber \hspace{2cm}\\
 && \frac{u+t}{su^2}
\int [d\x d\y] \frac{1}{\sqrt{\lambda^2}}
\frac{1}{\x - \frac{t}{u} \y} \, \frac{1}{\x + \frac{t}{u}\y+1}
\, \Gamma_{i}\times[\ldots] 
\end{eqnarray}
\begin{eqnarray}
\int d_{PS}  \!&&\!\!\!\!\frac{1}{((\ell+k_{1}-k_{2})^{2}-m^{2})}
\frac{1}{((\ell-k_{2}+k_{4})^{2}-m^{2})} \Gamma_{i}\times[\ldots] =
\frac{1}{8 \pi^{2}} \frac{\sqrt{s t u}}{4 (u+t)}
\nonumber \hspace{0cm}\\
&& 
\frac{u+t}{su^2}
\int [d\x d\y]\frac{1}{\sqrt{\lambda^2}}
\frac{1}{\x - \frac{t}{u}\y - \frac{u+t}{u}} \, \frac{1}{\x + \frac{t}{u} (\y-1)}
\, \Gamma_{i}\times[\ldots]
\end{eqnarray}
\begin{eqnarray}
\int d_{PS}  \!&&\!\!\!\!\frac{1}{((\ell+k_{1}-k_{2})^{2}-m^{2})} \frac{1}{((\ell+k_{1}-k_{4})^{2}-m^{2})}
\Gamma_{i}\times[\ldots] =-\frac{1}{8 \pi^{2}} \frac{\sqrt{s t u}}{4 (u+t)}
\nonumber \hspace{0cm}\\
&& 
\frac{u+t}{su^2}
\int [d\x d\y]\frac{1}{\sqrt{\lambda^2}}
 \frac{1}{\x - \frac{t}{u}\y - \frac{u+t}{u}} \, \frac{1}{\x + \frac{t}{u}\y+1}
\, \Gamma_{i}\times[\ldots],
\end{eqnarray}
with $\lambda=\pm \sqrt{-\frac{su}{u+t}(\x-\x_{1}(\y))(\x-\x_{2}(\y)) }$ in $[\ldots]$.

Let us introduce a short-hand notation for the denominators, the vertex functions and the
jacobian $\frac{1}{\sqrt{\lambda^2}}$:
\begin{eqnarray}\label{main}
 D_3(d_{i},e_{j},\Gamma_{k})&\equiv&
\frac{1}{\sqrt{\lambda^2}}
 \frac{1}{\x + d_{i}} \, \frac{1}{\x + e_{j}} \Gamma_k
\end{eqnarray}
with 
\begin{eqnarray}
&& d_{1}  = - \frac{t}{u} \y    \, ,  \\
&& d_{2}  = - \frac{t}{u}\y + \frac{u+t}{u} \, , \\
&& e_{1}  = + \frac{t}{u} (\y-1)   \, , \\
&& e_{2}  = + \frac{t}{u}\y+1 \, .
\end{eqnarray}
defined so that
\begin{eqnarray}\label{eq:def_d_ie_j}
 \frac{1}{(\ell^{2}-m^{2})} &=& \frac{u+t}{s u}\, \, \frac{1}{\x +d_{1}}
 \, , \nn \\
 \frac{1}{((\ell+k_{1}-k_{2})^{2}-m^{2})} &=&
- \frac{u+t}{s u}\, \, \frac{1}{\x +d_{2}}
 \, ,\nn\\
  \frac{1}{((\ell-k_{2}+k_{4})^{2}-m^{2})} &=&
- \frac{1}{u}\, \, \frac{1}{\x + e_{1 }}
\, ,  \nn \\
 \frac{1}{((\ell+k_{1}-k_{4})^{2}-m^{2})} &=&
 \frac{1}{u}\, \, \frac{1}{\x + e_{2}}
\, ,
\end{eqnarray}

The amplitudes can then be written as\footnote{ the $+$ or $-$ signs before $\cal C$ simply
come from \ce{eq:def_d_ie_j}.}:
\begin{eqnarray}\label{eq:expr_final_pol_ampl}
{ \bf [I]}  { \bf [III]}(p,q,r,s) &=&
     -{\cal C} \int[d\x d\y] D_3(d_{1},e_{1},\Gamma_{1})\mbox{tr}_{I-III}^{pqrs} \\
{ \bf [I]}  { \bf [IV]}(p,q,r,s) &=&
     +{\cal C}\int[d\x d\y] D_3(d_{1},e_{2},\Gamma_{2})\mbox{tr}_{I-IV}^{pqrs} \\
{ \bf [II]}  { \bf [III]}(p,q,r,s) &=&
     +{\cal C}\int[d\x d\y] D_3(d_{2},e_{1},\Gamma_{1})\mbox{tr}_{II-III}^{pqrs} \\
{ \bf [II]}  { \bf [IV]}(p,q,r,s) &=&
     -{\cal C}\int[d\x d\y] D_3(d_{2},e_{2},\Gamma_{2})\mbox{tr}_{I-IV}^{pqrs} \\
{ \bf [I]}  { \bf [V_{1}]}(p,q,r,s) &=&
     +{\cal C}\int[d\x d\y] \left[ D_3(d_{1},e_{2},\Gamma_{1}) - D_3(d_{1},e_{2},\Gamma_{2})\right] 
\mbox{tr}_{I-V_{1}}^{pqrs} \\
{ \bf [I]}  { \bf [V_{2}]}(p,q,r,s) &=&
     -{\cal C}\int[d\x d\y] \left[ D_3(d_{1},e_{1},\Gamma_{1}) - D_3(d_{1},e_{1},\Gamma_{2})\right] 
\mbox{tr}_{I-V_{2}}^{pqrs}\\
{ \bf [II]}  { \bf [V_{1}]}(p,q,r,s) &=&
    -{\cal C}\int[d\x d\y] \left[ D_3(d_{2},e_{2},\Gamma_{1}) - D_3(d_{2},e_{2},\Gamma_{2})\right] 
\mbox{tr}_{II-V_{1}}^{pqrs} \\
{ \bf [II]}  { \bf [V_{2}]}(p,q,r,s) &=&
    +{\cal C}\int[d\x d\y] \left[ D_3(d_{2},e_{1},\Gamma_{1}) - D_3(d_{2},e_{1},\Gamma_{2})\right] 
\mbox{tr}_{II-V_{2}}^{pqrs}
\end{eqnarray}
with
\begin{eqnarray}%
   {\cal C} = \frac{1}{32 \pi^{2}}\frac{\sqrt{s t u}}{s u^2}
\end{eqnarray}

Apart from the denominators and the vertex function in $D_{3}$, 
the only dependence  on $\left[\x\y\right]$ left
is  contained in the $tr_{X-Y}$ -- through scalar products involving $\ell$. 
Indeed, as we have seen, the decomposition of $\ell$ will introduce
$\x$, $\y$, $\z$ and $\lambda$ factors. The
 $\z$ factors are dropped in favour of $\x$ and $\y$. On the other
hand, as $\lambda$ has non-linear dependence on $\x$ and $\y$, we shall keep it explicitly.

It is also easy to convince oneself that we can decompose any $tr_{X-Y}$ as:
\eqs{\label{trace_decomp}
tr_{X-Y} = \sum_{l,m,n} K_{l,m,n}^{tr_{X-Y}} \lambda^l \x^m \y^n,
}
where $l,m,n = (0,1,2,3,4)$ and $l+m+n \leq 4$. This follows simply from the number of $\ell$ in
each $tr_{X-Y}$. 
Nothing here will be stated about $K_{l,m,n}^{tr_{X-Y}}$, except 
that they do not contain $\x$ and $\y$, {\it i.e.} the integration
variables.

Therefore, we are simply to perform one type of integrals
\begin{eqnarray}\label{ai_def}
 I(d_{i},e_{j},\Gamma_{k}; l,m,n)&\equiv&
\int[d\x d\y]\lambda^{l}\x^{m} \y^{n} D_3(d_{i},e_{j},\Gamma_{k})\\&=&
\int[d\x d\y]\frac{\lambda^{l}\x^{m} \y^{n}}{\sqrt{-\frac{su}{u+t}(\x-\x_{1}(\y))(\x-\x_{2}(\y)) }}
 \frac{1}{\x + d_{i}} \, \frac{1}{\x + e_{j}}
\, \Gamma_k \, \nonumber\\
&=&\int[d\x d\y]\frac{\lambda^{l}\x^{m} \y^{n}}{\sqrt{\lambda^2}}
\frac{1}{\x + d_{i}} \, \frac{1}{\x + e_{j}}
\, \Gamma_k \, .
\end{eqnarray}
We already know that the integrals will vanish for 
$l$ odd, as the two branches of the integrals over $\lambda^2$ cancel each other.

From this feature, we can also draw some conclusions on some helicity amplitudes. Indeed, the only
way to make $\lambda$  appear and not $\lambda^2$ is through the scalar product 
$\ell.\epsilon_T=\lambda$; all other possible scalar products with $\ell$ (see \ce{eq:scal_prod_ell})
do not involve an odd power of $\lambda$. Since $\epsilon_T$ 
appears solely in $\ep^{T_1}_i$, 
we can check that all amplitudes with an odd number of particles polarised
along $\epsilon_T$ (vector orthogonal to the collision plane) vanish, as should be.

\section{Polarised cross sections}

Combining \ce{trace_decomp} with  the expressions for the polarised amplitudes as 
in~\ce{eq:expr_final_pol_ampl}
and the generic expression for the integrals on the internal phase space~\ce{ai_def}, we have
\eqs{
{\cal M}(p,q,r,s)= \sum_{X,Y}{ \bf [X]}  { \bf [Y]}(p,q,r,s)= \sum_{X,Y} 
\sum_{l,m,n} K_{l,m,n}^{tr_{X-Y}^{pqrs}}I(d_{i},e_{j},\Gamma_{k}; l,m,n),
}
the coefficients $K_{l,m,n}^{tr_{X-Y}^{pqrs}}$ are determined by FORM~\cite{Vermaseren:2000nd} 
and are functions\footnote{As now, we are dealing back with hadronic and partonic processes, 
we put the circumflexes back for the partonic Mandelstam variables.}
 of $\hat s$, $\hat t$ and $\hat u$ whereas the integrals 
$I(d_{i},e_{j},\Gamma_{k}; l,m,n)$ are performed numerically for each sets 
of $\hat s$, $\hat t$ and $\hat u$ after having chosen the parameter
defining $\Gamma$.

Squaring the sum of the expressions ${ \bf [X]}  { \bf [Y]}(p,q,r,s)$, 
multiplying them by $\frac{1}{64}$ for 
the averaging on initial colour, by $\frac{1}{4}$ for  the averaging on initial spin, we get 
$|\overline{\cal M}(p,q,r,s)|^2$ to be plugged into~\ce{eq:sigmadt}. This in turn gives the polarised 
differential cross section $\frac{d \sigma(p,q,r,s)}{d\hat t}$.

As we are concerned with the polarisation effects only for the vector meson, we shall sum the 
various helicity contributions for the gluons and define the following polarised cross sections:
\eqs{
\frac{d \sigma_{T_1}}{d\hat t}=\sum_{p,q,s=T_1,T_2}\frac{d \sigma(p,q,T_1,s)}{d\hat t} \\
\frac{d \sigma_{T_2}}{d\hat t}=\sum_{p,q,s=T_1,T_2}\frac{d \sigma(p,q,T_2,s)}{d\hat t} \\
\frac{d \sigma_L}{d\hat t}=\sum_{p,q,s=T_1,T_2}\frac{d \sigma(p,q,L,s)}{d\hat t} \\
}
The hadronic cross sections are therefore obtained with 
\eqs{
\frac{d\sigma_h}{dydP_T}=\int_{x_1^{min}}^1 dx_1 \frac{2 \hat s P_T G_1(x_1) G_2(x_2(x_1))}
{\sqrt{s}(\sqrt{s}x_1-E_T e^{y})}\frac{d\sigma_h}{d\hat t},
}
for $h=T_1,T_2$ or $L$. Specifying $s$, $y$ and $P_T$, we get the final result.

\section{Numerical results and comparison with the CSM}

The most natural comparison of our calculation is with the LO CSM predictions. Indeed, the structure 
of that calculation is similar to that of our model with
solely the imaginary part of the amplitude. We recall here that the two quarks forming the meson
are put on their mass shell in the CSM.

The comparison will be done with the unpolarised partonic cross section for $gg\to\!\!\ ^3S_1g$
 obtained in~\cite{CSM_hadron}: 
\begin{eqnarray}%\label{eq:CSM_dsdt}
&\ & \hspace{-0.5cm} \frac{d\sigma}{d\hat t}= \frac{20\pi^2 M 
\alpha_s^3|\psi(0)|^2}{9\hat s^2} 
 \frac{\left[\hat s^2
(\hat s-M^2)^2\right]+\left[\hat t^2(\hat t-M^2)^2\right]+
\left[\hat u^2(\hat u-M^2)^2\right]}
{(\hat s-M^2)^2(\hat t-M^2)^2(\hat u-M^2)^2}. \nonumber
\end{eqnarray}

We shall use the gluon pdf's from CTEQ6LO fit~\cite{Pumplin:2002vw} and MRST2001LO~\cite{Martin:2002dr}
evaluated at the scale $Q=\sqrt{M^2+P_T^2}$. The coupling constant is computed as 
\eqs{
\alpha_s(Q)=\frac{12\pi}{23\log(\frac{Q^2}{0.165^2})} 
}
with 5 active flavours and $\Lambda_{QCD}=0.165$ GeV~\cite{Pumplin:2002vw}. Indeed,
$Q$ is above $m_b$ as soon as $P_T$ reaches 3-3.5 GeV for the $J/\psi$ .

\subsection{$J/\psi$ at the Tevatron}\index{jpsi@$J/\psi$|(}

Setting $\sqrt{s}$ to 1800 GeV and considering the cross section in the range $|\eta|<0.6$, 
we get the following results for $J/\psi$ production at the Tevatron in the experimental
conditions of the CDF collaboration whose measurements are detailed in chapter~\ref{ch:experim}.
The first plot~(see \cf{fig:g_m187_l183_mrst}) shows our prediction 
($\sigma_{TOT}$, $\sigma_T$ and $\sigma_L$) for $m_c=1.87$ GeV and $\Lambda=1.8$ GeV to be 
compared with the LO CSM (which does not depend on $m_c$) and the CSM fragmentation. We see that
our prediction for the imaginary part of the amplitude
is lower than the LO CSM, except at large $P_T$ and is mostly longitudinal.

On the other hand, these results include the gauge-invariance restoring vertex whose choice is arbitrary.
The conclusion drawn here might be only relevant for the GIPA, \ie~the standard 
pQCD set of diagrams
with a phenomenological vertex for $q \bar q\to {\cal Q}$ and the gauge-invariance restoring 
term of~\ce{eq:v_eff_vect_4}. Differences between possible choices of gauge-invariance 
restoring vertices are described by what we call autonomous contributions, their effects 
are analysed in the next section.

\begin{figure}[h]
\centering
\includegraphics[height=9cm]{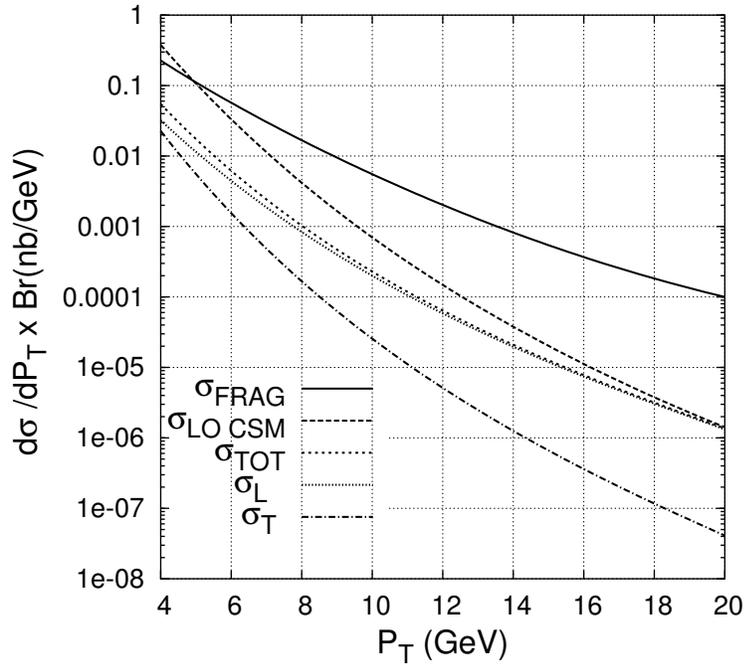}
\caption{Polarised ($\sigma_T$ and $\sigma_L$) and total ($\sigma_{TOT}$) cross sections obtained 
with a gaussian vertex functions, $m_c=1.87$ GeV, $\Lambda=1.8$ GeV and the MRST gluon distribution, 
to be compared with LO CSM and fragmentation CSM contributions.
}
\label{fig:g_m187_l183_mrst}
\end{figure}

Then, we illustrate (see~\cf{fig:GIPA_jpsi_comp}~(a)) the effect due to a change in the form 
of the vertex functions, \ie~from a dipole to a gaussian. The effects due to a variation
in $\Lambda$ (see \cf{fig:GIPA_jpsi_comp}~(b)), $m_c$ and $\Lambda$ (see \cf{fig:GIPA_jpsi_comp}~(c)),
due to a change in the scale $Q=\sqrt{M^2+P^2_T}$ (see \cf{fig:GIPA_jpsi_comp}~(d)) are also shown.

\begin{figure}[H]
\centering
\mbox{\subfigure[Vertex function]{\includegraphics[height=6.5cm]{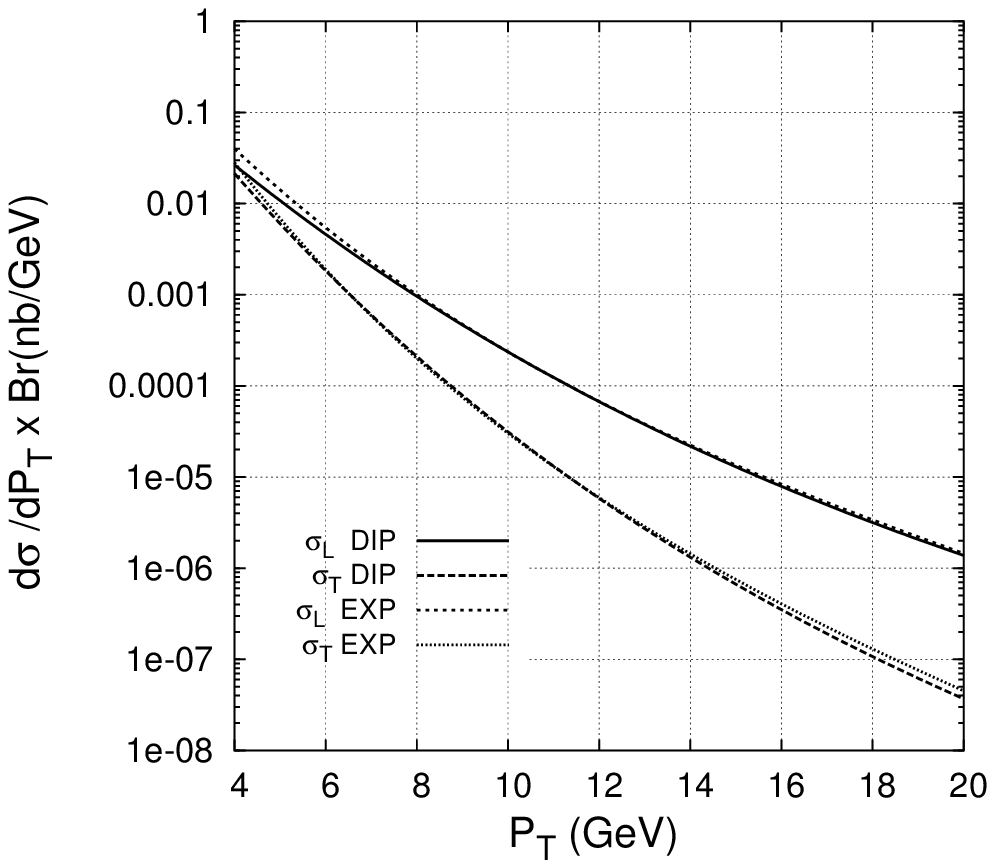}}\quad\quad
      \subfigure[$\Lambda$]{\includegraphics[height=6.5cm]{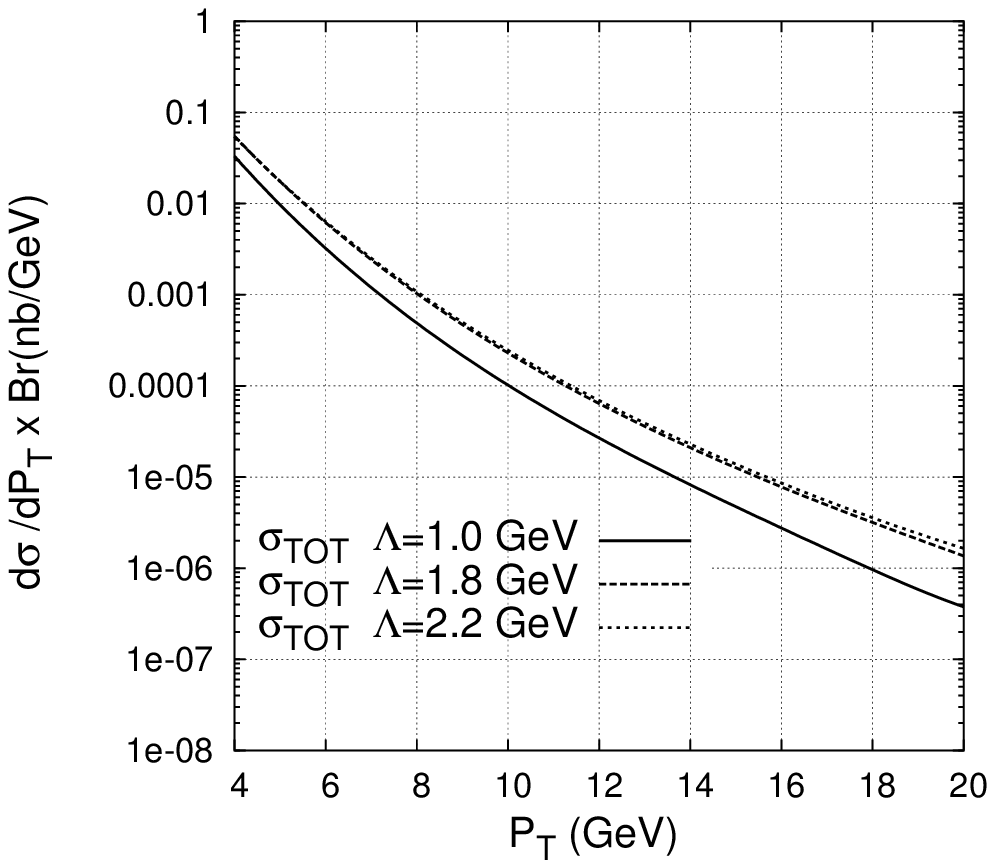}}} 
\mbox{\subfigure[$\Lambda$ and $m_c$]{\includegraphics[height=6.5cm]{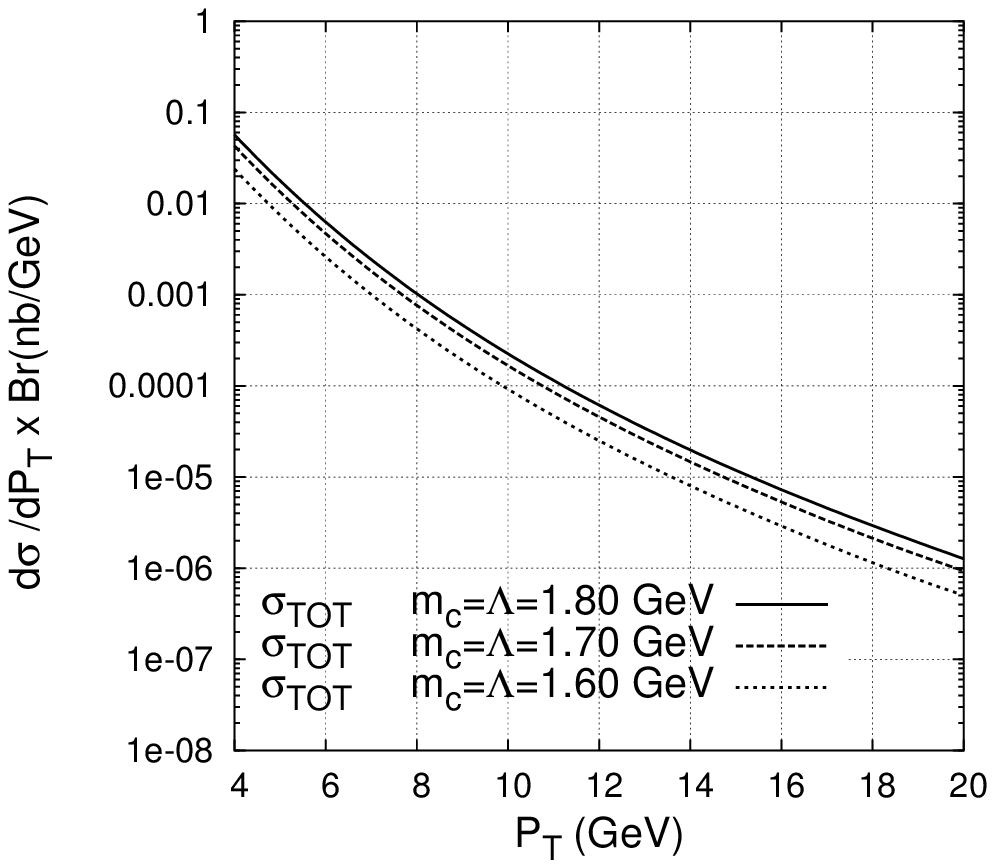}}\quad\quad
      \subfigure[Scale]{\includegraphics[height=6.5cm]{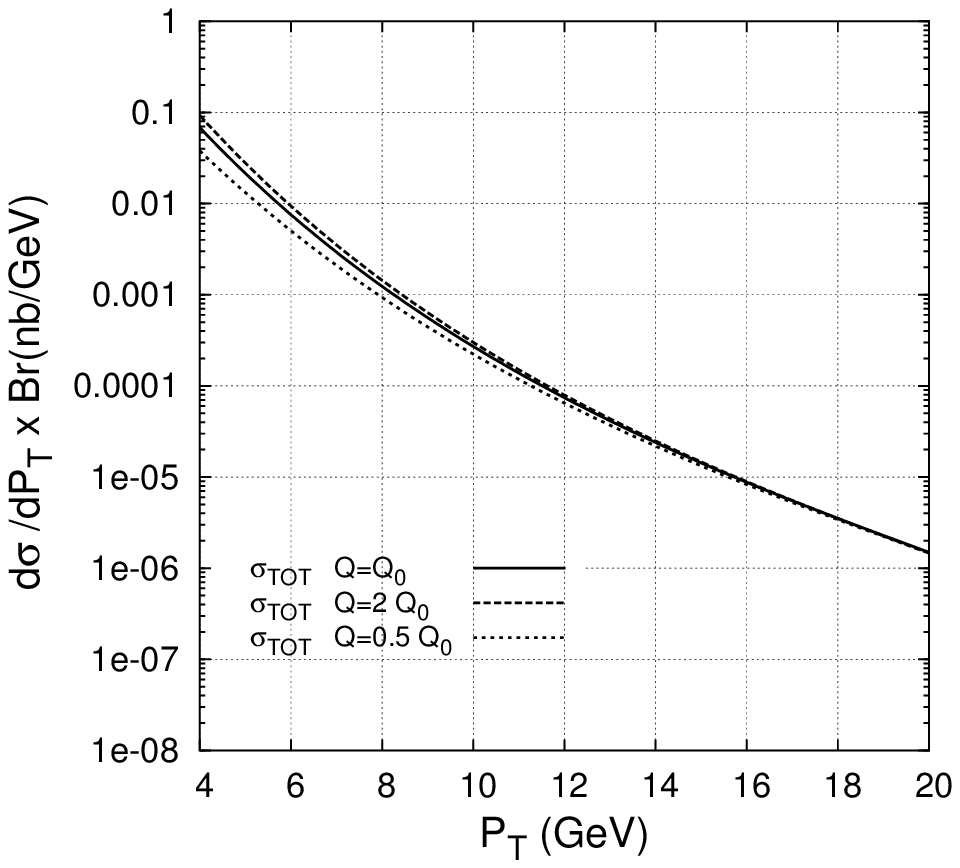}}
}
\caption{(a) Comparison between the cross section obtained with the dipole and the gaussian
vertex functions (MRST); (b) Variation of the cross section due to a change in $\Lambda$ for a fixed value
of the quark mass (MRST); (c) Variation of the cross section due to a change in $m_c$ and $\Lambda$ (CTEQ);
(d)  Variation of the cross section due to a change of a factor two in the scale $Q=\sqrt{M^2+P^2_T}$ 
(CTEQ).}
\label{fig:GIPA_jpsi_comp}
\end{figure}

As the changes in the vertex function illustrated above are of little influence on the final 
results, all the following plots for the $J/\psi$, unless specified differently, will 
be made for $m_c=1.87$ GeV, $\Lambda=$ 1.8 GeV, $Q=\sqrt{M^2+P^2_T}$  and a gaussian vertex function.
\index{jpsi@$J/\psi$|)}

\subsection{$\Upsilon(1S)$ at the Tevatron}\index{upsi@$\Upsilon$|(}

Here are similar plots for the $\Upsilon(1S)$ at the Tevatron ($\eta=0$). The deviation
of the GIPA with the CSM is slightly more marked, perhaps indicating that the deviation behaves like
a power of $\frac{m_Q}{P_T}$. In the following, all plots for the $\Upsilon(1S)$, 
unless specified differently, will  be made for $m_b=5.28$ GeV, $\Lambda=$~6.0 GeV, 
$Q=\sqrt{M^2+P^2_T}$  and a gaussian vertex function.

\begin{figure}[H]
\centerline{\includegraphics[width=8cm]{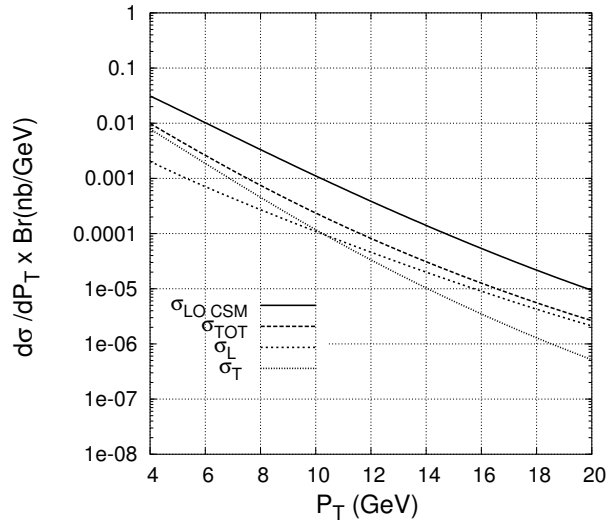}}
\caption{Polarised ($\sigma_T$ and $\sigma_L$) and total ($\sigma_{TOT}$) cross sections obtained 
with a gaussian vertex functions, $m_b=5.28$~GeV, $\Lambda=6.0$~GeV and the CTEQ gluon distribution, 
to be compared with LO CSM and fragmentation CSM contributions}
\label{fig:g_m528_l600_cteq}
\end{figure}

\begin{figure}[H]
\centerline{\includegraphics[width=8cm]{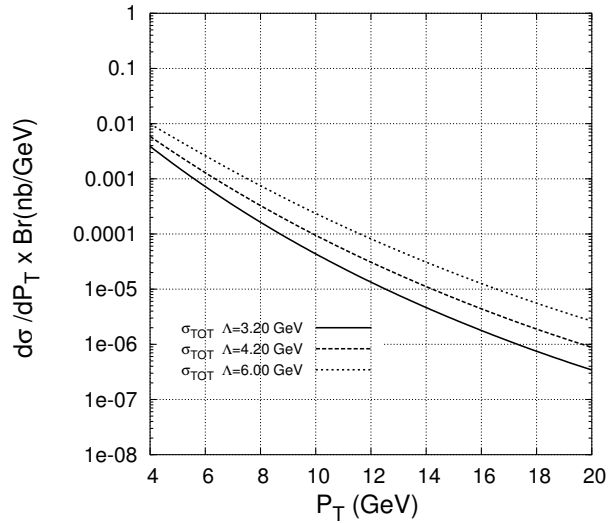}}
\caption{Variation of the cross section due to a change in $\Lambda$ for a fixed value
of the quark mass (CTEQ)}
\label{fig:g_m528_lvar_cteq}
\end{figure}\index{upsi@$\Upsilon$|)}

\subsection{$J/\psi$ at RHIC}\index{jpsi@$J/\psi$|(}

Here are  plots for the $J/\psi$ at RHIC ($\sqrt{s}=200$ GeV) (for $-0.5<y<0.5$).

\begin{figure}[H]
\centerline{\includegraphics[width=8.0cm]{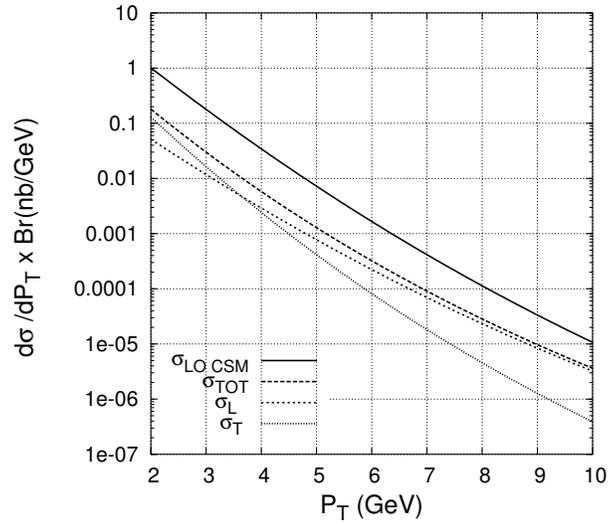}}
\caption{Polarised ($\sigma_T$ and $\sigma_L$) and total ($\sigma_{TOT}$) cross sections obtained 
at $\sqrt{s}=200$~GeV with a gaussian vertex functions, $m_c=1.87$~GeV, 
$\Lambda=1.8$~GeV and the CTEQ gluon distribution, to be compared with LO CSM.}
\label{fig:g_m187_l183_cteq_RHIC200}
\end{figure}

\begin{figure}[H]
\centerline{\includegraphics[width=8.0cm]{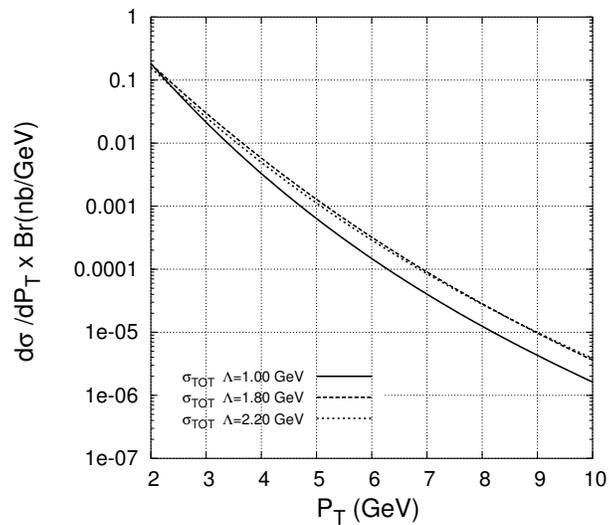}}
\caption{Polarised ($\sigma_T$ and $\sigma_L$) and total ($\sigma_{TOT}$) cross sections obtained 
at $\sqrt{s}=200$~GeV with a gaussian vertex functions, $m_c=1.87$~GeV, 
$\Lambda=1.8$~GeV and the CTEQ gluon distribution, to be compared with LO CSM.}
\label{fig:g_m187_lvar_cteq_RHIC200}
\end{figure}\index{jpsi@$J/\psi$|)}

\subsection{$\psi'$ at the Tevatron}\index{psip@$\psi'$|(}\index{node}

Here we plot the cross section for the $\psi'$ at  $\sqrt{s}=1800$~GeV with  $m_c=1.87$~GeV, 
$\Lambda=1.8$~GeV and for three values of the node parameter $a_{node}$: 0.8, 1.46, 2.0 GeV
(see \ce{eq:gam2s_exp} and \ct{tab:N_psip})
We see that this parameter can have a significant influence. This comes from the fact 
that the normalisation strongly
depends on the node position, whereas the production process is not very sensitive to this parameter.

In the following, we shall use $a_{node}=1.46$~GeV,
$m_c=1.87$~GeV, $\Lambda=1.8$~GeV and a gaussian vertex function.

\begin{figure}[H]
\centerline{\includegraphics[height=9cm]{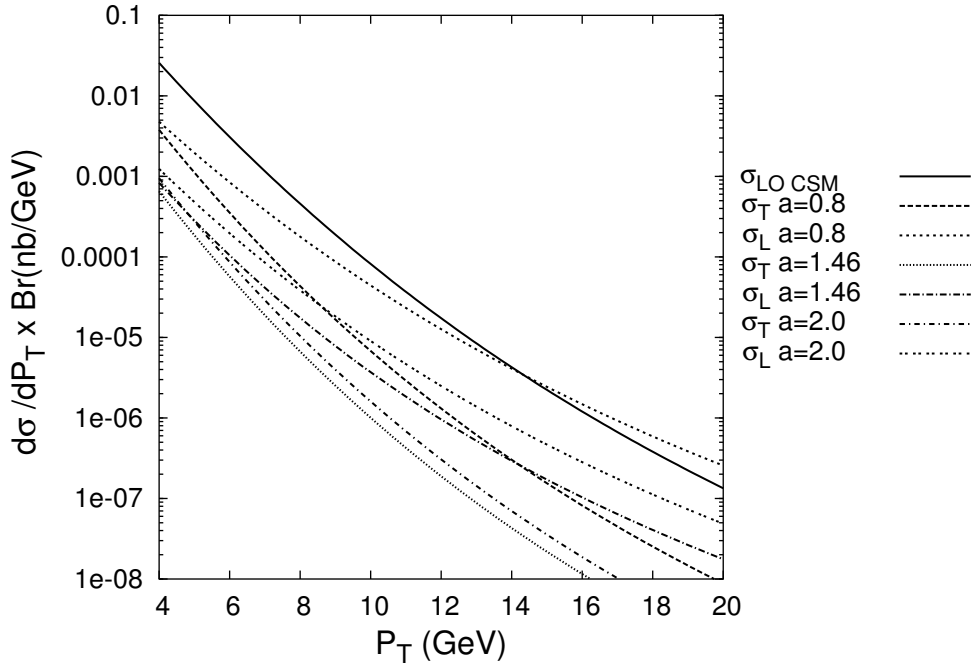}}
\caption{Polarised ($\sigma_T$ and $\sigma_L$) cross sections obtained 
at $\sqrt{s}=1800$~GeV with a gaussian vertex functions, $m_c=1.87$ GeV, 
$\Lambda=1.8$ GeV and the CTEQ gluon distribution for various values of $a_{node}$.}
\label{fig:g_psip_avar_0-0}
\end{figure}
\index{psip@$\psi'$|)}

\subsection{Taking stock of the situation}

Following the discussion concerning the $J/\psi$ at the Tevatron and RHIC, the $\Upsilon$ 
at the Tevatron and the $\psi'$ at the Tevatron, we may draw some conclusions concerning
the effects of changing the energy involved in the collision, the mass of the quarks and
the internal structure of the quarkonium. We can first say that the main features
of our predictions are not drastically affected by these changes.

To what concerns the deviation between the GIPA and the LO CSM, it seems that it is a permanent
feature at low $P_T$. As we have said before, it indicates that the real part can be large. 
One important result of our study of non-static effects -- or relativistic effects -- 
is the existence of a large uncertainty in $\psi'$ production linked with the 
position of the node in the vertex function.
\index{static!non-static}

We have shown that it is possible to go beyond the static approximation and that our 
procedure does not show any instabilities and uncontrollable behaviour due to changes in 
$m_q$ and $\Lambda$. Our procedure to normalise the vertex function also seems 
to work and to provide sensible 
values.\index{static!approximation}

We have the hope that tuning the internal dynamics
of the quarkonium in production processes could be applied to their study  in media; for instance,
to study their suppression due to an evolution in their size or due to an alteration of the binding
of their constituting quarks.

However, numerous choices of gauge-invariance restoring vertices are possible. No 
firm conclusions should be drawn before we have an idea of the effects due to a change in them.
The easiest way to get insights on this issue comes from the analysis of contributions that
were derived in section~\ref{sec:auton_contrib}. This is the subject of the following study.

\section{Analysis of autonomous vertices}

We now consider the contributions from autonomous vertices as 
in~\ce{eq:general_form_vmunu_contrib_qqVg}. We have simply to introduce
sub-amplitude ${ \bf [VI]}$, ${ \bf [VII]}$, \dots, whose expressions are
obtained from the vertex chosen. 

We limit ourselves to choices for the vertex coming 
from~\ce{eq:general_form_vmunu_contrib_qqVgb} with constant factors, we have therefore three 
possibilities, that we rewrite in the conventions of this chapter:
\begin{enumerate}
\item $(\Gamma_1-\Gamma_2)\alpha \gmu q^\nu \to (\Gamma_1-\Gamma_2) \alpha  \gamma^\sigma k_4^\rho$
\hspace*{\fill} the ``$\alpha$-term''
\item $(\Gamma_1-\Gamma_2)\beta \, (p+p')^\mu (p+p')^\nu \to (\Gamma_1-\Gamma_2) \beta \,
(2\ell+k_1-k_2)^\rho (2\ell+k_1-k_2)^\sigma$ 
the ``$\beta$-term''
\item $(\Gamma_1-\Gamma_2)\xi \, g^{\mu\nu} \to (\Gamma_1-\Gamma_2) \xi \,g^{\rho\sigma}$
\hspace*{\fill} the ``$\xi$-term''
\end{enumerate}

Let us consider the first one. If we include it in our calculation, it will induce
the following sub-amplitude to be added to the set ${ \bf [III]+[IV]+[V]}$,
\begin{eqnarray}
{ \bf [VI]}\!\!\!&=&\!\!\!
-i \alpha \left( \Gamma_{1}-\Gamma_{2} \right)(-i g_s)T_{ki}^c \bar u(\ell-k_2)
i \gamma^{\sigma} k_4^\rho
u(\ell+k_1)\ep^{r\star}_\rho(k_3)\ep^{s\star}_{\sigma}(k_4) .
\end{eqnarray}

Let us now have a closer look at the product $k_4^\rho\ep^{r\star}_\rho(k_3)$.
The latter is zero for $r=T$ (\ce{eq:pol-orthtr2}), whereas for $r=L$ it is (\ce{eq:pol-final})
\eqs{
\ep^{L}(k_3).k_4=\frac{\hat s-M^2}{2M}.
}

This contribution will thus produce solely longitudinal vector mesons. It is the ingredient
needed for a possible solution of the polarisation issue. We shall come back to this fact 
when we analyse the numerical results.

In practice, the introduction of these contributions does not modify much the procedure 
to get the polarised cross section as explained in the previous sections. 
The main difference is the introduction of a new quantity
\begin{eqnarray}
 D_2(d_{i},\Gamma_{j})&\equiv&
\frac{1}{\sqrt{\lambda^2}}
 \frac{1}{\x + d_{i}} \,  \Gamma_k
\end{eqnarray}
which appears in the following new integrals,
\begin{eqnarray}
{ \bf [I]}  { \bf [VI]}(p,q,r,s) &=&
     +{\cal C'}\int[d\x d\y] \left[ D_2(d_{1},\Gamma_{1}) - D_2(d_{1},\Gamma_{2})\right] 
\mbox{tr}_{I-VI}^{pqrs}\\
{ \bf [II]}  { \bf [VI]}(p,q,r,s) &=&
    -{\cal C'}\int[d\x d\y] \left[ D_2(d_{2},\Gamma_{1}) - D_2(d_{2},\Gamma_{2})\right] 
\mbox{tr}_{II-VI}^{pqrs},
\end{eqnarray}
where $\cal C'$ is $\frac{1}{32 \pi^2}\frac{\sqrt{\hat s\hat t\hat u}}{\hat s\hat u}$ and the quantities $\mbox{tr}_{I-VI}^{pqrs}$ and $\mbox{tr}_{II-VI}^{pqrs}$  are
\eqs{
\mbox{tr}^{pqrs}_{I-VI}=&
-\alpha g_s^3\mbox{Tr}\left[T^aT^bT^c\right]
\mbox{Tr} \left[
 \gamma^{\mu} \left(\ks{\ell} + m \right) \gamma^{\nu} 
(   ( \ks{\ell}-\ks{k}_{2} ) + m  )
\right.\hspace{3.5cm}\ \\&\left.\times
\gamma^{\sigma}
(  ( \ks{\ell}+\ks{k}_{1}) + m  )
\right]
k_4^\rho\ep_{\mu}^p(k_1)\ep_{\nu}^q(k_2)\ep^{r\star}_{\rho}(k_3)\ep^{s\star}_{\sigma}(k_4)
} 
\eqs{
\mbox{tr}^{pqrs}_{II-VI}=&
-\alpha g_s^3\mbox{Tr}\left[T^bT^aT^c\right]
\mbox{Tr} \left[
 \gamma^{\nu} \left( ( \ks{\ell}+\ks{k}_{1}-\ks{k}_{2}) + m \right) \gamma^{\mu} 
(   ( \ks{\ell}-\ks{k}_{2} ) + m  )
\right.\hspace{1cm}\ \\&\left.\times
\gamma^{\sigma}
(  ( \ks{\ell}+\ks{k}_{1}) + m  )
\right]
k_4^\rho\ep_{\mu}^p(k_1)\ep_{\nu}^q(k_2)\ep^{r\star}_{\rho}(k_3)\ep^{s\star}_{\sigma}(k_4)
}

\subsection{Numerical results}

To get a first idea, we have introduced these vertices in the amplitude
one by one with $\alpha$, $\beta$ or $\xi$ equal\footnote{Note that in this first application
the coefficients  $\alpha$, $\beta$ and $\xi$ do not have the same dimension.} to 1. 
In \cf{fig:g_m187_l183_cteq_xi}, we show the individual contributions of these terms. Note that some of 
these produce only one helicity.
One example already foreseen concerns the $\alpha$-term  whose transverse
cross section is zero. However, if we want to fit the data, the autonomous vertices
will need to be dominant, hence the GIPA and any interferences terms 
will not be important.

\begin{figure}[H]
\centering
\includegraphics[width=10cm]{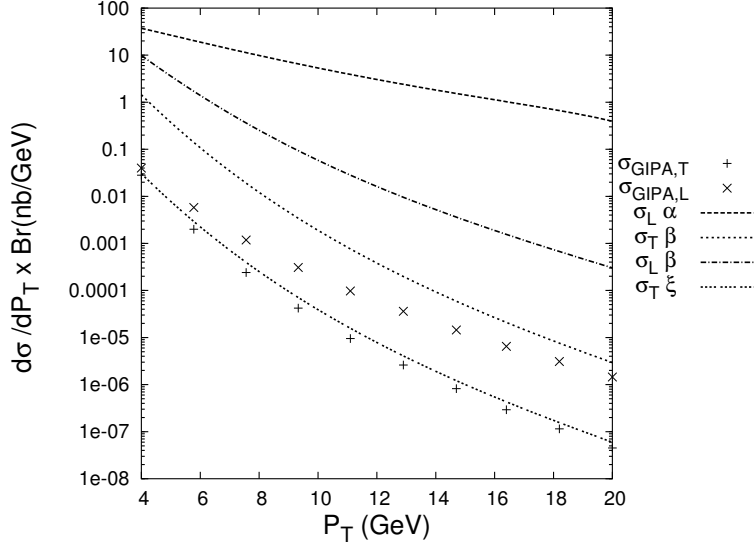}
\caption{Polarised ($\sigma_T$ and $\sigma_L$)  cross sections for the $J/\psi$ at the Tevatron obtained 
with three different autonomous vertices alone.  Note that $\sigma_T^\alpha=\sigma_L^\xi=0$.}
\label{fig:g_m187_l183_cteq_xi}
\end{figure}

The most striking result is that this latter contribution ($\alpha$-term) has a
slope compatible with the data. The contribution from 
the $\xi$-term has no longitudinal contribution\footnote{This can be also foreseen
since $g^{\rho\sigma}\ep^{r\star}_\rho(k_3)\ep^{s\star}_{\sigma}(k_4)= \ep^{r\star}(k_3).\ep^{s\star}(k_4)$ is zero for $r=L$ (see section~\ref{subsec:polvec}).} and its transverse component is 
slightly bigger than the GIPA. Yet, one might expect, since its 
dimension is lower, that the actual value of $\xi$ would enhance it.
The contribution of the $\beta$-term has no remarkable characteristic and considering 
its dimension, we do not expect $\beta$ to  enhance its contribution.

\subsection{Fitting the data}

We are then going to make use of these two unconstrained structures ($\alpha$ and $\xi$-terms).
One is purely  transverse whereas the other is purely longitudinal. 
This is a motivation to try to reproduce the behaviour of the polarisation parameter 
$\alpha_{pol}$ as measured by CDF (see section~\ref{sec:polarisation}).

First we shall rescale  the multiplicative factors $\alpha$ and $\xi$ to get 
dimensionless parameters:
\eqs{
      \alpha &\to \frac{\alpha}{\sqrt{\hat s}m_Q}, \\
      \xi &\to \frac{\xi}{m_Q}.
}
Note that this rescaling is purely arbitrary. It is far from obvious which quantity among  
$m_{\cal Q}$ (mass of the quarkonium), 
$m_Q$ , $\Lambda$, $\Lambda_{QCD}$, $\sqrt{\hat s}$, \dots is the most suitable.
However, except  for $\Lambda_{QCD}$, they are all of the same order. Besides, 
the uncertainties coming from
the missing real part may modify the final values obtained in the fit. 

\subsubsection{$J/\psi$ at the Tevatron}\index{jpsi@$J/\psi$|(}

Setting $\alpha$ to 8 and $\xi$ to 37.5,  we reproduce the cross section for $J/\psi$ at 
$\sqrt{s}=1800$~GeV~\cite{CDF7997b} as displayed in~\cf{fig:g_m187_l183_cteq_37_5-8}.  
However since $\sigma_L$ is much larger than $\sigma_T$
for $P_T \gtrsim 10$ GeV, we predict that $\alpha_{pol}$ will be negative for $P_T \gtrsim 6$~GeV
at variance with COM fragmentation predictions, for which $\alpha_{pol}$ is significantly positive. 
This fact can easily traced back to the steeper
slope of the $\xi$-term compared to the $\alpha$-term 
(see~\cf{fig:g_m187_l183_cteq_xi} (b)).\index{jpsi@$J/\psi$|)}

\subsubsection{$\Upsilon(1S)$ at the Tevatron}\index{upsi@$\Upsilon$|(}

Setting $\alpha$ to 8 and $\xi$ to 10,  we reproduce the cross section for $\Upsilon(1S)$  at 
$\sqrt{s}=1800$~GeV~\cite{Acosta:2001gv} as shown in~\cf{fig:g_m528_l600_cteq_10-8}. 
We observe the same behaviour as in the $J/\psi$ case
to what concerns  $\sigma_L$ relative to $\sigma_T$ and  we predict that $\alpha_{pol}$ 
will be negative for $P_T \gtrsim 8$~GeV.\index{upsi@$\Upsilon$|)}

\subsubsection{$J/\psi$ at RHIC}\index{jpsi@$J/\psi$|(}

Since the $P_T$ range of the data~\cite{Adler:2003qs} is not very wide and the error bars 
are rather large,
we use the same values for $\alpha$ and $\xi$ as for the Tevatron energy. Furthermore, 
our calculation (\cf{fig:g_m187_l183_cteq_37_5-8-RHIC}) agrees with the 
data, even though slightly
improved agreement could be obtained with a different set of values for $\alpha$ and $\xi$.
To what concerns polarisation, we predict that $\alpha_{pol}$ will be positive for 
$P_T \lesssim 4$~GeV and negative for larger values of $P_T$.

Note that our LO CSM curve includes solely gluon fusion at variance with curves of 
Nayak~\etal~\cite{Nayak:2003jp} which include all parton-fusion processes.\index{jpsi@$J/\psi$|)}

\subsubsection{$\psi'$ at the Tevatron}\index{psip@$\psi'$|(}\index{node}

Setting $a_{node}$  to 1.46 GeV and $\alpha$ to 27.5, we get the cross section for $\psi'$  at 
$\sqrt{s}=1800$~GeV as shown in~\cf{g_psip_a146_0-27_5}. This fits perfectly the
data obtained by CDF~\cite{CDF7997b}. There is no need to introduce any contribution from
the $\xi$-term and we predict that the cross section is longitudinally polarised.

\begin{figure}[H]
\centerline{\includegraphics[height=8.5cm]{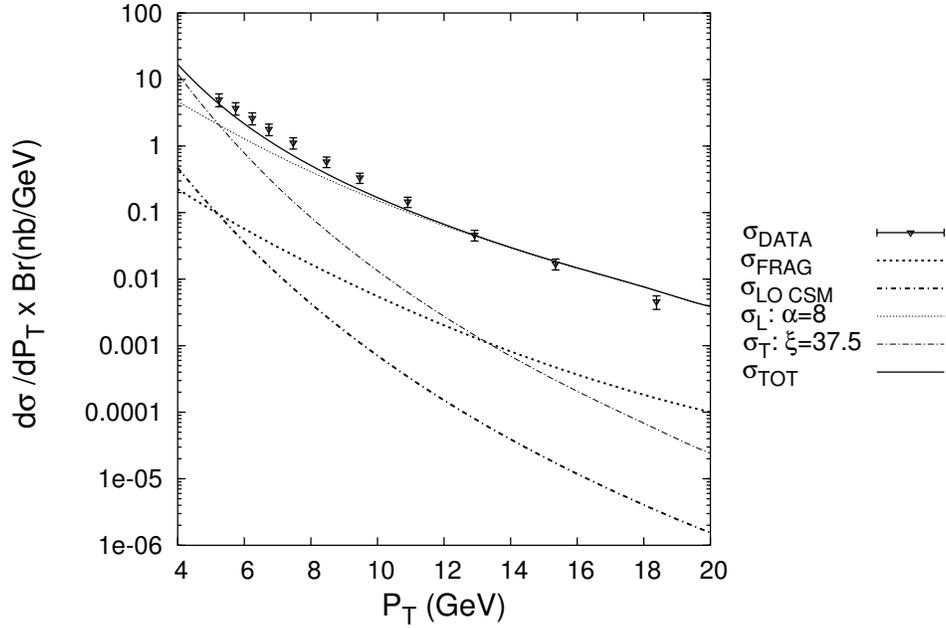}}
\caption{Polarised ($\sigma_T$ and $\sigma_L$) and total ($\sigma_{TOT}$) cross sections obtained
for $J/\psi$ at $\sqrt{s}=1800$~GeV 
with $\alpha=8$ and $\xi=37.5$ to be compared with LO CSM, the fragmentation CSM and to the data of 
CDF~\cite{CDF7997b}.}
\label{fig:g_m187_l183_cteq_37_5-8}
\end{figure}

\begin{figure}[H]
\centerline{\includegraphics[height=8.5cm]{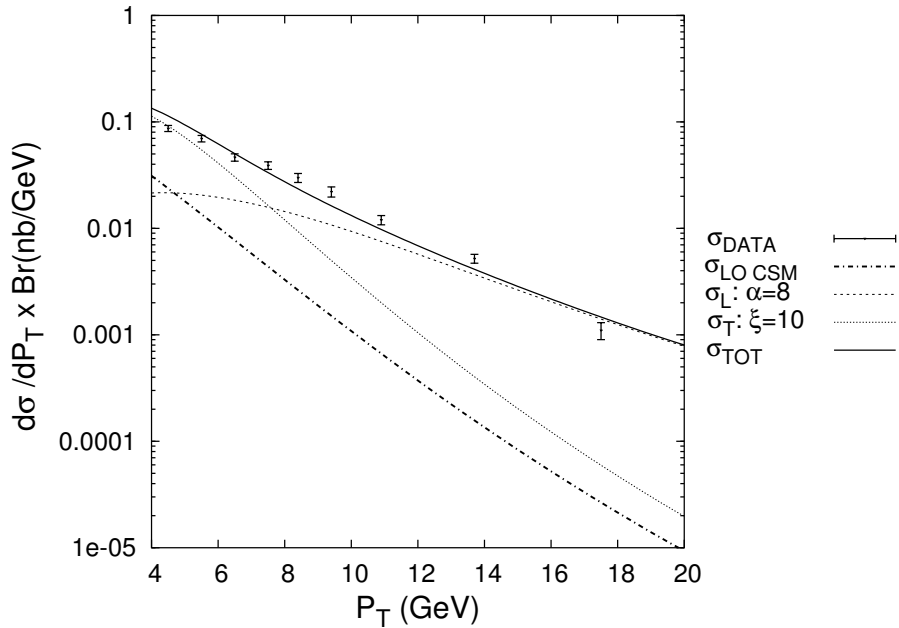}}
\caption{Polarised ($\sigma_T$ and $\sigma_L$) and total ($\sigma_{TOT}$) cross sections obtained 
for $\Upsilon(1S)$  at 
$\sqrt{s}=1800$~GeV with $\alpha=8$ and $\xi=10$ to be compared with LO CSM and to the data of 
CDF~\cite{Acosta:2001gv}.}
\label{fig:g_m528_l600_cteq_10-8}
\end{figure}

\begin{figure}[H]
\centerline{\includegraphics[height=8.5cm]{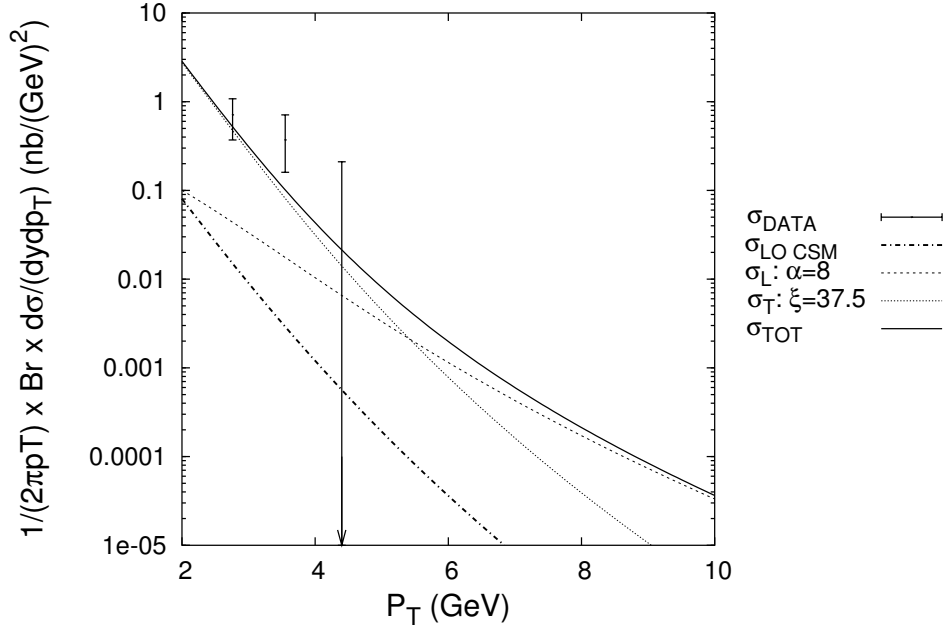}}
\caption{Polarised ($\sigma_T$ and $\sigma_L$) and total ($\sigma_{TOT}$) cross sections obtained 
for $J/\psi$ at $\sqrt{s}=200$~GeV
with $\alpha=8$ and $\xi=37.5$ to be compared with LO CSM and to the data of 
PHENIX~\cite{Adler:2003qs}.}
\label{fig:g_m187_l183_cteq_37_5-8-RHIC}
\end{figure}

\begin{figure}[H]
\centerline{\includegraphics[height=8.5cm]{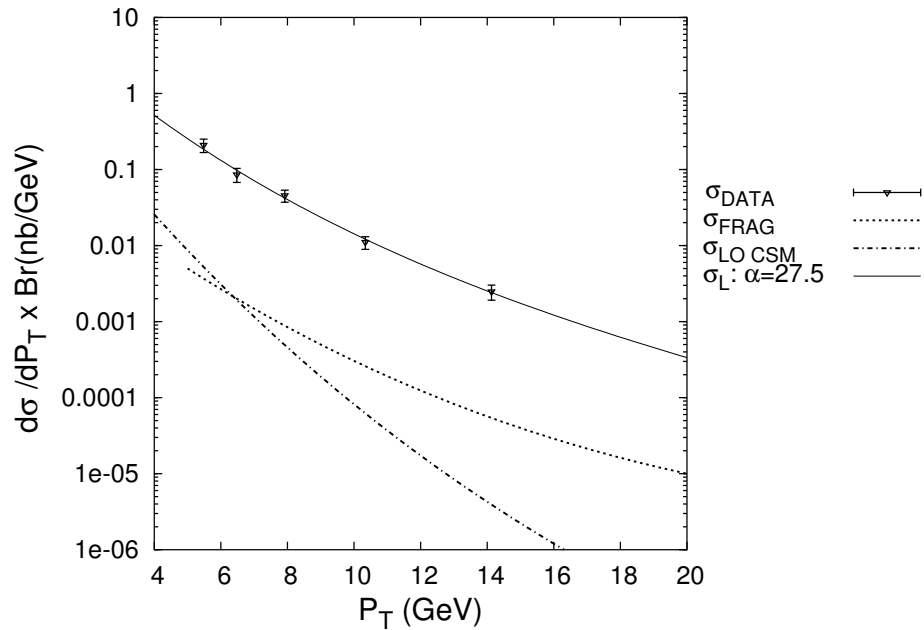}}
\caption{Polarised ($\sigma_T$ and $\sigma_L$) and total ($\sigma_{TOT}$) cross sections obtained
for $\psi'$ at $\sqrt{s}=1800$~GeV 
with $\alpha=27.5$ to be compared with LO CSM, the fragmentation CSM and to the data of 
CDF~\cite{CDF7997b}.}
\label{g_psip_a146_0-27_5}
\end{figure}
\index{psip@$\psi'$|)}

\subsection{Refining for the polarisation}\index{polarisation|(}

Following the discussion of~\ct{fig:summary_CSM_COM_pt_depend}, we know that fragmentation-like
processes should dominate for sufficiently large $P_T$. We have not yet included these  in our
calculation. Since the COM contribution 
is mainly transverse\footnote{We shall not consider its longitudinal part 
in plots, even  though we included it in our calculation for the sake of completeness, plotting 
them would have diminished the readability of the plots.} and can account with appropriate LDME 
(Long-Distance Matrix Element) value 
for the CDF data~\cite{Cho:1995vh},  we are going to combine it with our contribution 
from the $\alpha$-term  which is purely longitudinal and which can also account for the data.

The amplitude to produce a $^3S_1$ quarkonium ${\cal Q}$ by fragmentation of a colour octet state
$^3S_1^{(8)}$ is given by\index{colour-octet mechanism|(}\index{NRQCD}
\eqs{
|{\cal  M}(gg\to \!\!\ ^3S_1^{(8)}g\to \!\! \ ^3S_1g)|^2=
\overline{\sum_{T}} |{\cal  A}(gg\to \!\!\ ^3S_1^{(8)}g)|^2 \langle{\cal O}^{\cal Q}[^{3}S_1^{(8)}]\rangle
}
with for a transverse polarisation state~\cite{Cho:1995vh}
\eqs{
\overline{\sum_{T}} |{\cal  A}(gg\to \!\!\ ^3S_1^{(8)}g)|^2=-&\frac{(4\pi \alpha_s)^3}{144 M^3}\frac{\hat s^2}{(\hat s-M^2)^2}
\left[
(\hat s-M^2)^4+\hat t^4+\hat u^4 +2 M^4 \left(\frac{\hat t \hat u}{\hat s} \right)^2
\right] \times \\
\times 
&\frac{27 (\hat s \hat t+ \hat t \hat u+\hat u \hat s) -19 M^4}{[(\hat s-M^2)(\hat t-M^2)(\hat u-M^2)]^2}
}
and for a longitudinal polarisation state~\cite{Cho:1995vh}
\eqs{
\overline{\sum_{L}} |{\cal  A}(gg\to \!\!\ ^3S_1^{(8)}g)|^2=-&\frac{(4\pi \alpha_s)^3}{144 M^3}\frac{2 M^2\hat s}{(\hat s-M^2)^2}
(\hat t^2+\hat u^2)\hat t \hat u
\frac{27 (\hat s \hat t+ \hat t \hat u+\hat u \hat s) -19 M^4}{[(\hat s-M^2)(\hat t-M^2)(\hat u-M^2)]^2}
}
and where the LDME $\langle{\cal O}^{\cal Q}[^{3}S_1^{(8)}]\rangle$ depends on ${\cal Q}$. The 
longitudinal cross section is suppressed by $\frac{\hat s}{M^2}$ and therefore always small
compared to its transverse counterpart.

\index{fragmentation}
Combining our contribution and COM fragmentation with smaller coefficients (roughly 
decreasing the cross section of each contribution of a factor of 
2), we should recover a good description of the measured cross section as a function of $P_T$ but also
a reasonable description of the evolution of the polarisation parameter 
$\alpha_{pol}=\frac{\sigma_T-2\sigma_L}{\sigma_T+2\sigma_L}$ as a function of $P_T$ .

\subsubsection{$J/\psi$ at the Tevatron}\index{jpsi@$J/\psi$|(}

Setting $\alpha$ to 4.2, $\xi$ to 26.5 and $\langle {\cal O}^{J/\psi}({}^3S^{(8)}_1)\rangle$ 
to $4\times10^{-3}\ {\rm GeV}^3$, we got the cross section for the $J/\psi$  at 
$\sqrt{s}=1800$~GeV shown in~\cf{fig:g_m187_l183_cteq_8_26_5-4_2}. This fits rather well the
cross section measured by CDF~\cite{Acosta:2001gv}. To what concerns polarisation 
data~\cite{Affolder:2000nn}, they are only available for prompt $J/\psi$ and not for the direct
component of the signal for which our prediction are relevant. However, we can say that we obtain
the same trend (see \cf{fig:alpha_g_m187_l183_cteq_8_26_5-4_2}). 

\begin{figure}[H]
\centerline{\includegraphics[height=9cm]{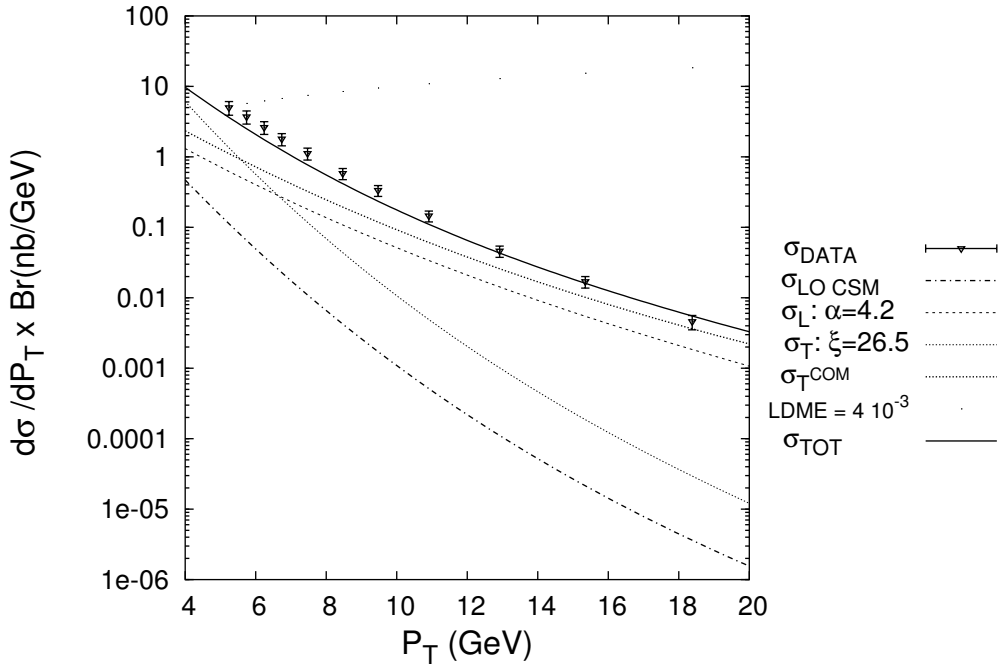}}
\caption{Various cross sections (see legend) obtained 
for the $J/\psi$ at $\sqrt{s}=1800$~GeV with $\alpha=4.2$ and $\xi=26.5$  and with a COM contribution with $\langle {\cal O}^{J/\psi}({}^3S^{(8)}_1)\rangle=4\times10^{-3}\ {\rm GeV}^3$ 
to be compared with LO CSM and to the data of CDF~\cite{CDF7997b}.}
\label{fig:g_m187_l183_cteq_8_26_5-4_2}
\end{figure}

\begin{figure}[H]
\centerline{\includegraphics[width=6cm,angle=-90]{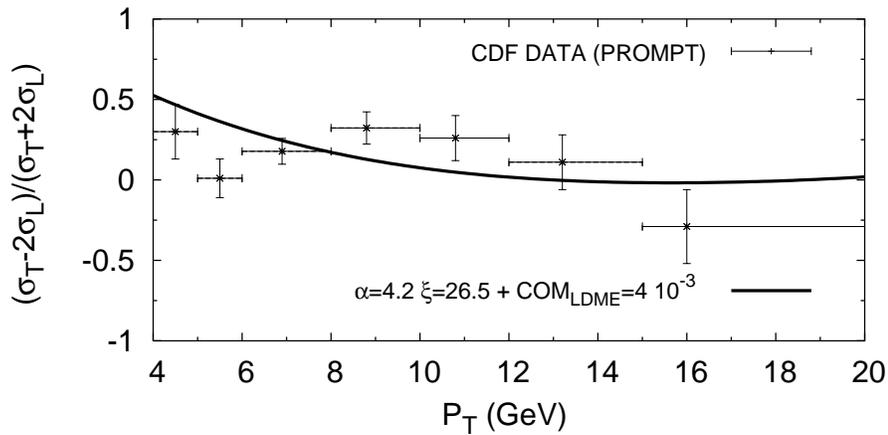}}
\caption{The polarisation parameter $\alpha_{pol}=\frac{\sigma_T-2\sigma_L}{\sigma_T+2\sigma_L}$
obtained for the $J/\psi$ at $\sqrt{s}=1800$~GeV with $\alpha=4.2$ and $\xi=26.5$  and with a COM contribution with $\langle {\cal O}^{J/\psi}({}^3S^{(8)}_1)\rangle = 4\times10^{-3}\ {\rm GeV}^3$.}
\label{fig:alpha_g_m187_l183_cteq_8_26_5-4_2}
\end{figure}\index{jpsi@$J/\psi$|)}

\subsubsection{$\psi'$ at the Tevatron}\index{psip@$\psi'$|(}\index{node}

With $a_{node}=1.46$ GeV, setting $\alpha$ to 17.5 and 
$\langle {\cal O}^{\psi'}({}^3S^{(8)}_1)\rangle$ to $4.5\times10^{-3}\ {\rm GeV}^3$, 
we get the cross section for the $\psi'$  at 
$\sqrt{s}=1800$~GeV shown in~\cf{fig:g_psip_a146_0-17_5} and the polarisation parameter 
$\alpha_{pol}$, in~\cf{fig:alpha_g_psip_a146_0-17_5}. The agreement is quite good both for
 cross section and for the polarisation.
\begin{figure}[H]
\centerline{\includegraphics[height=8cm]{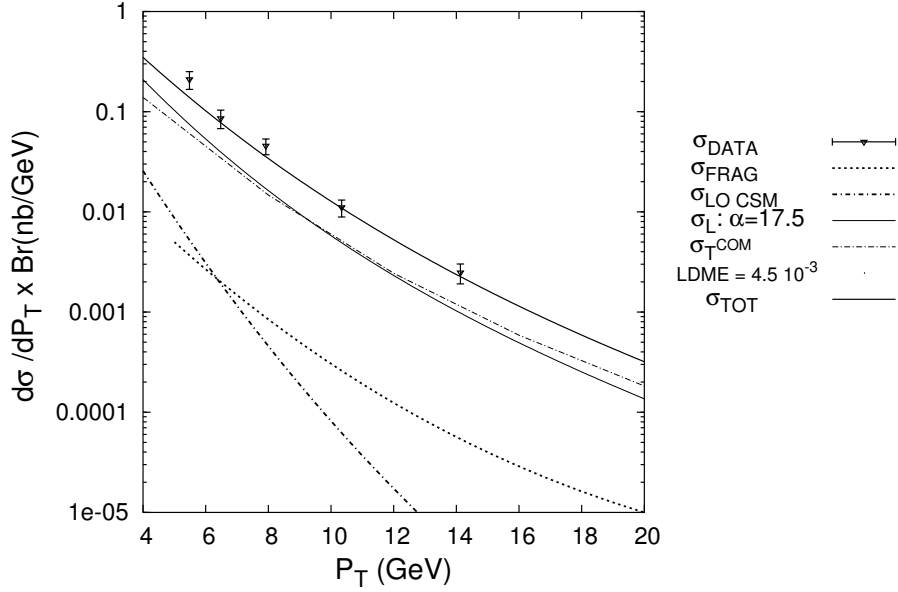}}
\caption{Various cross sections (see legend) obtained 
for $\psi'$ at $\sqrt{s}=1800$~GeV with $\alpha=17.5$ and with a COM contribution with $\langle {\cal O}^{\psi'}({}^3S^{(8)}_1)\rangle=4.5\times10^{-3}\ {\rm GeV}^3$ 
to be compared with LO CSM and to the data of CDF~\cite{CDF7997b}.}
\label{fig:g_psip_a146_0-17_5}
\end{figure}
\begin{figure}[H]
\centerline{\includegraphics[width=6cm,angle=-90]{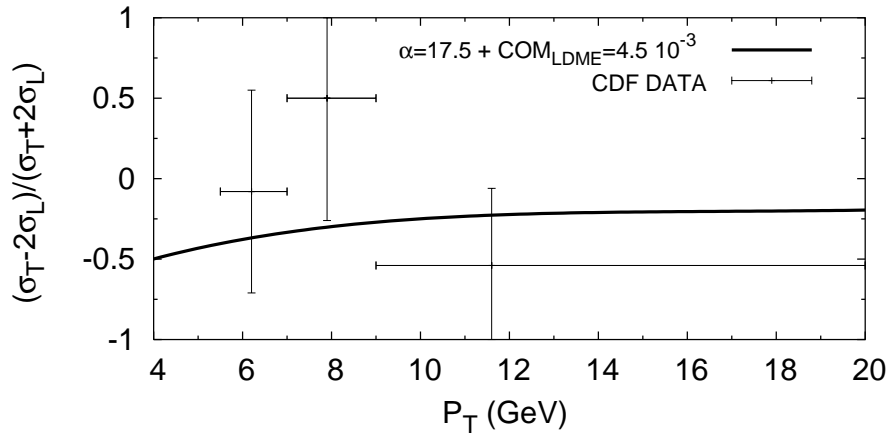}}
\caption{The polarisation parameter $\alpha_{pol}=\frac{\sigma_T-2\sigma_L}{\sigma_T+2\sigma_L}$
obtained for the $\psi'$ at $\sqrt{s}=1800$~GeV with $\alpha=17.5$ and with a COM contribution with $\langle {\cal O}^{\psi'}({}^3S^{(8)}_1)\rangle=4.5\times10^{-3}\ {\rm GeV}^3$, to be compared
with the CDF measurements~\cite{Affolder:2000nn}.}
\label{fig:alpha_g_psip_a146_0-17_5}
\end{figure}
\index{psip@$\psi'$|)}\index{colour-octet mechanism|)}\index{polarisation|)}

\subsubsection{In brief}

Our fitted values for $\alpha$ and $\xi$ are:

\begin{enumerate}
\item Without COM contribution:
\begin{itemize}
\item for the $J/\psi$ at the Tevatron:  $\alpha$=8 --  $\xi$=37.5
\item for the $J/\psi$ at RHIC: cf. Tevatron
\item for the $\Upsilon(1S)$ at the Tevatron: $\alpha$=8 --  $\xi$=10
\item for the $\psi'$ at the Tevatron:  $\alpha$=27 --  $\xi$:  $0<\xi\lesssim 30$ (not constrained)
\end{itemize}
\item With COM contribution:
\begin{itemize}
\item for the $J/\psi$ at the Tevatron: $\alpha$=4.2 --  $\xi$=26.5 -- $\langle {\cal O}^{J/\psi}({}^3S^{(8)}_1)\rangle=4.0\times10^{-3}\ {\rm GeV}^3$ 
\item for the $\psi'$ at the Tevatron:  $\alpha$=17.5  --  $\xi$: $0<\xi\lesssim 30$ (not constrained)  -- $\langle {\cal O}^{\psi'}({}^3S^{(8)}_1)\rangle=4.5\times10^{-3}\ {\rm GeV}^3$
\end{itemize}
\end{enumerate}

\clearemptydoublepage
\chapter{Conclusions and outlook}

``Use NRQCD !''. I shall never forget this wise and humoristic reply of Eric Braaten 
to me when I asked him
what to do to go beyond the CSM -- and its static approximation -- without having problems with 
gauge-invariance. Indeed, that approximation implying that whatever the hard part could be, the 
non-relativistic nature of the quarkonium should reduce the problem to a static one, was seriously 
disturbing me. Of course, NRQCD was a way to probe the internal dynamics of the quarkonium, but
still, the quark in the bound state were always considered at rest.

Now, are we in position to reply that we could go beyond
this static approximation and that we included relativistic effects in the calculation ? 
On the one hand, yes. We have introduced 
a relative momentum in the problem, we have tackled the issue of gauge 
invariance, we have constrained our model and we have applied it to processes for which 
the CSM was failing.

On the other hand, no. We have not calculated the real part of the amplitude and 
the contribution from the imaginary part is lower than the LO CSM at low $P_T$.  
However, we have shown that the choice of the 
gauge-invariance restoring vertex is not unique: this is the price to
pay to introduce non-static effects. And, then, things are getting very interesting since we
can reproduce the data, both from the Tevatron and from RHIC. This is the first time
that a model can reproduce both the rate and the polarisation of the data.

\subsubsection{What we did}

Let us recall one by one the achievements of this thesis. As we have just said, our aim 
was to introduce non-static relativistic effects in the problem 
of hadroproduction of quarkonium, especially for
the $J/\psi$, $\psi'$ and the $\Upsilon$ for which data are numerous, reliable but in evident 
contradiction with the models derived from pQCD.

In chapter~\ref{ch:build_model}, we studied the theoretical uncertainties affecting 
the LO CSM, as an indication of those affecting all models. These uncertainties being a factor $ 3$, 
our conclusion was that something
new was to be introduced, like the colour-octet contributions brought in by NRQCD. 

The way that we have chosen to introduce non-static effects was via
a phenomenological 3-particle vertex, made of a Dirac structure inferred from the quantum numbers of 
the considered quarkonium and a vertex function inspired by phenomenological
arguments. Of course, the vertex function was to be constrained. Firstly, its size parameter 
 -- telling us how rapidly it falls when the relative momentum 
of the quarks inside the quarkonium increases -- was chosen to be compatible 
with existing theoretical studies. Secondly, the
factor multiplying  the vertex -- the normalisation $N$ -- was fixed to reproduce 
the leptonic decay width. 

In the case of the $\psi'$, we showed that the normalisation, or equivalently the 
decay width, was strongly affected by a change
in the position of the node in the vertex function. 

Whereas the choice of the vertex function was driven by phenomenological considerations, the method, 
which we have proposed to restore gauge invariance 
was based on a rigorous and {\it ab initio} derivation exposed in chapter~\ref{ch:vertex_derivation}. 
First, we stated how gauge invariance was broken by the introduction of a 3-particle vertex function 
in the case of a pseudoscalar bound state. This was followed by the description of how 
to restore gauge invariance, keeping in mind that this procedure, 
which introduces new 4-particle vertices in the calculation, was not unique.

We considered the same problem for vector particles associated 
with a gluon. The same conclusions were drawn, and we also studied in details how to 
parametrise the freedom left in the 4-particle vertex. Indeed, we
have given simple and easily manageable Feynman rules for what we have called ``autonomous'' 
contributions, which correspond to different choices in the gauge-invariance 
restoring vertices. Note that the new contributions disappear in the case
of point-like particles, for which there is no problem with gauge invariance.

In chapter~\ref{ch:application}, we have applied our model to the hadroproduction of quarkonia associated
with a gluon, with the sole consideration of the imaginary part of the amplitude. This constraint comes
from the necessity of having internal quark on-shell to derive the new vertices.  
We made the calculations, first, for one choice of a gauge-invariance restoring vertex, then,  
for the same contribution with autonomous vertices whose weights were fitted to reproduce 
the cross-sections and, 
finally, with the further addition of colour-octet fragmentation contributions to 
reproduce the polarisation data.

\subsubsection{What we have obtained}

The main result that we have obtained is that one autonomous contribution is naturally dominant
and produces only longitudinally polarised quarkonia. Moreover, the $P_T$ slope  of this 
contribution is compatible with the data. Indeed,
it is expected that any non vanishing scalar product with a  helicity vector of a longitudinally
polarised quarkonium should scale like $\hat s$, which, at threshold, is proportional to $P_T^2$.
This explains why longitudinal components have a less steep slope. 

If we combine this contribution with colour-octet fragmentation, which is also naturally
large and produces only transversally polarised quarkonia, we can reach a perfect
agreement with all existing data and with a minimal number of free parameters. 

This combination with
colour-octet fragmentation was actually motivated by the fact that we were missing
fragmentation contributions, which are expected to be significant, due to
their $P_T$ slope (see~\cf{fig:summary_CSM_COM_pt_depend}).

Our new contribution is the only one, 
which is able to produce only longitudinally polarised quarkonium. 

Of course, we obtained other results. We have managed to introduce relativistic effects in the sense
that we took account of contributions with nonzero relative momentum -- non-static configurations -- 
between the two constituent quarks. Our estimate of the imaginary part is lower than the LO CSM
at low $P_T$, but of the same order at large $P_T$. 
We also got some interesting insights on the influence of the node of the $2S$ states.

More generally we have provided a formalism to study effects 
that come from the internal structure of the quarkonia and that were never 
considered before. One very interesting application would be to study
the effect of a variation of the binding potential on the production 
cross-section of quarkonia. This variation of potential could for instance be 
introduced to study the influence of the quark-gluon plasma.

\subsubsection{What we still miss}

Even though these results are new, there are still some drawbacks. 
The first is the missing real part. If we take the LO CSM as a benchmark, the real part seems to be 
dominant over the imaginary part at low $P_T$. One way to calculate it would be to use
dispersion relations. However, due to the expected asymptotic $\log \hat s$ behaviour
of the partonic amplitude, we have to introduce a subtraction constant in the dispersion relation.  
Furthermore, this constant is a function of $P_T$. We could fix it by fitting
the data at a given energy and make predictions for different energies. In the present situation, 
the data from RHIC have too small a $P_T$ range to proceed through the analysis. 
The second missing point it that our description of the quarkonia does not take 
into account cuts in the vertices. 

These missing contributions make us believe that the fits which we have made are 
reliable only up to a factor of order 3.

Another thing that we miss is an understanding of how the autonomous contributions arise. 
One of the three autonomous contributions that we have considered, namely the $\xi$-term 
($g^{\mu\nu}$), could 
come from off-shell contributions. This can be seen easily by setting a different quark mass
in the off-shell quark propagator in \ce{eq:ampl_v_gen} and performing the classification 
of the different Dirac structures with their factors. However, for the other ones, we have currently no
simple argument to motivate their presence, except that they exist in the standard contribution and
that their slope in $P_T$ is expected to be quite low, explaining why they dominate at large 
$P_T$.

\subsubsection{What is expedient to do}

A first necessary test of our formalism would be to apply it to production regimes
other  than hadroproduction. The application to photoproduction at HERA is the first that comes
to mind. Moreover, results on photoproduction are compatible with NLO calculations within the
CSM~\cite{Kramer:1995nb} and no contribution from colour-octet mechanisms are needed. 
In that context, a smaller value of the LDME's fixed by the fits on hadroproduction data would be
welcome, while our model provides with smaller ones. For all these reasons, we do plan to consider 
photoproduction in a close future.\index{HERA}

Besides photoproduction, electroproduction is also expected to provide us with interesting tests
and constraints. Compared to photoproduction, the kinematics would be slightly modified.
There exist data from LEP and a comparison with those would also be useful.\index{LEP}

Finally, in the case of $B$-factories, the situation is even more appealing,
since both the CSM and the COM diagree with Belle data. For some 
observables, the discrepancy amounts to more than one order of magnitude. Let us mention the double-charm 
production, \eg~$e^+e^-\to J/\psi +\eta_c$, and the inclusive production rate. However, a confirmation
from {\sc BaBar} of theses discrepancies would be greatly welcome.

In a totally different context, another attractive application is the modelisation of Generalised
Parton Distributions (GPD). In previous works, we have worked out these for the pion in a simple but
rough model where the restoration of gauge invariance was achieved by symmetrising the vertex function
$\Gamma_{(1,2)} \to \Gamma_{1}+\Gamma_{2}$. The result were encouraging, however we have now 
the tools to greatly refine this model. An important asset of our approach is the possibility to cast off
the obligation to work on the light cone. \index{GPD}

Moreover, as mentioned a couple of times, the application to production in media is opportune. For some
time now, $J/\psi$ has been thought to be an ideal probe for quark-gluon plasma through its suppression. 
Besides the fact that the absolute rates are not completely understood -- as we have seen 
throughout this work --, the way to introduce modification in the quark binding induced 
by the hadronic medium are not completely settled. Our formalism may be used to study these processes.

If we now turn back to the formalism rather than to its possible applications and tests, we are thinking to
try to justify the presence of these autonomous 4-particle vertices from first principles, \eg~from 
standard perturbative Feynman rules applied to some typical 
sets of diagrams. We would then obtain an estimate of the free parameters $\alpha$,
$\beta$ and $\xi$, possibly in a similar fashion than the velocity scaling rules in NRQCD.
\index{velocity scaling rules}

Another direct improvement of our formalism would be of course the consideration of the real part
of the amplitude. We have
already addressed the possibility of making use of dispersion relation with the drawback which is 
the presence of arbitrary subtraction constants. Another way may be to generalise our gauge-invariance
restoring vertices to off-shell quarks. Then, we would be in position
to consider the real part at any $\hat s$ or to determine it at a precise value of $\hat s$, 
providing us with  the value of the subtraction constants of the dispersion relations.

By showing that it was possible 
to {\it build} a model reproducing all available data on high-energy hadroproduction
of quarkonia, we do believe that we have provided a new subject of 
discussion and research and, hopefully, shed some light on what may be considered to be
the path to a resolution of this issue. 

\clearemptydoublepage
%---------------------------------------------------------------------
\appendix
\newpage
\chapter{Acronyms and abbreviations}\label{ch:acronyms}
\begin{tabular}{ll}
AGS & Alternating Gradient Synchrotron \\
BNL & Brookhaven National Laboratory \\
%BS & Bound State\\
BSE& Bethe-Salpeter Equation\\
CDF & Collider Detector at Fermilab \\
CERN & Conseil Europ\'een pour la Recherche Nucl\'eaire \\
CES & Central Electromagnetic Strip chambers \\
c.m. & center-of-momentum \\
CMP & Central Muon Upgrade \\
CEM & Colour Evaporation Model \\
CLEO & CLEO(patra)\protect\footnote{Named suggested by Chris Day: short for Cleopatra} 
 or  Cornell's Largest Experimental Object (or Operation or Organization) \\
COM & Colour Octet Mechanism \\
CSM & Colour Singlet Model \\
CTC & Central Tracking System \\
DESY & Deutsches Elektronen-Synchrotron\\
DSE & Dyson-Schwinger Equation\\
EMC & ElectroMagnetic Calorimeter \\
FNAL & Fermi National Accelerator Laboratory (Fermilab)\\
GIPA & Gauge-Invariant Phenomenological Approach \\
GPD  & Generalised Parton Distribution \\
HERA & Hadron Elektron Ring Anlage (Hadron Electron Ring Accelerator)\\
ISR & Intersecting Storage Rings \\
KEK &Kou Enerugii Butsurigaku Kenkyusho  \\
LDME & Long-Distance Matrix Element\\
LEP & Large Electron - Positron (storage ring) \\
LO  & Leading order \\

\end{tabular}
\footnotetext{Named sugested by Chris Day: short for Cleopatra}

\begin{tabular}{ll}
MuID& Muon IDentifier  \\
MuTr& Muon Tracker  \\
NLO & Next-to-Leading Order \\
NRQCD & Non Relativistic Quantum ChromoDynamics \\
pdf & parton distribution function\\
pQCD & perturbative Quantum ChromoDynamics \\
QCD & Quantum ChromoDynamics \\
QED & Quantum ElectroDynamics \\
RHIC & Relativistic Heavy Ion Collider \\
RICH & Ring Imaging \v C(h)erenkov \\
SCI & Soft Colour Interactions \\
SLAC & Stanford Linear Accelerator \\
SPEAR & Stanford Positron Electron Accelerating Ring \\
SVX & Silicon Vertex Detector\\
\end{tabular}

\clearemptydoublepage
\chapter{Normalisation}
\section{Results relative to the normalisation}\label{sec:res_norm_tab}
\index{normalisation|(}

\begin{table}[H]
{\tiny%%%%%%%%%%%%%%%%%%%%%%%%%%%%%%%%%%%%%%%%%%%%%%%%%%%%%%%%%%%%%%%%%%%%%%
\def\ifundefined#1{\expandafter\ifx\csname#1\endcsname\relax}

%%  Check for the \def token for inputed files. If it is not        %%
%%  defined, the file will be processed as a standalone and the     %%
%%  preamble will be used.                                          %%
\ifundefined{inputGnumericTable}

%%  End of the preamble for the standalone. The next section is for %%
%%  documents which are included into other LaTeX2e files.          %%
\else

%%  We are not a stand alone document. For a regular table, we will %%
%%  have no preamble and only define the closing to mean nothing.   %%
    \def\gnumericTableEnd{}

%%  If we want landscape mode in an embedded document, comment out  %%
%%  the line above and uncomment the two below. The table will      %%
%%  begin on a new page and run in landscape mode.                  %%
%       \def\gnumericTableEnd{\end{landscape}}
%       \begin{landscape}

%%  End of the else clause for this file being \input.              %%
\fi

%%%%%%%%%%%%%%%%%%%%%%%%%%%%%%%%%%%%%%%%%%%%%%%%%%%%%%%%%%%%%%%%%%%%%%
%%                                                                  %%
%%  The rest is the gnumeric table, except for the closing          %%
%%  statement. Changes below will alter the table's appearance.     %%
%%                                                                  %%
%%%%%%%%%%%%%%%%%%%%%%%%%%%%%%%%%%%%%%%%%%%%%%%%%%%%%%%%%%%%%%%%%%%%%%

\providecommand{\gnumericPB}[1]%
{\let\gnumericTemp=\\#1\let\\=\gnumericTemp\hspace{0pt}}

%%  The default table format retains the relative column widths of  %%
%%  gnumeric. They can easily be changed to c, r or l. In that case %%
%%  you may want to comment out the next line and uncomment the one %%
%%  thereafter                                                      %%
\providecommand\gnumbox{\makebox[0pt]}
%%\providecommand\gnumbox[1][]{\makebox}

%% to adjust positions in multirow situations                       %%
\setlength{\bigstrutjot}{\jot}
\setlength{\extrarowheight}{\doublerulesep}

%%  The \setlongtables command keeps column widths the same across %%
%%  pages. Simply comment out next line for varying column widths.  %%
\setlongtables

\setlength\gnumericTableWidth{%
	30pt+%
	30pt+%
	30pt+%
	30pt+%
0pt}
\def\gumericNumCols{4}
\setlength\gnumericTableWidthComplete{\gnumericTableWidth+\tabcolsep*\gumericNumCols*2+\arrayrulewidth*\gumericNumCols}
\ifthenelse{\lengthtest{\gnumericTableWidthComplete > \textwidth}}%
{\def\gnumericScale{\ratio{\textwidth-\tabcolsep*\gumericNumCols*2-\arrayrulewidth*\gumericNumCols}%
{\gnumericTableWidth}}}%
{\def\gnumericScale{1}}

%%%%%%%%%%%%%%%%%%%%%%%%%%%%%%%%%%%%%%%%%%%%%%%%%%%%%%%%%%%%%%%%%%%%%%
%%                                                                  %%
%% The following are the widths of the various columns. We are      %%
%% defining them here because then they are easier to change.       %%
%% Depending on the cell formats we may use them more than once.    %%
%%                                                                  %%
%%%%%%%%%%%%%%%%%%%%%%%%%%%%%%%%%%%%%%%%%%%%%%%%%%%%%%%%%%%%%%%%%%%%%%

\def\gnumericColB{30pt*\gnumericScale}
\def\gnumericColC{30pt*\gnumericScale}
\def\gnumericColD{30pt*\gnumericScale}
\def\gnumericColE{30pt*\gnumericScale}

\begin{longtable}[c]{%
	b{\gnumericColB}%
	b{\gnumericColC}%
	b{\gnumericColD}%
	b{\gnumericColE}%
	}

%%%%%%%%%%%%%%%%%%%%%%%%%%%%%%%%%%%%%%%%%%%%%%%%%%%%%%%%%%%%%%%%%%%%%%
%%  The longtable options. (Caption. headers... see Goosens. p.124) %%
%	\caption{The Table Caption.}             \\	%
% \hline	% Across the top of the table.
%%  The rest of these options are table rows which are placed on    %%
%%  the first. last or every page. Use \multicolumn if you want.    %%

%%  Header for the first page.                                      %%
%	\multicolumn{4}{c}{The First Header} \\ \hline 
%	\multicolumn{1}{c}{colTag}	%Column 1
%	&\multicolumn{1}{c}{colTag}	%Column 1
%	&\multicolumn{1}{c}{colTag}	%Column 2
%	&\multicolumn{1}{c}{colTag}	%Column 3
%	&\multicolumn{1}{c}{colTag}	\\ \hline %Last column
%	\endfirsthead

%%  The running header definition.                                  %%
%	\hline
%	\multicolumn{4}{l}{\ldots\small\slshape continued} \\ \hline
%	\multicolumn{1}{c}{colTag}	%Column 1
%	&\multicolumn{1}{c}{colTag}	%Column 1
%	&\multicolumn{1}{c}{colTag}	%Column 2
%	&\multicolumn{1}{c}{colTag}	%Column 3
%	&\multicolumn{1}{c}{colTag}	\\ \hline %Last column
%	\endhead

%%  The running footer definition.                                  %%
%	\hline
%	\multicolumn{4}{r}{\small\slshape continued\ldots} \\
%	\endfoot

%%  The ending footer definition.                                   %%
%	\multicolumn{4}{c}{That's all folks} \\ \hline 
%	\endlastfoot
%%%%%%%%%%%%%%%%%%%%%%%%%%%%%%%%%%%%%%%%%%%%%%%%%%%%%%%%%%%%%%%%%%%%%%

\hhline{|-|-|-|-|}
	 \multicolumn{1}{|p{\gnumericColB}|}%
	{\gnumericPB{\raggedleft}\gnumbox[r]{\textsf{$\Lambda$}}}
	&\multicolumn{1}{p{\gnumericColC}|}%
	{\gnumericPB{\raggedleft}\gnumbox[r]{\textsf{$N$}}}
	&\multicolumn{1}{p{\gnumericColD}|}%
	{\gnumericPB{\raggedleft}\gnumbox[r]{\textsf{$\Lambda$}}}
	&\multicolumn{1}{p{\gnumericColE}|}%
	{\gnumericPB{\raggedleft}\gnumbox[r]{\textsf{$N$}}}
\\
\hhline{|----|}
	 \multicolumn{1}{|p{\gnumericColB}|}%
	{\gnumericPB{\raggedleft}\gnumbox[r]{\textsf{$1.00$}}}
	&\multicolumn{1}{p{\gnumericColC}|}%
	{\gnumericPB{\raggedleft}\gnumbox[r]{\textsf{$5.95$}}}
	&\multicolumn{1}{p{\gnumericColD}|}%
	{\gnumericPB{\raggedleft}\gnumbox[r]{\textsf{$1.63$}}}
	&\multicolumn{1}{p{\gnumericColE}|}%
	{\gnumericPB{\raggedleft}\gnumbox[r]{\textsf{$2.64$}}}
\\
\hhline{|----|}
	 \multicolumn{1}{|p{\gnumericColB}|}%
	{\gnumericPB{\raggedleft}\gnumbox[r]{\textsf{$1.06$}}}
	&\multicolumn{1}{p{\gnumericColC}|}%
	{\gnumericPB{\raggedleft}\gnumbox[r]{\textsf{$5.36$}}}
	&\multicolumn{1}{p{\gnumericColD}|}%
	{\gnumericPB{\raggedleft}\gnumbox[r]{\textsf{$1.69$}}}
	&\multicolumn{1}{p{\gnumericColE}|}%
	{\gnumericPB{\raggedleft}\gnumbox[r]{\textsf{$2.48$}}}
\\
\hhline{|----|}
	 \multicolumn{1}{|p{\gnumericColB}|}%
	{\gnumericPB{\raggedleft}\gnumbox[r]{\textsf{$1.13$}}}
	&\multicolumn{1}{p{\gnumericColC}|}%
	{\gnumericPB{\raggedleft}\gnumbox[r]{\textsf{$4.86$}}}
	&\multicolumn{1}{p{\gnumericColD}|}%
	{\gnumericPB{\raggedleft}\gnumbox[r]{\textsf{$1.76$}}}
	&\multicolumn{1}{p{\gnumericColE}|}%
	{\gnumericPB{\raggedleft}\gnumbox[r]{\textsf{$2.34$}}}
\\
\hhline{|----|}
	 \multicolumn{1}{|p{\gnumericColB}|}%
	{\gnumericPB{\raggedleft}\gnumbox[r]{\textsf{$1.19$}}}
	&\multicolumn{1}{p{\gnumericColC}|}%
	{\gnumericPB{\raggedleft}\gnumbox[r]{\textsf{$4.44$}}}
	&\multicolumn{1}{p{\gnumericColD}|}%
	{\gnumericPB{\raggedleft}\gnumbox[r]{\textsf{$1.82$}}}
	&\multicolumn{1}{p{\gnumericColE}|}%
	{\gnumericPB{\raggedleft}\gnumbox[r]{\textsf{$2.21$}}}
\\
\hhline{|----|}
	 \multicolumn{1}{|p{\gnumericColB}|}%
	{\gnumericPB{\raggedleft}\gnumbox[r]{\textsf{$1.25$}}}
	&\multicolumn{1}{p{\gnumericColC}|}%
	{\gnumericPB{\raggedleft}\gnumbox[r]{\textsf{$4.07$}}}
	&\multicolumn{1}{p{\gnumericColD}|}%
	{\gnumericPB{\raggedleft}\gnumbox[r]{\textsf{$1.88$}}}
	&\multicolumn{1}{p{\gnumericColE}|}%
	{\gnumericPB{\raggedleft}\gnumbox[r]{\textsf{$2.09$}}}
\\
\hhline{|----|}
	 \multicolumn{1}{|p{\gnumericColB}|}%
	{\gnumericPB{\raggedleft}\gnumbox[r]{\textsf{$1.32$}}}
	&\multicolumn{1}{p{\gnumericColC}|}%
	{\gnumericPB{\raggedleft}\gnumbox[r]{\textsf{$3.76$}}}
	&\multicolumn{1}{p{\gnumericColD}|}%
	{\gnumericPB{\raggedleft}\gnumbox[r]{\textsf{$1.95$}}}
	&\multicolumn{1}{p{\gnumericColE}|}%
	{\gnumericPB{\raggedleft}\gnumbox[r]{\textsf{$1.98$}}}
\\
\hhline{|----|}
	 \multicolumn{1}{|p{\gnumericColB}|}%
	{\gnumericPB{\raggedleft}\gnumbox[r]{\textsf{$1.38$}}}
	&\multicolumn{1}{p{\gnumericColC}|}%
	{\gnumericPB{\raggedleft}\gnumbox[r]{\textsf{$3.48$}}}
	&\multicolumn{1}{p{\gnumericColD}|}%
	{\gnumericPB{\raggedleft}\gnumbox[r]{\textsf{$2.01$}}}
	&\multicolumn{1}{p{\gnumericColE}|}%
	{\gnumericPB{\raggedleft}\gnumbox[r]{\textsf{$1.88$}}}
\\
\hhline{|----|}
	 \multicolumn{1}{|p{\gnumericColB}|}%
	{\gnumericPB{\raggedleft}\gnumbox[r]{\textsf{$1.44$}}}
	&\multicolumn{1}{p{\gnumericColC}|}%
	{\gnumericPB{\raggedleft}\gnumbox[r]{\textsf{$3.23$}}}
	&\multicolumn{1}{p{\gnumericColD}|}%
	{\gnumericPB{\raggedleft}\gnumbox[r]{\textsf{$2.07$}}}
	&\multicolumn{1}{p{\gnumericColE}|}%
	{\gnumericPB{\raggedleft}\gnumbox[r]{\textsf{$1.79$}}}
\\
\hhline{|----|}
	 \multicolumn{1}{|p{\gnumericColB}|}%
	{\gnumericPB{\raggedleft}\gnumbox[r]{\textsf{$1.51$}}}
	&\multicolumn{1}{p{\gnumericColC}|}%
	{\gnumericPB{\raggedleft}\gnumbox[r]{\textsf{$3.01$}}}
	&\multicolumn{1}{p{\gnumericColD}|}%
	{\gnumericPB{\raggedleft}\gnumbox[r]{\textsf{$2.14$}}}
	&\multicolumn{1}{p{\gnumericColE}|}%
	{\gnumericPB{\raggedleft}\gnumbox[r]{\textsf{$1.70$}}}
\\
\hhline{|----|}
	 \multicolumn{1}{|p{\gnumericColB}|}%
	{\gnumericPB{\raggedleft}\gnumbox[r]{\textsf{$1.57$}}}
	&\multicolumn{1}{p{\gnumericColC}|}%
	{\gnumericPB{\raggedleft}\gnumbox[r]{\textsf{$2.82$}}}
	&\multicolumn{1}{p{\gnumericColD}|}%
	{\gnumericPB{\raggedleft}\gnumbox[r]{\textsf{$2.20$}}}
	&\multicolumn{1}{p{\gnumericColE}|}%
	{\gnumericPB{\raggedleft}\gnumbox[r]{\textsf{$1.62$}}}
\\
\hhline{|-|-|-|-|}
\end{longtable}

\gnumericTableEnd
}
\caption{Normalisation for a dipolar form for $J/\psi$ as a function of $\Lambda$ for $m_c=1.60$~GeV.}\label{tab:N_dip_jpsi_lam1}
\end{table}
\begin{table}[H]
{\tiny%%%%%%%%%%%%%%%%%%%%%%%%%%%%%%%%%%%%%%%%%%%%%%%%%%%%%%%%%%%%%%%%%%%%%%
\def\ifundefined#1{\expandafter\ifx\csname#1\endcsname\relax}

%%  Check for the \def token for inputed files. If it is not        %%
%%  defined, the file will be processed as a standalone and the     %%
%%  preamble will be used.                                          %%
\ifundefined{inputGnumericTable}

%%  End of the preamble for the standalone. The next section is for %%
%%  documents which are included into other LaTeX2e files.          %%
\else

%%  We are not a stand alone document. For a regular table, we will %%
%%  have no preamble and only define the closing to mean nothing.   %%
    \def\gnumericTableEnd{}

%%  If we want landscape mode in an embedded document, comment out  %%
%%  the line above and uncomment the two below. The table will      %%
%%  begin on a new page and run in landscape mode.                  %%
%       \def\gnumericTableEnd{\end{landscape}}
%       \begin{landscape}

%%  End of the else clause for this file being \input.              %%
\fi

%%%%%%%%%%%%%%%%%%%%%%%%%%%%%%%%%%%%%%%%%%%%%%%%%%%%%%%%%%%%%%%%%%%%%%
%%                                                                  %%
%%  The rest is the gnumeric table, except for the closing          %%
%%  statement. Changes below will alter the table's appearance.     %%
%%                                                                  %%
%%%%%%%%%%%%%%%%%%%%%%%%%%%%%%%%%%%%%%%%%%%%%%%%%%%%%%%%%%%%%%%%%%%%%%

\providecommand{\gnumericPB}[1]%
{\let\gnumericTemp=\\#1\let\\=\gnumericTemp\hspace{0pt}}

%%  The default table format retains the relative column widths of  %%
%%  gnumeric. They can easily be changed to c, r or l. In that case %%
%%  you may want to comment out the next line and uncomment the one %%
%%  thereafter                                                      %%
\providecommand\gnumbox{\makebox[0pt]}
%%\providecommand\gnumbox[1][]{\makebox}

%% to adjust positions in multirow situations                       %%
\setlength{\bigstrutjot}{\jot}
\setlength{\extrarowheight}{\doublerulesep}

%%  The \setlongtables command keeps column widths the same across %%
%%  pages. Simply comment out next line for varying column widths.  %%
\setlongtables

\setlength\gnumericTableWidth{%
	30pt+%
	30pt+%
	30pt+%
	30pt+%
0pt}
\def\gumericNumCols{4}
\setlength\gnumericTableWidthComplete{\gnumericTableWidth+\tabcolsep*\gumericNumCols*2+\arrayrulewidth*\gumericNumCols}
\ifthenelse{\lengthtest{\gnumericTableWidthComplete > \textwidth}}%
{\def\gnumericScale{\ratio{\textwidth-\tabcolsep*\gumericNumCols*2-\arrayrulewidth*\gumericNumCols}%
{\gnumericTableWidth}}}%
{\def\gnumericScale{1}}

%%%%%%%%%%%%%%%%%%%%%%%%%%%%%%%%%%%%%%%%%%%%%%%%%%%%%%%%%%%%%%%%%%%%%%
%%                                                                  %%
%% The following are the widths of the various columns. We are      %%
%% defining them here because then they are easier to change.       %%
%% Depending on the cell formats we may use them more than once.    %%
%%                                                                  %%
%%%%%%%%%%%%%%%%%%%%%%%%%%%%%%%%%%%%%%%%%%%%%%%%%%%%%%%%%%%%%%%%%%%%%%

\def\gnumericColB{30pt*\gnumericScale}
\def\gnumericColC{30pt*\gnumericScale}
\def\gnumericColD{30pt*\gnumericScale}
\def\gnumericColE{30pt*\gnumericScale}

\begin{longtable}[c]{%
	b{\gnumericColB}%
	b{\gnumericColC}%
	b{\gnumericColD}%
	b{\gnumericColE}%
	}

%%%%%%%%%%%%%%%%%%%%%%%%%%%%%%%%%%%%%%%%%%%%%%%%%%%%%%%%%%%%%%%%%%%%%%
%%  The longtable options. (Caption. headers... see Goosens. p.124) %%
%	\caption{The Table Caption.}             \\	%
% \hline	% Across the top of the table.
%%  The rest of these options are table rows which are placed on    %%
%%  the first. last or every page. Use \multicolumn if you want.    %%

%%  Header for the first page.                                      %%
%	\multicolumn{4}{c}{The First Header} \\ \hline 
%	\multicolumn{1}{c}{colTag}	%Column 1
%	&\multicolumn{1}{c}{colTag}	%Column 1
%	&\multicolumn{1}{c}{colTag}	%Column 2
%	&\multicolumn{1}{c}{colTag}	%Column 3
%	&\multicolumn{1}{c}{colTag}	\\ \hline %Last column
%	\endfirsthead

%%  The running header definition.                                  %%
%	\hline
%	\multicolumn{4}{l}{\ldots\small\slshape continued} \\ \hline
%	\multicolumn{1}{c}{colTag}	%Column 1
%	&\multicolumn{1}{c}{colTag}	%Column 1
%	&\multicolumn{1}{c}{colTag}	%Column 2
%	&\multicolumn{1}{c}{colTag}	%Column 3
%	&\multicolumn{1}{c}{colTag}	\\ \hline %Last column
%	\endhead

%%  The running footer definition.                                  %%
%	\hline
%	\multicolumn{4}{r}{\small\slshape continued\ldots} \\
%	\endfoot

%%  The ending footer definition.                                   %%
%	\multicolumn{4}{c}{That's all folks} \\ \hline 
%	\endlastfoot
%%%%%%%%%%%%%%%%%%%%%%%%%%%%%%%%%%%%%%%%%%%%%%%%%%%%%%%%%%%%%%%%%%%%%%

\hhline{|-|-|-|-|}
	 \multicolumn{1}{|p{\gnumericColB}|}%
	{\gnumericPB{\raggedleft}\gnumbox[r]{\textsf{$1.00$}}}
	&\multicolumn{1}{p{\gnumericColC}|}%
	{\gnumericPB{\raggedleft}\gnumbox[r]{\textsf{$7.93$}}}
	&\multicolumn{1}{p{\gnumericColD}|}%
	{\gnumericPB{\raggedleft}\gnumbox[r]{\textsf{$1.63$}}}
	&\multicolumn{1}{p{\gnumericColE}|}%
	{\gnumericPB{\raggedleft}\gnumbox[r]{\textsf{$3.17$}}}
\\
\hhline{|----|}
	 \multicolumn{1}{|p{\gnumericColB}|}%
	{\gnumericPB{\raggedleft}\gnumbox[r]{\textsf{$1.06$}}}
	&\multicolumn{1}{p{\gnumericColC}|}%
	{\gnumericPB{\raggedleft}\gnumbox[r]{\textsf{$7.04$}}}
	&\multicolumn{1}{p{\gnumericColD}|}%
	{\gnumericPB{\raggedleft}\gnumbox[r]{\textsf{$1.69$}}}
	&\multicolumn{1}{p{\gnumericColE}|}%
	{\gnumericPB{\raggedleft}\gnumbox[r]{\textsf{$2.96$}}}
\\
\hhline{|----|}
	 \multicolumn{1}{|p{\gnumericColB}|}%
	{\gnumericPB{\raggedleft}\gnumbox[r]{\textsf{$1.13$}}}
	&\multicolumn{1}{p{\gnumericColC}|}%
	{\gnumericPB{\raggedleft}\gnumbox[r]{\textsf{$6.31$}}}
	&\multicolumn{1}{p{\gnumericColD}|}%
	{\gnumericPB{\raggedleft}\gnumbox[r]{\textsf{$1.76$}}}
	&\multicolumn{1}{p{\gnumericColE}|}%
	{\gnumericPB{\raggedleft}\gnumbox[r]{\textsf{$2.77$}}}
\\
\hhline{|----|}
	 \multicolumn{1}{|p{\gnumericColB}|}%
	{\gnumericPB{\raggedleft}\gnumbox[r]{\textsf{$1.19$}}}
	&\multicolumn{1}{p{\gnumericColC}|}%
	{\gnumericPB{\raggedleft}\gnumbox[r]{\textsf{$5.69$}}}
	&\multicolumn{1}{p{\gnumericColD}|}%
	{\gnumericPB{\raggedleft}\gnumbox[r]{\textsf{$1.82$}}}
	&\multicolumn{1}{p{\gnumericColE}|}%
	{\gnumericPB{\raggedleft}\gnumbox[r]{\textsf{$2.60$}}}
\\
\hhline{|----|}
	 \multicolumn{1}{|p{\gnumericColB}|}%
	{\gnumericPB{\raggedleft}\gnumbox[r]{\textsf{$1.25$}}}
	&\multicolumn{1}{p{\gnumericColC}|}%
	{\gnumericPB{\raggedleft}\gnumbox[r]{\textsf{$5.16$}}}
	&\multicolumn{1}{p{\gnumericColD}|}%
	{\gnumericPB{\raggedleft}\gnumbox[r]{\textsf{$1.88$}}}
	&\multicolumn{1}{p{\gnumericColE}|}%
	{\gnumericPB{\raggedleft}\gnumbox[r]{\textsf{$2.44$}}}
\\
\hhline{|----|}
	 \multicolumn{1}{|p{\gnumericColB}|}%
	{\gnumericPB{\raggedleft}\gnumbox[r]{\textsf{$1.32$}}}
	&\multicolumn{1}{p{\gnumericColC}|}%
	{\gnumericPB{\raggedleft}\gnumbox[r]{\textsf{$4.71$}}}
	&\multicolumn{1}{p{\gnumericColD}|}%
	{\gnumericPB{\raggedleft}\gnumbox[r]{\textsf{$1.95$}}}
	&\multicolumn{1}{p{\gnumericColE}|}%
	{\gnumericPB{\raggedleft}\gnumbox[r]{\textsf{$2.30$}}}
\\
\hhline{|----|}
	 \multicolumn{1}{|p{\gnumericColB}|}%
	{\gnumericPB{\raggedleft}\gnumbox[r]{\textsf{$1.38$}}}
	&\multicolumn{1}{p{\gnumericColC}|}%
	{\gnumericPB{\raggedleft}\gnumbox[r]{\textsf{$4.31$}}}
	&\multicolumn{1}{p{\gnumericColD}|}%
	{\gnumericPB{\raggedleft}\gnumbox[r]{\textsf{$2.01$}}}
	&\multicolumn{1}{p{\gnumericColE}|}%
	{\gnumericPB{\raggedleft}\gnumbox[r]{\textsf{$2.17$}}}
\\
\hhline{|----|}
	 \multicolumn{1}{|p{\gnumericColB}|}%
	{\gnumericPB{\raggedleft}\gnumbox[r]{\textsf{$1.44$}}}
	&\multicolumn{1}{p{\gnumericColC}|}%
	{\gnumericPB{\raggedleft}\gnumbox[r]{\textsf{$3.97$}}}
	&\multicolumn{1}{p{\gnumericColD}|}%
	{\gnumericPB{\raggedleft}\gnumbox[r]{\textsf{$2.07$}}}
	&\multicolumn{1}{p{\gnumericColE}|}%
	{\gnumericPB{\raggedleft}\gnumbox[r]{\textsf{$2.06$}}}
\\
\hhline{|----|}
	 \multicolumn{1}{|p{\gnumericColB}|}%
	{\gnumericPB{\raggedleft}\gnumbox[r]{\textsf{$1.51$}}}
	&\multicolumn{1}{p{\gnumericColC}|}%
	{\gnumericPB{\raggedleft}\gnumbox[r]{\textsf{$3.67$}}}
	&\multicolumn{1}{p{\gnumericColD}|}%
	{\gnumericPB{\raggedleft}\gnumbox[r]{\textsf{$2.14$}}}
	&\multicolumn{1}{p{\gnumericColE}|}%
	{\gnumericPB{\raggedleft}\gnumbox[r]{\textsf{$1.95$}}}
\\
\hhline{|----|}
	 \multicolumn{1}{|p{\gnumericColB}|}%
	{\gnumericPB{\raggedleft}\gnumbox[r]{\textsf{$1.57$}}}
	&\multicolumn{1}{p{\gnumericColC}|}%
	{\gnumericPB{\raggedleft}\gnumbox[r]{\textsf{$3.40$}}}
	&\multicolumn{1}{p{\gnumericColD}|}%
	{\gnumericPB{\raggedleft}\gnumbox[r]{\textsf{$2.20$}}}
	&\multicolumn{1}{p{\gnumericColE}|}%
	{\gnumericPB{\raggedleft}\gnumbox[r]{\textsf{$1.85$}}}
\\
\hhline{|-|-|-|-|}
\end{longtable}

\gnumericTableEnd
}
\caption{Normalisation for a dipolar form for $J/\psi$ as a function of $\Lambda$ for $m_c=1.69$~GeV.}\label{tab:N_dip_jpsi_lam2}
\end{table}
\begin{table}[H]
{\tiny%%%%%%%%%%%%%%%%%%%%%%%%%%%%%%%%%%%%%%%%%%%%%%%%%%%%%%%%%%%%%%%%%%%%%%
\def\ifundefined#1{\expandafter\ifx\csname#1\endcsname\relax}

%%  Check for the \def token for inputed files. If it is not        %%
%%  defined, the file will be processed as a standalone and the     %%
%%  preamble will be used.                                          %%
\ifundefined{inputGnumericTable}

%%  End of the preamble for the standalone. The next section is for %%
%%  documents which are included into other LaTeX2e files.          %%
\else

%%  We are not a stand alone document. For a regular table, we will %%
%%  have no preamble and only define the closing to mean nothing.   %%
    \def\gnumericTableEnd{}

%%  If we want landscape mode in an embedded document, comment out  %%
%%  the line above and uncomment the two below. The table will      %%
%%  begin on a new page and run in landscape mode.                  %%
%       \def\gnumericTableEnd{\end{landscape}}
%       \begin{landscape}

%%  End of the else clause for this file being \input.              %%
\fi

%%%%%%%%%%%%%%%%%%%%%%%%%%%%%%%%%%%%%%%%%%%%%%%%%%%%%%%%%%%%%%%%%%%%%%
%%                                                                  %%
%%  The rest is the gnumeric table, except for the closing          %%
%%  statement. Changes below will alter the table's appearance.     %%
%%                                                                  %%
%%%%%%%%%%%%%%%%%%%%%%%%%%%%%%%%%%%%%%%%%%%%%%%%%%%%%%%%%%%%%%%%%%%%%%

\providecommand{\gnumericPB}[1]%
{\let\gnumericTemp=\\#1\let\\=\gnumericTemp\hspace{0pt}}

%%  The default table format retains the relative column widths of  %%
%%  gnumeric. They can easily be changed to c, r or l. In that case %%
%%  you may want to comment out the next line and uncomment the one %%
%%  thereafter                                                      %%
\providecommand\gnumbox{\makebox[0pt]}
%%\providecommand\gnumbox[1][]{\makebox}

%% to adjust positions in multirow situations                       %%
\setlength{\bigstrutjot}{\jot}
\setlength{\extrarowheight}{\doublerulesep}

%%  The \setlongtables command keeps column widths the same across %%
%%  pages. Simply comment out next line for varying column widths.  %%
\setlongtables

\setlength\gnumericTableWidth{%
	30pt+%
	30pt+%
	30pt+%
	30pt+%
0pt}
\def\gumericNumCols{4}
\setlength\gnumericTableWidthComplete{\gnumericTableWidth+\tabcolsep*\gumericNumCols*2+\arrayrulewidth*\gumericNumCols}
\ifthenelse{\lengthtest{\gnumericTableWidthComplete > \textwidth}}%
{\def\gnumericScale{\ratio{\textwidth-\tabcolsep*\gumericNumCols*2-\arrayrulewidth*\gumericNumCols}%
{\gnumericTableWidth}}}%
{\def\gnumericScale{1}}

%%%%%%%%%%%%%%%%%%%%%%%%%%%%%%%%%%%%%%%%%%%%%%%%%%%%%%%%%%%%%%%%%%%%%%
%%                                                                  %%
%% The following are the widths of the various columns. We are      %%
%% defining them here because then they are easier to change.       %%
%% Depending on the cell formats we may use them more than once.    %%
%%                                                                  %%
%%%%%%%%%%%%%%%%%%%%%%%%%%%%%%%%%%%%%%%%%%%%%%%%%%%%%%%%%%%%%%%%%%%%%%

\def\gnumericColB{30pt*\gnumericScale}
\def\gnumericColC{30pt*\gnumericScale}
\def\gnumericColD{30pt*\gnumericScale}
\def\gnumericColE{30pt*\gnumericScale}

\begin{longtable}[c]{%
	b{\gnumericColB}%
	b{\gnumericColC}%
	b{\gnumericColD}%
	b{\gnumericColE}%
	}

%%%%%%%%%%%%%%%%%%%%%%%%%%%%%%%%%%%%%%%%%%%%%%%%%%%%%%%%%%%%%%%%%%%%%%
%%  The longtable options. (Caption. headers... see Goosens. p.124) %%
%	\caption{The Table Caption.}             \\	%
% \hline	% Across the top of the table.
%%  The rest of these options are table rows which are placed on    %%
%%  the first. last or every page. Use \multicolumn if you want.    %%

%%  Header for the first page.                                      %%
%	\multicolumn{4}{c}{The First Header} \\ \hline 
%	\multicolumn{1}{c}{colTag}	%Column 1
%	&\multicolumn{1}{c}{colTag}	%Column 1
%	&\multicolumn{1}{c}{colTag}	%Column 2
%	&\multicolumn{1}{c}{colTag}	%Column 3
%	&\multicolumn{1}{c}{colTag}	\\ \hline %Last column
%	\endfirsthead

%%  The running header definition.                                  %%
%	\hline
%	\multicolumn{4}{l}{\ldots\small\slshape continued} \\ \hline
%	\multicolumn{1}{c}{colTag}	%Column 1
%	&\multicolumn{1}{c}{colTag}	%Column 1
%	&\multicolumn{1}{c}{colTag}	%Column 2
%	&\multicolumn{1}{c}{colTag}	%Column 3
%	&\multicolumn{1}{c}{colTag}	\\ \hline %Last column
%	\endhead

%%  The running footer definition.                                  %%
%	\hline
%	\multicolumn{4}{r}{\small\slshape continued\ldots} \\
%	\endfoot

%%  The ending footer definition.                                   %%
%	\multicolumn{4}{c}{That's all folks} \\ \hline 
%	\endlastfoot
%%%%%%%%%%%%%%%%%%%%%%%%%%%%%%%%%%%%%%%%%%%%%%%%%%%%%%%%%%%%%%%%%%%%%%

\hhline{|-|-|-|-|}
	 \multicolumn{1}{|p{\gnumericColB}|}%
	{\gnumericPB{\raggedleft}\gnumbox[r]{\textsf{$1.00$}}}
	&\multicolumn{1}{p{\gnumericColC}|}%
	{\gnumericPB{\raggedleft}\gnumbox[r]{\textsf{$9.39$}}}
	&\multicolumn{1}{p{\gnumericColD}|}%
	{\gnumericPB{\raggedleft}\gnumbox[r]{\textsf{$1.63$}}}
	&\multicolumn{1}{p{\gnumericColE}|}%
	{\gnumericPB{\raggedleft}\gnumbox[r]{\textsf{$3.54$}}}
\\
\hhline{|----|}
	 \multicolumn{1}{|p{\gnumericColB}|}%
	{\gnumericPB{\raggedleft}\gnumbox[r]{\textsf{$1.06$}}}
	&\multicolumn{1}{p{\gnumericColC}|}%
	{\gnumericPB{\raggedleft}\gnumbox[r]{\textsf{$8.28$}}}
	&\multicolumn{1}{p{\gnumericColD}|}%
	{\gnumericPB{\raggedleft}\gnumbox[r]{\textsf{$1.69$}}}
	&\multicolumn{1}{p{\gnumericColE}|}%
	{\gnumericPB{\raggedleft}\gnumbox[r]{\textsf{$3.29$}}}
\\
\hhline{|----|}
	 \multicolumn{1}{|p{\gnumericColB}|}%
	{\gnumericPB{\raggedleft}\gnumbox[r]{\textsf{$1.13$}}}
	&\multicolumn{1}{p{\gnumericColC}|}%
	{\gnumericPB{\raggedleft}\gnumbox[r]{\textsf{$7.36$}}}
	&\multicolumn{1}{p{\gnumericColD}|}%
	{\gnumericPB{\raggedleft}\gnumbox[r]{\textsf{$1.76$}}}
	&\multicolumn{1}{p{\gnumericColE}|}%
	{\gnumericPB{\raggedleft}\gnumbox[r]{\textsf{$3.07$}}}
\\
\hhline{|----|}
	 \multicolumn{1}{|p{\gnumericColB}|}%
	{\gnumericPB{\raggedleft}\gnumbox[r]{\textsf{$1.19$}}}
	&\multicolumn{1}{p{\gnumericColC}|}%
	{\gnumericPB{\raggedleft}\gnumbox[r]{\textsf{$6.59$}}}
	&\multicolumn{1}{p{\gnumericColD}|}%
	{\gnumericPB{\raggedleft}\gnumbox[r]{\textsf{$1.82$}}}
	&\multicolumn{1}{p{\gnumericColE}|}%
	{\gnumericPB{\raggedleft}\gnumbox[r]{\textsf{$2.87$}}}
\\
\hhline{|----|}
	 \multicolumn{1}{|p{\gnumericColB}|}%
	{\gnumericPB{\raggedleft}\gnumbox[r]{\textsf{$1.25$}}}
	&\multicolumn{1}{p{\gnumericColC}|}%
	{\gnumericPB{\raggedleft}\gnumbox[r]{\textsf{$5.94$}}}
	&\multicolumn{1}{p{\gnumericColD}|}%
	{\gnumericPB{\raggedleft}\gnumbox[r]{\textsf{$1.88$}}}
	&\multicolumn{1}{p{\gnumericColE}|}%
	{\gnumericPB{\raggedleft}\gnumbox[r]{\textsf{$2.69$}}}
\\
\hhline{|----|}
	 \multicolumn{1}{|p{\gnumericColB}|}%
	{\gnumericPB{\raggedleft}\gnumbox[r]{\textsf{$1.32$}}}
	&\multicolumn{1}{p{\gnumericColC}|}%
	{\gnumericPB{\raggedleft}\gnumbox[r]{\textsf{$5.39$}}}
	&\multicolumn{1}{p{\gnumericColD}|}%
	{\gnumericPB{\raggedleft}\gnumbox[r]{\textsf{$1.95$}}}
	&\multicolumn{1}{p{\gnumericColE}|}%
	{\gnumericPB{\raggedleft}\gnumbox[r]{\textsf{$2.52$}}}
\\
\hhline{|----|}
	 \multicolumn{1}{|p{\gnumericColB}|}%
	{\gnumericPB{\raggedleft}\gnumbox[r]{\textsf{$1.38$}}}
	&\multicolumn{1}{p{\gnumericColC}|}%
	{\gnumericPB{\raggedleft}\gnumbox[r]{\textsf{$4.91$}}}
	&\multicolumn{1}{p{\gnumericColD}|}%
	{\gnumericPB{\raggedleft}\gnumbox[r]{\textsf{$2.01$}}}
	&\multicolumn{1}{p{\gnumericColE}|}%
	{\gnumericPB{\raggedleft}\gnumbox[r]{\textsf{$2.38$}}}
\\
\hhline{|----|}
	 \multicolumn{1}{|p{\gnumericColB}|}%
	{\gnumericPB{\raggedleft}\gnumbox[r]{\textsf{$1.44$}}}
	&\multicolumn{1}{p{\gnumericColC}|}%
	{\gnumericPB{\raggedleft}\gnumbox[r]{\textsf{$4.50$}}}
	&\multicolumn{1}{p{\gnumericColD}|}%
	{\gnumericPB{\raggedleft}\gnumbox[r]{\textsf{$2.07$}}}
	&\multicolumn{1}{p{\gnumericColE}|}%
	{\gnumericPB{\raggedleft}\gnumbox[r]{\textsf{$2.24$}}}
\\
\hhline{|----|}
	 \multicolumn{1}{|p{\gnumericColB}|}%
	{\gnumericPB{\raggedleft}\gnumbox[r]{\textsf{$1.51$}}}
	&\multicolumn{1}{p{\gnumericColC}|}%
	{\gnumericPB{\raggedleft}\gnumbox[r]{\textsf{$4.14$}}}
	&\multicolumn{1}{p{\gnumericColD}|}%
	{\gnumericPB{\raggedleft}\gnumbox[r]{\textsf{$2.14$}}}
	&\multicolumn{1}{p{\gnumericColE}|}%
	{\gnumericPB{\raggedleft}\gnumbox[r]{\textsf{$2.12$}}}
\\
\hhline{|----|}
	 \multicolumn{1}{|p{\gnumericColB}|}%
	{\gnumericPB{\raggedleft}\gnumbox[r]{\textsf{$1.57$}}}
	&\multicolumn{1}{p{\gnumericColC}|}%
	{\gnumericPB{\raggedleft}\gnumbox[r]{\textsf{$3.82$}}}
	&\multicolumn{1}{p{\gnumericColD}|}%
	{\gnumericPB{\raggedleft}\gnumbox[r]{\textsf{$2.20$}}}
	&\multicolumn{1}{p{\gnumericColE}|}%
	{\gnumericPB{\raggedleft}\gnumbox[r]{\textsf{$2.01$}}}
\\
\hhline{|-|-|-|-|}
\end{longtable}

\gnumericTableEnd
}
\caption{Normalisation for a dipolar form for $J/\psi$ as a function of $\Lambda$ for $m_c=1.78$~GeV.}\label{tab:N_dip_jpsi_lam3}
\end{table}
\begin{table}[H]
{\tiny%%%%%%%%%%%%%%%%%%%%%%%%%%%%%%%%%%%%%%%%%%%%%%%%%%%%%%%%%%%%%%%%%%%%%%
\def\ifundefined#1{\expandafter\ifx\csname#1\endcsname\relax}

%%  Check for the \def token for inputed files. If it is not        %%
%%  defined, the file will be processed as a standalone and the     %%
%%  preamble will be used.                                          %%
\ifundefined{inputGnumericTable}

%%  End of the preamble for the standalone. The next section is for %%
%%  documents which are included into other LaTeX2e files.          %%
\else

%%  We are not a stand alone document. For a regular table, we will %%
%%  have no preamble and only define the closing to mean nothing.   %%
    \def\gnumericTableEnd{}

%%  If we want landscape mode in an embedded document, comment out  %%
%%  the line above and uncomment the two below. The table will      %%
%%  begin on a new page and run in landscape mode.                  %%
%       \def\gnumericTableEnd{\end{landscape}}
%       \begin{landscape}

%%  End of the else clause for this file being \input.              %%
\fi

%%%%%%%%%%%%%%%%%%%%%%%%%%%%%%%%%%%%%%%%%%%%%%%%%%%%%%%%%%%%%%%%%%%%%%
%%                                                                  %%
%%  The rest is the gnumeric table, except for the closing          %%
%%  statement. Changes below will alter the table's appearance.     %%
%%                                                                  %%
%%%%%%%%%%%%%%%%%%%%%%%%%%%%%%%%%%%%%%%%%%%%%%%%%%%%%%%%%%%%%%%%%%%%%%

\providecommand{\gnumericPB}[1]%
{\let\gnumericTemp=\\#1\let\\=\gnumericTemp\hspace{0pt}}

%%  The default table format retains the relative column widths of  %%
%%  gnumeric. They can easily be changed to c, r or l. In that case %%
%%  you may want to comment out the next line and uncomment the one %%
%%  thereafter                                                      %%
\providecommand\gnumbox{\makebox[0pt]}
%%\providecommand\gnumbox[1][]{\makebox}

%% to adjust positions in multirow situations                       %%
\setlength{\bigstrutjot}{\jot}
\setlength{\extrarowheight}{\doublerulesep}

%%  The \setlongtables command keeps column widths the same across %%
%%  pages. Simply comment out next line for varying column widths.  %%
\setlongtables

\setlength\gnumericTableWidth{%
	30pt+%
	30pt+%
	30pt+%
	30pt+%
0pt}
\def\gumericNumCols{4}
\setlength\gnumericTableWidthComplete{\gnumericTableWidth+\tabcolsep*\gumericNumCols*2+\arrayrulewidth*\gumericNumCols}
\ifthenelse{\lengthtest{\gnumericTableWidthComplete > \textwidth}}%
{\def\gnumericScale{\ratio{\textwidth-\tabcolsep*\gumericNumCols*2-\arrayrulewidth*\gumericNumCols}%
{\gnumericTableWidth}}}%
{\def\gnumericScale{1}}

%%%%%%%%%%%%%%%%%%%%%%%%%%%%%%%%%%%%%%%%%%%%%%%%%%%%%%%%%%%%%%%%%%%%%%
%%                                                                  %%
%% The following are the widths of the various columns. We are      %%
%% defining them here because then they are easier to change.       %%
%% Depending on the cell formats we may use them more than once.    %%
%%                                                                  %%
%%%%%%%%%%%%%%%%%%%%%%%%%%%%%%%%%%%%%%%%%%%%%%%%%%%%%%%%%%%%%%%%%%%%%%

\def\gnumericColB{30pt*\gnumericScale}
\def\gnumericColC{30pt*\gnumericScale}
\def\gnumericColD{30pt*\gnumericScale}
\def\gnumericColE{30pt*\gnumericScale}

\begin{longtable}[c]{%
	b{\gnumericColB}%
	b{\gnumericColC}%
	b{\gnumericColD}%
	b{\gnumericColE}%
	}

%%%%%%%%%%%%%%%%%%%%%%%%%%%%%%%%%%%%%%%%%%%%%%%%%%%%%%%%%%%%%%%%%%%%%%
%%  The longtable options. (Caption. headers... see Goosens. p.124) %%
%	\caption{The Table Caption.}             \\	%
% \hline	% Across the top of the table.
%%  The rest of these options are table rows which are placed on    %%
%%  the first. last or every page. Use \multicolumn if you want.    %%

%%  Header for the first page.                                      %%
%	\multicolumn{4}{c}{The First Header} \\ \hline 
%	\multicolumn{1}{c}{colTag}	%Column 1
%	&\multicolumn{1}{c}{colTag}	%Column 1
%	&\multicolumn{1}{c}{colTag}	%Column 2
%	&\multicolumn{1}{c}{colTag}	%Column 3
%	&\multicolumn{1}{c}{colTag}	\\ \hline %Last column
%	\endfirsthead

%%  The running header definition.                                  %%
%	\hline
%	\multicolumn{4}{l}{\ldots\small\slshape continued} \\ \hline
%	\multicolumn{1}{c}{colTag}	%Column 1
%	&\multicolumn{1}{c}{colTag}	%Column 1
%	&\multicolumn{1}{c}{colTag}	%Column 2
%	&\multicolumn{1}{c}{colTag}	%Column 3
%	&\multicolumn{1}{c}{colTag}	\\ \hline %Last column
%	\endhead

%%  The running footer definition.                                  %%
%	\hline
%	\multicolumn{4}{r}{\small\slshape continued\ldots} \\
%	\endfoot

%%  The ending footer definition.                                   %%
%	\multicolumn{4}{c}{That's all folks} \\ \hline 
%	\endlastfoot
%%%%%%%%%%%%%%%%%%%%%%%%%%%%%%%%%%%%%%%%%%%%%%%%%%%%%%%%%%%%%%%%%%%%%%

\hhline{|-|-|-|-|}
	 \multicolumn{1}{|p{\gnumericColB}|}%
	{\gnumericPB{\raggedleft}\gnumbox[r]{\textsf{$1.00$}}}
	&\multicolumn{1}{p{\gnumericColC}|}%
	{\gnumericPB{\raggedleft}\gnumbox[r]{\textsf{$10.61$}}}
	&\multicolumn{1}{p{\gnumericColD}|}%
	{\gnumericPB{\raggedleft}\gnumbox[r]{\textsf{$1.63$}}}
	&\multicolumn{1}{p{\gnumericColE}|}%
	{\gnumericPB{\raggedleft}\gnumbox[r]{\textsf{$3.84$}}}
\\
\hhline{|----|}
	 \multicolumn{1}{|p{\gnumericColB}|}%
	{\gnumericPB{\raggedleft}\gnumbox[r]{\textsf{$1.06$}}}
	&\multicolumn{1}{p{\gnumericColC}|}%
	{\gnumericPB{\raggedleft}\gnumbox[r]{\textsf{$9.31$}}}
	&\multicolumn{1}{p{\gnumericColD}|}%
	{\gnumericPB{\raggedleft}\gnumbox[r]{\textsf{$1.69$}}}
	&\multicolumn{1}{p{\gnumericColE}|}%
	{\gnumericPB{\raggedleft}\gnumbox[r]{\textsf{$3.56$}}}
\\
\hhline{|----|}
	 \multicolumn{1}{|p{\gnumericColB}|}%
	{\gnumericPB{\raggedleft}\gnumbox[r]{\textsf{$1.13$}}}
	&\multicolumn{1}{p{\gnumericColC}|}%
	{\gnumericPB{\raggedleft}\gnumbox[r]{\textsf{$8.23$}}}
	&\multicolumn{1}{p{\gnumericColD}|}%
	{\gnumericPB{\raggedleft}\gnumbox[r]{\textsf{$1.76$}}}
	&\multicolumn{1}{p{\gnumericColE}|}%
	{\gnumericPB{\raggedleft}\gnumbox[r]{\textsf{$3.31$}}}
\\
\hhline{|----|}
	 \multicolumn{1}{|p{\gnumericColB}|}%
	{\gnumericPB{\raggedleft}\gnumbox[r]{\textsf{$1.19$}}}
	&\multicolumn{1}{p{\gnumericColC}|}%
	{\gnumericPB{\raggedleft}\gnumbox[r]{\textsf{$7.34$}}}
	&\multicolumn{1}{p{\gnumericColD}|}%
	{\gnumericPB{\raggedleft}\gnumbox[r]{\textsf{$1.82$}}}
	&\multicolumn{1}{p{\gnumericColE}|}%
	{\gnumericPB{\raggedleft}\gnumbox[r]{\textsf{$3.09$}}}
\\
\hhline{|----|}
	 \multicolumn{1}{|p{\gnumericColB}|}%
	{\gnumericPB{\raggedleft}\gnumbox[r]{\textsf{$1.25$}}}
	&\multicolumn{1}{p{\gnumericColC}|}%
	{\gnumericPB{\raggedleft}\gnumbox[r]{\textsf{$6.59$}}}
	&\multicolumn{1}{p{\gnumericColD}|}%
	{\gnumericPB{\raggedleft}\gnumbox[r]{\textsf{$1.88$}}}
	&\multicolumn{1}{p{\gnumericColE}|}%
	{\gnumericPB{\raggedleft}\gnumbox[r]{\textsf{$2.88$}}}
\\
\hhline{|----|}
	 \multicolumn{1}{|p{\gnumericColB}|}%
	{\gnumericPB{\raggedleft}\gnumbox[r]{\textsf{$1.32$}}}
	&\multicolumn{1}{p{\gnumericColC}|}%
	{\gnumericPB{\raggedleft}\gnumbox[r]{\textsf{$5.95$}}}
	&\multicolumn{1}{p{\gnumericColD}|}%
	{\gnumericPB{\raggedleft}\gnumbox[r]{\textsf{$1.95$}}}
	&\multicolumn{1}{p{\gnumericColE}|}%
	{\gnumericPB{\raggedleft}\gnumbox[r]{\textsf{$2.70$}}}
\\
\hhline{|----|}
	 \multicolumn{1}{|p{\gnumericColB}|}%
	{\gnumericPB{\raggedleft}\gnumbox[r]{\textsf{$1.38$}}}
	&\multicolumn{1}{p{\gnumericColC}|}%
	{\gnumericPB{\raggedleft}\gnumbox[r]{\textsf{$5.40$}}}
	&\multicolumn{1}{p{\gnumericColD}|}%
	{\gnumericPB{\raggedleft}\gnumbox[r]{\textsf{$2.01$}}}
	&\multicolumn{1}{p{\gnumericColE}|}%
	{\gnumericPB{\raggedleft}\gnumbox[r]{\textsf{$2.54$}}}
\\
\hhline{|----|}
	 \multicolumn{1}{|p{\gnumericColB}|}%
	{\gnumericPB{\raggedleft}\gnumbox[r]{\textsf{$1.44$}}}
	&\multicolumn{1}{p{\gnumericColC}|}%
	{\gnumericPB{\raggedleft}\gnumbox[r]{\textsf{$4.93$}}}
	&\multicolumn{1}{p{\gnumericColD}|}%
	{\gnumericPB{\raggedleft}\gnumbox[r]{\textsf{$2.07$}}}
	&\multicolumn{1}{p{\gnumericColE}|}%
	{\gnumericPB{\raggedleft}\gnumbox[r]{\textsf{$2.39$}}}
\\
\hhline{|----|}
	 \multicolumn{1}{|p{\gnumericColB}|}%
	{\gnumericPB{\raggedleft}\gnumbox[r]{\textsf{$1.51$}}}
	&\multicolumn{1}{p{\gnumericColC}|}%
	{\gnumericPB{\raggedleft}\gnumbox[r]{\textsf{$4.52$}}}
	&\multicolumn{1}{p{\gnumericColD}|}%
	{\gnumericPB{\raggedleft}\gnumbox[r]{\textsf{$2.14$}}}
	&\multicolumn{1}{p{\gnumericColE}|}%
	{\gnumericPB{\raggedleft}\gnumbox[r]{\textsf{$2.25$}}}
\\
\hhline{|----|}
	 \multicolumn{1}{|p{\gnumericColB}|}%
	{\gnumericPB{\raggedleft}\gnumbox[r]{\textsf{$1.57$}}}
	&\multicolumn{1}{p{\gnumericColC}|}%
	{\gnumericPB{\raggedleft}\gnumbox[r]{\textsf{$4.16$}}}
	&\multicolumn{1}{p{\gnumericColD}|}%
	{\gnumericPB{\raggedleft}\gnumbox[r]{\textsf{$2.20$}}}
	&\multicolumn{1}{p{\gnumericColE}|}%
	{\gnumericPB{\raggedleft}\gnumbox[r]{\textsf{$2.13$}}}
\\
\hhline{|-|-|-|-|}
\end{longtable}

\gnumericTableEnd
}
\caption{Normalisation for a dipolar form for $J/\psi$ as a function of $\Lambda$ for $m_c=1.87$~GeV.}\label{tab:N_dip_jpsi_lam4}
\end{table}

\begin{table}[H]
{\tiny%%%%%%%%%%%%%%%%%%%%%%%%%%%%%%%%%%%%%%%%%%%%%%%%%%%%%%%%%%%%%%%%%%%%%%
\def\ifundefined#1{\expandafter\ifx\csname#1\endcsname\relax}

%%  Check for the \def token for inputed files. If it is not        %%
%%  defined, the file will be processed as a standalone and the     %%
%%  preamble will be used.                                          %%
\ifundefined{inputGnumericTable}

%%  End of the preamble for the standalone. The next section is for %%
%%  documents which are included into other LaTeX2e files.          %%
\else

%%  We are not a stand alone document. For a regular table, we will %%
%%  have no preamble and only define the closing to mean nothing.   %%
    \def\gnumericTableEnd{}

%%  If we want landscape mode in an embedded document, comment out  %%
%%  the line above and uncomment the two below. The table will      %%
%%  begin on a new page and run in landscape mode.                  %%
%       \def\gnumericTableEnd{\end{landscape}}
%       \begin{landscape}

%%  End of the else clause for this file being \input.              %%
\fi

%%%%%%%%%%%%%%%%%%%%%%%%%%%%%%%%%%%%%%%%%%%%%%%%%%%%%%%%%%%%%%%%%%%%%%
%%                                                                  %%
%%  The rest is the gnumeric table, except for the closing          %%
%%  statement. Changes below will alter the table's appearance.     %%
%%                                                                  %%
%%%%%%%%%%%%%%%%%%%%%%%%%%%%%%%%%%%%%%%%%%%%%%%%%%%%%%%%%%%%%%%%%%%%%%

\providecommand{\gnumericPB}[1]%
{\let\gnumericTemp=\\#1\let\\=\gnumericTemp\hspace{0pt}}

%%  The default table format retains the relative column widths of  %%
%%  gnumeric. They can easily be changed to c, r or l. In that case %%
%%  you may want to comment out the next line and uncomment the one %%
%%  thereafter                                                      %%
\providecommand\gnumbox{\makebox[0pt]}
%%\providecommand\gnumbox[1][]{\makebox}

%% to adjust positions in multirow situations                       %%
\setlength{\bigstrutjot}{\jot}
\setlength{\extrarowheight}{\doublerulesep}

%%  The \setlongtables command keeps column widths the same across %%
%%  pages. Simply comment out next line for varying column widths.  %%
\setlongtables

\setlength\gnumericTableWidth{%
	30pt+%
	30pt+%
	30pt+%
	30pt+%
0pt}
\def\gumericNumCols{4}
\setlength\gnumericTableWidthComplete{\gnumericTableWidth+\tabcolsep*\gumericNumCols*2+\arrayrulewidth*\gumericNumCols}
\ifthenelse{\lengthtest{\gnumericTableWidthComplete > \textwidth}}%
{\def\gnumericScale{\ratio{\textwidth-\tabcolsep*\gumericNumCols*2-\arrayrulewidth*\gumericNumCols}%
{\gnumericTableWidth}}}%
{\def\gnumericScale{1}}

%%%%%%%%%%%%%%%%%%%%%%%%%%%%%%%%%%%%%%%%%%%%%%%%%%%%%%%%%%%%%%%%%%%%%%
%%                                                                  %%
%% The following are the widths of the various columns. We are      %%
%% defining them here because then they are easier to change.       %%
%% Depending on the cell formats we may use them more than once.    %%
%%                                                                  %%
%%%%%%%%%%%%%%%%%%%%%%%%%%%%%%%%%%%%%%%%%%%%%%%%%%%%%%%%%%%%%%%%%%%%%%

\def\gnumericColB{30pt*\gnumericScale}
\def\gnumericColC{30pt*\gnumericScale}
\def\gnumericColD{30pt*\gnumericScale}
\def\gnumericColE{30pt*\gnumericScale}

\begin{longtable}[c]{%
	b{\gnumericColB}%
	b{\gnumericColC}%
	b{\gnumericColD}%
	b{\gnumericColE}%
	}

%%%%%%%%%%%%%%%%%%%%%%%%%%%%%%%%%%%%%%%%%%%%%%%%%%%%%%%%%%%%%%%%%%%%%%
%%  The longtable options. (Caption. headers... see Goosens. p.124) %%
%	\caption{The Table Caption.}             \\	%
% \hline	% Across the top of the table.
%%  The rest of these options are table rows which are placed on    %%
%%  the first. last or every page. Use \multicolumn if you want.    %%

%%  Header for the first page.                                      %%
%	\multicolumn{4}{c}{The First Header} \\ \hline 
%	\multicolumn{1}{c}{colTag}	%Column 1
%	&\multicolumn{1}{c}{colTag}	%Column 1
%	&\multicolumn{1}{c}{colTag}	%Column 2
%	&\multicolumn{1}{c}{colTag}	%Column 3
%	&\multicolumn{1}{c}{colTag}	\\ \hline %Last column
%	\endfirsthead

%%  The running header definition.                                  %%
%	\hline
%	\multicolumn{4}{l}{\ldots\small\slshape continued} \\ \hline
%	\multicolumn{1}{c}{colTag}	%Column 1
%	&\multicolumn{1}{c}{colTag}	%Column 1
%	&\multicolumn{1}{c}{colTag}	%Column 2
%	&\multicolumn{1}{c}{colTag}	%Column 3
%	&\multicolumn{1}{c}{colTag}	\\ \hline %Last column
%	\endhead

%%  The running footer definition.                                  %%
%	\hline
%	\multicolumn{4}{r}{\small\slshape continued\ldots} \\
%	\endfoot

%%  The ending footer definition.                                   %%
%	\multicolumn{4}{c}{That's all folks} \\ \hline 
%	\endlastfoot
%%%%%%%%%%%%%%%%%%%%%%%%%%%%%%%%%%%%%%%%%%%%%%%%%%%%%%%%%%%%%%%%%%%%%%

\hhline{|-|-|-|-|}
	 \multicolumn{1}{|p{\gnumericColB}|}%
	{\gnumericPB{\raggedleft}\gnumbox[r]{\textsf{$m_c$}}}
	&\multicolumn{1}{p{\gnumericColC}|}%
	{\gnumericPB{\raggedleft}\gnumbox[r]{\textsf{$N$}}}
	&\multicolumn{1}{p{\gnumericColD}|}%
	{\gnumericPB{\raggedleft}\gnumbox[r]{\textsf{$m_c$}}}
	&\multicolumn{1}{p{\gnumericColE}|}%
	{\gnumericPB{\raggedleft}\gnumbox[r]{\textsf{$N$}}}
\\
\hhline{|----|}
	 \multicolumn{1}{|p{\gnumericColB}|}%
	{\gnumericPB{\raggedleft}\gnumbox[r]{\textsf{$1.60$}}}
	&\multicolumn{1}{p{\gnumericColC}|}%
	{\gnumericPB{\raggedleft}\gnumbox[r]{\textsf{$5.95$}}}
	&\multicolumn{1}{p{\gnumericColD}|}%
	{\gnumericPB{\raggedleft}\gnumbox[r]{\textsf{$1.74$}}}
	&\multicolumn{1}{p{\gnumericColE}|}%
	{\gnumericPB{\raggedleft}\gnumbox[r]{\textsf{$8.81$}}}
\\
\hhline{|----|}
	 \multicolumn{1}{|p{\gnumericColB}|}%
	{\gnumericPB{\raggedleft}\gnumbox[r]{\textsf{$1.61$}}}
	&\multicolumn{1}{p{\gnumericColC}|}%
	{\gnumericPB{\raggedleft}\gnumbox[r]{\textsf{$6.33$}}}
	&\multicolumn{1}{p{\gnumericColD}|}%
	{\gnumericPB{\raggedleft}\gnumbox[r]{\textsf{$1.76$}}}
	&\multicolumn{1}{p{\gnumericColE}|}%
	{\gnumericPB{\raggedleft}\gnumbox[r]{\textsf{$9.03$}}}
\\
\hhline{|----|}
	 \multicolumn{1}{|p{\gnumericColB}|}%
	{\gnumericPB{\raggedleft}\gnumbox[r]{\textsf{$1.63$}}}
	&\multicolumn{1}{p{\gnumericColC}|}%
	{\gnumericPB{\raggedleft}\gnumbox[r]{\textsf{$6.67$}}}
	&\multicolumn{1}{p{\gnumericColD}|}%
	{\gnumericPB{\raggedleft}\gnumbox[r]{\textsf{$1.77$}}}
	&\multicolumn{1}{p{\gnumericColE}|}%
	{\gnumericPB{\raggedleft}\gnumbox[r]{\textsf{$9.25$}}}
\\
\hhline{|----|}
	 \multicolumn{1}{|p{\gnumericColB}|}%
	{\gnumericPB{\raggedleft}\gnumbox[r]{\textsf{$1.64$}}}
	&\multicolumn{1}{p{\gnumericColC}|}%
	{\gnumericPB{\raggedleft}\gnumbox[r]{\textsf{$6.99$}}}
	&\multicolumn{1}{p{\gnumericColD}|}%
	{\gnumericPB{\raggedleft}\gnumbox[r]{\textsf{$1.78$}}}
	&\multicolumn{1}{p{\gnumericColE}|}%
	{\gnumericPB{\raggedleft}\gnumbox[r]{\textsf{$9.46$}}}
\\
\hhline{|----|}
	 \multicolumn{1}{|p{\gnumericColB}|}%
	{\gnumericPB{\raggedleft}\gnumbox[r]{\textsf{$1.66$}}}
	&\multicolumn{1}{p{\gnumericColC}|}%
	{\gnumericPB{\raggedleft}\gnumbox[r]{\textsf{$7.29$}}}
	&\multicolumn{1}{p{\gnumericColD}|}%
	{\gnumericPB{\raggedleft}\gnumbox[r]{\textsf{$1.80$}}}
	&\multicolumn{1}{p{\gnumericColE}|}%
	{\gnumericPB{\raggedleft}\gnumbox[r]{\textsf{$9.66$}}}
\\
\hhline{|----|}
	 \multicolumn{1}{|p{\gnumericColB}|}%
	{\gnumericPB{\raggedleft}\gnumbox[r]{\textsf{$1.67$}}}
	&\multicolumn{1}{p{\gnumericColC}|}%
	{\gnumericPB{\raggedleft}\gnumbox[r]{\textsf{$7.57$}}}
	&\multicolumn{1}{p{\gnumericColD}|}%
	{\gnumericPB{\raggedleft}\gnumbox[r]{\textsf{$1.81$}}}
	&\multicolumn{1}{p{\gnumericColE}|}%
	{\gnumericPB{\raggedleft}\gnumbox[r]{\textsf{$9.86$}}}
\\
\hhline{|----|}
	 \multicolumn{1}{|p{\gnumericColB}|}%
	{\gnumericPB{\raggedleft}\gnumbox[r]{\textsf{$1.69$}}}
	&\multicolumn{1}{p{\gnumericColC}|}%
	{\gnumericPB{\raggedleft}\gnumbox[r]{\textsf{$7.84$}}}
	&\multicolumn{1}{p{\gnumericColD}|}%
	{\gnumericPB{\raggedleft}\gnumbox[r]{\textsf{$1.83$}}}
	&\multicolumn{1}{p{\gnumericColE}|}%
	{\gnumericPB{\raggedleft}\gnumbox[r]{\textsf{$10.05$}}}
\\
\hhline{|----|}
	 \multicolumn{1}{|p{\gnumericColB}|}%
	{\gnumericPB{\raggedleft}\gnumbox[r]{\textsf{$1.70$}}}
	&\multicolumn{1}{p{\gnumericColC}|}%
	{\gnumericPB{\raggedleft}\gnumbox[r]{\textsf{$8.10$}}}
	&\multicolumn{1}{p{\gnumericColD}|}%
	{\gnumericPB{\raggedleft}\gnumbox[r]{\textsf{$1.84$}}}
	&\multicolumn{1}{p{\gnumericColE}|}%
	{\gnumericPB{\raggedleft}\gnumbox[r]{\textsf{$10.24$}}}
\\
\hhline{|----|}
	 \multicolumn{1}{|p{\gnumericColB}|}%
	{\gnumericPB{\raggedleft}\gnumbox[r]{\textsf{$1.71$}}}
	&\multicolumn{1}{p{\gnumericColC}|}%
	{\gnumericPB{\raggedleft}\gnumbox[r]{\textsf{$8.34$}}}
	&\multicolumn{1}{p{\gnumericColD}|}%
	{\gnumericPB{\raggedleft}\gnumbox[r]{\textsf{$1.86$}}}
	&\multicolumn{1}{p{\gnumericColE}|}%
	{\gnumericPB{\raggedleft}\gnumbox[r]{\textsf{$10.43$}}}
\\
\hhline{|----|}
	 \multicolumn{1}{|p{\gnumericColB}|}%
	{\gnumericPB{\raggedleft}\gnumbox[r]{\textsf{$1.73$}}}
	&\multicolumn{1}{p{\gnumericColC}|}%
	{\gnumericPB{\raggedleft}\gnumbox[r]{\textsf{$8.58$}}}
	&\multicolumn{1}{p{\gnumericColD}|}%
	{\gnumericPB{\raggedleft}\gnumbox[r]{\textsf{$1.87$}}}
	&\multicolumn{1}{p{\gnumericColE}|}%
	{\gnumericPB{\raggedleft}\gnumbox[r]{\textsf{$10.61$}}}
\\
\hhline{|-|-|-|-|}
\end{longtable}

\gnumericTableEnd
}
\caption{Normalisation for a dipolar form for $J/\psi$ as a function of  the $c$ quark mass
for $\Lambda=1.0$~GeV.}\label{tab:N_dip_jpsi_mq1}
\end{table}

\begin{table}[H]
{\tiny%%%%%%%%%%%%%%%%%%%%%%%%%%%%%%%%%%%%%%%%%%%%%%%%%%%%%%%%%%%%%%%%%%%%%%
\def\ifundefined#1{\expandafter\ifx\csname#1\endcsname\relax}

%%  Check for the \def token for inputed files. If it is not        %%
%%  defined, the file will be processed as a standalone and the     %%
%%  preamble will be used.                                          %%
\ifundefined{inputGnumericTable}

%%  End of the preamble for the standalone. The next section is for %%
%%  documents which are included into other LaTeX2e files.          %%
\else

%%  We are not a stand alone document. For a regular table, we will %%
%%  have no preamble and only define the closing to mean nothing.   %%
    \def\gnumericTableEnd{}

%%  If we want landscape mode in an embedded document, comment out  %%
%%  the line above and uncomment the two below. The table will      %%
%%  begin on a new page and run in landscape mode.                  %%
%       \def\gnumericTableEnd{\end{landscape}}
%       \begin{landscape}

%%  End of the else clause for this file being \input.              %%
\fi

%%%%%%%%%%%%%%%%%%%%%%%%%%%%%%%%%%%%%%%%%%%%%%%%%%%%%%%%%%%%%%%%%%%%%%
%%                                                                  %%
%%  The rest is the gnumeric table, except for the closing          %%
%%  statement. Changes below will alter the table's appearance.     %%
%%                                                                  %%
%%%%%%%%%%%%%%%%%%%%%%%%%%%%%%%%%%%%%%%%%%%%%%%%%%%%%%%%%%%%%%%%%%%%%%

\providecommand{\gnumericPB}[1]%
{\let\gnumericTemp=\\#1\let\\=\gnumericTemp\hspace{0pt}}

%%  The default table format retains the relative column widths of  %%
%%  gnumeric. They can easily be changed to c, r or l. In that case %%
%%  you may want to comment out the next line and uncomment the one %%
%%  thereafter                                                      %%
\providecommand\gnumbox{\makebox[0pt]}
%%\providecommand\gnumbox[1][]{\makebox}

%% to adjust positions in multirow situations                       %%
\setlength{\bigstrutjot}{\jot}
\setlength{\extrarowheight}{\doublerulesep}

%%  The \setlongtables command keeps column widths the same across %%
%%  pages. Simply comment out next line for varying column widths.  %%
\setlongtables

\setlength\gnumericTableWidth{%
	30pt+%
	30pt+%
	30pt+%
	30pt+%
0pt}
\def\gumericNumCols{4}
\setlength\gnumericTableWidthComplete{\gnumericTableWidth+\tabcolsep*\gumericNumCols*2+\arrayrulewidth*\gumericNumCols}
\ifthenelse{\lengthtest{\gnumericTableWidthComplete > \textwidth}}%
{\def\gnumericScale{\ratio{\textwidth-\tabcolsep*\gumericNumCols*2-\arrayrulewidth*\gumericNumCols}%
{\gnumericTableWidth}}}%
{\def\gnumericScale{1}}

%%%%%%%%%%%%%%%%%%%%%%%%%%%%%%%%%%%%%%%%%%%%%%%%%%%%%%%%%%%%%%%%%%%%%%
%%                                                                  %%
%% The following are the widths of the various columns. We are      %%
%% defining them here because then they are easier to change.       %%
%% Depending on the cell formats we may use them more than once.    %%
%%                                                                  %%
%%%%%%%%%%%%%%%%%%%%%%%%%%%%%%%%%%%%%%%%%%%%%%%%%%%%%%%%%%%%%%%%%%%%%%

\def\gnumericColB{30pt*\gnumericScale}
\def\gnumericColC{30pt*\gnumericScale}
\def\gnumericColD{30pt*\gnumericScale}
\def\gnumericColE{30pt*\gnumericScale}

\begin{longtable}[c]{%
	b{\gnumericColB}%
	b{\gnumericColC}%
	b{\gnumericColD}%
	b{\gnumericColE}%
	}

%%%%%%%%%%%%%%%%%%%%%%%%%%%%%%%%%%%%%%%%%%%%%%%%%%%%%%%%%%%%%%%%%%%%%%
%%  The longtable options. (Caption. headers... see Goosens. p.124) %%
%	\caption{The Table Caption.}             \\	%
% \hline	% Across the top of the table.
%%  The rest of these options are table rows which are placed on    %%
%%  the first. last or every page. Use \multicolumn if you want.    %%

%%  Header for the first page.                                      %%
%	\multicolumn{4}{c}{The First Header} \\ \hline 
%	\multicolumn{1}{c}{colTag}	%Column 1
%	&\multicolumn{1}{c}{colTag}	%Column 1
%	&\multicolumn{1}{c}{colTag}	%Column 2
%	&\multicolumn{1}{c}{colTag}	%Column 3
%	&\multicolumn{1}{c}{colTag}	\\ \hline %Last column
%	\endfirsthead

%%  The running header definition.                                  %%
%	\hline
%	\multicolumn{4}{l}{\ldots\small\slshape continued} \\ \hline
%	\multicolumn{1}{c}{colTag}	%Column 1
%	&\multicolumn{1}{c}{colTag}	%Column 1
%	&\multicolumn{1}{c}{colTag}	%Column 2
%	&\multicolumn{1}{c}{colTag}	%Column 3
%	&\multicolumn{1}{c}{colTag}	\\ \hline %Last column
%	\endhead

%%  The running footer definition.                                  %%
%	\hline
%	\multicolumn{4}{r}{\small\slshape continued\ldots} \\
%	\endfoot

%%  The ending footer definition.                                   %%
%	\multicolumn{4}{c}{That's all folks} \\ \hline 
%	\endlastfoot
%%%%%%%%%%%%%%%%%%%%%%%%%%%%%%%%%%%%%%%%%%%%%%%%%%%%%%%%%%%%%%%%%%%%%%

\hhline{|-|-|-|-|}
	 \multicolumn{1}{|p{\gnumericColB}|}%
	{\gnumericPB{\raggedleft}\gnumbox[r]{\textsf{$1.60$}}}
	&\multicolumn{1}{p{\gnumericColC}|}%
	{\gnumericPB{\raggedleft}\gnumbox[r]{\textsf{$3.39$}}}
	&\multicolumn{1}{p{\gnumericColD}|}%
	{\gnumericPB{\raggedleft}\gnumbox[r]{\textsf{$1.74$}}}
	&\multicolumn{1}{p{\gnumericColE}|}%
	{\gnumericPB{\raggedleft}\gnumbox[r]{\textsf{$4.54$}}}
\\
\hhline{|----|}
	 \multicolumn{1}{|p{\gnumericColB}|}%
	{\gnumericPB{\raggedleft}\gnumbox[r]{\textsf{$1.61$}}}
	&\multicolumn{1}{p{\gnumericColC}|}%
	{\gnumericPB{\raggedleft}\gnumbox[r]{\textsf{$3.55$}}}
	&\multicolumn{1}{p{\gnumericColD}|}%
	{\gnumericPB{\raggedleft}\gnumbox[r]{\textsf{$1.76$}}}
	&\multicolumn{1}{p{\gnumericColE}|}%
	{\gnumericPB{\raggedleft}\gnumbox[r]{\textsf{$4.63$}}}
\\
\hhline{|----|}
	 \multicolumn{1}{|p{\gnumericColB}|}%
	{\gnumericPB{\raggedleft}\gnumbox[r]{\textsf{$1.63$}}}
	&\multicolumn{1}{p{\gnumericColC}|}%
	{\gnumericPB{\raggedleft}\gnumbox[r]{\textsf{$3.69$}}}
	&\multicolumn{1}{p{\gnumericColD}|}%
	{\gnumericPB{\raggedleft}\gnumbox[r]{\textsf{$1.77$}}}
	&\multicolumn{1}{p{\gnumericColE}|}%
	{\gnumericPB{\raggedleft}\gnumbox[r]{\textsf{$4.71$}}}
\\
\hhline{|----|}
	 \multicolumn{1}{|p{\gnumericColB}|}%
	{\gnumericPB{\raggedleft}\gnumbox[r]{\textsf{$1.64$}}}
	&\multicolumn{1}{p{\gnumericColC}|}%
	{\gnumericPB{\raggedleft}\gnumbox[r]{\textsf{$3.82$}}}
	&\multicolumn{1}{p{\gnumericColD}|}%
	{\gnumericPB{\raggedleft}\gnumbox[r]{\textsf{$1.78$}}}
	&\multicolumn{1}{p{\gnumericColE}|}%
	{\gnumericPB{\raggedleft}\gnumbox[r]{\textsf{$4.79$}}}
\\
\hhline{|----|}
	 \multicolumn{1}{|p{\gnumericColB}|}%
	{\gnumericPB{\raggedleft}\gnumbox[r]{\textsf{$1.66$}}}
	&\multicolumn{1}{p{\gnumericColC}|}%
	{\gnumericPB{\raggedleft}\gnumbox[r]{\textsf{$3.94$}}}
	&\multicolumn{1}{p{\gnumericColD}|}%
	{\gnumericPB{\raggedleft}\gnumbox[r]{\textsf{$1.80$}}}
	&\multicolumn{1}{p{\gnumericColE}|}%
	{\gnumericPB{\raggedleft}\gnumbox[r]{\textsf{$4.87$}}}
\\
\hhline{|----|}
	 \multicolumn{1}{|p{\gnumericColB}|}%
	{\gnumericPB{\raggedleft}\gnumbox[r]{\textsf{$1.67$}}}
	&\multicolumn{1}{p{\gnumericColC}|}%
	{\gnumericPB{\raggedleft}\gnumbox[r]{\textsf{$4.05$}}}
	&\multicolumn{1}{p{\gnumericColD}|}%
	{\gnumericPB{\raggedleft}\gnumbox[r]{\textsf{$1.81$}}}
	&\multicolumn{1}{p{\gnumericColE}|}%
	{\gnumericPB{\raggedleft}\gnumbox[r]{\textsf{$4.95$}}}
\\
\hhline{|----|}
	 \multicolumn{1}{|p{\gnumericColB}|}%
	{\gnumericPB{\raggedleft}\gnumbox[r]{\textsf{$1.69$}}}
	&\multicolumn{1}{p{\gnumericColC}|}%
	{\gnumericPB{\raggedleft}\gnumbox[r]{\textsf{$4.16$}}}
	&\multicolumn{1}{p{\gnumericColD}|}%
	{\gnumericPB{\raggedleft}\gnumbox[r]{\textsf{$1.83$}}}
	&\multicolumn{1}{p{\gnumericColE}|}%
	{\gnumericPB{\raggedleft}\gnumbox[r]{\textsf{$5.02$}}}
\\
\hhline{|----|}
	 \multicolumn{1}{|p{\gnumericColB}|}%
	{\gnumericPB{\raggedleft}\gnumbox[r]{\textsf{$1.70$}}}
	&\multicolumn{1}{p{\gnumericColC}|}%
	{\gnumericPB{\raggedleft}\gnumbox[r]{\textsf{$4.26$}}}
	&\multicolumn{1}{p{\gnumericColD}|}%
	{\gnumericPB{\raggedleft}\gnumbox[r]{\textsf{$1.84$}}}
	&\multicolumn{1}{p{\gnumericColE}|}%
	{\gnumericPB{\raggedleft}\gnumbox[r]{\textsf{$5.09$}}}
\\
\hhline{|----|}
	 \multicolumn{1}{|p{\gnumericColB}|}%
	{\gnumericPB{\raggedleft}\gnumbox[r]{\textsf{$1.71$}}}
	&\multicolumn{1}{p{\gnumericColC}|}%
	{\gnumericPB{\raggedleft}\gnumbox[r]{\textsf{$4.36$}}}
	&\multicolumn{1}{p{\gnumericColD}|}%
	{\gnumericPB{\raggedleft}\gnumbox[r]{\textsf{$1.86$}}}
	&\multicolumn{1}{p{\gnumericColE}|}%
	{\gnumericPB{\raggedleft}\gnumbox[r]{\textsf{$5.17$}}}
\\
\hhline{|----|}
	 \multicolumn{1}{|p{\gnumericColB}|}%
	{\gnumericPB{\raggedleft}\gnumbox[r]{\textsf{$1.73$}}}
	&\multicolumn{1}{p{\gnumericColC}|}%
	{\gnumericPB{\raggedleft}\gnumbox[r]{\textsf{$4.45$}}}
	&\multicolumn{1}{p{\gnumericColD}|}%
	{\gnumericPB{\raggedleft}\gnumbox[r]{\textsf{$1.87$}}}
	&\multicolumn{1}{p{\gnumericColE}|}%
	{\gnumericPB{\raggedleft}\gnumbox[r]{\textsf{$5.24$}}}
\\
\hhline{|-|-|-|-|}
\end{longtable}

\gnumericTableEnd
}
\caption{Normalisation for a dipolar form for $J/\psi$ as a function of the $c$ quark mass
for $\Lambda=1.4$~GeV.}\label{tab:N_dip_jpsi_mq2}
\end{table}

\begin{table}[H]
{\tiny%%%%%%%%%%%%%%%%%%%%%%%%%%%%%%%%%%%%%%%%%%%%%%%%%%%%%%%%%%%%%%%%%%%%%%
\def\ifundefined#1{\expandafter\ifx\csname#1\endcsname\relax}

%%  Check for the \def token for inputed files. If it is not        %%
%%  defined, the file will be processed as a standalone and the     %%
%%  preamble will be used.                                          %%
\ifundefined{inputGnumericTable}

%%  End of the preamble for the standalone. The next section is for %%
%%  documents which are included into other LaTeX2e files.          %%
\else

%%  We are not a stand alone document. For a regular table, we will %%
%%  have no preamble and only define the closing to mean nothing.   %%
    \def\gnumericTableEnd{}

%%  If we want landscape mode in an embedded document, comment out  %%
%%  the line above and uncomment the two below. The table will      %%
%%  begin on a new page and run in landscape mode.                  %%
%       \def\gnumericTableEnd{\end{landscape}}
%       \begin{landscape}

%%  End of the else clause for this file being \input.              %%
\fi

%%%%%%%%%%%%%%%%%%%%%%%%%%%%%%%%%%%%%%%%%%%%%%%%%%%%%%%%%%%%%%%%%%%%%%
%%                                                                  %%
%%  The rest is the gnumeric table, except for the closing          %%
%%  statement. Changes below will alter the table's appearance.     %%
%%                                                                  %%
%%%%%%%%%%%%%%%%%%%%%%%%%%%%%%%%%%%%%%%%%%%%%%%%%%%%%%%%%%%%%%%%%%%%%%

\providecommand{\gnumericPB}[1]%
{\let\gnumericTemp=\\#1\let\\=\gnumericTemp\hspace{0pt}}

%%  The default table format retains the relative column widths of  %%
%%  gnumeric. They can easily be changed to c, r or l. In that case %%
%%  you may want to comment out the next line and uncomment the one %%
%%  thereafter                                                      %%
\providecommand\gnumbox{\makebox[0pt]}
%%\providecommand\gnumbox[1][]{\makebox}

%% to adjust positions in multirow situations                       %%
\setlength{\bigstrutjot}{\jot}
\setlength{\extrarowheight}{\doublerulesep}

%%  The \setlongtables command keeps column widths the same across %%
%%  pages. Simply comment out next line for varying column widths.  %%
\setlongtables

\setlength\gnumericTableWidth{%
	30pt+%
	30pt+%
	30pt+%
	30pt+%
0pt}
\def\gumericNumCols{4}
\setlength\gnumericTableWidthComplete{\gnumericTableWidth+\tabcolsep*\gumericNumCols*2+\arrayrulewidth*\gumericNumCols}
\ifthenelse{\lengthtest{\gnumericTableWidthComplete > \textwidth}}%
{\def\gnumericScale{\ratio{\textwidth-\tabcolsep*\gumericNumCols*2-\arrayrulewidth*\gumericNumCols}%
{\gnumericTableWidth}}}%
{\def\gnumericScale{1}}

%%%%%%%%%%%%%%%%%%%%%%%%%%%%%%%%%%%%%%%%%%%%%%%%%%%%%%%%%%%%%%%%%%%%%%
%%                                                                  %%
%% The following are the widths of the various columns. We are      %%
%% defining them here because then they are easier to change.       %%
%% Depending on the cell formats we may use them more than once.    %%
%%                                                                  %%
%%%%%%%%%%%%%%%%%%%%%%%%%%%%%%%%%%%%%%%%%%%%%%%%%%%%%%%%%%%%%%%%%%%%%%

\def\gnumericColB{30pt*\gnumericScale}
\def\gnumericColC{30pt*\gnumericScale}
\def\gnumericColD{30pt*\gnumericScale}
\def\gnumericColE{30pt*\gnumericScale}

\begin{longtable}[c]{%
	b{\gnumericColB}%
	b{\gnumericColC}%
	b{\gnumericColD}%
	b{\gnumericColE}%
	}

%%%%%%%%%%%%%%%%%%%%%%%%%%%%%%%%%%%%%%%%%%%%%%%%%%%%%%%%%%%%%%%%%%%%%%
%%  The longtable options. (Caption. headers... see Goosens. p.124) %%
%	\caption{The Table Caption.}             \\	%
% \hline	% Across the top of the table.
%%  The rest of these options are table rows which are placed on    %%
%%  the first. last or every page. Use \multicolumn if you want.    %%

%%  Header for the first page.                                      %%
%	\multicolumn{4}{c}{The First Header} \\ \hline 
%	\multicolumn{1}{c}{colTag}	%Column 1
%	&\multicolumn{1}{c}{colTag}	%Column 1
%	&\multicolumn{1}{c}{colTag}	%Column 2
%	&\multicolumn{1}{c}{colTag}	%Column 3
%	&\multicolumn{1}{c}{colTag}	\\ \hline %Last column
%	\endfirsthead

%%  The running header definition.                                  %%
%	\hline
%	\multicolumn{4}{l}{\ldots\small\slshape continued} \\ \hline
%	\multicolumn{1}{c}{colTag}	%Column 1
%	&\multicolumn{1}{c}{colTag}	%Column 1
%	&\multicolumn{1}{c}{colTag}	%Column 2
%	&\multicolumn{1}{c}{colTag}	%Column 3
%	&\multicolumn{1}{c}{colTag}	\\ \hline %Last column
%	\endhead

%%  The running footer definition.                                  %%
%	\hline
%	\multicolumn{4}{r}{\small\slshape continued\ldots} \\
%	\endfoot

%%  The ending footer definition.                                   %%
%	\multicolumn{4}{c}{That's all folks} \\ \hline 
%	\endlastfoot
%%%%%%%%%%%%%%%%%%%%%%%%%%%%%%%%%%%%%%%%%%%%%%%%%%%%%%%%%%%%%%%%%%%%%%

\hhline{|-|-|-|-|}
	 \multicolumn{1}{|p{\gnumericColB}|}%
	{\gnumericPB{\raggedleft}\gnumbox[r]{\textsf{$1.60$}}}
	&\multicolumn{1}{p{\gnumericColC}|}%
	{\gnumericPB{\raggedleft}\gnumbox[r]{\textsf{$2.25$}}}
	&\multicolumn{1}{p{\gnumericColD}|}%
	{\gnumericPB{\raggedleft}\gnumbox[r]{\textsf{$1.74$}}}
	&\multicolumn{1}{p{\gnumericColE}|}%
	{\gnumericPB{\raggedleft}\gnumbox[r]{\textsf{$2.82$}}}
\\
\hhline{|----|}
	 \multicolumn{1}{|p{\gnumericColB}|}%
	{\gnumericPB{\raggedleft}\gnumbox[r]{\textsf{$1.61$}}}
	&\multicolumn{1}{p{\gnumericColC}|}%
	{\gnumericPB{\raggedleft}\gnumbox[r]{\textsf{$2.33$}}}
	&\multicolumn{1}{p{\gnumericColD}|}%
	{\gnumericPB{\raggedleft}\gnumbox[r]{\textsf{$1.76$}}}
	&\multicolumn{1}{p{\gnumericColE}|}%
	{\gnumericPB{\raggedleft}\gnumbox[r]{\textsf{$2.86$}}}
\\
\hhline{|----|}
	 \multicolumn{1}{|p{\gnumericColB}|}%
	{\gnumericPB{\raggedleft}\gnumbox[r]{\textsf{$1.63$}}}
	&\multicolumn{1}{p{\gnumericColC}|}%
	{\gnumericPB{\raggedleft}\gnumbox[r]{\textsf{$2.40$}}}
	&\multicolumn{1}{p{\gnumericColD}|}%
	{\gnumericPB{\raggedleft}\gnumbox[r]{\textsf{$1.77$}}}
	&\multicolumn{1}{p{\gnumericColE}|}%
	{\gnumericPB{\raggedleft}\gnumbox[r]{\textsf{$2.90$}}}
\\
\hhline{|----|}
	 \multicolumn{1}{|p{\gnumericColB}|}%
	{\gnumericPB{\raggedleft}\gnumbox[r]{\textsf{$1.64$}}}
	&\multicolumn{1}{p{\gnumericColC}|}%
	{\gnumericPB{\raggedleft}\gnumbox[r]{\textsf{$2.47$}}}
	&\multicolumn{1}{p{\gnumericColD}|}%
	{\gnumericPB{\raggedleft}\gnumbox[r]{\textsf{$1.78$}}}
	&\multicolumn{1}{p{\gnumericColE}|}%
	{\gnumericPB{\raggedleft}\gnumbox[r]{\textsf{$2.94$}}}
\\
\hhline{|----|}
	 \multicolumn{1}{|p{\gnumericColB}|}%
	{\gnumericPB{\raggedleft}\gnumbox[r]{\textsf{$1.66$}}}
	&\multicolumn{1}{p{\gnumericColC}|}%
	{\gnumericPB{\raggedleft}\gnumbox[r]{\textsf{$2.53$}}}
	&\multicolumn{1}{p{\gnumericColD}|}%
	{\gnumericPB{\raggedleft}\gnumbox[r]{\textsf{$1.80$}}}
	&\multicolumn{1}{p{\gnumericColE}|}%
	{\gnumericPB{\raggedleft}\gnumbox[r]{\textsf{$2.98$}}}
\\
\hhline{|----|}
	 \multicolumn{1}{|p{\gnumericColB}|}%
	{\gnumericPB{\raggedleft}\gnumbox[r]{\textsf{$1.67$}}}
	&\multicolumn{1}{p{\gnumericColC}|}%
	{\gnumericPB{\raggedleft}\gnumbox[r]{\textsf{$2.58$}}}
	&\multicolumn{1}{p{\gnumericColD}|}%
	{\gnumericPB{\raggedleft}\gnumbox[r]{\textsf{$1.81$}}}
	&\multicolumn{1}{p{\gnumericColE}|}%
	{\gnumericPB{\raggedleft}\gnumbox[r]{\textsf{$3.02$}}}
\\
\hhline{|----|}
	 \multicolumn{1}{|p{\gnumericColB}|}%
	{\gnumericPB{\raggedleft}\gnumbox[r]{\textsf{$1.69$}}}
	&\multicolumn{1}{p{\gnumericColC}|}%
	{\gnumericPB{\raggedleft}\gnumbox[r]{\textsf{$2.63$}}}
	&\multicolumn{1}{p{\gnumericColD}|}%
	{\gnumericPB{\raggedleft}\gnumbox[r]{\textsf{$1.83$}}}
	&\multicolumn{1}{p{\gnumericColE}|}%
	{\gnumericPB{\raggedleft}\gnumbox[r]{\textsf{$3.05$}}}
\\
\hhline{|----|}
	 \multicolumn{1}{|p{\gnumericColB}|}%
	{\gnumericPB{\raggedleft}\gnumbox[r]{\textsf{$1.70$}}}
	&\multicolumn{1}{p{\gnumericColC}|}%
	{\gnumericPB{\raggedleft}\gnumbox[r]{\textsf{$2.68$}}}
	&\multicolumn{1}{p{\gnumericColD}|}%
	{\gnumericPB{\raggedleft}\gnumbox[r]{\textsf{$1.84$}}}
	&\multicolumn{1}{p{\gnumericColE}|}%
	{\gnumericPB{\raggedleft}\gnumbox[r]{\textsf{$3.09$}}}
\\
\hhline{|----|}
	 \multicolumn{1}{|p{\gnumericColB}|}%
	{\gnumericPB{\raggedleft}\gnumbox[r]{\textsf{$1.71$}}}
	&\multicolumn{1}{p{\gnumericColC}|}%
	{\gnumericPB{\raggedleft}\gnumbox[r]{\textsf{$2.73$}}}
	&\multicolumn{1}{p{\gnumericColD}|}%
	{\gnumericPB{\raggedleft}\gnumbox[r]{\textsf{$1.86$}}}
	&\multicolumn{1}{p{\gnumericColE}|}%
	{\gnumericPB{\raggedleft}\gnumbox[r]{\textsf{$3.12$}}}
\\
\hhline{|----|}
	 \multicolumn{1}{|p{\gnumericColB}|}%
	{\gnumericPB{\raggedleft}\gnumbox[r]{\textsf{$1.73$}}}
	&\multicolumn{1}{p{\gnumericColC}|}%
	{\gnumericPB{\raggedleft}\gnumbox[r]{\textsf{$2.78$}}}
	&\multicolumn{1}{p{\gnumericColD}|}%
	{\gnumericPB{\raggedleft}\gnumbox[r]{\textsf{$1.87$}}}
	&\multicolumn{1}{p{\gnumericColE}|}%
	{\gnumericPB{\raggedleft}\gnumbox[r]{\textsf{$3.16$}}}
\\
\hhline{|-|-|-|-|}
\end{longtable}

\gnumericTableEnd
}
\caption{Normalisation for a dipolar form for $J/\psi$ as a function of the $c$ quark mass
for $\Lambda=1.8$~GeV.}\label{tab:N_dip_jpsi_mq3}
\end{table}

\begin{table}[H]
{\tiny%%%%%%%%%%%%%%%%%%%%%%%%%%%%%%%%%%%%%%%%%%%%%%%%%%%%%%%%%%%%%%%%%%%%%%
\def\ifundefined#1{\expandafter\ifx\csname#1\endcsname\relax}

%%  Check for the \def token for inputed files. If it is not        %%
%%  defined, the file will be processed as a standalone and the     %%
%%  preamble will be used.                                          %%
\ifundefined{inputGnumericTable}

%%  End of the preamble for the standalone. The next section is for %%
%%  documents which are included into other LaTeX2e files.          %%
\else

%%  We are not a stand alone document. For a regular table, we will %%
%%  have no preamble and only define the closing to mean nothing.   %%
    \def\gnumericTableEnd{}

%%  If we want landscape mode in an embedded document, comment out  %%
%%  the line above and uncomment the two below. The table will      %%
%%  begin on a new page and run in landscape mode.                  %%
%       \def\gnumericTableEnd{\end{landscape}}
%       \begin{landscape}

%%  End of the else clause for this file being \input.              %%
\fi

%%%%%%%%%%%%%%%%%%%%%%%%%%%%%%%%%%%%%%%%%%%%%%%%%%%%%%%%%%%%%%%%%%%%%%
%%                                                                  %%
%%  The rest is the gnumeric table, except for the closing          %%
%%  statement. Changes below will alter the table's appearance.     %%
%%                                                                  %%
%%%%%%%%%%%%%%%%%%%%%%%%%%%%%%%%%%%%%%%%%%%%%%%%%%%%%%%%%%%%%%%%%%%%%%

\providecommand{\gnumericPB}[1]%
{\let\gnumericTemp=\\#1\let\\=\gnumericTemp\hspace{0pt}}

%%  The default table format retains the relative column widths of  %%
%%  gnumeric. They can easily be changed to c, r or l. In that case %%
%%  you may want to comment out the next line and uncomment the one %%
%%  thereafter                                                      %%
\providecommand\gnumbox{\makebox[0pt]}
%%\providecommand\gnumbox[1][]{\makebox}

%% to adjust positions in multirow situations                       %%
\setlength{\bigstrutjot}{\jot}
\setlength{\extrarowheight}{\doublerulesep}

%%  The \setlongtables command keeps column widths the same across %%
%%  pages. Simply comment out next line for varying column widths.  %%
\setlongtables

\setlength\gnumericTableWidth{%
	30pt+%
	30pt+%
	30pt+%
	30pt+%
0pt}
\def\gumericNumCols{4}
\setlength\gnumericTableWidthComplete{\gnumericTableWidth+\tabcolsep*\gumericNumCols*2+\arrayrulewidth*\gumericNumCols}
\ifthenelse{\lengthtest{\gnumericTableWidthComplete > \textwidth}}%
{\def\gnumericScale{\ratio{\textwidth-\tabcolsep*\gumericNumCols*2-\arrayrulewidth*\gumericNumCols}%
{\gnumericTableWidth}}}%
{\def\gnumericScale{1}}

%%%%%%%%%%%%%%%%%%%%%%%%%%%%%%%%%%%%%%%%%%%%%%%%%%%%%%%%%%%%%%%%%%%%%%
%%                                                                  %%
%% The following are the widths of the various columns. We are      %%
%% defining them here because then they are easier to change.       %%
%% Depending on the cell formats we may use them more than once.    %%
%%                                                                  %%
%%%%%%%%%%%%%%%%%%%%%%%%%%%%%%%%%%%%%%%%%%%%%%%%%%%%%%%%%%%%%%%%%%%%%%

\def\gnumericColB{30pt*\gnumericScale}
\def\gnumericColC{30pt*\gnumericScale}
\def\gnumericColD{30pt*\gnumericScale}
\def\gnumericColE{30pt*\gnumericScale}

\begin{longtable}[c]{%
	b{\gnumericColB}%
	b{\gnumericColC}%
	b{\gnumericColD}%
	b{\gnumericColE}%
	}

%%%%%%%%%%%%%%%%%%%%%%%%%%%%%%%%%%%%%%%%%%%%%%%%%%%%%%%%%%%%%%%%%%%%%%
%%  The longtable options. (Caption. headers... see Goosens. p.124) %%
%	\caption{The Table Caption.}             \\	%
% \hline	% Across the top of the table.
%%  The rest of these options are table rows which are placed on    %%
%%  the first. last or every page. Use \multicolumn if you want.    %%

%%  Header for the first page.                                      %%
%	\multicolumn{4}{c}{The First Header} \\ \hline 
%	\multicolumn{1}{c}{colTag}	%Column 1
%	&\multicolumn{1}{c}{colTag}	%Column 1
%	&\multicolumn{1}{c}{colTag}	%Column 2
%	&\multicolumn{1}{c}{colTag}	%Column 3
%	&\multicolumn{1}{c}{colTag}	\\ \hline %Last column
%	\endfirsthead

%%  The running header definition.                                  %%
%	\hline
%	\multicolumn{4}{l}{\ldots\small\slshape continued} \\ \hline
%	\multicolumn{1}{c}{colTag}	%Column 1
%	&\multicolumn{1}{c}{colTag}	%Column 1
%	&\multicolumn{1}{c}{colTag}	%Column 2
%	&\multicolumn{1}{c}{colTag}	%Column 3
%	&\multicolumn{1}{c}{colTag}	\\ \hline %Last column
%	\endhead

%%  The running footer definition.                                  %%
%	\hline
%	\multicolumn{4}{r}{\small\slshape continued\ldots} \\
%	\endfoot

%%  The ending footer definition.                                   %%
%	\multicolumn{4}{c}{That's all folks} \\ \hline 
%	\endlastfoot
%%%%%%%%%%%%%%%%%%%%%%%%%%%%%%%%%%%%%%%%%%%%%%%%%%%%%%%%%%%%%%%%%%%%%%

\hhline{|-|-|-|-|}
	 \multicolumn{1}{|p{\gnumericColB}|}%
	{\gnumericPB{\raggedleft}\gnumbox[r]{\textsf{$1.60$}}}
	&\multicolumn{1}{p{\gnumericColC}|}%
	{\gnumericPB{\raggedleft}\gnumbox[r]{\textsf{$1.62$}}}
	&\multicolumn{1}{p{\gnumericColD}|}%
	{\gnumericPB{\raggedleft}\gnumbox[r]{\textsf{$1.74$}}}
	&\multicolumn{1}{p{\gnumericColE}|}%
	{\gnumericPB{\raggedleft}\gnumbox[r]{\textsf{$1.94$}}}
\\
\hhline{|----|}
	 \multicolumn{1}{|p{\gnumericColB}|}%
	{\gnumericPB{\raggedleft}\gnumbox[r]{\textsf{$1.61$}}}
	&\multicolumn{1}{p{\gnumericColC}|}%
	{\gnumericPB{\raggedleft}\gnumbox[r]{\textsf{$1.67$}}}
	&\multicolumn{1}{p{\gnumericColD}|}%
	{\gnumericPB{\raggedleft}\gnumbox[r]{\textsf{$1.76$}}}
	&\multicolumn{1}{p{\gnumericColE}|}%
	{\gnumericPB{\raggedleft}\gnumbox[r]{\textsf{$1.97$}}}
\\
\hhline{|----|}
	 \multicolumn{1}{|p{\gnumericColB}|}%
	{\gnumericPB{\raggedleft}\gnumbox[r]{\textsf{$1.63$}}}
	&\multicolumn{1}{p{\gnumericColC}|}%
	{\gnumericPB{\raggedleft}\gnumbox[r]{\textsf{$1.71$}}}
	&\multicolumn{1}{p{\gnumericColD}|}%
	{\gnumericPB{\raggedleft}\gnumbox[r]{\textsf{$1.77$}}}
	&\multicolumn{1}{p{\gnumericColE}|}%
	{\gnumericPB{\raggedleft}\gnumbox[r]{\textsf{$1.99$}}}
\\
\hhline{|----|}
	 \multicolumn{1}{|p{\gnumericColB}|}%
	{\gnumericPB{\raggedleft}\gnumbox[r]{\textsf{$1.64$}}}
	&\multicolumn{1}{p{\gnumericColC}|}%
	{\gnumericPB{\raggedleft}\gnumbox[r]{\textsf{$1.75$}}}
	&\multicolumn{1}{p{\gnumericColD}|}%
	{\gnumericPB{\raggedleft}\gnumbox[r]{\textsf{$1.78$}}}
	&\multicolumn{1}{p{\gnumericColE}|}%
	{\gnumericPB{\raggedleft}\gnumbox[r]{\textsf{$2.01$}}}
\\
\hhline{|----|}
	 \multicolumn{1}{|p{\gnumericColB}|}%
	{\gnumericPB{\raggedleft}\gnumbox[r]{\textsf{$1.66$}}}
	&\multicolumn{1}{p{\gnumericColC}|}%
	{\gnumericPB{\raggedleft}\gnumbox[r]{\textsf{$1.78$}}}
	&\multicolumn{1}{p{\gnumericColD}|}%
	{\gnumericPB{\raggedleft}\gnumbox[r]{\textsf{$1.80$}}}
	&\multicolumn{1}{p{\gnumericColE}|}%
	{\gnumericPB{\raggedleft}\gnumbox[r]{\textsf{$2.03$}}}
\\
\hhline{|----|}
	 \multicolumn{1}{|p{\gnumericColB}|}%
	{\gnumericPB{\raggedleft}\gnumbox[r]{\textsf{$1.67$}}}
	&\multicolumn{1}{p{\gnumericColC}|}%
	{\gnumericPB{\raggedleft}\gnumbox[r]{\textsf{$1.81$}}}
	&\multicolumn{1}{p{\gnumericColD}|}%
	{\gnumericPB{\raggedleft}\gnumbox[r]{\textsf{$1.81$}}}
	&\multicolumn{1}{p{\gnumericColE}|}%
	{\gnumericPB{\raggedleft}\gnumbox[r]{\textsf{$2.05$}}}
\\
\hhline{|----|}
	 \multicolumn{1}{|p{\gnumericColB}|}%
	{\gnumericPB{\raggedleft}\gnumbox[r]{\textsf{$1.69$}}}
	&\multicolumn{1}{p{\gnumericColC}|}%
	{\gnumericPB{\raggedleft}\gnumbox[r]{\textsf{$1.84$}}}
	&\multicolumn{1}{p{\gnumericColD}|}%
	{\gnumericPB{\raggedleft}\gnumbox[r]{\textsf{$1.83$}}}
	&\multicolumn{1}{p{\gnumericColE}|}%
	{\gnumericPB{\raggedleft}\gnumbox[r]{\textsf{$2.07$}}}
\\
\hhline{|----|}
	 \multicolumn{1}{|p{\gnumericColB}|}%
	{\gnumericPB{\raggedleft}\gnumbox[r]{\textsf{$1.70$}}}
	&\multicolumn{1}{p{\gnumericColC}|}%
	{\gnumericPB{\raggedleft}\gnumbox[r]{\textsf{$1.87$}}}
	&\multicolumn{1}{p{\gnumericColD}|}%
	{\gnumericPB{\raggedleft}\gnumbox[r]{\textsf{$1.84$}}}
	&\multicolumn{1}{p{\gnumericColE}|}%
	{\gnumericPB{\raggedleft}\gnumbox[r]{\textsf{$2.09$}}}
\\
\hhline{|----|}
	 \multicolumn{1}{|p{\gnumericColB}|}%
	{\gnumericPB{\raggedleft}\gnumbox[r]{\textsf{$1.71$}}}
	&\multicolumn{1}{p{\gnumericColC}|}%
	{\gnumericPB{\raggedleft}\gnumbox[r]{\textsf{$1.90$}}}
	&\multicolumn{1}{p{\gnumericColD}|}%
	{\gnumericPB{\raggedleft}\gnumbox[r]{\textsf{$1.86$}}}
	&\multicolumn{1}{p{\gnumericColE}|}%
	{\gnumericPB{\raggedleft}\gnumbox[r]{\textsf{$2.11$}}}
\\
\hhline{|----|}
	 \multicolumn{1}{|p{\gnumericColB}|}%
	{\gnumericPB{\raggedleft}\gnumbox[r]{\textsf{$1.73$}}}
	&\multicolumn{1}{p{\gnumericColC}|}%
	{\gnumericPB{\raggedleft}\gnumbox[r]{\textsf{$1.92$}}}
	&\multicolumn{1}{p{\gnumericColD}|}%
	{\gnumericPB{\raggedleft}\gnumbox[r]{\textsf{$1.87$}}}
	&\multicolumn{1}{p{\gnumericColE}|}%
	{\gnumericPB{\raggedleft}\gnumbox[r]{\textsf{$2.13$}}}
\\
\hhline{|-|-|-|-|}
\end{longtable}

\gnumericTableEnd
}
\caption{Normalisation for a dipolar form for $J/\psi$ as a function of  the $c$ quark mass
for $\Lambda=2.2$~GeV.}\label{tab:N_dip_jpsi_mq4}
\end{table}

\begin{table}[H]
{\tiny%%%%%%%%%%%%%%%%%%%%%%%%%%%%%%%%%%%%%%%%%%%%%%%%%%%%%%%%%%%%%%%%%%%%%%
\def\ifundefined#1{\expandafter\ifx\csname#1\endcsname\relax}

%%  Check for the \def token for inputed files. If it is not        %%
%%  defined, the file will be processed as a standalone and the     %%
%%  preamble will be used.                                          %%
\ifundefined{inputGnumericTable}

%%  End of the preamble for the standalone. The next section is for %%
%%  documents which are included into other LaTeX2e files.          %%
\else

%%  We are not a stand alone document. For a regular table, we will %%
%%  have no preamble and only define the closing to mean nothing.   %%
    \def\gnumericTableEnd{}

%%  If we want landscape mode in an embedded document, comment out  %%
%%  the line above and uncomment the two below. The table will      %%
%%  begin on a new page and run in landscape mode.                  %%
%       \def\gnumericTableEnd{\end{landscape}}
%       \begin{landscape}

%%  End of the else clause for this file being \input.              %%
\fi

%%%%%%%%%%%%%%%%%%%%%%%%%%%%%%%%%%%%%%%%%%%%%%%%%%%%%%%%%%%%%%%%%%%%%%
%%                                                                  %%
%%  The rest is the gnumeric table, except for the closing          %%
%%  statement. Changes below will alter the table's appearance.     %%
%%                                                                  %%
%%%%%%%%%%%%%%%%%%%%%%%%%%%%%%%%%%%%%%%%%%%%%%%%%%%%%%%%%%%%%%%%%%%%%%

\providecommand{\gnumericPB}[1]%
{\let\gnumericTemp=\\#1\let\\=\gnumericTemp\hspace{0pt}}

%%  The default table format retains the relative column widths of  %%
%%  gnumeric. They can easily be changed to c, r or l. In that case %%
%%  you may want to comment out the next line and uncomment the one %%
%%  thereafter                                                      %%
\providecommand\gnumbox{\makebox[0pt]}
%%\providecommand\gnumbox[1][]{\makebox}

%% to adjust positions in multirow situations                       %%
\setlength{\bigstrutjot}{\jot}
\setlength{\extrarowheight}{\doublerulesep}

%%  The \setlongtables command keeps column widths the same across %%
%%  pages. Simply comment out next line for varying column widths.  %%
\setlongtables

\setlength\gnumericTableWidth{%
	30pt+%
	30pt+%
	30pt+%
	30pt+%
0pt}
\def\gumericNumCols{4}
\setlength\gnumericTableWidthComplete{\gnumericTableWidth+\tabcolsep*\gumericNumCols*2+\arrayrulewidth*\gumericNumCols}
\ifthenelse{\lengthtest{\gnumericTableWidthComplete > \textwidth}}%
{\def\gnumericScale{\ratio{\textwidth-\tabcolsep*\gumericNumCols*2-\arrayrulewidth*\gumericNumCols}%
{\gnumericTableWidth}}}%
{\def\gnumericScale{1}}

%%%%%%%%%%%%%%%%%%%%%%%%%%%%%%%%%%%%%%%%%%%%%%%%%%%%%%%%%%%%%%%%%%%%%%
%%                                                                  %%
%% The following are the widths of the various columns. We are      %%
%% defining them here because then they are easier to change.       %%
%% Depending on the cell formats we may use them more than once.    %%
%%                                                                  %%
%%%%%%%%%%%%%%%%%%%%%%%%%%%%%%%%%%%%%%%%%%%%%%%%%%%%%%%%%%%%%%%%%%%%%%

\def\gnumericColB{30pt*\gnumericScale}
\def\gnumericColC{30pt*\gnumericScale}
\def\gnumericColD{30pt*\gnumericScale}
\def\gnumericColE{30pt*\gnumericScale}

\begin{longtable}[c]{%
	b{\gnumericColB}%
	b{\gnumericColC}%
	b{\gnumericColD}%
	b{\gnumericColE}%
	}

%%%%%%%%%%%%%%%%%%%%%%%%%%%%%%%%%%%%%%%%%%%%%%%%%%%%%%%%%%%%%%%%%%%%%%
%%  The longtable options. (Caption. headers... see Goosens. p.124) %%
%	\caption{The Table Caption.}             \\	%
% \hline	% Across the top of the table.
%%  The rest of these options are table rows which are placed on    %%
%%  the first. last or every page. Use \multicolumn if you want.    %%

%%  Header for the first page.                                      %%
%	\multicolumn{4}{c}{The First Header} \\ \hline 
%	\multicolumn{1}{c}{colTag}	%Column 1
%	&\multicolumn{1}{c}{colTag}	%Column 1
%	&\multicolumn{1}{c}{colTag}	%Column 2
%	&\multicolumn{1}{c}{colTag}	%Column 3
%	&\multicolumn{1}{c}{colTag}	\\ \hline %Last column
%	\endfirsthead

%%  The running header definition.                                  %%
%	\hline
%	\multicolumn{4}{l}{\ldots\small\slshape continued} \\ \hline
%	\multicolumn{1}{c}{colTag}	%Column 1
%	&\multicolumn{1}{c}{colTag}	%Column 1
%	&\multicolumn{1}{c}{colTag}	%Column 2
%	&\multicolumn{1}{c}{colTag}	%Column 3
%	&\multicolumn{1}{c}{colTag}	\\ \hline %Last column
%	\endhead

%%  The running footer definition.                                  %%
%	\hline
%	\multicolumn{4}{r}{\small\slshape continued\ldots} \\
%	\endfoot

%%  The ending footer definition.                                   %%
%	\multicolumn{4}{c}{That's all folks} \\ \hline 
%	\endlastfoot
%%%%%%%%%%%%%%%%%%%%%%%%%%%%%%%%%%%%%%%%%%%%%%%%%%%%%%%%%%%%%%%%%%%%%%

\hhline{|-|-|-|-|}
	 \multicolumn{1}{|p{\gnumericColB}|}%
	{\gnumericPB{\raggedleft}\gnumbox[r]{\textsf{$\Lambda$}}}
	&\multicolumn{1}{p{\gnumericColC}|}%
	{\gnumericPB{\raggedleft}\gnumbox[r]{\textsf{$N$}}}
	&\multicolumn{1}{p{\gnumericColD}|}%
	{\gnumericPB{\raggedleft}\gnumbox[r]{\textsf{$\Lambda$}}}
	&\multicolumn{1}{p{\gnumericColE}|}%
	{\gnumericPB{\raggedleft}\gnumbox[r]{\textsf{$N$}}}
\\
\hhline{|----|}
	 \multicolumn{1}{|p{\gnumericColB}|}%
	{\gnumericPB{\raggedleft}\gnumbox[r]{\textsf{$1.00$}}}
	&\multicolumn{1}{p{\gnumericColC}|}%
	{\gnumericPB{\raggedleft}\gnumbox[r]{\textsf{$5.36$}}}
	&\multicolumn{1}{p{\gnumericColD}|}%
	{\gnumericPB{\raggedleft}\gnumbox[r]{\textsf{$1.63$}}}
	&\multicolumn{1}{p{\gnumericColE}|}%
	{\gnumericPB{\raggedleft}\gnumbox[r]{\textsf{$2.43$}}}
\\
\hhline{|----|}
	 \multicolumn{1}{|p{\gnumericColB}|}%
	{\gnumericPB{\raggedleft}\gnumbox[r]{\textsf{$1.06$}}}
	&\multicolumn{1}{p{\gnumericColC}|}%
	{\gnumericPB{\raggedleft}\gnumbox[r]{\textsf{$4.84$}}}
	&\multicolumn{1}{p{\gnumericColD}|}%
	{\gnumericPB{\raggedleft}\gnumbox[r]{\textsf{$1.69$}}}
	&\multicolumn{1}{p{\gnumericColE}|}%
	{\gnumericPB{\raggedleft}\gnumbox[r]{\textsf{$2.29$}}}
\\
\hhline{|----|}
	 \multicolumn{1}{|p{\gnumericColB}|}%
	{\gnumericPB{\raggedleft}\gnumbox[r]{\textsf{$1.13$}}}
	&\multicolumn{1}{p{\gnumericColC}|}%
	{\gnumericPB{\raggedleft}\gnumbox[r]{\textsf{$4.40$}}}
	&\multicolumn{1}{p{\gnumericColD}|}%
	{\gnumericPB{\raggedleft}\gnumbox[r]{\textsf{$1.76$}}}
	&\multicolumn{1}{p{\gnumericColE}|}%
	{\gnumericPB{\raggedleft}\gnumbox[r]{\textsf{$2.16$}}}
\\
\hhline{|----|}
	 \multicolumn{1}{|p{\gnumericColB}|}%
	{\gnumericPB{\raggedleft}\gnumbox[r]{\textsf{$1.19$}}}
	&\multicolumn{1}{p{\gnumericColC}|}%
	{\gnumericPB{\raggedleft}\gnumbox[r]{\textsf{$4.02$}}}
	&\multicolumn{1}{p{\gnumericColD}|}%
	{\gnumericPB{\raggedleft}\gnumbox[r]{\textsf{$1.82$}}}
	&\multicolumn{1}{p{\gnumericColE}|}%
	{\gnumericPB{\raggedleft}\gnumbox[r]{\textsf{$2.04$}}}
\\
\hhline{|----|}
	 \multicolumn{1}{|p{\gnumericColB}|}%
	{\gnumericPB{\raggedleft}\gnumbox[r]{\textsf{$1.25$}}}
	&\multicolumn{1}{p{\gnumericColC}|}%
	{\gnumericPB{\raggedleft}\gnumbox[r]{\textsf{$3.70$}}}
	&\multicolumn{1}{p{\gnumericColD}|}%
	{\gnumericPB{\raggedleft}\gnumbox[r]{\textsf{$1.88$}}}
	&\multicolumn{1}{p{\gnumericColE}|}%
	{\gnumericPB{\raggedleft}\gnumbox[r]{\textsf{$1.94$}}}
\\
\hhline{|----|}
	 \multicolumn{1}{|p{\gnumericColB}|}%
	{\gnumericPB{\raggedleft}\gnumbox[r]{\textsf{$1.32$}}}
	&\multicolumn{1}{p{\gnumericColC}|}%
	{\gnumericPB{\raggedleft}\gnumbox[r]{\textsf{$3.42$}}}
	&\multicolumn{1}{p{\gnumericColD}|}%
	{\gnumericPB{\raggedleft}\gnumbox[r]{\textsf{$1.95$}}}
	&\multicolumn{1}{p{\gnumericColE}|}%
	{\gnumericPB{\raggedleft}\gnumbox[r]{\textsf{$1.84$}}}
\\
\hhline{|----|}
	 \multicolumn{1}{|p{\gnumericColB}|}%
	{\gnumericPB{\raggedleft}\gnumbox[r]{\textsf{$1.38$}}}
	&\multicolumn{1}{p{\gnumericColC}|}%
	{\gnumericPB{\raggedleft}\gnumbox[r]{\textsf{$3.17$}}}
	&\multicolumn{1}{p{\gnumericColD}|}%
	{\gnumericPB{\raggedleft}\gnumbox[r]{\textsf{$2.01$}}}
	&\multicolumn{1}{p{\gnumericColE}|}%
	{\gnumericPB{\raggedleft}\gnumbox[r]{\textsf{$1.75$}}}
\\
\hhline{|----|}
	 \multicolumn{1}{|p{\gnumericColB}|}%
	{\gnumericPB{\raggedleft}\gnumbox[r]{\textsf{$1.44$}}}
	&\multicolumn{1}{p{\gnumericColC}|}%
	{\gnumericPB{\raggedleft}\gnumbox[r]{\textsf{$2.95$}}}
	&\multicolumn{1}{p{\gnumericColD}|}%
	{\gnumericPB{\raggedleft}\gnumbox[r]{\textsf{$2.07$}}}
	&\multicolumn{1}{p{\gnumericColE}|}%
	{\gnumericPB{\raggedleft}\gnumbox[r]{\textsf{$1.67$}}}
\\
\hhline{|----|}
	 \multicolumn{1}{|p{\gnumericColB}|}%
	{\gnumericPB{\raggedleft}\gnumbox[r]{\textsf{$1.51$}}}
	&\multicolumn{1}{p{\gnumericColC}|}%
	{\gnumericPB{\raggedleft}\gnumbox[r]{\textsf{$2.76$}}}
	&\multicolumn{1}{p{\gnumericColD}|}%
	{\gnumericPB{\raggedleft}\gnumbox[r]{\textsf{$2.14$}}}
	&\multicolumn{1}{p{\gnumericColE}|}%
	{\gnumericPB{\raggedleft}\gnumbox[r]{\textsf{$1.59$}}}
\\
\hhline{|----|}
	 \multicolumn{1}{|p{\gnumericColB}|}%
	{\gnumericPB{\raggedleft}\gnumbox[r]{\textsf{$1.57$}}}
	&\multicolumn{1}{p{\gnumericColC}|}%
	{\gnumericPB{\raggedleft}\gnumbox[r]{\textsf{$2.58$}}}
	&\multicolumn{1}{p{\gnumericColD}|}%
	{\gnumericPB{\raggedleft}\gnumbox[r]{\textsf{$2.20$}}}
	&\multicolumn{1}{p{\gnumericColE}|}%
	{\gnumericPB{\raggedleft}\gnumbox[r]{\textsf{$1.52$}}}
\\
\hhline{|-|-|-|-|}
\end{longtable}

\gnumericTableEnd
}
\caption{Normalisation for a gaussian form for $J/\psi$ as a function of $\Lambda$ for $m_c=1.60$~GeV.}\label{tab:N_exp_jpsi_lam1}
\end{table}

\begin{table}[H]
{\tiny%%%%%%%%%%%%%%%%%%%%%%%%%%%%%%%%%%%%%%%%%%%%%%%%%%%%%%%%%%%%%%%%%%%%%%
\def\ifundefined#1{\expandafter\ifx\csname#1\endcsname\relax}

%%  Check for the \def token for inputed files. If it is not        %%
%%  defined, the file will be processed as a standalone and the     %%
%%  preamble will be used.                                          %%
\ifundefined{inputGnumericTable}

%%  End of the preamble for the standalone. The next section is for %%
%%  documents which are included into other LaTeX2e files.          %%
\else

%%  We are not a stand alone document. For a regular table, we will %%
%%  have no preamble and only define the closing to mean nothing.   %%
    \def\gnumericTableEnd{}

%%  If we want landscape mode in an embedded document, comment out  %%
%%  the line above and uncomment the two below. The table will      %%
%%  begin on a new page and run in landscape mode.                  %%
%       \def\gnumericTableEnd{\end{landscape}}
%       \begin{landscape}

%%  End of the else clause for this file being \input.              %%
\fi

%%%%%%%%%%%%%%%%%%%%%%%%%%%%%%%%%%%%%%%%%%%%%%%%%%%%%%%%%%%%%%%%%%%%%%
%%                                                                  %%
%%  The rest is the gnumeric table, except for the closing          %%
%%  statement. Changes below will alter the table's appearance.     %%
%%                                                                  %%
%%%%%%%%%%%%%%%%%%%%%%%%%%%%%%%%%%%%%%%%%%%%%%%%%%%%%%%%%%%%%%%%%%%%%%

\providecommand{\gnumericPB}[1]%
{\let\gnumericTemp=\\#1\let\\=\gnumericTemp\hspace{0pt}}

%%  The default table format retains the relative column widths of  %%
%%  gnumeric. They can easily be changed to c, r or l. In that case %%
%%  you may want to comment out the next line and uncomment the one %%
%%  thereafter                                                      %%
\providecommand\gnumbox{\makebox[0pt]}
%%\providecommand\gnumbox[1][]{\makebox}

%% to adjust positions in multirow situations                       %%
\setlength{\bigstrutjot}{\jot}
\setlength{\extrarowheight}{\doublerulesep}

%%  The \setlongtables command keeps column widths the same across %%
%%  pages. Simply comment out next line for varying column widths.  %%
\setlongtables

\setlength\gnumericTableWidth{%
	30pt+%
	30pt+%
	30pt+%
	30pt+%
0pt}
\def\gumericNumCols{4}
\setlength\gnumericTableWidthComplete{\gnumericTableWidth+\tabcolsep*\gumericNumCols*2+\arrayrulewidth*\gumericNumCols}
\ifthenelse{\lengthtest{\gnumericTableWidthComplete > \textwidth}}%
{\def\gnumericScale{\ratio{\textwidth-\tabcolsep*\gumericNumCols*2-\arrayrulewidth*\gumericNumCols}%
{\gnumericTableWidth}}}%
{\def\gnumericScale{1}}

%%%%%%%%%%%%%%%%%%%%%%%%%%%%%%%%%%%%%%%%%%%%%%%%%%%%%%%%%%%%%%%%%%%%%%
%%                                                                  %%
%% The following are the widths of the various columns. We are      %%
%% defining them here because then they are easier to change.       %%
%% Depending on the cell formats we may use them more than once.    %%
%%                                                                  %%
%%%%%%%%%%%%%%%%%%%%%%%%%%%%%%%%%%%%%%%%%%%%%%%%%%%%%%%%%%%%%%%%%%%%%%

\def\gnumericColB{30pt*\gnumericScale}
\def\gnumericColC{30pt*\gnumericScale}
\def\gnumericColD{30pt*\gnumericScale}
\def\gnumericColE{30pt*\gnumericScale}

\begin{longtable}[c]{%
	b{\gnumericColB}%
	b{\gnumericColC}%
	b{\gnumericColD}%
	b{\gnumericColE}%
	}

%%%%%%%%%%%%%%%%%%%%%%%%%%%%%%%%%%%%%%%%%%%%%%%%%%%%%%%%%%%%%%%%%%%%%%
%%  The longtable options. (Caption. headers... see Goosens. p.124) %%
%	\caption{The Table Caption.}             \\	%
% \hline	% Across the top of the table.
%%  The rest of these options are table rows which are placed on    %%
%%  the first. last or every page. Use \multicolumn if you want.    %%

%%  Header for the first page.                                      %%
%	\multicolumn{4}{c}{The First Header} \\ \hline 
%	\multicolumn{1}{c}{colTag}	%Column 1
%	&\multicolumn{1}{c}{colTag}	%Column 1
%	&\multicolumn{1}{c}{colTag}	%Column 2
%	&\multicolumn{1}{c}{colTag}	%Column 3
%	&\multicolumn{1}{c}{colTag}	\\ \hline %Last column
%	\endfirsthead

%%  The running header definition.                                  %%
%	\hline
%	\multicolumn{4}{l}{\ldots\small\slshape continued} \\ \hline
%	\multicolumn{1}{c}{colTag}	%Column 1
%	&\multicolumn{1}{c}{colTag}	%Column 1
%	&\multicolumn{1}{c}{colTag}	%Column 2
%	&\multicolumn{1}{c}{colTag}	%Column 3
%	&\multicolumn{1}{c}{colTag}	\\ \hline %Last column
%	\endhead

%%  The running footer definition.                                  %%
%	\hline
%	\multicolumn{4}{r}{\small\slshape continued\ldots} \\
%	\endfoot

%%  The ending footer definition.                                   %%
%	\multicolumn{4}{c}{That's all folks} \\ \hline 
%	\endlastfoot
%%%%%%%%%%%%%%%%%%%%%%%%%%%%%%%%%%%%%%%%%%%%%%%%%%%%%%%%%%%%%%%%%%%%%%

\hhline{|-|-|-|-|}
	 \multicolumn{1}{|p{\gnumericColB}|}%
	{\gnumericPB{\raggedleft}\gnumbox[r]{\textsf{$1.00$}}}
	&\multicolumn{1}{p{\gnumericColC}|}%
	{\gnumericPB{\raggedleft}\gnumbox[r]{\textsf{$7.38$}}}
	&\multicolumn{1}{p{\gnumericColD}|}%
	{\gnumericPB{\raggedleft}\gnumbox[r]{\textsf{$1.63$}}}
	&\multicolumn{1}{p{\gnumericColE}|}%
	{\gnumericPB{\raggedleft}\gnumbox[r]{\textsf{$2.95$}}}
\\
\hhline{|----|}
	 \multicolumn{1}{|p{\gnumericColB}|}%
	{\gnumericPB{\raggedleft}\gnumbox[r]{\textsf{$1.06$}}}
	&\multicolumn{1}{p{\gnumericColC}|}%
	{\gnumericPB{\raggedleft}\gnumbox[r]{\textsf{$6.54$}}}
	&\multicolumn{1}{p{\gnumericColD}|}%
	{\gnumericPB{\raggedleft}\gnumbox[r]{\textsf{$1.69$}}}
	&\multicolumn{1}{p{\gnumericColE}|}%
	{\gnumericPB{\raggedleft}\gnumbox[r]{\textsf{$2.76$}}}
\\
\hhline{|----|}
	 \multicolumn{1}{|p{\gnumericColB}|}%
	{\gnumericPB{\raggedleft}\gnumbox[r]{\textsf{$1.13$}}}
	&\multicolumn{1}{p{\gnumericColC}|}%
	{\gnumericPB{\raggedleft}\gnumbox[r]{\textsf{$5.86$}}}
	&\multicolumn{1}{p{\gnumericColD}|}%
	{\gnumericPB{\raggedleft}\gnumbox[r]{\textsf{$1.76$}}}
	&\multicolumn{1}{p{\gnumericColE}|}%
	{\gnumericPB{\raggedleft}\gnumbox[r]{\textsf{$2.58$}}}
\\
\hhline{|----|}
	 \multicolumn{1}{|p{\gnumericColB}|}%
	{\gnumericPB{\raggedleft}\gnumbox[r]{\textsf{$1.19$}}}
	&\multicolumn{1}{p{\gnumericColC}|}%
	{\gnumericPB{\raggedleft}\gnumbox[r]{\textsf{$5.28$}}}
	&\multicolumn{1}{p{\gnumericColD}|}%
	{\gnumericPB{\raggedleft}\gnumbox[r]{\textsf{$1.82$}}}
	&\multicolumn{1}{p{\gnumericColE}|}%
	{\gnumericPB{\raggedleft}\gnumbox[r]{\textsf{$2.43$}}}
\\
\hhline{|----|}
	 \multicolumn{1}{|p{\gnumericColB}|}%
	{\gnumericPB{\raggedleft}\gnumbox[r]{\textsf{$1.25$}}}
	&\multicolumn{1}{p{\gnumericColC}|}%
	{\gnumericPB{\raggedleft}\gnumbox[r]{\textsf{$4.79$}}}
	&\multicolumn{1}{p{\gnumericColD}|}%
	{\gnumericPB{\raggedleft}\gnumbox[r]{\textsf{$1.88$}}}
	&\multicolumn{1}{p{\gnumericColE}|}%
	{\gnumericPB{\raggedleft}\gnumbox[r]{\textsf{$2.29$}}}
\\
\hhline{|----|}
	 \multicolumn{1}{|p{\gnumericColB}|}%
	{\gnumericPB{\raggedleft}\gnumbox[r]{\textsf{$1.32$}}}
	&\multicolumn{1}{p{\gnumericColC}|}%
	{\gnumericPB{\raggedleft}\gnumbox[r]{\textsf{$4.37$}}}
	&\multicolumn{1}{p{\gnumericColD}|}%
	{\gnumericPB{\raggedleft}\gnumbox[r]{\textsf{$1.95$}}}
	&\multicolumn{1}{p{\gnumericColE}|}%
	{\gnumericPB{\raggedleft}\gnumbox[r]{\textsf{$2.16$}}}
\\
\hhline{|----|}
	 \multicolumn{1}{|p{\gnumericColB}|}%
	{\gnumericPB{\raggedleft}\gnumbox[r]{\textsf{$1.38$}}}
	&\multicolumn{1}{p{\gnumericColC}|}%
	{\gnumericPB{\raggedleft}\gnumbox[r]{\textsf{$4.01$}}}
	&\multicolumn{1}{p{\gnumericColD}|}%
	{\gnumericPB{\raggedleft}\gnumbox[r]{\textsf{$2.01$}}}
	&\multicolumn{1}{p{\gnumericColE}|}%
	{\gnumericPB{\raggedleft}\gnumbox[r]{\textsf{$2.04$}}}
\\
\hhline{|----|}
	 \multicolumn{1}{|p{\gnumericColB}|}%
	{\gnumericPB{\raggedleft}\gnumbox[r]{\textsf{$1.44$}}}
	&\multicolumn{1}{p{\gnumericColC}|}%
	{\gnumericPB{\raggedleft}\gnumbox[r]{\textsf{$3.69$}}}
	&\multicolumn{1}{p{\gnumericColD}|}%
	{\gnumericPB{\raggedleft}\gnumbox[r]{\textsf{$2.07$}}}
	&\multicolumn{1}{p{\gnumericColE}|}%
	{\gnumericPB{\raggedleft}\gnumbox[r]{\textsf{$1.93$}}}
\\
\hhline{|----|}
	 \multicolumn{1}{|p{\gnumericColB}|}%
	{\gnumericPB{\raggedleft}\gnumbox[r]{\textsf{$1.51$}}}
	&\multicolumn{1}{p{\gnumericColC}|}%
	{\gnumericPB{\raggedleft}\gnumbox[r]{\textsf{$3.41$}}}
	&\multicolumn{1}{p{\gnumericColD}|}%
	{\gnumericPB{\raggedleft}\gnumbox[r]{\textsf{$2.14$}}}
	&\multicolumn{1}{p{\gnumericColE}|}%
	{\gnumericPB{\raggedleft}\gnumbox[r]{\textsf{$1.83$}}}
\\
\hhline{|----|}
	 \multicolumn{1}{|p{\gnumericColB}|}%
	{\gnumericPB{\raggedleft}\gnumbox[r]{\textsf{$1.57$}}}
	&\multicolumn{1}{p{\gnumericColC}|}%
	{\gnumericPB{\raggedleft}\gnumbox[r]{\textsf{$3.17$}}}
	&\multicolumn{1}{p{\gnumericColD}|}%
	{\gnumericPB{\raggedleft}\gnumbox[r]{\textsf{$2.20$}}}
	&\multicolumn{1}{p{\gnumericColE}|}%
	{\gnumericPB{\raggedleft}\gnumbox[r]{\textsf{$1.74$}}}
\\
\hhline{|-|-|-|-|}
\end{longtable}

\gnumericTableEnd
}
\caption{Normalisation for a gaussian form for $J/\psi$ as a function of $\Lambda$ for $m_c=1.69$~GeV.}\label{tab:N_exp_jpsi_lam2}
\end{table}

\begin{table}[H]
{\tiny%%%%%%%%%%%%%%%%%%%%%%%%%%%%%%%%%%%%%%%%%%%%%%%%%%%%%%%%%%%%%%%%%%%%%%
\def\ifundefined#1{\expandafter\ifx\csname#1\endcsname\relax}

%%  Check for the \def token for inputed files. If it is not        %%
%%  defined, the file will be processed as a standalone and the     %%
%%  preamble will be used.                                          %%
\ifundefined{inputGnumericTable}

%%  End of the preamble for the standalone. The next section is for %%
%%  documents which are included into other LaTeX2e files.          %%
\else

%%  We are not a stand alone document. For a regular table, we will %%
%%  have no preamble and only define the closing to mean nothing.   %%
    \def\gnumericTableEnd{}

%%  If we want landscape mode in an embedded document, comment out  %%
%%  the line above and uncomment the two below. The table will      %%
%%  begin on a new page and run in landscape mode.                  %%
%       \def\gnumericTableEnd{\end{landscape}}
%       \begin{landscape}

%%  End of the else clause for this file being \input.              %%
\fi

%%%%%%%%%%%%%%%%%%%%%%%%%%%%%%%%%%%%%%%%%%%%%%%%%%%%%%%%%%%%%%%%%%%%%%
%%                                                                  %%
%%  The rest is the gnumeric table, except for the closing          %%
%%  statement. Changes below will alter the table's appearance.     %%
%%                                                                  %%
%%%%%%%%%%%%%%%%%%%%%%%%%%%%%%%%%%%%%%%%%%%%%%%%%%%%%%%%%%%%%%%%%%%%%%

\providecommand{\gnumericPB}[1]%
{\let\gnumericTemp=\\#1\let\\=\gnumericTemp\hspace{0pt}}

%%  The default table format retains the relative column widths of  %%
%%  gnumeric. They can easily be changed to c, r or l. In that case %%
%%  you may want to comment out the next line and uncomment the one %%
%%  thereafter                                                      %%
\providecommand\gnumbox{\makebox[0pt]}
%%\providecommand\gnumbox[1][]{\makebox}

%% to adjust positions in multirow situations                       %%
\setlength{\bigstrutjot}{\jot}
\setlength{\extrarowheight}{\doublerulesep}

%%  The \setlongtables command keeps column widths the same across %%
%%  pages. Simply comment out next line for varying column widths.  %%
\setlongtables

\setlength\gnumericTableWidth{%
	30pt+%
	30pt+%
	30pt+%
	30pt+%
0pt}
\def\gumericNumCols{4}
\setlength\gnumericTableWidthComplete{\gnumericTableWidth+\tabcolsep*\gumericNumCols*2+\arrayrulewidth*\gumericNumCols}
\ifthenelse{\lengthtest{\gnumericTableWidthComplete > \textwidth}}%
{\def\gnumericScale{\ratio{\textwidth-\tabcolsep*\gumericNumCols*2-\arrayrulewidth*\gumericNumCols}%
{\gnumericTableWidth}}}%
{\def\gnumericScale{1}}

%%%%%%%%%%%%%%%%%%%%%%%%%%%%%%%%%%%%%%%%%%%%%%%%%%%%%%%%%%%%%%%%%%%%%%
%%                                                                  %%
%% The following are the widths of the various columns. We are      %%
%% defining them here because then they are easier to change.       %%
%% Depending on the cell formats we may use them more than once.    %%
%%                                                                  %%
%%%%%%%%%%%%%%%%%%%%%%%%%%%%%%%%%%%%%%%%%%%%%%%%%%%%%%%%%%%%%%%%%%%%%%

\def\gnumericColB{30pt*\gnumericScale}
\def\gnumericColC{30pt*\gnumericScale}
\def\gnumericColD{30pt*\gnumericScale}
\def\gnumericColE{30pt*\gnumericScale}

\begin{longtable}[c]{%
	b{\gnumericColB}%
	b{\gnumericColC}%
	b{\gnumericColD}%
	b{\gnumericColE}%
	}

%%%%%%%%%%%%%%%%%%%%%%%%%%%%%%%%%%%%%%%%%%%%%%%%%%%%%%%%%%%%%%%%%%%%%%
%%  The longtable options. (Caption. headers... see Goosens. p.124) %%
%	\caption{The Table Caption.}             \\	%
% \hline	% Across the top of the table.
%%  The rest of these options are table rows which are placed on    %%
%%  the first. last or every page. Use \multicolumn if you want.    %%

%%  Header for the first page.                                      %%
%	\multicolumn{4}{c}{The First Header} \\ \hline 
%	\multicolumn{1}{c}{colTag}	%Column 1
%	&\multicolumn{1}{c}{colTag}	%Column 1
%	&\multicolumn{1}{c}{colTag}	%Column 2
%	&\multicolumn{1}{c}{colTag}	%Column 3
%	&\multicolumn{1}{c}{colTag}	\\ \hline %Last column
%	\endfirsthead

%%  The running header definition.                                  %%
%	\hline
%	\multicolumn{4}{l}{\ldots\small\slshape continued} \\ \hline
%	\multicolumn{1}{c}{colTag}	%Column 1
%	&\multicolumn{1}{c}{colTag}	%Column 1
%	&\multicolumn{1}{c}{colTag}	%Column 2
%	&\multicolumn{1}{c}{colTag}	%Column 3
%	&\multicolumn{1}{c}{colTag}	\\ \hline %Last column
%	\endhead

%%  The running footer definition.                                  %%
%	\hline
%	\multicolumn{4}{r}{\small\slshape continued\ldots} \\
%	\endfoot

%%  The ending footer definition.                                   %%
%	\multicolumn{4}{c}{That's all folks} \\ \hline 
%	\endlastfoot
%%%%%%%%%%%%%%%%%%%%%%%%%%%%%%%%%%%%%%%%%%%%%%%%%%%%%%%%%%%%%%%%%%%%%%

\hhline{|-|-|-|-|}
	 \multicolumn{1}{|p{\gnumericColB}|}%
	{\gnumericPB{\raggedleft}\gnumbox[r]{\textsf{$1.00$}}}
	&\multicolumn{1}{p{\gnumericColC}|}%
	{\gnumericPB{\raggedleft}\gnumbox[r]{\textsf{$9.00$}}}
	&\multicolumn{1}{p{\gnumericColD}|}%
	{\gnumericPB{\raggedleft}\gnumbox[r]{\textsf{$1.63$}}}
	&\multicolumn{1}{p{\gnumericColE}|}%
	{\gnumericPB{\raggedleft}\gnumbox[r]{\textsf{$3.34$}}}
\\
\hhline{|----|}
	 \multicolumn{1}{|p{\gnumericColB}|}%
	{\gnumericPB{\raggedleft}\gnumbox[r]{\textsf{$1.06$}}}
	&\multicolumn{1}{p{\gnumericColC}|}%
	{\gnumericPB{\raggedleft}\gnumbox[r]{\textsf{$7.90$}}}
	&\multicolumn{1}{p{\gnumericColD}|}%
	{\gnumericPB{\raggedleft}\gnumbox[r]{\textsf{$1.69$}}}
	&\multicolumn{1}{p{\gnumericColE}|}%
	{\gnumericPB{\raggedleft}\gnumbox[r]{\textsf{$3.11$}}}
\\
\hhline{|----|}
	 \multicolumn{1}{|p{\gnumericColB}|}%
	{\gnumericPB{\raggedleft}\gnumbox[r]{\textsf{$1.13$}}}
	&\multicolumn{1}{p{\gnumericColC}|}%
	{\gnumericPB{\raggedleft}\gnumbox[r]{\textsf{$7.00$}}}
	&\multicolumn{1}{p{\gnumericColD}|}%
	{\gnumericPB{\raggedleft}\gnumbox[r]{\textsf{$1.76$}}}
	&\multicolumn{1}{p{\gnumericColE}|}%
	{\gnumericPB{\raggedleft}\gnumbox[r]{\textsf{$2.90$}}}
\\
\hhline{|----|}
	 \multicolumn{1}{|p{\gnumericColB}|}%
	{\gnumericPB{\raggedleft}\gnumbox[r]{\textsf{$1.19$}}}
	&\multicolumn{1}{p{\gnumericColC}|}%
	{\gnumericPB{\raggedleft}\gnumbox[r]{\textsf{$6.26$}}}
	&\multicolumn{1}{p{\gnumericColD}|}%
	{\gnumericPB{\raggedleft}\gnumbox[r]{\textsf{$1.82$}}}
	&\multicolumn{1}{p{\gnumericColE}|}%
	{\gnumericPB{\raggedleft}\gnumbox[r]{\textsf{$2.71$}}}
\\
\hhline{|----|}
	 \multicolumn{1}{|p{\gnumericColB}|}%
	{\gnumericPB{\raggedleft}\gnumbox[r]{\textsf{$1.25$}}}
	&\multicolumn{1}{p{\gnumericColC}|}%
	{\gnumericPB{\raggedleft}\gnumbox[r]{\textsf{$5.63$}}}
	&\multicolumn{1}{p{\gnumericColD}|}%
	{\gnumericPB{\raggedleft}\gnumbox[r]{\textsf{$1.88$}}}
	&\multicolumn{1}{p{\gnumericColE}|}%
	{\gnumericPB{\raggedleft}\gnumbox[r]{\textsf{$2.54$}}}
\\
\hhline{|----|}
	 \multicolumn{1}{|p{\gnumericColB}|}%
	{\gnumericPB{\raggedleft}\gnumbox[r]{\textsf{$1.32$}}}
	&\multicolumn{1}{p{\gnumericColC}|}%
	{\gnumericPB{\raggedleft}\gnumbox[r]{\textsf{$5.10$}}}
	&\multicolumn{1}{p{\gnumericColD}|}%
	{\gnumericPB{\raggedleft}\gnumbox[r]{\textsf{$1.95$}}}
	&\multicolumn{1}{p{\gnumericColE}|}%
	{\gnumericPB{\raggedleft}\gnumbox[r]{\textsf{$2.39$}}}
\\
\hhline{|----|}
	 \multicolumn{1}{|p{\gnumericColB}|}%
	{\gnumericPB{\raggedleft}\gnumbox[r]{\textsf{$1.38$}}}
	&\multicolumn{1}{p{\gnumericColC}|}%
	{\gnumericPB{\raggedleft}\gnumbox[r]{\textsf{$4.65$}}}
	&\multicolumn{1}{p{\gnumericColD}|}%
	{\gnumericPB{\raggedleft}\gnumbox[r]{\textsf{$2.01$}}}
	&\multicolumn{1}{p{\gnumericColE}|}%
	{\gnumericPB{\raggedleft}\gnumbox[r]{\textsf{$2.25$}}}
\\
\hhline{|----|}
	 \multicolumn{1}{|p{\gnumericColB}|}%
	{\gnumericPB{\raggedleft}\gnumbox[r]{\textsf{$1.44$}}}
	&\multicolumn{1}{p{\gnumericColC}|}%
	{\gnumericPB{\raggedleft}\gnumbox[r]{\textsf{$4.25$}}}
	&\multicolumn{1}{p{\gnumericColD}|}%
	{\gnumericPB{\raggedleft}\gnumbox[r]{\textsf{$2.07$}}}
	&\multicolumn{1}{p{\gnumericColE}|}%
	{\gnumericPB{\raggedleft}\gnumbox[r]{\textsf{$2.13$}}}
\\
\hhline{|----|}
	 \multicolumn{1}{|p{\gnumericColB}|}%
	{\gnumericPB{\raggedleft}\gnumbox[r]{\textsf{$1.51$}}}
	&\multicolumn{1}{p{\gnumericColC}|}%
	{\gnumericPB{\raggedleft}\gnumbox[r]{\textsf{$3.91$}}}
	&\multicolumn{1}{p{\gnumericColD}|}%
	{\gnumericPB{\raggedleft}\gnumbox[r]{\textsf{$2.14$}}}
	&\multicolumn{1}{p{\gnumericColE}|}%
	{\gnumericPB{\raggedleft}\gnumbox[r]{\textsf{$2.01$}}}
\\
\hhline{|----|}
	 \multicolumn{1}{|p{\gnumericColB}|}%
	{\gnumericPB{\raggedleft}\gnumbox[r]{\textsf{$1.57$}}}
	&\multicolumn{1}{p{\gnumericColC}|}%
	{\gnumericPB{\raggedleft}\gnumbox[r]{\textsf{$3.61$}}}
	&\multicolumn{1}{p{\gnumericColD}|}%
	{\gnumericPB{\raggedleft}\gnumbox[r]{\textsf{$2.20$}}}
	&\multicolumn{1}{p{\gnumericColE}|}%
	{\gnumericPB{\raggedleft}\gnumbox[r]{\textsf{$1.91$}}}
\\
\hhline{|-|-|-|-|}
\end{longtable}

\gnumericTableEnd
}
\caption{Normalisation for a gaussian form for $J/\psi$ as a function of $\Lambda$ for $m_c=1.78$~GeV.}\label{tab:N_exp_jpsi_lam3}
\end{table}

\begin{table}[H]
{\tiny%%%%%%%%%%%%%%%%%%%%%%%%%%%%%%%%%%%%%%%%%%%%%%%%%%%%%%%%%%%%%%%%%%%%%%
\def\ifundefined#1{\expandafter\ifx\csname#1\endcsname\relax}

%%  Check for the \def token for inputed files. If it is not        %%
%%  defined, the file will be processed as a standalone and the     %%
%%  preamble will be used.                                          %%
\ifundefined{inputGnumericTable}

%%  End of the preamble for the standalone. The next section is for %%
%%  documents which are included into other LaTeX2e files.          %%
\else

%%  We are not a stand alone document. For a regular table, we will %%
%%  have no preamble and only define the closing to mean nothing.   %%
    \def\gnumericTableEnd{}

%%  If we want landscape mode in an embedded document, comment out  %%
%%  the line above and uncomment the two below. The table will      %%
%%  begin on a new page and run in landscape mode.                  %%
%       \def\gnumericTableEnd{\end{landscape}}
%       \begin{landscape}

%%  End of the else clause for this file being \input.              %%
\fi

%%%%%%%%%%%%%%%%%%%%%%%%%%%%%%%%%%%%%%%%%%%%%%%%%%%%%%%%%%%%%%%%%%%%%%
%%                                                                  %%
%%  The rest is the gnumeric table, except for the closing          %%
%%  statement. Changes below will alter the table's appearance.     %%
%%                                                                  %%
%%%%%%%%%%%%%%%%%%%%%%%%%%%%%%%%%%%%%%%%%%%%%%%%%%%%%%%%%%%%%%%%%%%%%%

\providecommand{\gnumericPB}[1]%
{\let\gnumericTemp=\\#1\let\\=\gnumericTemp\hspace{0pt}}

%%  The default table format retains the relative column widths of  %%
%%  gnumeric. They can easily be changed to c, r or l. In that case %%
%%  you may want to comment out the next line and uncomment the one %%
%%  thereafter                                                      %%
\providecommand\gnumbox{\makebox[0pt]}
%%\providecommand\gnumbox[1][]{\makebox}

%% to adjust positions in multirow situations                       %%
\setlength{\bigstrutjot}{\jot}
\setlength{\extrarowheight}{\doublerulesep}

%%  The \setlongtables command keeps column widths the same across %%
%%  pages. Simply comment out next line for varying column widths.  %%
\setlongtables

\setlength\gnumericTableWidth{%
	30pt+%
	30pt+%
	30pt+%
	30pt+%
0pt}
\def\gumericNumCols{4}
\setlength\gnumericTableWidthComplete{\gnumericTableWidth+\tabcolsep*\gumericNumCols*2+\arrayrulewidth*\gumericNumCols}
\ifthenelse{\lengthtest{\gnumericTableWidthComplete > \textwidth}}%
{\def\gnumericScale{\ratio{\textwidth-\tabcolsep*\gumericNumCols*2-\arrayrulewidth*\gumericNumCols}%
{\gnumericTableWidth}}}%
{\def\gnumericScale{1}}

%%%%%%%%%%%%%%%%%%%%%%%%%%%%%%%%%%%%%%%%%%%%%%%%%%%%%%%%%%%%%%%%%%%%%%
%%                                                                  %%
%% The following are the widths of the various columns. We are      %%
%% defining them here because then they are easier to change.       %%
%% Depending on the cell formats we may use them more than once.    %%
%%                                                                  %%
%%%%%%%%%%%%%%%%%%%%%%%%%%%%%%%%%%%%%%%%%%%%%%%%%%%%%%%%%%%%%%%%%%%%%%

\def\gnumericColB{30pt*\gnumericScale}
\def\gnumericColC{30pt*\gnumericScale}
\def\gnumericColD{30pt*\gnumericScale}
\def\gnumericColE{30pt*\gnumericScale}

\begin{longtable}[c]{%
	b{\gnumericColB}%
	b{\gnumericColC}%
	b{\gnumericColD}%
	b{\gnumericColE}%
	}

%%%%%%%%%%%%%%%%%%%%%%%%%%%%%%%%%%%%%%%%%%%%%%%%%%%%%%%%%%%%%%%%%%%%%%
%%  The longtable options. (Caption. headers... see Goosens. p.124) %%
%	\caption{The Table Caption.}             \\	%
% \hline	% Across the top of the table.
%%  The rest of these options are table rows which are placed on    %%
%%  the first. last or every page. Use \multicolumn if you want.    %%

%%  Header for the first page.                                      %%
%	\multicolumn{4}{c}{The First Header} \\ \hline 
%	\multicolumn{1}{c}{colTag}	%Column 1
%	&\multicolumn{1}{c}{colTag}	%Column 1
%	&\multicolumn{1}{c}{colTag}	%Column 2
%	&\multicolumn{1}{c}{colTag}	%Column 3
%	&\multicolumn{1}{c}{colTag}	\\ \hline %Last column
%	\endfirsthead

%%  The running header definition.                                  %%
%	\hline
%	\multicolumn{4}{l}{\ldots\small\slshape continued} \\ \hline
%	\multicolumn{1}{c}{colTag}	%Column 1
%	&\multicolumn{1}{c}{colTag}	%Column 1
%	&\multicolumn{1}{c}{colTag}	%Column 2
%	&\multicolumn{1}{c}{colTag}	%Column 3
%	&\multicolumn{1}{c}{colTag}	\\ \hline %Last column
%	\endhead

%%  The running footer definition.                                  %%
%	\hline
%	\multicolumn{4}{r}{\small\slshape continued\ldots} \\
%	\endfoot

%%  The ending footer definition.                                   %%
%	\multicolumn{4}{c}{That's all folks} \\ \hline 
%	\endlastfoot
%%%%%%%%%%%%%%%%%%%%%%%%%%%%%%%%%%%%%%%%%%%%%%%%%%%%%%%%%%%%%%%%%%%%%%

\hhline{|-|-|-|-|}
	 \multicolumn{1}{|p{\gnumericColB}|}%
	{\gnumericPB{\raggedleft}\gnumbox[r]{\textsf{$1.00$}}}
	&\multicolumn{1}{p{\gnumericColC}|}%
	{\gnumericPB{\raggedleft}\gnumbox[r]{\textsf{$10.44$}}}
	&\multicolumn{1}{p{\gnumericColD}|}%
	{\gnumericPB{\raggedleft}\gnumbox[r]{\textsf{$1.63$}}}
	&\multicolumn{1}{p{\gnumericColE}|}%
	{\gnumericPB{\raggedleft}\gnumbox[r]{\textsf{$3.68$}}}
\\
\hhline{|----|}
	 \multicolumn{1}{|p{\gnumericColB}|}%
	{\gnumericPB{\raggedleft}\gnumbox[r]{\textsf{$1.06$}}}
	&\multicolumn{1}{p{\gnumericColC}|}%
	{\gnumericPB{\raggedleft}\gnumbox[r]{\textsf{$9.10$}}}
	&\multicolumn{1}{p{\gnumericColD}|}%
	{\gnumericPB{\raggedleft}\gnumbox[r]{\textsf{$1.69$}}}
	&\multicolumn{1}{p{\gnumericColE}|}%
	{\gnumericPB{\raggedleft}\gnumbox[r]{\textsf{$3.41$}}}
\\
\hhline{|----|}
	 \multicolumn{1}{|p{\gnumericColB}|}%
	{\gnumericPB{\raggedleft}\gnumbox[r]{\textsf{$1.13$}}}
	&\multicolumn{1}{p{\gnumericColC}|}%
	{\gnumericPB{\raggedleft}\gnumbox[r]{\textsf{$8.02$}}}
	&\multicolumn{1}{p{\gnumericColD}|}%
	{\gnumericPB{\raggedleft}\gnumbox[r]{\textsf{$1.76$}}}
	&\multicolumn{1}{p{\gnumericColE}|}%
	{\gnumericPB{\raggedleft}\gnumbox[r]{\textsf{$3.17$}}}
\\
\hhline{|----|}
	 \multicolumn{1}{|p{\gnumericColB}|}%
	{\gnumericPB{\raggedleft}\gnumbox[r]{\textsf{$1.19$}}}
	&\multicolumn{1}{p{\gnumericColC}|}%
	{\gnumericPB{\raggedleft}\gnumbox[r]{\textsf{$7.12$}}}
	&\multicolumn{1}{p{\gnumericColD}|}%
	{\gnumericPB{\raggedleft}\gnumbox[r]{\textsf{$1.82$}}}
	&\multicolumn{1}{p{\gnumericColE}|}%
	{\gnumericPB{\raggedleft}\gnumbox[r]{\textsf{$2.95$}}}
\\
\hhline{|----|}
	 \multicolumn{1}{|p{\gnumericColB}|}%
	{\gnumericPB{\raggedleft}\gnumbox[r]{\textsf{$1.25$}}}
	&\multicolumn{1}{p{\gnumericColC}|}%
	{\gnumericPB{\raggedleft}\gnumbox[r]{\textsf{$6.37$}}}
	&\multicolumn{1}{p{\gnumericColD}|}%
	{\gnumericPB{\raggedleft}\gnumbox[r]{\textsf{$1.88$}}}
	&\multicolumn{1}{p{\gnumericColE}|}%
	{\gnumericPB{\raggedleft}\gnumbox[r]{\textsf{$2.76$}}}
\\
\hhline{|----|}
	 \multicolumn{1}{|p{\gnumericColB}|}%
	{\gnumericPB{\raggedleft}\gnumbox[r]{\textsf{$1.32$}}}
	&\multicolumn{1}{p{\gnumericColC}|}%
	{\gnumericPB{\raggedleft}\gnumbox[r]{\textsf{$5.74$}}}
	&\multicolumn{1}{p{\gnumericColD}|}%
	{\gnumericPB{\raggedleft}\gnumbox[r]{\textsf{$1.95$}}}
	&\multicolumn{1}{p{\gnumericColE}|}%
	{\gnumericPB{\raggedleft}\gnumbox[r]{\textsf{$2.59$}}}
\\
\hhline{|----|}
	 \multicolumn{1}{|p{\gnumericColB}|}%
	{\gnumericPB{\raggedleft}\gnumbox[r]{\textsf{$1.38$}}}
	&\multicolumn{1}{p{\gnumericColC}|}%
	{\gnumericPB{\raggedleft}\gnumbox[r]{\textsf{$5.20$}}}
	&\multicolumn{1}{p{\gnumericColD}|}%
	{\gnumericPB{\raggedleft}\gnumbox[r]{\textsf{$2.01$}}}
	&\multicolumn{1}{p{\gnumericColE}|}%
	{\gnumericPB{\raggedleft}\gnumbox[r]{\textsf{$2.43$}}}
\\
\hhline{|----|}
	 \multicolumn{1}{|p{\gnumericColB}|}%
	{\gnumericPB{\raggedleft}\gnumbox[r]{\textsf{$1.44$}}}
	&\multicolumn{1}{p{\gnumericColC}|}%
	{\gnumericPB{\raggedleft}\gnumbox[r]{\textsf{$4.74$}}}
	&\multicolumn{1}{p{\gnumericColD}|}%
	{\gnumericPB{\raggedleft}\gnumbox[r]{\textsf{$2.07$}}}
	&\multicolumn{1}{p{\gnumericColE}|}%
	{\gnumericPB{\raggedleft}\gnumbox[r]{\textsf{$2.29$}}}
\\
\hhline{|----|}
	 \multicolumn{1}{|p{\gnumericColB}|}%
	{\gnumericPB{\raggedleft}\gnumbox[r]{\textsf{$1.51$}}}
	&\multicolumn{1}{p{\gnumericColC}|}%
	{\gnumericPB{\raggedleft}\gnumbox[r]{\textsf{$4.34$}}}
	&\multicolumn{1}{p{\gnumericColD}|}%
	{\gnumericPB{\raggedleft}\gnumbox[r]{\textsf{$2.14$}}}
	&\multicolumn{1}{p{\gnumericColE}|}%
	{\gnumericPB{\raggedleft}\gnumbox[r]{\textsf{$2.16$}}}
\\
\hhline{|----|}
	 \multicolumn{1}{|p{\gnumericColB}|}%
	{\gnumericPB{\raggedleft}\gnumbox[r]{\textsf{$1.57$}}}
	&\multicolumn{1}{p{\gnumericColC}|}%
	{\gnumericPB{\raggedleft}\gnumbox[r]{\textsf{$3.99$}}}
	&\multicolumn{1}{p{\gnumericColD}|}%
	{\gnumericPB{\raggedleft}\gnumbox[r]{\textsf{$2.20$}}}
	&\multicolumn{1}{p{\gnumericColE}|}%
	{\gnumericPB{\raggedleft}\gnumbox[r]{\textsf{$2.04$}}}
\\
\hhline{|-|-|-|-|}
\end{longtable}

\gnumericTableEnd
}
\caption{Normalisation for a gaussian form for $J/\psi$ as a function of $\Lambda$ for $m_c=1.87$~GeV.}\label{tab:N_exp_jpsi_lam4}
\end{table}

\begin{table}[H]
{\tiny%%%%%%%%%%%%%%%%%%%%%%%%%%%%%%%%%%%%%%%%%%%%%%%%%%%%%%%%%%%%%%%%%%%%%%
\def\ifundefined#1{\expandafter\ifx\csname#1\endcsname\relax}

%%  Check for the \def token for inputed files. If it is not        %%
%%  defined, the file will be processed as a standalone and the     %%
%%  preamble will be used.                                          %%
\ifundefined{inputGnumericTable}

%%  End of the preamble for the standalone. The next section is for %%
%%  documents which are included into other LaTeX2e files.          %%
\else

%%  We are not a stand alone document. For a regular table, we will %%
%%  have no preamble and only define the closing to mean nothing.   %%
    \def\gnumericTableEnd{}

%%  If we want landscape mode in an embedded document, comment out  %%
%%  the line above and uncomment the two below. The table will      %%
%%  begin on a new page and run in landscape mode.                  %%
%       \def\gnumericTableEnd{\end{landscape}}
%       \begin{landscape}

%%  End of the else clause for this file being \input.              %%
\fi

%%%%%%%%%%%%%%%%%%%%%%%%%%%%%%%%%%%%%%%%%%%%%%%%%%%%%%%%%%%%%%%%%%%%%%
%%                                                                  %%
%%  The rest is the gnumeric table, except for the closing          %%
%%  statement. Changes below will alter the table's appearance.     %%
%%                                                                  %%
%%%%%%%%%%%%%%%%%%%%%%%%%%%%%%%%%%%%%%%%%%%%%%%%%%%%%%%%%%%%%%%%%%%%%%

\providecommand{\gnumericPB}[1]%
{\let\gnumericTemp=\\#1\let\\=\gnumericTemp\hspace{0pt}}

%%  The default table format retains the relative column widths of  %%
%%  gnumeric. They can easily be changed to c, r or l. In that case %%
%%  you may want to comment out the next line and uncomment the one %%
%%  thereafter                                                      %%
\providecommand\gnumbox{\makebox[0pt]}
%%\providecommand\gnumbox[1][]{\makebox}

%% to adjust positions in multirow situations                       %%
\setlength{\bigstrutjot}{\jot}
\setlength{\extrarowheight}{\doublerulesep}

%%  The \setlongtables command keeps column widths the same across %%
%%  pages. Simply comment out next line for varying column widths.  %%
\setlongtables

\setlength\gnumericTableWidth{%
	30pt+%
	30pt+%
	30pt+%
	30pt+%
0pt}
\def\gumericNumCols{4}
\setlength\gnumericTableWidthComplete{\gnumericTableWidth+\tabcolsep*\gumericNumCols*2+\arrayrulewidth*\gumericNumCols}
\ifthenelse{\lengthtest{\gnumericTableWidthComplete > \textwidth}}%
{\def\gnumericScale{\ratio{\textwidth-\tabcolsep*\gumericNumCols*2-\arrayrulewidth*\gumericNumCols}%
{\gnumericTableWidth}}}%
{\def\gnumericScale{1}}

%%%%%%%%%%%%%%%%%%%%%%%%%%%%%%%%%%%%%%%%%%%%%%%%%%%%%%%%%%%%%%%%%%%%%%
%%                                                                  %%
%% The following are the widths of the various columns. We are      %%
%% defining them here because then they are easier to change.       %%
%% Depending on the cell formats we may use them more than once.    %%
%%                                                                  %%
%%%%%%%%%%%%%%%%%%%%%%%%%%%%%%%%%%%%%%%%%%%%%%%%%%%%%%%%%%%%%%%%%%%%%%

\def\gnumericColB{30pt*\gnumericScale}
\def\gnumericColC{30pt*\gnumericScale}
\def\gnumericColD{30pt*\gnumericScale}
\def\gnumericColE{30pt*\gnumericScale}

\begin{longtable}[c]{%
	b{\gnumericColB}%
	b{\gnumericColC}%
	b{\gnumericColD}%
	b{\gnumericColE}%
	}

%%%%%%%%%%%%%%%%%%%%%%%%%%%%%%%%%%%%%%%%%%%%%%%%%%%%%%%%%%%%%%%%%%%%%%
%%  The longtable options. (Caption. headers... see Goosens. p.124) %%
%	\caption{The Table Caption.}             \\	%
% \hline	% Across the top of the table.
%%  The rest of these options are table rows which are placed on    %%
%%  the first. last or every page. Use \multicolumn if you want.    %%

%%  Header for the first page.                                      %%
%	\multicolumn{4}{c}{The First Header} \\ \hline 
%	\multicolumn{1}{c}{colTag}	%Column 1
%	&\multicolumn{1}{c}{colTag}	%Column 1
%	&\multicolumn{1}{c}{colTag}	%Column 2
%	&\multicolumn{1}{c}{colTag}	%Column 3
%	&\multicolumn{1}{c}{colTag}	\\ \hline %Last column
%	\endfirsthead

%%  The running header definition.                                  %%
%	\hline
%	\multicolumn{4}{l}{\ldots\small\slshape continued} \\ \hline
%	\multicolumn{1}{c}{colTag}	%Column 1
%	&\multicolumn{1}{c}{colTag}	%Column 1
%	&\multicolumn{1}{c}{colTag}	%Column 2
%	&\multicolumn{1}{c}{colTag}	%Column 3
%	&\multicolumn{1}{c}{colTag}	\\ \hline %Last column
%	\endhead

%%  The running footer definition.                                  %%
%	\hline
%	\multicolumn{4}{r}{\small\slshape continued\ldots} \\
%	\endfoot

%%  The ending footer definition.                                   %%
%	\multicolumn{4}{c}{That's all folks} \\ \hline 
%	\endlastfoot
%%%%%%%%%%%%%%%%%%%%%%%%%%%%%%%%%%%%%%%%%%%%%%%%%%%%%%%%%%%%%%%%%%%%%%

\hhline{|-|-|-|-|}
	 \multicolumn{1}{|p{\gnumericColB}|}%
	{\gnumericPB{\raggedleft}\gnumbox[r]{\textsf{$m_c$}}}
	&\multicolumn{1}{p{\gnumericColC}|}%
	{\gnumericPB{\raggedleft}\gnumbox[r]{\textsf{$N$}}}
	&\multicolumn{1}{p{\gnumericColD}|}%
	{\gnumericPB{\raggedleft}\gnumbox[r]{\textsf{$m_c$}}}
	&\multicolumn{1}{p{\gnumericColE}|}%
	{\gnumericPB{\raggedleft}\gnumbox[r]{\textsf{$N$}}}
\\
\hhline{|----|}
	 \multicolumn{1}{|p{\gnumericColB}|}%
	{\gnumericPB{\raggedleft}\gnumbox[r]{\textsf{$1.60$}}}
	&\multicolumn{1}{p{\gnumericColC}|}%
	{\gnumericPB{\raggedleft}\gnumbox[r]{\textsf{$5.36$}}}
	&\multicolumn{1}{p{\gnumericColD}|}%
	{\gnumericPB{\raggedleft}\gnumbox[r]{\textsf{$1.74$}}}
	&\multicolumn{1}{p{\gnumericColE}|}%
	{\gnumericPB{\raggedleft}\gnumbox[r]{\textsf{$8.34$}}}
\\
\hhline{|----|}
	 \multicolumn{1}{|p{\gnumericColB}|}%
	{\gnumericPB{\raggedleft}\gnumbox[r]{\textsf{$1.61$}}}
	&\multicolumn{1}{p{\gnumericColC}|}%
	{\gnumericPB{\raggedleft}\gnumbox[r]{\textsf{$5.73$}}}
	&\multicolumn{1}{p{\gnumericColD}|}%
	{\gnumericPB{\raggedleft}\gnumbox[r]{\textsf{$1.76$}}}
	&\multicolumn{1}{p{\gnumericColE}|}%
	{\gnumericPB{\raggedleft}\gnumbox[r]{\textsf{$8.59$}}}
\\
\hhline{|----|}
	 \multicolumn{1}{|p{\gnumericColB}|}%
	{\gnumericPB{\raggedleft}\gnumbox[r]{\textsf{$1.63$}}}
	&\multicolumn{1}{p{\gnumericColC}|}%
	{\gnumericPB{\raggedleft}\gnumbox[r]{\textsf{$6.07$}}}
	&\multicolumn{1}{p{\gnumericColD}|}%
	{\gnumericPB{\raggedleft}\gnumbox[r]{\textsf{$1.77$}}}
	&\multicolumn{1}{p{\gnumericColE}|}%
	{\gnumericPB{\raggedleft}\gnumbox[r]{\textsf{$8.84$}}}
\\
\hhline{|----|}
	 \multicolumn{1}{|p{\gnumericColB}|}%
	{\gnumericPB{\raggedleft}\gnumbox[r]{\textsf{$1.64$}}}
	&\multicolumn{1}{p{\gnumericColC}|}%
	{\gnumericPB{\raggedleft}\gnumbox[r]{\textsf{$6.40$}}}
	&\multicolumn{1}{p{\gnumericColD}|}%
	{\gnumericPB{\raggedleft}\gnumbox[r]{\textsf{$1.78$}}}
	&\multicolumn{1}{p{\gnumericColE}|}%
	{\gnumericPB{\raggedleft}\gnumbox[r]{\textsf{$9.08$}}}
\\
\hhline{|----|}
	 \multicolumn{1}{|p{\gnumericColB}|}%
	{\gnumericPB{\raggedleft}\gnumbox[r]{\textsf{$1.66$}}}
	&\multicolumn{1}{p{\gnumericColC}|}%
	{\gnumericPB{\raggedleft}\gnumbox[r]{\textsf{$6.70$}}}
	&\multicolumn{1}{p{\gnumericColD}|}%
	{\gnumericPB{\raggedleft}\gnumbox[r]{\textsf{$1.80$}}}
	&\multicolumn{1}{p{\gnumericColE}|}%
	{\gnumericPB{\raggedleft}\gnumbox[r]{\textsf{$9.31$}}}
\\
\hhline{|----|}
	 \multicolumn{1}{|p{\gnumericColB}|}%
	{\gnumericPB{\raggedleft}\gnumbox[r]{\textsf{$1.67$}}}
	&\multicolumn{1}{p{\gnumericColC}|}%
	{\gnumericPB{\raggedleft}\gnumbox[r]{\textsf{$7.00$}}}
	&\multicolumn{1}{p{\gnumericColD}|}%
	{\gnumericPB{\raggedleft}\gnumbox[r]{\textsf{$1.81$}}}
	&\multicolumn{1}{p{\gnumericColE}|}%
	{\gnumericPB{\raggedleft}\gnumbox[r]{\textsf{$9.54$}}}
\\
\hhline{|----|}
	 \multicolumn{1}{|p{\gnumericColB}|}%
	{\gnumericPB{\raggedleft}\gnumbox[r]{\textsf{$1.69$}}}
	&\multicolumn{1}{p{\gnumericColC}|}%
	{\gnumericPB{\raggedleft}\gnumbox[r]{\textsf{$7.28$}}}
	&\multicolumn{1}{p{\gnumericColD}|}%
	{\gnumericPB{\raggedleft}\gnumbox[r]{\textsf{$1.83$}}}
	&\multicolumn{1}{p{\gnumericColE}|}%
	{\gnumericPB{\raggedleft}\gnumbox[r]{\textsf{$9.77$}}}
\\
\hhline{|----|}
	 \multicolumn{1}{|p{\gnumericColB}|}%
	{\gnumericPB{\raggedleft}\gnumbox[r]{\textsf{$1.70$}}}
	&\multicolumn{1}{p{\gnumericColC}|}%
	{\gnumericPB{\raggedleft}\gnumbox[r]{\textsf{$7.56$}}}
	&\multicolumn{1}{p{\gnumericColD}|}%
	{\gnumericPB{\raggedleft}\gnumbox[r]{\textsf{$1.84$}}}
	&\multicolumn{1}{p{\gnumericColE}|}%
	{\gnumericPB{\raggedleft}\gnumbox[r]{\textsf{$10.00$}}}
\\
\hhline{|----|}
	 \multicolumn{1}{|p{\gnumericColB}|}%
	{\gnumericPB{\raggedleft}\gnumbox[r]{\textsf{$1.71$}}}
	&\multicolumn{1}{p{\gnumericColC}|}%
	{\gnumericPB{\raggedleft}\gnumbox[r]{\textsf{$7.83$}}}
	&\multicolumn{1}{p{\gnumericColD}|}%
	{\gnumericPB{\raggedleft}\gnumbox[r]{\textsf{$1.86$}}}
	&\multicolumn{1}{p{\gnumericColE}|}%
	{\gnumericPB{\raggedleft}\gnumbox[r]{\textsf{$10.22$}}}
\\
\hhline{|----|}
	 \multicolumn{1}{|p{\gnumericColB}|}%
	{\gnumericPB{\raggedleft}\gnumbox[r]{\textsf{$1.73$}}}
	&\multicolumn{1}{p{\gnumericColC}|}%
	{\gnumericPB{\raggedleft}\gnumbox[r]{\textsf{$8.09$}}}
	&\multicolumn{1}{p{\gnumericColD}|}%
	{\gnumericPB{\raggedleft}\gnumbox[r]{\textsf{$1.87$}}}
	&\multicolumn{1}{p{\gnumericColE}|}%
	{\gnumericPB{\raggedleft}\gnumbox[r]{\textsf{$10.44$}}}
\\
\hhline{|-|-|-|-|}
\end{longtable}

\gnumericTableEnd
}
\caption{Normalisation for a gaussian form for $J/\psi$ as a function of  the $c$ quark mass
for $\Lambda=1.0$~GeV.}\label{tab:N_exp_jpsi_mq1}
\end{table}

\begin{table}[H]
{\tiny%%%%%%%%%%%%%%%%%%%%%%%%%%%%%%%%%%%%%%%%%%%%%%%%%%%%%%%%%%%%%%%%%%%%%%
\def\ifundefined#1{\expandafter\ifx\csname#1\endcsname\relax}

%%  Check for the \def token for inputed files. If it is not        %%
%%  defined, the file will be processed as a standalone and the     %%
%%  preamble will be used.                                          %%
\ifundefined{inputGnumericTable}

%%  End of the preamble for the standalone. The next section is for %%
%%  documents which are included into other LaTeX2e files.          %%
\else

%%  We are not a stand alone document. For a regular table, we will %%
%%  have no preamble and only define the closing to mean nothing.   %%
    \def\gnumericTableEnd{}

%%  If we want landscape mode in an embedded document, comment out  %%
%%  the line above and uncomment the two below. The table will      %%
%%  begin on a new page and run in landscape mode.                  %%
%       \def\gnumericTableEnd{\end{landscape}}
%       \begin{landscape}

%%  End of the else clause for this file being \input.              %%
\fi

%%%%%%%%%%%%%%%%%%%%%%%%%%%%%%%%%%%%%%%%%%%%%%%%%%%%%%%%%%%%%%%%%%%%%%
%%                                                                  %%
%%  The rest is the gnumeric table, except for the closing          %%
%%  statement. Changes below will alter the table's appearance.     %%
%%                                                                  %%
%%%%%%%%%%%%%%%%%%%%%%%%%%%%%%%%%%%%%%%%%%%%%%%%%%%%%%%%%%%%%%%%%%%%%%

\providecommand{\gnumericPB}[1]%
{\let\gnumericTemp=\\#1\let\\=\gnumericTemp\hspace{0pt}}

%%  The default table format retains the relative column widths of  %%
%%  gnumeric. They can easily be changed to c, r or l. In that case %%
%%  you may want to comment out the next line and uncomment the one %%
%%  thereafter                                                      %%
\providecommand\gnumbox{\makebox[0pt]}
%%\providecommand\gnumbox[1][]{\makebox}

%% to adjust positions in multirow situations                       %%
\setlength{\bigstrutjot}{\jot}
\setlength{\extrarowheight}{\doublerulesep}

%%  The \setlongtables command keeps column widths the same across %%
%%  pages. Simply comment out next line for varying column widths.  %%
\setlongtables

\setlength\gnumericTableWidth{%
	30pt+%
	30pt+%
	30pt+%
	30pt+%
0pt}
\def\gumericNumCols{4}
\setlength\gnumericTableWidthComplete{\gnumericTableWidth+\tabcolsep*\gumericNumCols*2+\arrayrulewidth*\gumericNumCols}
\ifthenelse{\lengthtest{\gnumericTableWidthComplete > \textwidth}}%
{\def\gnumericScale{\ratio{\textwidth-\tabcolsep*\gumericNumCols*2-\arrayrulewidth*\gumericNumCols}%
{\gnumericTableWidth}}}%
{\def\gnumericScale{1}}

%%%%%%%%%%%%%%%%%%%%%%%%%%%%%%%%%%%%%%%%%%%%%%%%%%%%%%%%%%%%%%%%%%%%%%
%%                                                                  %%
%% The following are the widths of the various columns. We are      %%
%% defining them here because then they are easier to change.       %%
%% Depending on the cell formats we may use them more than once.    %%
%%                                                                  %%
%%%%%%%%%%%%%%%%%%%%%%%%%%%%%%%%%%%%%%%%%%%%%%%%%%%%%%%%%%%%%%%%%%%%%%

\def\gnumericColB{30pt*\gnumericScale}
\def\gnumericColC{30pt*\gnumericScale}
\def\gnumericColD{30pt*\gnumericScale}
\def\gnumericColE{30pt*\gnumericScale}

\begin{longtable}[c]{%
	b{\gnumericColB}%
	b{\gnumericColC}%
	b{\gnumericColD}%
	b{\gnumericColE}%
	}

%%%%%%%%%%%%%%%%%%%%%%%%%%%%%%%%%%%%%%%%%%%%%%%%%%%%%%%%%%%%%%%%%%%%%%
%%  The longtable options. (Caption. headers... see Goosens. p.124) %%
%	\caption{The Table Caption.}             \\	%
% \hline	% Across the top of the table.
%%  The rest of these options are table rows which are placed on    %%
%%  the first. last or every page. Use \multicolumn if you want.    %%

%%  Header for the first page.                                      %%
%	\multicolumn{4}{c}{The First Header} \\ \hline 
%	\multicolumn{1}{c}{colTag}	%Column 1
%	&\multicolumn{1}{c}{colTag}	%Column 1
%	&\multicolumn{1}{c}{colTag}	%Column 2
%	&\multicolumn{1}{c}{colTag}	%Column 3
%	&\multicolumn{1}{c}{colTag}	\\ \hline %Last column
%	\endfirsthead

%%  The running header definition.                                  %%
%	\hline
%	\multicolumn{4}{l}{\ldots\small\slshape continued} \\ \hline
%	\multicolumn{1}{c}{colTag}	%Column 1
%	&\multicolumn{1}{c}{colTag}	%Column 1
%	&\multicolumn{1}{c}{colTag}	%Column 2
%	&\multicolumn{1}{c}{colTag}	%Column 3
%	&\multicolumn{1}{c}{colTag}	\\ \hline %Last column
%	\endhead

%%  The running footer definition.                                  %%
%	\hline
%	\multicolumn{4}{r}{\small\slshape continued\ldots} \\
%	\endfoot

%%  The ending footer definition.                                   %%
%	\multicolumn{4}{c}{That's all folks} \\ \hline 
%	\endlastfoot
%%%%%%%%%%%%%%%%%%%%%%%%%%%%%%%%%%%%%%%%%%%%%%%%%%%%%%%%%%%%%%%%%%%%%%

\hhline{|-|-|-|-|}
	 \multicolumn{1}{|p{\gnumericColB}|}%
	{\gnumericPB{\raggedleft}\gnumbox[r]{\textsf{$1.60$}}}
	&\multicolumn{1}{p{\gnumericColC}|}%
	{\gnumericPB{\raggedleft}\gnumbox[r]{\textsf{$3.09$}}}
	&\multicolumn{1}{p{\gnumericColD}|}%
	{\gnumericPB{\raggedleft}\gnumbox[r]{\textsf{$1.74$}}}
	&\multicolumn{1}{p{\gnumericColE}|}%
	{\gnumericPB{\raggedleft}\gnumbox[r]{\textsf{$4.26$}}}
\\
\hhline{|----|}
	 \multicolumn{1}{|p{\gnumericColB}|}%
	{\gnumericPB{\raggedleft}\gnumbox[r]{\textsf{$1.61$}}}
	&\multicolumn{1}{p{\gnumericColC}|}%
	{\gnumericPB{\raggedleft}\gnumbox[r]{\textsf{$3.25$}}}
	&\multicolumn{1}{p{\gnumericColD}|}%
	{\gnumericPB{\raggedleft}\gnumbox[r]{\textsf{$1.76$}}}
	&\multicolumn{1}{p{\gnumericColE}|}%
	{\gnumericPB{\raggedleft}\gnumbox[r]{\textsf{$4.36$}}}
\\
\hhline{|----|}
	 \multicolumn{1}{|p{\gnumericColB}|}%
	{\gnumericPB{\raggedleft}\gnumbox[r]{\textsf{$1.63$}}}
	&\multicolumn{1}{p{\gnumericColC}|}%
	{\gnumericPB{\raggedleft}\gnumbox[r]{\textsf{$3.38$}}}
	&\multicolumn{1}{p{\gnumericColD}|}%
	{\gnumericPB{\raggedleft}\gnumbox[r]{\textsf{$1.77$}}}
	&\multicolumn{1}{p{\gnumericColE}|}%
	{\gnumericPB{\raggedleft}\gnumbox[r]{\textsf{$4.45$}}}
\\
\hhline{|----|}
	 \multicolumn{1}{|p{\gnumericColB}|}%
	{\gnumericPB{\raggedleft}\gnumbox[r]{\textsf{$1.64$}}}
	&\multicolumn{1}{p{\gnumericColC}|}%
	{\gnumericPB{\raggedleft}\gnumbox[r]{\textsf{$3.51$}}}
	&\multicolumn{1}{p{\gnumericColD}|}%
	{\gnumericPB{\raggedleft}\gnumbox[r]{\textsf{$1.78$}}}
	&\multicolumn{1}{p{\gnumericColE}|}%
	{\gnumericPB{\raggedleft}\gnumbox[r]{\textsf{$4.54$}}}
\\
\hhline{|----|}
	 \multicolumn{1}{|p{\gnumericColB}|}%
	{\gnumericPB{\raggedleft}\gnumbox[r]{\textsf{$1.66$}}}
	&\multicolumn{1}{p{\gnumericColC}|}%
	{\gnumericPB{\raggedleft}\gnumbox[r]{\textsf{$3.63$}}}
	&\multicolumn{1}{p{\gnumericColD}|}%
	{\gnumericPB{\raggedleft}\gnumbox[r]{\textsf{$1.80$}}}
	&\multicolumn{1}{p{\gnumericColE}|}%
	{\gnumericPB{\raggedleft}\gnumbox[r]{\textsf{$4.63$}}}
\\
\hhline{|----|}
	 \multicolumn{1}{|p{\gnumericColB}|}%
	{\gnumericPB{\raggedleft}\gnumbox[r]{\textsf{$1.67$}}}
	&\multicolumn{1}{p{\gnumericColC}|}%
	{\gnumericPB{\raggedleft}\gnumbox[r]{\textsf{$3.75$}}}
	&\multicolumn{1}{p{\gnumericColD}|}%
	{\gnumericPB{\raggedleft}\gnumbox[r]{\textsf{$1.81$}}}
	&\multicolumn{1}{p{\gnumericColE}|}%
	{\gnumericPB{\raggedleft}\gnumbox[r]{\textsf{$4.71$}}}
\\
\hhline{|----|}
	 \multicolumn{1}{|p{\gnumericColB}|}%
	{\gnumericPB{\raggedleft}\gnumbox[r]{\textsf{$1.69$}}}
	&\multicolumn{1}{p{\gnumericColC}|}%
	{\gnumericPB{\raggedleft}\gnumbox[r]{\textsf{$3.86$}}}
	&\multicolumn{1}{p{\gnumericColD}|}%
	{\gnumericPB{\raggedleft}\gnumbox[r]{\textsf{$1.83$}}}
	&\multicolumn{1}{p{\gnumericColE}|}%
	{\gnumericPB{\raggedleft}\gnumbox[r]{\textsf{$4.79$}}}
\\
\hhline{|----|}
	 \multicolumn{1}{|p{\gnumericColB}|}%
	{\gnumericPB{\raggedleft}\gnumbox[r]{\textsf{$1.70$}}}
	&\multicolumn{1}{p{\gnumericColC}|}%
	{\gnumericPB{\raggedleft}\gnumbox[r]{\textsf{$3.97$}}}
	&\multicolumn{1}{p{\gnumericColD}|}%
	{\gnumericPB{\raggedleft}\gnumbox[r]{\textsf{$1.84$}}}
	&\multicolumn{1}{p{\gnumericColE}|}%
	{\gnumericPB{\raggedleft}\gnumbox[r]{\textsf{$4.88$}}}
\\
\hhline{|----|}
	 \multicolumn{1}{|p{\gnumericColB}|}%
	{\gnumericPB{\raggedleft}\gnumbox[r]{\textsf{$1.71$}}}
	&\multicolumn{1}{p{\gnumericColC}|}%
	{\gnumericPB{\raggedleft}\gnumbox[r]{\textsf{$4.07$}}}
	&\multicolumn{1}{p{\gnumericColD}|}%
	{\gnumericPB{\raggedleft}\gnumbox[r]{\textsf{$1.86$}}}
	&\multicolumn{1}{p{\gnumericColE}|}%
	{\gnumericPB{\raggedleft}\gnumbox[r]{\textsf{$4.96$}}}
\\
\hhline{|----|}
	 \multicolumn{1}{|p{\gnumericColB}|}%
	{\gnumericPB{\raggedleft}\gnumbox[r]{\textsf{$1.73$}}}
	&\multicolumn{1}{p{\gnumericColC}|}%
	{\gnumericPB{\raggedleft}\gnumbox[r]{\textsf{$4.17$}}}
	&\multicolumn{1}{p{\gnumericColD}|}%
	{\gnumericPB{\raggedleft}\gnumbox[r]{\textsf{$1.87$}}}
	&\multicolumn{1}{p{\gnumericColE}|}%
	{\gnumericPB{\raggedleft}\gnumbox[r]{\textsf{$5.04$}}}
\\
\hhline{|-|-|-|-|}
\end{longtable}

\gnumericTableEnd
}
\caption{Normalisation for a gaussian form for $J/\psi$ as a function of  the $c$ quark mass
for $\Lambda=1.4$~GeV.}\label{tab:N_exp_jpsi_mq2}
\end{table}

\begin{table}[H]
{\tiny%%%%%%%%%%%%%%%%%%%%%%%%%%%%%%%%%%%%%%%%%%%%%%%%%%%%%%%%%%%%%%%%%%%%%%
\def\ifundefined#1{\expandafter\ifx\csname#1\endcsname\relax}

%%  Check for the \def token for inputed files. If it is not        %%
%%  defined, the file will be processed as a standalone and the     %%
%%  preamble will be used.                                          %%
\ifundefined{inputGnumericTable}

%%  End of the preamble for the standalone. The next section is for %%
%%  documents which are included into other LaTeX2e files.          %%
\else

%%  We are not a stand alone document. For a regular table, we will %%
%%  have no preamble and only define the closing to mean nothing.   %%
    \def\gnumericTableEnd{}

%%  If we want landscape mode in an embedded document, comment out  %%
%%  the line above and uncomment the two below. The table will      %%
%%  begin on a new page and run in landscape mode.                  %%
%       \def\gnumericTableEnd{\end{landscape}}
%       \begin{landscape}

%%  End of the else clause for this file being \input.              %%
\fi

%%%%%%%%%%%%%%%%%%%%%%%%%%%%%%%%%%%%%%%%%%%%%%%%%%%%%%%%%%%%%%%%%%%%%%
%%                                                                  %%
%%  The rest is the gnumeric table, except for the closing          %%
%%  statement. Changes below will alter the table's appearance.     %%
%%                                                                  %%
%%%%%%%%%%%%%%%%%%%%%%%%%%%%%%%%%%%%%%%%%%%%%%%%%%%%%%%%%%%%%%%%%%%%%%

\providecommand{\gnumericPB}[1]%
{\let\gnumericTemp=\\#1\let\\=\gnumericTemp\hspace{0pt}}

%%  The default table format retains the relative column widths of  %%
%%  gnumeric. They can easily be changed to c, r or l. In that case %%
%%  you may want to comment out the next line and uncomment the one %%
%%  thereafter                                                      %%
\providecommand\gnumbox{\makebox[0pt]}
%%\providecommand\gnumbox[1][]{\makebox}

%% to adjust positions in multirow situations                       %%
\setlength{\bigstrutjot}{\jot}
\setlength{\extrarowheight}{\doublerulesep}

%%  The \setlongtables command keeps column widths the same across %%
%%  pages. Simply comment out next line for varying column widths.  %%
\setlongtables

\setlength\gnumericTableWidth{%
	30pt+%
	30pt+%
	30pt+%
	30pt+%
0pt}
\def\gumericNumCols{4}
\setlength\gnumericTableWidthComplete{\gnumericTableWidth+\tabcolsep*\gumericNumCols*2+\arrayrulewidth*\gumericNumCols}
\ifthenelse{\lengthtest{\gnumericTableWidthComplete > \textwidth}}%
{\def\gnumericScale{\ratio{\textwidth-\tabcolsep*\gumericNumCols*2-\arrayrulewidth*\gumericNumCols}%
{\gnumericTableWidth}}}%
{\def\gnumericScale{1}}

%%%%%%%%%%%%%%%%%%%%%%%%%%%%%%%%%%%%%%%%%%%%%%%%%%%%%%%%%%%%%%%%%%%%%%
%%                                                                  %%
%% The following are the widths of the various columns. We are      %%
%% defining them here because then they are easier to change.       %%
%% Depending on the cell formats we may use them more than once.    %%
%%                                                                  %%
%%%%%%%%%%%%%%%%%%%%%%%%%%%%%%%%%%%%%%%%%%%%%%%%%%%%%%%%%%%%%%%%%%%%%%

\def\gnumericColB{30pt*\gnumericScale}
\def\gnumericColC{30pt*\gnumericScale}
\def\gnumericColD{30pt*\gnumericScale}
\def\gnumericColE{30pt*\gnumericScale}

\begin{longtable}[c]{%
	b{\gnumericColB}%
	b{\gnumericColC}%
	b{\gnumericColD}%
	b{\gnumericColE}%
	}

%%%%%%%%%%%%%%%%%%%%%%%%%%%%%%%%%%%%%%%%%%%%%%%%%%%%%%%%%%%%%%%%%%%%%%
%%  The longtable options. (Caption. headers... see Goosens. p.124) %%
%	\caption{The Table Caption.}             \\	%
% \hline	% Across the top of the table.
%%  The rest of these options are table rows which are placed on    %%
%%  the first. last or every page. Use \multicolumn if you want.    %%

%%  Header for the first page.                                      %%
%	\multicolumn{4}{c}{The First Header} \\ \hline 
%	\multicolumn{1}{c}{colTag}	%Column 1
%	&\multicolumn{1}{c}{colTag}	%Column 1
%	&\multicolumn{1}{c}{colTag}	%Column 2
%	&\multicolumn{1}{c}{colTag}	%Column 3
%	&\multicolumn{1}{c}{colTag}	\\ \hline %Last column
%	\endfirsthead

%%  The running header definition.                                  %%
%	\hline
%	\multicolumn{4}{l}{\ldots\small\slshape continued} \\ \hline
%	\multicolumn{1}{c}{colTag}	%Column 1
%	&\multicolumn{1}{c}{colTag}	%Column 1
%	&\multicolumn{1}{c}{colTag}	%Column 2
%	&\multicolumn{1}{c}{colTag}	%Column 3
%	&\multicolumn{1}{c}{colTag}	\\ \hline %Last column
%	\endhead

%%  The running footer definition.                                  %%
%	\hline
%	\multicolumn{4}{r}{\small\slshape continued\ldots} \\
%	\endfoot

%%  The ending footer definition.                                   %%
%	\multicolumn{4}{c}{That's all folks} \\ \hline 
%	\endlastfoot
%%%%%%%%%%%%%%%%%%%%%%%%%%%%%%%%%%%%%%%%%%%%%%%%%%%%%%%%%%%%%%%%%%%%%%

\hhline{|-|-|-|-|}
	 \multicolumn{1}{|p{\gnumericColB}|}%
	{\gnumericPB{\raggedleft}\gnumbox[r]{\textsf{$1.60$}}}
	&\multicolumn{1}{p{\gnumericColC}|}%
	{\gnumericPB{\raggedleft}\gnumbox[r]{\textsf{$2.08$}}}
	&\multicolumn{1}{p{\gnumericColD}|}%
	{\gnumericPB{\raggedleft}\gnumbox[r]{\textsf{$1.74$}}}
	&\multicolumn{1}{p{\gnumericColE}|}%
	{\gnumericPB{\raggedleft}\gnumbox[r]{\textsf{$2.66$}}}
\\
\hhline{|----|}
	 \multicolumn{1}{|p{\gnumericColB}|}%
	{\gnumericPB{\raggedleft}\gnumbox[r]{\textsf{$1.61$}}}
	&\multicolumn{1}{p{\gnumericColC}|}%
	{\gnumericPB{\raggedleft}\gnumbox[r]{\textsf{$2.16$}}}
	&\multicolumn{1}{p{\gnumericColD}|}%
	{\gnumericPB{\raggedleft}\gnumbox[r]{\textsf{$1.76$}}}
	&\multicolumn{1}{p{\gnumericColE}|}%
	{\gnumericPB{\raggedleft}\gnumbox[r]{\textsf{$2.70$}}}
\\
\hhline{|----|}
	 \multicolumn{1}{|p{\gnumericColB}|}%
	{\gnumericPB{\raggedleft}\gnumbox[r]{\textsf{$1.63$}}}
	&\multicolumn{1}{p{\gnumericColC}|}%
	{\gnumericPB{\raggedleft}\gnumbox[r]{\textsf{$2.23$}}}
	&\multicolumn{1}{p{\gnumericColD}|}%
	{\gnumericPB{\raggedleft}\gnumbox[r]{\textsf{$1.77$}}}
	&\multicolumn{1}{p{\gnumericColE}|}%
	{\gnumericPB{\raggedleft}\gnumbox[r]{\textsf{$2.74$}}}
\\
\hhline{|----|}
	 \multicolumn{1}{|p{\gnumericColB}|}%
	{\gnumericPB{\raggedleft}\gnumbox[r]{\textsf{$1.64$}}}
	&\multicolumn{1}{p{\gnumericColC}|}%
	{\gnumericPB{\raggedleft}\gnumbox[r]{\textsf{$2.29$}}}
	&\multicolumn{1}{p{\gnumericColD}|}%
	{\gnumericPB{\raggedleft}\gnumbox[r]{\textsf{$1.78$}}}
	&\multicolumn{1}{p{\gnumericColE}|}%
	{\gnumericPB{\raggedleft}\gnumbox[r]{\textsf{$2.79$}}}
\\
\hhline{|----|}
	 \multicolumn{1}{|p{\gnumericColB}|}%
	{\gnumericPB{\raggedleft}\gnumbox[r]{\textsf{$1.66$}}}
	&\multicolumn{1}{p{\gnumericColC}|}%
	{\gnumericPB{\raggedleft}\gnumbox[r]{\textsf{$2.35$}}}
	&\multicolumn{1}{p{\gnumericColD}|}%
	{\gnumericPB{\raggedleft}\gnumbox[r]{\textsf{$1.80$}}}
	&\multicolumn{1}{p{\gnumericColE}|}%
	{\gnumericPB{\raggedleft}\gnumbox[r]{\textsf{$2.83$}}}
\\
\hhline{|----|}
	 \multicolumn{1}{|p{\gnumericColB}|}%
	{\gnumericPB{\raggedleft}\gnumbox[r]{\textsf{$1.67$}}}
	&\multicolumn{1}{p{\gnumericColC}|}%
	{\gnumericPB{\raggedleft}\gnumbox[r]{\textsf{$2.41$}}}
	&\multicolumn{1}{p{\gnumericColD}|}%
	{\gnumericPB{\raggedleft}\gnumbox[r]{\textsf{$1.81$}}}
	&\multicolumn{1}{p{\gnumericColE}|}%
	{\gnumericPB{\raggedleft}\gnumbox[r]{\textsf{$2.87$}}}
\\
\hhline{|----|}
	 \multicolumn{1}{|p{\gnumericColB}|}%
	{\gnumericPB{\raggedleft}\gnumbox[r]{\textsf{$1.69$}}}
	&\multicolumn{1}{p{\gnumericColC}|}%
	{\gnumericPB{\raggedleft}\gnumbox[r]{\textsf{$2.46$}}}
	&\multicolumn{1}{p{\gnumericColD}|}%
	{\gnumericPB{\raggedleft}\gnumbox[r]{\textsf{$1.83$}}}
	&\multicolumn{1}{p{\gnumericColE}|}%
	{\gnumericPB{\raggedleft}\gnumbox[r]{\textsf{$2.91$}}}
\\
\hhline{|----|}
	 \multicolumn{1}{|p{\gnumericColB}|}%
	{\gnumericPB{\raggedleft}\gnumbox[r]{\textsf{$1.70$}}}
	&\multicolumn{1}{p{\gnumericColC}|}%
	{\gnumericPB{\raggedleft}\gnumbox[r]{\textsf{$2.51$}}}
	&\multicolumn{1}{p{\gnumericColD}|}%
	{\gnumericPB{\raggedleft}\gnumbox[r]{\textsf{$1.84$}}}
	&\multicolumn{1}{p{\gnumericColE}|}%
	{\gnumericPB{\raggedleft}\gnumbox[r]{\textsf{$2.95$}}}
\\
\hhline{|----|}
	 \multicolumn{1}{|p{\gnumericColB}|}%
	{\gnumericPB{\raggedleft}\gnumbox[r]{\textsf{$1.71$}}}
	&\multicolumn{1}{p{\gnumericColC}|}%
	{\gnumericPB{\raggedleft}\gnumbox[r]{\textsf{$2.56$}}}
	&\multicolumn{1}{p{\gnumericColD}|}%
	{\gnumericPB{\raggedleft}\gnumbox[r]{\textsf{$1.86$}}}
	&\multicolumn{1}{p{\gnumericColE}|}%
	{\gnumericPB{\raggedleft}\gnumbox[r]{\textsf{$2.99$}}}
\\
\hhline{|----|}
	 \multicolumn{1}{|p{\gnumericColB}|}%
	{\gnumericPB{\raggedleft}\gnumbox[r]{\textsf{$1.73$}}}
	&\multicolumn{1}{p{\gnumericColC}|}%
	{\gnumericPB{\raggedleft}\gnumbox[r]{\textsf{$2.61$}}}
	&\multicolumn{1}{p{\gnumericColD}|}%
	{\gnumericPB{\raggedleft}\gnumbox[r]{\textsf{$1.87$}}}
	&\multicolumn{1}{p{\gnumericColE}|}%
	{\gnumericPB{\raggedleft}\gnumbox[r]{\textsf{$3.02$}}}
\\
\hhline{|-|-|-|-|}
\end{longtable}

\gnumericTableEnd
}
\caption{Normalisation for a gaussian form for $J/\psi$ as a function of  the $c$ quark mass
for $\Lambda=1.8$~GeV.}\label{tab:N_exp_jpsi_mq3}
\end{table}

\begin{table}[H]
{\tiny%%%%%%%%%%%%%%%%%%%%%%%%%%%%%%%%%%%%%%%%%%%%%%%%%%%%%%%%%%%%%%%%%%%%%%
\def\ifundefined#1{\expandafter\ifx\csname#1\endcsname\relax}

%%  Check for the \def token for inputed files. If it is not        %%
%%  defined, the file will be processed as a standalone and the     %%
%%  preamble will be used.                                          %%
\ifundefined{inputGnumericTable}

%%  End of the preamble for the standalone. The next section is for %%
%%  documents which are included into other LaTeX2e files.          %%
\else

%%  We are not a stand alone document. For a regular table, we will %%
%%  have no preamble and only define the closing to mean nothing.   %%
    \def\gnumericTableEnd{}

%%  If we want landscape mode in an embedded document, comment out  %%
%%  the line above and uncomment the two below. The table will      %%
%%  begin on a new page and run in landscape mode.                  %%
%       \def\gnumericTableEnd{\end{landscape}}
%       \begin{landscape}

%%  End of the else clause for this file being \input.              %%
\fi

%%%%%%%%%%%%%%%%%%%%%%%%%%%%%%%%%%%%%%%%%%%%%%%%%%%%%%%%%%%%%%%%%%%%%%
%%                                                                  %%
%%  The rest is the gnumeric table, except for the closing          %%
%%  statement. Changes below will alter the table's appearance.     %%
%%                                                                  %%
%%%%%%%%%%%%%%%%%%%%%%%%%%%%%%%%%%%%%%%%%%%%%%%%%%%%%%%%%%%%%%%%%%%%%%

\providecommand{\gnumericPB}[1]%
{\let\gnumericTemp=\\#1\let\\=\gnumericTemp\hspace{0pt}}

%%  The default table format retains the relative column widths of  %%
%%  gnumeric. They can easily be changed to c, r or l. In that case %%
%%  you may want to comment out the next line and uncomment the one %%
%%  thereafter                                                      %%
\providecommand\gnumbox{\makebox[0pt]}
%%\providecommand\gnumbox[1][]{\makebox}

%% to adjust positions in multirow situations                       %%
\setlength{\bigstrutjot}{\jot}
\setlength{\extrarowheight}{\doublerulesep}

%%  The \setlongtables command keeps column widths the same across %%
%%  pages. Simply comment out next line for varying column widths.  %%
\setlongtables

\setlength\gnumericTableWidth{%
	30pt+%
	30pt+%
	30pt+%
	30pt+%
0pt}
\def\gumericNumCols{4}
\setlength\gnumericTableWidthComplete{\gnumericTableWidth+\tabcolsep*\gumericNumCols*2+\arrayrulewidth*\gumericNumCols}
\ifthenelse{\lengthtest{\gnumericTableWidthComplete > \textwidth}}%
{\def\gnumericScale{\ratio{\textwidth-\tabcolsep*\gumericNumCols*2-\arrayrulewidth*\gumericNumCols}%
{\gnumericTableWidth}}}%
{\def\gnumericScale{1}}

%%%%%%%%%%%%%%%%%%%%%%%%%%%%%%%%%%%%%%%%%%%%%%%%%%%%%%%%%%%%%%%%%%%%%%
%%                                                                  %%
%% The following are the widths of the various columns. We are      %%
%% defining them here because then they are easier to change.       %%
%% Depending on the cell formats we may use them more than once.    %%
%%                                                                  %%
%%%%%%%%%%%%%%%%%%%%%%%%%%%%%%%%%%%%%%%%%%%%%%%%%%%%%%%%%%%%%%%%%%%%%%

\def\gnumericColB{30pt*\gnumericScale}
\def\gnumericColC{30pt*\gnumericScale}
\def\gnumericColD{30pt*\gnumericScale}
\def\gnumericColE{30pt*\gnumericScale}

\begin{longtable}[c]{%
	b{\gnumericColB}%
	b{\gnumericColC}%
	b{\gnumericColD}%
	b{\gnumericColE}%
	}

%%%%%%%%%%%%%%%%%%%%%%%%%%%%%%%%%%%%%%%%%%%%%%%%%%%%%%%%%%%%%%%%%%%%%%
%%  The longtable options. (Caption. headers... see Goosens. p.124) %%
%	\caption{The Table Caption.}             \\	%
% \hline	% Across the top of the table.
%%  The rest of these options are table rows which are placed on    %%
%%  the first. last or every page. Use \multicolumn if you want.    %%

%%  Header for the first page.                                      %%
%	\multicolumn{4}{c}{The First Header} \\ \hline 
%	\multicolumn{1}{c}{colTag}	%Column 1
%	&\multicolumn{1}{c}{colTag}	%Column 1
%	&\multicolumn{1}{c}{colTag}	%Column 2
%	&\multicolumn{1}{c}{colTag}	%Column 3
%	&\multicolumn{1}{c}{colTag}	\\ \hline %Last column
%	\endfirsthead

%%  The running header definition.                                  %%
%	\hline
%	\multicolumn{4}{l}{\ldots\small\slshape continued} \\ \hline
%	\multicolumn{1}{c}{colTag}	%Column 1
%	&\multicolumn{1}{c}{colTag}	%Column 1
%	&\multicolumn{1}{c}{colTag}	%Column 2
%	&\multicolumn{1}{c}{colTag}	%Column 3
%	&\multicolumn{1}{c}{colTag}	\\ \hline %Last column
%	\endhead

%%  The running footer definition.                                  %%
%	\hline
%	\multicolumn{4}{r}{\small\slshape continued\ldots} \\
%	\endfoot

%%  The ending footer definition.                                   %%
%	\multicolumn{4}{c}{That's all folks} \\ \hline 
%	\endlastfoot
%%%%%%%%%%%%%%%%%%%%%%%%%%%%%%%%%%%%%%%%%%%%%%%%%%%%%%%%%%%%%%%%%%%%%%

\hhline{|-|-|-|-|}
	 \multicolumn{1}{|p{\gnumericColB}|}%
	{\gnumericPB{\raggedleft}\gnumbox[r]{\textsf{$1.60$}}}
	&\multicolumn{1}{p{\gnumericColC}|}%
	{\gnumericPB{\raggedleft}\gnumbox[r]{\textsf{$1.52$}}}
	&\multicolumn{1}{p{\gnumericColD}|}%
	{\gnumericPB{\raggedleft}\gnumbox[r]{\textsf{$1.74$}}}
	&\multicolumn{1}{p{\gnumericColE}|}%
	{\gnumericPB{\raggedleft}\gnumbox[r]{\textsf{$1.84$}}}
\\
\hhline{|----|}
	 \multicolumn{1}{|p{\gnumericColB}|}%
	{\gnumericPB{\raggedleft}\gnumbox[r]{\textsf{$1.61$}}}
	&\multicolumn{1}{p{\gnumericColC}|}%
	{\gnumericPB{\raggedleft}\gnumbox[r]{\textsf{$1.56$}}}
	&\multicolumn{1}{p{\gnumericColD}|}%
	{\gnumericPB{\raggedleft}\gnumbox[r]{\textsf{$1.76$}}}
	&\multicolumn{1}{p{\gnumericColE}|}%
	{\gnumericPB{\raggedleft}\gnumbox[r]{\textsf{$1.87$}}}
\\
\hhline{|----|}
	 \multicolumn{1}{|p{\gnumericColB}|}%
	{\gnumericPB{\raggedleft}\gnumbox[r]{\textsf{$1.63$}}}
	&\multicolumn{1}{p{\gnumericColC}|}%
	{\gnumericPB{\raggedleft}\gnumbox[r]{\textsf{$1.60$}}}
	&\multicolumn{1}{p{\gnumericColD}|}%
	{\gnumericPB{\raggedleft}\gnumbox[r]{\textsf{$1.77$}}}
	&\multicolumn{1}{p{\gnumericColE}|}%
	{\gnumericPB{\raggedleft}\gnumbox[r]{\textsf{$1.89$}}}
\\
\hhline{|----|}
	 \multicolumn{1}{|p{\gnumericColB}|}%
	{\gnumericPB{\raggedleft}\gnumbox[r]{\textsf{$1.64$}}}
	&\multicolumn{1}{p{\gnumericColC}|}%
	{\gnumericPB{\raggedleft}\gnumbox[r]{\textsf{$1.64$}}}
	&\multicolumn{1}{p{\gnumericColD}|}%
	{\gnumericPB{\raggedleft}\gnumbox[r]{\textsf{$1.78$}}}
	&\multicolumn{1}{p{\gnumericColE}|}%
	{\gnumericPB{\raggedleft}\gnumbox[r]{\textsf{$1.91$}}}
\\
\hhline{|----|}
	 \multicolumn{1}{|p{\gnumericColB}|}%
	{\gnumericPB{\raggedleft}\gnumbox[r]{\textsf{$1.66$}}}
	&\multicolumn{1}{p{\gnumericColC}|}%
	{\gnumericPB{\raggedleft}\gnumbox[r]{\textsf{$1.67$}}}
	&\multicolumn{1}{p{\gnumericColD}|}%
	{\gnumericPB{\raggedleft}\gnumbox[r]{\textsf{$1.80$}}}
	&\multicolumn{1}{p{\gnumericColE}|}%
	{\gnumericPB{\raggedleft}\gnumbox[r]{\textsf{$1.94$}}}
\\
\hhline{|----|}
	 \multicolumn{1}{|p{\gnumericColB}|}%
	{\gnumericPB{\raggedleft}\gnumbox[r]{\textsf{$1.67$}}}
	&\multicolumn{1}{p{\gnumericColC}|}%
	{\gnumericPB{\raggedleft}\gnumbox[r]{\textsf{$1.70$}}}
	&\multicolumn{1}{p{\gnumericColD}|}%
	{\gnumericPB{\raggedleft}\gnumbox[r]{\textsf{$1.81$}}}
	&\multicolumn{1}{p{\gnumericColE}|}%
	{\gnumericPB{\raggedleft}\gnumbox[r]{\textsf{$1.96$}}}
\\
\hhline{|----|}
	 \multicolumn{1}{|p{\gnumericColB}|}%
	{\gnumericPB{\raggedleft}\gnumbox[r]{\textsf{$1.69$}}}
	&\multicolumn{1}{p{\gnumericColC}|}%
	{\gnumericPB{\raggedleft}\gnumbox[r]{\textsf{$1.73$}}}
	&\multicolumn{1}{p{\gnumericColD}|}%
	{\gnumericPB{\raggedleft}\gnumbox[r]{\textsf{$1.83$}}}
	&\multicolumn{1}{p{\gnumericColE}|}%
	{\gnumericPB{\raggedleft}\gnumbox[r]{\textsf{$1.98$}}}
\\
\hhline{|----|}
	 \multicolumn{1}{|p{\gnumericColB}|}%
	{\gnumericPB{\raggedleft}\gnumbox[r]{\textsf{$1.70$}}}
	&\multicolumn{1}{p{\gnumericColC}|}%
	{\gnumericPB{\raggedleft}\gnumbox[r]{\textsf{$1.76$}}}
	&\multicolumn{1}{p{\gnumericColD}|}%
	{\gnumericPB{\raggedleft}\gnumbox[r]{\textsf{$1.84$}}}
	&\multicolumn{1}{p{\gnumericColE}|}%
	{\gnumericPB{\raggedleft}\gnumbox[r]{\textsf{$2.00$}}}
\\
\hhline{|----|}
	 \multicolumn{1}{|p{\gnumericColB}|}%
	{\gnumericPB{\raggedleft}\gnumbox[r]{\textsf{$1.71$}}}
	&\multicolumn{1}{p{\gnumericColC}|}%
	{\gnumericPB{\raggedleft}\gnumbox[r]{\textsf{$1.79$}}}
	&\multicolumn{1}{p{\gnumericColD}|}%
	{\gnumericPB{\raggedleft}\gnumbox[r]{\textsf{$1.86$}}}
	&\multicolumn{1}{p{\gnumericColE}|}%
	{\gnumericPB{\raggedleft}\gnumbox[r]{\textsf{$2.02$}}}
\\
\hhline{|----|}
	 \multicolumn{1}{|p{\gnumericColB}|}%
	{\gnumericPB{\raggedleft}\gnumbox[r]{\textsf{$1.73$}}}
	&\multicolumn{1}{p{\gnumericColC}|}%
	{\gnumericPB{\raggedleft}\gnumbox[r]{\textsf{$1.82$}}}
	&\multicolumn{1}{p{\gnumericColD}|}%
	{\gnumericPB{\raggedleft}\gnumbox[r]{\textsf{$1.87$}}}
	&\multicolumn{1}{p{\gnumericColE}|}%
	{\gnumericPB{\raggedleft}\gnumbox[r]{\textsf{$2.04$}}}
\\
\hhline{|-|-|-|-|}
\end{longtable}

\gnumericTableEnd
}
\caption{Normalisation for a gaussian form for $J/\psi$ as a function of  the $c$ quark mass
for $\Lambda=2.2$~GeV.}\label{tab:N_exp_jpsi_mq4}
\end{table}

%*******************************************
%*******************************************

\begin{table}[H]
{\tiny%%%%%%%%%%%%%%%%%%%%%%%%%%%%%%%%%%%%%%%%%%%%%%%%%%%%%%%%%%%%%%%%%%%%%%
\def\ifundefined#1{\expandafter\ifx\csname#1\endcsname\relax}

%%  Check for the \def token for inputed files. If it is not        %%
%%  defined, the file will be processed as a standalone and the     %%
%%  preamble will be used.                                          %%
\ifundefined{inputGnumericTable}

%%  End of the preamble for the standalone. The next section is for %%
%%  documents which are included into other LaTeX2e files.          %%
\else

%%  We are not a stand alone document. For a regular table, we will %%
%%  have no preamble and only define the closing to mean nothing.   %%
    \def\gnumericTableEnd{}

%%  If we want landscape mode in an embedded document, comment out  %%
%%  the line above and uncomment the two below. The table will      %%
%%  begin on a new page and run in landscape mode.                  %%
%       \def\gnumericTableEnd{\end{landscape}}
%       \begin{landscape}

%%  End of the else clause for this file being \input.              %%
\fi

%%%%%%%%%%%%%%%%%%%%%%%%%%%%%%%%%%%%%%%%%%%%%%%%%%%%%%%%%%%%%%%%%%%%%%
%%                                                                  %%
%%  The rest is the gnumeric table, except for the closing          %%
%%  statement. Changes below will alter the table's appearance.     %%
%%                                                                  %%
%%%%%%%%%%%%%%%%%%%%%%%%%%%%%%%%%%%%%%%%%%%%%%%%%%%%%%%%%%%%%%%%%%%%%%

\providecommand{\gnumericPB}[1]%
{\let\gnumericTemp=\\#1\let\\=\gnumericTemp\hspace{0pt}}

%%  The default table format retains the relative column widths of  %%
%%  gnumeric. They can easily be changed to c, r or l. In that case %%
%%  you may want to comment out the next line and uncomment the one %%
%%  thereafter                                                      %%
\providecommand\gnumbox{\makebox[0pt]}
%%\providecommand\gnumbox[1][]{\makebox}

%% to adjust positions in multirow situations                       %%
\setlength{\bigstrutjot}{\jot}
\setlength{\extrarowheight}{\doublerulesep}

%%  The \setlongtables command keeps column widths the same across %%
%%  pages. Simply comment out next line for varying column widths.  %%
\setlongtables

\setlength\gnumericTableWidth{%
	30pt+%
	30pt+%
	30pt+%
	30pt+%
0pt}
\def\gumericNumCols{4}
\setlength\gnumericTableWidthComplete{\gnumericTableWidth+\tabcolsep*\gumericNumCols*2+\arrayrulewidth*\gumericNumCols}
\ifthenelse{\lengthtest{\gnumericTableWidthComplete > \textwidth}}%
{\def\gnumericScale{\ratio{\textwidth-\tabcolsep*\gumericNumCols*2-\arrayrulewidth*\gumericNumCols}%
{\gnumericTableWidth}}}%
{\def\gnumericScale{1}}

%%%%%%%%%%%%%%%%%%%%%%%%%%%%%%%%%%%%%%%%%%%%%%%%%%%%%%%%%%%%%%%%%%%%%%
%%                                                                  %%
%% The following are the widths of the various columns. We are      %%
%% defining them here because then they are easier to change.       %%
%% Depending on the cell formats we may use them more than once.    %%
%%                                                                  %%
%%%%%%%%%%%%%%%%%%%%%%%%%%%%%%%%%%%%%%%%%%%%%%%%%%%%%%%%%%%%%%%%%%%%%%

\def\gnumericColB{30pt*\gnumericScale}
\def\gnumericColC{30pt*\gnumericScale}
\def\gnumericColD{30pt*\gnumericScale}
\def\gnumericColE{30pt*\gnumericScale}

\begin{longtable}[c]{%
	b{\gnumericColB}%
	b{\gnumericColC}%
	b{\gnumericColD}%
	b{\gnumericColE}%
	}

%%%%%%%%%%%%%%%%%%%%%%%%%%%%%%%%%%%%%%%%%%%%%%%%%%%%%%%%%%%%%%%%%%%%%%
%%  The longtable options. (Caption. headers... see Goosens. p.124) %%
%	\caption{The Table Caption.}             \\	%
% \hline	% Across the top of the table.
%%  The rest of these options are table rows which are placed on    %%
%%  the first. last or every page. Use \multicolumn if you want.    %%

%%  Header for the first page.                                      %%
%	\multicolumn{4}{c}{The First Header} \\ \hline 
%	\multicolumn{1}{c}{colTag}	%Column 1
%	&\multicolumn{1}{c}{colTag}	%Column 1
%	&\multicolumn{1}{c}{colTag}	%Column 2
%	&\multicolumn{1}{c}{colTag}	%Column 3
%	&\multicolumn{1}{c}{colTag}	\\ \hline %Last column
%	\endfirsthead

%%  The running header definition.                                  %%
%	\hline
%	\multicolumn{4}{l}{\ldots\small\slshape continued} \\ \hline
%	\multicolumn{1}{c}{colTag}	%Column 1
%	&\multicolumn{1}{c}{colTag}	%Column 1
%	&\multicolumn{1}{c}{colTag}	%Column 2
%	&\multicolumn{1}{c}{colTag}	%Column 3
%	&\multicolumn{1}{c}{colTag}	\\ \hline %Last column
%	\endhead

%%  The running footer definition.                                  %%
%	\hline
%	\multicolumn{4}{r}{\small\slshape continued\ldots} \\
%	\endfoot

%%  The ending footer definition.                                   %%
%	\multicolumn{4}{c}{That's all folks} \\ \hline 
%	\endlastfoot
%%%%%%%%%%%%%%%%%%%%%%%%%%%%%%%%%%%%%%%%%%%%%%%%%%%%%%%%%%%%%%%%%%%%%%

\hhline{|-|-|-|-|}
	 \multicolumn{1}{|p{\gnumericColB}|}%
	{\gnumericPB{\raggedleft}\gnumbox[r]{\textsf{$\Lambda$}}}
	&\multicolumn{1}{p{\gnumericColC}|}%
	{\gnumericPB{\raggedleft}\gnumbox[r]{\textsf{$N$}}}
	&\multicolumn{1}{p{\gnumericColD}|}%
	{\gnumericPB{\raggedleft}\gnumbox[r]{\textsf{$\Lambda$}}}
	&\multicolumn{1}{p{\gnumericColE}|}%
	{\gnumericPB{\raggedleft}\gnumbox[r]{\textsf{$N$}}}
\\
\hhline{|----|}
	 \multicolumn{1}{|p{\gnumericColB}|}%
	{\gnumericPB{\raggedleft}\gnumbox[r]{\textsf{$3.00$}}}
	&\multicolumn{1}{p{\gnumericColC}|}%
	{\gnumericPB{\raggedleft}\gnumbox[r]{\textsf{$2.84$}}}
	&\multicolumn{1}{p{\gnumericColD}|}%
	{\gnumericPB{\raggedleft}\gnumbox[r]{\textsf{$4.71$}}}
	&\multicolumn{1}{p{\gnumericColE}|}%
	{\gnumericPB{\raggedleft}\gnumbox[r]{\textsf{$1.43$}}}
\\
\hhline{|----|}
	 \multicolumn{1}{|p{\gnumericColB}|}%
	{\gnumericPB{\raggedleft}\gnumbox[r]{\textsf{$3.17$}}}
	&\multicolumn{1}{p{\gnumericColC}|}%
	{\gnumericPB{\raggedleft}\gnumbox[r]{\textsf{$2.61$}}}
	&\multicolumn{1}{p{\gnumericColD}|}%
	{\gnumericPB{\raggedleft}\gnumbox[r]{\textsf{$4.88$}}}
	&\multicolumn{1}{p{\gnumericColE}|}%
	{\gnumericPB{\raggedleft}\gnumbox[r]{\textsf{$1.35$}}}
\\
\hhline{|----|}
	 \multicolumn{1}{|p{\gnumericColB}|}%
	{\gnumericPB{\raggedleft}\gnumbox[r]{\textsf{$3.34$}}}
	&\multicolumn{1}{p{\gnumericColC}|}%
	{\gnumericPB{\raggedleft}\gnumbox[r]{\textsf{$2.40$}}}
	&\multicolumn{1}{p{\gnumericColD}|}%
	{\gnumericPB{\raggedleft}\gnumbox[r]{\textsf{$5.05$}}}
	&\multicolumn{1}{p{\gnumericColE}|}%
	{\gnumericPB{\raggedleft}\gnumbox[r]{\textsf{$1.29$}}}
\\
\hhline{|----|}
	 \multicolumn{1}{|p{\gnumericColB}|}%
	{\gnumericPB{\raggedleft}\gnumbox[r]{\textsf{$3.51$}}}
	&\multicolumn{1}{p{\gnumericColC}|}%
	{\gnumericPB{\raggedleft}\gnumbox[r]{\textsf{$2.23$}}}
	&\multicolumn{1}{p{\gnumericColD}|}%
	{\gnumericPB{\raggedleft}\gnumbox[r]{\textsf{$5.22$}}}
	&\multicolumn{1}{p{\gnumericColE}|}%
	{\gnumericPB{\raggedleft}\gnumbox[r]{\textsf{$1.22$}}}
\\
\hhline{|----|}
	 \multicolumn{1}{|p{\gnumericColB}|}%
	{\gnumericPB{\raggedleft}\gnumbox[r]{\textsf{$3.68$}}}
	&\multicolumn{1}{p{\gnumericColC}|}%
	{\gnumericPB{\raggedleft}\gnumbox[r]{\textsf{$2.07$}}}
	&\multicolumn{1}{p{\gnumericColD}|}%
	{\gnumericPB{\raggedleft}\gnumbox[r]{\textsf{$5.39$}}}
	&\multicolumn{1}{p{\gnumericColE}|}%
	{\gnumericPB{\raggedleft}\gnumbox[r]{\textsf{$1.16$}}}
\\
\hhline{|----|}
	 \multicolumn{1}{|p{\gnumericColB}|}%
	{\gnumericPB{\raggedleft}\gnumbox[r]{\textsf{$3.86$}}}
	&\multicolumn{1}{p{\gnumericColC}|}%
	{\gnumericPB{\raggedleft}\gnumbox[r]{\textsf{$1.93$}}}
	&\multicolumn{1}{p{\gnumericColD}|}%
	{\gnumericPB{\raggedleft}\gnumbox[r]{\textsf{$5.57$}}}
	&\multicolumn{1}{p{\gnumericColE}|}%
	{\gnumericPB{\raggedleft}\gnumbox[r]{\textsf{$1.11$}}}
\\
\hhline{|----|}
	 \multicolumn{1}{|p{\gnumericColB}|}%
	{\gnumericPB{\raggedleft}\gnumbox[r]{\textsf{$4.03$}}}
	&\multicolumn{1}{p{\gnumericColC}|}%
	{\gnumericPB{\raggedleft}\gnumbox[r]{\textsf{$1.81$}}}
	&\multicolumn{1}{p{\gnumericColD}|}%
	{\gnumericPB{\raggedleft}\gnumbox[r]{\textsf{$5.74$}}}
	&\multicolumn{1}{p{\gnumericColE}|}%
	{\gnumericPB{\raggedleft}\gnumbox[r]{\textsf{$1.06$}}}
\\
\hhline{|----|}
	 \multicolumn{1}{|p{\gnumericColB}|}%
	{\gnumericPB{\raggedleft}\gnumbox[r]{\textsf{$4.20$}}}
	&\multicolumn{1}{p{\gnumericColC}|}%
	{\gnumericPB{\raggedleft}\gnumbox[r]{\textsf{$1.70$}}}
	&\multicolumn{1}{p{\gnumericColD}|}%
	{\gnumericPB{\raggedleft}\gnumbox[r]{\textsf{$5.91$}}}
	&\multicolumn{1}{p{\gnumericColE}|}%
	{\gnumericPB{\raggedleft}\gnumbox[r]{\textsf{$1.01$}}}
\\
\hhline{|----|}
	 \multicolumn{1}{|p{\gnumericColB}|}%
	{\gnumericPB{\raggedleft}\gnumbox[r]{\textsf{$4.37$}}}
	&\multicolumn{1}{p{\gnumericColC}|}%
	{\gnumericPB{\raggedleft}\gnumbox[r]{\textsf{$1.60$}}}
	&\multicolumn{1}{p{\gnumericColD}|}%
	{\gnumericPB{\raggedleft}\gnumbox[r]{\textsf{$6.08$}}}
	&\multicolumn{1}{p{\gnumericColE}|}%
	{\gnumericPB{\raggedleft}\gnumbox[r]{\textsf{$0.97$}}}
\\
\hhline{|----|}
	 \multicolumn{1}{|p{\gnumericColB}|}%
	{\gnumericPB{\raggedleft}\gnumbox[r]{\textsf{$4.54$}}}
	&\multicolumn{1}{p{\gnumericColC}|}%
	{\gnumericPB{\raggedleft}\gnumbox[r]{\textsf{$1.51$}}}
	&\multicolumn{1}{p{\gnumericColD}|}%
	{\gnumericPB{\raggedleft}\gnumbox[r]{\textsf{$6.25$}}}
	&\multicolumn{1}{p{\gnumericColE}|}%
	{\gnumericPB{\raggedleft}\gnumbox[r]{\textsf{$0.93$}}}
\\
\hhline{|-|-|-|-|}
\end{longtable}

\gnumericTableEnd
}
\caption{Normalisation for a dipolar form for $\Upsilon(1S)$ as a function of $\Lambda$ for $m_b=4.80$~GeV.}\label{tab:N_dip_Upsi_lam1}
\end{table}

\begin{table}[H]
{\tiny%%%%%%%%%%%%%%%%%%%%%%%%%%%%%%%%%%%%%%%%%%%%%%%%%%%%%%%%%%%%%%%%%%%%%%
\def\ifundefined#1{\expandafter\ifx\csname#1\endcsname\relax}

%%  Check for the \def token for inputed files. If it is not        %%
%%  defined, the file will be processed as a standalone and the     %%
%%  preamble will be used.                                          %%
\ifundefined{inputGnumericTable}

%%  End of the preamble for the standalone. The next section is for %%
%%  documents which are included into other LaTeX2e files.          %%
\else

%%  We are not a stand alone document. For a regular table, we will %%
%%  have no preamble and only define the closing to mean nothing.   %%
    \def\gnumericTableEnd{}

%%  If we want landscape mode in an embedded document, comment out  %%
%%  the line above and uncomment the two below. The table will      %%
%%  begin on a new page and run in landscape mode.                  %%
%       \def\gnumericTableEnd{\end{landscape}}
%       \begin{landscape}

%%  End of the else clause for this file being \input.              %%
\fi

%%%%%%%%%%%%%%%%%%%%%%%%%%%%%%%%%%%%%%%%%%%%%%%%%%%%%%%%%%%%%%%%%%%%%%
%%                                                                  %%
%%  The rest is the gnumeric table, except for the closing          %%
%%  statement. Changes below will alter the table's appearance.     %%
%%                                                                  %%
%%%%%%%%%%%%%%%%%%%%%%%%%%%%%%%%%%%%%%%%%%%%%%%%%%%%%%%%%%%%%%%%%%%%%%

\providecommand{\gnumericPB}[1]%
{\let\gnumericTemp=\\#1\let\\=\gnumericTemp\hspace{0pt}}

%%  The default table format retains the relative column widths of  %%
%%  gnumeric. They can easily be changed to c, r or l. In that case %%
%%  you may want to comment out the next line and uncomment the one %%
%%  thereafter                                                      %%
\providecommand\gnumbox{\makebox[0pt]}
%%\providecommand\gnumbox[1][]{\makebox}

%% to adjust positions in multirow situations                       %%
\setlength{\bigstrutjot}{\jot}
\setlength{\extrarowheight}{\doublerulesep}

%%  The \setlongtables command keeps column widths the same across %%
%%  pages. Simply comment out next line for varying column widths.  %%
\setlongtables

\setlength\gnumericTableWidth{%
	30pt+%
	30pt+%
	30pt+%
	30pt+%
0pt}
\def\gumericNumCols{4}
\setlength\gnumericTableWidthComplete{\gnumericTableWidth+\tabcolsep*\gumericNumCols*2+\arrayrulewidth*\gumericNumCols}
\ifthenelse{\lengthtest{\gnumericTableWidthComplete > \textwidth}}%
{\def\gnumericScale{\ratio{\textwidth-\tabcolsep*\gumericNumCols*2-\arrayrulewidth*\gumericNumCols}%
{\gnumericTableWidth}}}%
{\def\gnumericScale{1}}

%%%%%%%%%%%%%%%%%%%%%%%%%%%%%%%%%%%%%%%%%%%%%%%%%%%%%%%%%%%%%%%%%%%%%%
%%                                                                  %%
%% The following are the widths of the various columns. We are      %%
%% defining them here because then they are easier to change.       %%
%% Depending on the cell formats we may use them more than once.    %%
%%                                                                  %%
%%%%%%%%%%%%%%%%%%%%%%%%%%%%%%%%%%%%%%%%%%%%%%%%%%%%%%%%%%%%%%%%%%%%%%

\def\gnumericColB{30pt*\gnumericScale}
\def\gnumericColC{30pt*\gnumericScale}
\def\gnumericColD{30pt*\gnumericScale}
\def\gnumericColE{30pt*\gnumericScale}

\begin{longtable}[c]{%
	b{\gnumericColB}%
	b{\gnumericColC}%
	b{\gnumericColD}%
	b{\gnumericColE}%
	}

%%%%%%%%%%%%%%%%%%%%%%%%%%%%%%%%%%%%%%%%%%%%%%%%%%%%%%%%%%%%%%%%%%%%%%
%%  The longtable options. (Caption. headers... see Goosens. p.124) %%
%	\caption{The Table Caption.}             \\	%
% \hline	% Across the top of the table.
%%  The rest of these options are table rows which are placed on    %%
%%  the first. last or every page. Use \multicolumn if you want.    %%

%%  Header for the first page.                                      %%
%	\multicolumn{4}{c}{The First Header} \\ \hline 
%	\multicolumn{1}{c}{colTag}	%Column 1
%	&\multicolumn{1}{c}{colTag}	%Column 1
%	&\multicolumn{1}{c}{colTag}	%Column 2
%	&\multicolumn{1}{c}{colTag}	%Column 3
%	&\multicolumn{1}{c}{colTag}	\\ \hline %Last column
%	\endfirsthead

%%  The running header definition.                                  %%
%	\hline
%	\multicolumn{4}{l}{\ldots\small\slshape continued} \\ \hline
%	\multicolumn{1}{c}{colTag}	%Column 1
%	&\multicolumn{1}{c}{colTag}	%Column 1
%	&\multicolumn{1}{c}{colTag}	%Column 2
%	&\multicolumn{1}{c}{colTag}	%Column 3
%	&\multicolumn{1}{c}{colTag}	\\ \hline %Last column
%	\endhead

%%  The running footer definition.                                  %%
%	\hline
%	\multicolumn{4}{r}{\small\slshape continued\ldots} \\
%	\endfoot

%%  The ending footer definition.                                   %%
%	\multicolumn{4}{c}{That's all folks} \\ \hline 
%	\endlastfoot
%%%%%%%%%%%%%%%%%%%%%%%%%%%%%%%%%%%%%%%%%%%%%%%%%%%%%%%%%%%%%%%%%%%%%%

\hhline{|-|-|-|-|}
	 \multicolumn{1}{|p{\gnumericColB}|}%
	{\gnumericPB{\raggedleft}\gnumbox[r]{\textsf{$3.00$}}}
	&\multicolumn{1}{p{\gnumericColC}|}%
	{\gnumericPB{\raggedleft}\gnumbox[r]{\textsf{$4.06$}}}
	&\multicolumn{1}{p{\gnumericColD}|}%
	{\gnumericPB{\raggedleft}\gnumbox[r]{\textsf{$4.71$}}}
	&\multicolumn{1}{p{\gnumericColE}|}%
	{\gnumericPB{\raggedleft}\gnumbox[r]{\textsf{$1.80$}}}
\\
\hhline{|----|}
	 \multicolumn{1}{|p{\gnumericColB}|}%
	{\gnumericPB{\raggedleft}\gnumbox[r]{\textsf{$3.17$}}}
	&\multicolumn{1}{p{\gnumericColC}|}%
	{\gnumericPB{\raggedleft}\gnumbox[r]{\textsf{$3.66$}}}
	&\multicolumn{1}{p{\gnumericColD}|}%
	{\gnumericPB{\raggedleft}\gnumbox[r]{\textsf{$4.88$}}}
	&\multicolumn{1}{p{\gnumericColE}|}%
	{\gnumericPB{\raggedleft}\gnumbox[r]{\textsf{$1.69$}}}
\\
\hhline{|----|}
	 \multicolumn{1}{|p{\gnumericColB}|}%
	{\gnumericPB{\raggedleft}\gnumbox[r]{\textsf{$3.34$}}}
	&\multicolumn{1}{p{\gnumericColC}|}%
	{\gnumericPB{\raggedleft}\gnumbox[r]{\textsf{$3.33$}}}
	&\multicolumn{1}{p{\gnumericColD}|}%
	{\gnumericPB{\raggedleft}\gnumbox[r]{\textsf{$5.05$}}}
	&\multicolumn{1}{p{\gnumericColE}|}%
	{\gnumericPB{\raggedleft}\gnumbox[r]{\textsf{$1.59$}}}
\\
\hhline{|----|}
	 \multicolumn{1}{|p{\gnumericColB}|}%
	{\gnumericPB{\raggedleft}\gnumbox[r]{\textsf{$3.51$}}}
	&\multicolumn{1}{p{\gnumericColC}|}%
	{\gnumericPB{\raggedleft}\gnumbox[r]{\textsf{$3.04$}}}
	&\multicolumn{1}{p{\gnumericColD}|}%
	{\gnumericPB{\raggedleft}\gnumbox[r]{\textsf{$5.22$}}}
	&\multicolumn{1}{p{\gnumericColE}|}%
	{\gnumericPB{\raggedleft}\gnumbox[r]{\textsf{$1.50$}}}
\\
\hhline{|----|}
	 \multicolumn{1}{|p{\gnumericColB}|}%
	{\gnumericPB{\raggedleft}\gnumbox[r]{\textsf{$3.68$}}}
	&\multicolumn{1}{p{\gnumericColC}|}%
	{\gnumericPB{\raggedleft}\gnumbox[r]{\textsf{$2.79$}}}
	&\multicolumn{1}{p{\gnumericColD}|}%
	{\gnumericPB{\raggedleft}\gnumbox[r]{\textsf{$5.39$}}}
	&\multicolumn{1}{p{\gnumericColE}|}%
	{\gnumericPB{\raggedleft}\gnumbox[r]{\textsf{$1.42$}}}
\\
\hhline{|----|}
	 \multicolumn{1}{|p{\gnumericColB}|}%
	{\gnumericPB{\raggedleft}\gnumbox[r]{\textsf{$3.86$}}}
	&\multicolumn{1}{p{\gnumericColC}|}%
	{\gnumericPB{\raggedleft}\gnumbox[r]{\textsf{$2.57$}}}
	&\multicolumn{1}{p{\gnumericColD}|}%
	{\gnumericPB{\raggedleft}\gnumbox[r]{\textsf{$5.57$}}}
	&\multicolumn{1}{p{\gnumericColE}|}%
	{\gnumericPB{\raggedleft}\gnumbox[r]{\textsf{$1.35$}}}
\\
\hhline{|----|}
	 \multicolumn{1}{|p{\gnumericColB}|}%
	{\gnumericPB{\raggedleft}\gnumbox[r]{\textsf{$4.03$}}}
	&\multicolumn{1}{p{\gnumericColC}|}%
	{\gnumericPB{\raggedleft}\gnumbox[r]{\textsf{$2.38$}}}
	&\multicolumn{1}{p{\gnumericColD}|}%
	{\gnumericPB{\raggedleft}\gnumbox[r]{\textsf{$5.74$}}}
	&\multicolumn{1}{p{\gnumericColE}|}%
	{\gnumericPB{\raggedleft}\gnumbox[r]{\textsf{$1.28$}}}
\\
\hhline{|----|}
	 \multicolumn{1}{|p{\gnumericColB}|}%
	{\gnumericPB{\raggedleft}\gnumbox[r]{\textsf{$4.20$}}}
	&\multicolumn{1}{p{\gnumericColC}|}%
	{\gnumericPB{\raggedleft}\gnumbox[r]{\textsf{$2.21$}}}
	&\multicolumn{1}{p{\gnumericColD}|}%
	{\gnumericPB{\raggedleft}\gnumbox[r]{\textsf{$5.91$}}}
	&\multicolumn{1}{p{\gnumericColE}|}%
	{\gnumericPB{\raggedleft}\gnumbox[r]{\textsf{$1.21$}}}
\\
\hhline{|----|}
	 \multicolumn{1}{|p{\gnumericColB}|}%
	{\gnumericPB{\raggedleft}\gnumbox[r]{\textsf{$4.37$}}}
	&\multicolumn{1}{p{\gnumericColC}|}%
	{\gnumericPB{\raggedleft}\gnumbox[r]{\textsf{$2.06$}}}
	&\multicolumn{1}{p{\gnumericColD}|}%
	{\gnumericPB{\raggedleft}\gnumbox[r]{\textsf{$6.08$}}}
	&\multicolumn{1}{p{\gnumericColE}|}%
	{\gnumericPB{\raggedleft}\gnumbox[r]{\textsf{$1.16$}}}
\\
\hhline{|----|}
	 \multicolumn{1}{|p{\gnumericColB}|}%
	{\gnumericPB{\raggedleft}\gnumbox[r]{\textsf{$4.54$}}}
	&\multicolumn{1}{p{\gnumericColC}|}%
	{\gnumericPB{\raggedleft}\gnumbox[r]{\textsf{$1.92$}}}
	&\multicolumn{1}{p{\gnumericColD}|}%
	{\gnumericPB{\raggedleft}\gnumbox[r]{\textsf{$6.25$}}}
	&\multicolumn{1}{p{\gnumericColE}|}%
	{\gnumericPB{\raggedleft}\gnumbox[r]{\textsf{$1.10$}}}
\\
\hhline{|-|-|-|-|}
\end{longtable}

\gnumericTableEnd
}
\caption{Normalisation for a dipolar form for $\Upsilon(1S)$ as a function of $\Lambda$ for $m_b=5.00$~GeV.}\label{tab:N_dip_Upsi_lam2}
\end{table}

\begin{table}[H]
{\tiny%%%%%%%%%%%%%%%%%%%%%%%%%%%%%%%%%%%%%%%%%%%%%%%%%%%%%%%%%%%%%%%%%%%%%%
\def\ifundefined#1{\expandafter\ifx\csname#1\endcsname\relax}

%%  Check for the \def token for inputed files. If it is not        %%
%%  defined, the file will be processed as a standalone and the     %%
%%  preamble will be used.                                          %%
\ifundefined{inputGnumericTable}

%%  End of the preamble for the standalone. The next section is for %%
%%  documents which are included into other LaTeX2e files.          %%
\else

%%  We are not a stand alone document. For a regular table, we will %%
%%  have no preamble and only define the closing to mean nothing.   %%
    \def\gnumericTableEnd{}

%%  If we want landscape mode in an embedded document, comment out  %%
%%  the line above and uncomment the two below. The table will      %%
%%  begin on a new page and run in landscape mode.                  %%
%       \def\gnumericTableEnd{\end{landscape}}
%       \begin{landscape}

%%  End of the else clause for this file being \input.              %%
\fi

%%%%%%%%%%%%%%%%%%%%%%%%%%%%%%%%%%%%%%%%%%%%%%%%%%%%%%%%%%%%%%%%%%%%%%
%%                                                                  %%
%%  The rest is the gnumeric table, except for the closing          %%
%%  statement. Changes below will alter the table's appearance.     %%
%%                                                                  %%
%%%%%%%%%%%%%%%%%%%%%%%%%%%%%%%%%%%%%%%%%%%%%%%%%%%%%%%%%%%%%%%%%%%%%%

\providecommand{\gnumericPB}[1]%
{\let\gnumericTemp=\\#1\let\\=\gnumericTemp\hspace{0pt}}

%%  The default table format retains the relative column widths of  %%
%%  gnumeric. They can easily be changed to c, r or l. In that case %%
%%  you may want to comment out the next line and uncomment the one %%
%%  thereafter                                                      %%
\providecommand\gnumbox{\makebox[0pt]}
%%\providecommand\gnumbox[1][]{\makebox}

%% to adjust positions in multirow situations                       %%
\setlength{\bigstrutjot}{\jot}
\setlength{\extrarowheight}{\doublerulesep}

%%  The \setlongtables command keeps column widths the same across %%
%%  pages. Simply comment out next line for varying column widths.  %%
\setlongtables

\setlength\gnumericTableWidth{%
	30pt+%
	30pt+%
	30pt+%
	30pt+%
0pt}
\def\gumericNumCols{4}
\setlength\gnumericTableWidthComplete{\gnumericTableWidth+\tabcolsep*\gumericNumCols*2+\arrayrulewidth*\gumericNumCols}
\ifthenelse{\lengthtest{\gnumericTableWidthComplete > \textwidth}}%
{\def\gnumericScale{\ratio{\textwidth-\tabcolsep*\gumericNumCols*2-\arrayrulewidth*\gumericNumCols}%
{\gnumericTableWidth}}}%
{\def\gnumericScale{1}}

%%%%%%%%%%%%%%%%%%%%%%%%%%%%%%%%%%%%%%%%%%%%%%%%%%%%%%%%%%%%%%%%%%%%%%
%%                                                                  %%
%% The following are the widths of the various columns. We are      %%
%% defining them here because then they are easier to change.       %%
%% Depending on the cell formats we may use them more than once.    %%
%%                                                                  %%
%%%%%%%%%%%%%%%%%%%%%%%%%%%%%%%%%%%%%%%%%%%%%%%%%%%%%%%%%%%%%%%%%%%%%%

\def\gnumericColB{30pt*\gnumericScale}
\def\gnumericColC{30pt*\gnumericScale}
\def\gnumericColD{30pt*\gnumericScale}
\def\gnumericColE{30pt*\gnumericScale}

\begin{longtable}[c]{%
	b{\gnumericColB}%
	b{\gnumericColC}%
	b{\gnumericColD}%
	b{\gnumericColE}%
	}

%%%%%%%%%%%%%%%%%%%%%%%%%%%%%%%%%%%%%%%%%%%%%%%%%%%%%%%%%%%%%%%%%%%%%%
%%  The longtable options. (Caption. headers... see Goosens. p.124) %%
%	\caption{The Table Caption.}             \\	%
% \hline	% Across the top of the table.
%%  The rest of these options are table rows which are placed on    %%
%%  the first. last or every page. Use \multicolumn if you want.    %%

%%  Header for the first page.                                      %%
%	\multicolumn{4}{c}{The First Header} \\ \hline 
%	\multicolumn{1}{c}{colTag}	%Column 1
%	&\multicolumn{1}{c}{colTag}	%Column 1
%	&\multicolumn{1}{c}{colTag}	%Column 2
%	&\multicolumn{1}{c}{colTag}	%Column 3
%	&\multicolumn{1}{c}{colTag}	\\ \hline %Last column
%	\endfirsthead

%%  The running header definition.                                  %%
%	\hline
%	\multicolumn{4}{l}{\ldots\small\slshape continued} \\ \hline
%	\multicolumn{1}{c}{colTag}	%Column 1
%	&\multicolumn{1}{c}{colTag}	%Column 1
%	&\multicolumn{1}{c}{colTag}	%Column 2
%	&\multicolumn{1}{c}{colTag}	%Column 3
%	&\multicolumn{1}{c}{colTag}	\\ \hline %Last column
%	\endhead

%%  The running footer definition.                                  %%
%	\hline
%	\multicolumn{4}{r}{\small\slshape continued\ldots} \\
%	\endfoot

%%  The ending footer definition.                                   %%
%	\multicolumn{4}{c}{That's all folks} \\ \hline 
%	\endlastfoot
%%%%%%%%%%%%%%%%%%%%%%%%%%%%%%%%%%%%%%%%%%%%%%%%%%%%%%%%%%%%%%%%%%%%%%

\hhline{|-|-|-|-|}
	 \multicolumn{1}{|p{\gnumericColB}|}%
	{\gnumericPB{\raggedleft}\gnumbox[r]{\textsf{$3.00$}}}
	&\multicolumn{1}{p{\gnumericColC}|}%
	{\gnumericPB{\raggedleft}\gnumbox[r]{\textsf{$4.88$}}}
	&\multicolumn{1}{p{\gnumericColD}|}%
	{\gnumericPB{\raggedleft}\gnumbox[r]{\textsf{$4.71$}}}
	&\multicolumn{1}{p{\gnumericColE}|}%
	{\gnumericPB{\raggedleft}\gnumbox[r]{\textsf{$2.04$}}}
\\
\hhline{|----|}
	 \multicolumn{1}{|p{\gnumericColB}|}%
	{\gnumericPB{\raggedleft}\gnumbox[r]{\textsf{$3.17$}}}
	&\multicolumn{1}{p{\gnumericColC}|}%
	{\gnumericPB{\raggedleft}\gnumbox[r]{\textsf{$4.37$}}}
	&\multicolumn{1}{p{\gnumericColD}|}%
	{\gnumericPB{\raggedleft}\gnumbox[r]{\textsf{$4.88$}}}
	&\multicolumn{1}{p{\gnumericColE}|}%
	{\gnumericPB{\raggedleft}\gnumbox[r]{\textsf{$1.91$}}}
\\
\hhline{|----|}
	 \multicolumn{1}{|p{\gnumericColB}|}%
	{\gnumericPB{\raggedleft}\gnumbox[r]{\textsf{$3.34$}}}
	&\multicolumn{1}{p{\gnumericColC}|}%
	{\gnumericPB{\raggedleft}\gnumbox[r]{\textsf{$3.94$}}}
	&\multicolumn{1}{p{\gnumericColD}|}%
	{\gnumericPB{\raggedleft}\gnumbox[r]{\textsf{$5.05$}}}
	&\multicolumn{1}{p{\gnumericColE}|}%
	{\gnumericPB{\raggedleft}\gnumbox[r]{\textsf{$1.79$}}}
\\
\hhline{|----|}
	 \multicolumn{1}{|p{\gnumericColB}|}%
	{\gnumericPB{\raggedleft}\gnumbox[r]{\textsf{$3.51$}}}
	&\multicolumn{1}{p{\gnumericColC}|}%
	{\gnumericPB{\raggedleft}\gnumbox[r]{\textsf{$3.57$}}}
	&\multicolumn{1}{p{\gnumericColD}|}%
	{\gnumericPB{\raggedleft}\gnumbox[r]{\textsf{$5.22$}}}
	&\multicolumn{1}{p{\gnumericColE}|}%
	{\gnumericPB{\raggedleft}\gnumbox[r]{\textsf{$1.68$}}}
\\
\hhline{|----|}
	 \multicolumn{1}{|p{\gnumericColB}|}%
	{\gnumericPB{\raggedleft}\gnumbox[r]{\textsf{$3.68$}}}
	&\multicolumn{1}{p{\gnumericColC}|}%
	{\gnumericPB{\raggedleft}\gnumbox[r]{\textsf{$3.26$}}}
	&\multicolumn{1}{p{\gnumericColD}|}%
	{\gnumericPB{\raggedleft}\gnumbox[r]{\textsf{$5.39$}}}
	&\multicolumn{1}{p{\gnumericColE}|}%
	{\gnumericPB{\raggedleft}\gnumbox[r]{\textsf{$1.58$}}}
\\
\hhline{|----|}
	 \multicolumn{1}{|p{\gnumericColB}|}%
	{\gnumericPB{\raggedleft}\gnumbox[r]{\textsf{$3.86$}}}
	&\multicolumn{1}{p{\gnumericColC}|}%
	{\gnumericPB{\raggedleft}\gnumbox[r]{\textsf{$2.98$}}}
	&\multicolumn{1}{p{\gnumericColD}|}%
	{\gnumericPB{\raggedleft}\gnumbox[r]{\textsf{$5.57$}}}
	&\multicolumn{1}{p{\gnumericColE}|}%
	{\gnumericPB{\raggedleft}\gnumbox[r]{\textsf{$1.49$}}}
\\
\hhline{|----|}
	 \multicolumn{1}{|p{\gnumericColB}|}%
	{\gnumericPB{\raggedleft}\gnumbox[r]{\textsf{$4.03$}}}
	&\multicolumn{1}{p{\gnumericColC}|}%
	{\gnumericPB{\raggedleft}\gnumbox[r]{\textsf{$2.75$}}}
	&\multicolumn{1}{p{\gnumericColD}|}%
	{\gnumericPB{\raggedleft}\gnumbox[r]{\textsf{$5.74$}}}
	&\multicolumn{1}{p{\gnumericColE}|}%
	{\gnumericPB{\raggedleft}\gnumbox[r]{\textsf{$1.41$}}}
\\
\hhline{|----|}
	 \multicolumn{1}{|p{\gnumericColB}|}%
	{\gnumericPB{\raggedleft}\gnumbox[r]{\textsf{$4.20$}}}
	&\multicolumn{1}{p{\gnumericColC}|}%
	{\gnumericPB{\raggedleft}\gnumbox[r]{\textsf{$2.54$}}}
	&\multicolumn{1}{p{\gnumericColD}|}%
	{\gnumericPB{\raggedleft}\gnumbox[r]{\textsf{$5.91$}}}
	&\multicolumn{1}{p{\gnumericColE}|}%
	{\gnumericPB{\raggedleft}\gnumbox[r]{\textsf{$1.34$}}}
\\
\hhline{|----|}
	 \multicolumn{1}{|p{\gnumericColB}|}%
	{\gnumericPB{\raggedleft}\gnumbox[r]{\textsf{$4.37$}}}
	&\multicolumn{1}{p{\gnumericColC}|}%
	{\gnumericPB{\raggedleft}\gnumbox[r]{\textsf{$2.35$}}}
	&\multicolumn{1}{p{\gnumericColD}|}%
	{\gnumericPB{\raggedleft}\gnumbox[r]{\textsf{$6.08$}}}
	&\multicolumn{1}{p{\gnumericColE}|}%
	{\gnumericPB{\raggedleft}\gnumbox[r]{\textsf{$1.27$}}}
\\
\hhline{|----|}
	 \multicolumn{1}{|p{\gnumericColB}|}%
	{\gnumericPB{\raggedleft}\gnumbox[r]{\textsf{$4.54$}}}
	&\multicolumn{1}{p{\gnumericColC}|}%
	{\gnumericPB{\raggedleft}\gnumbox[r]{\textsf{$2.19$}}}
	&\multicolumn{1}{p{\gnumericColD}|}%
	{\gnumericPB{\raggedleft}\gnumbox[r]{\textsf{$6.25$}}}
	&\multicolumn{1}{p{\gnumericColE}|}%
	{\gnumericPB{\raggedleft}\gnumbox[r]{\textsf{$1.21$}}}
\\
\hhline{|-|-|-|-|}
\end{longtable}

\gnumericTableEnd
}
\caption{Normalisation for a dipolar form for $\Upsilon(1S)$ as a function of $\Lambda$ for $m_b=5.28$~GeV.}\label{tab:N_dip_Upsi_lam3}
\end{table}

\begin{table}[H]
{\tiny%%%%%%%%%%%%%%%%%%%%%%%%%%%%%%%%%%%%%%%%%%%%%%%%%%%%%%%%%%%%%%%%%%%%%%
\def\ifundefined#1{\expandafter\ifx\csname#1\endcsname\relax}

%%  Check for the \def token for inputed files. If it is not        %%
%%  defined, the file will be processed as a standalone and the     %%
%%  preamble will be used.                                          %%
\ifundefined{inputGnumericTable}

%%  End of the preamble for the standalone. The next section is for %%
%%  documents which are included into other LaTeX2e files.          %%
\else

%%  We are not a stand alone document. For a regular table, we will %%
%%  have no preamble and only define the closing to mean nothing.   %%
    \def\gnumericTableEnd{}

%%  If we want landscape mode in an embedded document, comment out  %%
%%  the line above and uncomment the two below. The table will      %%
%%  begin on a new page and run in landscape mode.                  %%
%       \def\gnumericTableEnd{\end{landscape}}
%       \begin{landscape}

%%  End of the else clause for this file being \input.              %%
\fi

%%%%%%%%%%%%%%%%%%%%%%%%%%%%%%%%%%%%%%%%%%%%%%%%%%%%%%%%%%%%%%%%%%%%%%
%%                                                                  %%
%%  The rest is the gnumeric table, except for the closing          %%
%%  statement. Changes below will alter the table's appearance.     %%
%%                                                                  %%
%%%%%%%%%%%%%%%%%%%%%%%%%%%%%%%%%%%%%%%%%%%%%%%%%%%%%%%%%%%%%%%%%%%%%%

\providecommand{\gnumericPB}[1]%
{\let\gnumericTemp=\\#1\let\\=\gnumericTemp\hspace{0pt}}

%%  The default table format retains the relative column widths of  %%
%%  gnumeric. They can easily be changed to c, r or l. In that case %%
%%  you may want to comment out the next line and uncomment the one %%
%%  thereafter                                                      %%
\providecommand\gnumbox{\makebox[0pt]}
%%\providecommand\gnumbox[1][]{\makebox}

%% to adjust positions in multirow situations                       %%
\setlength{\bigstrutjot}{\jot}
\setlength{\extrarowheight}{\doublerulesep}

%%  The \setlongtables command keeps column widths the same across %%
%%  pages. Simply comment out next line for varying column widths.  %%
\setlongtables

\setlength\gnumericTableWidth{%
	30pt+%
	30pt+%
	30pt+%
	30pt+%
0pt}
\def\gumericNumCols{4}
\setlength\gnumericTableWidthComplete{\gnumericTableWidth+\tabcolsep*\gumericNumCols*2+\arrayrulewidth*\gumericNumCols}
\ifthenelse{\lengthtest{\gnumericTableWidthComplete > \textwidth}}%
{\def\gnumericScale{\ratio{\textwidth-\tabcolsep*\gumericNumCols*2-\arrayrulewidth*\gumericNumCols}%
{\gnumericTableWidth}}}%
{\def\gnumericScale{1}}

%%%%%%%%%%%%%%%%%%%%%%%%%%%%%%%%%%%%%%%%%%%%%%%%%%%%%%%%%%%%%%%%%%%%%%
%%                                                                  %%
%% The following are the widths of the various columns. We are      %%
%% defining them here because then they are easier to change.       %%
%% Depending on the cell formats we may use them more than once.    %%
%%                                                                  %%
%%%%%%%%%%%%%%%%%%%%%%%%%%%%%%%%%%%%%%%%%%%%%%%%%%%%%%%%%%%%%%%%%%%%%%

\def\gnumericColB{30pt*\gnumericScale}
\def\gnumericColC{30pt*\gnumericScale}
\def\gnumericColD{30pt*\gnumericScale}
\def\gnumericColE{30pt*\gnumericScale}

\begin{longtable}[c]{%
	b{\gnumericColB}%
	b{\gnumericColC}%
	b{\gnumericColD}%
	b{\gnumericColE}%
	}

%%%%%%%%%%%%%%%%%%%%%%%%%%%%%%%%%%%%%%%%%%%%%%%%%%%%%%%%%%%%%%%%%%%%%%
%%  The longtable options. (Caption. headers... see Goosens. p.124) %%
%	\caption{The Table Caption.}             \\	%
% \hline	% Across the top of the table.
%%  The rest of these options are table rows which are placed on    %%
%%  the first. last or every page. Use \multicolumn if you want.    %%

%%  Header for the first page.                                      %%
%	\multicolumn{4}{c}{The First Header} \\ \hline 
%	\multicolumn{1}{c}{colTag}	%Column 1
%	&\multicolumn{1}{c}{colTag}	%Column 1
%	&\multicolumn{1}{c}{colTag}	%Column 2
%	&\multicolumn{1}{c}{colTag}	%Column 3
%	&\multicolumn{1}{c}{colTag}	\\ \hline %Last column
%	\endfirsthead

%%  The running header definition.                                  %%
%	\hline
%	\multicolumn{4}{l}{\ldots\small\slshape continued} \\ \hline
%	\multicolumn{1}{c}{colTag}	%Column 1
%	&\multicolumn{1}{c}{colTag}	%Column 1
%	&\multicolumn{1}{c}{colTag}	%Column 2
%	&\multicolumn{1}{c}{colTag}	%Column 3
%	&\multicolumn{1}{c}{colTag}	\\ \hline %Last column
%	\endhead

%%  The running footer definition.                                  %%
%	\hline
%	\multicolumn{4}{r}{\small\slshape continued\ldots} \\
%	\endfoot

%%  The ending footer definition.                                   %%
%	\multicolumn{4}{c}{That's all folks} \\ \hline 
%	\endlastfoot
%%%%%%%%%%%%%%%%%%%%%%%%%%%%%%%%%%%%%%%%%%%%%%%%%%%%%%%%%%%%%%%%%%%%%%

\hhline{|-|-|-|-|}
	 \multicolumn{1}{|p{\gnumericColB}|}%
	{\gnumericPB{\raggedleft}\gnumbox[r]{\textsf{$3.00$}}}
	&\multicolumn{1}{p{\gnumericColC}|}%
	{\gnumericPB{\raggedleft}\gnumbox[r]{\textsf{$5.56$}}}
	&\multicolumn{1}{p{\gnumericColD}|}%
	{\gnumericPB{\raggedleft}\gnumbox[r]{\textsf{$4.71$}}}
	&\multicolumn{1}{p{\gnumericColE}|}%
	{\gnumericPB{\raggedleft}\gnumbox[r]{\textsf{$2.23$}}}
\\
\hhline{|----|}
	 \multicolumn{1}{|p{\gnumericColB}|}%
	{\gnumericPB{\raggedleft}\gnumbox[r]{\textsf{$3.17$}}}
	&\multicolumn{1}{p{\gnumericColC}|}%
	{\gnumericPB{\raggedleft}\gnumbox[r]{\textsf{$4.95$}}}
	&\multicolumn{1}{p{\gnumericColD}|}%
	{\gnumericPB{\raggedleft}\gnumbox[r]{\textsf{$4.88$}}}
	&\multicolumn{1}{p{\gnumericColE}|}%
	{\gnumericPB{\raggedleft}\gnumbox[r]{\textsf{$2.08$}}}
\\
\hhline{|----|}
	 \multicolumn{1}{|p{\gnumericColB}|}%
	{\gnumericPB{\raggedleft}\gnumbox[r]{\textsf{$3.34$}}}
	&\multicolumn{1}{p{\gnumericColC}|}%
	{\gnumericPB{\raggedleft}\gnumbox[r]{\textsf{$4.44$}}}
	&\multicolumn{1}{p{\gnumericColD}|}%
	{\gnumericPB{\raggedleft}\gnumbox[r]{\textsf{$5.05$}}}
	&\multicolumn{1}{p{\gnumericColE}|}%
	{\gnumericPB{\raggedleft}\gnumbox[r]{\textsf{$1.94$}}}
\\
\hhline{|----|}
	 \multicolumn{1}{|p{\gnumericColB}|}%
	{\gnumericPB{\raggedleft}\gnumbox[r]{\textsf{$3.51$}}}
	&\multicolumn{1}{p{\gnumericColC}|}%
	{\gnumericPB{\raggedleft}\gnumbox[r]{\textsf{$4.01$}}}
	&\multicolumn{1}{p{\gnumericColD}|}%
	{\gnumericPB{\raggedleft}\gnumbox[r]{\textsf{$5.22$}}}
	&\multicolumn{1}{p{\gnumericColE}|}%
	{\gnumericPB{\raggedleft}\gnumbox[r]{\textsf{$1.82$}}}
\\
\hhline{|----|}
	 \multicolumn{1}{|p{\gnumericColB}|}%
	{\gnumericPB{\raggedleft}\gnumbox[r]{\textsf{$3.68$}}}
	&\multicolumn{1}{p{\gnumericColC}|}%
	{\gnumericPB{\raggedleft}\gnumbox[r]{\textsf{$3.64$}}}
	&\multicolumn{1}{p{\gnumericColD}|}%
	{\gnumericPB{\raggedleft}\gnumbox[r]{\textsf{$5.39$}}}
	&\multicolumn{1}{p{\gnumericColE}|}%
	{\gnumericPB{\raggedleft}\gnumbox[r]{\textsf{$1.71$}}}
\\
\hhline{|----|}
	 \multicolumn{1}{|p{\gnumericColB}|}%
	{\gnumericPB{\raggedleft}\gnumbox[r]{\textsf{$3.86$}}}
	&\multicolumn{1}{p{\gnumericColC}|}%
	{\gnumericPB{\raggedleft}\gnumbox[r]{\textsf{$3.32$}}}
	&\multicolumn{1}{p{\gnumericColD}|}%
	{\gnumericPB{\raggedleft}\gnumbox[r]{\textsf{$5.57$}}}
	&\multicolumn{1}{p{\gnumericColE}|}%
	{\gnumericPB{\raggedleft}\gnumbox[r]{\textsf{$1.61$}}}
\\
\hhline{|----|}
	 \multicolumn{1}{|p{\gnumericColB}|}%
	{\gnumericPB{\raggedleft}\gnumbox[r]{\textsf{$4.03$}}}
	&\multicolumn{1}{p{\gnumericColC}|}%
	{\gnumericPB{\raggedleft}\gnumbox[r]{\textsf{$3.04$}}}
	&\multicolumn{1}{p{\gnumericColD}|}%
	{\gnumericPB{\raggedleft}\gnumbox[r]{\textsf{$5.74$}}}
	&\multicolumn{1}{p{\gnumericColE}|}%
	{\gnumericPB{\raggedleft}\gnumbox[r]{\textsf{$1.52$}}}
\\
\hhline{|----|}
	 \multicolumn{1}{|p{\gnumericColB}|}%
	{\gnumericPB{\raggedleft}\gnumbox[r]{\textsf{$4.20$}}}
	&\multicolumn{1}{p{\gnumericColC}|}%
	{\gnumericPB{\raggedleft}\gnumbox[r]{\textsf{$2.80$}}}
	&\multicolumn{1}{p{\gnumericColD}|}%
	{\gnumericPB{\raggedleft}\gnumbox[r]{\textsf{$5.91$}}}
	&\multicolumn{1}{p{\gnumericColE}|}%
	{\gnumericPB{\raggedleft}\gnumbox[r]{\textsf{$1.44$}}}
\\
\hhline{|----|}
	 \multicolumn{1}{|p{\gnumericColB}|}%
	{\gnumericPB{\raggedleft}\gnumbox[r]{\textsf{$4.37$}}}
	&\multicolumn{1}{p{\gnumericColC}|}%
	{\gnumericPB{\raggedleft}\gnumbox[r]{\textsf{$2.59$}}}
	&\multicolumn{1}{p{\gnumericColD}|}%
	{\gnumericPB{\raggedleft}\gnumbox[r]{\textsf{$6.08$}}}
	&\multicolumn{1}{p{\gnumericColE}|}%
	{\gnumericPB{\raggedleft}\gnumbox[r]{\textsf{$1.36$}}}
\\
\hhline{|----|}
	 \multicolumn{1}{|p{\gnumericColB}|}%
	{\gnumericPB{\raggedleft}\gnumbox[r]{\textsf{$4.54$}}}
	&\multicolumn{1}{p{\gnumericColC}|}%
	{\gnumericPB{\raggedleft}\gnumbox[r]{\textsf{$2.40$}}}
	&\multicolumn{1}{p{\gnumericColD}|}%
	{\gnumericPB{\raggedleft}\gnumbox[r]{\textsf{$6.25$}}}
	&\multicolumn{1}{p{\gnumericColE}|}%
	{\gnumericPB{\raggedleft}\gnumbox[r]{\textsf{$1.29$}}}
\\
\hhline{|-|-|-|-|}
\end{longtable}

\gnumericTableEnd
}
\caption{Normalisation for a dipolar form for $\Upsilon(1S)$ as a function of $\Lambda$ for $m_b=5.50$~GeV.}\label{tab:N_dip_Upsi_lam4}
\end{table}

\begin{table}[H]
{\tiny%%%%%%%%%%%%%%%%%%%%%%%%%%%%%%%%%%%%%%%%%%%%%%%%%%%%%%%%%%%%%%%%%%%%%%
\def\ifundefined#1{\expandafter\ifx\csname#1\endcsname\relax}

%%  Check for the \def token for inputed files. If it is not        %%
%%  defined, the file will be processed as a standalone and the     %%
%%  preamble will be used.                                          %%
\ifundefined{inputGnumericTable}

%%  End of the preamble for the standalone. The next section is for %%
%%  documents which are included into other LaTeX2e files.          %%
\else

%%  We are not a stand alone document. For a regular table, we will %%
%%  have no preamble and only define the closing to mean nothing.   %%
    \def\gnumericTableEnd{}

%%  If we want landscape mode in an embedded document, comment out  %%
%%  the line above and uncomment the two below. The table will      %%
%%  begin on a new page and run in landscape mode.                  %%
%       \def\gnumericTableEnd{\end{landscape}}
%       \begin{landscape}

%%  End of the else clause for this file being \input.              %%
\fi

%%%%%%%%%%%%%%%%%%%%%%%%%%%%%%%%%%%%%%%%%%%%%%%%%%%%%%%%%%%%%%%%%%%%%%
%%                                                                  %%
%%  The rest is the gnumeric table, except for the closing          %%
%%  statement. Changes below will alter the table's appearance.     %%
%%                                                                  %%
%%%%%%%%%%%%%%%%%%%%%%%%%%%%%%%%%%%%%%%%%%%%%%%%%%%%%%%%%%%%%%%%%%%%%%

\providecommand{\gnumericPB}[1]%
{\let\gnumericTemp=\\#1\let\\=\gnumericTemp\hspace{0pt}}

%%  The default table format retains the relative column widths of  %%
%%  gnumeric. They can easily be changed to c, r or l. In that case %%
%%  you may want to comment out the next line and uncomment the one %%
%%  thereafter                                                      %%
\providecommand\gnumbox{\makebox[0pt]}
%%\providecommand\gnumbox[1][]{\makebox}

%% to adjust positions in multirow situations                       %%
\setlength{\bigstrutjot}{\jot}
\setlength{\extrarowheight}{\doublerulesep}

%%  The \setlongtables command keeps column widths the same across %%
%%  pages. Simply comment out next line for varying column widths.  %%
\setlongtables

\setlength\gnumericTableWidth{%
	30pt+%
	30pt+%
	30pt+%
	30pt+%
0pt}
\def\gumericNumCols{4}
\setlength\gnumericTableWidthComplete{\gnumericTableWidth+\tabcolsep*\gumericNumCols*2+\arrayrulewidth*\gumericNumCols}
\ifthenelse{\lengthtest{\gnumericTableWidthComplete > \textwidth}}%
{\def\gnumericScale{\ratio{\textwidth-\tabcolsep*\gumericNumCols*2-\arrayrulewidth*\gumericNumCols}%
{\gnumericTableWidth}}}%
{\def\gnumericScale{1}}

%%%%%%%%%%%%%%%%%%%%%%%%%%%%%%%%%%%%%%%%%%%%%%%%%%%%%%%%%%%%%%%%%%%%%%
%%                                                                  %%
%% The following are the widths of the various columns. We are      %%
%% defining them here because then they are easier to change.       %%
%% Depending on the cell formats we may use them more than once.    %%
%%                                                                  %%
%%%%%%%%%%%%%%%%%%%%%%%%%%%%%%%%%%%%%%%%%%%%%%%%%%%%%%%%%%%%%%%%%%%%%%

\def\gnumericColB{30pt*\gnumericScale}
\def\gnumericColC{30pt*\gnumericScale}
\def\gnumericColD{30pt*\gnumericScale}
\def\gnumericColE{30pt*\gnumericScale}

\begin{longtable}[c]{%
	b{\gnumericColB}%
	b{\gnumericColC}%
	b{\gnumericColD}%
	b{\gnumericColE}%
	}

%%%%%%%%%%%%%%%%%%%%%%%%%%%%%%%%%%%%%%%%%%%%%%%%%%%%%%%%%%%%%%%%%%%%%%
%%  The longtable options. (Caption. headers... see Goosens. p.124) %%
%	\caption{The Table Caption.}             \\	%
% \hline	% Across the top of the table.
%%  The rest of these options are table rows which are placed on    %%
%%  the first. last or every page. Use \multicolumn if you want.    %%

%%  Header for the first page.                                      %%
%	\multicolumn{4}{c}{The First Header} \\ \hline 
%	\multicolumn{1}{c}{colTag}	%Column 1
%	&\multicolumn{1}{c}{colTag}	%Column 1
%	&\multicolumn{1}{c}{colTag}	%Column 2
%	&\multicolumn{1}{c}{colTag}	%Column 3
%	&\multicolumn{1}{c}{colTag}	\\ \hline %Last column
%	\endfirsthead

%%  The running header definition.                                  %%
%	\hline
%	\multicolumn{4}{l}{\ldots\small\slshape continued} \\ \hline
%	\multicolumn{1}{c}{colTag}	%Column 1
%	&\multicolumn{1}{c}{colTag}	%Column 1
%	&\multicolumn{1}{c}{colTag}	%Column 2
%	&\multicolumn{1}{c}{colTag}	%Column 3
%	&\multicolumn{1}{c}{colTag}	\\ \hline %Last column
%	\endhead

%%  The running footer definition.                                  %%
%	\hline
%	\multicolumn{4}{r}{\small\slshape continued\ldots} \\
%	\endfoot

%%  The ending footer definition.                                   %%
%	\multicolumn{4}{c}{That's all folks} \\ \hline 
%	\endlastfoot
%%%%%%%%%%%%%%%%%%%%%%%%%%%%%%%%%%%%%%%%%%%%%%%%%%%%%%%%%%%%%%%%%%%%%%

\hhline{|-|-|-|-|}
	 \multicolumn{1}{|p{\gnumericColB}|}%
	{\gnumericPB{\raggedleft}\gnumbox[r]{\textsf{$m_b$}}}
	&\multicolumn{1}{p{\gnumericColC}|}%
	{\gnumericPB{\raggedleft}\gnumbox[r]{\textsf{$N$}}}
	&\multicolumn{1}{p{\gnumericColD}|}%
	{\gnumericPB{\raggedleft}\gnumbox[r]{\textsf{$m_b$}}}
	&\multicolumn{1}{p{\gnumericColE}|}%
	{\gnumericPB{\raggedleft}\gnumbox[r]{\textsf{$N$}}}
\\
\hhline{|----|}
	 \multicolumn{1}{|p{\gnumericColB}|}%
	{\gnumericPB{\raggedleft}\gnumbox[r]{\textsf{$4.80$}}}
	&\multicolumn{1}{p{\gnumericColC}|}%
	{\gnumericPB{\raggedleft}\gnumbox[r]{\textsf{$2.84$}}}
	&\multicolumn{1}{p{\gnumericColD}|}%
	{\gnumericPB{\raggedleft}\gnumbox[r]{\textsf{$5.17$}}}
	&\multicolumn{1}{p{\gnumericColE}|}%
	{\gnumericPB{\raggedleft}\gnumbox[r]{\textsf{$4.56$}}}
\\
\hhline{|----|}
	 \multicolumn{1}{|p{\gnumericColB}|}%
	{\gnumericPB{\raggedleft}\gnumbox[r]{\textsf{$4.84$}}}
	&\multicolumn{1}{p{\gnumericColC}|}%
	{\gnumericPB{\raggedleft}\gnumbox[r]{\textsf{$3.10$}}}
	&\multicolumn{1}{p{\gnumericColD}|}%
	{\gnumericPB{\raggedleft}\gnumbox[r]{\textsf{$5.21$}}}
	&\multicolumn{1}{p{\gnumericColE}|}%
	{\gnumericPB{\raggedleft}\gnumbox[r]{\textsf{$4.68$}}}
\\
\hhline{|----|}
	 \multicolumn{1}{|p{\gnumericColB}|}%
	{\gnumericPB{\raggedleft}\gnumbox[r]{\textsf{$4.87$}}}
	&\multicolumn{1}{p{\gnumericColC}|}%
	{\gnumericPB{\raggedleft}\gnumbox[r]{\textsf{$3.31$}}}
	&\multicolumn{1}{p{\gnumericColD}|}%
	{\gnumericPB{\raggedleft}\gnumbox[r]{\textsf{$5.24$}}}
	&\multicolumn{1}{p{\gnumericColE}|}%
	{\gnumericPB{\raggedleft}\gnumbox[r]{\textsf{$4.80$}}}
\\
\hhline{|----|}
	 \multicolumn{1}{|p{\gnumericColB}|}%
	{\gnumericPB{\raggedleft}\gnumbox[r]{\textsf{$4.91$}}}
	&\multicolumn{1}{p{\gnumericColC}|}%
	{\gnumericPB{\raggedleft}\gnumbox[r]{\textsf{$3.51$}}}
	&\multicolumn{1}{p{\gnumericColD}|}%
	{\gnumericPB{\raggedleft}\gnumbox[r]{\textsf{$5.28$}}}
	&\multicolumn{1}{p{\gnumericColE}|}%
	{\gnumericPB{\raggedleft}\gnumbox[r]{\textsf{$4.92$}}}
\\
\hhline{|----|}
	 \multicolumn{1}{|p{\gnumericColB}|}%
	{\gnumericPB{\raggedleft}\gnumbox[r]{\textsf{$4.95$}}}
	&\multicolumn{1}{p{\gnumericColC}|}%
	{\gnumericPB{\raggedleft}\gnumbox[r]{\textsf{$3.69$}}}
	&\multicolumn{1}{p{\gnumericColD}|}%
	{\gnumericPB{\raggedleft}\gnumbox[r]{\textsf{$5.32$}}}
	&\multicolumn{1}{p{\gnumericColE}|}%
	{\gnumericPB{\raggedleft}\gnumbox[r]{\textsf{$5.03$}}}
\\
\hhline{|----|}
	 \multicolumn{1}{|p{\gnumericColB}|}%
	{\gnumericPB{\raggedleft}\gnumbox[r]{\textsf{$4.98$}}}
	&\multicolumn{1}{p{\gnumericColC}|}%
	{\gnumericPB{\raggedleft}\gnumbox[r]{\textsf{$3.85$}}}
	&\multicolumn{1}{p{\gnumericColD}|}%
	{\gnumericPB{\raggedleft}\gnumbox[r]{\textsf{$5.35$}}}
	&\multicolumn{1}{p{\gnumericColE}|}%
	{\gnumericPB{\raggedleft}\gnumbox[r]{\textsf{$5.14$}}}
\\
\hhline{|----|}
	 \multicolumn{1}{|p{\gnumericColB}|}%
	{\gnumericPB{\raggedleft}\gnumbox[r]{\textsf{$5.02$}}}
	&\multicolumn{1}{p{\gnumericColC}|}%
	{\gnumericPB{\raggedleft}\gnumbox[r]{\textsf{$4.01$}}}
	&\multicolumn{1}{p{\gnumericColD}|}%
	{\gnumericPB{\raggedleft}\gnumbox[r]{\textsf{$5.39$}}}
	&\multicolumn{1}{p{\gnumericColE}|}%
	{\gnumericPB{\raggedleft}\gnumbox[r]{\textsf{$5.25$}}}
\\
\hhline{|----|}
	 \multicolumn{1}{|p{\gnumericColB}|}%
	{\gnumericPB{\raggedleft}\gnumbox[r]{\textsf{$5.06$}}}
	&\multicolumn{1}{p{\gnumericColC}|}%
	{\gnumericPB{\raggedleft}\gnumbox[r]{\textsf{$4.16$}}}
	&\multicolumn{1}{p{\gnumericColD}|}%
	{\gnumericPB{\raggedleft}\gnumbox[r]{\textsf{$5.43$}}}
	&\multicolumn{1}{p{\gnumericColE}|}%
	{\gnumericPB{\raggedleft}\gnumbox[r]{\textsf{$5.36$}}}
\\
\hhline{|----|}
	 \multicolumn{1}{|p{\gnumericColB}|}%
	{\gnumericPB{\raggedleft}\gnumbox[r]{\textsf{$5.09$}}}
	&\multicolumn{1}{p{\gnumericColC}|}%
	{\gnumericPB{\raggedleft}\gnumbox[r]{\textsf{$4.30$}}}
	&\multicolumn{1}{p{\gnumericColD}|}%
	{\gnumericPB{\raggedleft}\gnumbox[r]{\textsf{$5.46$}}}
	&\multicolumn{1}{p{\gnumericColE}|}%
	{\gnumericPB{\raggedleft}\gnumbox[r]{\textsf{$5.46$}}}
\\
\hhline{|----|}
	 \multicolumn{1}{|p{\gnumericColB}|}%
	{\gnumericPB{\raggedleft}\gnumbox[r]{\textsf{$5.13$}}}
	&\multicolumn{1}{p{\gnumericColC}|}%
	{\gnumericPB{\raggedleft}\gnumbox[r]{\textsf{$4.43$}}}
	&\multicolumn{1}{p{\gnumericColD}|}%
	{\gnumericPB{\raggedleft}\gnumbox[r]{\textsf{$5.50$}}}
	&\multicolumn{1}{p{\gnumericColE}|}%
	{\gnumericPB{\raggedleft}\gnumbox[r]{\textsf{$5.56$}}}
\\
\hhline{|-|-|-|-|}
\end{longtable}

\gnumericTableEnd
}
\caption{Normalisation for a dipolar form for $\Upsilon(1S)$ as a function of  the $b$ quark mass
for $\Lambda=3.00$~GeV.}\label{tab:N_dip_Upsi_mq1}
\end{table}

\begin{table}[H]
{\tiny%%%%%%%%%%%%%%%%%%%%%%%%%%%%%%%%%%%%%%%%%%%%%%%%%%%%%%%%%%%%%%%%%%%%%%
\def\ifundefined#1{\expandafter\ifx\csname#1\endcsname\relax}

%%  Check for the \def token for inputed files. If it is not        %%
%%  defined, the file will be processed as a standalone and the     %%
%%  preamble will be used.                                          %%
\ifundefined{inputGnumericTable}

%%  End of the preamble for the standalone. The next section is for %%
%%  documents which are included into other LaTeX2e files.          %%
\else

%%  We are not a stand alone document. For a regular table, we will %%
%%  have no preamble and only define the closing to mean nothing.   %%
    \def\gnumericTableEnd{}

%%  If we want landscape mode in an embedded document, comment out  %%
%%  the line above and uncomment the two below. The table will      %%
%%  begin on a new page and run in landscape mode.                  %%
%       \def\gnumericTableEnd{\end{landscape}}
%       \begin{landscape}

%%  End of the else clause for this file being \input.              %%
\fi

%%%%%%%%%%%%%%%%%%%%%%%%%%%%%%%%%%%%%%%%%%%%%%%%%%%%%%%%%%%%%%%%%%%%%%
%%                                                                  %%
%%  The rest is the gnumeric table, except for the closing          %%
%%  statement. Changes below will alter the table's appearance.     %%
%%                                                                  %%
%%%%%%%%%%%%%%%%%%%%%%%%%%%%%%%%%%%%%%%%%%%%%%%%%%%%%%%%%%%%%%%%%%%%%%

\providecommand{\gnumericPB}[1]%
{\let\gnumericTemp=\\#1\let\\=\gnumericTemp\hspace{0pt}}

%%  The default table format retains the relative column widths of  %%
%%  gnumeric. They can easily be changed to c, r or l. In that case %%
%%  you may want to comment out the next line and uncomment the one %%
%%  thereafter                                                      %%
\providecommand\gnumbox{\makebox[0pt]}
%%\providecommand\gnumbox[1][]{\makebox}

%% to adjust positions in multirow situations                       %%
\setlength{\bigstrutjot}{\jot}
\setlength{\extrarowheight}{\doublerulesep}

%%  The \setlongtables command keeps column widths the same across %%
%%  pages. Simply comment out next line for varying column widths.  %%
\setlongtables

\setlength\gnumericTableWidth{%
	30pt+%
	30pt+%
	30pt+%
	30pt+%
0pt}
\def\gumericNumCols{4}
\setlength\gnumericTableWidthComplete{\gnumericTableWidth+\tabcolsep*\gumericNumCols*2+\arrayrulewidth*\gumericNumCols}
\ifthenelse{\lengthtest{\gnumericTableWidthComplete > \textwidth}}%
{\def\gnumericScale{\ratio{\textwidth-\tabcolsep*\gumericNumCols*2-\arrayrulewidth*\gumericNumCols}%
{\gnumericTableWidth}}}%
{\def\gnumericScale{1}}

%%%%%%%%%%%%%%%%%%%%%%%%%%%%%%%%%%%%%%%%%%%%%%%%%%%%%%%%%%%%%%%%%%%%%%
%%                                                                  %%
%% The following are the widths of the various columns. We are      %%
%% defining them here because then they are easier to change.       %%
%% Depending on the cell formats we may use them more than once.    %%
%%                                                                  %%
%%%%%%%%%%%%%%%%%%%%%%%%%%%%%%%%%%%%%%%%%%%%%%%%%%%%%%%%%%%%%%%%%%%%%%

\def\gnumericColB{30pt*\gnumericScale}
\def\gnumericColC{30pt*\gnumericScale}
\def\gnumericColD{30pt*\gnumericScale}
\def\gnumericColE{30pt*\gnumericScale}

\begin{longtable}[c]{%
	b{\gnumericColB}%
	b{\gnumericColC}%
	b{\gnumericColD}%
	b{\gnumericColE}%
	}

%%%%%%%%%%%%%%%%%%%%%%%%%%%%%%%%%%%%%%%%%%%%%%%%%%%%%%%%%%%%%%%%%%%%%%
%%  The longtable options. (Caption. headers... see Goosens. p.124) %%
%	\caption{The Table Caption.}             \\	%
% \hline	% Across the top of the table.
%%  The rest of these options are table rows which are placed on    %%
%%  the first. last or every page. Use \multicolumn if you want.    %%

%%  Header for the first page.                                      %%
%	\multicolumn{4}{c}{The First Header} \\ \hline 
%	\multicolumn{1}{c}{colTag}	%Column 1
%	&\multicolumn{1}{c}{colTag}	%Column 1
%	&\multicolumn{1}{c}{colTag}	%Column 2
%	&\multicolumn{1}{c}{colTag}	%Column 3
%	&\multicolumn{1}{c}{colTag}	\\ \hline %Last column
%	\endfirsthead

%%  The running header definition.                                  %%
%	\hline
%	\multicolumn{4}{l}{\ldots\small\slshape continued} \\ \hline
%	\multicolumn{1}{c}{colTag}	%Column 1
%	&\multicolumn{1}{c}{colTag}	%Column 1
%	&\multicolumn{1}{c}{colTag}	%Column 2
%	&\multicolumn{1}{c}{colTag}	%Column 3
%	&\multicolumn{1}{c}{colTag}	\\ \hline %Last column
%	\endhead

%%  The running footer definition.                                  %%
%	\hline
%	\multicolumn{4}{r}{\small\slshape continued\ldots} \\
%	\endfoot

%%  The ending footer definition.                                   %%
%	\multicolumn{4}{c}{That's all folks} \\ \hline 
%	\endlastfoot
%%%%%%%%%%%%%%%%%%%%%%%%%%%%%%%%%%%%%%%%%%%%%%%%%%%%%%%%%%%%%%%%%%%%%%

\hhline{|-|-|-|-|}
	 \multicolumn{1}{|p{\gnumericColB}|}%
	{\gnumericPB{\raggedleft}\gnumbox[r]{\textsf{$4.80$}}}
	&\multicolumn{1}{p{\gnumericColC}|}%
	{\gnumericPB{\raggedleft}\gnumbox[r]{\textsf{$1.77$}}}
	&\multicolumn{1}{p{\gnumericColD}|}%
	{\gnumericPB{\raggedleft}\gnumbox[r]{\textsf{$5.17$}}}
	&\multicolumn{1}{p{\gnumericColE}|}%
	{\gnumericPB{\raggedleft}\gnumbox[r]{\textsf{$2.53$}}}
\\
\hhline{|----|}
	 \multicolumn{1}{|p{\gnumericColB}|}%
	{\gnumericPB{\raggedleft}\gnumbox[r]{\textsf{$4.84$}}}
	&\multicolumn{1}{p{\gnumericColC}|}%
	{\gnumericPB{\raggedleft}\gnumbox[r]{\textsf{$1.89$}}}
	&\multicolumn{1}{p{\gnumericColD}|}%
	{\gnumericPB{\raggedleft}\gnumbox[r]{\textsf{$5.21$}}}
	&\multicolumn{1}{p{\gnumericColE}|}%
	{\gnumericPB{\raggedleft}\gnumbox[r]{\textsf{$2.59$}}}
\\
\hhline{|----|}
	 \multicolumn{1}{|p{\gnumericColB}|}%
	{\gnumericPB{\raggedleft}\gnumbox[r]{\textsf{$4.87$}}}
	&\multicolumn{1}{p{\gnumericColC}|}%
	{\gnumericPB{\raggedleft}\gnumbox[r]{\textsf{$1.99$}}}
	&\multicolumn{1}{p{\gnumericColD}|}%
	{\gnumericPB{\raggedleft}\gnumbox[r]{\textsf{$5.24$}}}
	&\multicolumn{1}{p{\gnumericColE}|}%
	{\gnumericPB{\raggedleft}\gnumbox[r]{\textsf{$2.64$}}}
\\
\hhline{|----|}
	 \multicolumn{1}{|p{\gnumericColB}|}%
	{\gnumericPB{\raggedleft}\gnumbox[r]{\textsf{$4.91$}}}
	&\multicolumn{1}{p{\gnumericColC}|}%
	{\gnumericPB{\raggedleft}\gnumbox[r]{\textsf{$2.08$}}}
	&\multicolumn{1}{p{\gnumericColD}|}%
	{\gnumericPB{\raggedleft}\gnumbox[r]{\textsf{$5.28$}}}
	&\multicolumn{1}{p{\gnumericColE}|}%
	{\gnumericPB{\raggedleft}\gnumbox[r]{\textsf{$2.69$}}}
\\
\hhline{|----|}
	 \multicolumn{1}{|p{\gnumericColB}|}%
	{\gnumericPB{\raggedleft}\gnumbox[r]{\textsf{$4.95$}}}
	&\multicolumn{1}{p{\gnumericColC}|}%
	{\gnumericPB{\raggedleft}\gnumbox[r]{\textsf{$2.16$}}}
	&\multicolumn{1}{p{\gnumericColD}|}%
	{\gnumericPB{\raggedleft}\gnumbox[r]{\textsf{$5.32$}}}
	&\multicolumn{1}{p{\gnumericColE}|}%
	{\gnumericPB{\raggedleft}\gnumbox[r]{\textsf{$2.74$}}}
\\
\hhline{|----|}
	 \multicolumn{1}{|p{\gnumericColB}|}%
	{\gnumericPB{\raggedleft}\gnumbox[r]{\textsf{$4.98$}}}
	&\multicolumn{1}{p{\gnumericColC}|}%
	{\gnumericPB{\raggedleft}\gnumbox[r]{\textsf{$2.23$}}}
	&\multicolumn{1}{p{\gnumericColD}|}%
	{\gnumericPB{\raggedleft}\gnumbox[r]{\textsf{$5.35$}}}
	&\multicolumn{1}{p{\gnumericColE}|}%
	{\gnumericPB{\raggedleft}\gnumbox[r]{\textsf{$2.78$}}}
\\
\hhline{|----|}
	 \multicolumn{1}{|p{\gnumericColB}|}%
	{\gnumericPB{\raggedleft}\gnumbox[r]{\textsf{$5.02$}}}
	&\multicolumn{1}{p{\gnumericColC}|}%
	{\gnumericPB{\raggedleft}\gnumbox[r]{\textsf{$2.30$}}}
	&\multicolumn{1}{p{\gnumericColD}|}%
	{\gnumericPB{\raggedleft}\gnumbox[r]{\textsf{$5.39$}}}
	&\multicolumn{1}{p{\gnumericColE}|}%
	{\gnumericPB{\raggedleft}\gnumbox[r]{\textsf{$2.83$}}}
\\
\hhline{|----|}
	 \multicolumn{1}{|p{\gnumericColB}|}%
	{\gnumericPB{\raggedleft}\gnumbox[r]{\textsf{$5.06$}}}
	&\multicolumn{1}{p{\gnumericColC}|}%
	{\gnumericPB{\raggedleft}\gnumbox[r]{\textsf{$2.36$}}}
	&\multicolumn{1}{p{\gnumericColD}|}%
	{\gnumericPB{\raggedleft}\gnumbox[r]{\textsf{$5.43$}}}
	&\multicolumn{1}{p{\gnumericColE}|}%
	{\gnumericPB{\raggedleft}\gnumbox[r]{\textsf{$2.87$}}}
\\
\hhline{|----|}
	 \multicolumn{1}{|p{\gnumericColB}|}%
	{\gnumericPB{\raggedleft}\gnumbox[r]{\textsf{$5.09$}}}
	&\multicolumn{1}{p{\gnumericColC}|}%
	{\gnumericPB{\raggedleft}\gnumbox[r]{\textsf{$2.42$}}}
	&\multicolumn{1}{p{\gnumericColD}|}%
	{\gnumericPB{\raggedleft}\gnumbox[r]{\textsf{$5.46$}}}
	&\multicolumn{1}{p{\gnumericColE}|}%
	{\gnumericPB{\raggedleft}\gnumbox[r]{\textsf{$2.92$}}}
\\
\hhline{|----|}
	 \multicolumn{1}{|p{\gnumericColB}|}%
	{\gnumericPB{\raggedleft}\gnumbox[r]{\textsf{$5.13$}}}
	&\multicolumn{1}{p{\gnumericColC}|}%
	{\gnumericPB{\raggedleft}\gnumbox[r]{\textsf{$2.48$}}}
	&\multicolumn{1}{p{\gnumericColD}|}%
	{\gnumericPB{\raggedleft}\gnumbox[r]{\textsf{$5.50$}}}
	&\multicolumn{1}{p{\gnumericColE}|}%
	{\gnumericPB{\raggedleft}\gnumbox[r]{\textsf{$2.96$}}}
\\
\hhline{|-|-|-|-|}
\end{longtable}

\gnumericTableEnd
}
\caption{Normalisation for a dipolar form for $\Upsilon(1S)$ as a function of the $b$ quark mass
for $\Lambda=4.10$~GeV.}\label{tab:N_dip_Upsi_mq2}
\end{table}

\begin{table}[H]
{\tiny%%%%%%%%%%%%%%%%%%%%%%%%%%%%%%%%%%%%%%%%%%%%%%%%%%%%%%%%%%%%%%%%%%%%%%
\def\ifundefined#1{\expandafter\ifx\csname#1\endcsname\relax}

%%  Check for the \def token for inputed files. If it is not        %%
%%  defined, the file will be processed as a standalone and the     %%
%%  preamble will be used.                                          %%
\ifundefined{inputGnumericTable}

%%  End of the preamble for the standalone. The next section is for %%
%%  documents which are included into other LaTeX2e files.          %%
\else

%%  We are not a stand alone document. For a regular table, we will %%
%%  have no preamble and only define the closing to mean nothing.   %%
    \def\gnumericTableEnd{}

%%  If we want landscape mode in an embedded document, comment out  %%
%%  the line above and uncomment the two below. The table will      %%
%%  begin on a new page and run in landscape mode.                  %%
%       \def\gnumericTableEnd{\end{landscape}}
%       \begin{landscape}

%%  End of the else clause for this file being \input.              %%
\fi

%%%%%%%%%%%%%%%%%%%%%%%%%%%%%%%%%%%%%%%%%%%%%%%%%%%%%%%%%%%%%%%%%%%%%%
%%                                                                  %%
%%  The rest is the gnumeric table, except for the closing          %%
%%  statement. Changes below will alter the table's appearance.     %%
%%                                                                  %%
%%%%%%%%%%%%%%%%%%%%%%%%%%%%%%%%%%%%%%%%%%%%%%%%%%%%%%%%%%%%%%%%%%%%%%

\providecommand{\gnumericPB}[1]%
{\let\gnumericTemp=\\#1\let\\=\gnumericTemp\hspace{0pt}}

%%  The default table format retains the relative column widths of  %%
%%  gnumeric. They can easily be changed to c, r or l. In that case %%
%%  you may want to comment out the next line and uncomment the one %%
%%  thereafter                                                      %%
\providecommand\gnumbox{\makebox[0pt]}
%%\providecommand\gnumbox[1][]{\makebox}

%% to adjust positions in multirow situations                       %%
\setlength{\bigstrutjot}{\jot}
\setlength{\extrarowheight}{\doublerulesep}

%%  The \setlongtables command keeps column widths the same across %%
%%  pages. Simply comment out next line for varying column widths.  %%
\setlongtables

\setlength\gnumericTableWidth{%
	30pt+%
	30pt+%
	30pt+%
	30pt+%
0pt}
\def\gumericNumCols{4}
\setlength\gnumericTableWidthComplete{\gnumericTableWidth+\tabcolsep*\gumericNumCols*2+\arrayrulewidth*\gumericNumCols}
\ifthenelse{\lengthtest{\gnumericTableWidthComplete > \textwidth}}%
{\def\gnumericScale{\ratio{\textwidth-\tabcolsep*\gumericNumCols*2-\arrayrulewidth*\gumericNumCols}%
{\gnumericTableWidth}}}%
{\def\gnumericScale{1}}

%%%%%%%%%%%%%%%%%%%%%%%%%%%%%%%%%%%%%%%%%%%%%%%%%%%%%%%%%%%%%%%%%%%%%%
%%                                                                  %%
%% The following are the widths of the various columns. We are      %%
%% defining them here because then they are easier to change.       %%
%% Depending on the cell formats we may use them more than once.    %%
%%                                                                  %%
%%%%%%%%%%%%%%%%%%%%%%%%%%%%%%%%%%%%%%%%%%%%%%%%%%%%%%%%%%%%%%%%%%%%%%

\def\gnumericColB{30pt*\gnumericScale}
\def\gnumericColC{30pt*\gnumericScale}
\def\gnumericColD{30pt*\gnumericScale}
\def\gnumericColE{30pt*\gnumericScale}

\begin{longtable}[c]{%
	b{\gnumericColB}%
	b{\gnumericColC}%
	b{\gnumericColD}%
	b{\gnumericColE}%
	}

%%%%%%%%%%%%%%%%%%%%%%%%%%%%%%%%%%%%%%%%%%%%%%%%%%%%%%%%%%%%%%%%%%%%%%
%%  The longtable options. (Caption. headers... see Goosens. p.124) %%
%	\caption{The Table Caption.}             \\	%
% \hline	% Across the top of the table.
%%  The rest of these options are table rows which are placed on    %%
%%  the first. last or every page. Use \multicolumn if you want.    %%

%%  Header for the first page.                                      %%
%	\multicolumn{4}{c}{The First Header} \\ \hline 
%	\multicolumn{1}{c}{colTag}	%Column 1
%	&\multicolumn{1}{c}{colTag}	%Column 1
%	&\multicolumn{1}{c}{colTag}	%Column 2
%	&\multicolumn{1}{c}{colTag}	%Column 3
%	&\multicolumn{1}{c}{colTag}	\\ \hline %Last column
%	\endfirsthead

%%  The running header definition.                                  %%
%	\hline
%	\multicolumn{4}{l}{\ldots\small\slshape continued} \\ \hline
%	\multicolumn{1}{c}{colTag}	%Column 1
%	&\multicolumn{1}{c}{colTag}	%Column 1
%	&\multicolumn{1}{c}{colTag}	%Column 2
%	&\multicolumn{1}{c}{colTag}	%Column 3
%	&\multicolumn{1}{c}{colTag}	\\ \hline %Last column
%	\endhead

%%  The running footer definition.                                  %%
%	\hline
%	\multicolumn{4}{r}{\small\slshape continued\ldots} \\
%	\endfoot

%%  The ending footer definition.                                   %%
%	\multicolumn{4}{c}{That's all folks} \\ \hline 
%	\endlastfoot
%%%%%%%%%%%%%%%%%%%%%%%%%%%%%%%%%%%%%%%%%%%%%%%%%%%%%%%%%%%%%%%%%%%%%%

\hhline{|-|-|-|-|}
	 \multicolumn{1}{|p{\gnumericColB}|}%
	{\gnumericPB{\raggedleft}\gnumbox[r]{\textsf{$4.80$}}}
	&\multicolumn{1}{p{\gnumericColC}|}%
	{\gnumericPB{\raggedleft}\gnumbox[r]{\textsf{$1.24$}}}
	&\multicolumn{1}{p{\gnumericColD}|}%
	{\gnumericPB{\raggedleft}\gnumbox[r]{\textsf{$5.17$}}}
	&\multicolumn{1}{p{\gnumericColE}|}%
	{\gnumericPB{\raggedleft}\gnumbox[r]{\textsf{$1.64$}}}
\\
\hhline{|----|}
	 \multicolumn{1}{|p{\gnumericColB}|}%
	{\gnumericPB{\raggedleft}\gnumbox[r]{\textsf{$4.84$}}}
	&\multicolumn{1}{p{\gnumericColC}|}%
	{\gnumericPB{\raggedleft}\gnumbox[r]{\textsf{$1.31$}}}
	&\multicolumn{1}{p{\gnumericColD}|}%
	{\gnumericPB{\raggedleft}\gnumbox[r]{\textsf{$5.21$}}}
	&\multicolumn{1}{p{\gnumericColE}|}%
	{\gnumericPB{\raggedleft}\gnumbox[r]{\textsf{$1.67$}}}
\\
\hhline{|----|}
	 \multicolumn{1}{|p{\gnumericColB}|}%
	{\gnumericPB{\raggedleft}\gnumbox[r]{\textsf{$4.87$}}}
	&\multicolumn{1}{p{\gnumericColC}|}%
	{\gnumericPB{\raggedleft}\gnumbox[r]{\textsf{$1.36$}}}
	&\multicolumn{1}{p{\gnumericColD}|}%
	{\gnumericPB{\raggedleft}\gnumbox[r]{\textsf{$5.24$}}}
	&\multicolumn{1}{p{\gnumericColE}|}%
	{\gnumericPB{\raggedleft}\gnumbox[r]{\textsf{$1.70$}}}
\\
\hhline{|----|}
	 \multicolumn{1}{|p{\gnumericColB}|}%
	{\gnumericPB{\raggedleft}\gnumbox[r]{\textsf{$4.91$}}}
	&\multicolumn{1}{p{\gnumericColC}|}%
	{\gnumericPB{\raggedleft}\gnumbox[r]{\textsf{$1.41$}}}
	&\multicolumn{1}{p{\gnumericColD}|}%
	{\gnumericPB{\raggedleft}\gnumbox[r]{\textsf{$5.28$}}}
	&\multicolumn{1}{p{\gnumericColE}|}%
	{\gnumericPB{\raggedleft}\gnumbox[r]{\textsf{$1.72$}}}
\\
\hhline{|----|}
	 \multicolumn{1}{|p{\gnumericColB}|}%
	{\gnumericPB{\raggedleft}\gnumbox[r]{\textsf{$4.95$}}}
	&\multicolumn{1}{p{\gnumericColC}|}%
	{\gnumericPB{\raggedleft}\gnumbox[r]{\textsf{$1.45$}}}
	&\multicolumn{1}{p{\gnumericColD}|}%
	{\gnumericPB{\raggedleft}\gnumbox[r]{\textsf{$5.32$}}}
	&\multicolumn{1}{p{\gnumericColE}|}%
	{\gnumericPB{\raggedleft}\gnumbox[r]{\textsf{$1.75$}}}
\\
\hhline{|----|}
	 \multicolumn{1}{|p{\gnumericColB}|}%
	{\gnumericPB{\raggedleft}\gnumbox[r]{\textsf{$4.98$}}}
	&\multicolumn{1}{p{\gnumericColC}|}%
	{\gnumericPB{\raggedleft}\gnumbox[r]{\textsf{$1.49$}}}
	&\multicolumn{1}{p{\gnumericColD}|}%
	{\gnumericPB{\raggedleft}\gnumbox[r]{\textsf{$5.35$}}}
	&\multicolumn{1}{p{\gnumericColE}|}%
	{\gnumericPB{\raggedleft}\gnumbox[r]{\textsf{$1.77$}}}
\\
\hhline{|----|}
	 \multicolumn{1}{|p{\gnumericColB}|}%
	{\gnumericPB{\raggedleft}\gnumbox[r]{\textsf{$5.02$}}}
	&\multicolumn{1}{p{\gnumericColC}|}%
	{\gnumericPB{\raggedleft}\gnumbox[r]{\textsf{$1.52$}}}
	&\multicolumn{1}{p{\gnumericColD}|}%
	{\gnumericPB{\raggedleft}\gnumbox[r]{\textsf{$5.39$}}}
	&\multicolumn{1}{p{\gnumericColE}|}%
	{\gnumericPB{\raggedleft}\gnumbox[r]{\textsf{$1.80$}}}
\\
\hhline{|----|}
	 \multicolumn{1}{|p{\gnumericColB}|}%
	{\gnumericPB{\raggedleft}\gnumbox[r]{\textsf{$5.06$}}}
	&\multicolumn{1}{p{\gnumericColC}|}%
	{\gnumericPB{\raggedleft}\gnumbox[r]{\textsf{$1.55$}}}
	&\multicolumn{1}{p{\gnumericColD}|}%
	{\gnumericPB{\raggedleft}\gnumbox[r]{\textsf{$5.43$}}}
	&\multicolumn{1}{p{\gnumericColE}|}%
	{\gnumericPB{\raggedleft}\gnumbox[r]{\textsf{$1.82$}}}
\\
\hhline{|----|}
	 \multicolumn{1}{|p{\gnumericColB}|}%
	{\gnumericPB{\raggedleft}\gnumbox[r]{\textsf{$5.09$}}}
	&\multicolumn{1}{p{\gnumericColC}|}%
	{\gnumericPB{\raggedleft}\gnumbox[r]{\textsf{$1.59$}}}
	&\multicolumn{1}{p{\gnumericColD}|}%
	{\gnumericPB{\raggedleft}\gnumbox[r]{\textsf{$5.46$}}}
	&\multicolumn{1}{p{\gnumericColE}|}%
	{\gnumericPB{\raggedleft}\gnumbox[r]{\textsf{$1.84$}}}
\\
\hhline{|----|}
	 \multicolumn{1}{|p{\gnumericColB}|}%
	{\gnumericPB{\raggedleft}\gnumbox[r]{\textsf{$5.13$}}}
	&\multicolumn{1}{p{\gnumericColC}|}%
	{\gnumericPB{\raggedleft}\gnumbox[r]{\textsf{$1.62$}}}
	&\multicolumn{1}{p{\gnumericColD}|}%
	{\gnumericPB{\raggedleft}\gnumbox[r]{\textsf{$5.50$}}}
	&\multicolumn{1}{p{\gnumericColE}|}%
	{\gnumericPB{\raggedleft}\gnumbox[r]{\textsf{$1.86$}}}
\\
\hhline{|-|-|-|-|}
\end{longtable}

\gnumericTableEnd
}
\caption{Normalisation for a dipolar form for $\Upsilon(1S)$ as a function of the $b$ quark mass
for $\Lambda=5.20$~GeV.}\label{tab:N_dip_Upsi_mq3}
\end{table}

\begin{table}[H]
{\tiny%%%%%%%%%%%%%%%%%%%%%%%%%%%%%%%%%%%%%%%%%%%%%%%%%%%%%%%%%%%%%%%%%%%%%%
\def\ifundefined#1{\expandafter\ifx\csname#1\endcsname\relax}

%%  Check for the \def token for inputed files. If it is not        %%
%%  defined, the file will be processed as a standalone and the     %%
%%  preamble will be used.                                          %%
\ifundefined{inputGnumericTable}

%%  End of the preamble for the standalone. The next section is for %%
%%  documents which are included into other LaTeX2e files.          %%
\else

%%  We are not a stand alone document. For a regular table, we will %%
%%  have no preamble and only define the closing to mean nothing.   %%
    \def\gnumericTableEnd{}

%%  If we want landscape mode in an embedded document, comment out  %%
%%  the line above and uncomment the two below. The table will      %%
%%  begin on a new page and run in landscape mode.                  %%
%       \def\gnumericTableEnd{\end{landscape}}
%       \begin{landscape}

%%  End of the else clause for this file being \input.              %%
\fi

%%%%%%%%%%%%%%%%%%%%%%%%%%%%%%%%%%%%%%%%%%%%%%%%%%%%%%%%%%%%%%%%%%%%%%
%%                                                                  %%
%%  The rest is the gnumeric table, except for the closing          %%
%%  statement. Changes below will alter the table's appearance.     %%
%%                                                                  %%
%%%%%%%%%%%%%%%%%%%%%%%%%%%%%%%%%%%%%%%%%%%%%%%%%%%%%%%%%%%%%%%%%%%%%%

\providecommand{\gnumericPB}[1]%
{\let\gnumericTemp=\\#1\let\\=\gnumericTemp\hspace{0pt}}

%%  The default table format retains the relative column widths of  %%
%%  gnumeric. They can easily be changed to c, r or l. In that case %%
%%  you may want to comment out the next line and uncomment the one %%
%%  thereafter                                                      %%
\providecommand\gnumbox{\makebox[0pt]}
%%\providecommand\gnumbox[1][]{\makebox}

%% to adjust positions in multirow situations                       %%
\setlength{\bigstrutjot}{\jot}
\setlength{\extrarowheight}{\doublerulesep}

%%  The \setlongtables command keeps column widths the same across %%
%%  pages. Simply comment out next line for varying column widths.  %%
\setlongtables

\setlength\gnumericTableWidth{%
	30pt+%
	30pt+%
	30pt+%
	30pt+%
0pt}
\def\gumericNumCols{4}
\setlength\gnumericTableWidthComplete{\gnumericTableWidth+\tabcolsep*\gumericNumCols*2+\arrayrulewidth*\gumericNumCols}
\ifthenelse{\lengthtest{\gnumericTableWidthComplete > \textwidth}}%
{\def\gnumericScale{\ratio{\textwidth-\tabcolsep*\gumericNumCols*2-\arrayrulewidth*\gumericNumCols}%
{\gnumericTableWidth}}}%
{\def\gnumericScale{1}}

%%%%%%%%%%%%%%%%%%%%%%%%%%%%%%%%%%%%%%%%%%%%%%%%%%%%%%%%%%%%%%%%%%%%%%
%%                                                                  %%
%% The following are the widths of the various columns. We are      %%
%% defining them here because then they are easier to change.       %%
%% Depending on the cell formats we may use them more than once.    %%
%%                                                                  %%
%%%%%%%%%%%%%%%%%%%%%%%%%%%%%%%%%%%%%%%%%%%%%%%%%%%%%%%%%%%%%%%%%%%%%%

\def\gnumericColB{30pt*\gnumericScale}
\def\gnumericColC{30pt*\gnumericScale}
\def\gnumericColD{30pt*\gnumericScale}
\def\gnumericColE{30pt*\gnumericScale}

\begin{longtable}[c]{%
	b{\gnumericColB}%
	b{\gnumericColC}%
	b{\gnumericColD}%
	b{\gnumericColE}%
	}

%%%%%%%%%%%%%%%%%%%%%%%%%%%%%%%%%%%%%%%%%%%%%%%%%%%%%%%%%%%%%%%%%%%%%%
%%  The longtable options. (Caption. headers... see Goosens. p.124) %%
%	\caption{The Table Caption.}             \\	%
% \hline	% Across the top of the table.
%%  The rest of these options are table rows which are placed on    %%
%%  the first. last or every page. Use \multicolumn if you want.    %%

%%  Header for the first page.                                      %%
%	\multicolumn{4}{c}{The First Header} \\ \hline 
%	\multicolumn{1}{c}{colTag}	%Column 1
%	&\multicolumn{1}{c}{colTag}	%Column 1
%	&\multicolumn{1}{c}{colTag}	%Column 2
%	&\multicolumn{1}{c}{colTag}	%Column 3
%	&\multicolumn{1}{c}{colTag}	\\ \hline %Last column
%	\endfirsthead

%%  The running header definition.                                  %%
%	\hline
%	\multicolumn{4}{l}{\ldots\small\slshape continued} \\ \hline
%	\multicolumn{1}{c}{colTag}	%Column 1
%	&\multicolumn{1}{c}{colTag}	%Column 1
%	&\multicolumn{1}{c}{colTag}	%Column 2
%	&\multicolumn{1}{c}{colTag}	%Column 3
%	&\multicolumn{1}{c}{colTag}	\\ \hline %Last column
%	\endhead

%%  The running footer definition.                                  %%
%	\hline
%	\multicolumn{4}{r}{\small\slshape continued\ldots} \\
%	\endfoot

%%  The ending footer definition.                                   %%
%	\multicolumn{4}{c}{That's all folks} \\ \hline 
%	\endlastfoot
%%%%%%%%%%%%%%%%%%%%%%%%%%%%%%%%%%%%%%%%%%%%%%%%%%%%%%%%%%%%%%%%%%%%%%

\hhline{|-|-|-|-|}
	 \multicolumn{1}{|p{\gnumericColB}|}%
	{\gnumericPB{\raggedleft}\gnumbox[r]{\textsf{$4.80$}}}
	&\multicolumn{1}{p{\gnumericColC}|}%
	{\gnumericPB{\raggedleft}\gnumbox[r]{\textsf{$0.93$}}}
	&\multicolumn{1}{p{\gnumericColD}|}%
	{\gnumericPB{\raggedleft}\gnumbox[r]{\textsf{$5.17$}}}
	&\multicolumn{1}{p{\gnumericColE}|}%
	{\gnumericPB{\raggedleft}\gnumbox[r]{\textsf{$1.17$}}}
\\
\hhline{|----|}
	 \multicolumn{1}{|p{\gnumericColB}|}%
	{\gnumericPB{\raggedleft}\gnumbox[r]{\textsf{$4.84$}}}
	&\multicolumn{1}{p{\gnumericColC}|}%
	{\gnumericPB{\raggedleft}\gnumbox[r]{\textsf{$0.97$}}}
	&\multicolumn{1}{p{\gnumericColD}|}%
	{\gnumericPB{\raggedleft}\gnumbox[r]{\textsf{$5.21$}}}
	&\multicolumn{1}{p{\gnumericColE}|}%
	{\gnumericPB{\raggedleft}\gnumbox[r]{\textsf{$1.18$}}}
\\
\hhline{|----|}
	 \multicolumn{1}{|p{\gnumericColB}|}%
	{\gnumericPB{\raggedleft}\gnumbox[r]{\textsf{$4.87$}}}
	&\multicolumn{1}{p{\gnumericColC}|}%
	{\gnumericPB{\raggedleft}\gnumbox[r]{\textsf{$1.00$}}}
	&\multicolumn{1}{p{\gnumericColD}|}%
	{\gnumericPB{\raggedleft}\gnumbox[r]{\textsf{$5.24$}}}
	&\multicolumn{1}{p{\gnumericColE}|}%
	{\gnumericPB{\raggedleft}\gnumbox[r]{\textsf{$1.20$}}}
\\
\hhline{|----|}
	 \multicolumn{1}{|p{\gnumericColB}|}%
	{\gnumericPB{\raggedleft}\gnumbox[r]{\textsf{$4.91$}}}
	&\multicolumn{1}{p{\gnumericColC}|}%
	{\gnumericPB{\raggedleft}\gnumbox[r]{\textsf{$1.03$}}}
	&\multicolumn{1}{p{\gnumericColD}|}%
	{\gnumericPB{\raggedleft}\gnumbox[r]{\textsf{$5.28$}}}
	&\multicolumn{1}{p{\gnumericColE}|}%
	{\gnumericPB{\raggedleft}\gnumbox[r]{\textsf{$1.21$}}}
\\
\hhline{|----|}
	 \multicolumn{1}{|p{\gnumericColB}|}%
	{\gnumericPB{\raggedleft}\gnumbox[r]{\textsf{$4.95$}}}
	&\multicolumn{1}{p{\gnumericColC}|}%
	{\gnumericPB{\raggedleft}\gnumbox[r]{\textsf{$1.05$}}}
	&\multicolumn{1}{p{\gnumericColD}|}%
	{\gnumericPB{\raggedleft}\gnumbox[r]{\textsf{$5.32$}}}
	&\multicolumn{1}{p{\gnumericColE}|}%
	{\gnumericPB{\raggedleft}\gnumbox[r]{\textsf{$1.23$}}}
\\
\hhline{|----|}
	 \multicolumn{1}{|p{\gnumericColB}|}%
	{\gnumericPB{\raggedleft}\gnumbox[r]{\textsf{$4.98$}}}
	&\multicolumn{1}{p{\gnumericColC}|}%
	{\gnumericPB{\raggedleft}\gnumbox[r]{\textsf{$1.07$}}}
	&\multicolumn{1}{p{\gnumericColD}|}%
	{\gnumericPB{\raggedleft}\gnumbox[r]{\textsf{$5.35$}}}
	&\multicolumn{1}{p{\gnumericColE}|}%
	{\gnumericPB{\raggedleft}\gnumbox[r]{\textsf{$1.24$}}}
\\
\hhline{|----|}
	 \multicolumn{1}{|p{\gnumericColB}|}%
	{\gnumericPB{\raggedleft}\gnumbox[r]{\textsf{$5.02$}}}
	&\multicolumn{1}{p{\gnumericColC}|}%
	{\gnumericPB{\raggedleft}\gnumbox[r]{\textsf{$1.10$}}}
	&\multicolumn{1}{p{\gnumericColD}|}%
	{\gnumericPB{\raggedleft}\gnumbox[r]{\textsf{$5.39$}}}
	&\multicolumn{1}{p{\gnumericColE}|}%
	{\gnumericPB{\raggedleft}\gnumbox[r]{\textsf{$1.25$}}}
\\
\hhline{|----|}
	 \multicolumn{1}{|p{\gnumericColB}|}%
	{\gnumericPB{\raggedleft}\gnumbox[r]{\textsf{$5.06$}}}
	&\multicolumn{1}{p{\gnumericColC}|}%
	{\gnumericPB{\raggedleft}\gnumbox[r]{\textsf{$1.11$}}}
	&\multicolumn{1}{p{\gnumericColD}|}%
	{\gnumericPB{\raggedleft}\gnumbox[r]{\textsf{$5.43$}}}
	&\multicolumn{1}{p{\gnumericColE}|}%
	{\gnumericPB{\raggedleft}\gnumbox[r]{\textsf{$1.27$}}}
\\
\hhline{|----|}
	 \multicolumn{1}{|p{\gnumericColB}|}%
	{\gnumericPB{\raggedleft}\gnumbox[r]{\textsf{$5.09$}}}
	&\multicolumn{1}{p{\gnumericColC}|}%
	{\gnumericPB{\raggedleft}\gnumbox[r]{\textsf{$1.13$}}}
	&\multicolumn{1}{p{\gnumericColD}|}%
	{\gnumericPB{\raggedleft}\gnumbox[r]{\textsf{$5.46$}}}
	&\multicolumn{1}{p{\gnumericColE}|}%
	{\gnumericPB{\raggedleft}\gnumbox[r]{\textsf{$1.28$}}}
\\
\hhline{|----|}
	 \multicolumn{1}{|p{\gnumericColB}|}%
	{\gnumericPB{\raggedleft}\gnumbox[r]{\textsf{$5.13$}}}
	&\multicolumn{1}{p{\gnumericColC}|}%
	{\gnumericPB{\raggedleft}\gnumbox[r]{\textsf{$1.15$}}}
	&\multicolumn{1}{p{\gnumericColD}|}%
	{\gnumericPB{\raggedleft}\gnumbox[r]{\textsf{$5.50$}}}
	&\multicolumn{1}{p{\gnumericColE}|}%
	{\gnumericPB{\raggedleft}\gnumbox[r]{\textsf{$1.29$}}}
\\
\hhline{|-|-|-|-|}
\end{longtable}

\gnumericTableEnd
}
\caption{Normalisation for a dipolar form for $\Upsilon(1S)$ as a function of  the $b$ quark mass
for $\Lambda=6.25$~GeV.}\label{tab:N_dip_Upsi_mq4}
\end{table}

\begin{table}[H]
{\tiny%%%%%%%%%%%%%%%%%%%%%%%%%%%%%%%%%%%%%%%%%%%%%%%%%%%%%%%%%%%%%%%%%%%%%%
\def\ifundefined#1{\expandafter\ifx\csname#1\endcsname\relax}

%%  Check for the \def token for inputed files. If it is not        %%
%%  defined, the file will be processed as a standalone and the     %%
%%  preamble will be used.                                          %%
\ifundefined{inputGnumericTable}

%%  End of the preamble for the standalone. The next section is for %%
%%  documents which are included into other LaTeX2e files.          %%
\else

%%  We are not a stand alone document. For a regular table, we will %%
%%  have no preamble and only define the closing to mean nothing.   %%
    \def\gnumericTableEnd{}

%%  If we want landscape mode in an embedded document, comment out  %%
%%  the line above and uncomment the two below. The table will      %%
%%  begin on a new page and run in landscape mode.                  %%
%       \def\gnumericTableEnd{\end{landscape}}
%       \begin{landscape}

%%  End of the else clause for this file being \input.              %%
\fi

%%%%%%%%%%%%%%%%%%%%%%%%%%%%%%%%%%%%%%%%%%%%%%%%%%%%%%%%%%%%%%%%%%%%%%
%%                                                                  %%
%%  The rest is the gnumeric table, except for the closing          %%
%%  statement. Changes below will alter the table's appearance.     %%
%%                                                                  %%
%%%%%%%%%%%%%%%%%%%%%%%%%%%%%%%%%%%%%%%%%%%%%%%%%%%%%%%%%%%%%%%%%%%%%%

\providecommand{\gnumericPB}[1]%
{\let\gnumericTemp=\\#1\let\\=\gnumericTemp\hspace{0pt}}

%%  The default table format retains the relative column widths of  %%
%%  gnumeric. They can easily be changed to c, r or l. In that case %%
%%  you may want to comment out the next line and uncomment the one %%
%%  thereafter                                                      %%
\providecommand\gnumbox{\makebox[0pt]}
%%\providecommand\gnumbox[1][]{\makebox}

%% to adjust positions in multirow situations                       %%
\setlength{\bigstrutjot}{\jot}
\setlength{\extrarowheight}{\doublerulesep}

%%  The \setlongtables command keeps column widths the same across %%
%%  pages. Simply comment out next line for varying column widths.  %%
\setlongtables

\setlength\gnumericTableWidth{%
	30pt+%
	30pt+%
	30pt+%
	30pt+%
0pt}
\def\gumericNumCols{4}
\setlength\gnumericTableWidthComplete{\gnumericTableWidth+\tabcolsep*\gumericNumCols*2+\arrayrulewidth*\gumericNumCols}
\ifthenelse{\lengthtest{\gnumericTableWidthComplete > \textwidth}}%
{\def\gnumericScale{\ratio{\textwidth-\tabcolsep*\gumericNumCols*2-\arrayrulewidth*\gumericNumCols}%
{\gnumericTableWidth}}}%
{\def\gnumericScale{1}}

%%%%%%%%%%%%%%%%%%%%%%%%%%%%%%%%%%%%%%%%%%%%%%%%%%%%%%%%%%%%%%%%%%%%%%
%%                                                                  %%
%% The following are the widths of the various columns. We are      %%
%% defining them here because then they are easier to change.       %%
%% Depending on the cell formats we may use them more than once.    %%
%%                                                                  %%
%%%%%%%%%%%%%%%%%%%%%%%%%%%%%%%%%%%%%%%%%%%%%%%%%%%%%%%%%%%%%%%%%%%%%%

\def\gnumericColB{30pt*\gnumericScale}
\def\gnumericColC{30pt*\gnumericScale}
\def\gnumericColD{30pt*\gnumericScale}
\def\gnumericColE{30pt*\gnumericScale}

\begin{longtable}[c]{%
	b{\gnumericColB}%
	b{\gnumericColC}%
	b{\gnumericColD}%
	b{\gnumericColE}%
	}

%%%%%%%%%%%%%%%%%%%%%%%%%%%%%%%%%%%%%%%%%%%%%%%%%%%%%%%%%%%%%%%%%%%%%%
%%  The longtable options. (Caption. headers... see Goosens. p.124) %%
%	\caption{The Table Caption.}             \\	%
% \hline	% Across the top of the table.
%%  The rest of these options are table rows which are placed on    %%
%%  the first. last or every page. Use \multicolumn if you want.    %%

%%  Header for the first page.                                      %%
%	\multicolumn{4}{c}{The First Header} \\ \hline 
%	\multicolumn{1}{c}{colTag}	%Column 1
%	&\multicolumn{1}{c}{colTag}	%Column 1
%	&\multicolumn{1}{c}{colTag}	%Column 2
%	&\multicolumn{1}{c}{colTag}	%Column 3
%	&\multicolumn{1}{c}{colTag}	\\ \hline %Last column
%	\endfirsthead

%%  The running header definition.                                  %%
%	\hline
%	\multicolumn{4}{l}{\ldots\small\slshape continued} \\ \hline
%	\multicolumn{1}{c}{colTag}	%Column 1
%	&\multicolumn{1}{c}{colTag}	%Column 1
%	&\multicolumn{1}{c}{colTag}	%Column 2
%	&\multicolumn{1}{c}{colTag}	%Column 3
%	&\multicolumn{1}{c}{colTag}	\\ \hline %Last column
%	\endhead

%%  The running footer definition.                                  %%
%	\hline
%	\multicolumn{4}{r}{\small\slshape continued\ldots} \\
%	\endfoot

%%  The ending footer definition.                                   %%
%	\multicolumn{4}{c}{That's all folks} \\ \hline 
%	\endlastfoot
%%%%%%%%%%%%%%%%%%%%%%%%%%%%%%%%%%%%%%%%%%%%%%%%%%%%%%%%%%%%%%%%%%%%%%

\hhline{|-|-|-|-|}
	 \multicolumn{1}{|p{\gnumericColB}|}%
	{\gnumericPB{\raggedleft}\gnumbox[r]{\textsf{$\Lambda$}}}
	&\multicolumn{1}{p{\gnumericColC}|}%
	{\gnumericPB{\raggedleft}\gnumbox[r]{\textsf{$N$}}}
	&\multicolumn{1}{p{\gnumericColD}|}%
	{\gnumericPB{\raggedleft}\gnumbox[r]{\textsf{$\Lambda$}}}
	&\multicolumn{1}{p{\gnumericColE}|}%
	{\gnumericPB{\raggedleft}\gnumbox[r]{\textsf{$N$}}}
\\
\hhline{|----|}
	 \multicolumn{1}{|p{\gnumericColB}|}%
	{\gnumericPB{\raggedleft}\gnumbox[r]{\textsf{$3.00$}}}
	&\multicolumn{1}{p{\gnumericColC}|}%
	{\gnumericPB{\raggedleft}\gnumbox[r]{\textsf{$2.53$}}}
	&\multicolumn{1}{p{\gnumericColD}|}%
	{\gnumericPB{\raggedleft}\gnumbox[r]{\textsf{$4.71$}}}
	&\multicolumn{1}{p{\gnumericColE}|}%
	{\gnumericPB{\raggedleft}\gnumbox[r]{\textsf{$1.31$}}}
\\
\hhline{|----|}
	 \multicolumn{1}{|p{\gnumericColB}|}%
	{\gnumericPB{\raggedleft}\gnumbox[r]{\textsf{$3.17$}}}
	&\multicolumn{1}{p{\gnumericColC}|}%
	{\gnumericPB{\raggedleft}\gnumbox[r]{\textsf{$2.33$}}}
	&\multicolumn{1}{p{\gnumericColD}|}%
	{\gnumericPB{\raggedleft}\gnumbox[r]{\textsf{$4.88$}}}
	&\multicolumn{1}{p{\gnumericColE}|}%
	{\gnumericPB{\raggedleft}\gnumbox[r]{\textsf{$1.24$}}}
\\
\hhline{|----|}
	 \multicolumn{1}{|p{\gnumericColB}|}%
	{\gnumericPB{\raggedleft}\gnumbox[r]{\textsf{$3.34$}}}
	&\multicolumn{1}{p{\gnumericColC}|}%
	{\gnumericPB{\raggedleft}\gnumbox[r]{\textsf{$2.16$}}}
	&\multicolumn{1}{p{\gnumericColD}|}%
	{\gnumericPB{\raggedleft}\gnumbox[r]{\textsf{$5.05$}}}
	&\multicolumn{1}{p{\gnumericColE}|}%
	{\gnumericPB{\raggedleft}\gnumbox[r]{\textsf{$1.18$}}}
\\
\hhline{|----|}
	 \multicolumn{1}{|p{\gnumericColB}|}%
	{\gnumericPB{\raggedleft}\gnumbox[r]{\textsf{$3.51$}}}
	&\multicolumn{1}{p{\gnumericColC}|}%
	{\gnumericPB{\raggedleft}\gnumbox[r]{\textsf{$2.00$}}}
	&\multicolumn{1}{p{\gnumericColD}|}%
	{\gnumericPB{\raggedleft}\gnumbox[r]{\textsf{$5.22$}}}
	&\multicolumn{1}{p{\gnumericColE}|}%
	{\gnumericPB{\raggedleft}\gnumbox[r]{\textsf{$1.12$}}}
\\
\hhline{|----|}
	 \multicolumn{1}{|p{\gnumericColB}|}%
	{\gnumericPB{\raggedleft}\gnumbox[r]{\textsf{$3.68$}}}
	&\multicolumn{1}{p{\gnumericColC}|}%
	{\gnumericPB{\raggedleft}\gnumbox[r]{\textsf{$1.87$}}}
	&\multicolumn{1}{p{\gnumericColD}|}%
	{\gnumericPB{\raggedleft}\gnumbox[r]{\textsf{$5.39$}}}
	&\multicolumn{1}{p{\gnumericColE}|}%
	{\gnumericPB{\raggedleft}\gnumbox[r]{\textsf{$1.07$}}}
\\
\hhline{|----|}
	 \multicolumn{1}{|p{\gnumericColB}|}%
	{\gnumericPB{\raggedleft}\gnumbox[r]{\textsf{$3.86$}}}
	&\multicolumn{1}{p{\gnumericColC}|}%
	{\gnumericPB{\raggedleft}\gnumbox[r]{\textsf{$1.75$}}}
	&\multicolumn{1}{p{\gnumericColD}|}%
	{\gnumericPB{\raggedleft}\gnumbox[r]{\textsf{$5.57$}}}
	&\multicolumn{1}{p{\gnumericColE}|}%
	{\gnumericPB{\raggedleft}\gnumbox[r]{\textsf{$1.02$}}}
\\
\hhline{|----|}
	 \multicolumn{1}{|p{\gnumericColB}|}%
	{\gnumericPB{\raggedleft}\gnumbox[r]{\textsf{$4.03$}}}
	&\multicolumn{1}{p{\gnumericColC}|}%
	{\gnumericPB{\raggedleft}\gnumbox[r]{\textsf{$1.64$}}}
	&\multicolumn{1}{p{\gnumericColD}|}%
	{\gnumericPB{\raggedleft}\gnumbox[r]{\textsf{$5.74$}}}
	&\multicolumn{1}{p{\gnumericColE}|}%
	{\gnumericPB{\raggedleft}\gnumbox[r]{\textsf{$0.98$}}}
\\
\hhline{|----|}
	 \multicolumn{1}{|p{\gnumericColB}|}%
	{\gnumericPB{\raggedleft}\gnumbox[r]{\textsf{$4.20$}}}
	&\multicolumn{1}{p{\gnumericColC}|}%
	{\gnumericPB{\raggedleft}\gnumbox[r]{\textsf{$1.54$}}}
	&\multicolumn{1}{p{\gnumericColD}|}%
	{\gnumericPB{\raggedleft}\gnumbox[r]{\textsf{$5.91$}}}
	&\multicolumn{1}{p{\gnumericColE}|}%
	{\gnumericPB{\raggedleft}\gnumbox[r]{\textsf{$0.94$}}}
\\
\hhline{|----|}
	 \multicolumn{1}{|p{\gnumericColB}|}%
	{\gnumericPB{\raggedleft}\gnumbox[r]{\textsf{$4.37$}}}
	&\multicolumn{1}{p{\gnumericColC}|}%
	{\gnumericPB{\raggedleft}\gnumbox[r]{\textsf{$1.46$}}}
	&\multicolumn{1}{p{\gnumericColD}|}%
	{\gnumericPB{\raggedleft}\gnumbox[r]{\textsf{$6.08$}}}
	&\multicolumn{1}{p{\gnumericColE}|}%
	{\gnumericPB{\raggedleft}\gnumbox[r]{\textsf{$0.90$}}}
\\
\hhline{|----|}
	 \multicolumn{1}{|p{\gnumericColB}|}%
	{\gnumericPB{\raggedleft}\gnumbox[r]{\textsf{$4.54$}}}
	&\multicolumn{1}{p{\gnumericColC}|}%
	{\gnumericPB{\raggedleft}\gnumbox[r]{\textsf{$1.38$}}}
	&\multicolumn{1}{p{\gnumericColD}|}%
	{\gnumericPB{\raggedleft}\gnumbox[r]{\textsf{$6.25$}}}
	&\multicolumn{1}{p{\gnumericColE}|}%
	{\gnumericPB{\raggedleft}\gnumbox[r]{\textsf{$0.87$}}}
\\
\hhline{|-|-|-|-|}
\end{longtable}

\gnumericTableEnd
}
\caption{Normalisation for a gaussian form for $\Upsilon(1S)$ as a function of $\Lambda$ for $m_b=4.80$~GeV.}\label{tab:N_exp_Upsi_lam1}
\end{table}

\begin{table}[H]
{\tiny%%%%%%%%%%%%%%%%%%%%%%%%%%%%%%%%%%%%%%%%%%%%%%%%%%%%%%%%%%%%%%%%%%%%%%
\def\ifundefined#1{\expandafter\ifx\csname#1\endcsname\relax}

%%  Check for the \def token for inputed files. If it is not        %%
%%  defined. the file will be processed as a standalone and the     %%
%%  preamble will be used.                                          %%
\ifundefined{inputGnumericTable}

%%  End of the preamble for the standalone. The next section is for %%
%%  documents which are included into other LaTeX2e files.          %%
\else

%%  We are not a stand alone document. For a regular table. we will %%
%%  have no preamble and only define the closing to mean nothing.   %%
    \def\gnumericTableEnd{}

%%  If we want landscape mode in an embedded document. comment out  %%
%%  the line above and uncomment the two below. The table will      %%
%%  begin on a new page and run in landscape mode.                  %%
%       \def\gnumericTableEnd{\end{landscape}}
%       \begin{landscape}

%%  End of the else clause for this file being \input.              %%
\fi

%%%%%%%%%%%%%%%%%%%%%%%%%%%%%%%%%%%%%%%%%%%%%%%%%%%%%%%%%%%%%%%%%%%%%%
%%                                                                  %%
%%  The rest is the gnumeric table. except for the closing          %%
%%  statement. Changes below will alter the table's appearance.     %%
%%                                                                  %%
%%%%%%%%%%%%%%%%%%%%%%%%%%%%%%%%%%%%%%%%%%%%%%%%%%%%%%%%%%%%%%%%%%%%%%

\providecommand{\gnumericPB}[1]%
{\let\gnumericTemp=\\#1\let\\=\gnumericTemp\hspace{0pt}}

%%  The default table format retains the relative column widths of  %%
%%  gnumeric. They can easily be changed to c. r or l. In that case %%
%%  you may want to comment out the next line and uncomment the one %%
%%  thereafter                                                      %%
\providecommand\gnumbox{\makebox[0pt]}
%%\providecommand\gnumbox[1][]{\makebox}

%% to adjust positions in multirow situations                       %%
\setlength{\bigstrutjot}{\jot}
\setlength{\extrarowheight}{\doublerulesep}

%%  The \setlongtables command keeps column widths the same across %%
%%  pages. Simply comment out next line for varying column widths.  %%
\setlongtables

\setlength\gnumericTableWidth{%
	30pt+%
	30pt+%
	30pt+%
	30pt+%
0pt}
\def\gumericNumCols{4}
\setlength\gnumericTableWidthComplete{\gnumericTableWidth+\tabcolsep*\gumericNumCols*2+\arrayrulewidth*\gumericNumCols}
\ifthenelse{\lengthtest{\gnumericTableWidthComplete > \textwidth}}%
{\def\gnumericScale{\ratio{\textwidth-\tabcolsep*\gumericNumCols*2-\arrayrulewidth*\gumericNumCols}%
{\gnumericTableWidth}}}%
{\def\gnumericScale{1}}

%%%%%%%%%%%%%%%%%%%%%%%%%%%%%%%%%%%%%%%%%%%%%%%%%%%%%%%%%%%%%%%%%%%%%%
%%                                                                  %%
%% The following are the widths of the various columns. We are      %%
%% defining them here because then they are easier to change.       %%
%% Depending on the cell formats we may use them more than once.    %%
%%                                                                  %%
%%%%%%%%%%%%%%%%%%%%%%%%%%%%%%%%%%%%%%%%%%%%%%%%%%%%%%%%%%%%%%%%%%%%%%

\def\gnumericColB{30pt*\gnumericScale}
\def\gnumericColC{30pt*\gnumericScale}
\def\gnumericColD{30pt*\gnumericScale}
\def\gnumericColE{30pt*\gnumericScale}

\begin{longtable}[c]{%
	b{\gnumericColB}%
	b{\gnumericColC}%
	b{\gnumericColD}%
	b{\gnumericColE}%
	}

%%%%%%%%%%%%%%%%%%%%%%%%%%%%%%%%%%%%%%%%%%%%%%%%%%%%%%%%%%%%%%%%%%%%%%
%%  The longtable options. (Caption. headers... see Goosens. p.124) %%
%	\caption{The Table Caption.}             \\	%
% \hline	% Across the top of the table.
%%  The rest of these options are table rows which are placed on    %%
%%  the first. last or every page. Use \multicolumn if you want.    %%

%%  Header for the first page.                                      %%
%	\multicolumn{4}{c}{The First Header} \\ \hline 
%	\multicolumn{1}{c}{colTag}	%Column 1
%	&\multicolumn{1}{c}{colTag}	%Column 1
%	&\multicolumn{1}{c}{colTag}	%Column 2
%	&\multicolumn{1}{c}{colTag}	%Column 3
%	&\multicolumn{1}{c}{colTag}	\\ \hline %Last column
%	\endfirsthead

%%  The running header definition.                                  %%
%	\hline
%	\multicolumn{4}{l}{\ldots\small\slshape continued} \\ \hline
%	\multicolumn{1}{c}{colTag}	%Column 1
%	&\multicolumn{1}{c}{colTag}	%Column 1
%	&\multicolumn{1}{c}{colTag}	%Column 2
%	&\multicolumn{1}{c}{colTag}	%Column 3
%	&\multicolumn{1}{c}{colTag}	\\ \hline %Last column
%	\endhead

%%  The running footer definition.                                  %%
%	\hline
%	\multicolumn{4}{r}{\small\slshape continued\ldots} \\
%	\endfoot

%%  The ending footer definition.                                   %%
%	\multicolumn{4}{c}{That's all folks} \\ \hline 
%	\endlastfoot
%%%%%%%%%%%%%%%%%%%%%%%%%%%%%%%%%%%%%%%%%%%%%%%%%%%%%%%%%%%%%%%%%%%%%%

\hhline{|-|-|-|-|}
	 \multicolumn{1}{|p{\gnumericColB}|}%
	{\gnumericPB{\raggedleft}\gnumbox[r]{\textsf{$3.00$}}}
	&\multicolumn{1}{p{\gnumericColC}|}%
	{\gnumericPB{\raggedleft}\gnumbox[r]{\textsf{$3.72$}}}
	&\multicolumn{1}{p{\gnumericColD}|}%
	{\gnumericPB{\raggedleft}\gnumbox[r]{\textsf{$4.71$}}}
	&\multicolumn{1}{p{\gnumericColE}|}%
	{\gnumericPB{\raggedleft}\gnumbox[r]{\textsf{$1.66$}}}
\\
\hhline{|----|}
	 \multicolumn{1}{|p{\gnumericColB}|}%
	{\gnumericPB{\raggedleft}\gnumbox[r]{\textsf{$3.17$}}}
	&\multicolumn{1}{p{\gnumericColC}|}%
	{\gnumericPB{\raggedleft}\gnumbox[r]{\textsf{$3.36$}}}
	&\multicolumn{1}{p{\gnumericColD}|}%
	{\gnumericPB{\raggedleft}\gnumbox[r]{\textsf{$4.88$}}}
	&\multicolumn{1}{p{\gnumericColE}|}%
	{\gnumericPB{\raggedleft}\gnumbox[r]{\textsf{$1.57$}}}
\\
\hhline{|----|}
	 \multicolumn{1}{|p{\gnumericColB}|}%
	{\gnumericPB{\raggedleft}\gnumbox[r]{\textsf{$3.34$}}}
	&\multicolumn{1}{p{\gnumericColC}|}%
	{\gnumericPB{\raggedleft}\gnumbox[r]{\textsf{$3.05$}}}
	&\multicolumn{1}{p{\gnumericColD}|}%
	{\gnumericPB{\raggedleft}\gnumbox[r]{\textsf{$5.05$}}}
	&\multicolumn{1}{p{\gnumericColE}|}%
	{\gnumericPB{\raggedleft}\gnumbox[r]{\textsf{$1.48$}}}
\\
\hhline{|----|}
	 \multicolumn{1}{|p{\gnumericColB}|}%
	{\gnumericPB{\raggedleft}\gnumbox[r]{\textsf{$3.51$}}}
	&\multicolumn{1}{p{\gnumericColC}|}%
	{\gnumericPB{\raggedleft}\gnumbox[r]{\textsf{$2.79$}}}
	&\multicolumn{1}{p{\gnumericColD}|}%
	{\gnumericPB{\raggedleft}\gnumbox[r]{\textsf{$5.22$}}}
	&\multicolumn{1}{p{\gnumericColE}|}%
	{\gnumericPB{\raggedleft}\gnumbox[r]{\textsf{$1.39$}}}
\\
\hhline{|----|}
	 \multicolumn{1}{|p{\gnumericColB}|}%
	{\gnumericPB{\raggedleft}\gnumbox[r]{\textsf{$3.68$}}}
	&\multicolumn{1}{p{\gnumericColC}|}%
	{\gnumericPB{\raggedleft}\gnumbox[r]{\textsf{$2.56$}}}
	&\multicolumn{1}{p{\gnumericColD}|}%
	{\gnumericPB{\raggedleft}\gnumbox[r]{\textsf{$5.39$}}}
	&\multicolumn{1}{p{\gnumericColE}|}%
	{\gnumericPB{\raggedleft}\gnumbox[r]{\textsf{$1.32$}}}
\\
\hhline{|----|}
	 \multicolumn{1}{|p{\gnumericColB}|}%
	{\gnumericPB{\raggedleft}\gnumbox[r]{\textsf{$3.86$}}}
	&\multicolumn{1}{p{\gnumericColC}|}%
	{\gnumericPB{\raggedleft}\gnumbox[r]{\textsf{$2.36$}}}
	&\multicolumn{1}{p{\gnumericColD}|}%
	{\gnumericPB{\raggedleft}\gnumbox[r]{\textsf{$5.57$}}}
	&\multicolumn{1}{p{\gnumericColE}|}%
	{\gnumericPB{\raggedleft}\gnumbox[r]{\textsf{$1.25$}}}
\\
\hhline{|----|}
	 \multicolumn{1}{|p{\gnumericColB}|}%
	{\gnumericPB{\raggedleft}\gnumbox[r]{\textsf{$4.03$}}}
	&\multicolumn{1}{p{\gnumericColC}|}%
	{\gnumericPB{\raggedleft}\gnumbox[r]{\textsf{$2.19$}}}
	&\multicolumn{1}{p{\gnumericColD}|}%
	{\gnumericPB{\raggedleft}\gnumbox[r]{\textsf{$5.74$}}}
	&\multicolumn{1}{p{\gnumericColE}|}%
	{\gnumericPB{\raggedleft}\gnumbox[r]{\textsf{$1.19$}}}
\\
\hhline{|----|}
	 \multicolumn{1}{|p{\gnumericColB}|}%
	{\gnumericPB{\raggedleft}\gnumbox[r]{\textsf{$4.20$}}}
	&\multicolumn{1}{p{\gnumericColC}|}%
	{\gnumericPB{\raggedleft}\gnumbox[r]{\textsf{$2.03$}}}
	&\multicolumn{1}{p{\gnumericColD}|}%
	{\gnumericPB{\raggedleft}\gnumbox[r]{\textsf{$5.91$}}}
	&\multicolumn{1}{p{\gnumericColE}|}%
	{\gnumericPB{\raggedleft}\gnumbox[r]{\textsf{$1.13$}}}
\\
\hhline{|----|}
	 \multicolumn{1}{|p{\gnumericColB}|}%
	{\gnumericPB{\raggedleft}\gnumbox[r]{\textsf{$4.37$}}}
	&\multicolumn{1}{p{\gnumericColC}|}%
	{\gnumericPB{\raggedleft}\gnumbox[r]{\textsf{$1.90$}}}
	&\multicolumn{1}{p{\gnumericColD}|}%
	{\gnumericPB{\raggedleft}\gnumbox[r]{\textsf{$6.08$}}}
	&\multicolumn{1}{p{\gnumericColE}|}%
	{\gnumericPB{\raggedleft}\gnumbox[r]{\textsf{$1.08$}}}
\\
\hhline{|----|}
	 \multicolumn{1}{|p{\gnumericColB}|}%
	{\gnumericPB{\raggedleft}\gnumbox[r]{\textsf{$4.54$}}}
	&\multicolumn{1}{p{\gnumericColC}|}%
	{\gnumericPB{\raggedleft}\gnumbox[r]{\textsf{$1.77$}}}
	&\multicolumn{1}{p{\gnumericColD}|}%
	{\gnumericPB{\raggedleft}\gnumbox[r]{\textsf{$6.25$}}}
	&\multicolumn{1}{p{\gnumericColE}|}%
	{\gnumericPB{\raggedleft}\gnumbox[r]{\textsf{$1.03$}}}
\\
\hhline{|-|-|-|-|}
\end{longtable}

\gnumericTableEnd
}
\caption{Normalisation for a gaussian form for $\Upsilon(1S)$ as a function of $\Lambda$ for $m_b=5.00$~GeV.}\label{tab:N_exp_Upsi_lam2}
\end{table}

\begin{table}[H]
{\tiny%%%%%%%%%%%%%%%%%%%%%%%%%%%%%%%%%%%%%%%%%%%%%%%%%%%%%%%%%%%%%%%%%%%%%%
\def\ifundefined#1{\expandafter\ifx\csname#1\endcsname\relax}

%%  Check for the \def token for inputed files. If it is not        %%
%%  defined. the file will be processed as a standalone and the     %%
%%  preamble will be used.                                          %%
\ifundefined{inputGnumericTable}

%%  End of the preamble for the standalone. The next section is for %%
%%  documents which are included into other LaTeX2e files.          %%
\else

%%  We are not a stand alone document. For a regular table. we will %%
%%  have no preamble and only define the closing to mean nothing.   %%
    \def\gnumericTableEnd{}

%%  If we want landscape mode in an embedded document. comment out  %%
%%  the line above and uncomment the two below. The table will      %%
%%  begin on a new page and run in landscape mode.                  %%
%       \def\gnumericTableEnd{\end{landscape}}
%       \begin{landscape}

%%  End of the else clause for this file being \input.              %%
\fi

%%%%%%%%%%%%%%%%%%%%%%%%%%%%%%%%%%%%%%%%%%%%%%%%%%%%%%%%%%%%%%%%%%%%%%
%%                                                                  %%
%%  The rest is the gnumeric table. except for the closing          %%
%%  statement. Changes below will alter the table's appearance.     %%
%%                                                                  %%
%%%%%%%%%%%%%%%%%%%%%%%%%%%%%%%%%%%%%%%%%%%%%%%%%%%%%%%%%%%%%%%%%%%%%%

\providecommand{\gnumericPB}[1]%
{\let\gnumericTemp=\\#1\let\\=\gnumericTemp\hspace{0pt}}

%%  The default table format retains the relative column widths of  %%
%%  gnumeric. They can easily be changed to c. r or l. In that case %%
%%  you may want to comment out the next line and uncomment the one %%
%%  thereafter                                                      %%
\providecommand\gnumbox{\makebox[0pt]}
%%\providecommand\gnumbox[1][]{\makebox}

%% to adjust positions in multirow situations                       %%
\setlength{\bigstrutjot}{\jot}
\setlength{\extrarowheight}{\doublerulesep}

%%  The \setlongtables command keeps column widths the same across %%
%%  pages. Simply comment out next line for varying column widths.  %%
\setlongtables

\setlength\gnumericTableWidth{%
	30pt+%
	30pt+%
	30pt+%
	30pt+%
0pt}
\def\gumericNumCols{4}
\setlength\gnumericTableWidthComplete{\gnumericTableWidth+\tabcolsep*\gumericNumCols*2+\arrayrulewidth*\gumericNumCols}
\ifthenelse{\lengthtest{\gnumericTableWidthComplete > \textwidth}}%
{\def\gnumericScale{\ratio{\textwidth-\tabcolsep*\gumericNumCols*2-\arrayrulewidth*\gumericNumCols}%
{\gnumericTableWidth}}}%
{\def\gnumericScale{1}}

%%%%%%%%%%%%%%%%%%%%%%%%%%%%%%%%%%%%%%%%%%%%%%%%%%%%%%%%%%%%%%%%%%%%%%
%%                                                                  %%
%% The following are the widths of the various columns. We are      %%
%% defining them here because then they are easier to change.       %%
%% Depending on the cell formats we may use them more than once.    %%
%%                                                                  %%
%%%%%%%%%%%%%%%%%%%%%%%%%%%%%%%%%%%%%%%%%%%%%%%%%%%%%%%%%%%%%%%%%%%%%%

\def\gnumericColB{30pt*\gnumericScale}
\def\gnumericColC{30pt*\gnumericScale}
\def\gnumericColD{30pt*\gnumericScale}
\def\gnumericColE{30pt*\gnumericScale}

\begin{longtable}[c]{%
	b{\gnumericColB}%
	b{\gnumericColC}%
	b{\gnumericColD}%
	b{\gnumericColE}%
	}

%%%%%%%%%%%%%%%%%%%%%%%%%%%%%%%%%%%%%%%%%%%%%%%%%%%%%%%%%%%%%%%%%%%%%%
%%  The longtable options. (Caption. headers... see Goosens. p.124) %%
%	\caption{The Table Caption.}             \\	%
% \hline	% Across the top of the table.
%%  The rest of these options are table rows which are placed on    %%
%%  the first. last or every page. Use \multicolumn if you want.    %%

%%  Header for the first page.                                      %%
%	\multicolumn{4}{c}{The First Header} \\ \hline 
%	\multicolumn{1}{c}{colTag}	%Column 1
%	&\multicolumn{1}{c}{colTag}	%Column 1
%	&\multicolumn{1}{c}{colTag}	%Column 2
%	&\multicolumn{1}{c}{colTag}	%Column 3
%	&\multicolumn{1}{c}{colTag}	\\ \hline %Last column
%	\endfirsthead

%%  The running header definition.                                  %%
%	\hline
%	\multicolumn{4}{l}{\ldots\small\slshape continued} \\ \hline
%	\multicolumn{1}{c}{colTag}	%Column 1
%	&\multicolumn{1}{c}{colTag}	%Column 1
%	&\multicolumn{1}{c}{colTag}	%Column 2
%	&\multicolumn{1}{c}{colTag}	%Column 3
%	&\multicolumn{1}{c}{colTag}	\\ \hline %Last column
%	\endhead

%%  The running footer definition.                                  %%
%	\hline
%	\multicolumn{4}{r}{\small\slshape continued\ldots} \\
%	\endfoot

%%  The ending footer definition.                                   %%
%	\multicolumn{4}{c}{That's all folks} \\ \hline 
%	\endlastfoot
%%%%%%%%%%%%%%%%%%%%%%%%%%%%%%%%%%%%%%%%%%%%%%%%%%%%%%%%%%%%%%%%%%%%%%

\hhline{|-|-|-|-|}
	 \multicolumn{1}{|p{\gnumericColB}|}%
	{\gnumericPB{\raggedleft}\gnumbox[r]{\textsf{$3.00$}}}
	&\multicolumn{1}{p{\gnumericColC}|}%
	{\gnumericPB{\raggedleft}\gnumbox[r]{\textsf{$4.60$}}}
	&\multicolumn{1}{p{\gnumericColD}|}%
	{\gnumericPB{\raggedleft}\gnumbox[r]{\textsf{$4.71$}}}
	&\multicolumn{1}{p{\gnumericColE}|}%
	{\gnumericPB{\raggedleft}\gnumbox[r]{\textsf{$1.91$}}}
\\
\hhline{|----|}
	 \multicolumn{1}{|p{\gnumericColB}|}%
	{\gnumericPB{\raggedleft}\gnumbox[r]{\textsf{$3.17$}}}
	&\multicolumn{1}{p{\gnumericColC}|}%
	{\gnumericPB{\raggedleft}\gnumbox[r]{\textsf{$4.11$}}}
	&\multicolumn{1}{p{\gnumericColD}|}%
	{\gnumericPB{\raggedleft}\gnumbox[r]{\textsf{$4.88$}}}
	&\multicolumn{1}{p{\gnumericColE}|}%
	{\gnumericPB{\raggedleft}\gnumbox[r]{\textsf{$1.79$}}}
\\
\hhline{|----|}
	 \multicolumn{1}{|p{\gnumericColB}|}%
	{\gnumericPB{\raggedleft}\gnumbox[r]{\textsf{$3.34$}}}
	&\multicolumn{1}{p{\gnumericColC}|}%
	{\gnumericPB{\raggedleft}\gnumbox[r]{\textsf{$3.70$}}}
	&\multicolumn{1}{p{\gnumericColD}|}%
	{\gnumericPB{\raggedleft}\gnumbox[r]{\textsf{$5.05$}}}
	&\multicolumn{1}{p{\gnumericColE}|}%
	{\gnumericPB{\raggedleft}\gnumbox[r]{\textsf{$1.68$}}}
\\
\hhline{|----|}
	 \multicolumn{1}{|p{\gnumericColB}|}%
	{\gnumericPB{\raggedleft}\gnumbox[r]{\textsf{$3.51$}}}
	&\multicolumn{1}{p{\gnumericColC}|}%
	{\gnumericPB{\raggedleft}\gnumbox[r]{\textsf{$3.35$}}}
	&\multicolumn{1}{p{\gnumericColD}|}%
	{\gnumericPB{\raggedleft}\gnumbox[r]{\textsf{$5.22$}}}
	&\multicolumn{1}{p{\gnumericColE}|}%
	{\gnumericPB{\raggedleft}\gnumbox[r]{\textsf{$1.58$}}}
\\
\hhline{|----|}
	 \multicolumn{1}{|p{\gnumericColB}|}%
	{\gnumericPB{\raggedleft}\gnumbox[r]{\textsf{$3.68$}}}
	&\multicolumn{1}{p{\gnumericColC}|}%
	{\gnumericPB{\raggedleft}\gnumbox[r]{\textsf{$3.05$}}}
	&\multicolumn{1}{p{\gnumericColD}|}%
	{\gnumericPB{\raggedleft}\gnumbox[r]{\textsf{$5.39$}}}
	&\multicolumn{1}{p{\gnumericColE}|}%
	{\gnumericPB{\raggedleft}\gnumbox[r]{\textsf{$1.49$}}}
\\
\hhline{|----|}
	 \multicolumn{1}{|p{\gnumericColB}|}%
	{\gnumericPB{\raggedleft}\gnumbox[r]{\textsf{$3.86$}}}
	&\multicolumn{1}{p{\gnumericColC}|}%
	{\gnumericPB{\raggedleft}\gnumbox[r]{\textsf{$2.79$}}}
	&\multicolumn{1}{p{\gnumericColD}|}%
	{\gnumericPB{\raggedleft}\gnumbox[r]{\textsf{$5.57$}}}
	&\multicolumn{1}{p{\gnumericColE}|}%
	{\gnumericPB{\raggedleft}\gnumbox[r]{\textsf{$1.40$}}}
\\
\hhline{|----|}
	 \multicolumn{1}{|p{\gnumericColB}|}%
	{\gnumericPB{\raggedleft}\gnumbox[r]{\textsf{$4.03$}}}
	&\multicolumn{1}{p{\gnumericColC}|}%
	{\gnumericPB{\raggedleft}\gnumbox[r]{\textsf{$2.57$}}}
	&\multicolumn{1}{p{\gnumericColD}|}%
	{\gnumericPB{\raggedleft}\gnumbox[r]{\textsf{$5.74$}}}
	&\multicolumn{1}{p{\gnumericColE}|}%
	{\gnumericPB{\raggedleft}\gnumbox[r]{\textsf{$1.33$}}}
\\
\hhline{|----|}
	 \multicolumn{1}{|p{\gnumericColB}|}%
	{\gnumericPB{\raggedleft}\gnumbox[r]{\textsf{$4.20$}}}
	&\multicolumn{1}{p{\gnumericColC}|}%
	{\gnumericPB{\raggedleft}\gnumbox[r]{\textsf{$2.37$}}}
	&\multicolumn{1}{p{\gnumericColD}|}%
	{\gnumericPB{\raggedleft}\gnumbox[r]{\textsf{$5.91$}}}
	&\multicolumn{1}{p{\gnumericColE}|}%
	{\gnumericPB{\raggedleft}\gnumbox[r]{\textsf{$1.26$}}}
\\
\hhline{|----|}
	 \multicolumn{1}{|p{\gnumericColB}|}%
	{\gnumericPB{\raggedleft}\gnumbox[r]{\textsf{$4.37$}}}
	&\multicolumn{1}{p{\gnumericColC}|}%
	{\gnumericPB{\raggedleft}\gnumbox[r]{\textsf{$2.20$}}}
	&\multicolumn{1}{p{\gnumericColD}|}%
	{\gnumericPB{\raggedleft}\gnumbox[r]{\textsf{$6.08$}}}
	&\multicolumn{1}{p{\gnumericColE}|}%
	{\gnumericPB{\raggedleft}\gnumbox[r]{\textsf{$1.20$}}}
\\
\hhline{|----|}
	 \multicolumn{1}{|p{\gnumericColB}|}%
	{\gnumericPB{\raggedleft}\gnumbox[r]{\textsf{$4.54$}}}
	&\multicolumn{1}{p{\gnumericColC}|}%
	{\gnumericPB{\raggedleft}\gnumbox[r]{\textsf{$2.05$}}}
	&\multicolumn{1}{p{\gnumericColD}|}%
	{\gnumericPB{\raggedleft}\gnumbox[r]{\textsf{$6.25$}}}
	&\multicolumn{1}{p{\gnumericColE}|}%
	{\gnumericPB{\raggedleft}\gnumbox[r]{\textsf{$1.14$}}}
\\
\hhline{|-|-|-|-|}
\end{longtable}

\gnumericTableEnd
}
\caption{Normalisation for a gaussian form for $\Upsilon(1S)$ as a function of $\Lambda$ for $m_b=5.28$~GeV.}\label{tab:N_exp_Upsi_lam3}
\end{table}

\begin{table}[H]
{\tiny%%%%%%%%%%%%%%%%%%%%%%%%%%%%%%%%%%%%%%%%%%%%%%%%%%%%%%%%%%%%%%%%%%%%%%
\def\ifundefined#1{\expandafter\ifx\csname#1\endcsname\relax}

%%  Check for the \def token for inputed files. If it is not        %%
%%  defined. the file will be processed as a standalone and the     %%
%%  preamble will be used.                                          %%
\ifundefined{inputGnumericTable}

%%  End of the preamble for the standalone. The next section is for %%
%%  documents which are included into other LaTeX2e files.          %%
\else

%%  We are not a stand alone document. For a regular table. we will %%
%%  have no preamble and only define the closing to mean nothing.   %%
    \def\gnumericTableEnd{}

%%  If we want landscape mode in an embedded document. comment out  %%
%%  the line above and uncomment the two below. The table will      %%
%%  begin on a new page and run in landscape mode.                  %%
%       \def\gnumericTableEnd{\end{landscape}}
%       \begin{landscape}

%%  End of the else clause for this file being \input.              %%
\fi

%%%%%%%%%%%%%%%%%%%%%%%%%%%%%%%%%%%%%%%%%%%%%%%%%%%%%%%%%%%%%%%%%%%%%%
%%                                                                  %%
%%  The rest is the gnumeric table. except for the closing          %%
%%  statement. Changes below will alter the table's appearance.     %%
%%                                                                  %%
%%%%%%%%%%%%%%%%%%%%%%%%%%%%%%%%%%%%%%%%%%%%%%%%%%%%%%%%%%%%%%%%%%%%%%

\providecommand{\gnumericPB}[1]%
{\let\gnumericTemp=\\#1\let\\=\gnumericTemp\hspace{0pt}}

%%  The default table format retains the relative column widths of  %%
%%  gnumeric. They can easily be changed to c. r or l. In that case %%
%%  you may want to comment out the next line and uncomment the one %%
%%  thereafter                                                      %%
\providecommand\gnumbox{\makebox[0pt]}
%%\providecommand\gnumbox[1][]{\makebox}

%% to adjust positions in multirow situations                       %%
\setlength{\bigstrutjot}{\jot}
\setlength{\extrarowheight}{\doublerulesep}

%%  The \setlongtables command keeps column widths the same across %%
%%  pages. Simply comment out next line for varying column widths.  %%
\setlongtables

\setlength\gnumericTableWidth{%
	30pt+%
	30pt+%
	30pt+%
	30pt+%
0pt}
\def\gumericNumCols{4}
\setlength\gnumericTableWidthComplete{\gnumericTableWidth+\tabcolsep*\gumericNumCols*2+\arrayrulewidth*\gumericNumCols}
\ifthenelse{\lengthtest{\gnumericTableWidthComplete > \textwidth}}%
{\def\gnumericScale{\ratio{\textwidth-\tabcolsep*\gumericNumCols*2-\arrayrulewidth*\gumericNumCols}%
{\gnumericTableWidth}}}%
{\def\gnumericScale{1}}

%%%%%%%%%%%%%%%%%%%%%%%%%%%%%%%%%%%%%%%%%%%%%%%%%%%%%%%%%%%%%%%%%%%%%%
%%                                                                  %%
%% The following are the widths of the various columns. We are      %%
%% defining them here because then they are easier to change.       %%
%% Depending on the cell formats we may use them more than once.    %%
%%                                                                  %%
%%%%%%%%%%%%%%%%%%%%%%%%%%%%%%%%%%%%%%%%%%%%%%%%%%%%%%%%%%%%%%%%%%%%%%

\def\gnumericColB{30pt*\gnumericScale}
\def\gnumericColC{30pt*\gnumericScale}
\def\gnumericColD{30pt*\gnumericScale}
\def\gnumericColE{30pt*\gnumericScale}

\begin{longtable}[c]{%
	b{\gnumericColB}%
	b{\gnumericColC}%
	b{\gnumericColD}%
	b{\gnumericColE}%
	}

%%%%%%%%%%%%%%%%%%%%%%%%%%%%%%%%%%%%%%%%%%%%%%%%%%%%%%%%%%%%%%%%%%%%%%
%%  The longtable options. (Caption. headers... see Goosens. p.124) %%
%	\caption{The Table Caption.}             \\	%
% \hline	% Across the top of the table.
%%  The rest of these options are table rows which are placed on    %%
%%  the first. last or every page. Use \multicolumn if you want.    %%

%%  Header for the first page.                                      %%
%	\multicolumn{4}{c}{The First Header} \\ \hline 
%	\multicolumn{1}{c}{colTag}	%Column 1
%	&\multicolumn{1}{c}{colTag}	%Column 1
%	&\multicolumn{1}{c}{colTag}	%Column 2
%	&\multicolumn{1}{c}{colTag}	%Column 3
%	&\multicolumn{1}{c}{colTag}	\\ \hline %Last column
%	\endfirsthead

%%  The running header definition.                                  %%
%	\hline
%	\multicolumn{4}{l}{\ldots\small\slshape continued} \\ \hline
%	\multicolumn{1}{c}{colTag}	%Column 1
%	&\multicolumn{1}{c}{colTag}	%Column 1
%	&\multicolumn{1}{c}{colTag}	%Column 2
%	&\multicolumn{1}{c}{colTag}	%Column 3
%	&\multicolumn{1}{c}{colTag}	\\ \hline %Last column
%	\endhead

%%  The running footer definition.                                  %%
%	\hline
%	\multicolumn{4}{r}{\small\slshape continued\ldots} \\
%	\endfoot

%%  The ending footer definition.                                   %%
%	\multicolumn{4}{c}{That's all folks} \\ \hline 
%	\endlastfoot
%%%%%%%%%%%%%%%%%%%%%%%%%%%%%%%%%%%%%%%%%%%%%%%%%%%%%%%%%%%%%%%%%%%%%%

\hhline{|-|-|-|-|}
	 \multicolumn{1}{|p{\gnumericColB}|}%
	{\gnumericPB{\raggedleft}\gnumbox[r]{\textsf{$3.00$}}}
	&\multicolumn{1}{p{\gnumericColC}|}%
	{\gnumericPB{\raggedleft}\gnumbox[r]{\textsf{$5.37$}}}
	&\multicolumn{1}{p{\gnumericColD}|}%
	{\gnumericPB{\raggedleft}\gnumbox[r]{\textsf{$4.71$}}}
	&\multicolumn{1}{p{\gnumericColE}|}%
	{\gnumericPB{\raggedleft}\gnumbox[r]{\textsf{$2.11$}}}
\\
\hhline{|----|}
	 \multicolumn{1}{|p{\gnumericColB}|}%
	{\gnumericPB{\raggedleft}\gnumbox[r]{\textsf{$3.17$}}}
	&\multicolumn{1}{p{\gnumericColC}|}%
	{\gnumericPB{\raggedleft}\gnumbox[r]{\textsf{$4.76$}}}
	&\multicolumn{1}{p{\gnumericColD}|}%
	{\gnumericPB{\raggedleft}\gnumbox[r]{\textsf{$4.88$}}}
	&\multicolumn{1}{p{\gnumericColE}|}%
	{\gnumericPB{\raggedleft}\gnumbox[r]{\textsf{$1.97$}}}
\\
\hhline{|----|}
	 \multicolumn{1}{|p{\gnumericColB}|}%
	{\gnumericPB{\raggedleft}\gnumbox[r]{\textsf{$3.34$}}}
	&\multicolumn{1}{p{\gnumericColC}|}%
	{\gnumericPB{\raggedleft}\gnumbox[r]{\textsf{$4.26$}}}
	&\multicolumn{1}{p{\gnumericColD}|}%
	{\gnumericPB{\raggedleft}\gnumbox[r]{\textsf{$5.05$}}}
	&\multicolumn{1}{p{\gnumericColE}|}%
	{\gnumericPB{\raggedleft}\gnumbox[r]{\textsf{$1.84$}}}
\\
\hhline{|----|}
	 \multicolumn{1}{|p{\gnumericColB}|}%
	{\gnumericPB{\raggedleft}\gnumbox[r]{\textsf{$3.51$}}}
	&\multicolumn{1}{p{\gnumericColC}|}%
	{\gnumericPB{\raggedleft}\gnumbox[r]{\textsf{$3.83$}}}
	&\multicolumn{1}{p{\gnumericColD}|}%
	{\gnumericPB{\raggedleft}\gnumbox[r]{\textsf{$5.22$}}}
	&\multicolumn{1}{p{\gnumericColE}|}%
	{\gnumericPB{\raggedleft}\gnumbox[r]{\textsf{$1.73$}}}
\\
\hhline{|----|}
	 \multicolumn{1}{|p{\gnumericColB}|}%
	{\gnumericPB{\raggedleft}\gnumbox[r]{\textsf{$3.68$}}}
	&\multicolumn{1}{p{\gnumericColC}|}%
	{\gnumericPB{\raggedleft}\gnumbox[r]{\textsf{$3.47$}}}
	&\multicolumn{1}{p{\gnumericColD}|}%
	{\gnumericPB{\raggedleft}\gnumbox[r]{\textsf{$5.39$}}}
	&\multicolumn{1}{p{\gnumericColE}|}%
	{\gnumericPB{\raggedleft}\gnumbox[r]{\textsf{$1.62$}}}
\\
\hhline{|----|}
	 \multicolumn{1}{|p{\gnumericColB}|}%
	{\gnumericPB{\raggedleft}\gnumbox[r]{\textsf{$3.86$}}}
	&\multicolumn{1}{p{\gnumericColC}|}%
	{\gnumericPB{\raggedleft}\gnumbox[r]{\textsf{$3.16$}}}
	&\multicolumn{1}{p{\gnumericColD}|}%
	{\gnumericPB{\raggedleft}\gnumbox[r]{\textsf{$5.57$}}}
	&\multicolumn{1}{p{\gnumericColE}|}%
	{\gnumericPB{\raggedleft}\gnumbox[r]{\textsf{$1.53$}}}
\\
\hhline{|----|}
	 \multicolumn{1}{|p{\gnumericColB}|}%
	{\gnumericPB{\raggedleft}\gnumbox[r]{\textsf{$4.03$}}}
	&\multicolumn{1}{p{\gnumericColC}|}%
	{\gnumericPB{\raggedleft}\gnumbox[r]{\textsf{$2.89$}}}
	&\multicolumn{1}{p{\gnumericColD}|}%
	{\gnumericPB{\raggedleft}\gnumbox[r]{\textsf{$5.74$}}}
	&\multicolumn{1}{p{\gnumericColE}|}%
	{\gnumericPB{\raggedleft}\gnumbox[r]{\textsf{$1.44$}}}
\\
\hhline{|----|}
	 \multicolumn{1}{|p{\gnumericColB}|}%
	{\gnumericPB{\raggedleft}\gnumbox[r]{\textsf{$4.20$}}}
	&\multicolumn{1}{p{\gnumericColC}|}%
	{\gnumericPB{\raggedleft}\gnumbox[r]{\textsf{$2.66$}}}
	&\multicolumn{1}{p{\gnumericColD}|}%
	{\gnumericPB{\raggedleft}\gnumbox[r]{\textsf{$5.91$}}}
	&\multicolumn{1}{p{\gnumericColE}|}%
	{\gnumericPB{\raggedleft}\gnumbox[r]{\textsf{$1.37$}}}
\\
\hhline{|----|}
	 \multicolumn{1}{|p{\gnumericColB}|}%
	{\gnumericPB{\raggedleft}\gnumbox[r]{\textsf{$4.37$}}}
	&\multicolumn{1}{p{\gnumericColC}|}%
	{\gnumericPB{\raggedleft}\gnumbox[r]{\textsf{$2.46$}}}
	&\multicolumn{1}{p{\gnumericColD}|}%
	{\gnumericPB{\raggedleft}\gnumbox[r]{\textsf{$6.08$}}}
	&\multicolumn{1}{p{\gnumericColE}|}%
	{\gnumericPB{\raggedleft}\gnumbox[r]{\textsf{$1.29$}}}
\\
\hhline{|----|}
	 \multicolumn{1}{|p{\gnumericColB}|}%
	{\gnumericPB{\raggedleft}\gnumbox[r]{\textsf{$4.54$}}}
	&\multicolumn{1}{p{\gnumericColC}|}%
	{\gnumericPB{\raggedleft}\gnumbox[r]{\textsf{$2.27$}}}
	&\multicolumn{1}{p{\gnumericColD}|}%
	{\gnumericPB{\raggedleft}\gnumbox[r]{\textsf{$6.25$}}}
	&\multicolumn{1}{p{\gnumericColE}|}%
	{\gnumericPB{\raggedleft}\gnumbox[r]{\textsf{$1.23$}}}
\\
\hhline{|-|-|-|-|}
\end{longtable}

\gnumericTableEnd
}
\caption{Normalisation for a gaussian form for $\Upsilon(1S)$ as a function of $\Lambda$ for $m_b=5.50$~GeV.}\label{tab:N_exp_Upsi_lam4}
\end{table}

\begin{table}[H]
{\tiny%%%%%%%%%%%%%%%%%%%%%%%%%%%%%%%%%%%%%%%%%%%%%%%%%%%%%%%%%%%%%%%%%%%%%%
\def\ifundefined#1{\expandafter\ifx\csname#1\endcsname\relax}

%%  Check for the \def token for inputed files. If it is not        %%
%%  defined. the file will be processed as a standalone and the     %%
%%  preamble will be used.                                          %%
\ifundefined{inputGnumericTable}

%%  End of the preamble for the standalone. The next section is for %%
%%  documents which are included into other LaTeX2e files.          %%
\else

%%  We are not a stand alone document. For a regular table. we will %%
%%  have no preamble and only define the closing to mean nothing.   %%
    \def\gnumericTableEnd{}

%%  If we want landscape mode in an embedded document. comment out  %%
%%  the line above and uncomment the two below. The table will      %%
%%  begin on a new page and run in landscape mode.                  %%
%       \def\gnumericTableEnd{\end{landscape}}
%       \begin{landscape}

%%  End of the else clause for this file being \input.              %%
\fi

%%%%%%%%%%%%%%%%%%%%%%%%%%%%%%%%%%%%%%%%%%%%%%%%%%%%%%%%%%%%%%%%%%%%%%
%%                                                                  %%
%%  The rest is the gnumeric table. except for the closing          %%
%%  statement. Changes below will alter the table's appearance.     %%
%%                                                                  %%
%%%%%%%%%%%%%%%%%%%%%%%%%%%%%%%%%%%%%%%%%%%%%%%%%%%%%%%%%%%%%%%%%%%%%%

\providecommand{\gnumericPB}[1]%
{\let\gnumericTemp=\\#1\let\\=\gnumericTemp\hspace{0pt}}

%%  The default table format retains the relative column widths of  %%
%%  gnumeric. They can easily be changed to c. r or l. In that case %%
%%  you may want to comment out the next line and uncomment the one %%
%%  thereafter                                                      %%
\providecommand\gnumbox{\makebox[0pt]}
%%\providecommand\gnumbox[1][]{\makebox}

%% to adjust positions in multirow situations                       %%
\setlength{\bigstrutjot}{\jot}
\setlength{\extrarowheight}{\doublerulesep}

%%  The \setlongtables command keeps column widths the same across %%
%%  pages. Simply comment out next line for varying column widths.  %%
\setlongtables

\setlength\gnumericTableWidth{%
	30pt+%
	30pt+%
	30pt+%
	30pt+%
0pt}
\def\gumericNumCols{4}
\setlength\gnumericTableWidthComplete{\gnumericTableWidth+\tabcolsep*\gumericNumCols*2+\arrayrulewidth*\gumericNumCols}
\ifthenelse{\lengthtest{\gnumericTableWidthComplete > \textwidth}}%
{\def\gnumericScale{\ratio{\textwidth-\tabcolsep*\gumericNumCols*2-\arrayrulewidth*\gumericNumCols}%
{\gnumericTableWidth}}}%
{\def\gnumericScale{1}}

%%%%%%%%%%%%%%%%%%%%%%%%%%%%%%%%%%%%%%%%%%%%%%%%%%%%%%%%%%%%%%%%%%%%%%
%%                                                                  %%
%% The following are the widths of the various columns. We are      %%
%% defining them here because then they are easier to change.       %%
%% Depending on the cell formats we may use them more than once.    %%
%%                                                                  %%
%%%%%%%%%%%%%%%%%%%%%%%%%%%%%%%%%%%%%%%%%%%%%%%%%%%%%%%%%%%%%%%%%%%%%%

\def\gnumericColB{30pt*\gnumericScale}
\def\gnumericColC{30pt*\gnumericScale}
\def\gnumericColD{30pt*\gnumericScale}
\def\gnumericColE{30pt*\gnumericScale}

\begin{longtable}[c]{%
	b{\gnumericColB}%
	b{\gnumericColC}%
	b{\gnumericColD}%
	b{\gnumericColE}%
	}

%%%%%%%%%%%%%%%%%%%%%%%%%%%%%%%%%%%%%%%%%%%%%%%%%%%%%%%%%%%%%%%%%%%%%%
%%  The longtable options. (Caption. headers... see Goosens. p.124) %%
%	\caption{The Table Caption.}             \\	%
% \hline	% Across the top of the table.
%%  The rest of these options are table rows which are placed on    %%
%%  the first. last or every page. Use \multicolumn if you want.    %%

%%  Header for the first page.                                      %%
%	\multicolumn{4}{c}{The First Header} \\ \hline 
%	\multicolumn{1}{c}{colTag}	%Column 1
%	&\multicolumn{1}{c}{colTag}	%Column 1
%	&\multicolumn{1}{c}{colTag}	%Column 2
%	&\multicolumn{1}{c}{colTag}	%Column 3
%	&\multicolumn{1}{c}{colTag}	\\ \hline %Last column
%	\endfirsthead

%%  The running header definition.                                  %%
%	\hline
%	\multicolumn{4}{l}{\ldots\small\slshape continued} \\ \hline
%	\multicolumn{1}{c}{colTag}	%Column 1
%	&\multicolumn{1}{c}{colTag}	%Column 1
%	&\multicolumn{1}{c}{colTag}	%Column 2
%	&\multicolumn{1}{c}{colTag}	%Column 3
%	&\multicolumn{1}{c}{colTag}	\\ \hline %Last column
%	\endhead

%%  The running footer definition.                                  %%
%	\hline
%	\multicolumn{4}{r}{\small\slshape continued\ldots} \\
%	\endfoot

%%  The ending footer definition.                                   %%
%	\multicolumn{4}{c}{That's all folks} \\ \hline 
%	\endlastfoot
%%%%%%%%%%%%%%%%%%%%%%%%%%%%%%%%%%%%%%%%%%%%%%%%%%%%%%%%%%%%%%%%%%%%%%

\hhline{|-|-|-|-|}
	 \multicolumn{1}{|p{\gnumericColB}|}%
	{\gnumericPB{\raggedleft}\gnumbox[r]{\textsf{$m_b$}}}
	&\multicolumn{1}{p{\gnumericColC}|}%
	{\gnumericPB{\raggedleft}\gnumbox[r]{\textsf{$N$}}}
	&\multicolumn{1}{p{\gnumericColD}|}%
	{\gnumericPB{\raggedleft}\gnumbox[r]{\textsf{$m_b$}}}
	&\multicolumn{1}{p{\gnumericColE}|}%
	{\gnumericPB{\raggedleft}\gnumbox[r]{\textsf{$N$}}}
\\
\hhline{|----|}
	 \multicolumn{1}{|p{\gnumericColB}|}%
	{\gnumericPB{\raggedleft}\gnumbox[r]{\textsf{$4.80$}}}
	&\multicolumn{1}{p{\gnumericColC}|}%
	{\gnumericPB{\raggedleft}\gnumbox[r]{\textsf{$2.53$}}}
	&\multicolumn{1}{p{\gnumericColD}|}%
	{\gnumericPB{\raggedleft}\gnumbox[r]{\textsf{$5.17$}}}
	&\multicolumn{1}{p{\gnumericColE}|}%
	{\gnumericPB{\raggedleft}\gnumbox[r]{\textsf{$4.25$}}}
\\
\hhline{|----|}
	 \multicolumn{1}{|p{\gnumericColB}|}%
	{\gnumericPB{\raggedleft}\gnumbox[r]{\textsf{$4.84$}}}
	&\multicolumn{1}{p{\gnumericColC}|}%
	{\gnumericPB{\raggedleft}\gnumbox[r]{\textsf{$2.77$}}}
	&\multicolumn{1}{p{\gnumericColD}|}%
	{\gnumericPB{\raggedleft}\gnumbox[r]{\textsf{$5.21$}}}
	&\multicolumn{1}{p{\gnumericColE}|}%
	{\gnumericPB{\raggedleft}\gnumbox[r]{\textsf{$4.38$}}}
\\
\hhline{|----|}
	 \multicolumn{1}{|p{\gnumericColB}|}%
	{\gnumericPB{\raggedleft}\gnumbox[r]{\textsf{$4.87$}}}
	&\multicolumn{1}{p{\gnumericColC}|}%
	{\gnumericPB{\raggedleft}\gnumbox[r]{\textsf{$2.98$}}}
	&\multicolumn{1}{p{\gnumericColD}|}%
	{\gnumericPB{\raggedleft}\gnumbox[r]{\textsf{$5.24$}}}
	&\multicolumn{1}{p{\gnumericColE}|}%
	{\gnumericPB{\raggedleft}\gnumbox[r]{\textsf{$4.51$}}}
\\
\hhline{|----|}
	 \multicolumn{1}{|p{\gnumericColB}|}%
	{\gnumericPB{\raggedleft}\gnumbox[r]{\textsf{$4.91$}}}
	&\multicolumn{1}{p{\gnumericColC}|}%
	{\gnumericPB{\raggedleft}\gnumbox[r]{\textsf{$3.17$}}}
	&\multicolumn{1}{p{\gnumericColD}|}%
	{\gnumericPB{\raggedleft}\gnumbox[r]{\textsf{$5.28$}}}
	&\multicolumn{1}{p{\gnumericColE}|}%
	{\gnumericPB{\raggedleft}\gnumbox[r]{\textsf{$4.64$}}}
\\
\hhline{|----|}
	 \multicolumn{1}{|p{\gnumericColB}|}%
	{\gnumericPB{\raggedleft}\gnumbox[r]{\textsf{$4.95$}}}
	&\multicolumn{1}{p{\gnumericColC}|}%
	{\gnumericPB{\raggedleft}\gnumbox[r]{\textsf{$3.35$}}}
	&\multicolumn{1}{p{\gnumericColD}|}%
	{\gnumericPB{\raggedleft}\gnumbox[r]{\textsf{$5.32$}}}
	&\multicolumn{1}{p{\gnumericColE}|}%
	{\gnumericPB{\raggedleft}\gnumbox[r]{\textsf{$4.77$}}}
\\
\hhline{|----|}
	 \multicolumn{1}{|p{\gnumericColB}|}%
	{\gnumericPB{\raggedleft}\gnumbox[r]{\textsf{$4.98$}}}
	&\multicolumn{1}{p{\gnumericColC}|}%
	{\gnumericPB{\raggedleft}\gnumbox[r]{\textsf{$3.51$}}}
	&\multicolumn{1}{p{\gnumericColD}|}%
	{\gnumericPB{\raggedleft}\gnumbox[r]{\textsf{$5.35$}}}
	&\multicolumn{1}{p{\gnumericColE}|}%
	{\gnumericPB{\raggedleft}\gnumbox[r]{\textsf{$4.89$}}}
\\
\hhline{|----|}
	 \multicolumn{1}{|p{\gnumericColB}|}%
	{\gnumericPB{\raggedleft}\gnumbox[r]{\textsf{$5.02$}}}
	&\multicolumn{1}{p{\gnumericColC}|}%
	{\gnumericPB{\raggedleft}\gnumbox[r]{\textsf{$3.67$}}}
	&\multicolumn{1}{p{\gnumericColD}|}%
	{\gnumericPB{\raggedleft}\gnumbox[r]{\textsf{$5.39$}}}
	&\multicolumn{1}{p{\gnumericColE}|}%
	{\gnumericPB{\raggedleft}\gnumbox[r]{\textsf{$5.01$}}}
\\
\hhline{|----|}
	 \multicolumn{1}{|p{\gnumericColB}|}%
	{\gnumericPB{\raggedleft}\gnumbox[r]{\textsf{$5.06$}}}
	&\multicolumn{1}{p{\gnumericColC}|}%
	{\gnumericPB{\raggedleft}\gnumbox[r]{\textsf{$3.82$}}}
	&\multicolumn{1}{p{\gnumericColD}|}%
	{\gnumericPB{\raggedleft}\gnumbox[r]{\textsf{$5.43$}}}
	&\multicolumn{1}{p{\gnumericColE}|}%
	{\gnumericPB{\raggedleft}\gnumbox[r]{\textsf{$5.13$}}}
\\
\hhline{|----|}
	 \multicolumn{1}{|p{\gnumericColB}|}%
	{\gnumericPB{\raggedleft}\gnumbox[r]{\textsf{$5.09$}}}
	&\multicolumn{1}{p{\gnumericColC}|}%
	{\gnumericPB{\raggedleft}\gnumbox[r]{\textsf{$3.97$}}}
	&\multicolumn{1}{p{\gnumericColD}|}%
	{\gnumericPB{\raggedleft}\gnumbox[r]{\textsf{$5.46$}}}
	&\multicolumn{1}{p{\gnumericColE}|}%
	{\gnumericPB{\raggedleft}\gnumbox[r]{\textsf{$5.25$}}}
\\
\hhline{|----|}
	 \multicolumn{1}{|p{\gnumericColB}|}%
	{\gnumericPB{\raggedleft}\gnumbox[r]{\textsf{$5.13$}}}
	&\multicolumn{1}{p{\gnumericColC}|}%
	{\gnumericPB{\raggedleft}\gnumbox[r]{\textsf{$4.11$}}}
	&\multicolumn{1}{p{\gnumericColD}|}%
	{\gnumericPB{\raggedleft}\gnumbox[r]{\textsf{$5.50$}}}
	&\multicolumn{1}{p{\gnumericColE}|}%
	{\gnumericPB{\raggedleft}\gnumbox[r]{\textsf{$5.37$}}}
\\
\hhline{|-|-|-|-|}
\end{longtable}

\gnumericTableEnd
}
\caption{Normalisation for a gaussian form for $\Upsilon(1S)$ as a function of  the $b$ quark mass
for $\Lambda=3.00$~GeV.}\label{tab:N_exp_Upsi_mq1}
\end{table}

\begin{table}[H]
{\tiny%%%%%%%%%%%%%%%%%%%%%%%%%%%%%%%%%%%%%%%%%%%%%%%%%%%%%%%%%%%%%%%%%%%%%%
\def\ifundefined#1{\expandafter\ifx\csname#1\endcsname\relax}

%%  Check for the \def token for inputed files. If it is not        %%
%%  defined. the file will be processed as a standalone and the     %%
%%  preamble will be used.                                          %%
\ifundefined{inputGnumericTable}

%%  End of the preamble for the standalone. The next section is for %%
%%  documents which are included into other LaTeX2e files.          %%
\else

%%  We are not a stand alone document. For a regular table. we will %%
%%  have no preamble and only define the closing to mean nothing.   %%
    \def\gnumericTableEnd{}

%%  If we want landscape mode in an embedded document. comment out  %%
%%  the line above and uncomment the two below. The table will      %%
%%  begin on a new page and run in landscape mode.                  %%
%       \def\gnumericTableEnd{\end{landscape}}
%       \begin{landscape}

%%  End of the else clause for this file being \input.              %%
\fi

%%%%%%%%%%%%%%%%%%%%%%%%%%%%%%%%%%%%%%%%%%%%%%%%%%%%%%%%%%%%%%%%%%%%%%
%%                                                                  %%
%%  The rest is the gnumeric table. except for the closing          %%
%%  statement. Changes below will alter the table's appearance.     %%
%%                                                                  %%
%%%%%%%%%%%%%%%%%%%%%%%%%%%%%%%%%%%%%%%%%%%%%%%%%%%%%%%%%%%%%%%%%%%%%%

\providecommand{\gnumericPB}[1]%
{\let\gnumericTemp=\\#1\let\\=\gnumericTemp\hspace{0pt}}

%%  The default table format retains the relative column widths of  %%
%%  gnumeric. They can easily be changed to c. r or l. In that case %%
%%  you may want to comment out the next line and uncomment the one %%
%%  thereafter                                                      %%
\providecommand\gnumbox{\makebox[0pt]}
%%\providecommand\gnumbox[1][]{\makebox}

%% to adjust positions in multirow situations                       %%
\setlength{\bigstrutjot}{\jot}
\setlength{\extrarowheight}{\doublerulesep}

%%  The \setlongtables command keeps column widths the same across %%
%%  pages. Simply comment out next line for varying column widths.  %%
\setlongtables

\setlength\gnumericTableWidth{%
	30pt+%
	30pt+%
	30pt+%
	30pt+%
0pt}
\def\gumericNumCols{4}
\setlength\gnumericTableWidthComplete{\gnumericTableWidth+\tabcolsep*\gumericNumCols*2+\arrayrulewidth*\gumericNumCols}
\ifthenelse{\lengthtest{\gnumericTableWidthComplete > \textwidth}}%
{\def\gnumericScale{\ratio{\textwidth-\tabcolsep*\gumericNumCols*2-\arrayrulewidth*\gumericNumCols}%
{\gnumericTableWidth}}}%
{\def\gnumericScale{1}}

%%%%%%%%%%%%%%%%%%%%%%%%%%%%%%%%%%%%%%%%%%%%%%%%%%%%%%%%%%%%%%%%%%%%%%
%%                                                                  %%
%% The following are the widths of the various columns. We are      %%
%% defining them here because then they are easier to change.       %%
%% Depending on the cell formats we may use them more than once.    %%
%%                                                                  %%
%%%%%%%%%%%%%%%%%%%%%%%%%%%%%%%%%%%%%%%%%%%%%%%%%%%%%%%%%%%%%%%%%%%%%%

\def\gnumericColB{30pt*\gnumericScale}
\def\gnumericColC{30pt*\gnumericScale}
\def\gnumericColD{30pt*\gnumericScale}
\def\gnumericColE{30pt*\gnumericScale}

\begin{longtable}[c]{%
	b{\gnumericColB}%
	b{\gnumericColC}%
	b{\gnumericColD}%
	b{\gnumericColE}%
	}

%%%%%%%%%%%%%%%%%%%%%%%%%%%%%%%%%%%%%%%%%%%%%%%%%%%%%%%%%%%%%%%%%%%%%%
%%  The longtable options. (Caption. headers... see Goosens. p.124) %%
%	\caption{The Table Caption.}             \\	%
% \hline	% Across the top of the table.
%%  The rest of these options are table rows which are placed on    %%
%%  the first. last or every page. Use \multicolumn if you want.    %%

%%  Header for the first page.                                      %%
%	\multicolumn{4}{c}{The First Header} \\ \hline 
%	\multicolumn{1}{c}{colTag}	%Column 1
%	&\multicolumn{1}{c}{colTag}	%Column 1
%	&\multicolumn{1}{c}{colTag}	%Column 2
%	&\multicolumn{1}{c}{colTag}	%Column 3
%	&\multicolumn{1}{c}{colTag}	\\ \hline %Last column
%	\endfirsthead

%%  The running header definition.                                  %%
%	\hline
%	\multicolumn{4}{l}{\ldots\small\slshape continued} \\ \hline
%	\multicolumn{1}{c}{colTag}	%Column 1
%	&\multicolumn{1}{c}{colTag}	%Column 1
%	&\multicolumn{1}{c}{colTag}	%Column 2
%	&\multicolumn{1}{c}{colTag}	%Column 3
%	&\multicolumn{1}{c}{colTag}	\\ \hline %Last column
%	\endhead

%%  The running footer definition.                                  %%
%	\hline
%	\multicolumn{4}{r}{\small\slshape continued\ldots} \\
%	\endfoot

%%  The ending footer definition.                                   %%
%	\multicolumn{4}{c}{That's all folks} \\ \hline 
%	\endlastfoot
%%%%%%%%%%%%%%%%%%%%%%%%%%%%%%%%%%%%%%%%%%%%%%%%%%%%%%%%%%%%%%%%%%%%%%

\hhline{|-|-|-|-|}
	 \multicolumn{1}{|p{\gnumericColB}|}%
	{\gnumericPB{\raggedleft}\gnumbox[r]{\textsf{$4.80$}}}
	&\multicolumn{1}{p{\gnumericColC}|}%
	{\gnumericPB{\raggedleft}\gnumbox[r]{\textsf{$1.61$}}}
	&\multicolumn{1}{p{\gnumericColD}|}%
	{\gnumericPB{\raggedleft}\gnumbox[r]{\textsf{$5.17$}}}
	&\multicolumn{1}{p{\gnumericColE}|}%
	{\gnumericPB{\raggedleft}\gnumbox[r]{\textsf{$2.35$}}}
\\
\hhline{|----|}
	 \multicolumn{1}{|p{\gnumericColB}|}%
	{\gnumericPB{\raggedleft}\gnumbox[r]{\textsf{$4.84$}}}
	&\multicolumn{1}{p{\gnumericColC}|}%
	{\gnumericPB{\raggedleft}\gnumbox[r]{\textsf{$1.72$}}}
	&\multicolumn{1}{p{\gnumericColD}|}%
	{\gnumericPB{\raggedleft}\gnumbox[r]{\textsf{$5.21$}}}
	&\multicolumn{1}{p{\gnumericColE}|}%
	{\gnumericPB{\raggedleft}\gnumbox[r]{\textsf{$2.41$}}}
\\
\hhline{|----|}
	 \multicolumn{1}{|p{\gnumericColB}|}%
	{\gnumericPB{\raggedleft}\gnumbox[r]{\textsf{$4.87$}}}
	&\multicolumn{1}{p{\gnumericColC}|}%
	{\gnumericPB{\raggedleft}\gnumbox[r]{\textsf{$1.81$}}}
	&\multicolumn{1}{p{\gnumericColD}|}%
	{\gnumericPB{\raggedleft}\gnumbox[r]{\textsf{$5.24$}}}
	&\multicolumn{1}{p{\gnumericColE}|}%
	{\gnumericPB{\raggedleft}\gnumbox[r]{\textsf{$2.46$}}}
\\
\hhline{|----|}
	 \multicolumn{1}{|p{\gnumericColB}|}%
	{\gnumericPB{\raggedleft}\gnumbox[r]{\textsf{$4.91$}}}
	&\multicolumn{1}{p{\gnumericColC}|}%
	{\gnumericPB{\raggedleft}\gnumbox[r]{\textsf{$1.89$}}}
	&\multicolumn{1}{p{\gnumericColD}|}%
	{\gnumericPB{\raggedleft}\gnumbox[r]{\textsf{$5.28$}}}
	&\multicolumn{1}{p{\gnumericColE}|}%
	{\gnumericPB{\raggedleft}\gnumbox[r]{\textsf{$2.52$}}}
\\
\hhline{|----|}
	 \multicolumn{1}{|p{\gnumericColB}|}%
	{\gnumericPB{\raggedleft}\gnumbox[r]{\textsf{$4.95$}}}
	&\multicolumn{1}{p{\gnumericColC}|}%
	{\gnumericPB{\raggedleft}\gnumbox[r]{\textsf{$1.97$}}}
	&\multicolumn{1}{p{\gnumericColD}|}%
	{\gnumericPB{\raggedleft}\gnumbox[r]{\textsf{$5.32$}}}
	&\multicolumn{1}{p{\gnumericColE}|}%
	{\gnumericPB{\raggedleft}\gnumbox[r]{\textsf{$2.57$}}}
\\
\hhline{|----|}
	 \multicolumn{1}{|p{\gnumericColB}|}%
	{\gnumericPB{\raggedleft}\gnumbox[r]{\textsf{$4.98$}}}
	&\multicolumn{1}{p{\gnumericColC}|}%
	{\gnumericPB{\raggedleft}\gnumbox[r]{\textsf{$2.04$}}}
	&\multicolumn{1}{p{\gnumericColD}|}%
	{\gnumericPB{\raggedleft}\gnumbox[r]{\textsf{$5.35$}}}
	&\multicolumn{1}{p{\gnumericColE}|}%
	{\gnumericPB{\raggedleft}\gnumbox[r]{\textsf{$2.62$}}}
\\
\hhline{|----|}
	 \multicolumn{1}{|p{\gnumericColB}|}%
	{\gnumericPB{\raggedleft}\gnumbox[r]{\textsf{$5.02$}}}
	&\multicolumn{1}{p{\gnumericColC}|}%
	{\gnumericPB{\raggedleft}\gnumbox[r]{\textsf{$2.11$}}}
	&\multicolumn{1}{p{\gnumericColD}|}%
	{\gnumericPB{\raggedleft}\gnumbox[r]{\textsf{$5.39$}}}
	&\multicolumn{1}{p{\gnumericColE}|}%
	{\gnumericPB{\raggedleft}\gnumbox[r]{\textsf{$2.67$}}}
\\
\hhline{|----|}
	 \multicolumn{1}{|p{\gnumericColB}|}%
	{\gnumericPB{\raggedleft}\gnumbox[r]{\textsf{$5.06$}}}
	&\multicolumn{1}{p{\gnumericColC}|}%
	{\gnumericPB{\raggedleft}\gnumbox[r]{\textsf{$2.17$}}}
	&\multicolumn{1}{p{\gnumericColD}|}%
	{\gnumericPB{\raggedleft}\gnumbox[r]{\textsf{$5.43$}}}
	&\multicolumn{1}{p{\gnumericColE}|}%
	{\gnumericPB{\raggedleft}\gnumbox[r]{\textsf{$2.72$}}}
\\
\hhline{|----|}
	 \multicolumn{1}{|p{\gnumericColB}|}%
	{\gnumericPB{\raggedleft}\gnumbox[r]{\textsf{$5.09$}}}
	&\multicolumn{1}{p{\gnumericColC}|}%
	{\gnumericPB{\raggedleft}\gnumbox[r]{\textsf{$2.24$}}}
	&\multicolumn{1}{p{\gnumericColD}|}%
	{\gnumericPB{\raggedleft}\gnumbox[r]{\textsf{$5.46$}}}
	&\multicolumn{1}{p{\gnumericColE}|}%
	{\gnumericPB{\raggedleft}\gnumbox[r]{\textsf{$2.77$}}}
\\
\hhline{|----|}
	 \multicolumn{1}{|p{\gnumericColB}|}%
	{\gnumericPB{\raggedleft}\gnumbox[r]{\textsf{$5.13$}}}
	&\multicolumn{1}{p{\gnumericColC}|}%
	{\gnumericPB{\raggedleft}\gnumbox[r]{\textsf{$2.30$}}}
	&\multicolumn{1}{p{\gnumericColD}|}%
	{\gnumericPB{\raggedleft}\gnumbox[r]{\textsf{$5.50$}}}
	&\multicolumn{1}{p{\gnumericColE}|}%
	{\gnumericPB{\raggedleft}\gnumbox[r]{\textsf{$2.81$}}}
\\
\hhline{|-|-|-|-|}
\end{longtable}

\gnumericTableEnd
}
\caption{Normalisation for a gaussian form for $\Upsilon(1S)$ as a function of  the $b$ quark mass
for $\Lambda=4.10$~GeV.}\label{tab:N_exp_Upsi_mq2}
\end{table}

\begin{table}[H]
{\tiny%%%%%%%%%%%%%%%%%%%%%%%%%%%%%%%%%%%%%%%%%%%%%%%%%%%%%%%%%%%%%%%%%%%%%%
\def\ifundefined#1{\expandafter\ifx\csname#1\endcsname\relax}

%%  Check for the \def token for inputed files. If it is not        %%
%%  defined. the file will be processed as a standalone and the     %%
%%  preamble will be used.                                          %%
\ifundefined{inputGnumericTable}

%%  End of the preamble for the standalone. The next section is for %%
%%  documents which are included into other LaTeX2e files.          %%
\else

%%  We are not a stand alone document. For a regular table. we will %%
%%  have no preamble and only define the closing to mean nothing.   %%
    \def\gnumericTableEnd{}

%%  If we want landscape mode in an embedded document. comment out  %%
%%  the line above and uncomment the two below. The table will      %%
%%  begin on a new page and run in landscape mode.                  %%
%       \def\gnumericTableEnd{\end{landscape}}
%       \begin{landscape}

%%  End of the else clause for this file being \input.              %%
\fi

%%%%%%%%%%%%%%%%%%%%%%%%%%%%%%%%%%%%%%%%%%%%%%%%%%%%%%%%%%%%%%%%%%%%%%
%%                                                                  %%
%%  The rest is the gnumeric table. except for the closing          %%
%%  statement. Changes below will alter the table's appearance.     %%
%%                                                                  %%
%%%%%%%%%%%%%%%%%%%%%%%%%%%%%%%%%%%%%%%%%%%%%%%%%%%%%%%%%%%%%%%%%%%%%%

\providecommand{\gnumericPB}[1]%
{\let\gnumericTemp=\\#1\let\\=\gnumericTemp\hspace{0pt}}

%%  The default table format retains the relative column widths of  %%
%%  gnumeric. They can easily be changed to c. r or l. In that case %%
%%  you may want to comment out the next line and uncomment the one %%
%%  thereafter                                                      %%
\providecommand\gnumbox{\makebox[0pt]}
%%\providecommand\gnumbox[1][]{\makebox}

%% to adjust positions in multirow situations                       %%
\setlength{\bigstrutjot}{\jot}
\setlength{\extrarowheight}{\doublerulesep}

%%  The \setlongtables command keeps column widths the same across %%
%%  pages. Simply comment out next line for varying column widths.  %%
\setlongtables

\setlength\gnumericTableWidth{%
	30pt+%
	30pt+%
	30pt+%
	30pt+%
0pt}
\def\gumericNumCols{4}
\setlength\gnumericTableWidthComplete{\gnumericTableWidth+\tabcolsep*\gumericNumCols*2+\arrayrulewidth*\gumericNumCols}
\ifthenelse{\lengthtest{\gnumericTableWidthComplete > \textwidth}}%
{\def\gnumericScale{\ratio{\textwidth-\tabcolsep*\gumericNumCols*2-\arrayrulewidth*\gumericNumCols}%
{\gnumericTableWidth}}}%
{\def\gnumericScale{1}}

%%%%%%%%%%%%%%%%%%%%%%%%%%%%%%%%%%%%%%%%%%%%%%%%%%%%%%%%%%%%%%%%%%%%%%
%%                                                                  %%
%% The following are the widths of the various columns. We are      %%
%% defining them here because then they are easier to change.       %%
%% Depending on the cell formats we may use them more than once.    %%
%%                                                                  %%
%%%%%%%%%%%%%%%%%%%%%%%%%%%%%%%%%%%%%%%%%%%%%%%%%%%%%%%%%%%%%%%%%%%%%%

\def\gnumericColB{30pt*\gnumericScale}
\def\gnumericColC{30pt*\gnumericScale}
\def\gnumericColD{30pt*\gnumericScale}
\def\gnumericColE{30pt*\gnumericScale}

\begin{longtable}[c]{%
	b{\gnumericColB}%
	b{\gnumericColC}%
	b{\gnumericColD}%
	b{\gnumericColE}%
	}

%%%%%%%%%%%%%%%%%%%%%%%%%%%%%%%%%%%%%%%%%%%%%%%%%%%%%%%%%%%%%%%%%%%%%%
%%  The longtable options. (Caption. headers... see Goosens. p.124) %%
%	\caption{The Table Caption.}             \\	%
% \hline	% Across the top of the table.
%%  The rest of these options are table rows which are placed on    %%
%%  the first. last or every page. Use \multicolumn if you want.    %%

%%  Header for the first page.                                      %%
%	\multicolumn{4}{c}{The First Header} \\ \hline 
%	\multicolumn{1}{c}{colTag}	%Column 1
%	&\multicolumn{1}{c}{colTag}	%Column 1
%	&\multicolumn{1}{c}{colTag}	%Column 2
%	&\multicolumn{1}{c}{colTag}	%Column 3
%	&\multicolumn{1}{c}{colTag}	\\ \hline %Last column
%	\endfirsthead

%%  The running header definition.                                  %%
%	\hline
%	\multicolumn{4}{l}{\ldots\small\slshape continued} \\ \hline
%	\multicolumn{1}{c}{colTag}	%Column 1
%	&\multicolumn{1}{c}{colTag}	%Column 1
%	&\multicolumn{1}{c}{colTag}	%Column 2
%	&\multicolumn{1}{c}{colTag}	%Column 3
%	&\multicolumn{1}{c}{colTag}	\\ \hline %Last column
%	\endhead

%%  The running footer definition.                                  %%
%	\hline
%	\multicolumn{4}{r}{\small\slshape continued\ldots} \\
%	\endfoot

%%  The ending footer definition.                                   %%
%	\multicolumn{4}{c}{That's all folks} \\ \hline 
%	\endlastfoot
%%%%%%%%%%%%%%%%%%%%%%%%%%%%%%%%%%%%%%%%%%%%%%%%%%%%%%%%%%%%%%%%%%%%%%

\hhline{|-|-|-|-|}
	 \multicolumn{1}{|p{\gnumericColB}|}%
	{\gnumericPB{\raggedleft}\gnumbox[r]{\textsf{$4.80$}}}
	&\multicolumn{1}{p{\gnumericColC}|}%
	{\gnumericPB{\raggedleft}\gnumbox[r]{\textsf{$1.14$}}}
	&\multicolumn{1}{p{\gnumericColD}|}%
	{\gnumericPB{\raggedleft}\gnumbox[r]{\textsf{$5.17$}}}
	&\multicolumn{1}{p{\gnumericColE}|}%
	{\gnumericPB{\raggedleft}\gnumbox[r]{\textsf{$1.54$}}}
\\
\hhline{|----|}
	 \multicolumn{1}{|p{\gnumericColB}|}%
	{\gnumericPB{\raggedleft}\gnumbox[r]{\textsf{$4.84$}}}
	&\multicolumn{1}{p{\gnumericColC}|}%
	{\gnumericPB{\raggedleft}\gnumbox[r]{\textsf{$1.20$}}}
	&\multicolumn{1}{p{\gnumericColD}|}%
	{\gnumericPB{\raggedleft}\gnumbox[r]{\textsf{$5.21$}}}
	&\multicolumn{1}{p{\gnumericColE}|}%
	{\gnumericPB{\raggedleft}\gnumbox[r]{\textsf{$1.56$}}}
\\
\hhline{|----|}
	 \multicolumn{1}{|p{\gnumericColB}|}%
	{\gnumericPB{\raggedleft}\gnumbox[r]{\textsf{$4.87$}}}
	&\multicolumn{1}{p{\gnumericColC}|}%
	{\gnumericPB{\raggedleft}\gnumbox[r]{\textsf{$1.25$}}}
	&\multicolumn{1}{p{\gnumericColD}|}%
	{\gnumericPB{\raggedleft}\gnumbox[r]{\textsf{$5.24$}}}
	&\multicolumn{1}{p{\gnumericColE}|}%
	{\gnumericPB{\raggedleft}\gnumbox[r]{\textsf{$1.59$}}}
\\
\hhline{|----|}
	 \multicolumn{1}{|p{\gnumericColB}|}%
	{\gnumericPB{\raggedleft}\gnumbox[r]{\textsf{$4.91$}}}
	&\multicolumn{1}{p{\gnumericColC}|}%
	{\gnumericPB{\raggedleft}\gnumbox[r]{\textsf{$1.30$}}}
	&\multicolumn{1}{p{\gnumericColD}|}%
	{\gnumericPB{\raggedleft}\gnumbox[r]{\textsf{$5.28$}}}
	&\multicolumn{1}{p{\gnumericColE}|}%
	{\gnumericPB{\raggedleft}\gnumbox[r]{\textsf{$1.62$}}}
\\
\hhline{|----|}
	 \multicolumn{1}{|p{\gnumericColB}|}%
	{\gnumericPB{\raggedleft}\gnumbox[r]{\textsf{$4.95$}}}
	&\multicolumn{1}{p{\gnumericColC}|}%
	{\gnumericPB{\raggedleft}\gnumbox[r]{\textsf{$1.34$}}}
	&\multicolumn{1}{p{\gnumericColD}|}%
	{\gnumericPB{\raggedleft}\gnumbox[r]{\textsf{$5.32$}}}
	&\multicolumn{1}{p{\gnumericColE}|}%
	{\gnumericPB{\raggedleft}\gnumbox[r]{\textsf{$1.64$}}}
\\
\hhline{|----|}
	 \multicolumn{1}{|p{\gnumericColB}|}%
	{\gnumericPB{\raggedleft}\gnumbox[r]{\textsf{$4.98$}}}
	&\multicolumn{1}{p{\gnumericColC}|}%
	{\gnumericPB{\raggedleft}\gnumbox[r]{\textsf{$1.37$}}}
	&\multicolumn{1}{p{\gnumericColD}|}%
	{\gnumericPB{\raggedleft}\gnumbox[r]{\textsf{$5.35$}}}
	&\multicolumn{1}{p{\gnumericColE}|}%
	{\gnumericPB{\raggedleft}\gnumbox[r]{\textsf{$1.67$}}}
\\
\hhline{|----|}
	 \multicolumn{1}{|p{\gnumericColB}|}%
	{\gnumericPB{\raggedleft}\gnumbox[r]{\textsf{$5.02$}}}
	&\multicolumn{1}{p{\gnumericColC}|}%
	{\gnumericPB{\raggedleft}\gnumbox[r]{\textsf{$1.41$}}}
	&\multicolumn{1}{p{\gnumericColD}|}%
	{\gnumericPB{\raggedleft}\gnumbox[r]{\textsf{$5.39$}}}
	&\multicolumn{1}{p{\gnumericColE}|}%
	{\gnumericPB{\raggedleft}\gnumbox[r]{\textsf{$1.69$}}}
\\
\hhline{|----|}
	 \multicolumn{1}{|p{\gnumericColB}|}%
	{\gnumericPB{\raggedleft}\gnumbox[r]{\textsf{$5.06$}}}
	&\multicolumn{1}{p{\gnumericColC}|}%
	{\gnumericPB{\raggedleft}\gnumbox[r]{\textsf{$1.44$}}}
	&\multicolumn{1}{p{\gnumericColD}|}%
	{\gnumericPB{\raggedleft}\gnumbox[r]{\textsf{$5.43$}}}
	&\multicolumn{1}{p{\gnumericColE}|}%
	{\gnumericPB{\raggedleft}\gnumbox[r]{\textsf{$1.72$}}}
\\
\hhline{|----|}
	 \multicolumn{1}{|p{\gnumericColB}|}%
	{\gnumericPB{\raggedleft}\gnumbox[r]{\textsf{$5.09$}}}
	&\multicolumn{1}{p{\gnumericColC}|}%
	{\gnumericPB{\raggedleft}\gnumbox[r]{\textsf{$1.47$}}}
	&\multicolumn{1}{p{\gnumericColD}|}%
	{\gnumericPB{\raggedleft}\gnumbox[r]{\textsf{$5.46$}}}
	&\multicolumn{1}{p{\gnumericColE}|}%
	{\gnumericPB{\raggedleft}\gnumbox[r]{\textsf{$1.74$}}}
\\
\hhline{|----|}
	 \multicolumn{1}{|p{\gnumericColB}|}%
	{\gnumericPB{\raggedleft}\gnumbox[r]{\textsf{$5.13$}}}
	&\multicolumn{1}{p{\gnumericColC}|}%
	{\gnumericPB{\raggedleft}\gnumbox[r]{\textsf{$1.51$}}}
	&\multicolumn{1}{p{\gnumericColD}|}%
	{\gnumericPB{\raggedleft}\gnumbox[r]{\textsf{$5.50$}}}
	&\multicolumn{1}{p{\gnumericColE}|}%
	{\gnumericPB{\raggedleft}\gnumbox[r]{\textsf{$1.77$}}}
\\
\hhline{|-|-|-|-|}
\end{longtable}

\gnumericTableEnd
}
\caption{Normalisation for a gaussian form for $\Upsilon(1S)$ as a function of  the $b$ quark mass
for $\Lambda=5.20$~GeV.}\label{tab:N_exp_Upsi_mq3}
\end{table}

\begin{table}[H]
{\tiny%%%%%%%%%%%%%%%%%%%%%%%%%%%%%%%%%%%%%%%%%%%%%%%%%%%%%%%%%%%%%%%%%%%%%%
\def\ifundefined#1{\expandafter\ifx\csname#1\endcsname\relax}

%%  Check for the \def token for inputed files. If it is not        %%
%%  defined. the file will be processed as a standalone and the     %%
%%  preamble will be used.                                          %%
\ifundefined{inputGnumericTable}

%%  End of the preamble for the standalone. The next section is for %%
%%  documents which are included into other LaTeX2e files.          %%
\else

%%  We are not a stand alone document. For a regular table. we will %%
%%  have no preamble and only define the closing to mean nothing.   %%
    \def\gnumericTableEnd{}

%%  If we want landscape mode in an embedded document. comment out  %%
%%  the line above and uncomment the two below. The table will      %%
%%  begin on a new page and run in landscape mode.                  %%
%       \def\gnumericTableEnd{\end{landscape}}
%       \begin{landscape}

%%  End of the else clause for this file being \input.              %%
\fi

%%%%%%%%%%%%%%%%%%%%%%%%%%%%%%%%%%%%%%%%%%%%%%%%%%%%%%%%%%%%%%%%%%%%%%
%%                                                                  %%
%%  The rest is the gnumeric table. except for the closing          %%
%%  statement. Changes below will alter the table's appearance.     %%
%%                                                                  %%
%%%%%%%%%%%%%%%%%%%%%%%%%%%%%%%%%%%%%%%%%%%%%%%%%%%%%%%%%%%%%%%%%%%%%%

\providecommand{\gnumericPB}[1]%
{\let\gnumericTemp=\\#1\let\\=\gnumericTemp\hspace{0pt}}

%%  The default table format retains the relative column widths of  %%
%%  gnumeric. They can easily be changed to c. r or l. In that case %%
%%  you may want to comment out the next line and uncomment the one %%
%%  thereafter                                                      %%
\providecommand\gnumbox{\makebox[0pt]}
%%\providecommand\gnumbox[1][]{\makebox}

%% to adjust positions in multirow situations                       %%
\setlength{\bigstrutjot}{\jot}
\setlength{\extrarowheight}{\doublerulesep}

%%  The \setlongtables command keeps column widths the same across %%
%%  pages. Simply comment out next line for varying column widths.  %%
\setlongtables

\setlength\gnumericTableWidth{%
	30pt+%
	30pt+%
	30pt+%
	30pt+%
0pt}
\def\gumericNumCols{4}
\setlength\gnumericTableWidthComplete{\gnumericTableWidth+\tabcolsep*\gumericNumCols*2+\arrayrulewidth*\gumericNumCols}
\ifthenelse{\lengthtest{\gnumericTableWidthComplete > \textwidth}}%
{\def\gnumericScale{\ratio{\textwidth-\tabcolsep*\gumericNumCols*2-\arrayrulewidth*\gumericNumCols}%
{\gnumericTableWidth}}}%
{\def\gnumericScale{1}}

%%%%%%%%%%%%%%%%%%%%%%%%%%%%%%%%%%%%%%%%%%%%%%%%%%%%%%%%%%%%%%%%%%%%%%
%%                                                                  %%
%% The following are the widths of the various columns. We are      %%
%% defining them here because then they are easier to change.       %%
%% Depending on the cell formats we may use them more than once.    %%
%%                                                                  %%
%%%%%%%%%%%%%%%%%%%%%%%%%%%%%%%%%%%%%%%%%%%%%%%%%%%%%%%%%%%%%%%%%%%%%%

\def\gnumericColB{30pt*\gnumericScale}
\def\gnumericColC{30pt*\gnumericScale}
\def\gnumericColD{30pt*\gnumericScale}
\def\gnumericColE{30pt*\gnumericScale}

\begin{longtable}[c]{%
	b{\gnumericColB}%
	b{\gnumericColC}%
	b{\gnumericColD}%
	b{\gnumericColE}%
	}

%%%%%%%%%%%%%%%%%%%%%%%%%%%%%%%%%%%%%%%%%%%%%%%%%%%%%%%%%%%%%%%%%%%%%%
%%  The longtable options. (Caption. headers... see Goosens. p.124) %%
%	\caption{The Table Caption.}             \\	%
% \hline	% Across the top of the table.
%%  The rest of these options are table rows which are placed on    %%
%%  the first. last or every page. Use \multicolumn if you want.    %%

%%  Header for the first page.                                      %%
%	\multicolumn{4}{c}{The First Header} \\ \hline 
%	\multicolumn{1}{c}{colTag}	%Column 1
%	&\multicolumn{1}{c}{colTag}	%Column 1
%	&\multicolumn{1}{c}{colTag}	%Column 2
%	&\multicolumn{1}{c}{colTag}	%Column 3
%	&\multicolumn{1}{c}{colTag}	\\ \hline %Last column
%	\endfirsthead

%%  The running header definition.                                  %%
%	\hline
%	\multicolumn{4}{l}{\ldots\small\slshape continued} \\ \hline
%	\multicolumn{1}{c}{colTag}	%Column 1
%	&\multicolumn{1}{c}{colTag}	%Column 1
%	&\multicolumn{1}{c}{colTag}	%Column 2
%	&\multicolumn{1}{c}{colTag}	%Column 3
%	&\multicolumn{1}{c}{colTag}	\\ \hline %Last column
%	\endhead

%%  The running footer definition.                                  %%
%	\hline
%	\multicolumn{4}{r}{\small\slshape continued\ldots} \\
%	\endfoot

%%  The ending footer definition.                                   %%
%	\multicolumn{4}{c}{That's all folks} \\ \hline 
%	\endlastfoot
%%%%%%%%%%%%%%%%%%%%%%%%%%%%%%%%%%%%%%%%%%%%%%%%%%%%%%%%%%%%%%%%%%%%%%

\hhline{|-|-|-|-|}
	 \multicolumn{1}{|p{\gnumericColB}|}%
	{\gnumericPB{\raggedleft}\gnumbox[r]{\textsf{$4.80$}}}
	&\multicolumn{1}{p{\gnumericColC}|}%
	{\gnumericPB{\raggedleft}\gnumbox[r]{\textsf{$0.87$}}}
	&\multicolumn{1}{p{\gnumericColD}|}%
	{\gnumericPB{\raggedleft}\gnumbox[r]{\textsf{$5.17$}}}
	&\multicolumn{1}{p{\gnumericColE}|}%
	{\gnumericPB{\raggedleft}\gnumbox[r]{\textsf{$1.10$}}}
\\
\hhline{|----|}
	 \multicolumn{1}{|p{\gnumericColB}|}%
	{\gnumericPB{\raggedleft}\gnumbox[r]{\textsf{$4.84$}}}
	&\multicolumn{1}{p{\gnumericColC}|}%
	{\gnumericPB{\raggedleft}\gnumbox[r]{\textsf{$0.90$}}}
	&\multicolumn{1}{p{\gnumericColD}|}%
	{\gnumericPB{\raggedleft}\gnumbox[r]{\textsf{$5.21$}}}
	&\multicolumn{1}{p{\gnumericColE}|}%
	{\gnumericPB{\raggedleft}\gnumbox[r]{\textsf{$1.11$}}}
\\
\hhline{|----|}
	 \multicolumn{1}{|p{\gnumericColB}|}%
	{\gnumericPB{\raggedleft}\gnumbox[r]{\textsf{$4.87$}}}
	&\multicolumn{1}{p{\gnumericColC}|}%
	{\gnumericPB{\raggedleft}\gnumbox[r]{\textsf{$0.93$}}}
	&\multicolumn{1}{p{\gnumericColD}|}%
	{\gnumericPB{\raggedleft}\gnumbox[r]{\textsf{$5.24$}}}
	&\multicolumn{1}{p{\gnumericColE}|}%
	{\gnumericPB{\raggedleft}\gnumbox[r]{\textsf{$1.13$}}}
\\
\hhline{|----|}
	 \multicolumn{1}{|p{\gnumericColB}|}%
	{\gnumericPB{\raggedleft}\gnumbox[r]{\textsf{$4.91$}}}
	&\multicolumn{1}{p{\gnumericColC}|}%
	{\gnumericPB{\raggedleft}\gnumbox[r]{\textsf{$0.96$}}}
	&\multicolumn{1}{p{\gnumericColD}|}%
	{\gnumericPB{\raggedleft}\gnumbox[r]{\textsf{$5.28$}}}
	&\multicolumn{1}{p{\gnumericColE}|}%
	{\gnumericPB{\raggedleft}\gnumbox[r]{\textsf{$1.14$}}}
\\
\hhline{|----|}
	 \multicolumn{1}{|p{\gnumericColB}|}%
	{\gnumericPB{\raggedleft}\gnumbox[r]{\textsf{$4.95$}}}
	&\multicolumn{1}{p{\gnumericColC}|}%
	{\gnumericPB{\raggedleft}\gnumbox[r]{\textsf{$0.98$}}}
	&\multicolumn{1}{p{\gnumericColD}|}%
	{\gnumericPB{\raggedleft}\gnumbox[r]{\textsf{$5.32$}}}
	&\multicolumn{1}{p{\gnumericColE}|}%
	{\gnumericPB{\raggedleft}\gnumbox[r]{\textsf{$1.16$}}}
\\
\hhline{|----|}
	 \multicolumn{1}{|p{\gnumericColB}|}%
	{\gnumericPB{\raggedleft}\gnumbox[r]{\textsf{$4.98$}}}
	&\multicolumn{1}{p{\gnumericColC}|}%
	{\gnumericPB{\raggedleft}\gnumbox[r]{\textsf{$1.00$}}}
	&\multicolumn{1}{p{\gnumericColD}|}%
	{\gnumericPB{\raggedleft}\gnumbox[r]{\textsf{$5.35$}}}
	&\multicolumn{1}{p{\gnumericColE}|}%
	{\gnumericPB{\raggedleft}\gnumbox[r]{\textsf{$1.17$}}}
\\
\hhline{|----|}
	 \multicolumn{1}{|p{\gnumericColB}|}%
	{\gnumericPB{\raggedleft}\gnumbox[r]{\textsf{$5.02$}}}
	&\multicolumn{1}{p{\gnumericColC}|}%
	{\gnumericPB{\raggedleft}\gnumbox[r]{\textsf{$1.02$}}}
	&\multicolumn{1}{p{\gnumericColD}|}%
	{\gnumericPB{\raggedleft}\gnumbox[r]{\textsf{$5.39$}}}
	&\multicolumn{1}{p{\gnumericColE}|}%
	{\gnumericPB{\raggedleft}\gnumbox[r]{\textsf{$1.19$}}}
\\
\hhline{|----|}
	 \multicolumn{1}{|p{\gnumericColB}|}%
	{\gnumericPB{\raggedleft}\gnumbox[r]{\textsf{$5.06$}}}
	&\multicolumn{1}{p{\gnumericColC}|}%
	{\gnumericPB{\raggedleft}\gnumbox[r]{\textsf{$1.04$}}}
	&\multicolumn{1}{p{\gnumericColD}|}%
	{\gnumericPB{\raggedleft}\gnumbox[r]{\textsf{$5.43$}}}
	&\multicolumn{1}{p{\gnumericColE}|}%
	{\gnumericPB{\raggedleft}\gnumbox[r]{\textsf{$1.20$}}}
\\
\hhline{|----|}
	 \multicolumn{1}{|p{\gnumericColB}|}%
	{\gnumericPB{\raggedleft}\gnumbox[r]{\textsf{$5.09$}}}
	&\multicolumn{1}{p{\gnumericColC}|}%
	{\gnumericPB{\raggedleft}\gnumbox[r]{\textsf{$1.06$}}}
	&\multicolumn{1}{p{\gnumericColD}|}%
	{\gnumericPB{\raggedleft}\gnumbox[r]{\textsf{$5.46$}}}
	&\multicolumn{1}{p{\gnumericColE}|}%
	{\gnumericPB{\raggedleft}\gnumbox[r]{\textsf{$1.21$}}}
\\
\hhline{|----|}
	 \multicolumn{1}{|p{\gnumericColB}|}%
	{\gnumericPB{\raggedleft}\gnumbox[r]{\textsf{$5.13$}}}
	&\multicolumn{1}{p{\gnumericColC}|}%
	{\gnumericPB{\raggedleft}\gnumbox[r]{\textsf{$1.08$}}}
	&\multicolumn{1}{p{\gnumericColD}|}%
	{\gnumericPB{\raggedleft}\gnumbox[r]{\textsf{$5.50$}}}
	&\multicolumn{1}{p{\gnumericColE}|}%
	{\gnumericPB{\raggedleft}\gnumbox[r]{\textsf{$1.23$}}}
\\
\hhline{|-|-|-|-|}
\end{longtable}

\gnumericTableEnd
}
\caption{Normalisation for a gaussian form for $\Upsilon(1S)$ as a function of  the $b$ quark mass
for $\Lambda=6.25$~GeV.}\label{tab:N_exp_Upsi_mq4}
\end{table}

\section{Determination of parameters: another approach}\label{sec:norm_alt_method}
\index{leptonic decay width|(}

Usually, the complete procedure to be followed is to obtain from the gap 
equation\footnote{which is another way to
write down the BSE.} a relation between the quark mass and the bound-state mass; 
$\Lambda$ -- the paramater size-- is determined from the 
leptonic decay width, whereas the normalisation $N$ is determined by 
the fact that the residue of the bound-state propagator be 1
at the bound-state mass. The value obtained for the quark mass from the gap 
equation depends unfortunately on $\Lambda$. This imposes the use of 
an iterative method to solve these coupled equations.

In the following, we limit ourselves to  the calculation 
of the leptonic decay width, to single out 
the few differences with our calculation but also to point out where 
there is a problem with the Wick rotation. \index{Wick rotation}

We present here the calculation of the decay width for the dipole vertex 
function. The same procedure can be applied for
the gaussian; instead of using Feynman-parameter method to gather the 
denominators, one has therefore to use
Schwinger ones, or equally the properties of the Laplace transform.

\subsection{Leptonic decay width}

From the Feynman rules and the vertex functions for the bound state, we have
\eqs{
i A^{\mu\nu}&= (-1) e_Q 3 \int \frac{d^4k}{(2 \pi)^4}
{\rm Tr} \left((i \Gamma(p_{rel}^2)\gmu) i \frac{\ks k -\frac{1}{2}\ks P +m}{(k-\frac{P}{2})^2-m^2_c}
(i e \gnu) i \frac{\ks k +\frac{1}{2}\ks P +m}{(k+\frac{P}{2})^2-m^2_c}
\right)\\
 &=-3e_Q\int \frac{d^4k}{(2 \pi)^4}
 \Gamma(p_{rel}^2)
\frac{g^{\mu\nu} (M^2+4m^2-4k^2)+8 k^\mu k^\nu-2P^\mu P^\nu}
{((k-\frac{P}{2})^2-m^2_c)((k+\frac{P}{2})^2-m^2_c)}
}
$e_Q$ is the heavy quark charge, $-1$ comes for the fermionic loop, $3$ is the colour factor.

Since $p_{rel}=k$, for the dipolar form of the vertex functions, we shall have
\eqs{
\Gamma(p_{rel}^2)=\frac{N}{(1-\frac{k^2}{\Lambda^2})^2}
}

Bearing in mind that $k^2$ is $g_{\mu\nu}k^\mu k^\nu$ and defining:
\eqs{
&I\equiv\int \frac{d^4k}{(2 \pi)^4}\Gamma(p_{rel}^2)\frac{1}
{((k-\frac{P}{2})^2-m^2_c)((k+\frac{P}{2})^2-m^2_c)},
\\
&I^{\mu\nu}\equiv\int \frac{d^4k}{(2 \pi)^4}\Gamma(p_{rel}^2)\frac{k^\mu k^\nu}
{((k-\frac{P}{2})^2-m^2_c)((k+\frac{P}{2})^2-m^2_c)},
}
we have
\eqs{
A^{\mu\nu}= -2eN\Lambda^4 \left[(g^{\mu\nu}(M^2+4 m^2)-8 P^\mu P^\nu)I-4g^{\mu\nu} g_{\mu'\nu'}I^{\mu'\nu'}
+8 I^{\mu\nu}\right]
}

Using the Feynman parametrisation for rational fractions, we have the following relation
\eqs{\frac{1}{A_1^2 A_2 A_3}=\int^1_0 dx dy dz \delta(x+y+z-1) \frac{6x}{(xA_1+yA_2+zA_3)^4},}
we may rewrite all the denominators as:
\eqs{\frac{1}{(k^2-\Lambda^2)^2((k-\frac{P}{2})^2-m^2_c)((k+\frac{P}{2})^2-m^2_c)}=
\int^1_0 dx dy dz \delta(x+y+z-1) \frac{6x}{D^4},
}
with 
\eqs{D&=x (k^2+\lambda^2)-y((k-\frac{P}{2})^2-m^2_c)+z((k+\frac{P}{2})^2-m^2_c)+(y+z)i\ep\\
&=k^2+(z-y) k.P-x\Lambda^2+(y+z) (\frac{M^2}{4}-m^2)+(y+z) i\ep}.

The aim now is to get a complete square. To achieve this, we make use of the following 
change of variable 
\eqs{\ell\equiv k +\frac{z-y}{2}P &\Rightarrow k=\ell-\frac{z-y}{2}P\\
&\Rightarrow k^2=\ell^2+\frac{(z-y)^2}{4} M^2-(z-y) \ell.P\\
&\Rightarrow k.P= \ell.P-\frac{z-y}{2} M^2,
}
which  cancels linear terms in the variable. Indeed, we have
\eqs{
D&=\ell^2-\frac{(z-y)^2}{4} M^2 -x\Lambda^2+(y+z) (\frac{M^2}{4}-m^2)+(y+z) i\ep\\
&= \ell^2-\Delta+(y+z) i\ep,
}
with 
\eqs{
\Delta= \underbrace{(z-y)^2 \frac{M^2}{4}}_{>0} + \underbrace{x\Lambda^2}_{>0}+(y+z)  
\underbrace{(m^2-\frac{M^2}{4})}_{>0 \hbox{ if } m^2>\frac{M^2}{4}}.
}

Now we have, 
\eqs{
&\int \frac{d^4k}{(2 \pi)^4}\frac{\Gamma(p_{rel}^2)}
{((k-\frac{P}{2})^2-m^2_c)((k+\frac{P}{2})^2-m^2_c)}
\to \int^1_0 dx dy dz \delta(..) \int \frac{d^4\ell}{(2 \pi)^4}
\frac{6x}{(\ell^2-\Delta)^4}
,
\\
&\int \frac{d^4k}{(2 \pi)^4}\frac{k^\mu k^\nu \Gamma(p_{rel}^2)}
{((k-\frac{P}{2})^2-m^2_c)((k+\frac{P}{2})^2-m^2_c)}
\to \int^1_0 dx dy dz \delta(..) \int \frac{d^4\ell}{(2 \pi)^4}
\frac{6x\ell^\mu \ell^\nu}{(\ell^2-\Delta)^4}
}

{\it The previous lines implicitly require to make the Wick rotation one cannot trust
as already explained}.\index{Wick rotation}

The integration over $d^4 \ell$ is carried out using these relations~\cite{peskin}:

\eqs{
&\int\frac{d^4 \ell}{(2 \pi)^4}\frac{1}{(\ell^2-\Delta)^4}=\frac{i}{(4\pi)^2} \frac{2!}{3!}
\left(\frac{1}{\Delta}\right)^2=\frac{i}{3(4\pi)^2} \left(\frac{1}{\Delta}\right)^2,\\
&\int\frac{d^4 \ell}{(2 \pi)^4}\frac{\ell^\mu \ell^\nu}{(\ell^2-\Delta)^4}=\frac{-1}{(4\pi)^2} i 
\frac{g^{\mu\nu}}{2} \frac{1!}{3!}\left(\frac{1}{\Delta}\right)=
\frac{-i}{3(4\pi)^2}\frac{g^{\mu\nu}}{4} \left(\frac{1}{\Delta}\right).
}

We are thus left with two types of integrals:
\eqs{&\int^1_0 dx dy dz \delta(..) \frac{x}{\Delta}\equiv A,\\
&\int^1_0 dx dy dz \delta(..) \frac{x}{\Delta^2}\equiv B,}
in terms of which  $A^{\mu\nu}$ reads :
\eqs{
A^{\mu\nu}=-2e N \Lambda^4 \left( \frac{6}{3(4 \pi)^2}\right)
&[
(g^{\mu\nu}(M^2+4m^2)-2 P^\mu P^\nu)B\\
&-(8 \left(\frac{g^{\mu\nu}A}{4}\right)
-4g^{\mu\nu}g_{\mu'\nu'}\left(\frac{g^{\mu'\nu'}A}{4}\right)]\\
=\frac{- e N \Lambda^4}{4 \pi^2}
[g^{\mu\nu}((M^2+&4m^2)B+2A)-2 P^\mu P^\nu B].
}

The integration on $z$ is done using the $\delta$-function, which gives $z=1-x-y$ and
the remaining domain of integration on $(x,y)$ is then restricted by $0\le1-x-y\le0$, that is $0\le x\le 1$ 
and $0\le y \le 1-x$.

First, let us concentrate on $A$. Defining $\gamma^2=m^2-\frac{M^2}{4}$, we have
\eqs{
\int_0^1 \frac{4 x dx}{M^2}\int_0^{1-x} \frac{dy}{\Delta_{z=1-x-y}}=&
\int_0^1 \frac{4 x dx}{M^2}\int_0^{1-x} \frac{dy}{((1-x)-2y)^2 +(x\Lambda^2+(1-x)\gamma^2)\frac{4}{M^2})}\\
=&\int_0^1 \frac{x dx}{M^2}\int_0^{1-x} \frac{dy}{y^2-y(1-x)+\frac{(1-x)^2}{4} +\frac{x\Lambda^2+(1-x)\gamma^2}{M^2})}
}

In order to simplify the denominator, we make the following change of variable:
\eqs{
y'=y-\frac{1-x}{2},\\
y'_{0}=0-\frac{1-x}{2},\\
y'_{1}=(1-x)-\frac{1-x}{2}=\frac{1-x}{2}.
}
We thus obtain:
\eqs{
&\int_0^1 \frac{x dx}{M^2}\int_{-\frac{1-x}{2}}^{\frac{1-x}{2}}
\frac{dy'}{y'^2+\frac{x\Lambda^2+(1-x)\gamma^2)}{M^2}}=
2\int_0^1 \frac{x dx}{M^2}\int_{0}^{\frac{1-x}{2}}
\frac{dy'}{y'^2+\frac{x\Lambda^2+(1-x)\gamma^2)}{M^2}}\\
&\Rightarrow A=2\int_0^1 \frac{x dx}{M^2}\frac{M}{\sqrt{\gamma^2+x(\Lambda^2-\gamma^2)}}
\arctan \left(\frac{(1-x)M}{2\sqrt{\gamma^2+x(\Lambda^2-\gamma^2)}}\right).
}

Acting in the same way with $B$, we may also write
\eqs{
&B=
2\int_0^1 \frac{x dx}{M^2}\int_{0}^{\frac{1-x}{2}}
\frac{dy'}{\left(y'^2+\frac{x\Lambda^2+(1-x)\gamma^2)}{M^2}\right)^2},
}
which, integrating by part gives $B=B_1+B_2$ with:
\eqs{
&B_1=\int_0^1 \frac{2 x (1-x) dx}{
\left(\gamma^2+x(\Lambda^2-\gamma^2)\right)
\left(4(\gamma^2+x(\Lambda^2-\gamma^2))+M^2(1-x)^2
\right),}\\
&B_2=\frac{1}{M}\int_0^1 \frac{x dx }{(\gamma^2+x(\Lambda^2-\gamma^2))^{\frac{3}{2}}}
\arctan \left(\frac{(1-x)M}{2\sqrt{\gamma^2+x(\Lambda^2-\gamma^2)}}\right).
}

The latter three integrals are not computable analytically. We shall therefore use 
numerical method to obtain their value as function of $\Lambda$ and $M^2$. In our approach, 
we recall that we have choosen to take $m$ as the mass of the lowest-lying meson composed
of a heavy quark $Q$.

On the other hand, we simply have 
\eqs{
C^{\nu'\mu'}=(A^{\mu\nu})^\dagger
=\frac{ -e N \Lambda^4}{4 \pi^2}
[g^{\mu\nu}((M^2+&4m^2)B+2A)-2 P^\mu P^\nu B].
}

From the Feynman rules, we have
\eqs{
B^{\rho\rho'}=& (ie)^2 \int d_2(PS) {\rm Tr} (\grho(-\ks a+m_\ell)\grhop (\ks b+m_\ell))\\
&(ie)^2 \int d_2(PS) 4(a.b g^{\rho\rho'}-a^\rho b^{\rho'}-a^{\rho'} b^{\rho}
+g^{\rho\rho'} m_\ell^2)
}
where $a$ and $b$ are the momenta of the lepton and antilepton ($a+b=P$), and $m_\ell$ their mass. 
As $m_\ell$ is much smaller than $M$ and $m$, we shall neglect it.

To perform the integration on the two-particle phase space, we shall use the following 
well known relations~\cite{barger}\footnote{$\lambda=
\lambda(1,a^2/M^2,b^2/M^2)=\lambda(1,m_\ell^2/M^2,m_\ell^2/M^2)=1-2 m_\ell^2/M^2\simeq 1$.}
\eqs{\int d_2(PS) a^\alpha b^\beta=&\frac{\pi M^2}{24}\sqrt{\lambda}
\left(g^{\alpha \beta}\lambda +2 \frac{P^\alpha P^\beta}{M^2}\left[1+\frac{a^2}{M^2}+
\frac{b^2}{M^2}-2 \frac{(a^2-b^2)^2}{M^4}\right]\right)\\
\simeq&\frac{\pi M^2}{24}\left(g^{\alpha\beta}+2 \frac{P^\alpha P^\beta}{M^2}\right),\\
\int d_2(PS) a.b=&\int d_2(PS)a^\alpha b^\beta g_{\alpha\beta}  =\frac{\pi M^2}{24}\left(4+2\right)=\frac{\pi M^2}{4}.}

Therefore, we have
\eqs{
B^{\rho\rho'}=& (ie)^2 \left[\pi M^2 g^{\rho \rho'}-8 \frac{\pi M^2}{24}\left(g^{\rho\rho'}+2 \frac{P^\rho P^{\rho'}}{M^2}\right) \right]=(ie)^2 \frac{2\pi}{3} M^2\underbrace{\left[g^{\rho\rho'}- \frac{P^\rho P^{\rho'}}{M^2}\right]}_{\Delta^{\rho\rho'}}.
}

Now that all quantities in \ce{eq:decomp_ampl_inv_decay} are determined, we can put everything
together, which gives\footnote{Recall the projector property of $\Delta_{\mu\nu}$: 
$\Delta_{\mu\nu}P^\mu=\Delta_{\mu\nu}P^\nu=0$.}:
\eqs{\int \left|\bar{\cal M}\right|^2 d_2(PS)=&\frac{1}{3}
\Delta_{\mu\mu'}\frac{-1}{M^4}\left(\frac{ e N \Lambda^4}{4 \pi^2}\right)^2\left((ie)^2 \frac{2\pi}{3} M^2\right)
\times\\
&\times [g^{\mu\nu}((M^2+4m^2)B+2A)]g_{\nu\rho}
\Delta^{\rho\rho'}g_{\rho'\nu'}
[g^{\nu'\mu'}((M^2+4m^2)B+2A)]\\
=&\frac{e^4 N^2 \Lambda^8}{18 \pi M^2}
(M^2+4m^2)B+2A)^2\Delta_{\mu\mu'}\Delta^{\mu\mu'}
=\frac{e^4 N^2 \Lambda^8}{6 \pi M^2}
((M^2+4m^2)B+2A)^2
}

\ce{eq:lept_width} then gives the width and provide with a relation between $\Lambda$ and the masses.

Combining this relation with the normalisation condition, it is therefore possible
to fully determine the vertex-function parameters. The result obtained by Costa~\etal~\cite{costa}
are given in~\ct{tab:res_norm_meth2}.

\index{normalisation|)}\index{leptonic decay width|)}

\addcontentsline{toc}{chapter}%
{\protect\numberline{}{List of Figures}}
\sffamily
\listoffigures
\clearemptydoublepage
\rmfamily

\addcontentsline{toc}{chapter}%
{\protect\numberline{}{List of Tables}}
\sffamily
\listoftables
\clearemptydoublepage
\rmfamily

\clearemptydoublepage

\newpage
\addcontentsline{toc}{chapter}%
{\protect\numberline{}{Index}}
\thispagestyle{empty}
\printindex
\clearemptydoublepage

\end{document}